\documentclass[12pt,a4paper]{report}
\let\openright=\clearpage

\usepackage[a-2u]{pdfx}

\usepackage[utf8]{inputenc}

\usepackage{lmodern}

\usepackage{amsmath}        
\usepackage{amsfonts}       
\usepackage{amsthm}         
\usepackage{bbding}         
\usepackage{bm}             
\usepackage{graphicx}       
\usepackage{fancyvrb}       
\usepackage{natbib}         
\usepackage[nottoc]{tocbibind} 
\usepackage{dcolumn}        
\usepackage{booktabs}       
\usepackage{paralist}       
\usepackage{xcolor}         
\usepackage{pdfpages}

\usepackage{epigraph}
\usepackage{etoolbox}
\usepackage{amssymb}  
\usepackage{aastex_hack}
\usepackage{pdflscape}
\usepackage{subcaption}
\usepackage{multirow}
\usepackage[font=footnotesize]{caption}
\usepackage[T1]{fontenc}
\usepackage[single]{accents}
\usepackage[ocgcolorlinks]{ocgx2}

\makeatletter
\newlength\epitextskip
\pretocmd{\@epitext}{\em}{}{}
\apptocmd{\@epitext}{\em}{}{}
\patchcmd{\epigraph}{\@epitext{#1}\\}{\@epitext{#1}\\[\epitextskip]}{}{}
\makeatother

\expandafter\ifcsname urlstyle\endcsname
  \else
  \fi

\def\NazevPrace{Polarizace rentgenového záření akreujících supermasivních černých děr}

\def\AutorPrace{Jakub Podgorný}

\def\Katedra{Astronomický ústav Univerzity Karlovy, Astronomický ústav Akademie věd České republiky, Strasbourg astronomical observatory}

\def\TypPracoviste{Ústav}

\def\Vedouci{RNDr. Michal Dovčiak, Ph.D.}
\def\VedouciFR{\mbox{Dr. Frédéric Marin}}

\def\KatedraVedouciho{Astronomický \mbox{ústav} Akademie věd České republiky}
\def\KatedraVedoucihoFR{Strasbourg astronomical \newline \mbox{observatory}}

\def\ThesisTitle{Polarisation properties of X-ray emission from accreting supermassive black holes}

\def\ThesisAuthor{Jakub Podgorný}

\def\YearSubmitted{2023}

\def\Department{Astronomical \mbox{Institute} of Charles University, Astronomical \newline \mbox{Institute}~of~the~Czech Academy of Sciences, Strasbourg astronomical observatory}

\def\DeptType{Department}

\def\Supervisor{RNDr. Michal Dovčiak, Ph.D.}
\def\SupervisorFrench{\mbox{Dr. Frédéric Marin}}

\def\SupervisorsDepartment{Astronomical Institute of the Czech Academy of Sciences}
\def\SupervisorsDepartmentFrench{Strasbourg astronomical observatory}

\def\StudyProgramme{Theoretical Physics, Astronomy \mbox{and Astrophysics}}
\def\StudyBranch{P4F1}

\def\Dedication{%
\subsection*{A personal dedication}
\vspace*{10px}
\thispagestyle{empty}
{\footnotesize I have always tried to think of my work within the broadest possible terms. My use of references -- as mentioned in the ``Preface'' -- is not intended, therefore, either to be pretentious or to~convey the impression that this dissertation carries greater weight than it actually does. Within any given lifetime -- a lifetime being a characteristic human measure --, a period of~three years’ work is one of considerable duration, though it amounts to almost nothing when considered from a~collective viewpoint. I am extremely grateful that, whilst possessing only the limited knowledge of an early-career researcher, I have been given the opportunity to add my own particular pine needle to what is a growing and collectively assembled anthill. Besides attempting to make certain parts of my chosen endeavor useful to the local scientific community -- as is only inevitable --, I have also sought to ensure that the work as a whole should serve my own personal purposes.

After familiarizing myself with the endless -- though non-routine -- cycle of initial insights and the ensuing development thereof through which results emerge and age, a choice of specialization had necessarily to be made. Although inquiry into the disciplines of physical cosmology, philosophy, (art) history, consciousness studies, or even mathematics, to cite several examples, might seem to offer a broader field of comprehension when it comes to providing a basis on~which to generalize, it is my firm belief that, in fact, there does not exist any true hierarchy as far as areas of research are concerned, even in the case of astrophysics. Once a detailed and solvable problem has been identified, a problem which possesses high potential in terms of its capacity to be instructive, all that remains is for one to work, during the constructive phase, with every ounce of one’s passion and honesty, as well as in accordance with one’s particular capacities, the recognition of all of which has a calming effect upon oneself. Optimization. From disorder to order.

While acknowledging all of its unavoidable associated dubieties, I am nevertheless convinced that science is indeed soothing both to its practitioner and to society as a whole, inasmuch as it provides long-term, objective sureties, pace postmodernist trends of thought. I chose the field of numerical simulations of black-hole environments for the reason that, to the extent that basic research itself allows, I found the implications useful -- in the event that such research should come to fruition -- and the object of study satisfying -- insofar as it made its way towards that fruition. Nevertheless, were I to have chosen any other phenomenon warranting the devotion of three years to its study, it might also, I believe, have provided me with just as much scope -- rather than insight per se --, not to mention preparation for upcoming challenges. I consider the greatest benefits of doctoral studies to be the following: they offer a person the opportunity to analyze one very specific problem in the deepest manner available; they oblige that person to exercise an overall mastery of efficient working practices; and they enable that same person to move towards a more productive period of their life once the learning phase has reached its~term, thereby enabling them creatively to explore the unknown.

~\

A large part of my doctoral dissertation was conducted in Strasbourg -- an imposing medieval town of many wonders -- and, most importantly, within the flourishing environment of~the~observatory, a workplace for which I shall probably struggle to find an equal at any time in~the~future. A comparable part was undertaken in Prague, the place of my birth and a~city to which I owe both the days of my youth and the majority of my schooling and university education. While spending much of the latter period in the Karlov district of Prague, I came to develop a sense of the genius loci associated with all the dark corners and pubs into which celebrated Czech and German writers -- e.g. Jaroslav Hašek (1883--1923) and Egon Erwin Kisch (1885--1948) -- must once have ventured. This genius loci also made its presence felt within the~university libraries themselves, as well as its lecture halls and corridors, through all of~which great thinkers, including none other than Albert Einstein (1879--1955), must have passed. These same streets, stairways and gardens had once been filled with revolutionary students, the effects of whom on 20\textsuperscript{th}-century (Czech) history is well known. The neighborhood of~Karlov likewise contained -- and continues to do so -- the aged hospitals where many have taken both their final and their first steps, -- as in the case of my most beloved grandmother, Margit, and my newborn niece, Olga -- and where many continue to devote their energies to~healing others -- as does my~inspirational sister, Gabriela.

I have always truly appreciated the degree of patience extended and of instruction freely given to me during such times by acknowledged authorities; and, accordingly, I feel obliged to~make use of it by virtue of producing fruitful research and to share my own acquired learning with those less fortunate or younger than myself. Alongside many others, of course, a~few teachers in particular laid down enduring paths of inspiration within my mind (for the sake of~brevity here mentioned without their academic titles): Jakub Haláček, who convinced me~to~opt for physics on completion of my high school education; Milan Pokorný, who showed me the beauty and power of mathematics during my undergraduate studies; and Oldřich Semerák, who not only reiterated to me at graduate level that same beauty and power, as found in General Relativity, but who also guided me towards the realization that art is of equal importance to~me as is science, and that many interesting possibilities arise at precisely those points where these two intersect. For the majority of the quotations presented here below, I am indebted to Oldřich and to Jiří Bičák, the latter being yet another distinguished figure who has exerted a~considerable degree of influence upon me. I should like to offer my foremost acknowledgement, however, to~my two supervisors, Michal Dovčiak and Frédéric Marin, who have shown incredible endurance, provided me with staggering levels of support throughout my PhD, and also offered crucial feedback on all of the results presented below. My personal thanks go to~you. Having been able to join the \textit{IXPE} team of NASA and, in that setting, to have seen in detail the work of the greatest exponents within our field, can only be classed as a once-in-a-lifetime experience. I should also like to register my appreciation of the contribution made by Marie Novotná, who lent me great assistance as regards the graphic components of this dissertation as well as producing a number of illustrations. Mention should also be made of Robert D.~Hughes (Prague), who kindly revised the English within the current section (viz. ``A~personal dedication''), and~Karolina Hughes, who translated the below quotation.

None of the above would have been possible, however, without the ongoing support of~my family and friends. I am well aware of the privilege I have enjoyed in terms of the fact that my~particular background has enabled me to conduct specialized studies for almost a full decade in relation to one subject alone, a period of time just long enough to enable me, with an adequate degree of professionalism, to make small, independent steps towards extending the boundaries of~human knowledge. In particular, I should like to thank my mother, Yvetta, who has had to make many sacrifices as I travelled down my chosen path and who herself has successfully confronted so many obstacles. My thanks next go to my girlfriend, Marie, whose support through the final and most important stages of the entire process, enabled me to finish my formal education. No less do I express my gratitude towards the friends and colleagues I have been able to make at both of the abovementioned institutions, as well as towards the Magnesium group in Prague and my flatmates in Strasbourg -- they all contributed towards the~fact that the three years I spent studying amounted to an experience of the greatest intensity, one of~significance to~me as much in terms of personal as of professional growth. Yet, despite the repeated upheavals our world has seen during the period in question, it was nevertheless for me -- it must be said thanks to them -- a fun time. Lastly, I should like to offer my thanks to those who have already died, yet remain with us in spirit. Most importantly so to my grandparents, Miroslav, Ludmila and Jiří, who all left strong roots from which to branch forth; to my mother's second husband, Václav, whose humor and wisdom still resonate in Podolí (a district of Prague); and to my own beloved father, Oleg, who can only have had the heart of a whale. I should like to devote this work to my Dad. And I hope that, despite his having left us so soon, his lingering presence remains somewhere near Vyšehrad Castle -- a place that means a lot to our family -- and that he is smiling with pride at the progress I have been making.

~\

}
\thispagestyle{empty}

\newpage
\thispagestyle{empty}

\vspace*{250px}
\epigraph{Vždy budou duchové..., kteří budou usilovat, aby spojenou mocí poznání a~snů, vědy a poezie, vytvořili jednotný obraz vesmírného dění, jenž by stejně odpovídal věčnému prahnutí lidského ducha po harmonii a kráse, i~žízni srdce po spravedlnosti.\\
\vspace*{40px}
\emph{[There shall always be spirits... who will strive, through the combined forces of knowledge and dreams, as of science and poetry, to create a~unified image of 
cosmic events, (one) which might reflect not only the~human spirit’s eternal longing for harmony and beauty, but also the~heart’s thirst for justice.]}
\vspace*{40px}
}{Otokar Březina [excerpt from a letter to Brno-based classical philologist František Novotný]}

}

\def\Abstrakt{%
Tato disertační práce se zabývá polarizací rentegenového záření charakteristickou pro astrofyzikální prostředí v blízkosti akreujících černých děr. Byť název a původní zadání vymezuje problematiku supermasivních černých děr v~aktivních galaktických jádrech, výsledky práce lze do velké míry aplikovat také na černé díry o hmotnosti hvězd uvnitř rentgenových binárních systémů. Představeno je vícero numerických modelů předpovídajících polarizaci emise z těchto zdrojů v rentgenovém oboru, včetně jejich přímé aplikace v rámci interpretace nejnovějších pozorování získaných díky misi \textit{Imaging X-ray Polarimetry Explorer} (\textit{IXPE}, v provozu od prosince 2021). Modelování pokrývá oblasti od přenosu záření v atmosférách akrečních disků, k obecně-relativistickým vlivům na rentgenové světlo ve vakuu v blízkosti centrálních černých děr, až k interakci záření se vzdálenými komponentami obklopujícími akreční jádro. Problematika je zkoumána na široké škále fyzikální a početní komplexity. Jednotícím prvkem práce je zaměření na odraz rentgenových paprsků od částečně ionizované hmoty.
}

\def\Abstract{%
This dissertation elaborates on X-ray polarisation features of astrophysical environments near accreting black holes. Although the work was originally assigned to supermassive black holes in active galactic nuclei, the results are also largely applicable to stellar-mass black holes in X-ray binary systems. Several numerical models predicting the X-ray polarisation from these sources are presented, including their immediate applications in the interpretation of the latest discoveries achieved thanks to the \textit{Imaging X-ray Polarimetry Explorer} (\textit{IXPE}) mission that began operating in December 2021. The modeling ranges from radiative transfer effects in atmospheres of accretion discs to general-relativistic signatures of X-rays travelling in vacuum near the central black holes to reprocessing events in distant, circumnuclear components. Various scales in physical and computational complexity are examined. A unifying element of this dissertation is the~focus on reflection of X-rays from partially ionized matter.
}

\def\KlicovaSlova{%
{rentgenová astrofyzika}, {aktivní galaktická jádra}, {relativistická \mbox{astrofyzika}}, {černé díry}, {polarizace}
}
\def\Keywords{%
{X-ray astrophysics}, {active galactic nuclei}, {relativistic astrophysics}, {black holes}, {polarisation}
}

\hypersetup{unicode}
\hypersetup{breaklinks=true}
\hypersetup{citecolor=blue}

\makeatletter
\def\@makechapterhead#1{
  {\parindent \z@ \raggedright \normalfont
   \Huge\bfseries \thechapter. #1
   \par\nobreak
   \vskip 20\p@
}}
\def\@makeschapterhead#1{
  {\parindent \z@ \raggedright \normalfont
   \Huge\bfseries #1
   \par\nobreak
   \vskip 20\p@
}}
\makeatother

\def\chapwithtoc#1{
\chapter*{#1}
\addcontentsline{toc}{chapter}{#1}
}

\theoremstyle{plain}

\theoremstyle{plain}

\theoremstyle{remark}

\DefineVerbatimEnvironment{code}{Verbatim}{fontsize=\small, frame=single}

\begin{document}
\includepdf{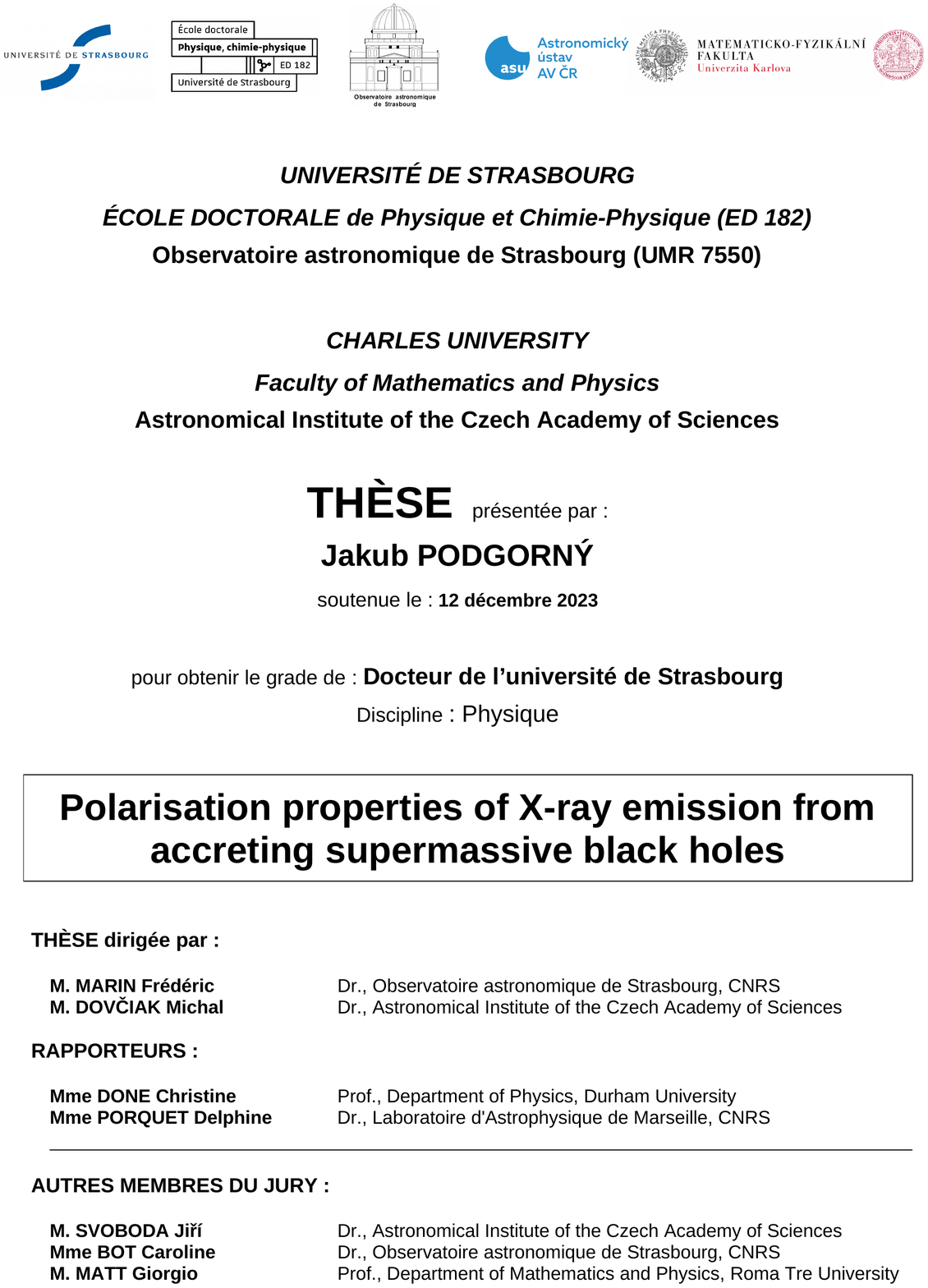}



\pagestyle{empty}
\hypersetup{pageanchor=false}
\begin{center}


\vspace{-8mm}
\vfill

{\bf\Large DOCTORAL THESIS at Charles University}

\vfill

{\LARGE\ThesisAuthor}

\vspace{15mm}

{\LARGE\bfseries\ThesisTitle}

\vfill

\Department

\vfill

{
\centerline{\vbox{\halign{\hbox to 0.5\hsize{\hfil #}&\hskip 0.5em\parbox[t]{0.5\hsize}{\raggedright #}\cr
Supervisors of the doctoral thesis:&\Supervisor \newline \SupervisorFrench \cr
\noalign{\vspace{2mm}}
Study programme at Charles University:&\StudyProgramme \cr
\noalign{\vspace{2mm}}
Study branch at Charles University:&\StudyBranch \cr
}}}}

\vfill

Prague and Strasbourg \YearSubmitted

\end{center}

\includepdf{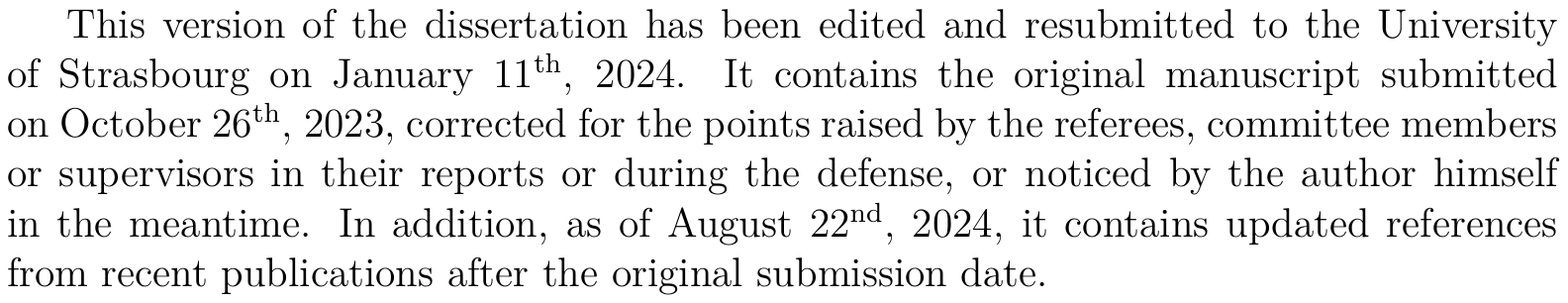}

\newpage


\openright
\hypersetup{pageanchor=true}
\pagestyle{plain}
\pagenumbering{roman}
\vglue 0pt plus 1fill
\thispagestyle{empty}
\noindent
I declare that I carried out this doctoral thesis independently, and only with the~cited
sources, literature and other professional sources. It has not been used to~obtain another
or the same degree. The content provided here, including all text and images, has been created without the use of any artificial intelligence tools. All material is the result of human creativity and effort.

\medskip\noindent
I understand that my work relates to the rights and obligations under the Act No.~121/2000 Sb.,
the Copyright Act, as amended, in particular the fact that the~Charles
University has the right to conclude a license agreement on the use of this
work as a school work pursuant to Section 60 subsection 1 of the Copyright~Act.

\vspace{10mm}

\hbox{\hbox to 0.5\hsize{%
In \hbox to 6em{\dotfill} date \hbox to 6em{\dotfill}
\hss}\hbox to 0.5\hsize{\dotfill\quad}}
\smallskip
\hbox{\hbox to 0.5\hsize{}\hbox to 0.5\hsize{\hfil Author's signature\hfil}}

\vspace{20mm}
\newpage


\openright

\noindent
\Dedication

\newpage


\openright

\vbox to 0.5\vsize{
\setlength\parindent{0mm}
\setlength\parskip{5mm}

\thispagestyle{empty}
Název práce:
\NazevPrace

Autor:
\AutorPrace

\TypPracoviste:
\Katedra

Vedoucí disertační práce:
\Vedouci, \KatedraVedouciho; \VedouciFR, \KatedraVedoucihoFR 

Abstrakt:
\Abstrakt

Klíčová slova:
\KlicovaSlova

\vss}\newpage\nobreak\vbox to 0.49\vsize{
\setlength\parindent{0mm}
\setlength\parskip{5mm}

\thispagestyle{empty}
Title:
\ThesisTitle

Author:
\ThesisAuthor

\DeptType:
\Department

Supervisors:
\Supervisor, \SupervisorsDepartment; \SupervisorFrench, \SupervisorsDepartmentFrench

Abstract:
\Abstract


\vss}

\newpage

\openright
\pagestyle{plain}
\pagenumbering{arabic}
\setcounter{page}{1}

\begin{sloppypar}
\tableofcontents

\chapwithtoc{List of publications}

Here below is a list of publications that this dissertation comprises of and that were coauthored by my supervisors, as well as many other brilliant colleagues that greatly affected my thoughts and whom I hold in high esteem. For this reason, it is my obligation to use the pronoun ``we'' for all results presented further, even though all results were carried with a necessary degree of independence and I am the author of this manuscript, which is written on the basis of these publications and puts them into larger context. If a particular credit deserves to be in addition given throughout the text, e.g. to my supervisors or to colleagues that contributed to parts of the publications, it will be acknowledged. All publications below were published in scientific journals and underwent a review process apart from one, which is still undergoing a journal revision on the day of submission. Preparatory works for this dissertation began across my master studies and were published in my master thesis \cite{Podgorny2020}, which I build upon, but do not include. This dissertation was only possible to carry out in time and sufficient detail due to full-time work schedule allowed by these resources acknowledged in~the publications: the Barrande Fellowship Programme of the Czech and French governments, the GAUK project No. 174121 from CU, the Czech-Polish mobility program (M\v{S}MT 8J20PL037 and PPN/BCZ/2019/1/00069), the conference support by GALHECOS team at~ObAS and ED 182, the Czech Science Foundation project GACR 21-06825X, the institutional support from ASU CAS RVO:67985815, and the computational facilities provided by both ASU CAS and ObAS.

\section*{As a main author}

\begin{list}{} {\leftmargin=2em \itemindent=-2em}
    \item[] J. Podgorn{\'y}, M. {Dov{\v{c}}iak}, F. Marin, R. W. Goosmann, and A. R{\'o}{\.z}a{\'n}ska. Spectral and polarization properties of reflected X-ray emission from black hole accretion discs. \textit{Monthly Notices of the Royal Astronomical Society}, 510(4):4723--4735, March 2022.
    \item[] J. Podgorn{\'y}, M. {Dov{\v{c}}iak}, R. W. Goosmann, F. Marin, G. Matt, A. R{\'o}{\.z}a{\'n}ska, and V. Karas. Spectral and polarization properties of reflected X-ray emission from black-hole accretion discs for a distant observer: the lamp-post model.
\textit{Monthly Notices of the Royal Astronomical Society}, 524(3):3853--3876, September 2023.
    \item[] J. Podgorn{\'y}, F. Marin, and M. {Dov{\v{c}}iak}. X-ray polarization properties of partially ionized equatorial obscurers around accreting compact objects. \textit{Monthly Notices of the Royal Astronomical Society}, 526(4):4929--4951, December 2023.
    \item[] J. Podgorn{\'y}, L. Marra, F. Muleri, N. Rodriguez Cavero, A. Ratheesh, M. {Dov{\v{c}}iak}, R. Miku{\v{s}}incov{\'a}, M. Brigitte, J. F. Steiner, A. Veledina, S. Bianchi, H. Krawczynski, J. Svoboda, P. Kaaret, G. Matt, J. A. Garc{\'i}a, P.-O. Petrucci, A. A. Lutovinov, A. N. Semena, A. Di Marco, M. Negro, M. C. Weisskopf, A. Ingram, J. Poutanen, B. Beheshtipour, S. Chun, K. Hu, T. Mizuno, Z. Sixuan, F. Tombesi, S. Zane, I. Agudo, L. A. Antonelli, M. Bachetti, L. Baldini, W. H. Baumgartner, R. Bellazzini, S. D. Bongiorno, R. Bonino, A. Brez, N. Bucciantini, F. Capitanio, S. Castellano, E. Cavazzuti, C.-T. Chen, S. Ciprini, E. Costa, A. De Rosa, E. Del Monte, L. Di Gesu, N. Di Lalla, I. Donnarumma, V. Doroshenko, S. R. Ehlert, T. Enoto, Y. Evangelista, S. Fabiani, R. Ferrazzoli, S. Gunji, K. Hayashida, J. Heyl, W. Iwakiri, S. G. Jorstad, V. Karas, F. Kislat, T. Kitaguchi, J. J. Kolodziejczak, F. La Monaca, L. Latronico, I. Liodakis, S. Maldera, A. Manfreda, F. Marin, A. Marinucci, A. P. Marscher, H. L. Marshall, F. Massaro, I. Mitsuishi, C.-Y. Ng, S. L. O'Dell, N. Omodei, C. Oppedisano, A. Papitto, G. G. Pavlov, A. L. Peirson, M. Perri, M. Pesce-Rollins, M. Pilia, A. Possenti, S. Puccetti, B. D. Ramsey, J. Rankin, O. J. Roberts, R. W. Romani, C. Sgro, P. Slane, P. Soffitta, G. Spandre, Doug A. Swartz, T. Tamagawa, F. Tavecchio, R. Taverna, Y. Tawara, A. F. Tennant, N. E. Thomas, A. Trois, S. S. Tsygankov, R. Turolla, J. Vink, K. Wu, and F. Xie. The first X-ray polarimetric observation of the black hole binary LMC X-1. \textit{Monthly Notices of the Royal Astronomical Society}, 526(4):5964--5975, December 2023.
    \item[] J. Podgorn{\'y}, F. Marin, and M. {Dov{\v{c}}iak}. X-ray polarization from parsec-scale components of active galactic nuclei: observational prospects. \textit{Monthly Notices of the Royal Astronomical Society}, 527(1):1114--1134, January 2024.
    \item[] J. Podgorn{\'y}, M. {Dov{\v{c}}iak} and F. Marin. Simple numerical X-ray polarization models of reflecting axially symmetric structures around accreting compact objects. \textit{Monthly Notices of the Royal Astronomical Society}, 530(3):2608--2626, May 2024.
\end{list}

\section*{As a substantially contributing coauthor}

\begin{list}{} {\leftmargin=2em \itemindent=-2em}
    \item[] H. Krawczynski, F. Muleri, M. Dov{\v{c}}iak, A. Veledina, N. Rodriguez Cavero, J. Svoboda, A. Ingram, G. Matt, J. A. Garc{\'i}a, V. Loktev, M. Negro, J. Poutanen, T. Kitaguchi,
J. Podgorn{\'y}, J. Rankin, W. Zhang, A. Berdyugin, S. V. Berdyugina, S. Bianchi, D. Blinov, F. Capitanio, N. Di Lalla, P. Draghis, S. Fabiani, M. Kagitani, V. Kravtsov, S. Kiehlmann, L. Latronico, A. A. Lutovinov, N. Mandarakas, F. Marin, A. Marinucci, J. M. Miller, T. Mizuno, S. V. Molkov, N. Omodei, P.-O. Petrucci, A. Ratheesh, T. Sakanoi, A. N. Semena, R. Skalidis, P. Soffitta, A. F. Tennant, P. Thalhammer, F. Tombesi, M. C. Weisskopf, J. Wilms, S. Zhang, I. Agudo, L. A. Antonelli, M. Bachetti, L. Baldini, W. H. Baumgartner, R. Bellazzini, S. D. Bongiorno, R. Bonino, A. Brez, N. Bucciantini, S. Castellano, E. Cavazzuti, S. Ciprini, E. Costa, A. De Rosa, E. Del Monte, L. Di Gesu, A. Di Marco, I. Donnarumma, V. Doroshenko, S. R. Ehlert, T. Enoto, Y. Evangelista, R. Ferrazzoli, S. Gunji, K. Hayashida, J. Heyl, W. Iwakiri, S. G. Jorstad, V. Karas, J. J. Kolodziejczak, F. La Monaca, I. Liodakis, S. Maldera, A. Manfreda, A. P. Marscher, H. L. Marshall, I. Mitsuishi, C.-Y. Ng, S. L. O'Dell, C. Oppedisano, A. Papitto, G. G. Pavlov, A. L. Peirson, M. Perri, M. Pesce-Rollins, M. Pilia, A. Possenti, S. Puccetti, B. D. Ramsey, R. W. Romani, C. Sgrò, P. Slane, G. Spandre, T. Tamagawa, F. Tavecchio, R. Taverna, Y. Tawara, N. E. Thomas, A. Trois, S. Tsygankov, R. Turolla, J. Vink, K. Wu, F. Xie, and S. Zane. Polarized x-rays constrain the disk-jet geometry in the black hole x-ray binary Cygnus X-1. \textit{Science}, 378(6620):650--654, November 2022.
    \item[] A. Ratheesh, M. Dov{\v{c}}iak, H. Krawczynski, J. Podgorn{\'y}, L. Marra, A. Veledina, V. Suleimanov, N. Rodriguez Cavero, J. F. Steiner, J. Svoboda
% ...
964(1):77, March 2024.
    \item[] A. Veledina, F. Muleri, J. Poutanen, J. Podgorn{\'y}, M. Dov{\v{c}}iak, F. Capitanio, E. Churazov, A. De Rosa, A. Di Marco, S. Forsblom
% ...
8:1031--1046, August 2024.
\end{list}

\newpage
\
\vspace{4.5cm} 
\begin{figure}[h]
        \centering
	\includegraphics[trim={0cm 0cm 0cm 0cm},clip,width=0.6\textwidth]{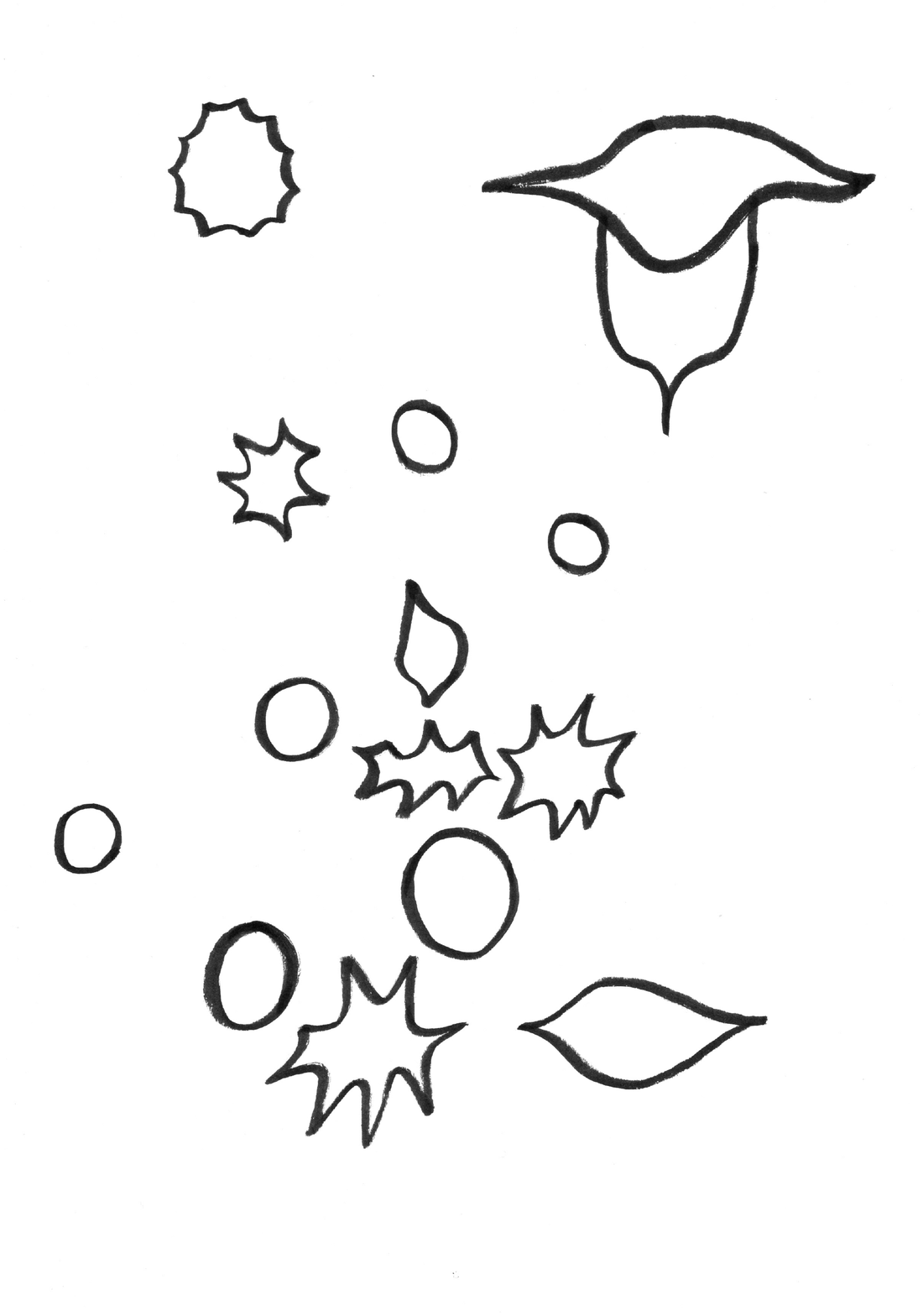}
\end{figure}
\chapter*{Preface}
\addcontentsline{toc}{chapter}{Preface}

This dissertation aims to broaden our knowledge of astrophysical black holes and~their surroundings. The idea that an object is so heavy and small that even light cannot escape its surface, today called the event horizon, is known since 1783 thanks to John Michell (1724--1793) \citep{Michell1784, Schaffer1979}. Without detailed scientific knowledge, a novelist named Gustav Meyrink (1868--1932) described in his 1903's  novel \textit{The Black Sphere} \citep{Meyrink1903} an imagination of ``mathematical nothing'', providing a somewhat brilliant intuition for the existence and properties of a directly unobservable attracting ball that was soon-after theoretically predicted by rigorous computations in Albert Einstein's theory of~gravity, the theory of General Relativity (GR), published in 1915 \citep{Einstein1915, Einstein1916}. 

The governing equations can be written in a compact tensorial notation
\begin{equation*}
	R_{\mu \nu} - \frac{1}{2}Rg_{\mu \nu} + \Lambda g_{\mu \nu} = \frac{8\pi G}{c^4} T_{\mu \nu} \, ,
\end{equation*}
where $R_{\mu \nu}$ and $R$ are the Ricci tensor and Ricci scalar, respectively, $g_{\mu \nu}$ is the~spacetime metric, $\Lambda$ is the cosmological constant, $G$ is the gravitational constant, $c$ is the speed of light, and $T_{\mu \nu}$ is the energy-momentum tensor. Max Born (1882--1970), another brilliant physicist coeval to Einstein, described a few decades later this ahead-of-time result: ``The foundation of general relativity appeared to me then, and it still does, the greatest feat of human thinking about Nature, the most amazing combination of philosophical penetration, physical intuition and mathematical skill. But its connections with experience were slender. It appealed to me like a great work of art, to be enjoyed and admired from a~distance,'' \citep{Born1956}. Nonetheless, Karl Schwarzschild (1873--1916) discovered already in 1916 an exact solution to the Einstein's equations that describes a~static black hole \citep{Schwarzschild1916}. Only in 1963, Roy Kerr (\text{*}1934) found another important analytical solution for an axially symmetric rotating black hole within the same theory \citep{Kerr1963b}, which was generalized by Ezra Newman (1929--2021) in 1965 to include the electric charge \citep{Newman1965a, Newman1965b}, building up on the decades older works of Hans Reissner (1874--1967) \citep{Reissner1916}, Hermann Weyl (1885--1955) \citep{Weyl1917}, Gunnar Nordström (1881--1923) \citep{Nordstrom1918} and George Barker Jeffery (1891--1957) \citep{Jeffery1921}. 

The research field rocketed from obscure sci-fi to also observationally well-established theory in the 1980s and 1990s. Although the black hole itself is impossible to be seen, and thus to be ever truly observationally confirmed via direct light signal by definition, the information observed from its surroundings by state-of-the-art telescopes gives us strong evidences  for the existence of such objects in both local and faraway universe and no astronomical observation disproved the long-standing tight theoretical predictions \citep[see e.g.][for further historical and contemporary references]{Misner1973, Seward2010}.

~\

Combining information from cosmology, galaxy and stellar studies, we are able to clearly distinguish two different populations of black holes existing in our universe. Firstly, supermassive black holes that are more than a hundred thousand, sometimes billion times heavier than our Sun. These are located in~the~center of almost every massive galaxy and, typically, only one of those giant attractors is present per galaxy. For example, the ``shadow'' near supermassive black hole in~our Galaxy and in the nearby M87 galaxy were recently captured in astonishing spatial resolution by the \textit{Event Horizon Telescope} (\textit{EHT}) project \citep{EHT2019} in the radio band. 

Of completely different origin and astrophysical context are stellar-mass black holes, which are frequent even inside the Milky Way and which are only a few times heavier than our Sun, at most a few hundred times heavier. This dissertation was originally assigned to study accreting -- meaning active, hoovering matter from its neighborhood and forming an accretion disc of nearly circular infall of this matter -- supermassive black holes, like the title suggests. But much of~the~theory of~these giants is applicable to their million times lighter counterparts, when located inside a multiple stellar system. This can be e.g. a close binary system of two stars mutually orbiting each other due to gravitational attraction: one of them so heavy and old that it already exploded as a supernova in the past and now is a dark remnant, the other one being still an ordinary star, yet in such an evolution phase that it passes material from its outer layers to~the~nearby black hole. Although we will point out important differences between the two species within black-hole taxonomy, if applicable, the dissertation will eventually be about both. The existence of intermediate mass black holes is, however, still until today speculative.

~\

By emphasizing the two discrete populations of black holes with infalling (and~simultaneously partly ejecting) matter in axially symmetric structures, we highlight the \textit{general} similarity of appearance in nature across orders-of-magnitude different scales. Not only that physical shapes of objects are repeating in nature -- this is due to optimization in energy, momentum and angular momentum conservation -- but the same mathematical rules often apply for completely independent phenomena. 

Although much of our understanding of physical behavior comes from variation principles and symmetries in different forms, let us name perhaps a more concrete example from the black-hole theory discovered in 1970s mostly by Jacob Beckenstein (1947--2015) and Stephen Hawking (1942--2018): the description of~basic black-hole attributes is reconciled with the laws of thermodynamics for ordinary laboratory systems \citep{Bekenstein1972, Bekenstein1973, Hawking1974, Hawking1975}. Based on this, taking the \textit{measurable} properties of a physical vacuum (thus not true nothing) leads to fascinating connections between microscopic and macroscopic theories and provides long-term cosmological predictions for evaporating black holes with fundamental consequences \citep{Fulling1973, Davies1975, Unruh1976}. 

Another example of universality from the history of science of compact (massive and small) objects is the prediction by Wilhelm Anderson (1880--1940), Edmund Stoner (1899--1968) and Subrahmanyan Chandrasekhar (1910--1995) in~1929 and 1930: according to the known principles of the micro-world, a star 1.4 times heavier than our Sun cannot support itself due to internal pressure and will collapse \citep[see][and references therein]{Nauenberg2008}. This macroscopic limit of $2.765\times10^{30}$ kilograms can then be directly expressed through combination of~known fundamental constants in nature, derived from the micro-world and being universally present at much larger scales. Until now, no ordinary star heavier than the Chandrasekhar's limit has been discovered. It collapses to either a neutron star or, if heavier, then to a black hole \citep{Tolman1939, Oppenheimer1939, Bombaci1996, Kalogera1996}.

~\

Alongside light emitted from black-hole vicinity, since 2015 we are able to measure a \textit{direct} signal from black holes through gravitational waves \citep{LIGO2016}, which have been first rigorously predicted by Einstein in 1916 \citep{Einstein1916b, Einstein1918}. Both theoretical and observational works on light and gravitational-wave messengers from black holes were awarded the 2017 and 2020 Nobel prizes, which is a noticeable acknowledgement for black-hole studies. The plain detection of gravitational waves is, similarly to~the~predictions of the existence of black holes, an additional illustrative case of~synergy between theory and observations that sometimes takes a century to be completed, but may become powerful in the end. Already in 1974 Russel Alan Hulse (\text{*}1950) and Joseph Hooton Taylor Jr. (\text{*}1941) discovered a binary system of a neutron star and a pulsar (a highly magnetized, rapidly rotating neutron star) orbiting each other, showing a gradual decay of orbital period due to loss of energy and angular momentum in gravitational radiation, which earned them a Nobel prize in 1993 \citep{Hulse1975, Taylor1979, Taylor1982, Taylor1989, Weisberg2010}. A combination of light and gravitational wave detection from merging compact objects in binary systems is a way to examine the fundamental parameters of our universe and its evolution, for~example by studying merging supermassive black holes, or through the so-called pulsar timing arrays \citep[see e.g.][]{Hellings1983, Hobbs2010, Amaro2017, Arzoumanian2018}.

~\

Much of modern astronomy that uses cutting-edge technology is conversely giving hard times to our theoretical understandings of the outer space. And~the~wider these gaps are, the more we see trends in dividing the contemporary researchers to (sometimes self-proclaimed) experimentalists and theorists, as if there was some inherent division necessary in the cornerstones of scientific methodology, which results in polarisation and lack of contact between the~two. Of course, the quantity of today's knowledge requires high specializations and~team work, but it is likely that the rise of artificial intelligence will soon transform the academic world and accelerate the literature search and cognitive practices, which could help to bridge the gaps \citep{Klutka2018, Bates2020}. Although this dissertation has been completed in the ``old fashioned way'', we will attempt to link both entities in the field of astrophysics as efficiently as possible. We will largely depart from numerical simulations with practical software applications, in order to interpret astronomical data. At all times, with such ambitions, one comes across the dilemma between a too sophisticated model that is of low usage when it comes to comparison with real measurement, or an efficient data-fitting tool, which is, however, too simplistic to fully capture the~intricate impacts of known processes that are occurring combined. The ultimate skill of~a~theorist is to describe in the most simple way the~most important characteristics of complex environments observed by far-sighted experimentalists, or to give rational predictions for phenomena yet to be observed with inventive instrumental design.

~\

Such noble objectives are touching the topic of beauty in cognition of nature, which causes high tension, for example, in modern theoretical physics, affecting actual methodology and executive decisions in research. Here we will not need to~go this far. Let us just note that throughout the dissertation, human desires for~understanding the mysteries of nature uniformly and egoless exploration of~the~edges of~knowledge -- while discovering the hidden aesthetics -- are also important motivations behind. This is besides the usually mentioned purposeful incentives for doing basic research, owing to long-term civilization benefits and technological advancement, which are nonetheless extremely important in today's world. We remind that to those that despise the latter while using the Internet in the middle of environmental and health crisis.

We quote from Einstein's \textit{Physics and Reality}: ``The eternal mystery of~the~world is its comprehensibility... The fact that it is comprehensible is a~miracle,'' \citep{Einstein1936}. And moreover, that it works so well, and on ranges far beyond our everyday experience where we can still progressively understand with human invented science (even in the admittedly shallower \textit{how}, not the deepest \textit{why}). Humbleness is accompanying the greatest scientists, rather than pride for~ruling the nature.

~\

Despite the successful gravitational-wave detections, and despite the fact that massive particles are frequently arriving to Earth and represent another useful messenger of the remote, we will focus on predictions for and from electromagnetic radiation. It provides still by far the most prevalent and detailed description of~the~sky, given the thousands of years of available astronomical records. 

We will primarily focus on one property of light: \textit{polarisation}. As polarisation is a purely geometrical property of light, much of this work will be about simple spatial symmetries and asymmetries generating the patterns beyond. Although throughout the dissertation we will try to avoid the textbook statement ``high asymmetry at the origin of emission means high polarisation of the observed light'', unless either implication of the equivalence brings more qualitative insight, it will always be silently present as a driver for discoveries. 

Symmetries and asymmetries are part of us since ancient civilizations. We~now even know that there must had been a delicate disproportion of matter and antimatter in initial conditions of the universe that kept us here. Much has been written by scientists, artists, or philosophers on the fascination of a human mind by geometry. Out of all, let us name the book \textit{Symmetry} by Hermann Weyl (1885--1955) \citep{Weyl1952}, or quote Francis Bacon (1561--1626): ``There is no excellent beauty that hath not some strangeness in proportion,'' which is further elaborated in Chandrasekhar's \textit{Beauty and the quest for beauty in science} \citep{Chandrasekhar1979}, or refer way to the past to the Pythagorean concepts and the \textit{Harmony of~spheres} by Johannes Kepler (1571--1630) \citep{Kepler1619, Vamvacas2009}. Despite much of contemporary philosophy or art is miles away from such basal needs and ideals, modern-day scientists still refer to the pure occurrence of order, objectivity, eternal principles, geometrical interpretations and invariances in~nature -- all expressed in the language of mathematics (the main and by far most powerful language of contemporary science, regardless of mocking by some) -- as one of their main motivations of everyday efforts. And to many, mathematical symmetry, and more generally any mathematical description is far more than a~surprisingly useful tool in cognition of nature, it seems to be inherent to nature.

~\

We will leave such platonic beliefs or their partly legitimate criticism above all the following and let us deeply focus on one particular astrophysical problem. Aiming at the informative potential of polarisation, we will study the interaction of light and matter in the hot and violent environments near black holes. Here the~interplay of known microscopic and macroscopic phenomena is so complex (some would say dirty) that one would not expect any noble geometrical principles to~arise from the muddle. Albeit black holes by definition are one of~the~simplest objects in the universe (they are fully described only by 3 numbers: mass, spin and charge), their interaction with the outside world is of infinite degrees of freedom. The choice of studying such environment forces us to be highly pragmatic in~linkage of theory and observations and is a challenge to the aforementioned desire for simplifications. Yet we will show that it is possible to depart from the first principles and then use fundamental (not only) geometrical elements in the global parametrization of emergent polarisation. Conversely, it is natural to assume that if, in the end, we study the geometrical property of light with telescopes, it will primarily tell us something about the geometrical structures of matter emitting, absorbing or reflecting this radiation within these faraway, hence angularly small systems, much like a microscope into an unresolved object.

~\

Although the presented dissertation is not rocket engineering, the character of the problem does not allow to use the so-called classical physics. In classical physics we operate at human scales, with intuitive deterministic behavior and~slow movement, time is separated from space, all is ``flat'', the stage play is independent of the auditorium and we study nature as it truly \textit{is}. But for more than a hundred years, physics is enriched by objectively more correct approaches: a)~the~Quantum Field Theory (QFT), which effectively describes particle properties and interactions at the smallest scales using probabilities, and~which takes into account the~effect of measurement on the studied system, and b) the aforementioned theory of GR, which effectively describes the gravitational field by curvature of~the~space and time itself (on the left side of the Einstein's equations, equalling the matter parameters on the right) and washes out the difference between large-scale attraction and distortion of geometrical background. Both of~these include the older principles of Special Relativity (SR), important when characteristic speeds approach the speed of light. SR sets the rules of causality through finite and maximal speed (that of the light, $c = 2.998 \times 10^{10}$ centimeters per second), states that determination of length, time, mass or energy depends on velocity with respect to frame of reference, that space and time are inherently connected, and that some quantities are absolutely invariant (unlike the~usual layman's understanding that relativistic physics has something to do with ignoble relativism).

~\

Quantum physics will be touched only lightly in this work, whenever microscopic phenomena on particle levels are believed to play a collective role on~orders-of-magnitude larger volumes, on at least day-averaged time periods that are polarimetrically interesting (i.e. technically observable), and at high, kilo-electronvolt characteristic energies. Contrary to small-scale interactions, gravity is a unique physical force, because (apart from repulsive cosmological dark energy effects), it does not have a neutralizing counterpart and the effects cumulate and emerge only macroscopically. The GR theory helped to predict black holes and describes strong-gravity effects in the proximity of black holes in sufficient detail. The previous theory of Isaac Newton does not give correct predictions in~this case \citep[see e.g.][]{Misner1973}. GR predicts non-negligible curvature of~spacetime still at distances ten or hundred times larger than the event horizon where much observable action is happening. Thus, GR will be omnipresent in~this dissertation, although we plan to stay within one simplistic vacuum solution to~the~governing Einstein's equations: the Kerr black hole. Even if we assume gravitational vacuum outside the black hole, the massive central object will affect the trajectories, frequency and polarisation of photons passing by that we primarily aim to interpret from Earth's perspective. Thus, bringing one other necessary \textit{geometrical} chess piece into the game, which might be a great servant or a bad master.

~\

We will study physical conditions so extreme that no laboratory on Earth can produce them. It is a realm where human intuition often fails, but we will be loyal to logical extrapolation and assumption that the same laws of nature apply across our universe, i.e. a variation of the Copernican principle \citep{Peacock1999}. We will not touch cosmology that studies the space-time edges (whether finite or infinite) of the entire physical universe, although some of these results might a cosmologist find useful \citep[as Karl Popper (1902--1994) says: ``All science is cosmology, I believe...'', \textit{The Logic of Scientific Discovery}, ][]{Popper1959}.

Understanding the central engines of black-hole accretion is important, because the fundamental process are seen on limits of operation where anomalies are perhaps more anticipated, which can be identified by disagreement of observations with models. The infall of matter onto black holes represents the most efficient energy release in nature, thus it is no surprise that accreting black holes are highly energetic systems luminous at all wavelengths (cf. the virial theorem) \citep{Clausius1870, Collins1978, Karas2021}. Bolometrically, i.e. integrated across all frequencies, accreting supermassive black holes are intrinsically the~most luminous objects we know \citep[more than ten million times luminous than our Sun,][]{Reynolds2003}. They can outshine thousand times an average galaxy from a region smaller than our Solar system. Thus, they are easily visible even if~thousands of light-years away from us, despite being ``black'' in the core.

~\

In this dissertation, we will restrict ourselves to the X-ray band, expecting to gain direct insight on the most important high-energy processes occurring in~the~system's nucleus. Because incoming photons, concurrently to polarisation, contain also information on direction, intensity, frequency and time variability with respect to the origin of emission, we will cover also other degrees of freedom of X-rays when relevant to polarisation properties. In fact, photon-demanding polarimetry is complementary to high-resolution spectroscopy, timing or photometry, which are typically more developed in both theory and instrumentation \citep{Seward2010, Karas2021}. This applies to other wavelengths too, but we will only summarize multi-wavelength information on accreting black holes relevant to X-ray polarisation output and in turn provide hints where high-resolution X-ray polarimetry can lift degeneracies known from other techniques.

~\

We will investigate X-ray polarimetric methods that allow determination of~black-hole spin and mass, which are independent to the more popular \mbox{X-ray} spectroscopic and timing techniques. Even within the simplified assumption of~aligned black-hole rotation with the disc, which we will keep for simplicity, the speed of rotation is a very useful quantity when one examines the evolution of~the~universe and attempts to answer whether some very old supermassive black holes were rotating and what made them rotate in the first place, if the universe is assumed to be of finite and relatively young in age. And for their mass: how did they grow? Are they primordial?

The statistical knowledge of spin and mass of much smaller and younger black holes located inside our Galaxy is interesting for the stellar evolution theorists, in~some cases predicting the properties of progenitors to black-hole binary mergers emitting detectable gravitational waves. The knowledge of the mechanism behind the observed massless and massive particle output, also in relation to~the~black-hole properties, is important for evaluating the feedback and evolution of~the~entire host galaxy or host stellar system. 

Apart from the black-hole attributes, we will attempt to provide tools for~observational interpretations and constraints on the surrounding material, such as the inclination, size and ionization structure of the accretion disc, the geometry of hot gaseous coronae nearby, and the structure, location and shape of more distant cold obscuring material around the accreting core.

~\

All of this modeling is timely, because after pioneering 1970s experiments, such as on board of the \textit{Orbiting Solar Observatory 8} (\textit{OSO-8}) mission \citep{Weisskopf1978}, and after nearly four decades of quietness, observational X-ray polarimetry is on its rise \citep{Fabiani2014}. Largely due to successful launch of the \textit{Imaging X-ray Polarimetry Explorer} \citep[\textit{IXPE}, ][]{Weisskopf2022} by NASA and ASI in December 2021, which provides the first high-precision \mbox{2--8~keV} linear polarisation window into the outer space. Once we discover the~X-ray polarisation output of accreting black holes and the data coincides with predictions, we may assume something on the faraway cosmic laboratories, depending with honesty on our best state of theoretical knowledge. If the data does not coincide with predictions at all, we are missing something fundamental, which for a scientist (a truly childish one) after a few minutes of silence is strangely a~moment of~felicity, because a new gate opens.

~\

\vspace{10mm}
\hspace*{\fill} written in Stružinec u Jistebnice, July 2023

\newpage
\ 
\vspace{4.5cm} 
\begin{figure}[h]
        \centering
	\includegraphics[trim={0cm 0cm 0cm 0cm},clip,width=0.6\textwidth]{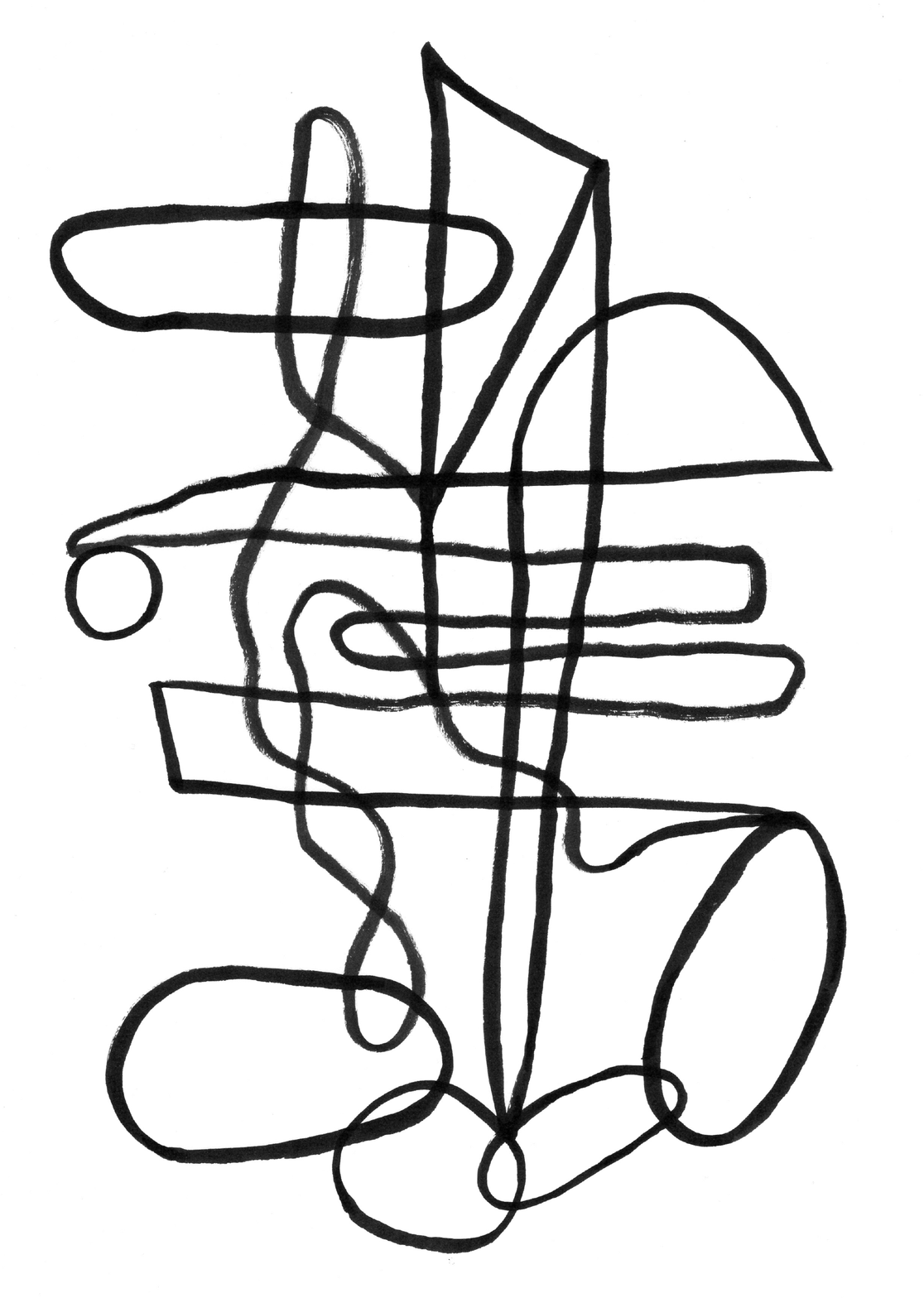}
\end{figure}
\chapter{Introduction}\label{introduction}

In this chapter, summary of literature, major definitions, and major conventions are given.

\section{Elementary processes}\label{elementary}

Let us begin with the most fundamental terminology and the overview of relevant physical processes. Throughout this dissertation, we will operate in~the~$cgs$ units apart from GR related calculations, where we will assume the standard geometrical unit system with $c = G = 1$. 

Because we will be interested in X-rays, we will often use keV units of energy and light will be mostly considered as photons, although introduction to polarisation is perhaps more intuitive within the equivalent wave description of~the~wave-particle duality of electromagnetic radiation. 

We express the energy of the radiation $\textrm{d}E$, in a specified frequency interval $(\nu, \nu + \textrm{d}\nu)$, which is transported across an element of area $\textrm{d}\sigma$ and in directions confined to an element of solid angle $\textrm{d}\omega$, during a time $\textrm{d}t$, in terms of the specific intensity $I$ by \citep{Chandrasekhar1960}
\begin{equation}\label{intensity}
    \textrm{d}E = I(x,y,z;l,m,n;t;\nu)\cos\vartheta \textrm{d}\nu \textrm{d}\sigma \textrm{d}\omega \textrm{d}t \textrm{ ,}
\end{equation}
where $\vartheta$ is the angle which the direction considered forms with the outward normal to $\textrm{d}\sigma$, $t$ is time, and $(x, y, z)$ and the direction cosines $(l, m, n)$ define
the point and the direction to which $I$ refers. Apart from intensity, we will often use scalar radiation flux $F_\textrm{E} \, [\textrm{erg} \cdot \textrm{cm}^{-2} \cdot \textrm{s}^{-1} \cdot \textrm{keV}^{-1}]$, which is the net flux of energy in all directions \citep{Castor2004}
\begin{equation}
    F_\textrm{E}(x,y,z;t;\nu) = \int I \cos\vartheta \textrm{d}\omega \textrm{ ,}
\end{equation}
where the integration holds over all solid angles $\textrm{d}\omega$. In order to provide simulation results regardless of distance towards the sources, it is more practical to~define flux not per element of detector area $\textrm{d}\sigma$, but per solid angle $\textrm{d}\Omega_0 = \textrm{d}\sigma / r_0^2$, by which the area $\textrm{d}\sigma$ is seen by an emitting point-like source at~a~distance $r_0$. In this case, we will use the notation $F_\textrm{E}^{\Omega_0} \, [\textrm{erg}~\cdot \textrm{sr}^{-1} \cdot~\textrm{s}^{-1} \cdot \textrm{keV}^{-1}] =~r_0^2F_\textrm{E}$. Sometimes a photon flux $N_\textrm{E} \, [\textrm{counts} \cdot \textrm{cm}^{-2} \cdot \textrm{s}^{-1} \cdot~\textrm{keV}^{-1}] =~F_\textrm{E}/(h_\textrm{P}\nu)$ or~$N_\textrm{E}^{\Omega_0} \, [\textrm{counts} \cdot~\textrm{sr}^{-1} \cdot~\textrm{s}^{-1} \cdot~\textrm{keV}^{-1}] =~F_\textrm{E}^{\Omega_0}/(h_\textrm{P}\nu)$ will be used, where $h_\textrm{P} =~6.626 \times~10^{-27} \, \textrm{cm}^2 \cdot \textrm{g} \cdot \textrm{s}^{-1}$ is the~Planck's constant.

\subsection{Polarisation of light}

We refer to e.g. \cite{Bachor2004} for the ``bra-ket'' notation of polarisation state $\vert P \rangle$ of individual photons, which is preferred for QFT literature. Polarisation is then explained by the spin angular momentum of the photon particles. Here we will define the basic microscopic quantities equivalently in~terms of~electromagnetic waves, which better illustrates the geometric terms used in~the~theory of polarisation. We will treat 3-dimensional space and time separately, because the governing Maxwell's equations and their solutions in this form are easily understood and naturally compatible with SR. We will reserve covariant 4-dimensional notation for GR calculations, if necessary. 

\begin{figure}[h]\centering
	\includegraphics[width=\textwidth]{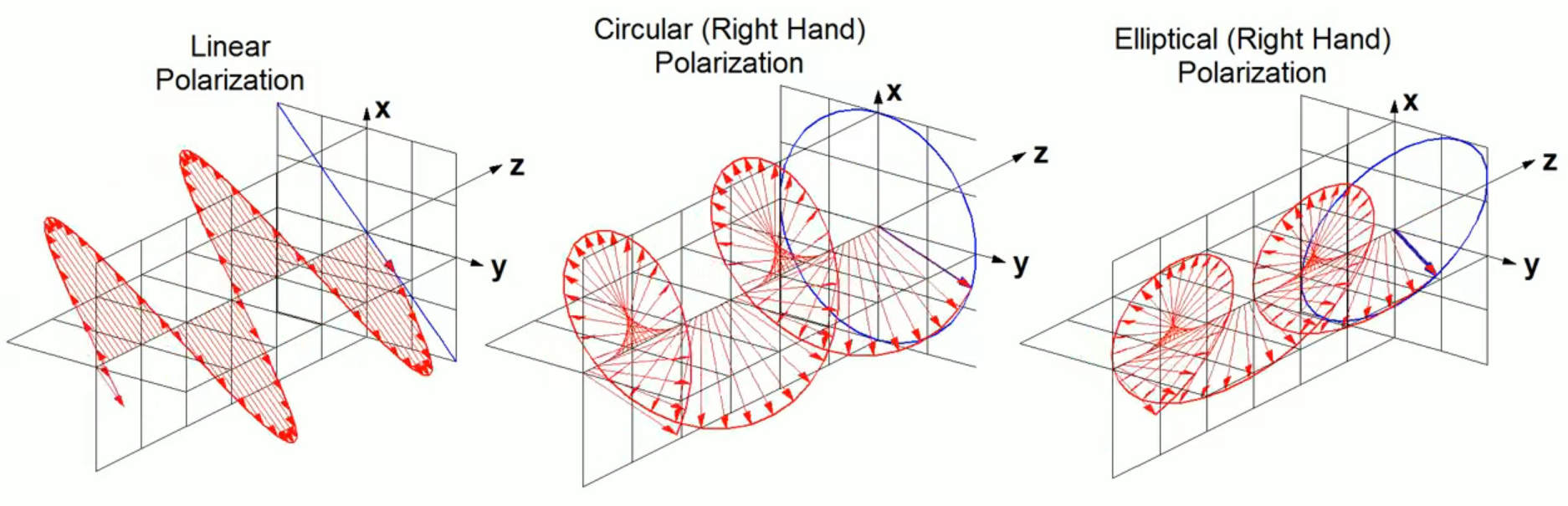}
	\caption{\footnotesize{Illustration of an electromagnetic wave that is linearly, circularly (right-handed) and elliptically (right-handed) polarized and traveling in the $z$-direction. Image from \cite{Yavuz2011}.}}
	\label{Elmag_polarization}
\end{figure}

Figure \ref{Elmag_polarization} shows three different types of polarisation in a Cartesian coordinate system ($x$,$y$,$z$). A single electromagnetic wave in vacuum is traveling with the~speed $c$ in $z$-direction. Maxwell's equations with the vacuum assumption then provide a solution of two non-trivial components of the electric field vector $\Vec{E}$:
\begin{equation}\label{harmonic_wave}
\begin{aligned}
E_x(t) =& E_x(0)cos(\omega t-\varphi_1)\textrm{ ,}\textrm{\space \space} \\
E_y(t) =& E_y(0)cos(\omega t-\varphi_2)\textrm{ ,}
\end{aligned}
\end{equation}
where $\omega = kc$ denotes the angular frequency, $k$ is the absolute value of the wave vector, and $\varphi_{1,2}$ denote the two –- a priori arbitrary –- phases (and analogically for the magnetic field).

Then, for the electric or magnetic field vector, in the plane perpendicular to~the~direction of propagation, if the ratio of the amplitudes and the difference in phases of the components in any two directions at right angles to each other are absolute constants, we call this wave (or a beam, if it holds for a superposition of~waves in this direction) to be ``elliptically polarized'' \citep{Chandrasekhar1960}. This condition guarantees that the field vector follows elliptical trajectory in~the~$xy$-plane, i.e. the ``polarisation plane'', because by eliminating $\omega t$ from equation (\ref{harmonic_wave}) we obtain equation of an ellipse
\begin{equation}
    \dfrac{E_x^2(t)}{E_x^2(0)} - \dfrac{2E_x(t)E_y(t)\cos\delta}{E_x(0)E_y(0)} - \dfrac{E_y^2(t)}{E_y^2(0)} = \sin^2\delta\textrm{ ,}
\end{equation}
where $\delta = \varphi_2 - \varphi_1$ is the relative phase between the two components of the electric field vector. 

The cases of ``linear polarisation'' and ``circular polarisation'' also shown in~Figure \ref{Elmag_polarization} are only subcases of elliptical polarisation when $\delta = 0$ and $\delta = \pm \pi / 2$, respectively. If $\delta > 0$, the tip motion from the point of view of the observer is in~counterclockwise direction and the polarisation is referred to as ``right-handed''. Otherwise, if $\delta < 0$, as ``left-handed''. The polarisation angle $\Psi$ is generally defined between the positive $x$-axis and the semi-major axis of the ellipse ($\Psi$ increases in~the~counterclockwise, right-handed direction from the observer's point of view) and is defined modulo $180^\circ$.

~\

Real astronomical detectors are sensitive only to a superposition of very large number of individual waves, which represents a macroscopic situation. Because all orientations of the electric field vector are in principle equally probable, the most natural state is unpolarized macroscopic signal. However, if a specific mechanism at the origin of radiation traveling in vacuum to the observer causes a preference in polarisation of a certain fraction of waves in a beam, we macroscopically characterize this fraction as the degree of polarisation $p$ of electromagnetic radiation
\begin{equation}
    p = \dfrac{I_P}{I}\textrm{ ,}
\end{equation}
where $I_\textrm{P}$ is the specific intensity of the polarized waves in the arriving beam, $I$ is the total specific intensity into this direction and at this point in space and time. For an individual wave described in the case above, the intensity $I$ representing a measurable energetic value, can be regarded as time average $\langle...\rangle$ of the electric field vector components over times $t$ much longer than $2\pi/\omega$ (it can still change with $t$ on large time scales for obtaining the apparent mean intensity directly from the electric field vibrations, as $2\pi/\omega$ is an extremely small scale for X-rays especially) \citep{Chandrasekhar1960}.

~\

In \cite{Stokes1851}, Sir George Gabriel Stokes (1819--1903) introduced a set of~four parameters, later called ``Stokes parameters'', which fully describe an~arbitrary polarisation state of radiation. One degree of freedom is represented by the~Stokes parameter $I$ defined by: 
\begin{equation}\label{Stokes_I}
    I = \langle E_x^2\rangle + \langle E_y^2\rangle\textrm{ .}
\end{equation}
The other three Stokes parameters $Q$, $U$ and $V$ are defined as
\begin{equation}
\begin{aligned}
Q &= \langle E_x^2\rangle - \langle E_y^2\rangle \textrm{ ,}\\
U &= 2\langle E_xE_y\cos\delta \rangle  \textrm{ ,}\\
V &= 2\langle E_xE_y\sin\delta \rangle   \textrm{ .}
\end{aligned}
\end{equation}
See e.g. \cite{Hamaker1996} for equivalent definition of Stokes parameters through complex exponential notation of the wave solution (\ref{harmonic_wave}). An important property of the~Stokes parameters is linearity, which allows that the~definitions hold for any macroscopic superposition of individual waves averaged over time.

~\

\begin{figure}[h]
	\centering
	\begin{minipage}[b]{0.49\textwidth}
		\includegraphics[width=\textwidth]{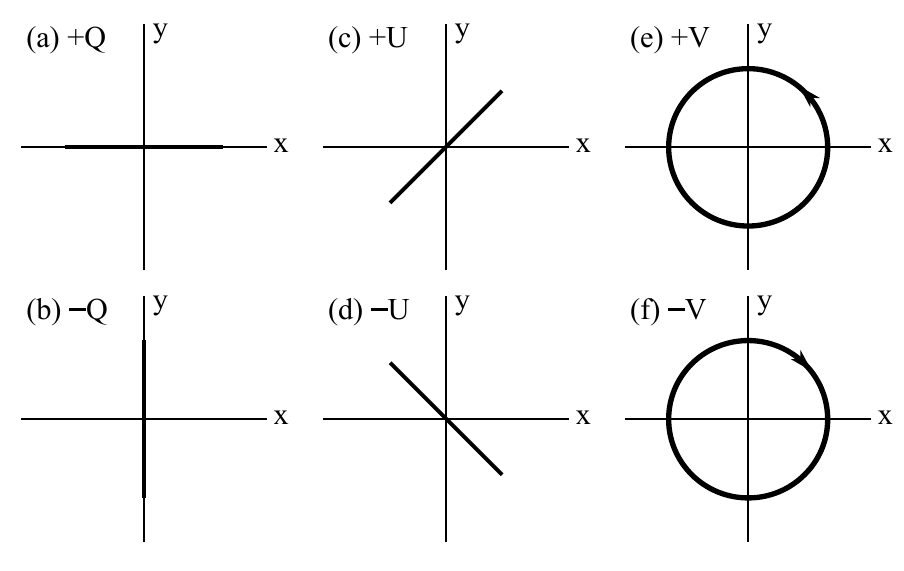}
		\caption{\footnotesize{Examples of polarisation directions in the polarisation plane for different values of~the~Stokes parameters. (a) $Q > 0$, $U = 0$, $V = 0$, then $\Psi = 0^\circ$; (b) $Q < 0$, $U = 0$, $V = 0$, then $\Psi = 90^\circ$; (c) $U > 0$, $Q = 0$, $V = 0$, then $\Psi = 45^\circ$; (d) $U < 0$, $Q = 0$, $V = 0$, then $\Psi = 135^\circ$; (e) $V > 0$, $Q = 0$, $U = 0$, then $\Psi$ is undefined; (f) $V < 0$, $Q = 0$, $U = 0$, then $\Psi$ is undefined. Image from \cite{Kislat2015}.}}
		\label{Stokes_params}
	\end{minipage}
	\hfill
	\begin{minipage}[b]{0.49\textwidth}
		\includegraphics[width=\textwidth]{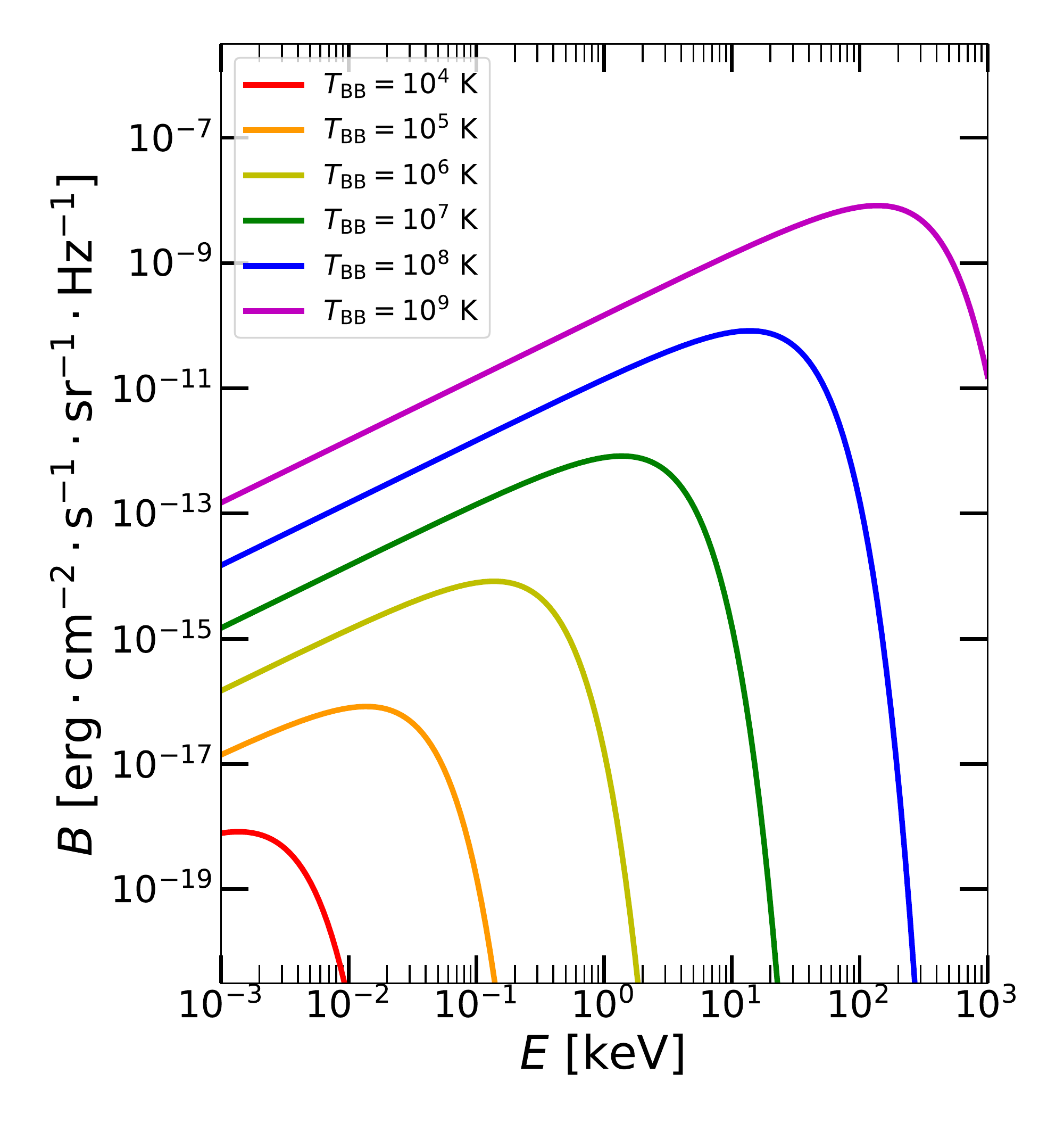}
		\caption{\footnotesize{The Planck's law of radiation (\ref{blackbody}) in X-rays.}}
		\label{bb_curves}
	\end{minipage}
\end{figure}

The meaning of Stokes parameters can be understood in the following way. The squared amplitude of $\Vec{E}$ in the $xy$-plane of the wave packet is Stokes $I$. Hence, it may be denoted by the same letter as the macroscopic specific intensity defined by equation (\ref{intensity}). Figure \ref{Stokes_params} illustrates the $Q$, $U$ and $V$ Stokes parameters and the definition of polarisation angle in the polarisation plane. A~difference in~intensities between the $x$ (``vertical'') and $y$ (``horizontal'') directions is quantified by $Q$. A~difference in intensities between the $45^\circ$ (``diagonal'') and~$135^\circ$ directions in $\Psi$ is quantified by $U$. Thus, $Q$ and $U$ provide information on~linear polarisation. The intensity of circularly polarized light is encoded in $V$. For~\textit{individual} waves, the Stokes parameters are related via
\begin{equation}
    I^2 = Q^2 + U^2 + V^2 \, ,
\end{equation}
which reduces the number of free parameters required for a full description to~three. If we superpose waves with the same polarisation adding to total intensity $I_\textrm{P} \leq I$, we may rewrite this equation as \citep{Chandrasekhar1960}
\begin{equation}
    I_\textrm{P}^2 = Q^2 + U^2 + V^2 \, ,
\end{equation}
which leads to the expected four degrees of freedom. 

~\

Two parameters can be replaced by the previously introduced degree of polarisation $p$ and polarisation angle $\Psi$. Let us redefine these for partial linear polarisation of a generic beam, because current and forthcoming X-ray polarimeters are not sensitive to circular polarisation \citep{Fabiani2014}. Therefore, for~the rest of this dissertation, we will use the linear\footnote{\,The degree of circular polarisation $p_\textrm{c}$ would then be defined analogically as $p_\textrm{c} = V/I$.} polarisation degree $p$ and~linear polarisation angle $\Psi$:
\begin{equation}\label{psi_def}
\begin{aligned}
p &= \dfrac{\sqrt{Q^2+U^2}}{I} \textrm{ ,}\\
\Psi &= \dfrac{1}{2}\textrm{\space}\textrm{arctan}_2\left(\dfrac{U}{Q}\right)   \textrm{ ,}
\end{aligned}
\end{equation}
where $\textrm{arctan}_2$ denotes the quadrant-preserving inverse of a tangent function. It~is often more intuitive to use $p$ and $\Psi$, but for e.g. superposition, energy binning and measurement uncertainties it is preferred to work in the $QU$-plane, where $p$ and $\Psi$ correspond to the length and orientation of a vector centered at the origin of this plane (the difference is mainly due to scalar versus vectorial approach and statistics of $Q$ and $U$, which are Ricean, see Section \ref{polarimetry}). Addition of two beams with equal intensities polarized with the same $p$ perpendicularly to each other in~terms of $\Psi$ results in net zero polarisation fraction.

~\

Numerical codes often require efficient transcription for physical elements that change the polarisation state of a macroscopic electromagnetic signal traveling otherwise in vacuum. If the Stokes parameters are held in 4-dimensional Stokes vector $\Vec{S} = (I, Q, U, V)$, then any suitable matrix transformation can represent the change of the polarisation state along the light trajectory. This is referred to as ``Müller formalism'' with ``Müller matrices'' \citep{Mueller1948}. This representation is equivalent to the ``Jones formalism'', which operates with the $x$- and $y$-amplitudes and phases of the electric field in complex exponential notation that are put into a vector and any series of optically active elements changing polarisation are
similarly the ``Jones matrices'' \citep{Jones1941}.

\subsection{X-ray polarigenesis}\label{processes}

Number of processes occurring in nature that alter the polarisation state have been observed and theoretically understood. They all involve internal asymmetry of the optically active element with respect to radiation. In astrophysical situations, these are typically scatterings (Thomson scattering, Compton scattering, Rayleigh scattering, resonant scattering, Raman scattering, or Mie scattering on dust grains), which cause linear polarisation along a well-defined direction on spherical particles or on collectively oriented grains with elongated shapes in optically thick media, e.g. dust clouds in interstellar medium (ISM) \citep{Chandrasekhar1960,Serkowski1975,Born1999,Draine2003,Trippe2014}. Then the linear or circular dichroic media, such as environments with long-chain molecules, helically shaped molecules or plasma in the presence of magnetic field, may cause that one component of the wave in the transverse plane experiences stronger attenuation while passing through \citep{Angel1974,Wolstencroft1974,Huard1997,Born1999}. The synchrotron radiation emits elliptically polarized light due to geometric projections of relativistically gyrating electric charges around magnetic field lines \citep{Ginzburg1965, Rybicky1979}. The radio observations of ISM often reveal the Zeeman effect on spectral lines in the presence of magnetic field (and~the~anomalous Zeeman effect, and the Goldreich-Kylafis effect for unresolved cases) that causes the resulting split spectral lines to be either linearly or circularly polarized, depending on the viewing angle towards the magnetic field lines \citep{Zeeman1897,Preston1898,Preston1899,Goldreich1981,Goldreich1982,Elitzur2002}. The~Hanle effect in fluorescent gases permeated by weak magnetic field results in linear polarisation genesis \citep{Hanle1924,Stenflo1982}. Linear polarisation emerges also due to~reflections on refractive boundary between two homogeneous media and linear or~circular polarisation arises in birefringence crystals and similarly in ISM plasma with external static homogeneous magnetic field, i.e. the Faraday rotation effect \citep{Chandrasekhar1953,Burn1966,Pacholczyk1970b,Rybicky1979,Fowles1989,Tribble1991,Huard1997,Born1999}. In high-energy astrophysics and QFT, we also encounter possible testing of axion-like particles through X-ray polarimetry and the quantum electrodynamic effect of vacuum birefringence \citep{Heisenberg1936,Heyl2000,Pavlov2000b,Heyl2002,Toma2012,Day2018,Caiazzo2022}. And as it was already mentioned, GR effects alter polarisation of light in vacuum too \citep{Connors1977, Connors1980}.

~\

We refer to \cite{Trippe2014} for a recent review on astrophysical polarisation and~further literature on the above. Here, we will elaborate slightly more on those processes that are important in the X-rays near black holes. Because unpolarized emission depolarizes (by dilution) any partially polarized beam of different origin, we will also briefly describe those elementary processes that produce unpolarized emission and are assumed to be important in the studied environments. Many discussed numerical codes implement some forms of absorption, which is also important for the net emerging polarisation state, if fraction of (un)polarized radiation is taken away in particular direction. 

Up to our knowledge at scales considered, no microscopic interactions are directly coupled with gravity effects that alter the entire space and time background through curvature, i.e. the curvature field is separable and depends only on~the~net energy and momentum tensor. Thus, we will describe the GR effects on~polarisation inside the follow-up section that provides the equations of radiative transfer, the global GR setup of this work and our additional ``linearized'' simplification (a given vacuum Kerr background, no self-gravitating elements typical for non-linear general-relativistic magneto-hydro-dynamical simulations) that will allow us to embed all in a simple separable relativistic radiative transfer treatment. 

We will omit the discussion of Faraday rotation and vacuum birefringence, as it will not be solved for magnetic fields in this dissertation, but we refer to~e.g. \cite{Caiazzo2018, Krawczynski2021} for estimations of possible effects in X-rays.

\subsubsection*{Thermal emission}

All matter with non-zero temperature emits thermal radiation, because any movements of particles in the material causes charge acceleration or dipole oscillation, which produces electromagnetic radiation. If the object approximates a black body (e.g. an optically thick accretion disc with sharp density drop in the photosphere) and thermal equilibrium is reached, the emission intensity profile per~unit frequency is isotropic and according to Planck's law \citep{Planck1901,Rybicky1979}:
\begin{equation}\label{blackbody}
    B = \frac{2h_\textrm{P}\nu^3}{c^2}\frac{1}{e^{\frac{h_\textrm{p}\nu}{k_\textrm{B}T_\textrm{BB}}}-1}\textrm{ ,}
\end{equation}
where $T_\textrm{BB}$ is the body's temperature and $k_\textrm{B} = 1.381 \times 10^{-16} \, \textrm{erg} \cdot \textrm{K}^{-1}$ is the~Boltzman's constant.

In the treatment of disc photospheres, we will sometimes use the effective temperature $T_\textrm{eff} = T_\textrm{BB}$ that characterizes radiation field of a quasi-thermal medium, i.e. it assigns the effective temperature value, as if it was a black body. For~simplicity, the radiative transfer codes often assume a local thermodynamic equilibrium (LTE), characterized by optically thick environment (photon's mean free path is shorter than characteristic length) and $I = B$, thus we also assume some mean local temperature. For the black-body emission to peak in the X-ray band, it requires extreme temperatures of $T_\textrm{BB} \gtrapprox 10^5$ K (see Figure \ref{bb_curves}). Black-body radiation is always unpolarized, as there is no internal asymmetry in the thermal pool.

\subsubsection*{Scattering on particles}

Thermal emission can, however, gain some polarisation through scattering on~electrons \citep[bound or free, although for X-rays above 1 keV the distinction does not matter for gas mostly composed of hydrogen,][]{Vainshtein1998} or dust grains. In the X-rays, we are mostly interested in Compton scattering of photons on free electrons, which is called Thomson scattering in the inelastic limit -- applicable when $h_\textrm{P}\nu$ of the incoming photon becomes comparable to the electron rest energy $m_\textrm{e}c^2$.

Let us list the formulae for unpolarized incident photon, which can be found in e.g. \cite{McMaster1961,Fernandez1993,Matt1996} more generalized. The photon's energy changes after the interaction as \citep{Compton1923}
\begin{equation}\label{cpt_energy}
    \varepsilon' = \dfrac{\varepsilon}{1+\dfrac{\varepsilon}{m_e c^2} (1-\cos \Theta)} \textrm{ ,}
\end{equation}
where $\varepsilon$ and $\varepsilon'$ are photon energies before and after the scattering at an angle $\Theta$, respectively. The cross-section $\sigma_\textrm{KN}$ of Compton scattering is expressed by the~Klein-Nishina formula \citep{Klein1929}
\begin{equation}\label{KN_formula}
    \sigma_\textrm{KN} = \dfrac{3}{4}\sigma_\textrm{T} \Big\{ \dfrac{1+u}{u^3}\Big[ \dfrac{2u(1+u)}{1+2u}-\log (1+2u) \Big] + \dfrac{\log (1+2u)}{2u} - \dfrac{1+3u}{(1+2u)^2}\Big\} \textrm{ ,}
\end{equation}
where $u = h_\textrm{P}\nu / m_\textrm{e} c^2$ and $\sigma_\textrm{T} = 8 \pi e_\textrm{C}^4 / 3 m_\textrm{e}^2 c^4$ is the classical Thomson cross section, where $e_\textrm{C}$ is the elementary charge. The differential cross section for electrons at~rest is \citep[e.g.][]{Matt1996}
\begin{equation}\label{KN_diff}
    \frac{\textrm{d}\sigma_\textrm{KN}}{\textrm{d}\omega} = \dfrac{1}{2}\frac{e_\textrm{C}^2}{m_\textrm{e} c^2}\left( \frac{\varepsilon'}{\varepsilon} \right)^2 \left[ 
  \left( \frac{\varepsilon'}{\varepsilon} \right) + \left( \frac{\varepsilon}{\varepsilon'} \right) - \sin^2 \Theta \right]  \textrm{ .}
\end{equation}

~\

\begin{figure}[h]
	\centering
	\begin{minipage}[t]{0.49\textwidth}
		\includegraphics[width=\textwidth]{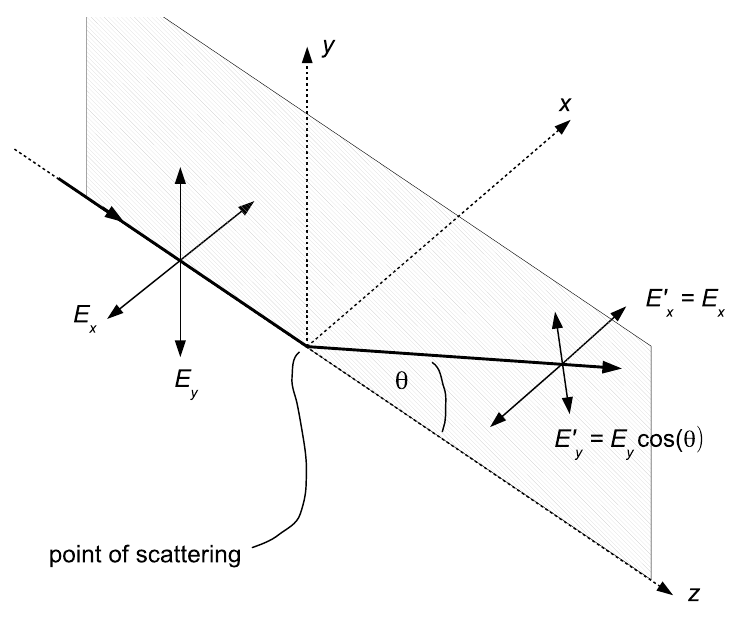}
		\caption{\footnotesize{Explanation of linear polarisation induced by scattering in particular geometry. Gray plane is called the~plane of scattering. Image from \cite{Trippe2014}.}}
		\label{Compton_scat}
	\end{minipage}
	\hfill
	\begin{minipage}[t]{0.49\textwidth}
		\includegraphics[width=\textwidth]{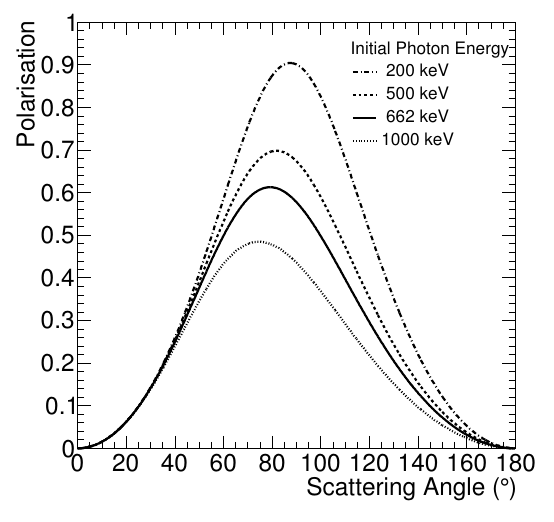}
		\caption{\footnotesize{Degree of linear polarisation $p_\textrm{C}$ induced by Compton scattering as a function of scattering angle $\Theta$ for a few cases of initial photon energy $\varepsilon$. Image from \cite{Knights2018}.}}
		\label{Compton_pol}
	\end{minipage}
\end{figure}

Figure \ref{Compton_scat} illustrates how the passing photon can gain linear polarisation, which is applicable also to other types of scatterings. The resulting polarisation is perpendicular to the ``scattering plane'' given by the incident and emission directions.

The resulting degree of polarisation for Compton scattering is given more pricesely for unpolarized incident photons by \citep{Hudson1968}
\begin{equation}\label{compton_polarization}
    p_\textrm{C} = \dfrac{\sin^2 \Theta}{\dfrac{\varepsilon}{\varepsilon'}+\dfrac{\varepsilon'}{\varepsilon}-\sin^2 \Theta} \textrm{ .}
\end{equation}
Thus it depends on energy and, for an initially unpolarized beam, the polarisation of~$p = 100\%$ is never achieved. The degree of polarisation as a~function of~the~scattering angle $\Theta$ is shown in Figure \ref{Compton_pol} for different initial photon energies. In simulations we then typically utilize these relations, combined with their generalization for polarized incident photons, and then for the fraction of~polarized photons $p$ the polarisation vector  $\Vec{P}$ (a unit vector parallel to $\Vec{E}$) of~the~scattered photon is \citep{Angel1969}
\begin{equation}
    \Vec{P} = \frac{(\Vec{P}_0 \times \Vec{e}_\textrm{z})  \times \Vec{e}_\textrm{z} }{\vert \Vec{P}\vert} \textrm{ ,}
\end{equation}
where $\Vec{P}_0$ is the~incident polarisation vector and $\Vec{e}_\textrm{z}$ is the unit vector in~the~scattered direction. The remaining fraction $1-p$ has the polarisation vector randomly distributed in the polarisation plane.
Multiple scatterings often result in~net depolarisation of the emergent radiation, because each scattering event occurs in~generally different direction with different photon energy change.

In the limit of low initial energies (i.e. the Thomson scattering), the curve shown in Figure \ref{Compton_pol} becomes symmetric around $\Theta = 90^\circ$, can reach $p = 100\%$, and follows a simple law:
\begin{equation}\label{thomson_polarization}
    p_\textrm{T} = \dfrac{1- \cos^2 \Theta}{1+\cos^2 \Theta} \textrm{ .}
\end{equation}

When the incident photon possesses higher energy than the electrons, we refer to ``Compton down-scattering'' and the photon looses energy after it scatters. In~extremely hot plasma in the vicinity of black holes, the situation can be reversed and photons can gain energy from ultra-relativistic electrons (with mean speeds comparable to $c$). Then we refer to ``Compton up-scattering'', ``inverse Compton scattering'', or ``Comptonization'' for the case of multiple scatterings. A~proper quantum treatment is required to fully describe this effect, but in~the~case of up-scattered thermal emission, it typically collectively leads to moderately polarized power-law emission in the case of coronae near accretion discs \citep{Rybicky1979, Sunyaev1980, Sunyaev1985}. We refer to \cite{Bonometto1970,Nagirner1993, Poutanen1993, Poutanen2010} for the analytical treatment of polarisation due to Compton up-scattering. For a relativistic isotropic electron gas, the resulting polarisation is zero \citep{Bonometto1970}.

\subsubsection*{Synchrotron emission}

Black-hole accreting systems often possess extreme magnetic fields \citep[$\approx 10^4$ G for~supermassive black holes, $\approx 10^8$ G for stellar-mass black holes, measured as accretion disc magnetic strength, see ][]{Daly2019}. Polarized synchrotron emission is particularly known to be a magnetic field structure tracer near the~black-hole horizon in the radio band \citep{EHT2021pol}. But the resulting polarized synchrotron emission, sometimes referred to~as ``magneto-bremsstrahlung'', of electrons gyrating around magnetic field lines can affect the~net output even in the X-rays \citep{Ginzburg1965,Weisskopf2022}. The emergent polarisation is orthogonal to the magnetic field and~the~details of polarisation computation, including absorption, can be found in e.g. \cite{Leung2011}.

The disordering of magnetic field by turbulences and shocks, and multiple scatterings can wash out the primary polarisation gain. For astrophysically realistic plasmas with a partially disordered magnetic field and $0 < \bar{\alpha} < 1$, where $\bar{\alpha}$ sets the power in the usually assumed power-law distribution of electron energies, the degree of linear polarisation is not expected to exceed $p \approx 16\%$ \citep{Pacholczyk1970}. For circular polarisation the realistic estimates are $p_\textrm{c} \lessapprox 1\%$ \citep{Angel1974}. 

In X-rays, we often encounter synchrotron self-Compton radiation, which are photons that get inverse Compton scattered by the same electrons that caused synchrotron emission due to acceleration in magnetic field \citep[see e.g.][]{Ghisellini2013}. Unpolarized synchrotron photons result in synchrotron self-Compton emission with small ($<5$\%) polarisation fraction for all but very low Lorentz factors describing the mean particle speeds with respect to $c$ \citep{Begelman1987,Celotti1994}.

\subsubsection*{Bremsstrahlung}

At soft X-rays, we typically encounter one other form of free-free radiation. This is the free-free absorption or emission of photons when free electron accelerates in the electric field of positively charged atomic nucleus \citep{Rybicky1979}. This effect is often called ``thermal bremsstrahlung'', although it is not truly \textit{thermal}, in comparison to black-body emission.

The cross-section exponentially decays towards high energies, and the effects are expected to be non-negligible in the X-ray band for hot gases with electron temperatures $T_\textrm{e} \gtrapprox 10^6 \textrm{ K}$ \citep{Seward2010}. It may impact the~net polarisation emerging nearby accreting black holes by means of~absorption, and~perhaps more importantly by means of emission, because the~cross-section for~bremsstrahlung is polarisation dependent \citep{Gluckstern1953}. For~the~mid (1--10 keV) and hard ($>10$ keV) X-rays, for example \cite{Haug1972,Komarov2016,Krawczynski2023} provide calculations of~polarized bremsstrahlung emission for pressure anisotropy of electrons in solar flares, galaxy clusters and black-hole accretion discs, respectively.  In the last case it is shown that in combination with Comptonization it may lead to surprisingly high (tens of \%) linear polarisation fraction.

\subsubsection*{Photoelectric effect}

An important class of processes in the X-rays is the photoionization of electrons and recombination, i.e. the bound-free absorption and emission. The cross-section for photoelectric absorption is affected by particular ionization edges, but in the X-rays generally declines as $\sim E^{-3}$ \citep[see e.g.][]{Garcia2013}. The~condition of hydrostatic equilibrium for e.g. accretion disc's plane-parallel atmosphere, can be replaced by even simpler assumption of constant pressure or constant density. In the latter case, a very useful quantity called ionization parameter $\xi$ $[\textrm{erg} \cdot \textrm{cm} \cdot \textrm{s}^{-1}]$ can be defined as \citep{tarter1969}
\begin{equation}\label{xi}
	\xi = \dfrac{4 \pi \int_0^\infty F_\textrm{E} \textrm{d}E}{n_\textrm{H}} \textrm{ ,}
\end{equation}
where $n_\textrm{H}$ is the neutral hydrogen density in the medium, $F_\textrm{E}$ is the radiation flux received at one location on the disc surface. It represents the competition between the photoionization rate ($\sim n_\textrm{H}$) and recombination rate ($\sim n_\textrm{H}^2$) inside the irradiated medium. Although recombination produces unpolarized photons, the photoelectric cross-section is dependent on the incident polarisation state, which is utilized in the gas-pixel detectors on board of \textit{IXPE} (see Section \ref{polarimetry}).

\subsubsection*{Spectral lines}

The surroundings of accreting supermassive and stellar-mass black holes produce many emission and absoprtion lines in the X-rays caused by bound-bound electron transitions inside atoms. Absorption lines typically emerge from cold gas irradiated by hotter thermalized medium behind from observer's point of view. Emission lines are typically a result of reflection. Fluorescent lines are unpolarized, but resonant line scattering may cause non-negligible polarisation \citep{Lee1994a, Lee1994b, Lee1997}, depending on the competing Auger effect. The polarisation of the scattered line emission depends on the~quantum numbers characterizing the electronic transition in each resonant line. It results in~a~wide range in strength, but may become as relevant as for Thomson scattering. 

Due to the ``iron peak'' in nuclear binding energy curve of chemical elements, iron is quite omnipresent in the universe, including nearby supermassive and~stellar-mass black holes. Therefore, in the X-rays, we typically encounter a photoelectric absorption by inner shells of partially ionized iron atoms, which results into excited state and immediate decay. This is done through either Auger de-excitation \citep[e.g. with $66\%$ probability for $n = 1$ shell of the iron atom,][]{Fabian2000} or characteristic fluorescent line X-ray emission, where the emitted photon may escape to the observer or encounter further interactions. The details of bound-bound processes require quantum treatment and are highly dependent on local physical conditions, similarly to ionization. 

All spectral lines originate at single energy values, but are broadened by vast number of physical effects. It is out of the scope of this dissertation to provide the theoretical background, but from non-relativistic effects we name the Doppler broadening due to thermal velocities, turbulences and large-scale velocities, natural broadening and pressure broadening. The relativistic line broadening will be discussed in the next section. We refer to e.g. \cite{Castor2004} for more details on~the~emergence and properties of spectral lines or ionization and recombination.

\subsection{Combined radiative transfer}\label{combined}

We briefly provide below the theoretical framework for numerical solving of radiative processes that is used in this work. In addition to these, we examine Monte Carlo (MC) methods \citep[see e.g.][]{Cashwell1959} that rely on injecting photons in the configuration space, ray tracing and calculating probabilities for~absorption and scattering.

\subsubsection*{Radiative transfer in astrophysical materials in local frame of reference}

When numerically solving for a radiation solution in interaction with partially ionized medium in plane-parallel static approximation, we typically discretize the medium in vertical layers, provide boundary conditions and iterate for temperature, pressure and fractional abundance of atomic species inside the medium the one-dimensional equation of radiative transfer \citep{Chandrasekhar1960}:
\begin{equation}\label{RTE}
   \mu_\textrm{e} \dfrac{\textrm{d} I}{\textrm{d} \tau} = I - \tilde{S} \, ,
\end{equation}
where $\mu_\textrm{e}$ is cosine of the~inclination angle from the slab's normal and $\tilde{S}$ is the~frequency dependent source function that combines all absorption, emission and~scattering processes contributions in an infinitely wide slab parametrized in~the~remaining dimension by optical depth $\tau$, again frequency dependent and proportional to geometrical depth $l$ measured from the side of the slab (in all cases for~disc atmospheres, we will assume this side to be a fictitious disc surface, below which the $\tau$ reaches infinity in optically thick regime). In non-LTE conditions, we have to solve the equation of statistical equilibrium \citep{Castor2004}
\begin{equation}\label{SE}
     n_{i} \sum_{j \neq i} P_{ij} - \sum_{j \neq i} n_{j} P_{ji} = 0 \, ,
\end{equation}
where $P_{ij}$ are all ionization and excitation rates for both radiative and collisional processes, $P_{ji}$ are all recombination and de-excitation rates and $n_{x}$ are the corresponding number densities for the species that we account for. Polarized radiative calculus in materials is well developed \citep[e.g.][]{Chandrasekhar1960,Sobolev1963,Stenflo1994, DeglInnocenti2004}, but will not be used in~this work, therefore we omit the theoretical background.

\subsubsection*{General-relativistic radiative transfer in vacuum}\label{relativistic}

We consider a single rotating black hole in this work and the rest of the universe is simply non-gravitating. Although this is a notable simplification and many times the axial and equatorial symmetry allows significant reductions in~calculation of~light trajectories and polarisation, analytical operations within the Kerr metric typically require extensive acrobatics with trigonometric functions. This is clear already from the derivation of the solution and e.g. Chandrasekhar writes on~perturbation studies of the Kerr metric \citep{Chandrasekhar1983}: ``... the analysis has led us into a realm of the rococo: splendorous, joyful, and immensely ornate.'' 

Despite challenging details, a Kerr black hole can be fully described only by mass $M$ and angular momentum $J$. We do not consider electric charge, because it is widely believed that in astrophysical black holes any excess of electric charge would autoneutralize, although this is still a matter of debate with the latest \textit{EHT} results \citep{EHT2021, EHT2022}. 

One of the simplest representations of the Kerr metric components from the length element $\textrm{d}s$ is in Boyer-Lindquist coordinates ($t$,$r$,$\theta$,$\phi$) \citep{Boyer1967}, being a natural generalization of the Schwarzschild coordinates: \citep[e.g.][]{Visser2007}
\begin{equation}\label{Kerr}
\begin{split}
    \textrm{d}s^2 = g_{\mu \nu}\textrm{d}x^{\mu}\textrm{d}x^{\nu} &= - \left(  1 - \frac{2Mr}{r^2 + a^2\cos^2{\theta}}  \right) \textrm{d}t^2 - \frac{4Mra\sin^2{\theta}}{r^2 + a^2\cos^2{\theta}}  \textrm{d}t \textrm{d}\phi 
    \\
    &\quad  + \left( \frac{r^2 + a^2 \cos^2{\theta}}{r^2 - 2Mr + a^2} \right) \textrm{d}r^2 + (r^2 + a^2 \cos^2{\theta} ) \textrm{d}\theta^2
    \\
    &\quad + \left( r^2 + a^2 + \frac{2Mra^2\sin^2{\theta}}{r^2+a^2\cos^2{\theta}}  \right) \sin^2{\theta} \textrm{d}\phi^2 \, ,
\end{split}
\end{equation}
where we used the specific angular momentum $a = J/M$, the ``black-hole spin'', which has a reduced range of $-1 \leq a \leq 1$ and is considered in this work only in one, positive direction of rotation, given by angular 3-velocity $\Vec{\Omega}$. In fact, \cite{Thorne1974} calculated a maximal theoretical rotation value to be $a = 0.998$ for~a~standard accretion disc (see below). 

We assume the surrounding disc to be corotating with the black hole and residing geometrically thin in the equatorial plane ($\theta = \pi/2$), but we note that this is another simplification that is in fact hard to justify. We note that there are many works studying misaligned rotation, counter rotation, or self-gravitating discs and~other departures from the Kerr metric \citep[e.g.][]{Bardeen1975,Laor1989,Collin1999,Karas2004,Collin2008,Nixon2012a,Nixon2012b,Krawczynski2012,Krawczynski2018,Tripathi2020, Hrubes2022}. Our estimates should thus serve as first-order approach. 

If $a = 0$, the solution (\ref{Kerr}) becomes a Schwarzschild solution and the inner-most circular orbit (ISCO) is located at 6 gravitational radii ($6 \, r_\textrm{g} = 6 \, GM/c^2$ in~non-geometrical units). If $a = 1$, the ISCO is much closer to the black-hole: at $1 \, r_\textrm{g}$. This is important for the allowed accretion disc's inner extension (see below). In this study, we will not consider any light escaping from the~infalling matter below ISCO, which is as well a debatable simplification \citep[e.g.][]{Dovciak2004}. The region between the (outer) event horizon and between the Killing horizon \citep[where Killing vectors are null,][]{Carroll2004}, is called the ``ergosphere'' and is important for explanation of the work necessary on observed ejected material in~polar directions through the Penrose process \citep{Penrose1969}. For a full theoretical background we refer to e.g. \cite{Misner1973}.

~\

\begin{figure}[h]
	\centering
	\begin{minipage}[t]{0.39\textwidth}
		\includegraphics[width=\textwidth]{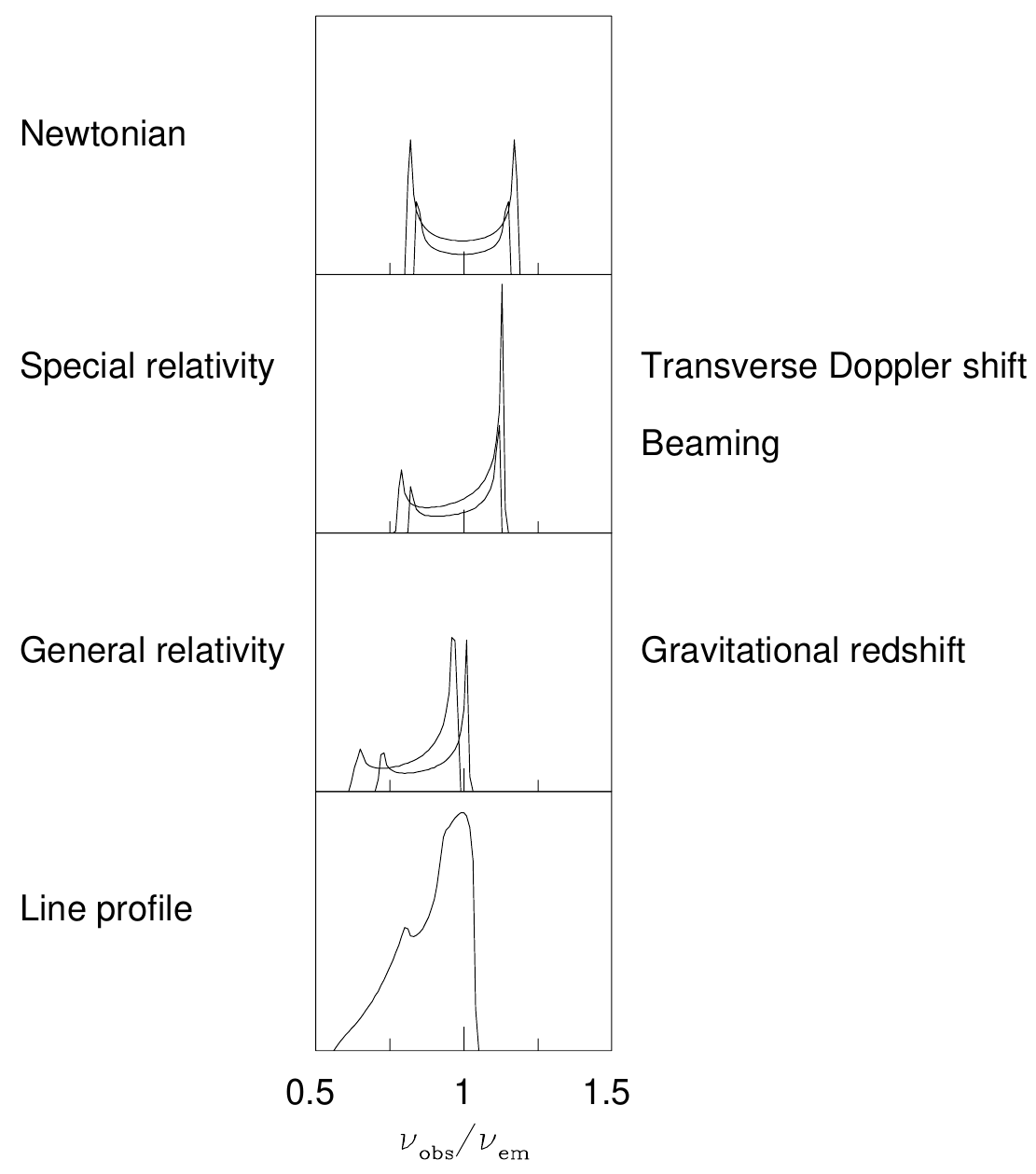}
		\caption{\footnotesize{Illustration of the spectral line broadening due to classical Dopler shift, SR Doppler effect and~aberration, GR energy shift and the net effect (from top to bottom). Image from \cite{Fabian2000}.}}
		\label{line_broad}
	\end{minipage}
	\hfill
	\begin{minipage}[t]{0.59\textwidth}
		\includegraphics[width=\textwidth]{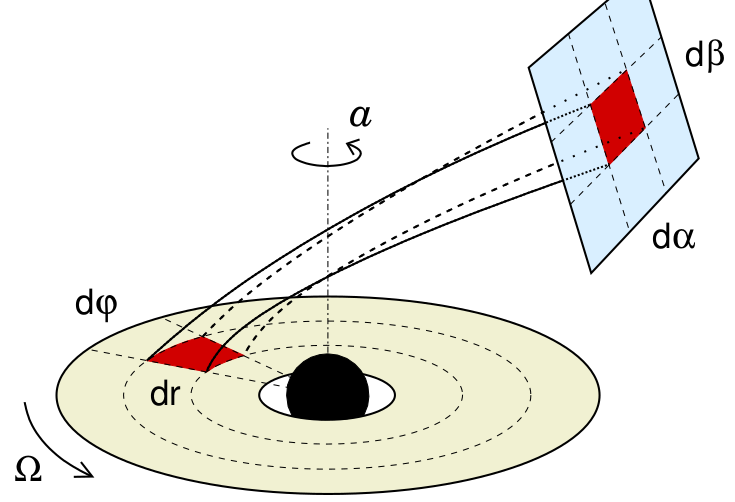}
		\caption{\footnotesize{Illustration of the transformation from local frame of reference, comoving with the accretion disc to the image on the sky. We define the disc area element typically in the Boyer-Lindquist coordinates $r$ and $\phi$ and the projected area that the observer sees in the impact parameters $\alpha$ and $\beta$. Image courtesy of Michal Dovčiak.}}
		\label{transfer_function}
	\end{minipage}
\end{figure}

The polarized radiative transfer calculations used in this dissertation are departing from the fundamental works of \cite{Cunningham1973, Cunningham1975, Connors1977, Connors1980}, which we summarize here below. Photon trajectories follow the geodesic equation
\begin{equation}\label{geodesic}
    \dfrac{dp^{\mu}}{d\lambda'}+\Gamma_{\sigma \tau}^{\mu} p^{\sigma} p^{\tau} = 0 \textrm{ ,}
\end{equation}
where $p^{\mu} = \frac{dx^{\mu}}{d\lambda'}$ is a four-momentum of light, $\lambda'$ is affine parameter of a~geodesic, and $\Gamma_{\mu \nu}^{\sigma}$ are Christoffel symbols for the Kerr metric. Brandon Carter (\text{*}1942) derived constants of motion that fully describe any test particle motion outside of~the~Kerr black hole \citep{Carter1968}. The lensing effects are captured by the~equation of geodesic deviation
\begin{equation}\label{deviation}
    \dfrac{d^2 Y_j^{\mu}}{d\lambda'^2}+2\Gamma_{\sigma \gamma}^{\mu} p^{\sigma} \dfrac{dY_j^{\gamma}}{d\lambda'} + \Gamma_{\sigma \tau , \gamma}^{\mu} p^{\sigma} p^{\tau} Y_j^{\gamma} = 0 \textrm{ , } j = 1,2 \textrm{ ,}
\end{equation}
where $Y_j^{\mu}$ are two vectors characterizing the distance between nearby geodesics. These equations are to be numerically integrated over the journey of a photon, in~order to obtain a distorted picture that the observer will obtain, since the~trajectories and energies of photons deviate from the classical Newtonian framework. All SR and GR effects are then automatically included, namely the~kinetic and gravitational frequency shifts, relativistic aberration, the emission direction changes, gravitational lensing and bending effects and relative time delays. The~relativistic effects on broadening of spectral lines are illustrated in~Figure \ref{line_broad} for~ratios of~observed and emitted radiation frequencies $\nu_\textrm{obs}$ and $\nu_\textrm{em}$, respectively. It serves as a tool for black-hole spin spectral fitting from X-ray radiation reflected from accretion disc, typically performed on the strong iron line complex at 6--7~keV, alongside spectral fits of the underlying continuum that also varies with black-hole spin.

~\

To obtain $F_\textrm{E}^{\Omega_0}$ at a distant detector that sees, as a point-source, an emitting accretion disc at inclination angle $\theta$ (coinciding with the Boyer-Lindquist coordinate $\theta$), measured from the rotational axis of the disc, with local intensity $I$, we integrate
\begin{equation}\label{cunningham}
    F_\textrm{E}^{\Omega_0} = \int_{r_\textrm{in}}^{r_\textrm{out}} \int_{g_\textrm{min}}^{g_\textrm{max}} g^3 I f \textrm{ d}g \textrm{d}r
\end{equation}
over the area of the disc with radial coordinate $r$ extending from $r_\textrm{in}$ in the outer vicinity of the black-hole horizon to arbitrary $r_\textrm{out}$. We only aim to cover one side of the disc and possible photons arriving from the far side of the disc due to non-zero curvature are neglected. And in most of the results, we also neglect reflections of photons that escape the disc and due to strong gravity arrive back to the disc (the returning radiation), i.e. neither to~the~event horizon, neither to~the~spatial infinity. Keplerian disc rotation is assumed, which is not needed for~the~local emission itself, but the assumed velocities $U^{(\phi)} \sim r^{-3/2}$ affect the~shifts upon locally emitted radiation. In (\ref{cunningham}), $f$ is the transfer function that represents the~Jacobian of the transformation from the detector's viewing position -- typically calculated in impact coordinates \citep{Cunningham1973} $\alpha$~and~$\beta$ measured perpendicular to and parallel with the spin axis of~the~black hole projected onto the observer’s sky, respectively -- towards the~disc area in~the~integration coordinates, here $g$ (see below) and $r$. The transfer function then includes all of~the~effects above upon the photon's journey and acts as an amplifier or damper of different disc sections. We define $n_\textrm{em}$ as the local emission angle measured from the principal system axis and $g=E_\textrm{obs}/E_\textrm{em}$ as the~energy shift from the emitted photon energy $E_\textrm{em}$ to the observed one $E_\textrm{obs}$. The energy shift is directly related to the Boyer-Lindquist coordinate $\phi$, being the azimuthal coordinate over the~disc. Therefore, $g$ may also be used as a distinct azimuthal coordinate. Both of~these have their minima and maxima, $g_\textrm{min}$, $\phi_\textrm{min}$ and $g_\textrm{max}$, $\phi_\textrm{max}$ respectively, for $g$ depending on the emission radius studied.

\begin{figure}[h]\centering
	\includegraphics[width=\textwidth]{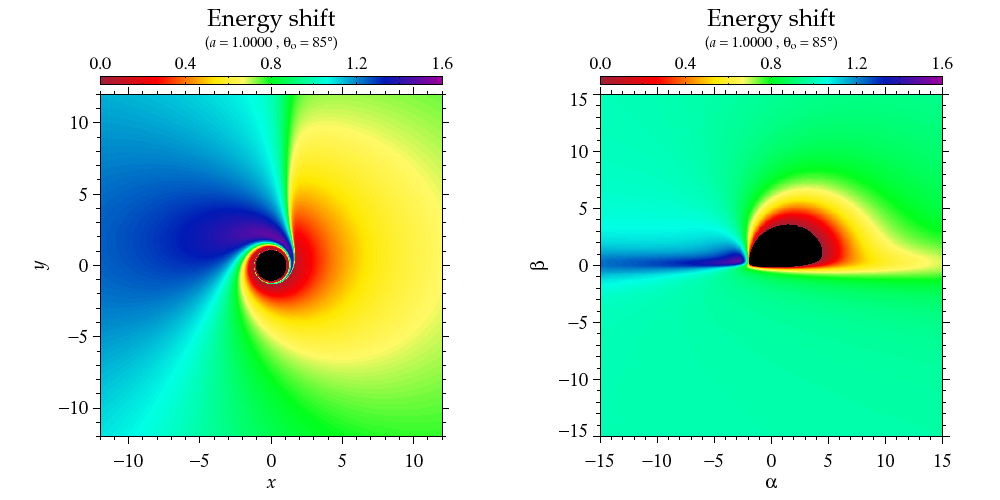}
	\caption{\footnotesize{The energy shift, $g$, in Kerr spacetime for the case of $a = 1$ and $\theta = 85^\circ$ in~the~equatorial $x$ and $y$ coordinates (left) and the impact parameters $\alpha$ and $\beta$ (right). Image courtesy of~Michal Dovčiak.}}
	\label{energy_shifts}
\end{figure}

Equation (\ref{cunningham}) may be rewritten in terms of observed photon flux $N_\textrm{E}^{\Omega_0}$ in~an~equally useful way \citep[see e.g.][]{Dovciak2004a}:
\begin{equation}\label{dov2.9}
	N_\textrm{E}^{\Omega_0} = N_0 \int_{r_\textrm{in}}^{r_\textrm{out}} \int_{\phi_\textrm{min}}^{\phi_\textrm{max}} I_\textrm{N} g \cos(n_\textrm{em}) \bar{l} r \textrm{ } \textrm{d}\phi \textrm{d} r \textrm{ ,}
\end{equation}
where $N_0$ is the normalization constant, $I_\textrm{N}=I/h_\textrm{P}\nu_\textrm{em}$ is the local photon intensity at $\nu_\textrm{em}$, $\bar{l}$ is the so-called lensing factor for far-away sources obtained from (\ref{deviation}), and the combination of parameters $g^2\cos(n_\textrm{em}) \bar{l}$ is an analogue of~the~transfer function $f$ from the former case. The transformation from local disc coordinates to the image that a distant observer would see is illustrated in Figure \ref{transfer_function}. The energy shift of a photon for an example of $a = 1$ and $\theta = 85^\circ$ in~the~equatorial $x = \sqrt{r^2+a^2}\cos{\phi}$ and $y=\sqrt{r^2+a^2}\sin{\phi}$ Cartesian coordinates and~the~impact parameters $\alpha$ and $\beta$ on the plane of the sky is shown in Figure \ref{energy_shifts}. For other examples and maps of the GR emission vector change, or the transfer function itself, see \cite{Dovciak2004a}.

~\

The previous paragraph developed for Stokes $I$ is equally valid for other Stokes parameters. It is, however, the polarisation angle that continuously rotates along a geodesic climbing up the potential well, which is apparent from parallel transport of the polarisation vector, residing in the polarisation plane perpendicular to the photon momentum. Then, for a distant observer, the total polarisation degree is also altered, if GR effects are included. We typically expect relativistic depolarisation on the total signal, because individual light rays are affected differently, although that is not always the case and in rare cases the polarisation can be amplified \citep{Dovciak2011}. We provide the integration formulae for~all Stokes parameters at once
\begin{equation}
\begin{aligned}\label{totalpol}
\begin{split}
\Delta I_\textrm{obs}(E_\textrm{obs},\Delta E_\textrm{obs}) &= N_0 \iint I r \cos(n_\textrm{em}) g^2 \bar{l}  \textrm{ } \textrm{d}S \textrm{ ,}\\
\Delta Q_\textrm{obs}(E_\textrm{obs},\Delta E_\textrm{obs}) &= N_0 \iint [Q \cos(2\Delta\Psi) \\&\quad - U \sin(2\Delta\Psi)] r \cos(n_\textrm{em}) g^2 \bar{l}  \textrm{ } \textrm{d}S  \textrm{ ,}\\
\Delta U_\textrm{obs}(E_\textrm{obs},\Delta E_\textrm{obs}) &= N_0 \iint [Q \sin(2\Delta\Psi) \\&\quad + U \cos(2\Delta\Psi)] r \cos(n_\textrm{em}) g^2 \bar{l}  \textrm{ } \textrm{d}S   \textrm{ ,}
\\
\Delta V_\textrm{obs}(E_\textrm{obs},\Delta E_\textrm{obs}) &= N_0 \iint  V r \cos(n_\textrm{em}) g^2 \bar{l}  \textrm{ } \textrm{d}S    \textrm{ .}
\end{split}
\end{aligned}
\end{equation}
where $\textrm{d}S$ denotes the planar integration across the disc in any coordinates and~$\Delta \Psi$ is the total change of the linear polarisation angle between the emission point (here at the disc) and the receiving point (here at the detector located at spatial infinity). In numerical codes that we will operate with, it is useful to also integrate the resulting photon flux $N_\textrm{E}^{\Omega_0}$ and analogic quantities for $Q$, $U$, $V$ over the final energy bin with size $\Delta E_\textrm{obs}$ at $E_\textrm{obs}$, in order to obtain the~Stokes parameters $\Delta I_\textrm{obs} (E_\textrm{obs}, \Delta E_\textrm{obs})$, $\Delta Q_\textrm{obs} (E_\textrm{obs}, \Delta E_\textrm{obs})$, $\Delta U_\textrm{obs} (E_\textrm{obs}, \Delta E_\textrm{obs})$, $\Delta V_\textrm{obs} (E_\textrm{obs}, \Delta E_\textrm{obs})$ in the units of $[\textrm{counts} \cdot \textrm{sr}^{-1} \cdot \textrm{s}^{-1}]$, which set the number of~photons measured per~energy bin (and inside the code we would set a distance to obtain results per detector unit area). Again, we will assume $\Delta V_\textrm{obs} = 0$ in~this work. We will often drop the ``obs'' and ``em'' subscripts when obvious from the~context.

~\

The acquisition of $\Delta \Psi$ between two spatial points is analytically possible in~many cases, unlike the solutions to the geodesic equation and equation of~geodetic deviation that are typically numerically integrated. This is because of~the~Walker-Penrose theorem \citep{Walker1970} that is derived in~the~Newman-Penrose formalism of GR \citep{Newman1962, Newman1963}, invoking complex calculus. A complex constant of motion $\kappa = \kappa_2 - i\kappa_1$, the~Walker-Penrose constant, can be expressed for the Kerr metric in Boyer-Lindquist coordinates:
\begin{equation}\label{KS}
\begin{split}
    \kappa = \frac{1}{r+ia\cos{\theta}} &  \{(r^2+a^2)(p_rf_t-p_tf_r) +a(p_rf_{\varphi}-p_{\varphi}f_r) \\
    &\quad + \frac{i}{\sin{\theta}}[p_\theta f_\varphi - p_\varphi f_\theta - a\sin^2{\!\theta}\,(p_t f_\theta - p_\theta f_t) ] \} \, ,
\end{split}
\end{equation}
where $f^\mu$ is the polarisation 4-vector parallelly transported along a null geodesic. Hence, in order to find the polarisation angle change, i.e. to find three components of the polarisation 4-vector $f^\mu$ in the local reference frame, these three unknown components can be expressed {\it at the receiving point} by means of~the~$\kappa_1$ and $\kappa_2$ quantities, while $\kappa_1$ and $\kappa_2$ can be expressed by means of~the~initial $p^\mu$ and~$f_0^\mu$ of~each photon {\it at the emission point} due to the conservation of~$\kappa$. As~last step, we construct the local polarisation angle $\Psi$ according to its definition at~the~receiving point and transform to this frame of reference. The GR effects on~polarisation near black holes have been studied extensively since the~pioneering 1970s and~1980s research, and the dependency of this effect on inclination, black-hole spin and other fundamental parameters enables independent fitting method via X-ray spectropolarimetry \citep[e.g.][]{Laor1990, Matt1993b, Agol2000, Dovciak2004, Dovciak2008, Schnittman2009, Schnittman2010}.

\section{Astronomical X-ray polarimetry}\label{all_polarimetry}

Before we introduce the astrophysical objects of study and estimate the efficiency of X-ray polarisation diagnostics, let us devote a few words to the current status of X-ray polarimetry in astronomy. The information provided in this section is selected from \cite{Fabiani2014, Kislat2015, DiLalla2019, Weisskopf2022, Muleri2022}, to which we also refer for further details.

\subsection{Detectors}\label{polarimetry}

In the literature, we find three classes of astronomical X-ray polarimeters: Bragg diffraction-based polarimeters, scattering-based polarimeters, and polarimeters on the basis of the photoelectric effect. The last category includes gas pixel detectors (GPD) and time projection chambers (TPC) that are most frequently used in~contemporary research. The \textit{IXPE} mission contains a GPD that traces a photo-excited electron path in a gas chamber. The direction of~emission of~the~photoelectron is dependent on the differential cross-section of the interaction, which is correlated to the absorbed photon's electric vector direction. The projected path of the electron in the sensitive volume is then analyzed using sophisticated methods to provide information on the initial linear polarisation state of each photon. Here we will provide just the basics of such analysis without technological details.

~\

\begin{figure}[h]\centering
	\includegraphics[trim={0cm 2.5cm 0cm 3cm},clip,width=\textwidth]{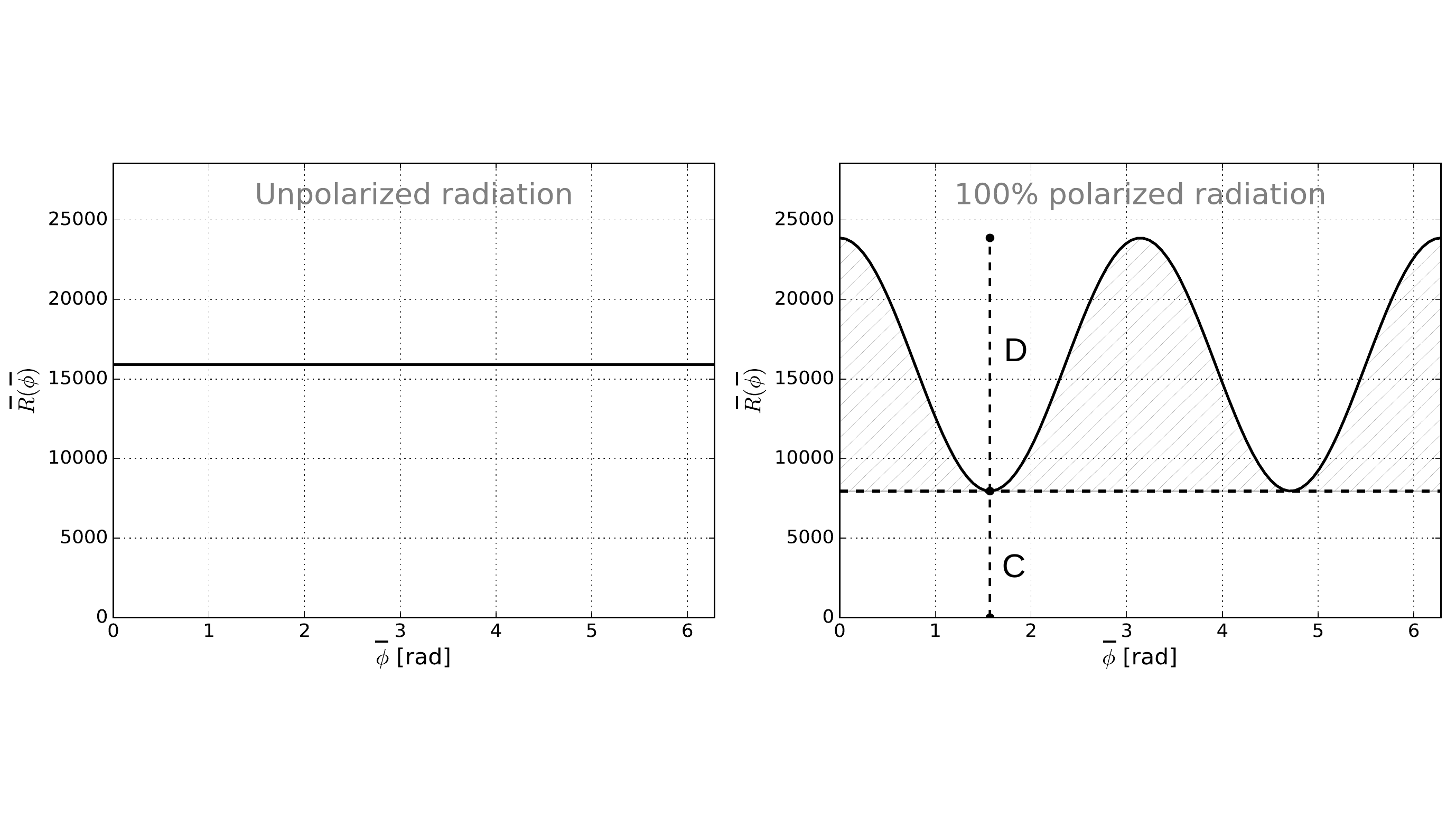}
	\caption{\footnotesize{Response curves from a fictitious linear polarimeter for fully unpolarized radiation (left) and fully polarized radiation (right). Image adapted from \cite{DiLalla2019}.}}
	\label{modulation}
\end{figure}

Any linear polarimeter measures the azimuthal modulation of some physical quantity $\bar{R}(\bar{\phi})$, around some direction of polarisation $\bar{\phi}_0$ in the polarisation plane. The response of a polarimeter can be characterized by two generic parameters $C$ and $D$:
\begin{equation}\label{response}
    \bar{R}(\bar{\phi}) = C + D\cos^2(\bar{\phi}-\bar{\phi}_0) \textrm{ .}
\end{equation}
Without systematic effects, unpolarized radiation results in statistically flat histogram of the response curve $\bar{R}(\bar{\phi})$ and $100\%$ polarized signal produces strongly modulated histogram of the response curve with some amplitude $D/2$. Figure \ref{modulation} illustrates the two extreme cases. For real polarimeters $C \neq 0$ and we introduce energy-dependent modulation factor $\mu$:
\begin{equation}
    \mu = \left. \dfrac{\bar{R}_\textrm{max}-\bar{R}_\textrm{min}}{\bar{R}_\textrm{max}+\bar{R}_\textrm{min}}\right|_{p=100\%} = \left. \dfrac{D}{D+2C}\right|_{p=100\%} \textrm{ ,}
\end{equation}
where $\bar{R}_\textrm{max}$ and $\bar{R}_\textrm{min}$ denote the maxima and minima of the response curve $\bar{R}(\bar{\phi})$, respectively. The modulation factor directly corresponds to the sensitivity of~a~given polarimeter. Ideal polarimeter with complete modulation means that $\mu = 1$. For good enough photon statistics, we can obtain polarisation degree from the measured modulation amplitude:
\begin{equation}\label{from_amplitude}
    p = \dfrac{1}{\mu} \dfrac{\bar{R}_\textrm{max}-\bar{R}_\textrm{min}}{\bar{R}_\textrm{max}+\bar{R}_\textrm{min}} \textrm{ .}
\end{equation}

~\

In case of low number of photons, which is the usual case for astronomical X-ray polarimetry, the polarisation fraction is, however, best obtained through Stokes parameters in the so-called event-by-event calculation \citep[see][for details and comparisons with the approach of fitting the modulation curve]{Muleri2022}. This approach also allows to use weights\footnote{\,The quality of a single photon track in a GPD in {\it IXPE} depends on many factors, e.g. inclination of the track with respect to the collecting plane, or due to Coulomb interactions near the generation point. The algorithm presented in \cite{DiMarco2022} assigns weights to each photo-electron track, given by the accuracy of the reconstruction of their path.} for photon trajectories in GPDs, which raises the overall sensitivity and reduces statistical errors \citep{DiMarco2022}. We define:
\begin{equation}
\begin{aligned}
i_k & = 1 \textrm{ ,}\\
q_k & = \cos(2\bar{\phi}_k) \textrm{ ,} \\
u_k & = \sin(2\bar{\phi}_k) \textrm{ ,}
\end{aligned}
\end{equation}
for each photon arrival (indexed by $k$) at the instrument-defined direction $\bar{\phi}_k$. The linearity of Stokes parameters allows to sum:
\begin{equation}
\begin{aligned}
\tilde{I}_\textrm{N} & = \dfrac{1}{T_\textrm{obs}}\sum_{k=1}^N i_k \textrm{ ,}\\
\tilde{Q}_\textrm{N} & = \dfrac{1}{T_\textrm{obs}}\sum_{k=1}^N q_k \textrm{ ,} \\
\tilde{U}_\textrm{N} & = \dfrac{1}{T_\textrm{obs}}\sum_{k=1}^N u_k \textrm{ ,}
\end{aligned}
\end{equation}
for a measurement in $[\textrm{counts} \cdot \textrm{s}^{-1}]$ of $N$ events per energy band over the exposition time $T_\textrm{obs}$ (here the Stokes parameters are not in the units per unit area, which is the reason for the tilde notation). Then Equations (\ref{response}) and (\ref{from_amplitude}) provide expressions for measured polarisation degree $p$ and polarisation angle $\Psi$
\begin{equation}
\begin{aligned}
p &= \dfrac{2}{\mu} \dfrac{\sqrt{\tilde{Q}_\textrm{N}^2+\tilde{U}_\textrm{N}^2}}{\tilde{I}_\textrm{N}} \textrm{ ,}\\
\Psi &= \dfrac{1}{2}\textrm{\space}\textrm{arctan}_2\left(\dfrac{\tilde{U}_\textrm{N}}{\tilde{Q}_\textrm{N}}\right) \, .
\end{aligned}
\end{equation}
One can equivalently define $\mu$-normalized Stokes parameters $Q$ and $U$ to account for the instrumental quality.

~\

Even for unpolarized photons, real polarimeters detect non-zero modulation amplitude due to Poisson-distributed statistical fluctuations. To quantify the~maximum polarisation, which can be produced by stochastic noise only (in~the~absence of true source polarisation), at a certain $X\%$ confidence level, we use the minimum detectable polarisation function $M\!D\!P_{X\%}$. It can be derived that for $99\%$ confidence level and in case of non-negligible background \citep{Elsner2012}:
\begin{equation}\label{MDP}
    M\!D\!P_{99\%} = \dfrac{4.29}{\mu}\dfrac{\sqrt{N_\textrm{S}+N_\textrm{B}}}{N_\textrm{S}} \textrm{ ,}
\end{equation}
where $N_\textrm{S}$ is the number of events received from the actual source, $N_\textrm{B}$ is the~number of events from the background contribution to the measurement. Thus, the~criterion of observed polarisation with $3\sigma$ statistical uncertainty above zero is not enough for an objectively credible polarisation detection. In addition, the~observed polarisation degree value is required to appear above $M\!D\!P_{99\%}$ to claim a statistically significant detection, according to the \textit{IXPE} standards \citep{Weisskopf2022}. Of course, both requirements are interdependent, as both depend on the modulation factor and number of collected counts.

~\

Multiple methods in obtaining statistical uncertainties for polarisation quantities exist. Although it is often more convenient to plot the polarisation degree and angle, the independence is peculiar to the Stokes parameters, which are typically assumed to be normally distributed (true when $\mu p \ll 1$). See \cite{Muleri2022} for derivation of the ``combined'' uncertainties for $p$ and $\Psi$, which is in many cases more appropriate than the classical 1-dimensional uncertainty. The systematic errors include the modulation factor $\mu$. A notable systematic in GPDs is spurious modulation.

A standard forward folding of models with instrumental response matrices is necessary for e.g. the $\chi^2$ analysis in spectral fitting software. The multiplication consists of matrix representing the effective mirror area, the redistribution matrix and for the Stokes parameters $Q$ and $U$ additionally the modulation factor. For example, \textit{IXPE} has its own simulation and analysis software developed: {\tt IXPEOBSSIM}\footnote{\,Available at \url{https://ixpeobssim.readthedocs.io/en/latest} on the day of submission, including user documentation.} \citep{Baldini2022}. The code serves for estimation of observational times, anticipation of measurement uncertainties, and trial of fitting procedures for sources yet to be observed with up-to-date response matrices, as well as for~processing of collected observations. Real data are often in addition analyzed with the \texttt{XSPEC}\footnote{\,Available at \url{https://heasarc.gsfc.nasa.gov/xanadu/xspec/}  on the day of submission, including user documentation.} software \citep{Arnaud1996}, a fitting tool commonly used in X-ray spectroscopy, which recently expanded to polarisation analysis (version \textit{12.13.0} 
 and onwards). Even an upper limit on polarisation may significantly reduce the theoretical parameter space and be of valuable insight on the source.

\subsection{Space observatories}\label{space_observatiories}

Due to absorption in the Earth's atmosphere, X-rays require space-based telescopes. But since the first measurement of cosmic X-rays in 1962 \citep{Giacconi1962}, X-ray astronomy has become one of the leading astronomy fields. The~sensitivity of X-ray instruments increased dramatically in the 1990s, for example with the launch of 
\begin{itemize}
    \item \textit{ROSAT} satellite \citep[0.1--2 keV, Germany, UK, USA, operating from 1990 to~1999,][]{Pfeffermann1987}.
\end{itemize}
Today, there are high-quality X-ray images and spectra from e.g.
\begin{itemize}
    \item \textit{Chandra} \citep[0.07--10 keV, SAO, NASA, launched
1999,][]{Pavlov2000},
    \item \textit{XMM-Newton} \citep[0.1--15 keV, ESA, launched 1999,][]{Jansen2001},
    \item NICER \citep[0.2--12 keV, NASA, launched 2017,][]{Gendreau2012},
    \item \textit{NuSTAR} \citep[3--79 keV, NASA, launched 2012,][]{Harrison2013},
\end{itemize}
among others. The X-ray community also benefits from all-sky monitoring campaigns by 
\begin{itemize}
    \item \textit{MAXI} \citep[0.5--30 keV, JAXA, launched 2009,][]{Matsuoka2009},
    \item \textit{Swift}/BAT \citep[15--50 keV, NASA, launched 2004,][]{Krimm2013}.
\end{itemize}
Ambitious projects are being launched and developed, such as 
\begin{itemize}
    \item \textit{XRISM} \citep[0.3--12 keV, JAXA, NASA, launched 2023,][]{xrism2020},
    \item new \textit{ATHENA} \citep[0.1--12 keV, ESA,][]{Barret2020}.
\end{itemize}
All of these are critical for simultaneous analysis of X-ray polarimetric observations.

~\

Although polarimetry from radio to optical bands has a long history, high-energy polarimetry faced number of technical difficulties alongside faint nature of the sources of interest. The pioneering instruments on board of 
\begin{itemize}
    \item \textit{OSO-8} \citep[2.4--2.8 keV and 4.8--5.6 keV, NASA, operating from 1975 to 1978][]{Novick1975, Weisskopf1976b}
\end{itemize}
in 1976 and 1977 confirmed X-ray polarisation from the Crab nebula \citep{Weisskopf1978}. And despite some success in the $\gamma$-rays and very hard X-rays on~board of 
\begin{itemize}
    \item \textit{INTEGRAL}/SPI \citep[130--8000 keV, ESA, launched 2002,][]{Vedrenne2003, Roques2003},
\end{itemize}
it took about four decades to launch the first X-ray polarimeters, when the GPDs or TPCs were developed, which improved the sensitivity from previous polarimeters by at least two orders of magnitude in $\mu$ \citep{Fabiani2014}. We name
\begin{itemize}
    \item \textit{X-Calibur} \citep[15--60 keV, NASA,][]{Krawczynski2011, Guo2013, Beilicke2014, Kislat2015, Kislat2017},
\end{itemize}
a balloon experiment already launched in 2014, 2016, 2018--2019 along with its successor the \begin{itemize}
    \item \textit{XL-Calibur} \citep[15--80 keV, USA, Japan, Sweden,][]{Abarr2021},
\end{itemize}
launched in May 2022 and planned to be relaunched. Then the forthcoming projects
\begin{itemize}
    \item \textit{REDSoX} \citep[0.2--0.8 keV, USA,][]{Marshall2018},
    \item \textit{XPP} \citep[0.2--60 keV, NASA, Europe,][]{Jahoda2019},
    \item \textit{XPoSat} \citep[8--30 keV, ISRO,][]{Paul2022},
    \item \textit{eXTP} \citep[2--12 keV, China, Europe,][]{Zhang2019}.
\end{itemize}
And most importantly the \textit{IXPE} mission, which lead to a significant breakthrough in the 2--8 keV band and operates successfully since its launch in December 2021. The \textit{XL-Calibur}'s launch faced technical difficulties and the \textit{X-Calibur} flights did not obtain a positive polarisation detection \citep[private communication with the~team and][]{Abarr2020}. Thus, noting also the \textit{INTEGRAL}/SPI observation of~Cygnus X-1 \citep[$76\% \pm 15\%$ at $230$--$850$ keV,][]{Jourdain2012} with debatable instrumental cross-validation with well-understood Crab nebula \citep{Forot2008}, \textit{IXPE} is the first X-ray polarimeter to begin bringing high-precision detections on weekly basis. Because the \textit{IXPE} observations of accreting black holes make large part of the dissertation, we will introduce this instrument in~more depth in the rest of this section. We refer to the recent reviews by \citet{Fabiani2018, Chattopadhyay2021} for other astronomical X-ray polarimeteres that are being developed.

~\

\begin{figure}[h]
	\centering
	\begin{minipage}[t]{0.49\textwidth}
		\includegraphics[width=\textwidth]{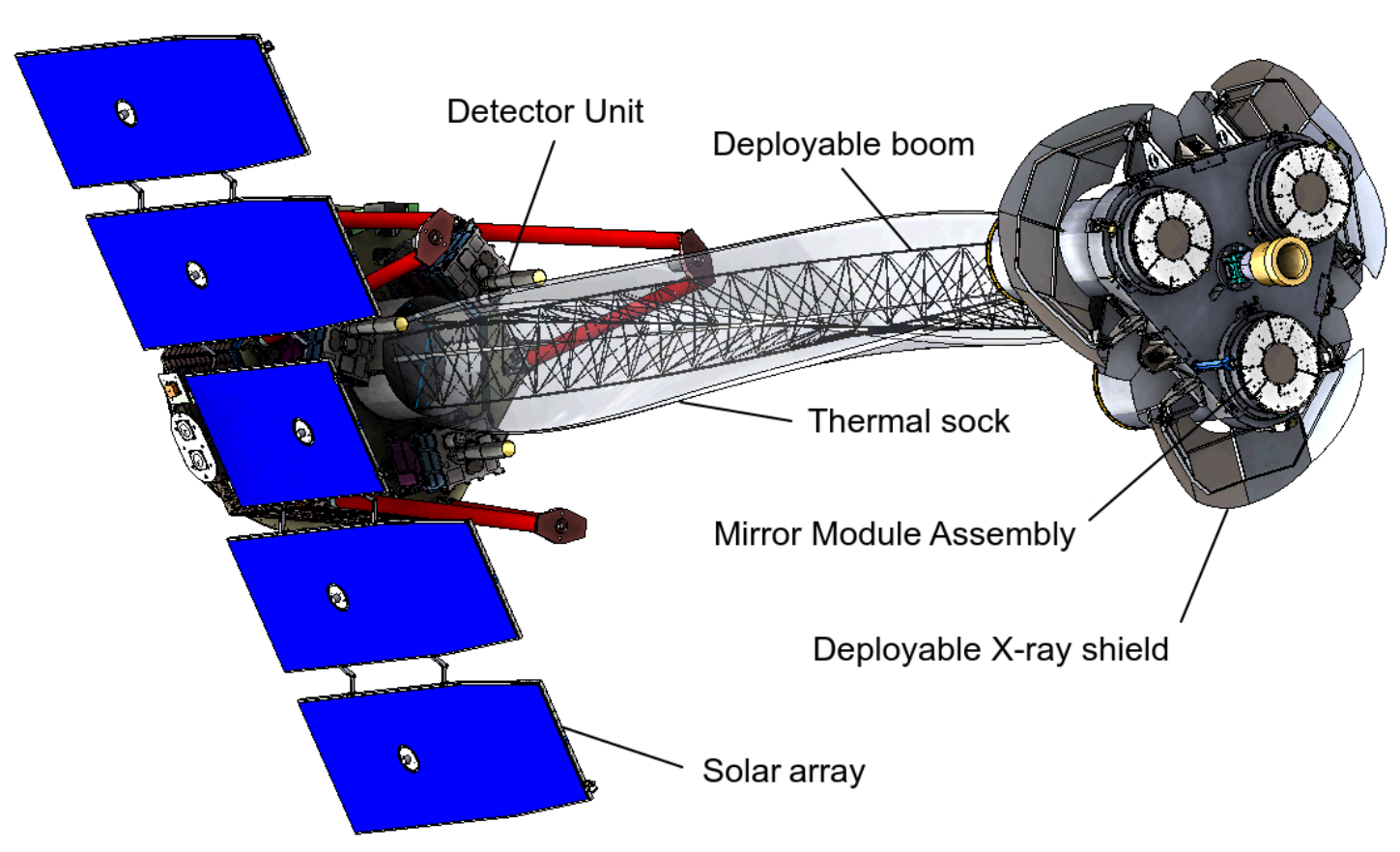}
		\caption{\footnotesize{Illustration of the \textit{IXPE} space observatory with the main instrumental parts described. Image from \cite{DiLalla2019}.}}
		\label{IXPEsketch}
	\end{minipage}
	\hfill
	\begin{minipage}[t]{0.49\textwidth}
		\includegraphics[width=\textwidth]{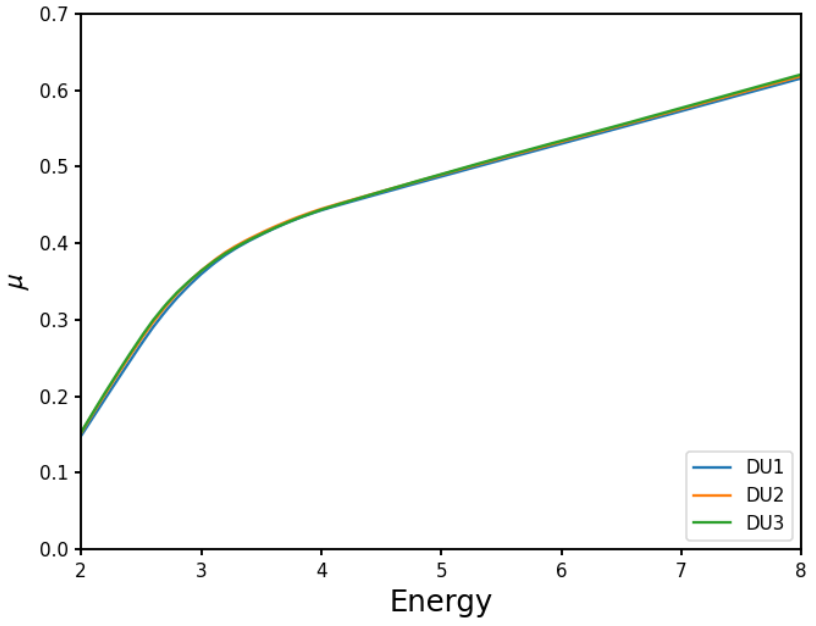}
		\caption{\footnotesize{The GPD modulation factor $\mu(E)$ of DUs on board of \textit{IXPE} at the time of launch. Image from \cite{Weisskopf2022}.}}
		\label{GPDmodulation}
	\end{minipage}
\end{figure}

\textit{IXPE} operates in point-and-stare regime for selected targets. Due to~the~recent technological advancement with GPDs, it surpasses the polarimetric sensitivity of \textit{OSO-8} by two orders of magnitude \citep{DiLalla2019}. \textit{IXPE} hosts three identical aligned telescopes on board. Each operates independently and consists of a Mirror Module Assembly (MMA) and a~ Detector Unit (DU). The~MMA is followed by a 4-m long focal-length in a~boom, and the~DUs have, apart from the~GPD, a~Filter Wheel Assembly, which is necessary for in-flight calibration and~source and~background attenuation and~assessments \citep{Weisskopf2018}. The~whole payload is illustrated in Figure \ref{IXPEsketch}. The~mirror effective area at single MMA is $\approx 240 \textrm{ cm}^2$ at $4.5 \textrm{ keV}$. Angular resolution is $< 30 \textrm{ arcsec}$, hence all Galactic and~extragalactic sources considered in this dissertation are observed as point-sources and theoretically studied for a distant observer as such. The overlapping field from all 3 DUs is $10 \textrm{ arcmin}$ in diameter. Timing accuracy is $\approx 20 \ \mu\textrm{s}$.  The \textit{IXPE} energy resolution is $23\%$ full width at half maximum at $2.6 \textrm{ keV}$, scaling as $1/\sqrt{E}$. The polarisation sensitivity can be illustrated by $M\!D\!P_{99\%} \lessapprox 5.5\%$ for~an exposure time of 10 days at $0.5 \textrm{ mCrab}$ source.\footnote{\,$\textrm{ Crab}$ is a flux unit designed for measuring X-ray astronomical sources and corresponds to the flux from a Crab nebula at a certain energy band \citep{Kirsch2005}. In the standard 2--10 keV range, the flux $F_\textrm{X,2--10} = 1 \, \textrm{Crab} = 2.4 \times 10^{-8} \, \textrm{erg} \cdot \textrm{cm}^{-2} \cdot \textrm{s}^{-1}$.} To be more precise, the~measured modulation factor $\mu$ with energy of each DU is depicted in Figure \ref{GPDmodulation}.

~\

We will elaborate on \textit{IXPE} observations as part of the results in this dissertation, but let us provide one more detailed example of its capabilities. One may assume the average count rate for the \textit{IXPE} observation of the X-ray bright accreting supermassive black hole in MCG-05-23-16 \citep[$(0.525 \pm 0.002) \, \textrm{counts}\cdot \textrm{s}^{-1}$ in 2--8 keV,][]{Marinucci2022}, where to obtain $M\!D\!P_{99\%} = 4.7\%$ took 482 ks. If compared with the X-ray bright accreting stellar-mass black hole in Cygnus X-1 \citep[$(40 \pm 20) \, \textrm{counts}\cdot \textrm{s}^{-1}$ in 2--8 keV,][]{Krawczynski2022}, then, if similar observational time was given for Cygnus X-1, the same confidence limit for~linear polarisation detection would be reduced to $M\!D\!P_{99\%} \approx 0.5\%$ for Cygnus X-1. These are extraordinary numbers, because polarimetry is in general photon-demanding when compared to spectroscopy or photometry. If the emission is e.g. $1\%$ polarized, we then require about 100 times more photons to process some information from the signal than a spectrograph counting each photon per the~same band width would need, in order to be, generally speaking, as informative.

\subsection{Science cases}\label{cases}

X-rays are produced by vast number of sources in the Universe, such as planets, stars, ISM, stellar remnants, galaxies, galaxy clusters and gamma ray bursts (GRB) \citep{Trumper2008}. Out of these, X-ray polarimeters typically target objects with powerful acceleration mechanism, strong magnetic fields, possible presence of vacuum birefringence, strong gravity, and scatterings in aspherical geometries. During the first year and a half of operation, \textit{IXPE} produced a number of interesting results when observing supernova remnants \citep{Vink2022, Ferrazzoli2023, Zhou2023}, pulsar wind nebulae \citep{Xie2022,Bucciantini2023,Romani2023}, the Sagittarius A* complex \citep{Marin2023}, GRBs \citep{Negro2023}, magnetars \citep{Taverna2022,Turolla2023,Zane2023}, accreting pulsars \citep{Doroshenko2022,Marshall2022,Tsygankov2022,Doroshenko2023,Forsblom2023,Malacaria2023,Mushtukov2023,Suleimanov2023b,Tsygankov2023} and weakly magnetized accreting neutron stars \citep{Capitanio2023,Chatterjee2023,Cocchi2023,Fabiani2023,Farinelli2023,Ursini2023b}. The first \textit{IXPE} observational results of accreting black holes will be summarized in Chapter \ref{chap05}.

\section{Active galactic nuclei}\label{AGN_intro}

From cosmological perspective, only a fraction of all galactic nuclei are active \citep[$\lessapprox 40\%$, but the selection criteria are key, see e.g. ][]{Chiang2019}, which means that there is a long-term, ongoing mass transfer from the surroundings onto the central supermassive black hole. Our Galaxy, for example, shows a~very low activity at this epoch, but perhaps was strongly accreting in the past \citep{Revnivtsev2004, Nobukawa2011, Chuard2018, Marin2023}. The~extreme power density, luminosity at all wavelengths, and rapid time variability of the active cores of galaxies caused a confusion in the 1950s on their origin, being first classified as ``quasi-stellar astronomical radio-sources'', shortened as  ``quasars'' \citep{Chiu1964}, by which we mean today a point-like UV or~optical emitting object associated with a radio-source. With the development of~radio astronomy, spectroscopy and interferometry in the 1960s, their true nature was revealed. Since 1970s and the development of gravitational lensing and high-resolution spectroscopy people are breaking records in finding active galactic nuclei (AGN) at~higher and higher cosmological redshifts. At the time of writing of~this dissertation, the furthest and oldest quasar is at redshift $z = 7.642$, i.e.~$\approx 670$~million years after the Big Bang, which challenges the structure formation theories within Big Bang cosmologies \citep{Wang2021}. Today we know that AGNs have diverse nature and dozens of classifications based on appearance at~all wavelengths exist \citep[see e.g.][]{Pringle1972, Shakura1973, Seward2010, Netzer2015}.

~\

If the host galaxy is identified, such galaxies with active nuclei are called ``Seyfert galaxies''. They are named after Carl Seyfert (1911--1960), who first described this class already in 1943 \citep{Seyfert1943}. Probably the most useful distinction comes from the presence or absence of broad optical emission spectral lines in~the~cores of Seyfert galaxies, defined as ``type-1'' or ``type-2'' AGNs, respectively \citep{RowanRobinson1977, Keel1980, Antonucci1993}. A~major breakthrough in~understanding that these are actually one and the same axially symmetric objects, viewed under different inclination with respect to~the~axis of symmetry, came (not surprisingly) from polarimetry mostly in the 1980s \citep{Dibai1966, Walker1966, Angel1976, Antonucci1982, Antonucci1984, Antonucci1985}. The so-called unification model \citep{Antonucci1993, Urry1995} assumes a nucleus with high orbital speeds of~the~infalling matter in equatorial disc structures, causing Doppler broadening of~the~spectral lines, that can be viewed bare from polar directions, but obscured from the sides. The cause of~obscuration is accounted to a parsec-scale cold dusty torus residing in the equatorial plane. Infrared and X-ray observations suggest clumpy structure \citep{Nenkova2002, Risaliti2002, Matt2003, Tristram2007} and elaborate spectral fitting shows large-scale geometries of the obscurer sometimes far from an actual geometrical torus \citep{Fritz2006, Suganuma2006, Namekata2014, Siebenmorgen2015, Buchner2019, Ricci2023}. Although the size, shape, clumpiness and~location of the obscuring material varies and is difficult to theoretically predict from the first principles, we know that its presence is related to~the~active state of inner black-hole accretion \citep{Hopkins2012a, Schawinski2015}.

~\

\begin{figure}[!htb]
	\centering
	\begin{minipage}[t]{0.49\textwidth}
		\includegraphics[width=\textwidth]{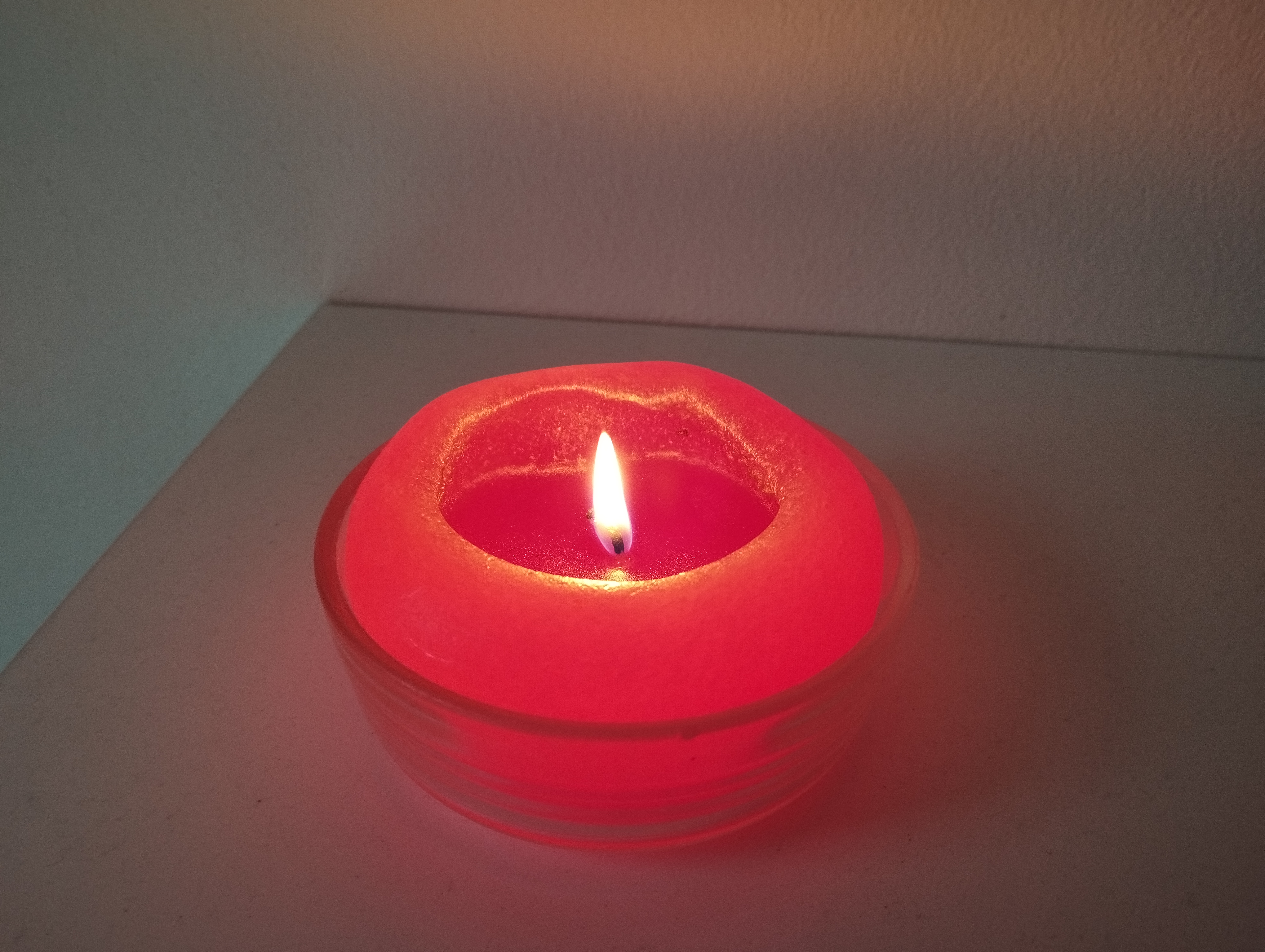}
		\caption{\footnotesize{Photograph of a candle illustrating a type-1 view of an AGN. The wick is the~central accretion engine, while the flame represents the polar jet that is often present.}}
		\label{candle_type1}
	\end{minipage}
	\hfill
	\begin{minipage}[t]{0.49\textwidth}
		\includegraphics[width=\textwidth]{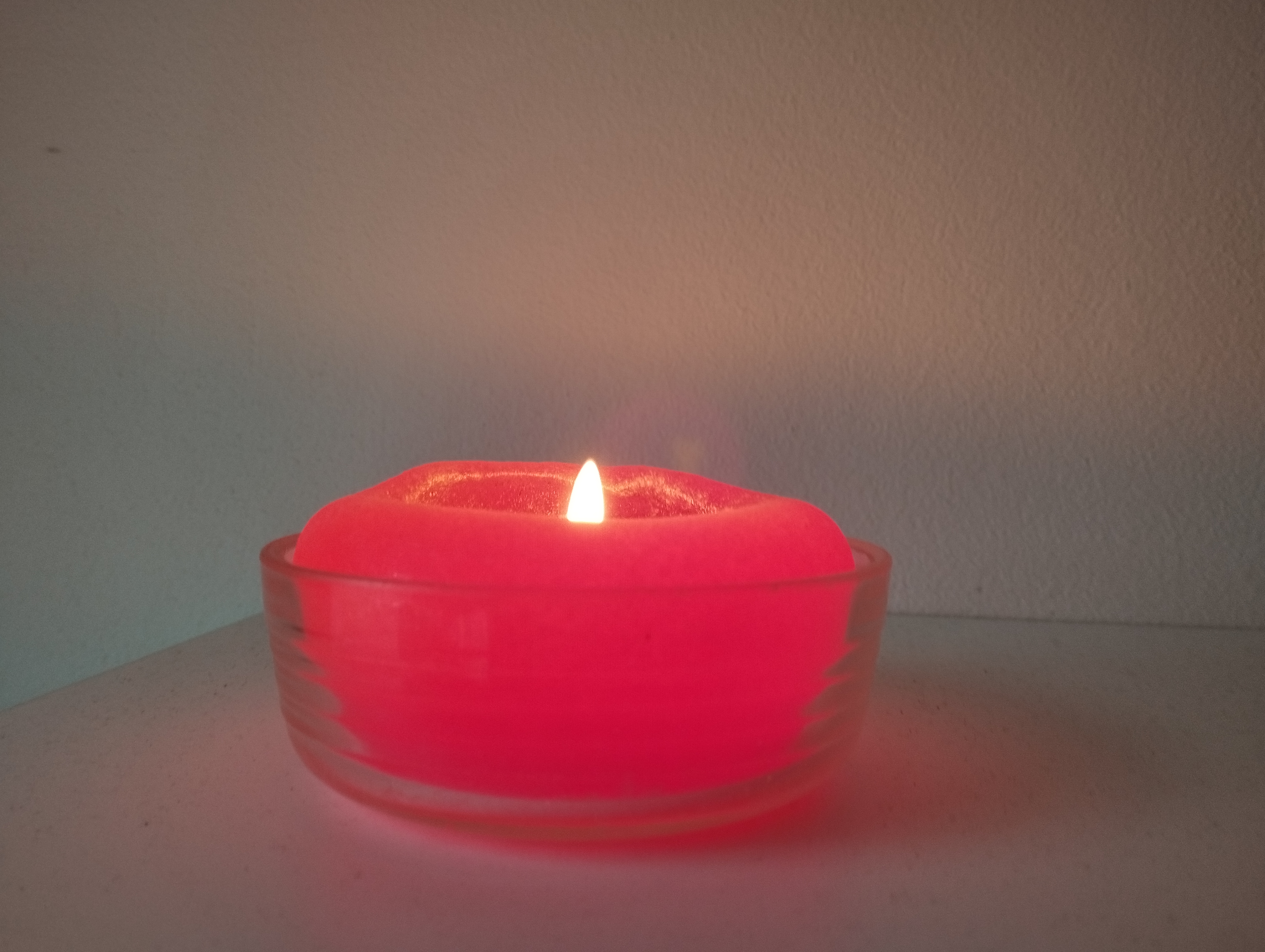}
		\caption{\footnotesize{The same as in Figure \ref{candle_type1}, but for a type-2 view. The central parts are obscured. Only for AGNs that are radio bright we identify polar ejecta of relativistic speeds.}}
		\label{candle_type2}
	\end{minipage}
\end{figure}

Strongly magnetized accretion disc and the ergosphere in close vicinity of~the~black hole, arising from full GR description, may cause that some infalling material is ejected with relativistic speeds in polar directions through the~so-called Blandford-Znajek or Blandford-Payne processes \citep[][]{Blandford1977,Blandford1982,Rees1982,Blandford2019}. The~rich structures that can form around the poles reflect the central emission, which can thus gain polarisation through first-order scattering. The presence of~broad lines in polarisation for type-2 AGNs unifies the picture \citep[see e.g.][]{Antonucci1993, Beckmann2012, Seward2010}. One can imagine the~entire structure like the candle in Figures \ref{candle_type1} and \ref{candle_type2} viewed from the~generic type-1 and type-2 directions, respectively. We also more specifically distinguish the clumpy equatorial broad line region (BLR) from the accretion disc located closer to the black hole and the torus itself, located further out. And if the clumpy regions extend at parsec scales vertically (in polar directions), we call them narrow line regions (NLR). It is also worth noting that perfect axial symmetry is often not the evolutionary case and some misalignment of the central black hole rotation axis from the accretion disc axis, misalignment of the disc plane from the~distant equatorial obscurers, or misalignment of the~dusty torus plane from the host galaxy plane, is considered \citep{Bardeen1975,Goosmann2011, Hopkins2012b,Tremain2014,Stalevski2017, Chatterjee2020}.

~\

\begin{figure}[h]\centering
	\includegraphics[width=1\textwidth]{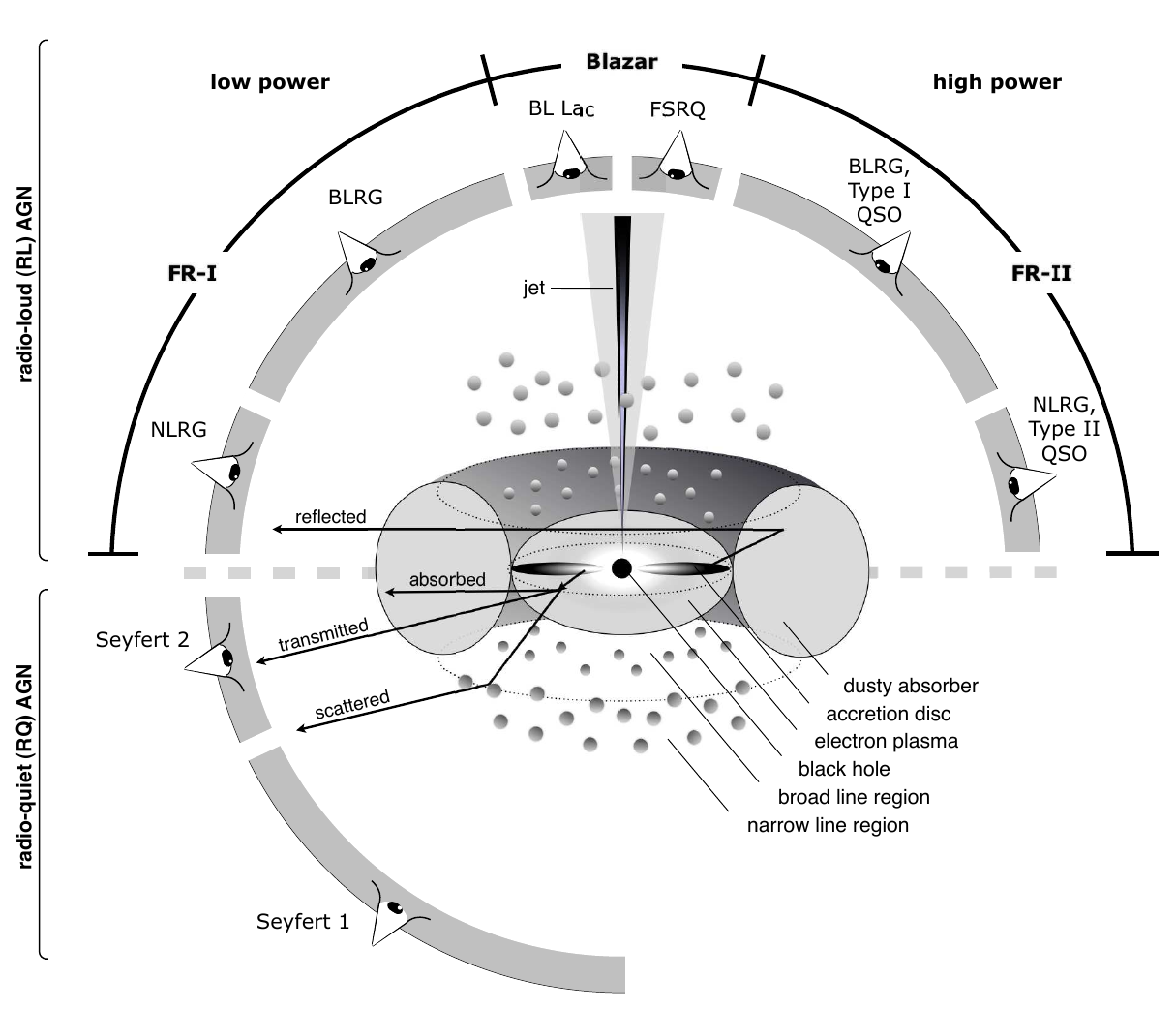}
	\caption{\footnotesize{Schematic drawing of the AGN unification model and different nomenclature. Image from \cite{Beckmann2012}.}}
	\label{AGN_scheme}
\end{figure}

Polar ejecta are often highly colimated (spanning only a few angular degrees from the black-hole poles) and represent perhaps the most energetic locations in~the~observable universe, hence called ``jets'', often surrounded by more angularly extended ionization cones \citep[see e.g.][]{Peterson1997, Trumper2008, Seward2010, Blandford2019}. Due to synchrotron emission, the jets are particularly bright in radio, leading to another very useful classification. If jets are present, which is in $10$--$20\%$ of AGNs \citep{Kellermann1989,Peterson1997}, we speak of ``radio-loud'' AGNs and ``radio-quiet'' AGNs are those with low jet activity. These relativistic streams of particles are remarkably stable and can outreach the entire host galaxy despite originating in the orders-of-magnitude smaller nucleus. Yet another theoretical challenge to~high-energy astrophysicists. If the observer is looking straight into the narrow jet, we speak of a ``blazar'', obviously showing very distinct observational traits from the rest of AGNs. In this dissertation, we will not cover blazars and we will focus on radio-quiet objects, not to complicate the already intricate story of~X-ray polarisation origin due to synchrotron emission on top of scattering and absorption effects. In addition, AGNs possess more inclined winds originating from the accretion disc at more than a few gravitational radii \citep[for recent reviews, see e.g.][]{Proga2007, Czerny2019, Karas2021}. These winds, significantly contributing to the total outflow rates and altering the central emission, are attributed to radiation, thermal or magneto-hydro-dynamical (MHD) driving mechanisms. Studying both jets and winds is also important for galaxy evolution theories that investigate the feedback of matter and radiation of AGNs with the host galaxies. One of the contemporary and ambitious unification diagrams \citep{Beckmann2012} with the major AGN constituents and some common classifications is shown in Figure \ref{AGN_scheme}. But we note that too detailed illustration can often cause a cognitive bias.

~\

The theory of black-hole accretion disc's formation, stability and radiation properties is one of the most challenging parts of today's astrophysics. Exact stable solutions to the governing Euler's and Maxwell's equations that would cover complicated microphysics are scarce \citep{Abramowicz2013}. A~practical analytical solution for a geometrically thin, optically thick disc was found by \cite{Shakura1973} and relativistically developed by \cite{Novikov1973, Page1974, Compere2017}. However, despite its remarkable utility for first-order estimates on radial structure and~emission profile, it is often criticized for neglect of microphysics (described via ad hoc parameter $\bar{\alpha}$), instability and lack of support from observations and MHD simulations departing from the first principles \citep[see e.g.][]{Shakura1976, Done1992, Abramowicz2005, Wang2014, Mishra2017, Zdziarski2022}. Other famous, and perhaps more realistic solutions include discs of~non-negligible size, which is related to the increased accretion rate and luminosity or increased advection to the detriment of radiative efficiency: the~slim discs \citep{Abramowicz1988}, the ``advection dominated accretion flows'' (ADAF) \citep{Narayan1994}, the thick discs or Polish donuts \citep{Jaroszynski1980}, the puffy discs \citep{Sadowski2016, Lancova2019}, or combined descriptions, such as the ``jet emitting disc standard accretion disc'' (JED-SAD) model \citep{Ferreira2006}. The results of sophisticated MHD simulations that include GR effects and simplified radiative transfer are providing reasonably similar predictions with different codes \citep{Porth2019}, but in the current state of development they are difficult to be directly observationally validated. The~intricate effects of thermal instabilities, convection and turbulences with respect to the vertical structure of the disc were studied, for example, by \cite{Agol2001, Turner2004, Zhu2013}.

For the sake of simplicity, we will be using the Novikov-Thorne model (i.e. the standard $\bar{\alpha}$-disc with relativistic corrections) for radial disc profiles, assuming Keplerian velocities, no torque on the inner boundary, and inner-disc radius extended to the inner-most circular orbit, if not stated otherwise. But we note that this might be one of the most critical model assumptions for the global emission predictions in Chapters \ref{chap02}, \ref{chap04} and \ref{chap05}, difficult to assess unless a completely new model based on different assumptions is built. 

In proper radiative transfer treatment of the accretion disc, we encounter another critical assumption for the result of mostly bound-bound and bound-free processes: the chemical composition and metallicity $A_\textrm{Fe}$, which sets the relative abundance of all ions with respect to some referential composition, typically that of our Sun \citep{Matteucci2001}. Various determinations of solar abundances relative to neutral hydrogen exist and are debated \citep{Grevesse1998, Asplund2005, Asplund2009, Bergemann2014}.

~\

We define the Eddington luminosity $L_\textrm{Edd}$ as \citep[e.g.][]{Karas2021}
\begin{equation}
    L_\textrm{Edd} = 1.3 \times 10^{46} \,  \frac{M}{10^8 \, M_\odot} \cdot \textrm{erg} \cdot \textrm{s}^{-1}
\end{equation}
and the Eddington mass accretion rate $\dot{M}_\textrm{Edd}$ as \citep[e.g.][]{Karas2021}
\begin{equation}
    \dot{M}_\textrm{Edd} = 1.39 \times 10^{18} \, \frac{M}{M_\odot} \cdot \textrm{g} \cdot \textrm{s}^{-1} \, .
\end{equation}
They effectively describe the accretion state when radial radiative force balances the gravitational pull by the central body, which is referred to as Eddington limit. In these black-hole mass dependent units, we can then more conveniently discuss the true bolometric (or X-ray) luminosity $L_\textrm{bol}$ (or $L_\textrm{X}$) and mass accretion rate $\dot{M}$, which are proportional and which can be also observationally determined \citep[see e.g.][]{Davis2011}. The geometrically thin and optically thick disc can be assumed to be locally thermally radiating as a black body, with a decreasing temperature profile with radius \citep{Shakura1973}
\begin{equation}
    T_\textrm{eff}^4 \sim \dfrac{\dot{M}}{\dot{M}_\textrm{Edd}} \left( \dfrac{r}{r_\textrm{g}}\right)^{-3} \dfrac{1}{M} \, ,
\end{equation}
hence forming a multicolor black-body spectrum, i.e. the ``big blue bump'' \citep{Malkan1983, Czerny1987}. Although state-of-the-art radiative transfer equation solvers are able to treat the disc's photosphere in great detail (see below), it appears difficult to capture the effects of Comptonization, which are to~be assessed with next generation of codes \citep{Shimura1995, Merloni2000, Davis2006, Done2012, Davis2019}. Therefore, the empirical color correction factor $f_\textrm{c}$ \citep{Ebisuzaki1984} is often used:
\begin{equation}
    I = f_\textrm{c}^{-4}B(f_\textrm{c}T_\textrm{eff})\textrm{ ,}
\end{equation}
where $B$ is from the Planck's law (\ref{blackbody}). Typically AGN discs exhibit \mbox{$n_\textrm{H} = 10^{13}$--$10^{18} \ \textrm{cm}^{-3}$} and $T_\textrm{eff} = 10^{3.5}$--$10^{6} \ \textrm{K}$, if the Novikov-Thorne model is assumed with observationally and theoretically motivated estimates of~\mbox{$1.5 \lessapprox f_\textrm{c} \lessapprox 1.8$}, $0.1 \lessapprox \bar{\alpha} \lessapprox 0.2$, $10^5 \lessapprox M/M_\odot \lessapprox 10^9$ and $0.01 \lessapprox \dot{M}/\dot{M}_\textrm{Edd} \lessapprox 1$. The spectrum then peaks at the UV band for AGNs, which is well confirmed by observations \citep[see e.g.][]{Seward2010}.

~\

\begin{figure}[h]\centering
	\includegraphics[width=\textwidth]{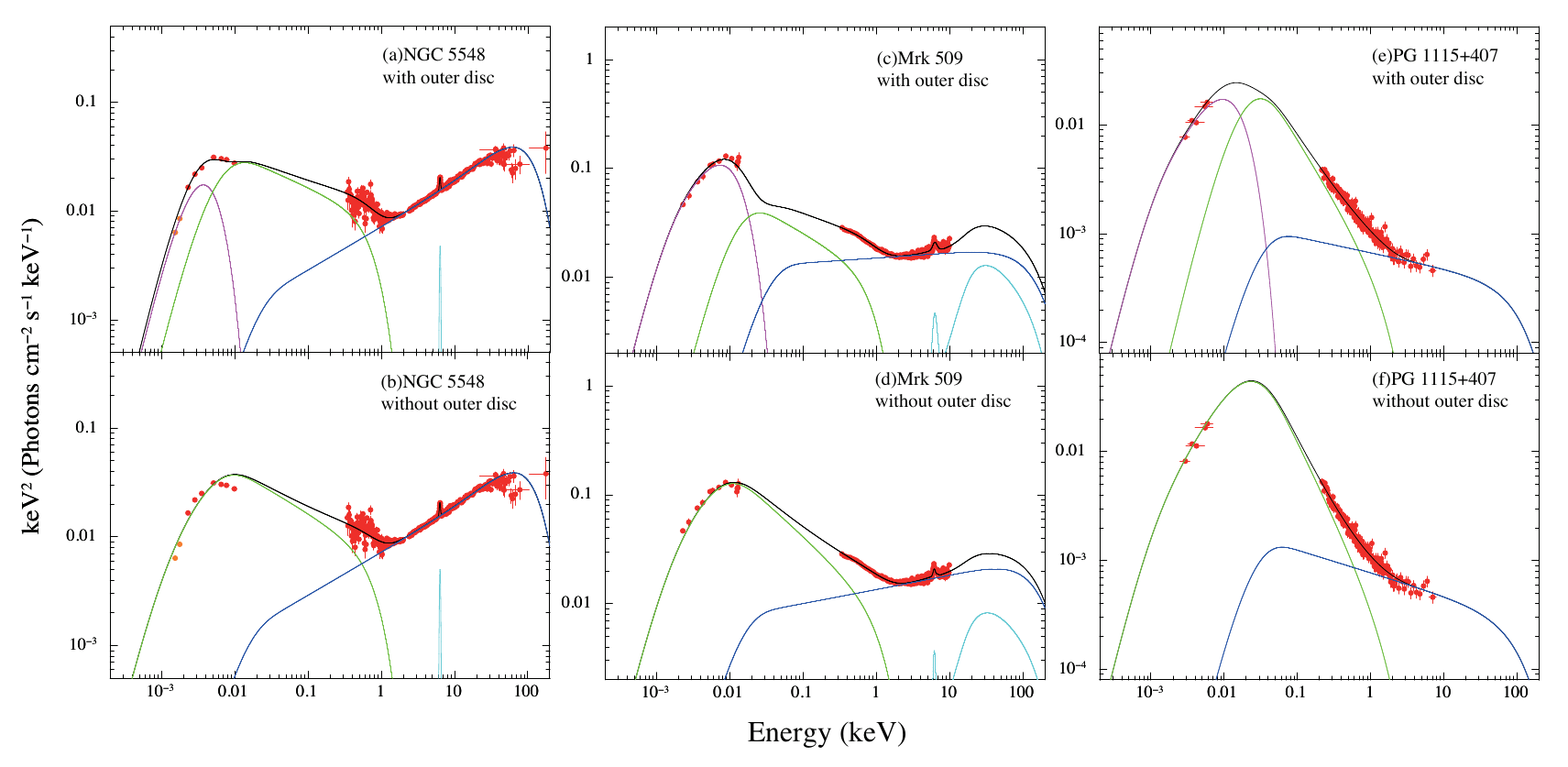}
	\caption{\footnotesize{Broadband spectral energy distribution fit of various AGNs by \cite{Kubota2018}. Details are provided therein. The fit consists of four distinct continua here: outer disc emission (in magenta), the warm Comptonized component (in green), and the hard Comptonized primary (in dark blue) and reflected (in light blue) components. Total unfolded spectrum is in black and the data from \cite{Mehdipour2011, Jin2012a, Jin2012b, Mehdipour2015} in red. Panels (a), (b), (c), (d), (e) and (f) show fits of NGC 5548 with and without outer disc, Mrk 509 with and without outer disc and PG 1115+407 with and without outer disc, respectively. Image from \cite{Kubota2018}.}}
	\label{AGN_SED}
\end{figure}

Lastly, let us focus more on the X-ray spectral properties of AGNs, for which they are undoubtedly famous. Examples of a broadband unfolded spectral energy distribution (SED) are shown in Figure \ref{AGN_SED}. In the X-rays, we predominantly observe a decreasing power-law with energy $N \sim E^{-\Gamma}$, described by a power-law index $\Gamma$. This is a natural consequence of thermal disc emission Compton up-scattered in hot ($T_\textrm{e} \gtrapprox 10^7$ K) gas of electrons, the corona, located nearby \citep{Shapiro1976,Liang1977,Bisnovatyi1977, Sunyaev1980, Haardt1991}. The measured spectral slope is thus also related to the accretion rate and luminosity \citep[e.g.][]{Zdziarski1999,Lu1999, Wang2004, Risaliti2009, Brightman2013, Yang2015, Trakhtenbrot2017, Huang2020}. The power-law is cut at hard energies (hundreds of keV) due to insufficient energy of the Comptonizing electrons. At the very hard (X$\gamma$) energy tail, electron-positron pair production and annihilation effects also play important role \citep{Coppi1990, Zdziarski1990, Coppi1999}. But we decided to omit the relevant theoretical background in Section \ref{processes}, because the models developed in this dissertation neglect these effects and operate at lower energy ranges. At soft energies (tenths of keV), the low-energy cut-off naturally appears due to the seed thermal photons energy, i.e. there are only a few scattered photons below some characteristic energy. Such coronae, which are inseparable part of the inner-most accreting regions, can be constrained up to some limit in~terms of optical depth, electron temperature and geometry from spectroscopic and timing techniques \citep[e.g.][]{Haardt1991,Haardt1993,Haardt1993b,Dove1997,Henri1997,Krolik1999,Ghisellini2004,Markoff2005,Dovciak2016,Kara2016,Fabian2017,Kara2017,Kubota2018, Ursini2020}. However, especially with geometry, which is a critical unknown, X-ray polarimetry has high potential in lifting the known degeneracies \citep{Haardt1993c,Poutanen1996, Marinucci2018, Poutanen2018,Ursini2022, Krawczynski2022a, Poutanen2023}.

~\

In the literature, one finds many realizations for the coronal geometry and bulk motion. Although non-stationarity of the primary source is sometimes considered \citep{Beloborodov1998, Beloberodov1999, Beloborodov1999b, Malzac2001, Poutanen2023}, we will assume a stationary situation in this dissertation, which affects the polarisation properties due to negelect of some potential aberration effects. The most simplified phenomenological model is the~``lamp-post model'' \citep[e.g.][]{Matt1991,Martocchia1996,Henri1997,Petrucci1997, Martocchia2000, Miniutti2004, Dovciak2004b, Dovciak2011, Parker2015, Furst2015, Miller2015, Niedzwiecki2016, Dovciak2016, Walton2017, Ursini2020} that assumes a point-like isotropic X-ray emitter located on~the~rotation axis at some height above the black hole. Although this model is difficult to support from photon conservation point of view and Comptonization codes always require at least a minimal spatial extension of the corona, it may serve as a prototype for an optically thin aborted jet or base of~a~jet with spherical, or~perhaps more favored conical shape. Of completely different emission properties are ``extended coronal models'' \citep[e.g.][]{Haardt1991, Haardt1993, Poutanen1996, Dabrowski2001,Malzac2005,Niedzwiecki2008, Schnittman2010, Marinucci2018, Poutanen2018}, where the corona is diffuse and elongated in equatorial direction. These range from sandwich models, where the corona forms only a~geometrically thin layer on the disc, to wedge models, to large spherical models engulfing the entire inner-accreting region. These are considered both patchy and with uniform density. Last, distinct archetype is the ``hot inner-accretion flow'' \citep{Poutanen1996, Dove1997, Esin1997, Esin1998, Zdziarski1998, Liu2007, Poutanen2009, Veledina2013, Poutanen2018} that does not assume the hot gas above and below the disc, or in the polar directions, but rather in wedge-like or cylindrical geometries inside the accretion disc, if truncated at significant distance from the~event horizon and with debatable role of synchrotron seed photons. Although the evidence and direct knowledge about X-ray emitting coronae is coming primarily from X-ray observations, some deeper theoretical support is provided from plasma simulations for the emergence of coronae and its heating, using the kinetic approach (and less from plasma fluid simulations), but determining the geometry anyhow would be a significant step forward for this field \citep[e.g.][]{Begelman1983a, Begelman1983b,DiMatteo1998, Wilkins2012, Beloborodov2017, Nattila2022, Liska2022, Mellah2022, Bambic2023}.

~\

\begin{figure}[h]
	\centering
	\begin{minipage}[t]{0.467\textwidth}
		\includegraphics[width=\textwidth]{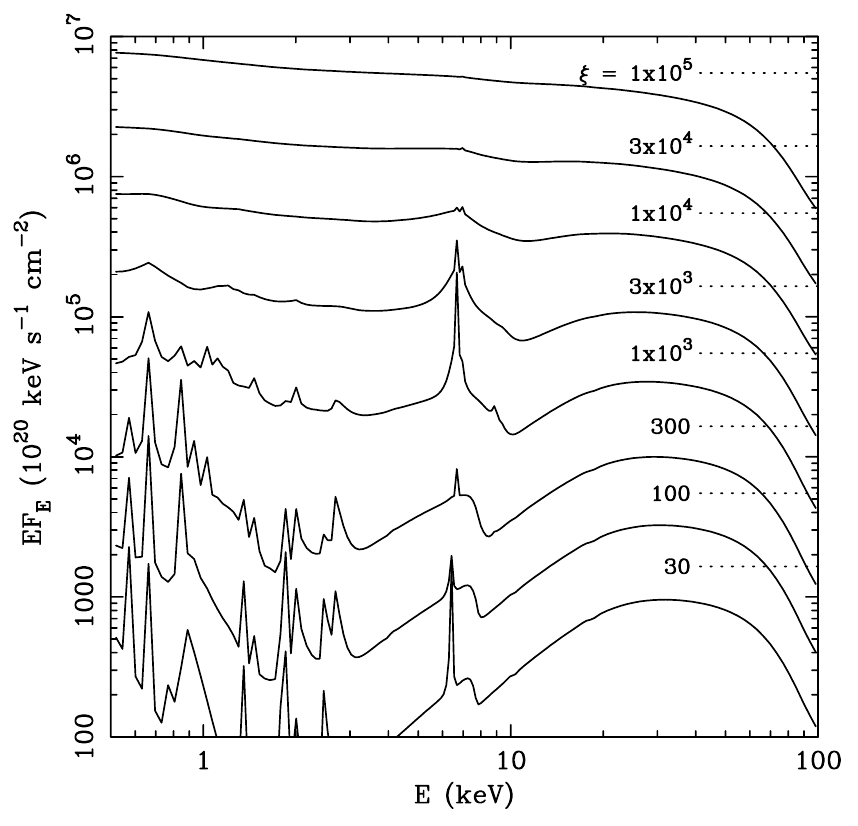}
		\caption{\footnotesize{Theoretical AGN X-ray reflection spectrum for various values of the ionization parameter $\xi$ defined by (\ref{xi}). The more the disc is ionized, the more the reflected spectrum resembles the original power-law. Image from \cite{Fabian2000}.}}
		\label{AGN_reflection}
	\end{minipage}
	\hfill
	\begin{minipage}[t]{0.513\textwidth}
		\includegraphics[width=\textwidth]{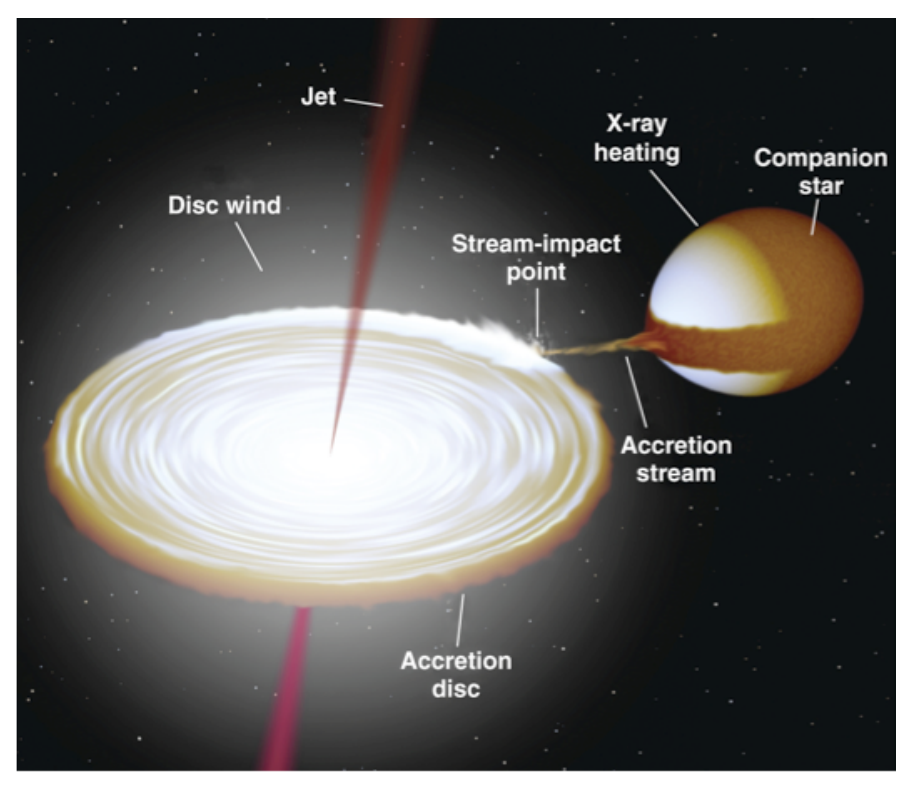}
		\caption{\footnotesize{Artist's impression of the XRB system, including the indication of main components. The accretion disc surrounding the stellar-mass black hole is formed by a stream of matter from the companion star. Image from \cite{Fender2012}, courtesy of Rob Hynes.}}
		\label{XRB_sketch}
	\end{minipage}
\end{figure}

X-ray spectra of radio-quiet AGNs may contain also a strong reflection component of the X-ray power-law originating in the hot corona \citep[e.g.][]{Guilbert1988,Lightman1988,Fabian1989,Matt1993,Fabian2000,Reynolds2003, Garcia2013}. In continuum, we typically see a Compton shoulder at about 20 keV that is caused by interplay of~Compton down-scattered photons with nearly constant \mbox{mid-X-ray} cross-section with energy and photoelectric absorption with comparable, but decreasing cross-section as $\sim E^{-3}$. Many emission lines appear on top of the continuum, especially below 2 keV and in the iron line complex near 6--7 keV. If these lines are narrow, we ascribe this reflection to the torus, distant disc or other components far-away from the central accretion engine. If broad, then these are assumed to~be relativistically smeared and originating in reflection off inner accretion disc close to~the~black hole or in the winds \citep[see][for~the~discussion of~degeneracy between the two]{Parker2022}. The relativistic reflection component in~both line and continuum features enables independent spectroscopic and timing determinations of~the~black-hole spin, coronal geometry, source orientation and~the~accretion disc's properties \citep[see][and references therein]{Reynolds2019}. The~anticipation is that X-ray polarimetry can independently constrain such properties \citep{Rees1975,Connors1977,Stark1977,Connors1980, Dovciak2004, Dovciak2008, Schnittman2009, Li2009, Marin2014, Cheng2016,Marin2016,Taverna2020, Taverna2021, Ursini2022, Krawczynski2022a, Mikusincova2023}. A~theoretical X-ray reflection spectrum from a~constant-density plane-parallel disc in the local comoving frame is shown in~Figure \ref{AGN_reflection} for different values of the ionization parameter, which is, regarding the presence of main spectral features, supported by larger number of AGN observations \citep[e.g.][]{Gottwald1995, Winter2009, Ng2010}. \cite{Fabian2000} discusses four distinct regimes of~the~spectral dependence on $\xi$. For~a~given density, the~higher the~impinging bolometric flux is, the~less continuum absorption and emission lines we see, and the disc acts as a highly reflective Compton mirror to the original power-law. The reflected power-law can also return to~the~disc due to strong gravity and reflect again, multiple times, which affects both spectra and polarisation, depending for example on the spin of the black hole and coronal parameters \citep{Schnittman2009, Schnittman2010, Connors2021, Dauser2022}.

~\

Disentangling the AGN X-ray spectra is a difficult task due to large multiplicity in the observed power-law spectra \citep{Trumper2008, Seward2010}. The soft X-rays of AGNs are further complicated by the~presence of the ``soft X-ray excess'', for which competing explanatory theories exist until today. The most promising being the warm corona model, or that this feature is a natural extension of the big blue bump located in the UV or that it appears due to a presence of broad spectral lines \citep[e.g.][]{Boroson1992,Boller1996,Pariev1998,BranduardiRaymont2001,Lee2002,Gierlinski2004, Mehdipour2011,Rozanska2015,Kubota2018, Petrucci2020}. Across all X-rays, we also observe variability in emission and absorption (on shorter time scales than viscous characteristic times) due to flares, cloud eclipses, warm absorbers and winds inside the~system, or due to absorption in the host galaxy or in the ISM of our Galaxy \citep[e.g.][]{Halpern1984,Nandra1994,Peterson1997, Porquet2000, Porquet2004, Trumper2008, Seward2010}. For data fitting, absorption is typically characterized through one parameter along the entire line of sight: the neutral hydrogen column density $N_\textrm{H} \ [\textrm{cm}^{-2}]$ \citep[see e.g. the {\tt TBABS} model,][]{Wilms2000}.

\section{X-ray binary systems}\label{all_XRBs}

Multiple gravitationally bound stellar systems are more common than solitary stars in the Universe \citep{Duchene2013}. Interesting configurations may occur over the life-time of a binary stellar system, resulting in increased X-ray brightness. We will focus on X-ray bright binary systems in stable mass transfer evolutionary phase with stellar-mass black hole as an acceptor, and a giant or supergiant ordinary star as a donor. These are called black-hole X-ray binary systems (XRB) and the brightness at all wavelengths (not only X-rays) is given much through the same mechanisms as in the quasars, including the presence of~the~disc-corona-jet system \citep{Rees1998, Done2007, Seward2010}. Hence, these are often nicknamed as ``microquasars'' \citep{Mirabel1992}. 
In XRBs, the mass transfer happens through Roche lobe (in XRBs with low mass of the donor) or additionally through stellar winds of the donor (with high-mass donor star). Sketch of the entire stellar binary system is shown in Figure \ref{XRB_sketch}. Due to radiation effects combined with the formed gravitational potential well, the accretion disc is formed with overall higher mean effective temperatures ($T_\textrm{eff} = 10^{5}$--$10^{7.5} \ \textrm{K}$) and neutral hydrogen densities ($n_\textrm{H} = 10^{19}$--$10^{25} \ \textrm{cm}^{-3}$) compared to AGNs. This again theoretically follows from the Novikov-Thorne disc solution, assuming orders-of-magnitude lighter central black hole \citep[\mbox{$M = 6$--$20 \, M_\odot$}, ][]{Reynolds2003}. This is important for our modeling, because the~peak of the multicolor black-body radiation is well within the soft X-rays, which is observationally confirmed \citep[see e.g.][]{Gierlinski1999}. The growing observational evidence of high-density accretion discs in XRBs is a recent success of~modern \mbox{X-ray} spectral fitting models of reflected disc emission \citep{Garcia2016,Tomsick2018, Jiang2019, Jiang2019b, Jiang2020, Connors2021, Liu2023}, although many of these studies assume a constant and extreme density even in the upper layers of the disc atmosphere, which is a debatable simplification. Because of the additional significant presence of the thermal component \mbox{in~the~X-rays} and because of the expected high disc densities, we will first develop models suitable for AGNs, which seems in principle a less ambitious task, and then try to adapt them to XRBs.

~\

\begin{figure}[h]\centering
	\includegraphics[trim={0cm 0.5cm 0cm 0cm},clip,width=1\textwidth]{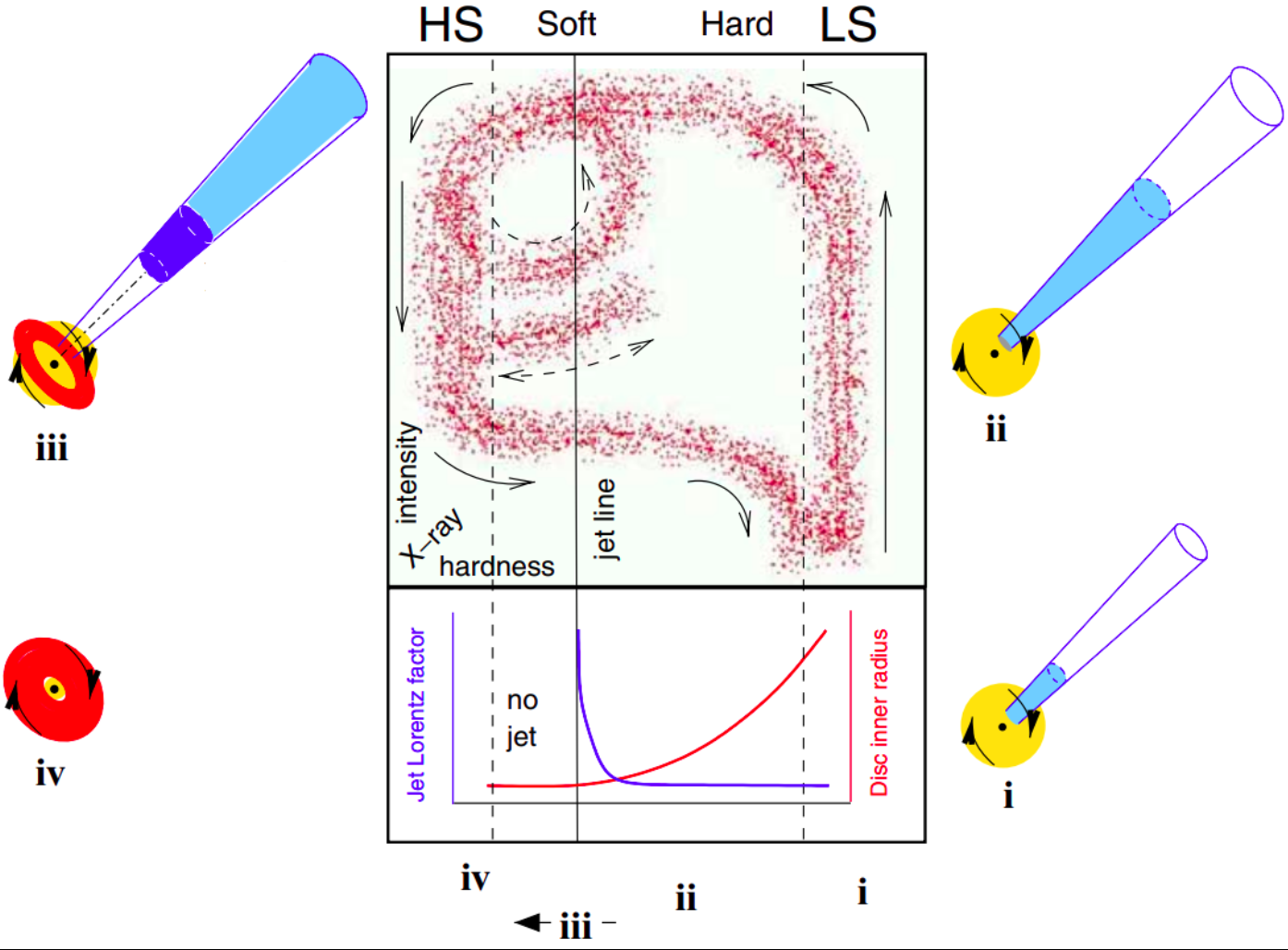}
	\caption{\footnotesize{Schematic drawing of the XRB spectral states in the X-ray hardness-intensity diagram. The evolution of a typical binary system follows a hysteresis behavior in the shape of~a~letter ``Q''. However, many sources exhibit irregularities and operate in just part of~the~configuration space. The two most extreme states, the low/hard state (LS on the picture) and~the~high/soft state (HS on the picture), vary by jet presence and theoretical inner disc radius. Image adapted from \cite{Fender2004}.}}
	\label{Q_diagram}
\end{figure}

\begin{figure}[h]
	\centering
	\begin{minipage}[b]{0.523\textwidth}
		\includegraphics[width=\textwidth]{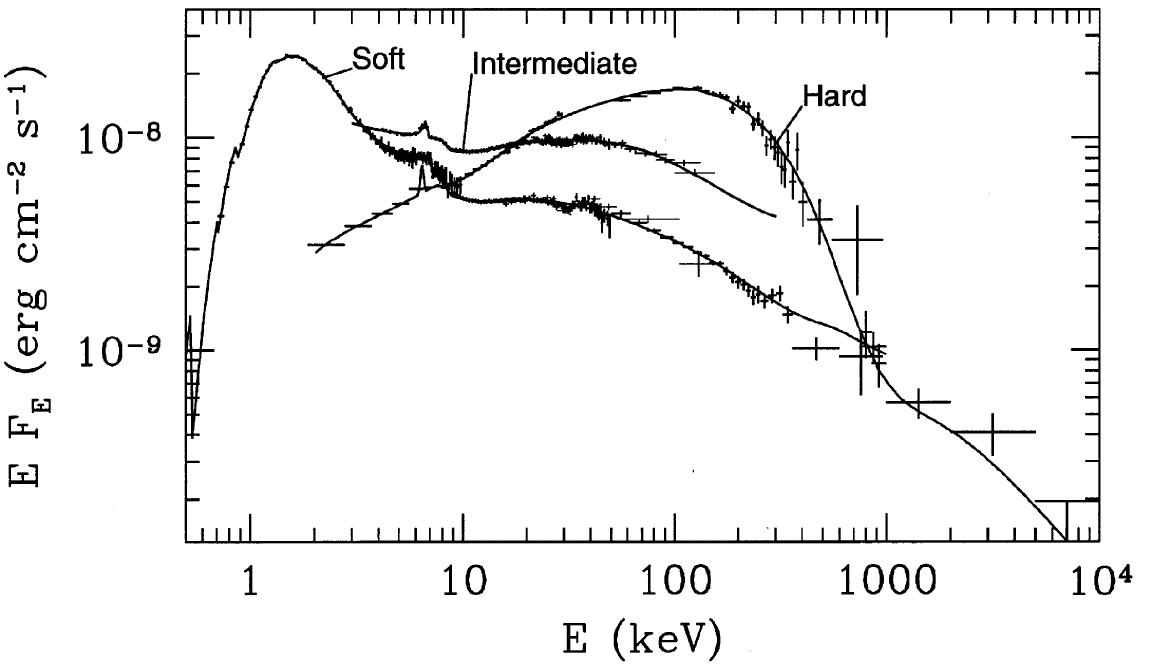}
		\caption{\footnotesize{Broadband fitted spectral energy distribution of the soft, intermediate and hard XRB states of Cygnus X-1 by \citet{Gierlinski1999}. The models are hybrid, consisting of the thermal disc component, a hard power-law continuum component and a broad power-law reflection component. The data were taken between 1991 and 1996 by the \textit{Ginga}, \textit{RXTE}, \textit{ASCA} and OSSE instruments. Image from \cite{Gierlinski1999}.}}
		\label{cygx1_spectrumall}
	\end{minipage}
	\hfill
	\begin{minipage}[b]{0.457\textwidth}
		\includegraphics[width=\textwidth]{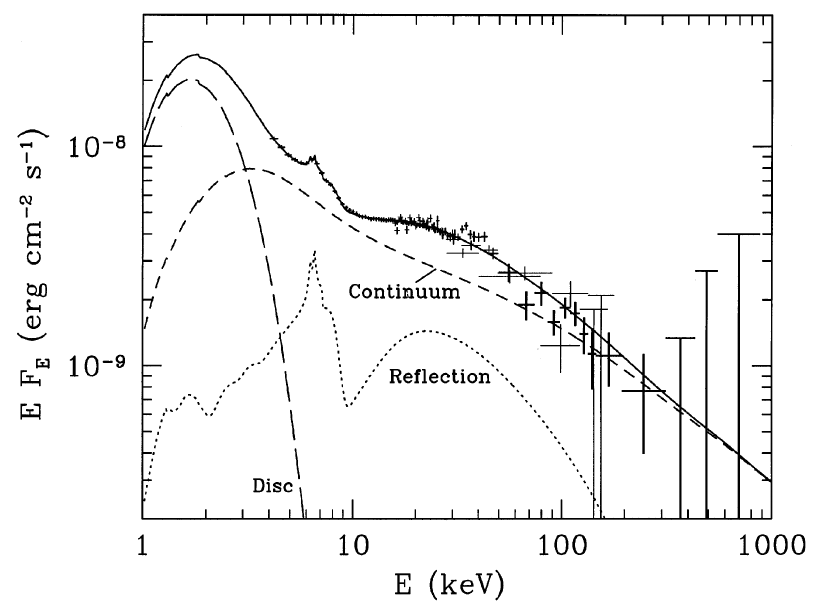}
		\caption{\footnotesize{The soft state from the~same study as in Figure \ref{cygx1_spectrumall} by \citet{Gierlinski1999} with spectral decomposition shown. A thermal black-body contribution in~the~X-rays is visible, alongside a harder reflection component accompanying the primary power-law coronal contribution. Image from \cite{Gierlinski1999}.}}
		\label{cygx1_spectrumdecomposed}
	\end{minipage}
\end{figure}

However, it may be easier to deduce observationally relevant information on~accretion physics with XRBs, not only due to their X-ray brightness compared to~AGNs, which is critical for photon-demanding polarimetry (see Section \ref{space_observatiories}). At~human time scales, the nature was perhaps more favorable to us with XRBs than AGNs, because stellar-mass black holes can change their accretion state in~periods of months or years, thus we can arguably infer more on their accreting nature \citep{Fender2004}. Some transient sources show unstable outbursts with eight to nine orders of magnitude difference in luminosity across day periods with most significant distinction in the X-rays \citep[see e.g. ][]{Connors2021}. The change of~state is best characterized in the so-called Q-diagram of \mbox{X-ray} flux ratios, which is depicted in Figure \ref{Q_diagram}. XRBs show hysteresis evolution in~hardness-intensity diagram, often evolving from the prototypical ``low/hard state'' to the ``high/soft state'' and back \citep[see][and references therein]{Zdziarski2004, Remillard2006}. The~former is characterized by bright coronal emission compared to thermal emission from the~disc, significant disc truncation and strong presence of jets. The latter is characterized by dominant thermal component, accretion disc reaching to~the~ISCO and~jets being switched off. An extensive nomenclature for intermediate or irregular states exists. A broadband spectral energy distribution fit of different X-ray spectral states of high-mass XRB Cyg X-1 is shown in Figure \ref{cygx1_spectrumall}. An example of modelled decomposition of the soft state of Cyg X-1 is shown in Figure \ref{cygx1_spectrumdecomposed}.

~\

The XRB spectral, timing and polarisation characteristics of the corona and~outflows related to accretion are similar to those of AGNs and we refer to~the~literature in Section \ref{AGN_intro}, which is in many cases applied to XRBs. The~characteristics of stellar winds and circumbinary obscuring material depend on~the~evolutionary phase of the binary system, which in turn is largely given by the initial mass of~both components, eccentricity, separation or orbital period \citep{Tauris2006}. The dusty torus as in AGNs is not forming, because the accreting matter is streamed directly from the nearby orbiting companion. Short-term variability in absorption and emission is observed similarly to~AGNs. If the binary orbital axis is misaligned with the black hole spin axis, the~inner accretion disc may become warped, which may be also linked to~the~Lense-Thirring precession phenomenon \citep{Lense1918,Bardeen1975,Stella1998,Fragile2007,Abarr2020b}. This effect is imprinted in light curves of many wavelengths as quasi-periodic oscillations (QPO) \citep{Ingram2009, Veledina2013b}.

People tried with moderate success to identify in AGNs the spectral states observed in XRBs, tied in theory to the accretion mechanism, although in AGNs they should occur at larger time scales, thus observable only statistically \citep{Kording2006,Svoboda2017, Fernandez2021, Moravec2022}. Without doubt, better understanding of the accretion disc itself would significantly move up all AGN and XRB branches of studies. Albeit decades of intensive research, it proved to be a challange. It is perhaps desirable to aim at particular subproblems, such as determination of the inner disc radius, which is important for both theoretical and observational consideration of accreting black holes \citep[see e.g.][]{Krolik2002, Seward2010}: particularly in understanding and putting constraints on the GR effects, disc theories, MHD simulations, multiwavelength spectroscopy, reverberation mapping, QPO studies, and last but not least, X-ray polarisation outlook of inner-accreting regions, which we will explore in this dissertation.

\section{Computational approach}

In Sections \ref{AGN_intro} and \ref{all_XRBs}, we focused on the observational and physical description of accreting black holes, integrating all obtained knowledge. Apart from X-ray polarisation, we focused on their X-ray spectral properties, because any polarisation simulation must adequately produce and cross-validate the Stokes parameter $I$ too, even if not in high precision.

Timing properties were not of key interest from X-ray polarimetric point of~view, thus we omitted the summary of relevant literature. Long-term changes expected in X-ray polarisation of accreting black holes are typically described through different slowly evolving configurations of static polarisation models. The~only exceptions being orbital-phase-resolved polarisation analysis of XRBs and a possibility of detecting QPOs in polarisation. The former is interesting from the point of view of changing geometry of reflection from the stellar wind, reflection from the stellar companion or periodic eclipses. The latter could be seen in associated quasi-periodic changes in the polarisation angle due to misalignment of the inner disc part, as it precesses. Regarding short-term variability caused by e.g. corotating flares or clouds with the accretion disc, one can estimate the typical Keplerian orbital periods around Kerr black holes from \cite{Bardeen1972}, which are negligibly short compared to typical observational time needed  with \textit{IXPE} to obtain a polarisation detection of XRBs or fainter AGNs (cf. Section \ref{space_observatiories}). We obtain $\approx10 \, \textrm{ms}$ orbital period at $10 \, r_\textrm{g}$ for a stellar-mass black hole with $M = 10 \, M_\odot$ and $\approx10 \, \textrm{ks}$ orbital period at $10 \, r_\textrm{g}$ for a supermassive black hole with $M = 10^7 \, M_\odot$. Thus, with instruments such as \textit{IXPE}, we observe a~rotationally smeared signal from the inner-accreting regions.

~\

The numerical methods used and developed in this work are built on~the~shoulders of decades of intensive modeling by various research groups. We will now introduce the literature on the X-ray spectropolarimetric modeling of the studied (static) systems. We will first summarize the main findings of the past, and then give technical background of this work, making use of the elementary processes outlined in Section \ref{elementary} and having in mind the practical motivations illustrated in Section \ref{all_polarimetry}.

\subsection{Modeling of environments near black holes}\label{modelling}

Typically, MHD or kinetic plasma simulations that are able to grasp a large region near black hole self-consistently \citep[see e.g.][]{Nattila2022, Liska2022, Wieglus2022, Liska2023} do not treat radiative transfer in sufficient depth or are too computationally expensive to provide trustworthy observational interpretation via direct data fitting and case-to-case interpolation. For this reason, we rather use for specific problems radiative transfer equation solvers, MC simulations, or fully analytical models that tend to be the most computationally efficient. We note that recently, machine learning techniques were used in X-ray studies of AGNs or XRBs not only for faster classification of~collected data \citep{Hebbar2023} or reducing spectral fitting time \citep{Parker2022c}, but also for increasing modeling resolution by means of~interpolating computationally expensive models computed in a few grid points at majority of the parameter space and in high density at a smaller region of the~parameter space that was used as a training set \citep{Matzeu2022}.

~\

Analytical approximations provided perhaps the most classical results in~the~second half of $20^\textrm{th}$ century. Chandrasekhar provided estimates on polarisation due to Rayleigh single-scattering in reflection from plane-parallel slab \citep{Chandrasekhar1960}, which we will often use. For unpolarized primary radiation \citep[in transcription from ][]{Dovciak2004a}
\begin{equation}\label{chandra1}
	\begin{aligned}
		I &=  \dfrac{I_\textrm{l}+I_\textrm{r}}{\langle I_\textrm{l} + I_\textrm{r} \rangle} I_\textrm{N}  \textrm{ ,}\\
		Q &=  \dfrac{I_\textrm{l}-I_\textrm{r}}{\langle I_\textrm{l} + I_\textrm{r} \rangle} I_\textrm{N}  \textrm{ ,}\\
		U &=  \dfrac{\hat{U}}{\langle I_\textrm{l} + I_\textrm{r} \rangle} I_\textrm{N}  \textrm{ ,}\\
		V &=  0  \textrm{ ,}
	\end{aligned}
\end{equation}
where
\begin{equation*}\label{chandra2}
	\begin{aligned}
		I_\textrm{l} &= \mu_\textrm{e}^2(1+\mu_\textrm{i}^2) + 2(1-\mu_\textrm{e}^2)(1-\mu_\textrm{i}^2) - 4\mu_\textrm{e}\mu_\textrm{i} \sqrt{(1-\mu_\textrm{e}^2)(1-\mu_\textrm{i}^2)} \cos(\Phi_\textrm{e}-\Phi_\textrm{i}) \textrm{ ,}\\ & \textrm{\space\space\space} - \mu_\textrm{e}^2(1-\mu_\textrm{i}^2)\cos[2(\Phi_\textrm{e}-\Phi_\textrm{i})] \textrm{ ,}\\
		I_\textrm{r} &=  1+ \mu_\textrm{i}^2 + (1-\mu_\textrm{i}^2) \cos[2(\Phi_\textrm{e}-\Phi_\textrm{i})] \textrm{ ,}\\
		\hat{U} &=  4\mu_\textrm{i} \sqrt{(1-\mu_\textrm{e}^2)(1-\mu_\textrm{i}^2)} \sin(\Phi_\textrm{e}-\Phi_\textrm{i}) -2\mu_\textrm{e}(1-\mu_\textrm{i}^2)  \sin[2(\Phi_\textrm{e}-\Phi_\textrm{i})] \textrm{ ,}
	\end{aligned}
\end{equation*}
where
\begin{equation*}
	\Phi_\textrm{i} = - R_\textrm{sgn}^\textrm{i} \arccos \left(  \dfrac{-1}{\sqrt{1-\mu_\textrm{i}^2}} \dfrac{p_{\textrm{i}\alpha}e_{(\phi)}^{\alpha}}{p_{\textrm{i}\mu}U^{\mu}} \right) + \dfrac{\pi}{2}
\end{equation*}
and $R_\textrm{sgn}^\textrm{i}$ is the sign of the radial component of the four-momentum $p_\textrm{i}^{\mu}$ of an~incident photon. It is positive if the incident photon travels outwards ($p_\textrm{i}^{(r)} > 0$), and~negative, if it travels inwards ($p_\textrm{i}^{(r)} < 0$) in the local rest frame of the disc. The~rest of the angles describe the reflection event with respect to the disc's normal and photon's incident and emergent direction: $\mu_\mathrm{i}$ is the cosine of incident inclination angle $\delta_\textrm{i}$ measured from the disc's normal, $\mu_\mathrm{e}$ is the cosine of~emergent inclination angle $\delta_\textrm{e}$ (in Section \ref{relativistic} considered as $n_\textrm{em}$, in Chapter \ref{chap01} will be sometimes considered as the global $\theta$) measured from the disc's normal, $\Phi_\mathrm{e}$ is the~relative azimuthal angle measured counter-clockwise at~the~disc plane between the~emission and the incident ray directions projected to the disc's surface. For~unpolarized primary, if the results are integrated in $\mu_\textrm{i}$ and $\Phi_\textrm{e}$, the~resulting polarisation fraction monotonically increases with inclination from $0\%$ up to~about $20\%$ and~the~polarisation angle is parallel with the slab. We refer the~reader to~\cite{Chandrasekhar1960} for prescriptions for polarized primary and~for~the~result for diffuse pure-scattering reflection with the assumption of~infinite scatterings.

\begin{figure}[h]
	\centering
	\begin{minipage}[b]{0.49\textwidth}
		\includegraphics[width=\textwidth]{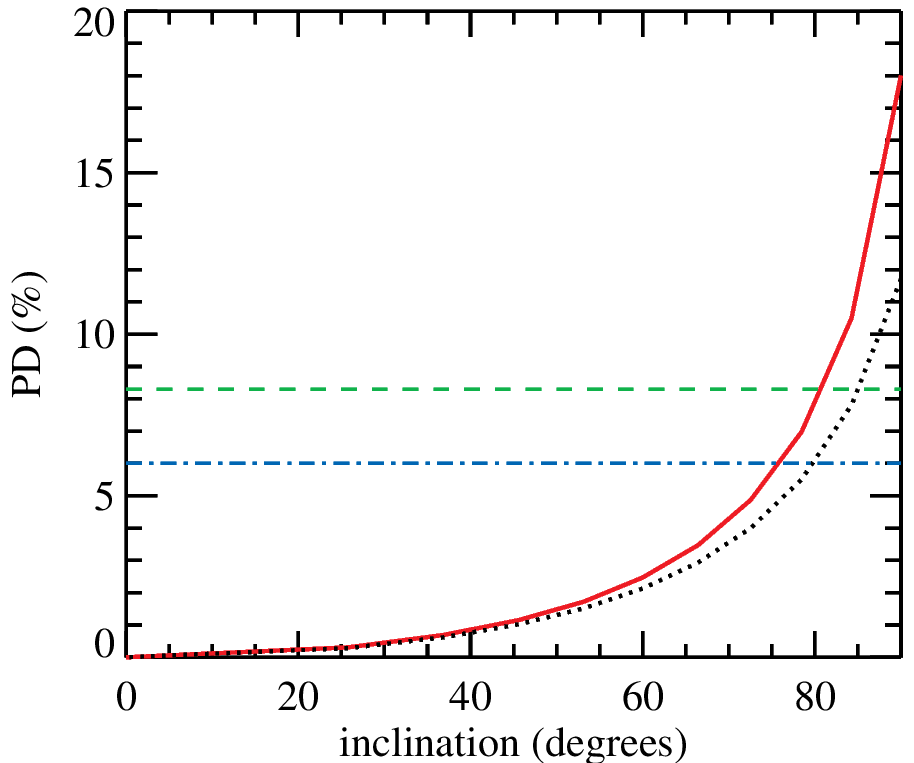}
		\caption{\footnotesize{Polarisation fraction versus observer's inclination for transmission through semi-infinite atmosphere in pure-scattering limit \citep[black dotted, ][]{Chandrasekhar1960} and~with absorption and internal sources of~radiation \citep[red solid, ][]{LoskutovSobolev1981}. The latter is also parametrized by~the~ratio of scattering coefficient to the sum of scattering and true absorption coefficients, which gives higher polarisation for smaller values of the ratio. The horizontal lines correspond to \textit{IXPE} data of XRB system 4U1630-47, see \citet{Ratheesh2023} for details. Image from \cite{Ratheesh2023}.}}
		\label{analytical_semiinfinite}
	\end{minipage}
	\hfill
	\begin{minipage}[b]{0.49\textwidth}
		\includegraphics[width=\textwidth]{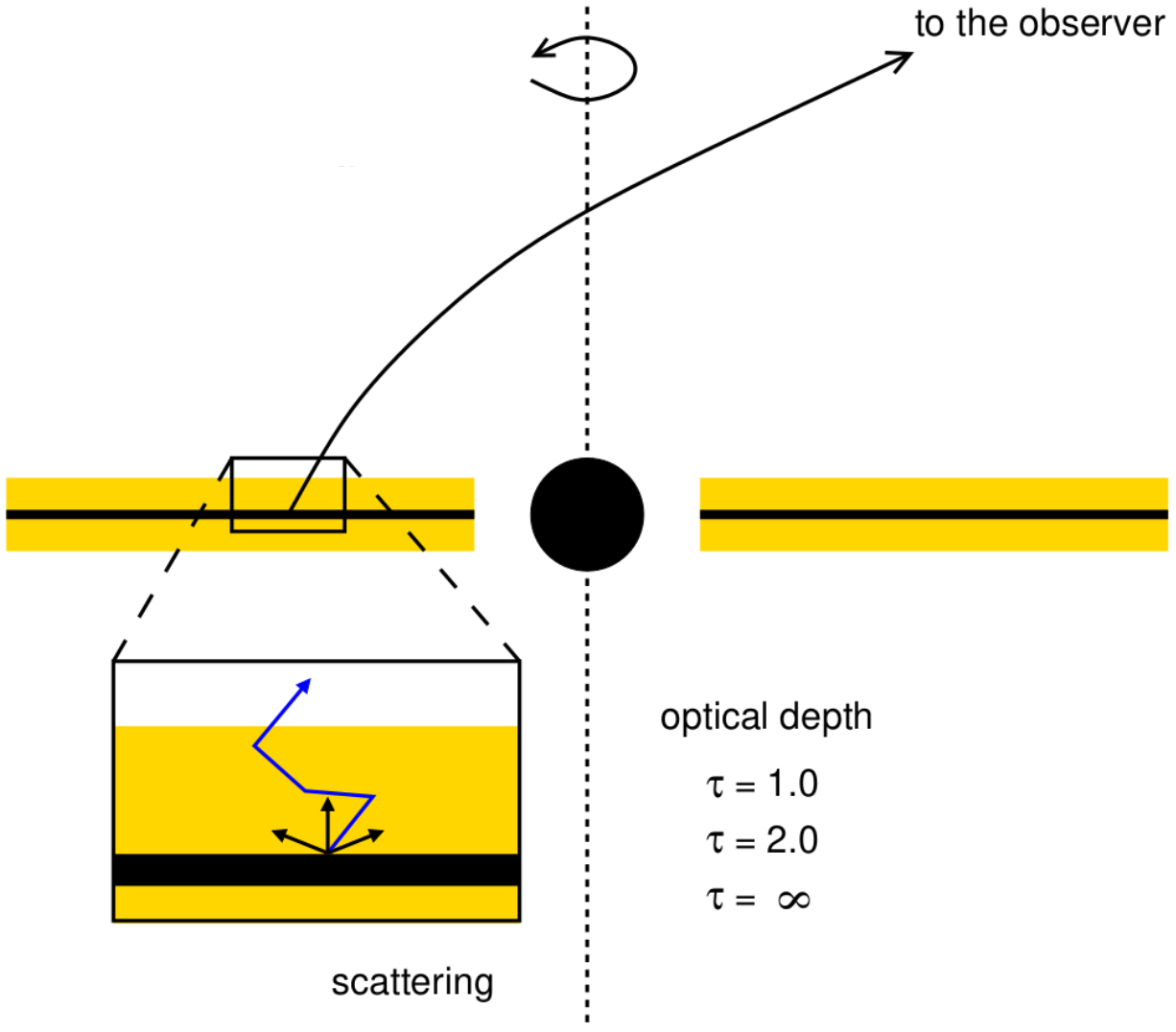}
		\caption{\footnotesize{Model of thermally radiating geometrically thin disc around Kerr black hole. The atmosphere is approximated in local comoving frame by slab of some optical depth $\tau$. In the most simple approach, we can assume the optical depth of $\tau \rightarrow \infty$ and only scattering events of thermal radiation reaching upwards. Once we precalculate the emission escaping the~semi-infinite atmosphere, we switch to GR ray tracing from equatorial plane towards an observer under general inclination $\theta$ with respect ot the principal axis, which coincides with the~Boyer-Lindquist coordinate $\theta$. Image courtesy of Michal Dovčiak.}}
		\label{thermal_illustration}
	\end{minipage}
\end{figure}

The semi-infinite plane-parallel atmosphere was also approached by Chandrasekhar and Sobolev in transmission problem \citep{Chandrasekhar1947, Sobolev1949, Chandrasekhar1960, Sobolev1963}, where the pure-scattering result is displayed in Figure \ref{analytical_semiinfinite}. The Loskutov-Sobolev approximation \citep{LoskutovSobolev1981} is also displayed for comparison, which takes into account additional effect of absorption, which increases polarisation due to effectively reduced symmetry: the photon paths are most likely vertical towards the slab surface, providing a preferred scattering angle for an observer at given inclination \citep{Lightman1979, Loskutov1979}. In addition, it accounts for internal sources within the atmosphere, which typically depolarize due to smaller mean free path through the atmosphere compared to the classical Milne problem, where the sources are located only at the bottom of the atmosphere \citep{Milne1921, LoskutovSobolev1981}. The details depend on~the~source function distribution with optical depth and the ratio of scattering to absorption opacities. In some cases, the vertical gradient of thermal source function can additionally polarize \citep{Harrington1969, Gnedin1978, Loskutov1979, Bochkarev1985}. And already \cite{Nagirner1962} pointed out that the~emergent polarisation angle may not always be parallel to~the~slab, if internal sources of radiation are considered and the result resembles an optically thin limb brightened atmosphere \citep{Gnedin1978,Matt1993,Davis2009}.

For the pure-scattering semi-infinite atmosphere in transmission, numerical solution to the guiding integro-differential equations is tabulated in e.g. \cite{Chandrasekhar1960}, but a useful empirical law can be used \citep{Viironen2004}:
\begin{equation}\label{viir}
	p = -\frac{1-\mu_\textrm{e}}{1+3.852\mu_\textrm{e}}11.71\% \, ,
\end{equation}
describing the monotonic increase with energy up to about $12\%$. The emergent polarisation angle is parallel to the disc in this case, as usual for scattering-dominated limb darkened atmospheres. The semi-infinite atmosphere with large absorption can be approximated by (\ref{compton_polarization}) or (\ref{thomson_polarization}) with polarisation angle parallel to the slab for photons emerging at large atmospheric depths and scattering once at the disc surface plane with scattering angle $\Theta$ coinciding with observer's inclination $\theta$ with respect to the disc's normal. Without the effects of absorption, the multiple scattering events for transmission through finite plane-parallel slab were studied by \cite{Sunyaev1985} with the resulting polarisation parallel to the slab for large optical thickness, but perpendicular for Thomson optical depth $\tau_\textrm{T} \lessapprox 3$, depending on the inclination angle. For optically thin case, the single-scattering approximation gives \citep{Sunyaev1985}
\begin{equation}\label{st85}
    p = \frac{1-\mu_\textrm{e}^2}{3-\mu_\textrm{e}^2} \, ,
\end{equation}
reaching $33\%$. The function (\ref{st85}) is monotonically decreasing with $\mu_\textrm{e}$ and less steeply than in (\ref{viir}). The corresponding polarisation is parallel to the projected slab normal, i.e. orthogonal to the scattering-induced polarisation vector from a~semi-infinite atmosphere. Polarisation by single scatterings from general axially symmetric structures was in exact form described by \cite{Brown1977}.

Modern analytical or semi-analytical codes focus for example on the X-ray polarisation arising from neutral or fully ionized disc, when illuminated by the coronal power-law, on two-phase coronae with relativistic electrons, or self-consistent treatment of the disc and corona, including multiple scattering orders and synchrotron radiation in particular geometries, or outflowing coronae with relativistic velocities that affects polarisation due to aberration effects \citep{Bonometto1970,Dolginov1970,Gnedin1978,Silantev1979,Williams1984a,Williams1984b,Sunyaev1985,Matt1993b,Haardt1993c,Poutanen1993, Nagirner1993, Poutanen1996, Poutanen1996a,Beloborodov1998,Nagirner2001,Silantev2007,Bianchi2010,Poutanen2010,Silantev2018,Silantev2020,Silantev2021,Silantev2022,veledina22,Loktev2022,Nagirner2022a,Nagirner2022b,Silantev2023,Poutanen2023,Loktev2023, Dexter2023}. The works of~\cite{Poutanen1996, veledina22} introduced a spectral fitting {\tt COMPPS} model, which has also a polarisation extension, but the extension is currently not available in interface with {\tt XSPEC} [private communication with the authors].

~\

The optically thick atmosphere for X-ray reflection or transmission with internal source distribution was numerically treated by many authors (either via radiative transfer equation solving or via MC methods) for the X-ray photospheres of black-hole accretion discs. For plain spectral output for example by \cite{Mitsuda1984} (the {\tt DISKBB} model for fitting of simple multicolor thermal disc emission), or \cite{Kriz1986, Hubeny1997,Hubeny1998, Davis2005, Davis2006b} (relativistically integrated thermal discs in hydrostatic equilibrium and the {\tt BHSPEC} model), or \cite{Suleymanov1992,Suleimanov2002,Suleimanov2007} (similarly), or \cite{Shimura1993,Shimura1995b,Shimura1995} (similarly), or \cite{Li2005} (the {\tt KERRBB} model for relativistic thermal disc emission), or \cite{Ross1993, Matt1993c, Ross1999, Ross2005, Ross2007} (the {\tt REFLIONX} models for power-law reflection from constant density discs), or \cite{Magdziarz1995, Zdziarski1996, Zycki1999} (the {\tt PEXRAV} and {\tt PEXRIV} models for power-law reflection from neutral and fully ionized constant-density disc, respectively, and the {\tt NTHCOMP} model for Comptonized power-law), or \cite{Kallman2001, Garcia2010, Garcia2011, Garcia2013, Garcia2016, Garcia2020} (the {\tt XSTAR} code and {\tt XILLVER} models for local power-law constant-density reflection), or \cite{Rozanska1996,Nayakshin2000,Nayakshin2001, Pequignot2001, Ballantyne2001, Dumont2002,Rozanska2002, Dumont2003, rozanska2008, Rozanska2011, vincent2016} (the {\tt NOAR}, {\tt TITAN} and {\tt ATM24} codes and comparisons of both constant-density and hydrostatic equilibrium power-law reflection), or \cite{Mehdipour2011, Rozanska2015, Petrucci2020, Gronkiewicz2023} (the radiative transfer treatment of warm coronal emission in AGNs), among others.

~\

Analytical or numerical treatment of locally radiating geometrically thin disc atmosphere is typically implemented into a GR ray tracing code, such as \cite{Dovciak2004, Dovciak2004b, Dovciak2008, Dovciak2011, Taverna2020,Taverna2021,Dovciak2022,Mikusincova2023} (the {\tt KY} codes applied to various different computations of emission in local comoving frame with the disc for various geometries, see below), or \cite{Garcia2014, Dauser2014, Dauser2022} (the {\tt RELXILL} code for relativistic lamp-post extension to {\tt XILLVER} tables, currently being developed also to treat various coronal geometries), or \cite{Niedzwiecki2016, Niedzwiecki2018, Niedzwiecki2019} (spectral relativistic lamp-post models), all building on the older works of e.g. \cite{Fabian1989,Laor1991,Matt1992a,Matt1992b, Matt1993d, Matt1996c, Agol1997,Martocchia2000,Beckwith2004}. Figure \ref{thermal_illustration} illustrates such relativistic embedding for thermally radiating geometrically thin discs, but similar sketch is applicable to the local disc reflection integration for geometrically thin discs. Slim disc models with relativistic extensions, suitable for observational fitting of e.g. ultra luminous X-ray sources (ULX) -- compact objects accreting in~super-Eddington regime --, were constructed by \cite{Kawaguchi2003, Straub2011, Vierdayanti2013, Caballero2017}, among others.

~\

We will compare our spectral results to some of the above studies to validate our more unique polarisation results, because numerical radiative transfer computations of X-ray polarisation near accreting black holes are scarce due to the complexity of the problem. We name the early works of \cite{Stark1981, Phillips1986, Nagendra1987, Matt1989, Laor1990, Coleman1990, Coleman1991, Matt1991, Matt1993, Matt1993b, Agol1996, Blaes1996, Matt1996, Hsu1998, Agol1998, Lee1999b, Agol2000, Silantev2002} and then more recently \cite{Dovciak2004, Gnedin2006, Silantev2008, Dovciak2008, Li2009, Davis2009, Schnittman2009, Dovciak2011, Marin2014c, Silantev2019, Silantev2019b, Taverna2020, Taverna2021} for the geometrically thin disc in transmission or reflection with various simplified assumptions, or combined with coronal contribution in selected geometries by \cite{Schnittman2010}, or for geometrically thick discs in the attempt by \cite{West2023} and warped discs by \cite{Cheng2016,Abarr2020b}. Particularly relevant to this dissertation (to be discussed in~depth later) is the lamp-post relativistic reflection of polarisation computed in~\cite{Dovciak2011} in simplified approach and the non-relativistic total polarisation gained by scattering and absorption of thermal disc emission computed by combined approach of radiative transfer solver {\tt CLOUDY} \citep{Ferland2013,Ferland2017} with MC simulator {\tt STOKES} \citep{Goosmann2007,Marin2012,Marin2015,Marin2018}, described in \cite{Taverna2021}. In the neighboring field of~accreting neutron stars, where often the role of magnetic fields is more critical, much theoretical progress motivated by the \textit{IXPE} observations is also seen \citep[e.g.][]{Sokolova2021, Suleimanov2023, Sokolova2023}. Lastly, there are two exceptional X-ray polarisation codes that focus on~the~Comptonization treatement in~coronae with MC method, including GR effects in Kerr spacetime. The {\tt MONK} code \citep{Zhang2019b, Zhang2022, Zhang2023} is predicting the~X-ray polarisation outcome for coronae of various geometries with seed photons from the disc and MC treatment of inverse Compton scattering in~the~corona. Perhaps of even more complexity, also including GR effects, is the~{\tt KerrC} code \citep{Krawczynski2022a}, which calculates also reflection from the accretion disc. Recently, another MC spectropolarimetric attempt for computation of coronal emission was obtained by \cite{Kumar2023}. Although some of these MC computations involve peculiar effects of partial ionization, or radiation returning to the disc, there are always some admittable simplifications to the detriment of detailed radiative transfer. A notable relativistic MHD simulation providing X-ray polarisation predictions was done by \cite{Schnittman2016} and in \cite{Moscibrodzka2023}, linked to a MC code.

~\

A specific category is the modeling of cold distant components of AGNs, where large libraries of torus X-ray spectral models exist \citep[see e.g. ][and references therein]{Liu2014, Furui2016, Buchner2019,Meulen2023}. These are typically separated from the inner disc-corona system computations and assume such external (central) source of emission. From polarisation perspective, these are easier to be treated by MC codes than radiative transfer equation solvers. We name the works of \cite{Ghisellini1994, Matt1996b, Ursini2023} and \cite{Goosmann2011, Marin2012, Marin2016b,Marin2018c,Marin2018b, Ursini2023} for dusty tori and polar outflows of uniform density, and pioneering attempts for X-ray polarisation signatures of~clumpy obscurers by \cite{Marin2012b, Marin2013, Marin2015b}. These could be perhaps accompanied in the future by results with the MC code {\tt SKIRT} \citep{Baes2011, Baes2015, Camps2015, Camps2020, Meulen2023} that is currently developing polarisation effects in the X-rays, similarly to~{\tt STOKES} [private communication with the authors]. However, some ambitious radiative transfer equation solvers that include X-ray polarisation estimate have been performed to evaluate the warm absorbers in AGNs \citep{Dorodnitsyn2010, Dorodnitsyn2011}. Many of these models can be also applied to the obscuring funnels of XRBs, ULXs, or accreting neutron stars or white dwarfs, which can be formed by vertically extended outflows or the accretion disc itself \citep[e.g.][]{Poutanen2007,Wielgus2016,Narayan2017,Koljonen2020,Neilsen2020,Miller2020,Ratheesh2021, Doroshenko2023}. Recently, analytical treatment of this problem also emerged in~\cite{Veledina2023} and \cite{Tomaru2023} attempted for X-ray polarisation MC modeling of polarisation of incident black-body emission of~soft-state XRBs gained by scattering in optically thin thermal-radiative winds, implementing polarisation into the {\tt MONACO} code \citep{Odaka2011}. The latter was followed by \cite{Tanimoto2023}, which modelled the X-ray polarisation specifically in the obscured Circinus Galaxy, using the MC {\tt MONACO} code linked to radiative hydrodynamic simulations.

\subsection{Codes explored in this dissertation}\label{codes_here}

We will now elaborate a bit more on the numerical codes and models adopted for~new results produced within this dissertation.

\subsubsection*{TITAN}

{\tt TITAN}\footnote{\,Public version not available at the time of writing this dissertation.} \citep{Dumont2003} is a non-LTE radiative transfer code suitable for~studying X-rays in hot photoionized media, such as atmospheres of accretion discs. It solves the energy balance, the ionization equilibria and the statistical equilibria (\ref{SE}) and the transfer equation in a plane-parallel geometry (\ref{RTE}), all for the lines and continuum up to about $26 \textrm{ keV}$. The iterative method is the~accelerated lambda iteration (ALI) \citep{Cannon1973, Scharmer1981, Olson1986, Hubeny2003}, ideal for fast convergence in spectral lines. In the latest version \textit{105v03f} that we use, {\tt TITAN} was updated for many atomic transitions \citep{Adhikari2015, Goosmann2016}. In total $\approx 4000$ X-ray spectral lines are considered. We will use the code for studying the radiative problem of~a~plane-parallel disc atmosphere in order to obtain the ionization structure, i.e. fractional abundance of each ion. Although {\tt TITAN} allows other enclosing relations, we will use it for a constant density atmosphere only. {\tt TITAN} does not have polarisation computations implemented and its treatment of Comptonization is simplistic. It should not be used for densities $n_\textrm{H} \gtrapprox 10^{18} \, \textrm{cm}^{-3}$ with the~atomic data and treatment of collisions that is available [private communication with the~authors].

\subsubsection*{STOKES}

{\tt STOKES}\footnote{\,Public versions in optical and UV energy bands available at \url{www.stokes-program.info} on the day of submission.} is a 3D MC simulator suitable for media where scattering and absorption are the dominant sources of opacity. The code is ideal for studying polarisation, because many critical line and continuum processes are implemented. Although various usage exists \citep{Marin2014b, Marin2017}, it was primarily developed to study AGNs in the optical and UV, and more recently the~code expanded to X-rays (version \textit{2.07} onwards). It includes the effects of~Compton down-scattering on free electrons, using the Klein-Nishina cross section (\ref{KN_formula}) and proper energy redistribution (\ref{cpt_energy}), but the angular redistribution is treated according to Thomson phase functions (see Section \ref{processes}). Then it includes scattering on bound electrons and dust (including dust absorption), the~bremsstrahlung absorption and emission (in emission for spectral lines only, i.e. the resonant bremsstrahlung; the free-free continuum emission is not implemented yet), and photoelectric absorption. It solves for several thousands of spectral lines. Polarisation due to resonant line scattering is treated according to \cite{Lee1994a, Lee1994b, Lee1997}. The~competing Auger effect is also implemented. On~the~other hand, as an MC code it cannot consistently solve for the~ionization structure, temperature and magnetic fields. The~densities and temperatures of the scattering and absorbing regions have to~be predefined and are used for various processes. But for example the~thermal emission or synchrotron radiation have to be included as external input emitters, in~case of~need. In the latest version (from \textit{v2.34} onwards), which was developed simultaneously with this dissertation and hence will not be used, {\tt STOKES} also includes the inverse Compton scattering according to \citet{Poutanen1993} with predefined temperature information. Although we will not use it, {\tt STOKES} also allows for~the~treatment of~circular polarisation, time variability and velocity of~scattering regions.

~\

The code allows for arbitrary polarisation of the~emission originating in predefined emission regions (directional or undirectional). We will use the notation $p_0$, $\Psi_0$ for the incident polarisation degree and angle of a photon that is to be reprocessed. The energy distribution of the primary radiation input can be self-defined or based on templates. A weighting algorithm is implemented to prevent cumulation of numerical noise in one part of the resulting spectrum. The scattering regions, which can coincide with emission regions, allow for non-trivial geometries (planar layers, cones, wedges etc.) and practical symmetry reductions, as MC simulations are known for large computational costs. Inside the scattering region, each photon can be either scattered (including true absorption and immediate reemission) or absorbed (and thermalized, although thermal reemission is neglected
) with some probability $P_\textrm{X}$. This is computed according to exponential probability law at distance $d$ travelled inside the scattering region with cross-section $\sigma_\textrm{X}$ for each process with particle species of uniform density $n_\textrm{X}$:
\begin{equation}\label{absorption_law}
	P_\textrm{X} = e^{-n_\textrm{X}\sigma_{X} d} \textrm{ .}
\end{equation}
The polarisation modification upon each reprocessing event is treated using the~Müller formalism. Although the photon cannot be emitted out of nothing outside of the emission region (e.g. through thermal radiation), the emission region can overlap with the scattering region. If the photon escapes the scattering medium and reaches a detector in particular predefined angular and energy binning, the polarisation information is transformed into the global reference frame and registered, including renormalized flux information. Additional routine {\tt ANALYZE} developed by the same authors is used to convert the raw output to conveniently readable files with the Stokes parameters per energy and angular bins of the detectors.

~\

We will use {\tt STOKES} for computations of reflection and transmission in plane-parallel atmospheres, where we will combine the MC approach with {\tt TITAN} to~precompute ionization structure of vertically stratified layers properly. This will be in depth described in Chapter \ref{chap01}. Apart from that, we will also use {\tt STOKES} alone for~computation of distant reflection in partially ionized equatorial and polar obscurers in Chapter \ref{chap04}. There we will assume homogeneous media in various simplified geometries and shrink the entire disc-corona emitting system to~a~point in~the~center of coordinates.

\subsubsection*{KY}

Lastly, let us focus again on the modeling of the inner-accreting regions and~the~necessary inclusion of relativistic effects. To solve the polarized radiative transfer in vacuum according to the layout sketched in Section \ref{relativistic}, we will use and adapt the well-tested {\tt KY} codes\footnote{\,Public versions available at \url{https://projects.asu.cas.cz/stronggravity/kyn} and \url{https://projects.asu.cas.cz/dovciak/kynsed} on the day of submission.}. There are numerous {\tt KY} routines available for solving of polarized radiative transfer in Kerr space-time: from solving of relativistically smeared spectral lines \citep{Dovciak2004, Dovciak2004b, Dovciak2004a} (including convolution models), to thermal radiation emitted from the disc \citep{Dovciak2008, Taverna2020, Taverna2021, Mikusincova2023}, to coronal emission added with simplified reflection from the disc \citep{Dovciak2011} and broadband spectra of lamp-post coronae with simplified reflection and thermalized continuum \citep{Dovciak2022}. All of these codes depart from the principles sketched in Section \ref{combined} and all relativistic effects are taken into account, apart from returning radiation. Returning radiation is not implemented yet to the codes apart from the {\tt KYNBBRR} routine \mbox{\citep{Taverna2020, Mikusincova2023}}, computing total disc thermal emission for a distant observer, which allows for simplified returning radiation effects, using albedo parameter $0 \leq A \leq 1$ and reflecting the same spectropolarimetric energy distribution of photons. Although we provided the main results in~Section \ref{relativistic} in the most commonly known Boyer-Lindquist coordinates, the~{\tt KY} codes solve (\ref{geodesic}), (\ref{deviation}) and (\ref{dov2.9}) in the so-called Kerr ingoing coordinates for numerical purposes \citep{Misner1973, Dovciak2004}. Symmetry of~the~coronal geometries adopted allow for largely simplified treatment of radiation in~between the corona and~the~disc, in between the corona and~the~observer and in between the disc and the observer. The code produces unfolded Stokes parameters for a distant observer (\ref{totalpol}), renormalized for a general source distance $r_0 = 1$ Mpc.

~\

\begin{figure}[h]\centering
	\includegraphics[width=0.43\textwidth]{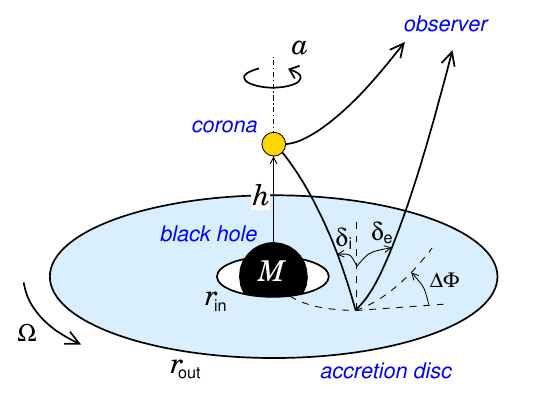}
        \put(-175,120){\scalebox{1.14}{a)}}
        \includegraphics[width=0.57\textwidth]{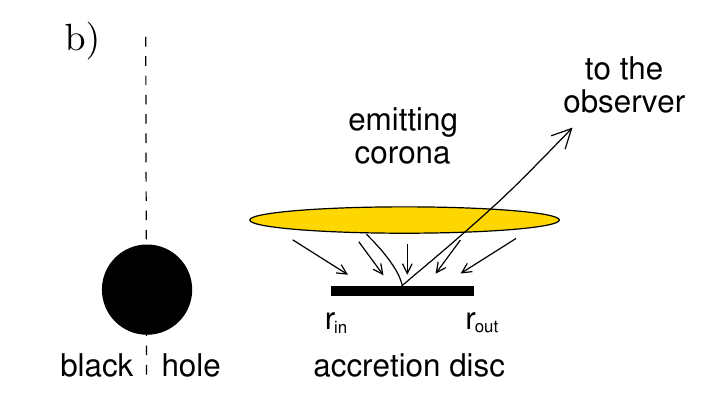}
	\caption{\footnotesize{Two model geometries of a corona above an accretion disc: a) the lamp-post model, b) the extended coronal model. The main model parameters are labeled. Images from \cite{Dovciak2004a, Dovciak2014}.}}
	\label{corona}
\end{figure}

In this dissertation, we will adapt the lamp-post and extended coronal geometries and the associated reflection routines {\tt KYNLPCR} and {\tt KYNXILLVER}, which use the local spectral reflection tables from {\tt NOAR} and {\tt XILLVER} codes and polarized reflection given by single-scattering Chandrasekhar's formulae (\ref{chandra1}) \citep{Dovciak2011, Marin2018c, Marin2018b}. A sketch of the two simplified disc-corona setups explored is in Figure \ref{corona}, including the parametrization. As previously described, the inner-accreting regions will be essentially computed in two different phases. First, we will compute locally the reflection or transmission with {\tt TITAN} and {\tt STOKES}, assuming a semi-infinite atmosphere, which will be described in Chapter \ref{chap01}. Then, we will assume a global view where GR radiative transfer in~vacuum with {\tt KY} is applied on a geometrically thin disc, which will be described in~Chapter \ref{chap02}. Hence, the precomputed semi-infinite atmosphere in tabular format will be used as infinitely geometrically thin in the second step, following the~essentials of the Novikov-Thorne accretion theorem. Assuming such two distinct local and global scales that can be computationally separated, we inexpensively combine detailed radiative transfer inside the critical top layers of the disc, simplified accretion physics and relativistic effects.

~\

For the lamp-post model, we will assume an isotropically emitting point-like corona at height $h$ above the center of the black hole (ranging in principle from the~event horizon all the way to relativistically nearly ``flat'' regions beyond $100 \, r_\textrm{g}$). Local reflection in comoving frame with the accretion disc can be then described by the previously defined three angles $\mu_\textrm{i}=\cos{\delta_\textrm{i}}$, $\mu_\textrm{e}=\cos{\delta_\textrm{e}}$, $\Phi_\textrm{e}$ [see (\ref{chandra1}) and Figure \ref{corona} for a sketch]. For the combined signal with the~primary emission we also have to specify the scaling of primary to reflected flux $N_\textrm{p}/N_\textrm{r}$, which in the model is unity for self-consistent treatment of corona with a given isotropic luminosity $L/L_\textrm{Edd}$ and reflection fraction according to the local reflection tables. The~extended corona is not parametrized by height, as we assume it is geometrically thin and glued at the top of geometrically thin disc, i.e. still within the black-hole equatorial plane. However, we have to define a radial profile of ionization, given by a power-law index $q$ and the ionization parameter normalization $\xi_0$:
\begin{equation}
	\xi = \xi_0 r^{-q} \, .
\end{equation}
The {\tt KY} codes allow integration of smaller sections of the disc given by $r_\textrm{in}$, $r_\textrm{out}$ and $\phi_\textrm{min}$, $\Delta \phi = \phi_\textrm{max} - \phi_\textrm{min}$ boundaries, in order to compute the effects of partial obscuration and flares. We will only work with radial disc truncation in this dissertation, due to the aforementioned insensitivity of current X-ray polarimeters to short-term variability in the accretion flow. The codes also allow for global redshift $z$ and time dependent treatment of emission, which we will not use in~this work. The same holds for non-constant radial profile of the disc's density.

\newpage
\ 
\vspace{4.5cm} 
\begin{figure}[h]
        \centering
	\includegraphics[trim={0cm 0cm 0cm 0cm},clip,width=0.6\textwidth]{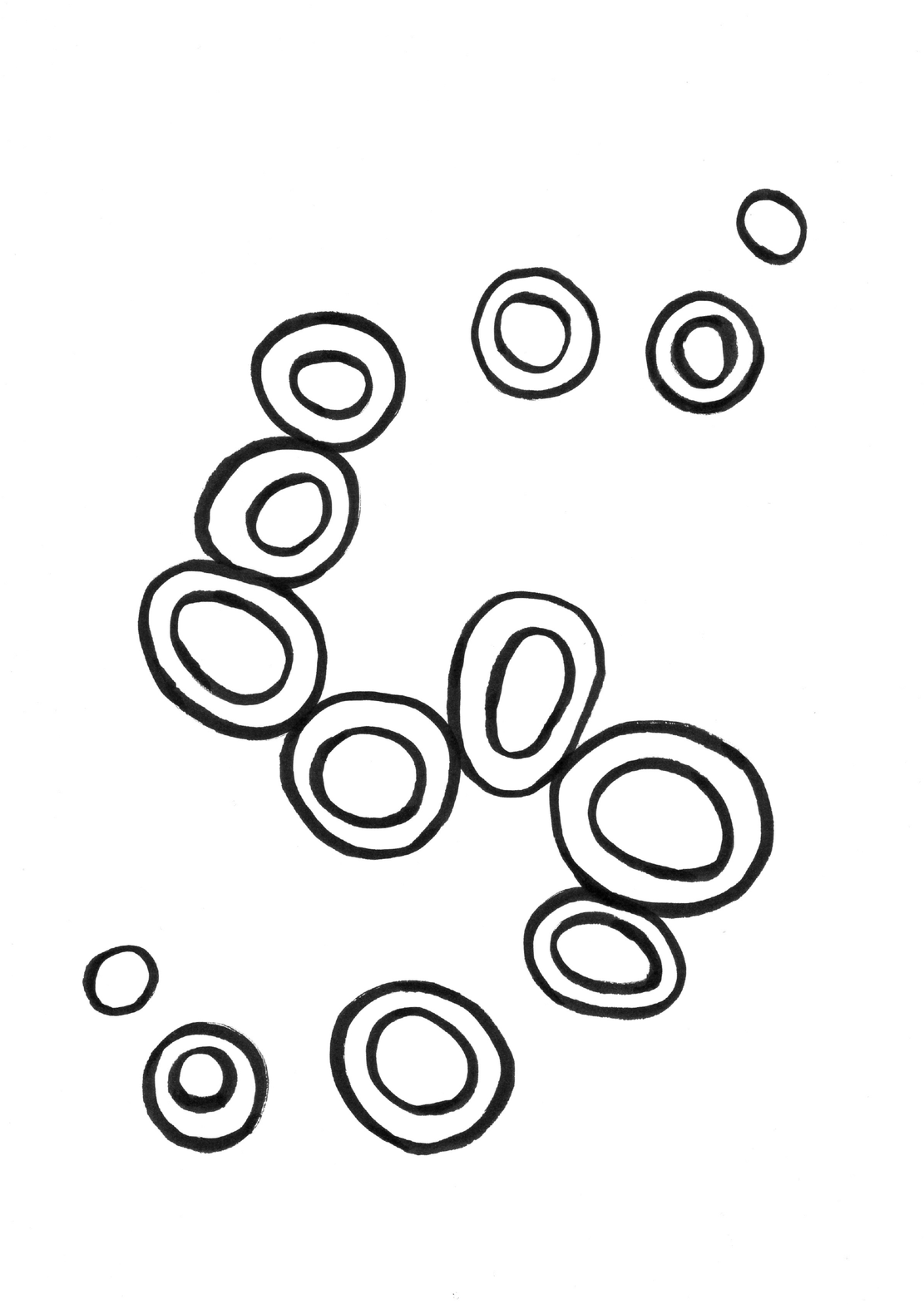}
\end{figure}
\chapter[Emission from a partially ionized accretion disc \newline in a local comoving frame]{Emission from a partially ionized accretion disc in a local comoving frame}\label{chap01}

This chapter describes modeling of local accretion disc emission in plane-parallel approximation, either through reflection or transmission. No SR or GR effects are assumed. We assume a passive plane-parallel atmosphere comoving with the~source of illuminating radiation.

\section{Power-law reflection off a disc's atmosphere}\label{reflection_tables_TS}

We begin with the results of reflection of a coronal power law from a constant-density slab of high optical thickness held in photoionization equilibrium. We use the combination of {\tt TITAN} and {\tt STOKES} (\textit{v2.32}) codes, which allows for accurate treatment of partial ionization effects. Because {\tt TITAN} is not equipped with polarisation computations, we couple it to {\tt STOKES}, which is built for accurate polarisation treatment. Conversely, the {\tt STOKES} MC simulation cannot solve self-consistently for the ionization structure, thus an iterative solver like {\tt TITAN} is required for detailed studies of hot X-ray photoionized media. The core of~the~presented model was already achieved as part of \cite{Podgorny2020}, including normalization and production of the final model tables in Flexible Image Transport System (FITS) format \citep{FITS} and a brief discussion. We examined and cross-validated the resulting spectropolarimetric tables as part of~this dissertation in greater detail, which is summarized in \cite{Podgorny2021, Podgorny2023}. In addition to these publications and the older tables used there, a~new version\footnote{\,Available as \textit{v2} to the original tables at \url{https://doi.org/10.6084/m9.figshare.16726207} on the day of submission, including documentation.} was recently created. This version contains only two changes: (i) a unified grid in $\xi$, which was in all previous publications not available (we worked with different $\xi$ values per each $\Gamma$) due to a technicality, and (ii) $100\%$ vertically polarized illumination for those tables that were previously computed for $100\%$ horizontally polarized illumination. The tabular model is now ready-to-use for~manipulation in {\tt XSPEC}, as the FITS files with unified $\xi$ conform to Office of~Guest Investigator Programs (OGIP) standard \citep{OGIP} of {\tt XSPEC}.

Table \ref{reflection_tables} describes the current parameter values, for which also the results below are drawn and which are physically motivated. Each model table contains one Stokes parameter and is computed for $100\%$ vertically polarized ($p_0 = 1$, $\Psi_0 = 0^{\circ}$), $100\%$ diagonally polarized ($p_0 = 1$, $\Psi_0 = 45^{\circ}$), and unpolarized ($p_0 = 0$) incident radiation. Three incident polarisation states allow immediate interpolation for~arbitrary polarisation state that will be used in Section \ref{chap02}. The source of~primary radiation is located directly above the slab. Each table was computed for $N_\textrm{tot} = 7 \times 10^8$ simulated photons, which is sufficient for investigating the elementary spectral and polarisation properties with negligible presence of~inevitable MC numerical noise.
\begin{table}
        \footnotesize
	\centering
	\caption{\footnotesize{Description of the FITS tables representing the local reflection tables computed, which were recently updated for ionization parameter grid that does not depend on $\Gamma$.}}
	\begin{tabular}{ll}
		\hline
		\hline
		Number of tables & 9 (vertical, diagonal, and no incident polarisation \\
		&  versus the $I$, $Q$, $U$ output)  \\
		Stokes parameters units            & $[\textrm{counts} \cdot \, \textrm{cm}^{-2} \cdot \, \textrm{s}^{-1}]$        \\
		Energy range     & $0.1 \textrm{ keV}$ to $100 \textrm{ keV}$                       \\
		Energy binning      & 300 bins, $\Delta \log E = 0.01$                                             \\
		$\Gamma$            & {\{}$1.4, 1.6, 1.8, 2.0, 2.2, 2.4, 2.6, 2.8, 3.0${\}}                                                             \\
		$\xi$ $[\textrm{erg} \cdot \textrm{cm} \cdot \textrm{s}^{-1}]$           & {\{}$5, 10, 20, 50, 100, 200, 500, 1000, 2000, 5000, 10000, $  \\
		&
		\multicolumn{1}{l}{$\ \ \ 20000${\}}}          \\
		$\mu_\mathrm{i}$            & {\{}$0.0,0.1,0.2,0.3,0.4,0.5,0.6,0.7,0.8,0.9,1.0${\}}                    \\
		$\mu_\mathrm{e}$            & {\{}$0.025,0.075,0.175,0.275,0.375,0.475,0.575,$   \\
		&
		\multicolumn{1}{l}{$\ \ \ 0.675,0.775,0.875,0.975${\}}}                        \\
		$\Phi_\mathrm{e}$           & {\{}$7.5^{\circ},22.5^{\circ},37.5^{\circ},52.5^{\circ},67.5^{\circ},82.5^{\circ},97.5^{\circ},$                               \\
		&
		\multicolumn{1}{l}{$\ \ \ 112.5^{\circ},127.5^{\circ},142.5^{\circ},157.5^{\circ},172.5^{\circ},187.5^{\circ},$}                                              \\
		& \multicolumn{1}{l}{$\ \ \ 202.5^{\circ},217.5^{\circ},232.5^{\circ},247.5^{\circ},262.5^{\circ},277.5^{\circ},$}                                \\
		& \multicolumn{1}{l}{$\ \ \ 292.5^{\circ},307.5^{\circ},322.5^{\circ},337.5^{\circ},352.5^{\circ}${\}}}         \\
		Extensions       & Primary Header  -- description of the tables                                             \\
		& `PARAMETERS’      -- parameter values                                                      \\
		& `ENERGIES’     -- low and high energy bin edges                                      \\
		& `SPECTRA’  -- values of the Stokes parameters and \\
		&
		\multicolumn{1}{l}{corresponding model parametric values}           \\ \hline \hline
	\end{tabular}%
	\label{reflection_tables}
\end{table}

~\

The results were obtained for neutral hydrogen density $n_\textrm{H} = 10^{15} \ \textrm{cm}^{-3}$; hence, the main application are accretion discs of AGNs. This is different from the~follow-up section on X-ray thermal radiation passing through a disc's atmosphere, which is clearly applicable to XRBs and which was studied for a~range of slab densities. The reflection is computed up to Thomson optical depth of~\mbox{$\tau_\textrm{T} = \sigma_\textrm{T} n_\textrm{H}H = \sigma_\textrm{T} \times 10^{25} \ \textrm{cm}^{-2} \approx 7$}, where $H$ is the total geometrical thickness of the slab studied. No radiation source is considered on the opposite side. The~solar abundance is from \cite{Asplund2005} with $A_\textrm{Fe} = 1$, neglecting the~presence of dust. The high and low-energy cut-offs of the primary power-law were considered exponential in {\tt TITAN} (for the computation of the ionization structure) and sharp in {\tt STOKES} (for the computation of spectropolarimetric properties), cut at $E_\textrm{min} = 10^{-1.1} \, \textrm{keV} \, \approx 0.079$ keV and $E_\textrm{max} = 10^{2.4}  \, \textrm{keV} \, \approx 251.189$ keV. We~used angle-averaged illumination in {\tt TITAN} for technical reasons, but we note that the vertical ionization structure should be incident angle dependent if pencil-like beams were studied, which leaves some deficiency on the results in~terms of~accuracy at high optical depths.

A much larger simplification with respect to real atmospheric conditions is the assumption of constant density. When studying the reflection in full hydrostatic equilibrium, well-known thermal ionization instability occurs along with narrow two-phase zones in the medium between the disc and the corona \citep[see e.g.][]{Begelman1983a,Rozanska1996}, even though the detailed geometry and boundary conditions of the illuminated atmosphere are unknown \citep{Ballantyne2001, Ross2005, Ross2007}. Reflection from slabs in~hydrostatic equilibrium were previously studied from spectroscopic point of~view, but comparisons to constant-density models are scarce, as there is no ionization parameter defined for the former case and it is difficult to define an~average atmospheric density \citep{Rozanska2002}. It is however clear that the~two classes of models vary in both continuum shapes and spectral line profiles \citep{Nayakshin2000, Nayakshin2001, Pequignot2001, Ballantyne2001, Rozanska2002, Dumont2002, Ross2007, rozanska2008, Rozanska2011, vincent2016}. It is unclear how refraining from the~constant-density assumption would quantitatively affect the~polarisation properties of the emergent spectra, which are of our prime interest, as they have not been numerically computed with detailed treatment of~partial ionization before neither in hydrostatic equilibrium, nor assuming constant density. Qualitatively, however, we know that if the upper atmospheric layers are heated, they expand and due to lower density their ionization level increases; and on the contrary the ionization level significantly drops below the~instability point \citep{Nayakshin2000,Done2007b,Done2010}. For a full hydrostatic equilibrium, we would thus expect a shift of polarisation towards the highly ionized computations shown below for a constant density case.

\subsection{Spectropolarimetric properties}\label{sp_props_reflection}

\begin{figure}[h]\centering
	\includegraphics[width=1\textwidth]{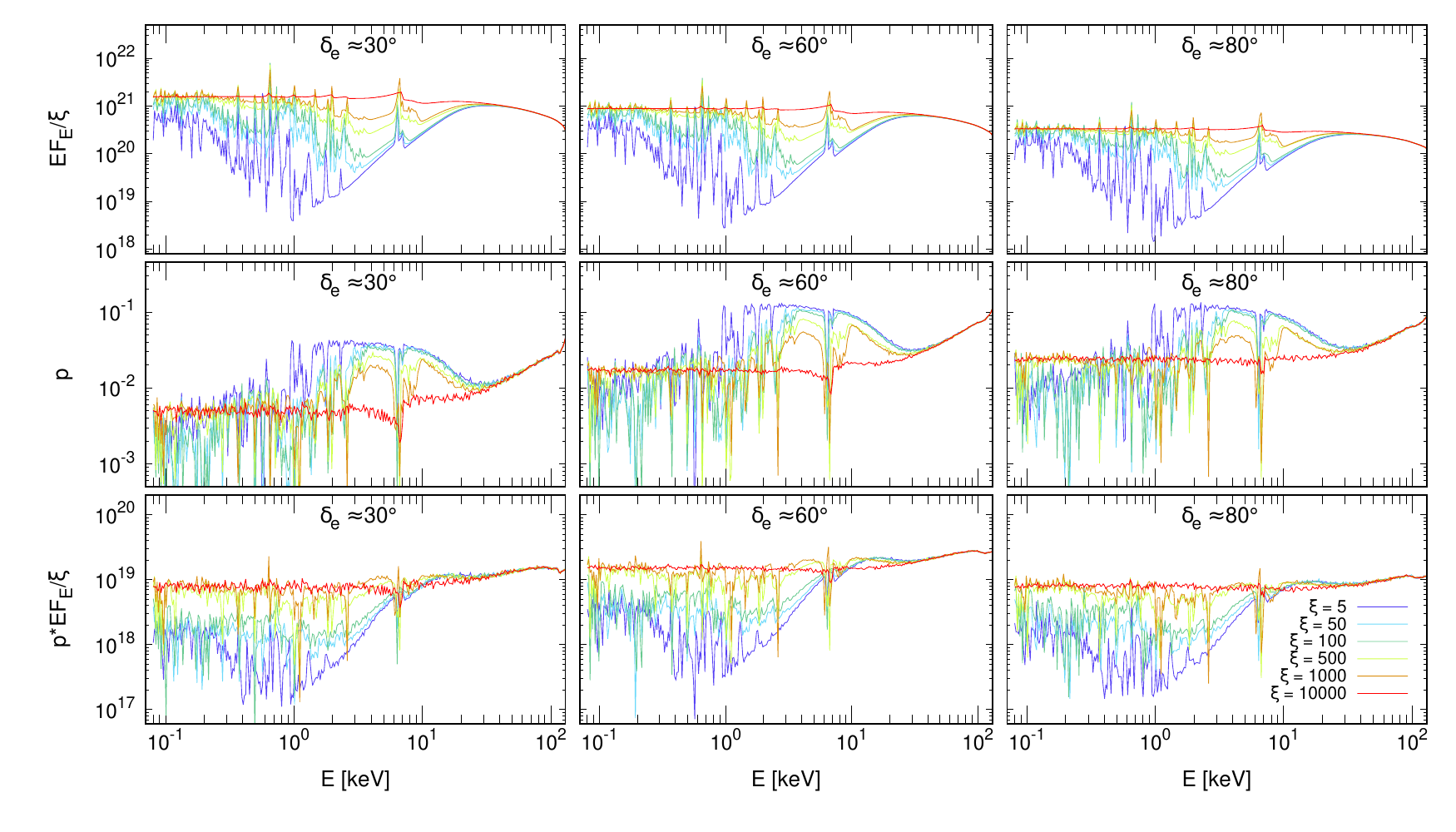}
	\caption{\footnotesize{The locally reprocessed emission of the unpolarized incident power-law in~the~constant-density atmosphere, as computed by {\tt TITAN} and {\tt STOKES} and angle-averaged in $\delta_\textrm{i}$ and~$\Phi_\textrm{e}$. The color code corresponds to different values of the ionization parameter $\xi$ $[\textrm{erg} \cdot \textrm{cm} \cdot \textrm{s}^{-1}]$. We show the values of $\mu_\mathrm{e} = 0.875, 0.475, 0.175$ (left to right panels, $\delta_\mathrm{e} = 28.96^\circ, 61.65^\circ, 79.92^\circ$). Top row: the spectra corrected for the slope of primary irradiation $\Gamma = 2$ and divided by $\xi$ to account for amplitude differences. Middle row: the corresponding polarisation degree versus energy. Bottom row: the corresponding polarized flux, i.e. the values in the top row multiplied by the values in the middle row.}}
	\label{local_not_in_mue}
\end{figure}

Figure \ref{local_not_in_mue} shows the reprocessed spectra, polarisation degree versus energy and~polarized flux for various $\xi$ values, three emission inclination angles $\delta_\textrm{e}$ and~for~unpolarized primary with $\Gamma = 2$. The results are integrated in incident inclination angles $\delta_\textrm{i}$ and emergent azimuthal angles $\Phi_\textrm{e}$, in order to gain higher energy resolution in polarisation and in order to approximate the situation of an~emitting disc observed under inclination $\delta_\textrm{e}$ with no GR or SR effects. In the spectra, we obtained the characteristic photon redistribution by disc reprocessing. We see the~Compton hump at~around 20 keV, the most prominent iron line complex at 6--7 keV and a forest of lines below 2 keV. The spectral shape changes with ionization according to the literature \citep[compare with e.g. Figure~\ref{AGN_reflection}, adopted from][]{Fabian2000}.

~\

In polarisation, we obtained depolarisation in spectral lines and the continuum changes according to the (multiple-)scattering and absorption opacity contributions. The higher the absorption, the higher the single-scattering contribution is and the geometry is effectively reduced to dominant single-scattering angles. We see almost energy-independent polarisation for the cases with highest ionization with lack of emission lines. The polarisation is generally order-of-magnitude lower from the tabular FITS values, which are not integrated in $\delta_\textrm{i}$ and $\Phi_\textrm{e}$. This is because in particular incident and emission angle combinations with $\Theta \approx 90^{\circ}$ as the~dominant single-scattering angle, the polarisation can easily reach values close to unity, while the angle-averaged results mean vectorial addition of~many individual polarisation directions, hence depolarisation. Multiple scatterings also generally depolarize, because any leading directionality is washed out. We observe a reverse imprint of the spectral Compton hump in polarisation due to~depolarisation by multiple scatterings. At the hard energy tail, we see increase in~polarisation, because the Compton down-scattering energy redistribution results in~more scatterings per emerging photon towards the lower energies. And~in~general, the~photons penetrate deeper in the slab for~smaller $\delta_\textrm{i}$. In deeper layers the~matter is less ionized and more absorbing, so the vertical stratification diminishes the depolarisation (and photon energy change) by multiple scatterings for those photons that reached deep inside and emerged unabsorbed. We would, however, require a combined simulation of~the~photosphere for thermally radiating inner disc (in the X-rays more relevant for XRBs than AGNs) and power-law reflection from the other side of the atmosphere \citep[see e.g.][]{Rozanska2011} to properly account for ionization, temperature and density structure closer to the disc mid plane. This is left for future studies and~the~produced reflection tables should be thus used rather for AGN accretion discs.

~\

Despite the accurate treatment of polarisation in resonantly scattered lines in {\tt STOKES} \citep{Lee1994a, Lee1994b, Lee1997}, we do not see any gain of polarisation in spectral lines, as is observed e.g. in \cite{Dorodnitsyn2010, Dorodnitsyn2011} for the problem of AGN warm absorbers. However, more detailed investigation of polarisation in individual spectral lines would require higher spatial\footnote{\,The spatial resolution was 500 vertically stratified layers in {\tt TITAN}, which was subsequently averaged up to 50 vertical layers in {\tt STOKES} for computational reasons.} and energy resolution.

~\

The flux and polarisation in hard X-rays would obtain a slightly different shape, if Comptonization was included. Qualitatively, we expect an effect of~smoothing out the hard-energy spectral tail and slight decrease of polarisation due to additional photon energy shifts \citep{Poutanen1993, Krawczynski2011, Garcia2020}. We estimate that for the studied slab temperatures ($T \approx 10^7$ K) this would begin to play a role at $\gtrapprox 10$ keV \citep{Prince2004} (i.e. above the \textit{IXPE} range), but a comparison with results using the latest version of~{\tt STOKES} \textit{v2.34} is still awaited. Polarisation through inverse Compton scattering is also related to the synchrotron emission, which emerges in optically thin media and is expected to be low for optically thick accretion discs.

~\

The \textit{angle-averaged} polarisation angle is in~all cases $\Psi \approx 90^{\circ}$ and constant in~energy (apart from the~unpolarized fluorescent spectral lines, for which the~default polarisation angle is undefined). The constancy with energy holds even for~individual local reflection tables, although, depending on the local disc normal, the energy-independent value may differ with respect to individual rays studied (see below).

~\

The dependency on inclination of the spectra follows from reflection limb darkening, as more photons are absorbed in the disc surface direction. The~polarisation is known to increase with inclination on the contrary, because if the~observer is looking at the reflecting surface pole-on, the individually polarized beams cancel out in polarisation from directions symmetric around the disc normal. The~most prominent rise in polarisation degree occurs up to $\theta \approx 60^{\circ}$
. An observer that would be further inclined sees similar polarisation in the Newtonian picture, while the average flux keeps decreasing towards the edge-on viewing angles. Hence, the~polarized flux reaches maximum at $\theta \approx 60^{\circ}$, which suggest the~most favorable inclination for detection of X-ray polarisation generated by plain disc reflection. The polarized flux panels not only support the antagonistic behavior of the flux and polarisation degree with energy, but symbolize the~original Stokes parameters $Q = Ip\cos{2\Psi}$ and $U = Ip\sin{2\Psi}$ without any sign oscillations (because the~continuum of $\Psi$ is constant in energy), again abstracted away from the slope and amplitude changes.

~\

\begin{figure}[h]
	\centering
	\begin{minipage}[t]{0.47\textwidth}
		\includegraphics[width=\textwidth]{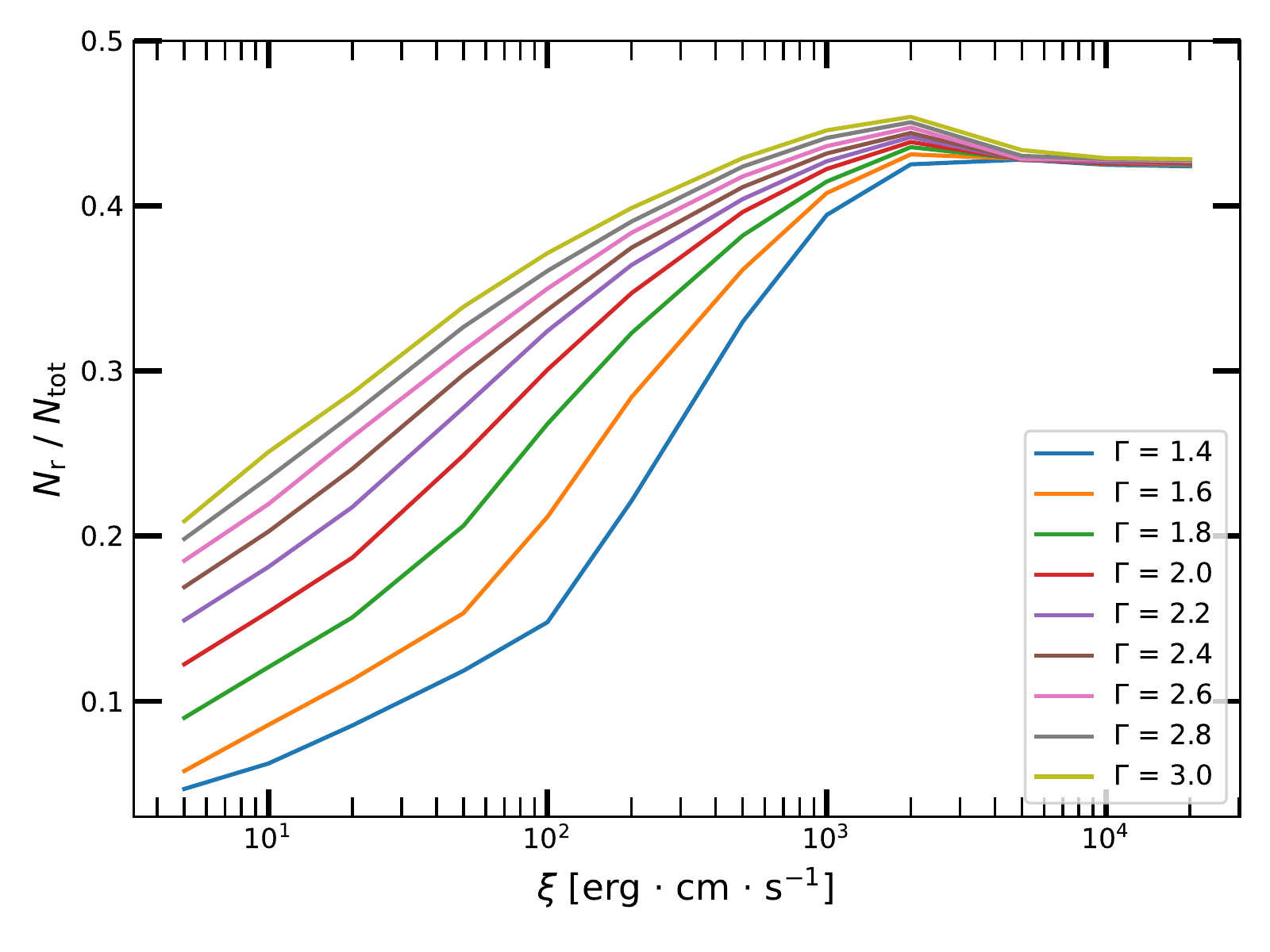}
		\caption{\footnotesize{The $\xi$-dependency of the fraction of reflected photons with respect to the number of incident photons in the total energy range studied, integrated in incident inclination angle $\delta_\textrm{i}$. The color code corresponds to~different values of $\Gamma$.}}
		\label{reflectivity}
	\end{minipage}
	\hfill
	\begin{minipage}[t]{0.51\textwidth}
		\includegraphics[width=\textwidth]{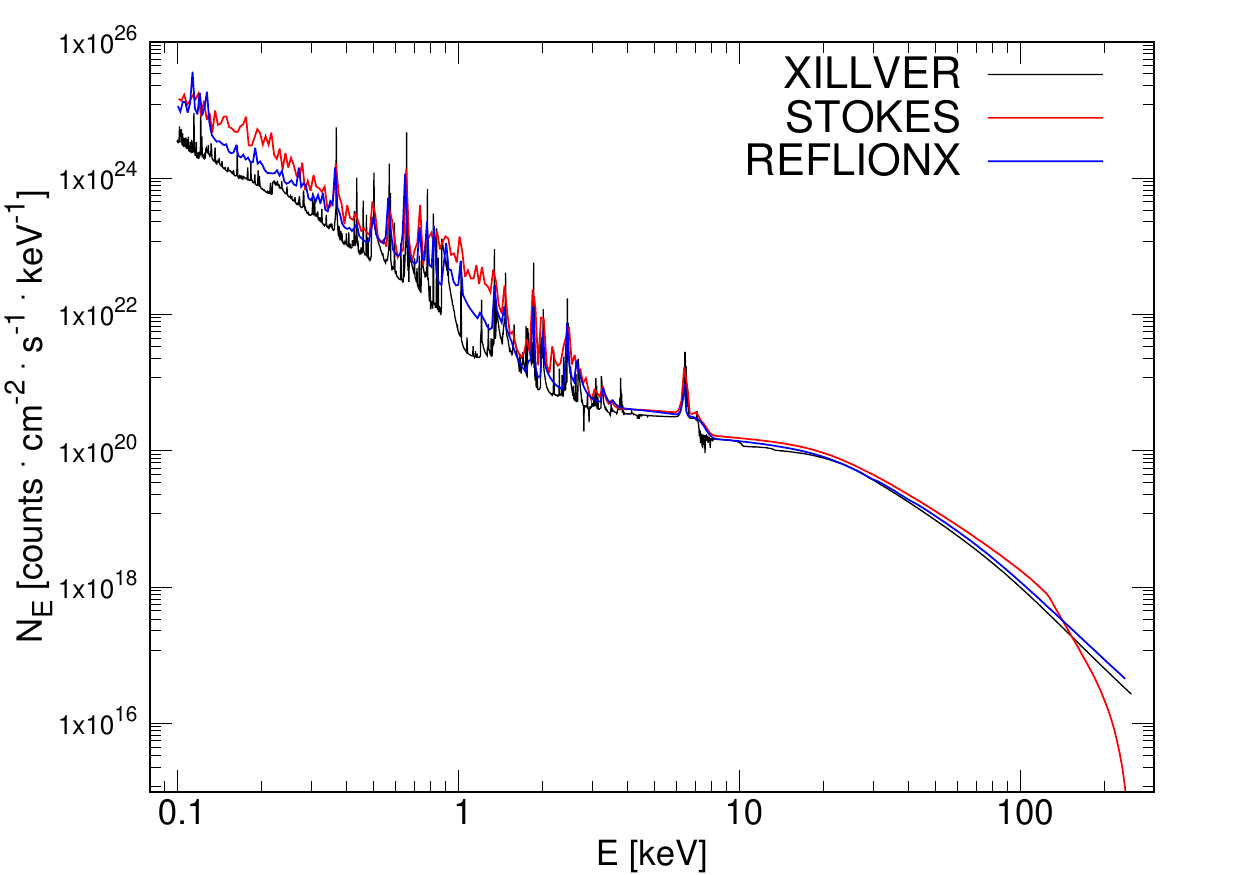}
		\caption{\footnotesize{Comparison of spectra for $\Gamma = 2.2$ and $\xi = 100 \, \, \textrm{erg} \cdot \textrm{cm} \cdot \textrm{s}^{-1}$, as computed by the~{\tt XILLVER} model (black), {\tt REFLIONX} model (blue) and {\tt TITAN} and {\tt STOKES} for unpolarized primary (red), all normalized accordingly and~angle-averaged.}}
		\label{xillver_reflionx_compare}
	\end{minipage}
\end{figure}

From spectroscopic point of view, an interesting quantity is the disc's reflectivity, which is considered in thermal reverberation disc-corona studies, where it is assumed that a fraction of X-ray coronal radiation is absorbed and thermalized \citep{Kammoun2019b, Dovciak2022}. We will plot it here as the~number of photons $N_\textrm{r}$, which are unabsorbed and reprocessed into all directions (i.e. integrated in $\delta_\textrm{e}$ and $\Phi_\textrm{e}$) divided by the total number of inserted photons $N_\textrm{tot}$ in~the~final {\tt STOKES} MC simulation. Apart from changing the slope of the reflected radiation, higher values of $\Gamma$ increase the presence of spectral lines in our reprocessed emission tables, similarly to the effects described in \cite{Garcia2013} for {\tt XILLVER} computations. If integrated in energy and in incident inclination angle $\delta_\textrm{i}$, which only slightly increases reflectivity due to higher fraction of~photons unabsorbed in the deeper atmospheric layers, the total reflectivity then rises with both $\Gamma$ and $\xi$, forming a characteristic ``S'' shape with $\xi$ for low $\Gamma$. This is displayed in Figure \ref{reflectivity} as another perspective of the well-known reflection spectral change with $\xi$.

~\

The dependency of the reflected spectra on incident polarisation is negligible. The dependency of the reflection-induced polarisation on the incident polarisation is substantial for individual combinations of \{$\delta_\textrm{i}, \delta_\textrm{e}, \Phi_\textrm{e}\}$, but if angle-averaged, it becomes of the order of incident polarisation degree. We will examine this dependency in detail altogether with GR effects for a distant observer in Section \ref{chap02}. The cases of $100\%$ polarized radiation modelled locally here are by itself ``unphysical'' in a sense that they were computed in order to subsequently interpolate for realistic low values \citep[$p_0 \lessapprox 10\%$,][]{Beheshtipour2017, BeheshtipourThesis, Tamborra2018, Ursini2022, Krawczynski2022a} of coronal polarisation.

\subsection{Comparison to other models}\label{local_comparison}

In this section, we directly juxtapose the spectra of {\tt XILLVER}, {\tt REFLIONX} and~our model for the same $\xi$ and $\Gamma$, although a thorough comparison between our model and the available constant-density spectral reflection tables would be also needed in terms of the used cut-offs, abundances or reflectivity. We overlap the three spectral models in Figure \ref{xillver_reflionx_compare} for one case of $\xi$ and $\Gamma$ and in Figure \ref{compare_all_xst} in Appendix \ref{supplementary} for a few other examples at various corners of the parameter space. The necessary angle averaging and renormalization of the three models was done according to the procedures described in \cite{Podgorny2020, Podgorny2021}, which proves the correct analytical transformation of the {\tt STOKES} output to physical units of our model stored in the FITS files described by Table \ref{reflection_tables}. The~results computed with {\tt TITAN} and {\tt STOKES} are generally closer to the {\tt REFLIONX} tables than to the {\tt XILLVER} tables. The same slab densities are assumed.

Without extensive investigation we can only state general reasons for possible discrepancies, but we note that even the differences between the widely used {\tt REFLIONX} and {\tt XILLVER} tables are poorly understood \citep{Garcia2013}. The~majority of the disagreement lies in the soft X-rays where many spectral lines occur and the spectra are more sensitive to the cut-off selection and abundances \citep[cf. the comparisons between {\tt XILLVER} and {\tt PEXRAV} and {\tt PEXRIV} in][where also the largest dissimilarities occur at soft X-rays]{Garcia2013}. The~abundances used for {\tt REFLIONX} are from \cite{Morrison1983} and for {\tt XILLVER} are from \cite{Grevesse1998}, i.e. more outdated to our model \citep[see the~discussion in][]{Grevesse2007}, but in \cite{Garcia2013} it is argued that the~choice of atomic data does not noticeably affect the result. There are, however, at~least two times more X-ray spectral lines considered in {\tt XSTAR}, forming the basis of~the~{\tt XILLVER} tables, than in {\tt TITAN}. Also, the {\tt XILLVER} tables were computed up to~$\tau_\textrm{T} = 10$, which may result in higher accuracy. We note that the~resolution of~current X-ray polarimeters is far beyond resolving individual spectral lines and our model is motivated by numerical simulation of polarisation, it should not be used for accurate spectral data fitting.

~\

At hard X-rays, the spectra are nearly identical. We stress that our MC results were obtained by a completely different numerical method than {\tt XILLVER} with {\tt XSTAR}, which simultaneously solves the equations of radiative transfer, energy balance, and ionization equilibrium in a Compton-thick, plane-parallel medium, using the Feautrier method \citep{Mihalas1978}, or {\tt REFLIONX}, which utilizes the Fokker-Planck diffusion equation, including a modified Kompaneets operator \citep{Ross1978, Ross1979}. The fact that in mid and hard X-rays we reach a~good agreement is favorable for the current and forthcoming X-rays polarimeters operating at these energy ranges, but less favorable for bringing more clarity in unveiling the nature of the spectral soft excess in AGNs. The inclusion of all three $\delta_\textrm{i}$, $\delta_\textrm{e}$ and $\Phi_\textrm{e}$ angles in our model should be an advantage over the other for~a~subsequent detailed disc integration in curved space-time. This is even more important for polarisation, which is more sensitive to geometry than pure spectra.

~\

An attempt was made to cross-validate the latest model by using the code {\tt CLOUDY} for ionization structure precomputations instead of {\tt TITAN}, and then to use {\tt STOKES} in the same manner. It was concluded that the {\tt CLOUDY} code, originally designed for much lower density environments and temperatures, is on its limits for the X-ray power-law reflection problem and the mechanisms inside the code are poorly understood to produce a trustworthy reference. The intermediate results of this experiment were forwarded to the group of Roberto Taverna who were more successful in constructing models of thermally radiating XRB disc atmospheres with {\tt CLOUDY} and {\tt STOKES} \citep{Taverna2021}, which we will discuss in the next section.

~\

\begin{figure}[h]
	\centering
	\begin{minipage}[t]{0.49\textwidth}
		\includegraphics[width=\textwidth]{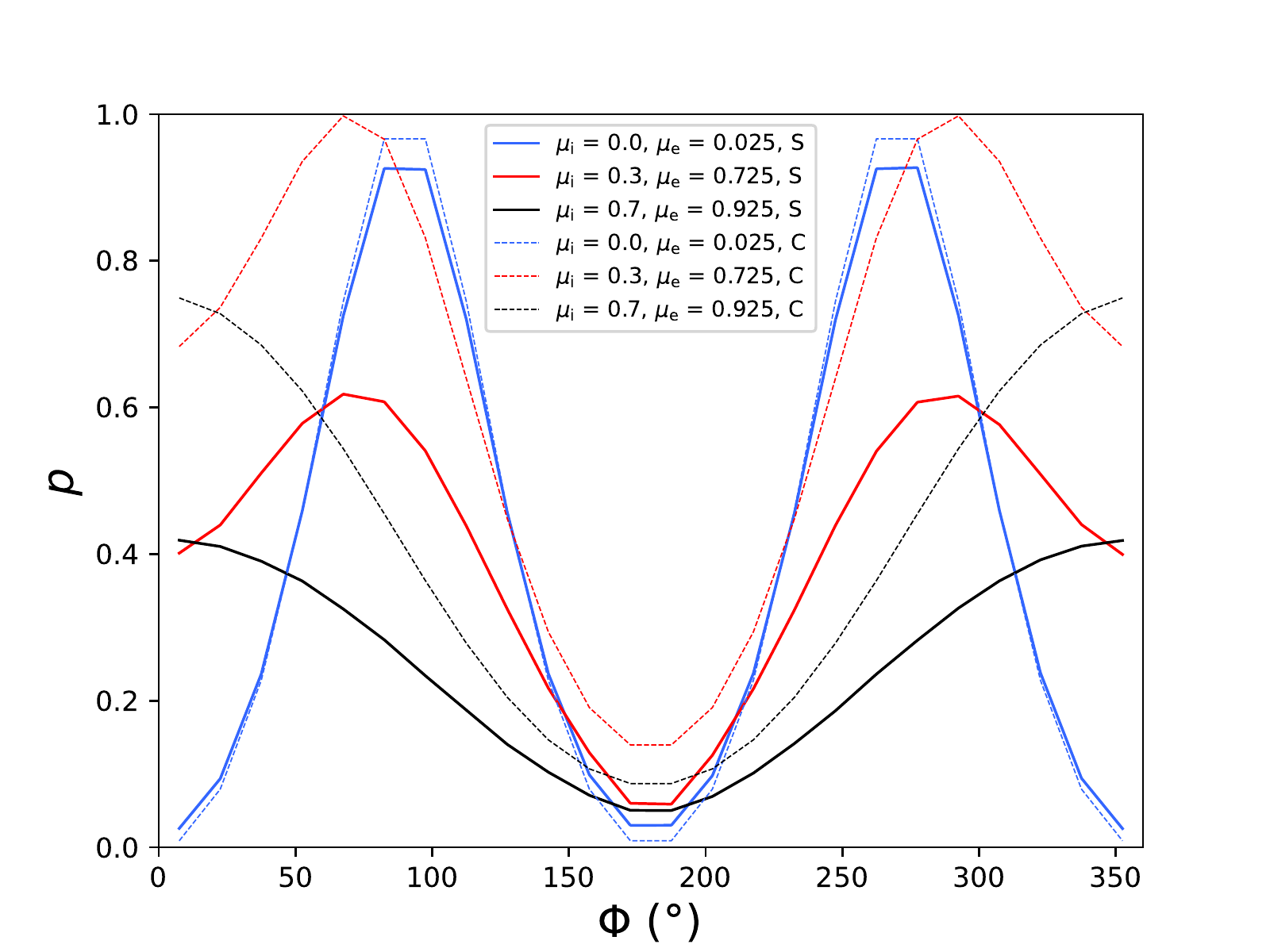}
	\end{minipage}
	\hfill
	\begin{minipage}[t]{0.49\textwidth}
		\includegraphics[width=\textwidth]{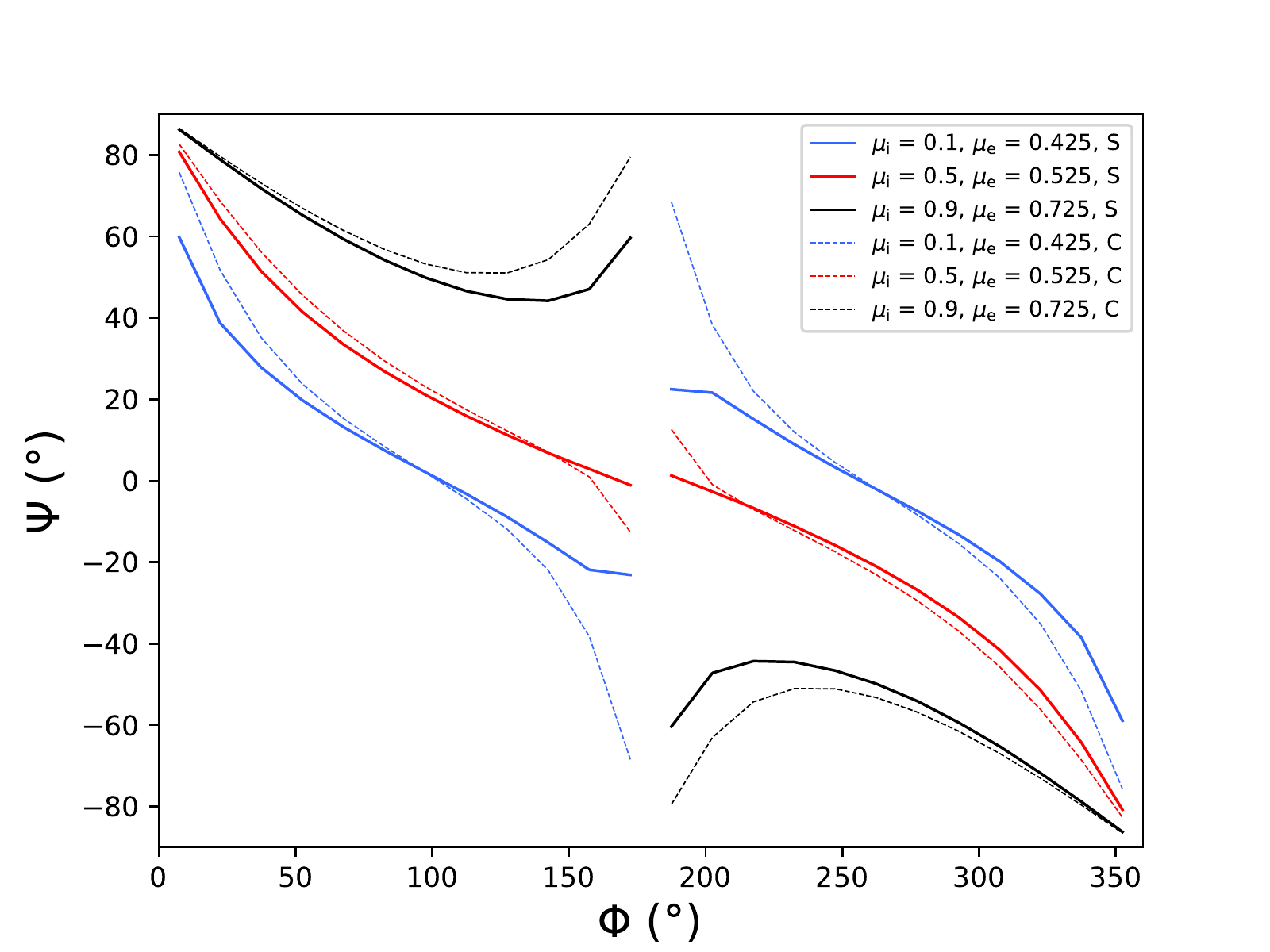}
	\end{minipage}
 \caption{\footnotesize{The polarisation degree (left) and the polarisation angle (right) versus azimuthal emission angle for various incident and emission inclination angles in the color code, as computed by {\tt TITAN} and {\tt STOKES} (solid lines) and as predicted by the Chandrasekhar's single-scattering approximation (dashed lines). The polarisation degree results are integrated in 10--100 keV, and computed for~unpolarized primary with $\Gamma = 2.2$ and $\xi = 100 \, \, \textrm{erg} \cdot \textrm{cm} \cdot \textrm{s}^{-1}$. The polarisation angle results are computed for unpolarized primary and integrated in $E$, $\Gamma$, $\xi$.}}
 \label{polcomp_chandra}
\end{figure}

To cross-validate the emergent polarisation from {\tt TITAN} and {\tt STOKES}, we used the Chandrasekhar's single-scattering analytical approximation (\ref{chandra1}), which was used e.g. in \cite{Dovciak2011} to estimate the net polarisation from disc reflection for a distant observer. We discovered however a sign error in these local reflection formulae when transcripted for \cite{Dovciak2011}, which was described and corrected in the appendix B of \cite{Podgorny2023}. To provide a direct comparison of the Chandrasekhar's approximation with our model at 10--100 keV (with no lines and polarisation gained almost entirely by Compton recoil), we plot in Figure \ref{polcomp_chandra} the polarisation degree and angle for unpolarized primary versus various angular combinations of \{$\delta_\textrm{i}, \delta_\textrm{e}, \Phi_\textrm{e}\}$. Compton multiple down-scatterings that are treated by {\tt STOKES} depolarize the output in~some cases by almost $50\%$ with respect to the value given by single Rayleigh scattering events. If only single-scattered photons are registered in {\tt STOKES}, the~results differ by just $\approx 1.5\%$ in the emergent polarisation degree. The results are symmetric in $\Phi_\textrm{e}$ around $180^{\circ}$, as expected. The angular dependency of our numerical simulation follows that of the Chandrasekhar's, which is reassuring from the code implementation point of view. Especially for the emergent polarisation angle, which is highly sensitive to the scattering geometry and which is suffering from coordinate degeneracy due to its definition. The angularly integrated value of~$\Psi \approx 90^\circ$ agrees with the angularly integrated Chandrasekhar's approximation.

\section{Thermal emission through a disc's atmosphere}\label{local_transmission}

With {\tt TITAN} and {\tt STOKES} (\textit{v2.33}, updated for external undirectional black-body irradiation) we performed numerous simulations for isotropic illumination of~a~single-temperature unpolarized black-body photon distribution (\ref{blackbody}), passing through a passive constant-density slab. The computational setup is essentially the same as in the previous section, although here we are interested in~transmitted spectropolarimetric result. If a constant-density assumption is used, we may operate with the ionization parameter (\ref{xi}), which here in combination with bolometric integration of (\ref{blackbody}) \citep[the Stefan-Boltzman law, ][]{Stefan1879,Boltzmann1884} gives
\begin{equation}\label{xibb}
    \xi_\textrm{BB} = \frac{4\pi \sigma T_\textrm{BB}^4}{n_\textrm{H}} \ ,  
\end{equation}
where $\sigma = (2 \pi^5 k_\textrm{B}^4)/(15 h_\textrm{P}^3 c^2)$ is the Stefan-Boltzman constant. This means that we use the natural normalization of the black body, as if the slab was glued directly to the source of this flux, which is convenient when thinking of the case of a colder photosphere directly on top of hotter dissipative disc layers. But in~a~few illustrative cases below we will also study the effects of a faraway reprocessing slab, e.g. a wind layer distant to a central thermally radiating accretion disc, which would have the flux normalization (and thus $\xi$) reduced by a factor of~squared distance to the source. Standard discs should be thermally radiating in the X-rays, if very hot, i.e. requiring a stellar-mass black hole and high mean densities (see Section \ref{all_XRBs}). So the main application will be the thermal emission from accretion discs of XRBs. We again neglect Compton up-scattering, which should be effective near the \textit{IXPE} band at higher slab temperatures than those considered \citep{Prince2004}, but a quantitative estimate on this assumption could be done with the latest version of {\tt STOKES} \textit{v2.34}. On the other hand Compton down-scatterings are important, as the main source of polarisation for individual photons, alongside absorption that effectively reduces the geometry of scattering and adds polarisation. The solar abundance is from \cite{Asplund2005} with $A_\textrm{Fe} = 1$.

~\

Essentials of this method were published in \cite{Taverna2021} with the~code {\tt CLOUDY} instead of {\tt TITAN} for disc structure precomputations: locally and then with relativistic integration across the disc. However, the recent advances in~X-ray polarimetric observations of XRBs in the thermal state (see Section \ref{chap05}) required a~more detailed theoretical exploration of the parametric space than in~\cite{Taverna2021}, especially in the local frame of reference. In \cite{Taverna2021}, the~slab was treated in collisional ionization equilibrium (CIE), assuming collisions are the most important process of ionization in dense discs, but a photon ionization equilibrium (PIE) is an equally (un)justifiable assumption for first-order investigations. Although {\tt TITAN} cannot cover CIE, it is an ideal photoionization code to cross-validate the results of {\tt CLOUDY} in PIE and, unlike {\tt CLOUDY}, it offers a) a possibility of double illumination of the slab, which we would use in the~future for a combined problem (power-law reflection and black-body transmission self-consistently), and b) a more realistic boundary condition option for optically thick regimes (see below), which we would use in the future for semi-infinite atmospheres with internal source distribution (i.e. towards a~more physical atmosphere). Therefore, having the procedures tested in reflection problem already, we explored the local transmission also with {\tt TITAN} and {\tt STOKES} in PIE, which is presented here.

~\

Additional local computations in both CIE and PIE, as a follow-up to \cite{Taverna2021}, were done simultaneously with {\tt CLOUDY} and {\tt STOKES} by Lorenzo Marra. The main principle difference is the inclusion of CIE possibility in {\tt CLOUDY} as opposed to {\tt TITAN}, and then obviously the codes' architecture and individual atomic data inclusion. Although a thorough comparison between {\tt TITAN} and {\tt CLOUDY} of the temperature and~ionization profiles and~spectra, depending on~the~slab density, irradiation properties and boundary conditions, still needs to be done, we obtained similar results in similar PIE conditions. Thus, not surprisingly, almost identical spectropolarimetric results with {\tt STOKES}.

This last step, however, confirms correct implementation of the ionization structure input to {\tt STOKES} from both {\tt CLOUDY} and {\tt TITAN}, which was done independently and which is technically not easy, as both photoionization codes provide diverse output information in intricate formats. The new computations with {\tt CLOUDY} and {\tt STOKES} in CIE and PIE will not be shown, as they were done and~privately communicated by Lorenzo Marra and because a detailed peer-reviewed comparative study is still awaited. More work is also needed to be able to present our results with {\tt TITAN} and {\tt STOKES} as a complete parametric grid explored and fully transformed to tabular format for relativistic integration and fitting of any X-ray source, similarly to the reflection tables discussed in~the~previous section. Hence, we will only show the intermediate discoveries, which lead to tabular summary of only a small (highly ionized) part of the irregular parameter space explored that was immediately used for relativistic integration and direct interpretation of~the~{\it IXPE} observations of 4U1630-47 and LMC X-3 (see Section \ref{chap05}). All of~this, including the local atmospheric modeling presented here, was summarized and~published in \cite{Ratheesh2023} and \cite{Svoboda2023}.

\subsection{Effects of absorption}

First, let us define the problem in terms of optical thickness. An atmosphere of~the~Novikov-Thorne disc solution is considered optically thick. One can imagine a nearly thermal equilibrium reached by viscous dissipation in deep layers of~the~disc, which then radiates as a black body and is transformed through colder layers up to some imaginary surface boundary (a density decrease below some negligible value), similarly to radiative transfer inside outer layers of stars. 

{\tt TITAN} offers two types of boundary conditions for PIE with external black-body illumination and a slab with constant density $n_\textrm{H}$ and geometrical thickness $H$. One option (\textit{irent} $ = 4$) keeps the same effective temperature as the temperature $T_\textrm{BB}$ of the single-color black body, i.e. no matter the slab thickness, the~temperature profile is roughly the same with monotonically increasing temperature $T$ with Thomson optical depth $\tau_\textrm{T} = n_\textrm{H}H\sigma_\textrm{T}$, where \mbox{$T(\tau_\textrm{T} = 1) = T_\textrm{eff} \approx T_\textrm{BB}$}. The~other option (\textit{irent} $ = 1$) keeps the same temperature at the bottom (illuminated) surface of the slab as the prescribed $T_\textrm{BB}$, meaning \mbox{$T(\tau_\textrm{T} = \tau_\textrm{T,max} = n_\textrm{H}H\sigma_\textrm{T} = N_\textrm{H}\sigma_\textrm{T}) \approx T_\textrm{BB}$}. The two available boundary conditions produce equal results for $\tau_\textrm{T,max} \approx 1$. It is the latter case, which is less physical, but used below, because it is identical to {\tt CLOUDY} treatment of PIE and~still valid in optically thin approximation of the upper atmospheric layers playing the largest role in altering the polarisation state of escaping radiation. 

In any more sophisticated approach in the future, we would use the first option to assess the entire semi-infinite atmosphere, but that would require also inclusion of internal sources in {\tt STOKES} (to keep the consistency between ionization structure precomputation and MC spectropolarimetric simulation, using a~source function $\tilde{S}$), which we do not have in the presented approach. Inside the slab scattering region, the external black-body photons are only scattered, absorbed and immediately reemitted or absorbed, but not thermalized with radiation and kinetic consequences, which the MC method cannot process. Also, in~case of~a~more realistic semi-infinite atmosphere a~hydrostatic equilibrium would have to~be used. 

The ignored phenomena were essentially included and altogether studied simultanously to our efforts by Valery Suleimanov using spectropolarimetric radiative transfer solver, adapted from \cite{Suleimanov2002, Suleimanov2007}. This model, including relativistic disc integration, was part of the data interpretation of \textit{IXPE} observation of 4U1630-47 and it is also summarized in \cite{Ratheesh2023} (see in particular figure B1 for local emission at two different rings and figure B2 for the net result with and without the relativistic effects). Even more physical computation, taking into account density inhomogeneities \citep{Coleman1990} and magnetic force support was done by \cite{Davis2009}, using MHD approach combined with 3D MC simulation for emergent polarisation.

~\

Self-consistent models of a semi-infinite local disc atmosphere serve as a good cross-validation to our simplified approach, but inevitably contain many internal parameters and are computationally costly, which makes them less flexible for data fitting. Our Milne approach with a passive slab sitting on external black-body source is less robust, but if proven to provide reasonably similar spectropolarimetric results with inclination, density and effective temperature, it introduces only the problem of height scaling and ionization structure, which shall be fine-tuned to represent the effects of a real photospheric transition medium between the matter in thermal equilibrium at high temperatures in the midplane and nearly vacuum on the other side. If the free parameter of our finite-slab model, the optical depth $\tau_\textrm{T,max}$, is too large, we depart from a more physical situation by neglecting internal sources of thermal radiation inside the slab. If it is too low, we may underestimate the absorption and scattering effects. As we will show below, in comparison with X-ray spectroscopic and energy-dependent polarisation data, we will validate and use our model for fitting in only part of the thinkable parameter space. The rest of the computations, perhaps too artificial, only help to~understand the role of elementary processes in the emergent polarisation.

~\

Apart from range of optical depths, we also tested several orders of magnitude in $n_\textrm{H}$, because a physical atmosphere would have a dominantly decreasing density profile with vertical height \citep[e.g. a gaussian profile, or more general ones are derived, ][]{Shakura1973, Davis2006, Taverna2020}. This was done also in order to test possible application of the model, assuming radial disc integration and various radial disc density profiles, $n_\textrm{H}(r)$, alongside various effective temperature profiles, $T_\textrm{eff}(r)$, that can appear according to~the~Novikov-Thorne estimates. After several simulation trials we proved that in the PIE conditions studied ($n_\textrm{H} = 10^{12}$--$10^{20.5} \ \textrm{cm}^{-3}$, $k_\textrm{B}T_\textrm{BB} = 0.16$--$1.4 \ \textrm{keV}$, $\tau_\textrm{T,max} \geq 0.4$, $\xi = \xi_\textrm{BB}$) a)~the~resulting polarisation angle is always constant with energy and~parallel with the slab (as expected for a passive atmosphere with scattering and absorption and moderate or high optical thickness) and b) the shape of the resulting polarisation degree with energy is heavily dependent on ionization structure (in turn dependent on $n_\textrm{H}$, $T_\textrm{BB}$ and $\tau_\textrm{T,max}$) -- no matter how the ionization state of the slab is prepared -- and then scaled at~all energies uniformly with $\mu_\textrm{e}$ and $\tau_\textrm{T,max}$.

~\

We will explain the results in depth in terms of absorption and scattering effects below, but in order to separate the effect of varying ionization with $\tau_\textrm{T,max}$ from the effects of pure $\tau_\textrm{T,max}$ on the emergent polarisation in the resulting figures, we made an additional modeling choice based on the preliminary parametric behavior. We decided to fix the ionization profile precomputed with {\tt TITAN} at~$\tau_\textrm{T,max,structure} = 1$ and then rescale this ionization structure when computing with {\tt STOKES} by extending or shrinking the height of the entire slab $H$, reaching different $\tau_\textrm{T,max}$ for the \textit{same} ionization profile. This goes certainly against the~original idea of keeping the two computational phases and their setups as equal as possible, but it is a reasonable trade-off for understanding of what causes what inside the slab. Nonetheless, for any thicker slabs we would not be able to keep the~{\tt STOKES} setup identical to the ionization structure precomputations due to lack of internal thermal sources and the constant-temperature boundary condition becomes invalid for a realistic ionization structure estimate. For data fitting later on, we will use only the highly ionized results where caveats with respect to partial ionization do not apply. In the partially ionized cases, if shrinking or extending vertically stratified ionization profile obtained with {\tt TITAN} for $\tau_\textrm{T,max,structure} = 1$, we overestimate or underestimate the real consistent ionization gradient for final thickness $\tau_\textrm{T,max}$, respectively.

We first assumed a stratification of 500 layers in {\tt TITAN} and then again averaging to up to 50 layers in {\tt STOKES} during the conversion of the ionization structure information. After comparing the results for $\tau_\textrm{T,max,structure} = 1$ with just averaging to 1 layer in {\tt STOKES} as the entire slab, there was negligible difference in~the~spectropolarimetric results and the temperature and ionization stratification obtained in {\tt TITAN} was low. Therefore, we continued with single-layer approximation for~computational efficiency in the MC simulation for all results presented below.

~\

Having laid down the procedure and giving away the main features, let us study the effects of partial ionization in detail. Figures \ref{absorbed_spectra} and \ref{absorbed_pd} show the~effects of changing density for particular single-temperature black-body illumination with $\xi = \xi_\textrm{BB}$ and $k_\textrm{B}T_\textrm{BB} = 0.16$ keV. We show the fractional abundances of~iron and oxygen ions from {\tt TITAN} (taken at intermediate layer with temperature closest to the averaged layer in {\tt STOKES}) to show that the lower the density, the~higher the ionization. For the highly ionized cases, the atomic nuclei are nearly stripped off all electrons. Next, we show the effective optical depth \citep{Rybicky1979}
\begin{equation}
    \tau_\mathrm{eff} = \sqrt{3 \tau_\mathrm{abs} ( \tau_\mathrm{abs} + \tau_\mathrm{es} )} \ ,
\end{equation}
where $\tau_\mathrm{abs}$ is the total (sum of bound-free and free-free) absorption optical depth and $\tau_\mathrm{es}$ is the electron scattering optical depth, which is the best information we can obtain on continuum opacities from {\tt TITAN}, including the provided line optical depths, which are plotted on top of the continuum. We divide by~the~total column density $N_\textrm{H}$ to compare to energy-independent Thomson cross section $\sigma_\textrm{T}$ for~pure scattering in elastic limit. Although this comparison is not very clarifying due to mixture of competing effects forming the effective optical depth, within the~peculiarities of {\tt TITAN} outputs it gives the only available estimate on~absorption importance (including ionization edges) and scattering. We then provide the corresponding density examples of the {\tt STOKES} output spectra for~$\tau_\textrm{T,max} = 1$ and a series of corresponding polarisation fractions with energy from {\tt STOKES} for~the~same densities and for $\tau_\textrm{T,max} = \{0.67, 1, 3\}$.
\begin{landscape}
\begin{figure}
    \centering
    \begin{picture}(137,99)
    \put(0,0){\includegraphics[width=0.33\textwidth]{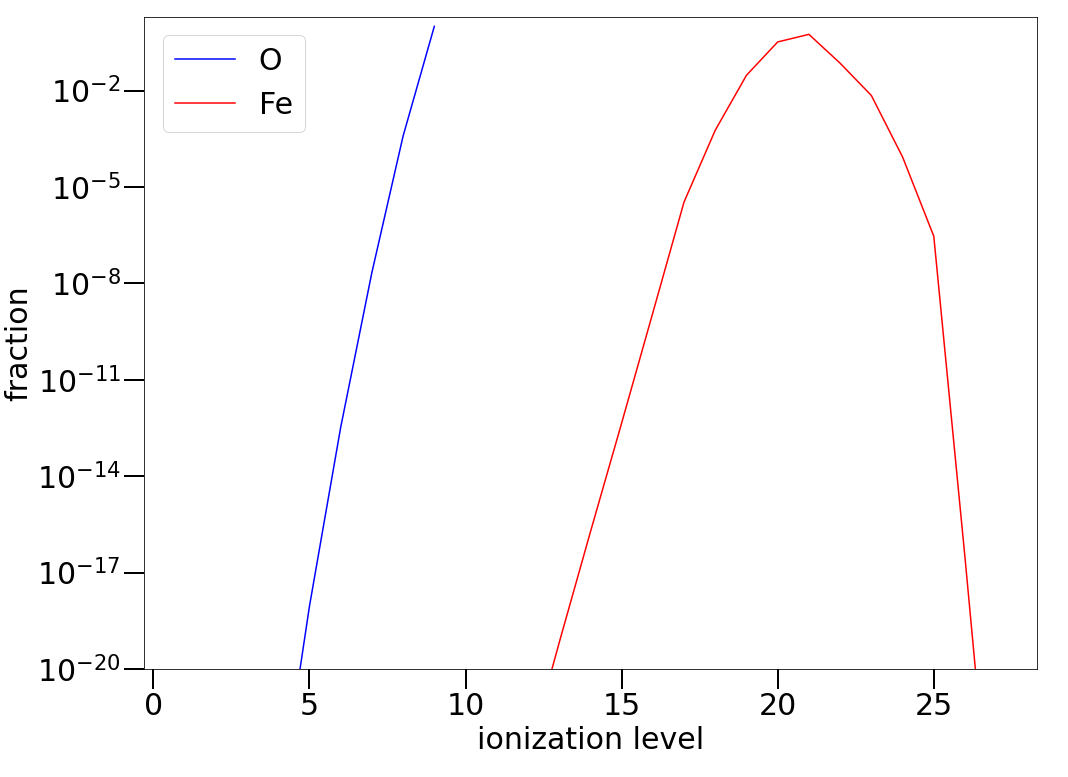}}
    \put(62,20){\scalebox{.55}{$n_\textrm{H} = 10^{18} \ \textrm{cm}^{-3}$}}
    \end{picture}
    \begin{picture}(137,99)
    \put(0,0){\includegraphics[width=0.33\textwidth]{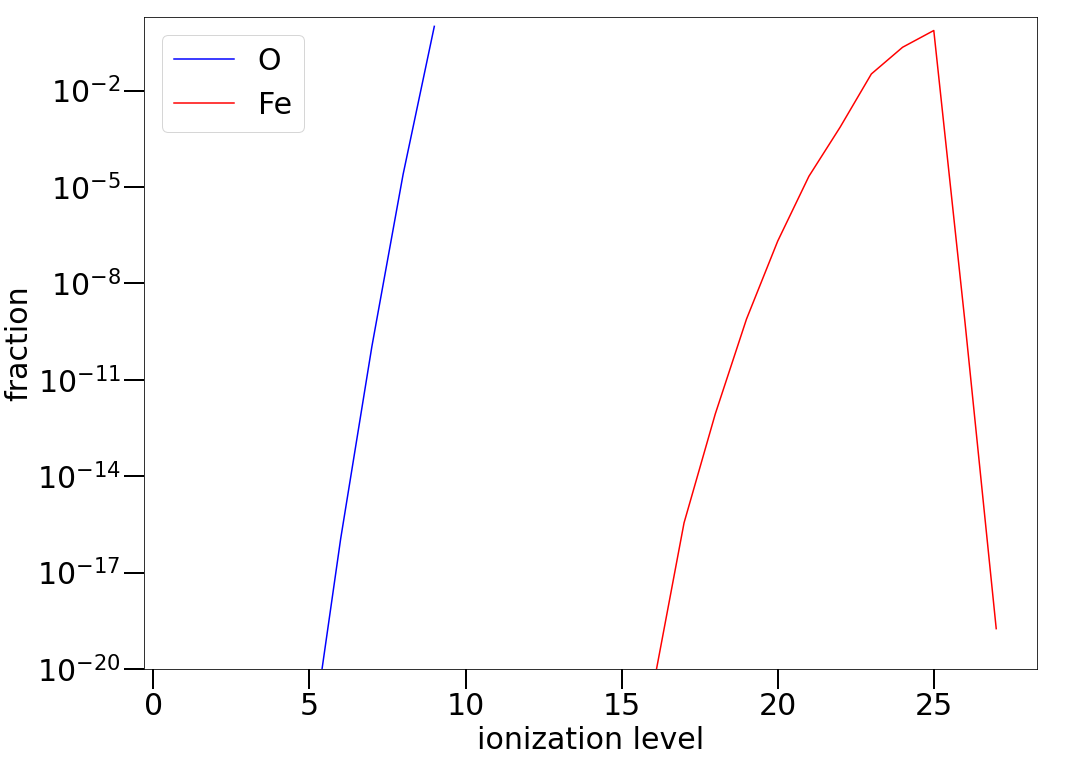}}
    \put(62,20){\scalebox{.55}{$n_\textrm{H} = 10^{17} \ \textrm{cm}^{-3}$}}
    \end{picture}
    \begin{picture}(137,99)
    \put(0,0){\includegraphics[width=0.33\textwidth]{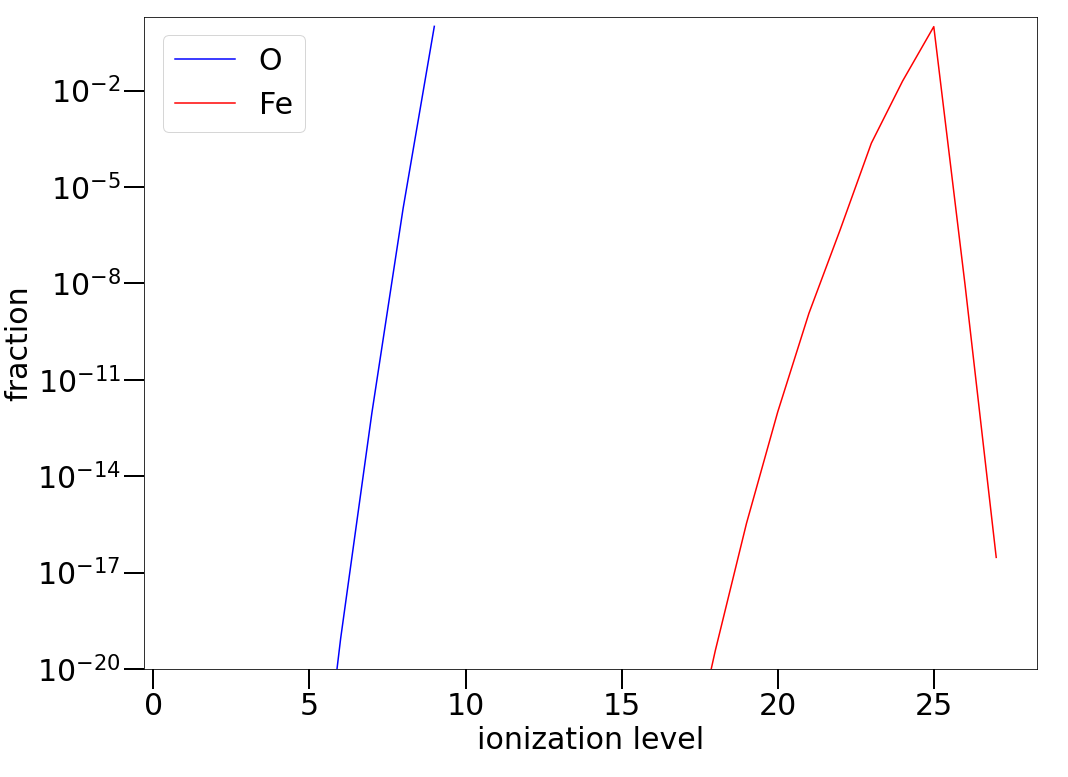}}
    \put(62,20){\scalebox{.55}{$n_\textrm{H} = 10^{16} \ \textrm{cm}^{-3}$}}
    \end{picture}
    \begin{picture}(137,99)
    \put(0,0){\includegraphics[width=0.33\textwidth]{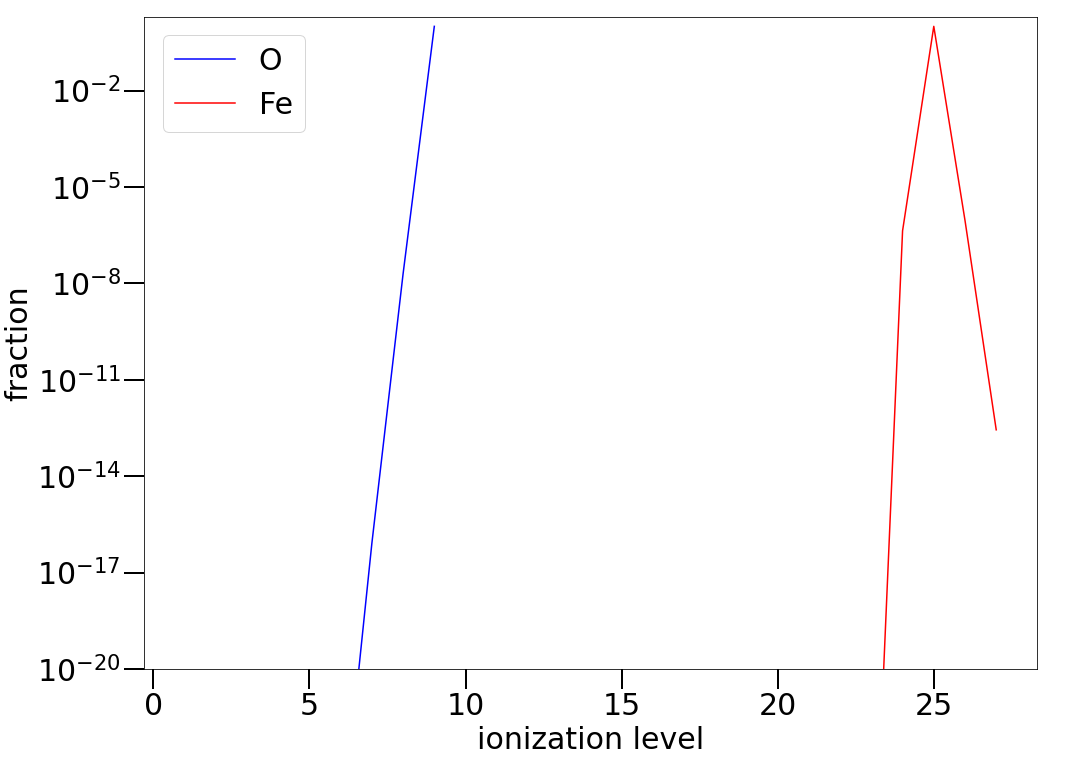}}
    \put(62,20){\scalebox{.55}{$n_\textrm{H} = 10^{15} \ \textrm{cm}^{-3}$}}
    \end{picture}
    \begin{picture}(137,99)
    \put(0,0){\includegraphics[width=0.33\textwidth]{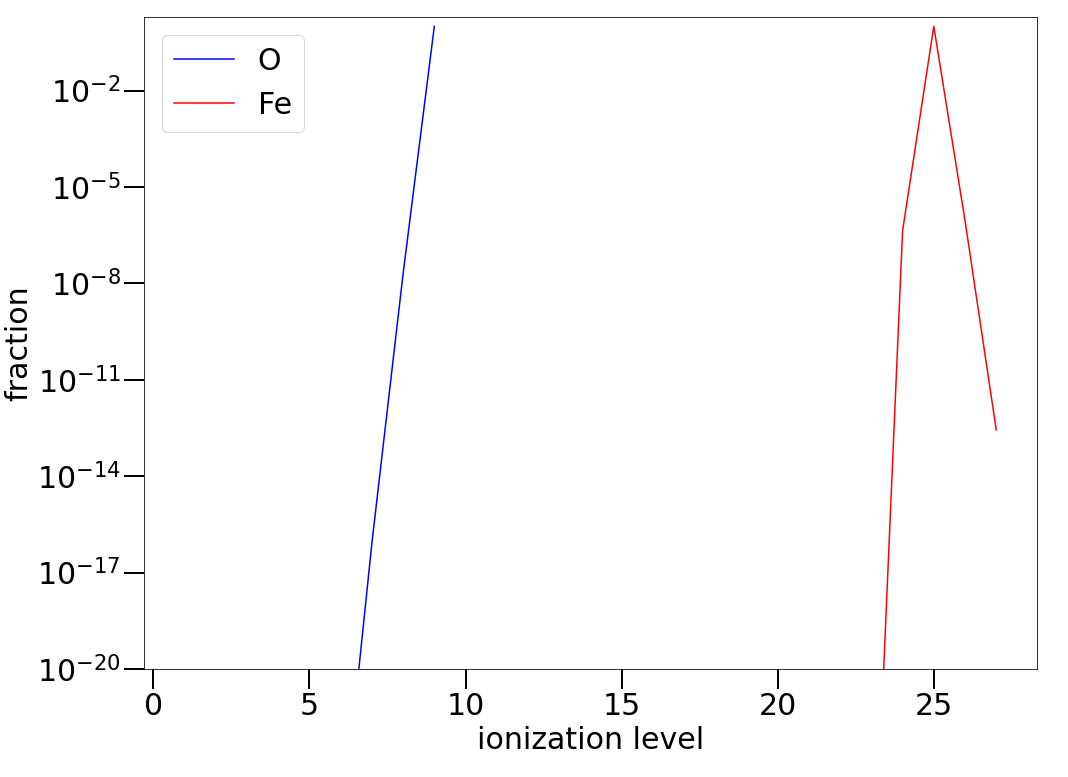}}
    \put(62,20){\scalebox{.55}{$n_\textrm{H} = 10^{14} \ \textrm{cm}^{-3}$}}
    \end{picture}\\
    \begin{picture}(137,99)
    \put(0,0){\includegraphics[width=0.33\textwidth]{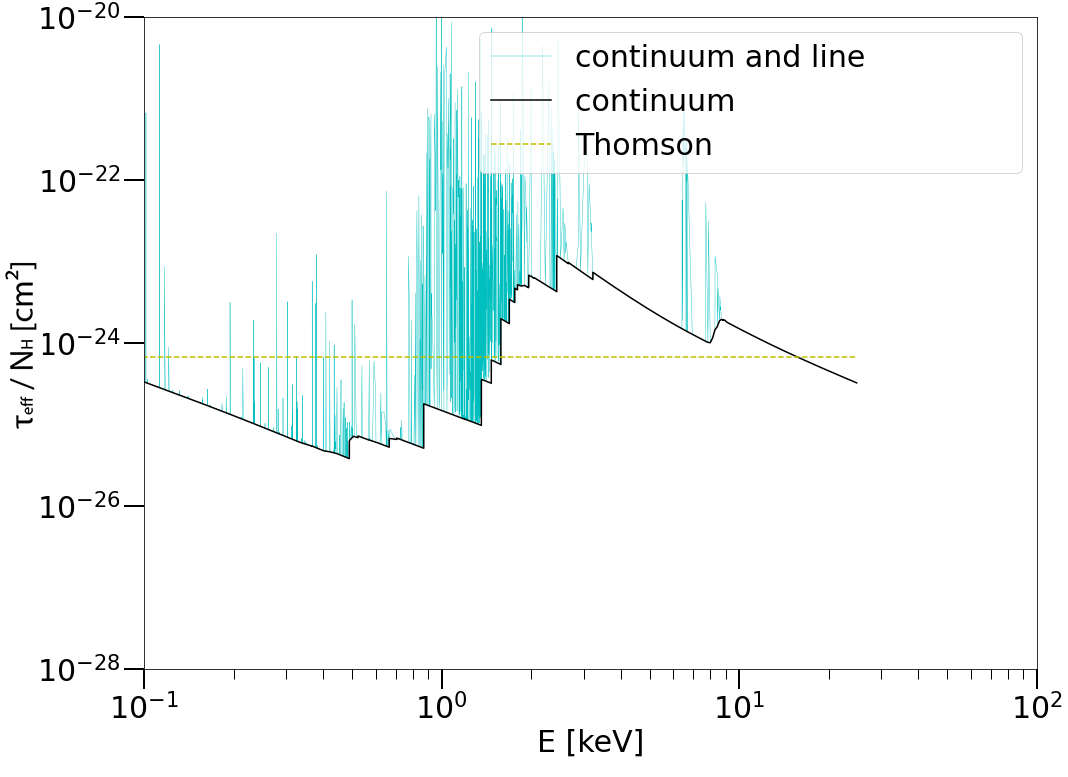}}
    \put(62,20){\scalebox{.55}{$n_\textrm{H} = 10^{18} \ \textrm{cm}^{-3}$}}
    \end{picture}
    \begin{picture}(137,99)
    \put(0,0){\includegraphics[width=0.33\textwidth]{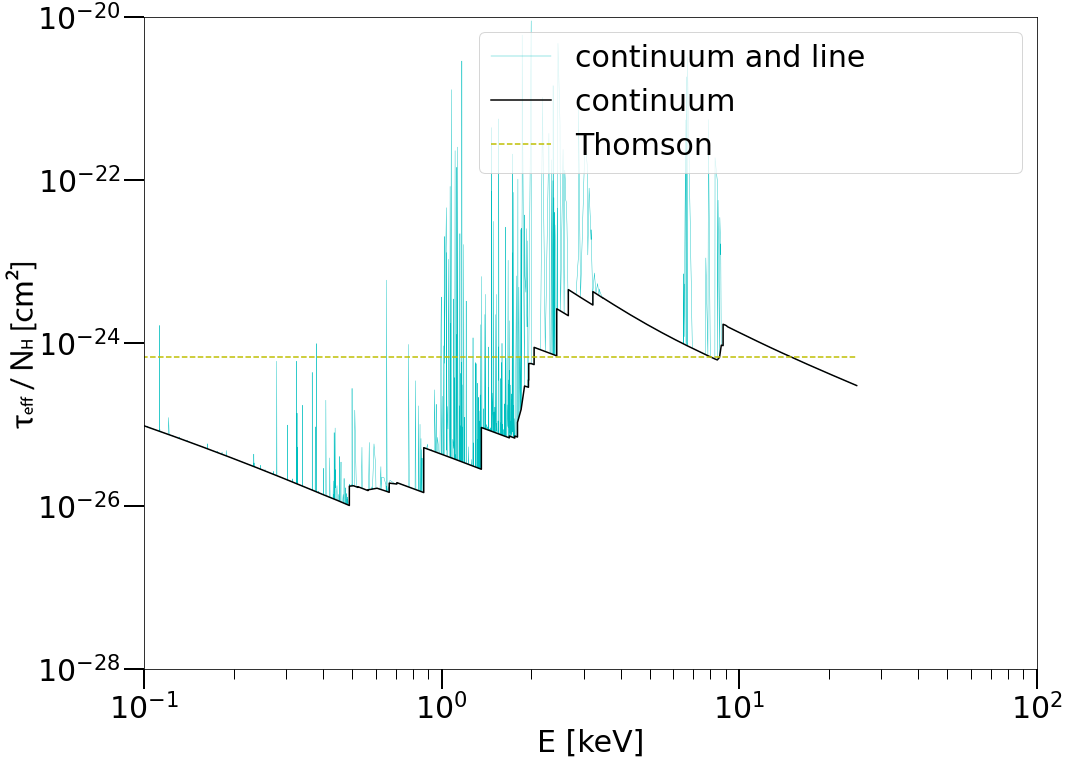}}
    \put(62,20){\scalebox{.55}{$n_\textrm{H} = 10^{17} \ \textrm{cm}^{-3}$}}
    \end{picture}
    \begin{picture}(137,99)
    \put(0,0){\includegraphics[width=0.33\textwidth]{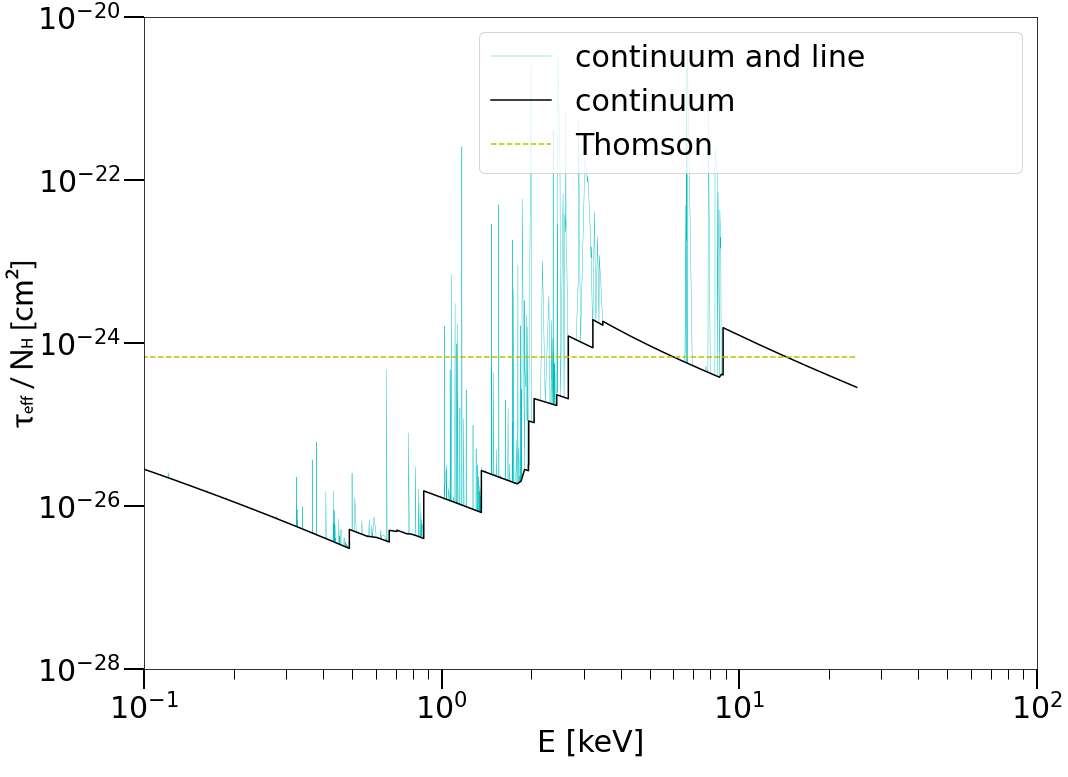}}
    \put(62,20){\scalebox{.55}{$n_\textrm{H} = 10^{16} \ \textrm{cm}^{-3}$}}
    \end{picture}
    \begin{picture}(137,99)
    \put(0,0){\includegraphics[width=0.33\textwidth]{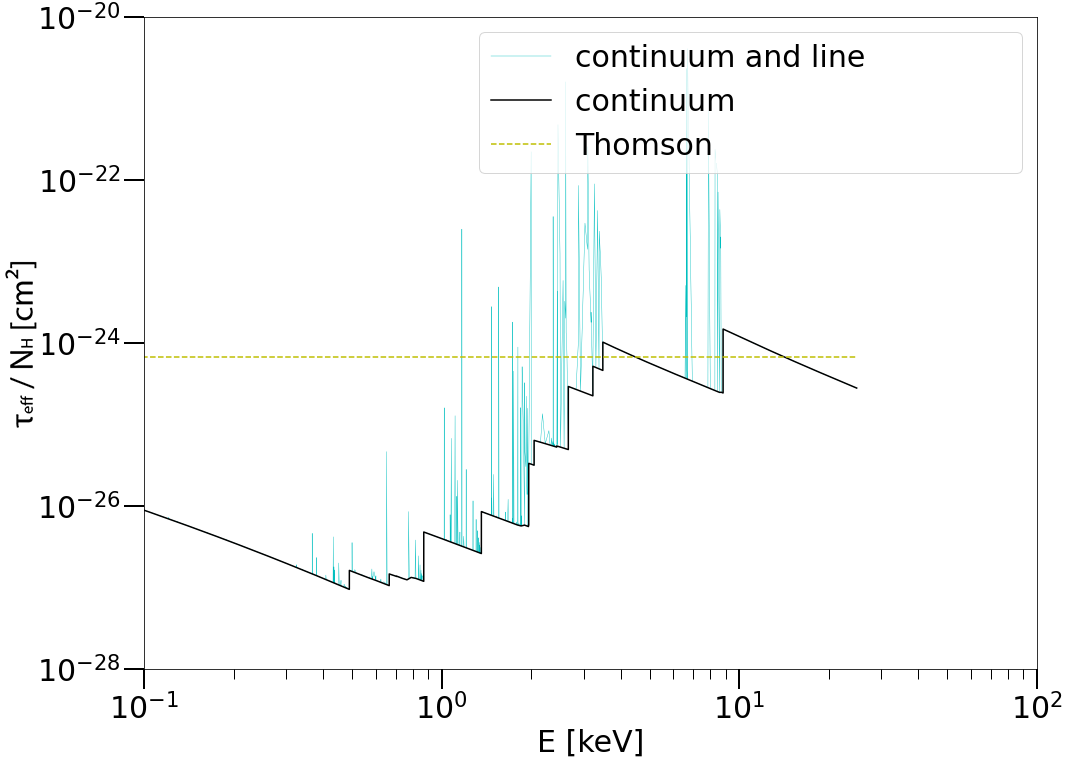}}
    \put(62,20){\scalebox{.55}{$n_\textrm{H} = 10^{15} \ \textrm{cm}^{-3}$}}
    \end{picture}
    \begin{picture}(137,99)
    \put(0,0){\includegraphics[width=0.33\textwidth]{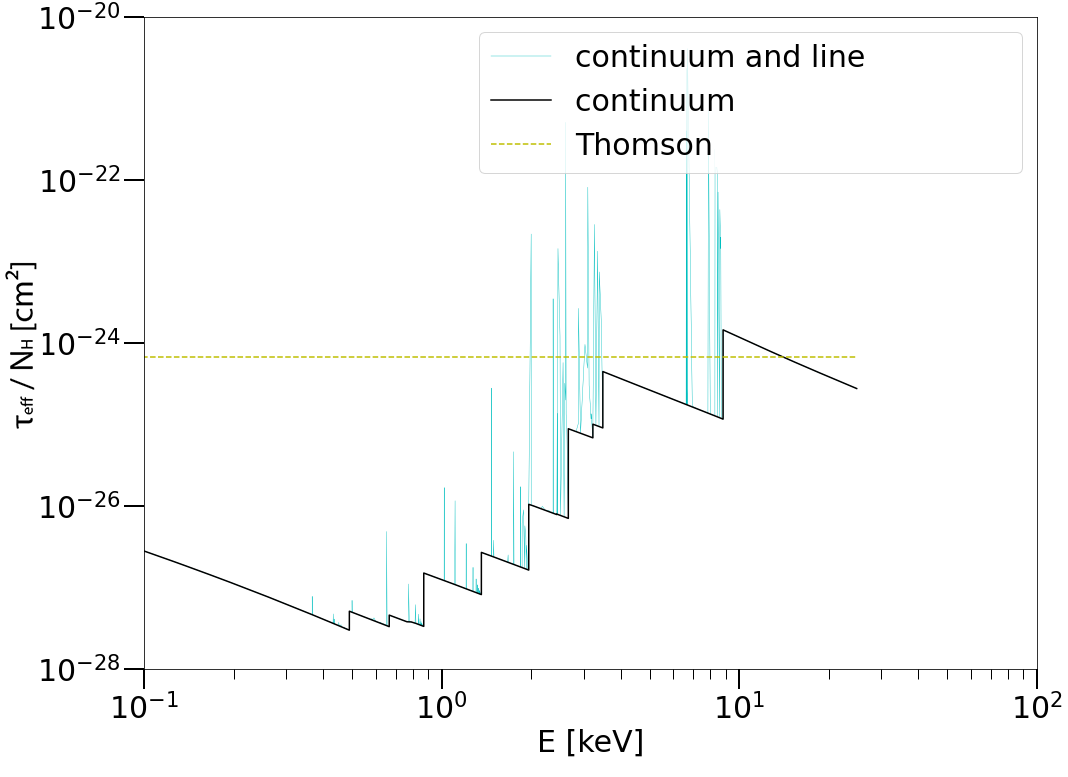}}
    \put(62,20){\scalebox{.55}{$n_\textrm{H} = 10^{14} \ \textrm{cm}^{-3}$}}
    \end{picture}\\
    \begin{picture}(137,99)
    \put(0,0){\includegraphics[width=0.33\textwidth]{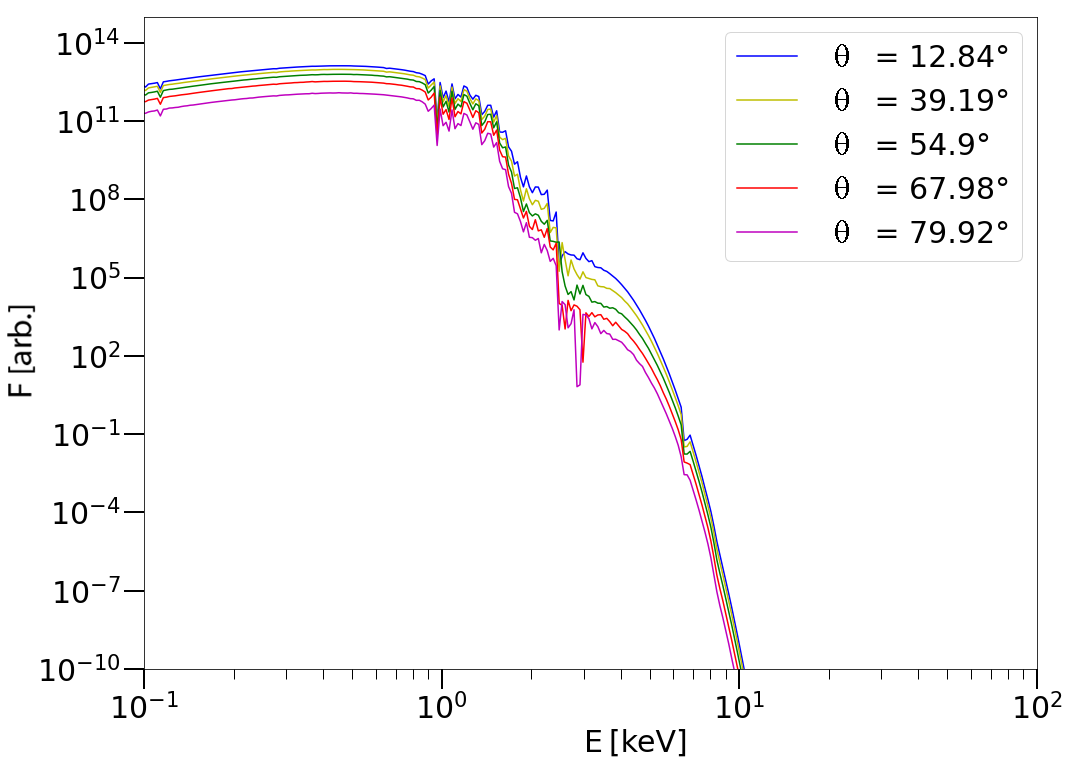}}
    \put(22,30){\scalebox{.55}{$\tau_\textrm{T,max} = 1$}}
    \put(22,20){\scalebox{.55}{$n_\textrm{H} = 10^{18} \ \textrm{cm}^{-3}$}}
    \end{picture}
    \begin{picture}(137,99)
    \put(0,0){\includegraphics[width=0.33\textwidth]{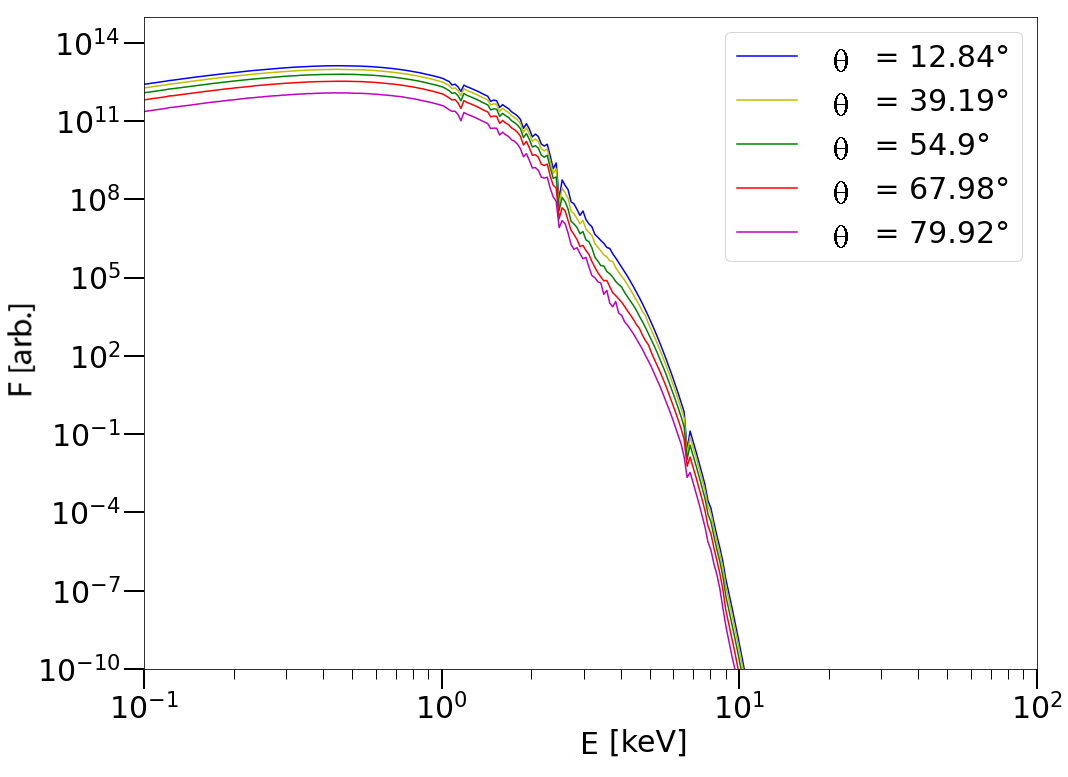}}
    \put(22,30){\scalebox{.55}{$\tau_\textrm{T,max} = 1$}}
    \put(22,20){\scalebox{.55}{$n_\textrm{H} = 10^{17} \ \textrm{cm}^{-3}$}}
    \end{picture}
    \begin{picture}(137,99)
    \put(0,0){\includegraphics[width=0.33\textwidth]{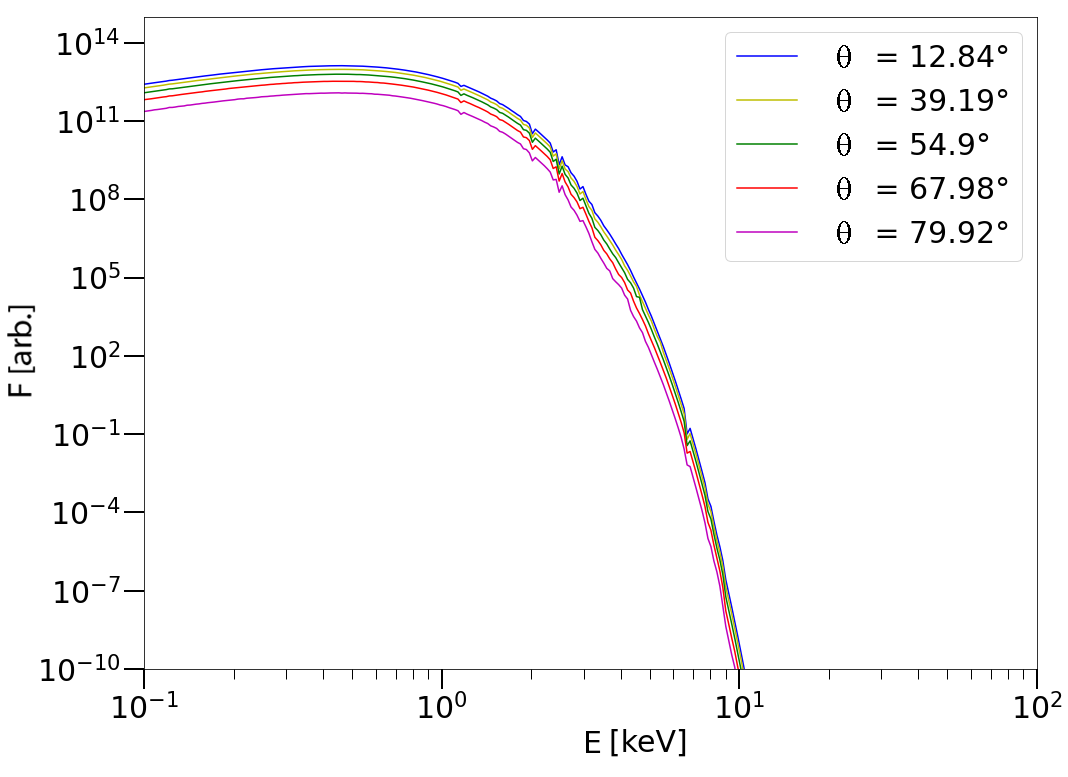}}
    \put(22,30){\scalebox{.55}{$\tau_\textrm{T,max} = 1$}}
    \put(22,20){\scalebox{.55}{$n_\textrm{H} = 10^{16} \ \textrm{cm}^{-3}$}}
    \end{picture}
    \begin{picture}(137,99)
    \put(0,0){\includegraphics[width=0.33\textwidth]{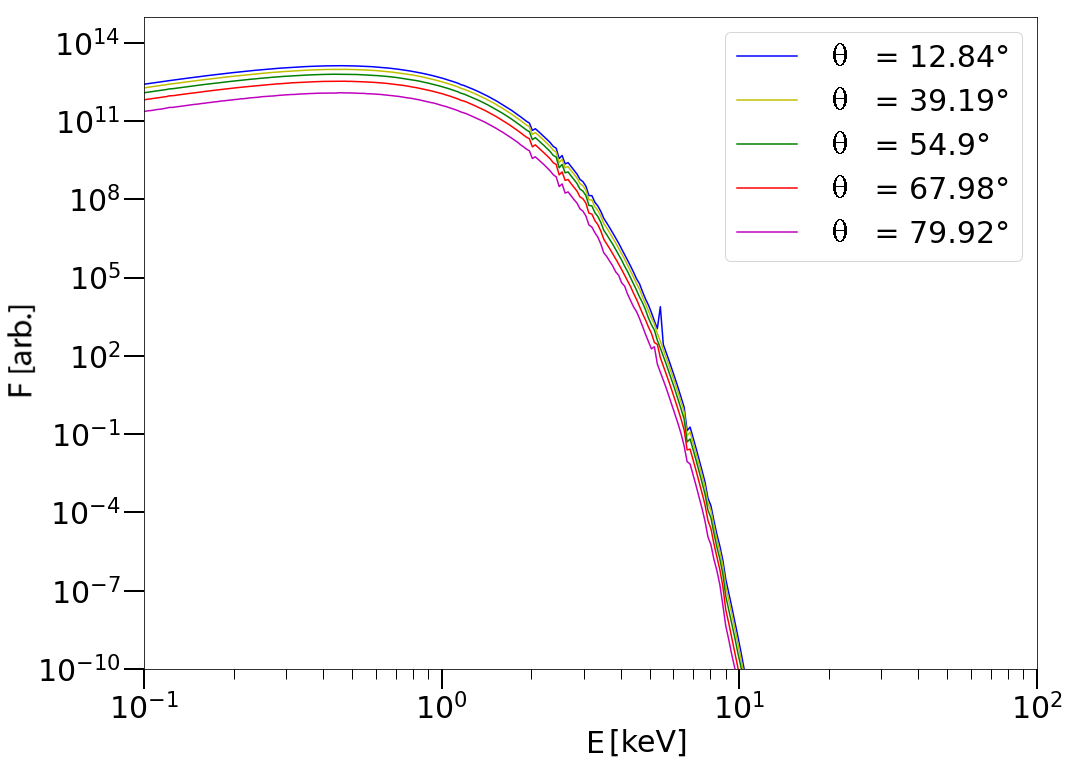}}
    \put(22,30){\scalebox{.55}{$\tau_\textrm{T,max} = 1$}}
    \put(22,20){\scalebox{.55}{$n_\textrm{H} = 10^{15} \ \textrm{cm}^{-3}$}}
    \end{picture}
    \begin{picture}(137,99)
    \put(0,0){\includegraphics[width=0.33\textwidth]{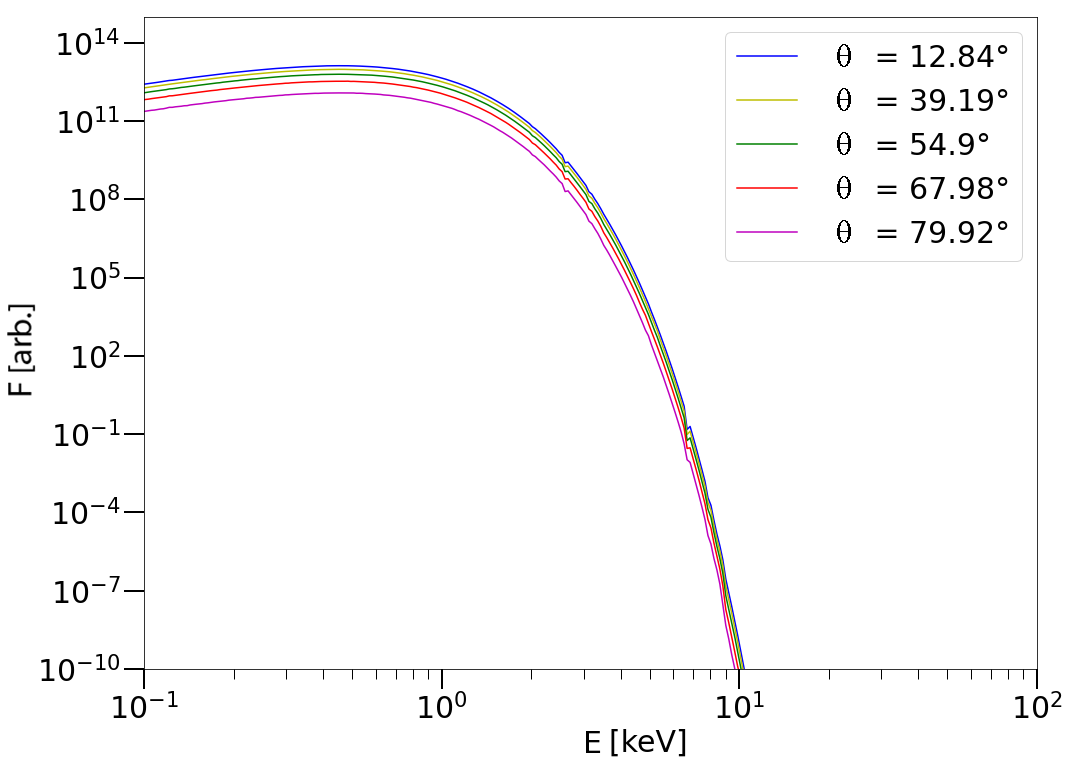}}
    \put(22,30){\scalebox{.55}{$\tau_\textrm{T,max} = 1$}}
    \put(22,20){\scalebox{.55}{$n_\textrm{H} = 10^{14} \ \textrm{cm}^{-3}$}}
    \end{picture}
    \caption{\footnotesize{Various cases of transmitted single-temperature unpolarized black-body radiation with $k_\textrm{B}T_\textrm{BB} = 0.16$ keV through a constant density slab, as computed with {\tt TITAN} and {\tt STOKES} in PIE. From left to right the density of the slab decreases ($n_\textrm{H} = 10^{18},10^{17},10^{16},10^{15},10^{14} \ \textrm{cm}^{-3}$, respectively) and the slab is more ionized for the same external illumination with $\xi = \xi_\textrm{BB}$. Top row: fractional abundance of ions of oxygen (blue) OI--OIX and iron (red) FeI--FeXXVII from {\tt TITAN} ionization structure precomputations. Middle row: the effective optical depth, $\tau_\textrm{eff}$, divided by column density, $N_\textrm{H} \, [\textrm{cm}^{-2}]$, versus energy for continuum processes (black) and including lines (cyan) from {\tt TITAN} ionization structure precomputations. Thomson cross-section $\sigma_\textrm{T}$ (yellow) represents pure scatterings. Bottom row: spectra, $F$, per energy bin for various observer's inclinations $\theta$ in the color code for $\tau_\textrm{T,max} = 1$, obtained from {\tt STOKES}.}}
    \label{absorbed_spectra}
\end{figure}
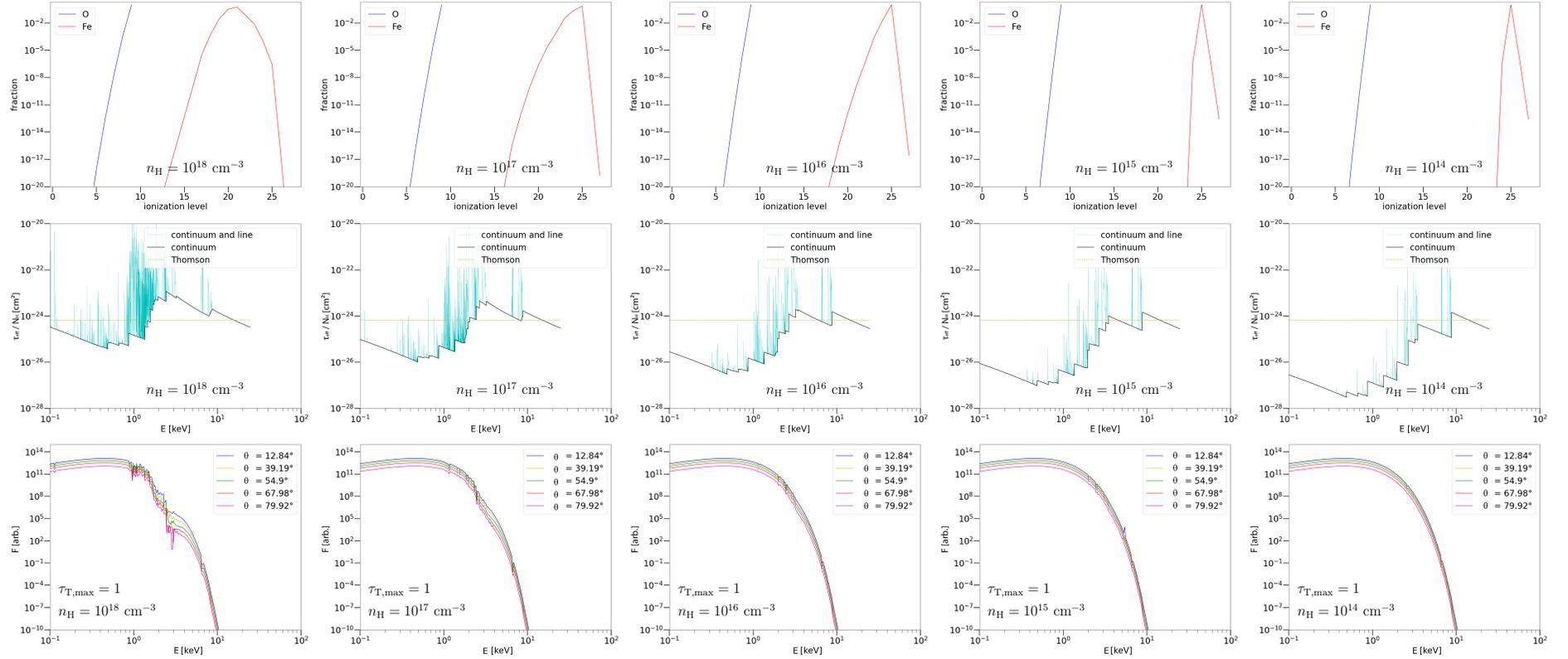
\end{landscape}
\begin{landscape}
\begin{figure}
    \centering
    \begin{picture}(137,99)
    \put(0,0){\includegraphics[width=0.33\textwidth]{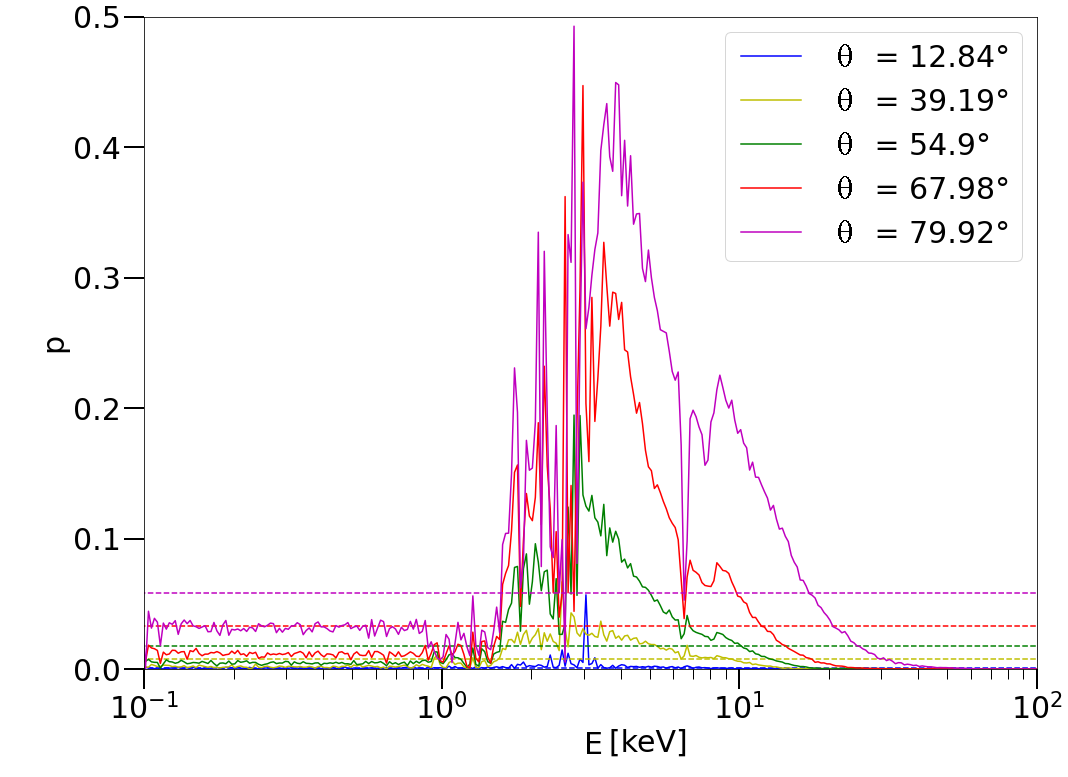}}
    \put(20,90){\scalebox{.55}{$\tau_\textrm{T,max} = 0.67$}}
    \put(20,80){\scalebox{.55}{$n_\textrm{H} = 10^{18} \ \textrm{cm}^{-3}$}}
    \end{picture}
    \begin{picture}(137,99)
    \put(0,0){\includegraphics[width=0.33\textwidth]{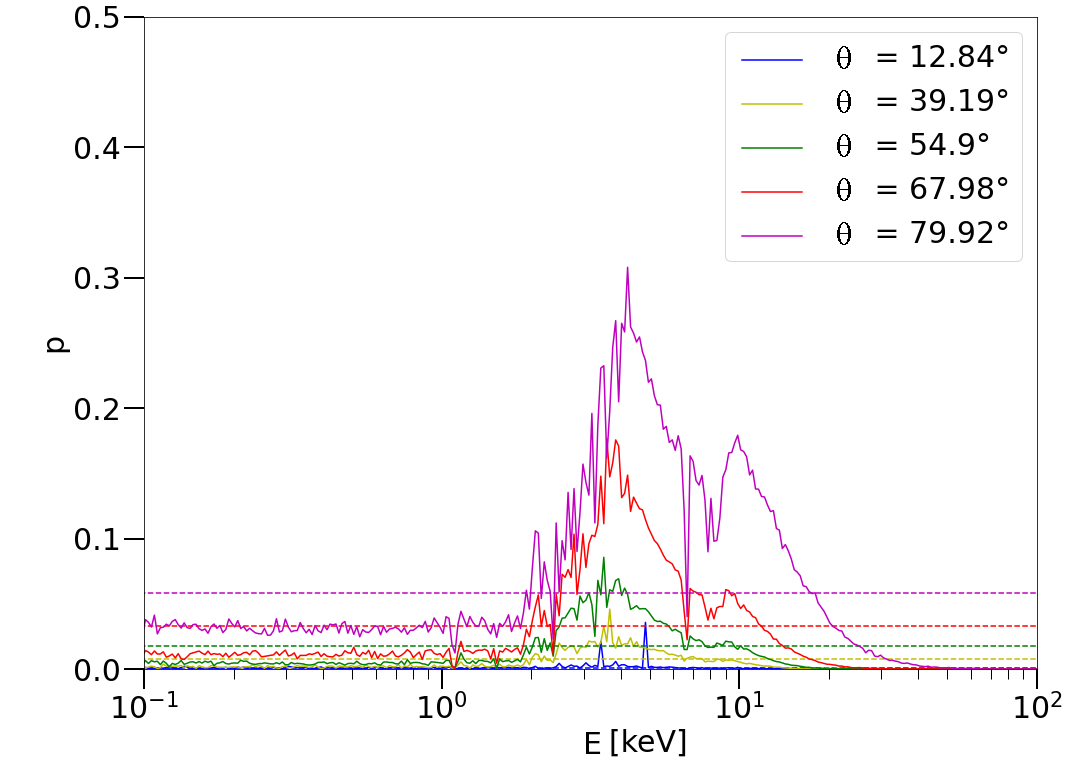}}
    \put(20,90){\scalebox{.55}{$\tau_\textrm{T,max} = 0.67$}}
    \put(20,80){\scalebox{.55}{$n_\textrm{H} = 10^{17} \ \textrm{cm}^{-3}$}}
    \end{picture}
    \begin{picture}(137,99)
    \put(0,0){\includegraphics[width=0.33\textwidth]{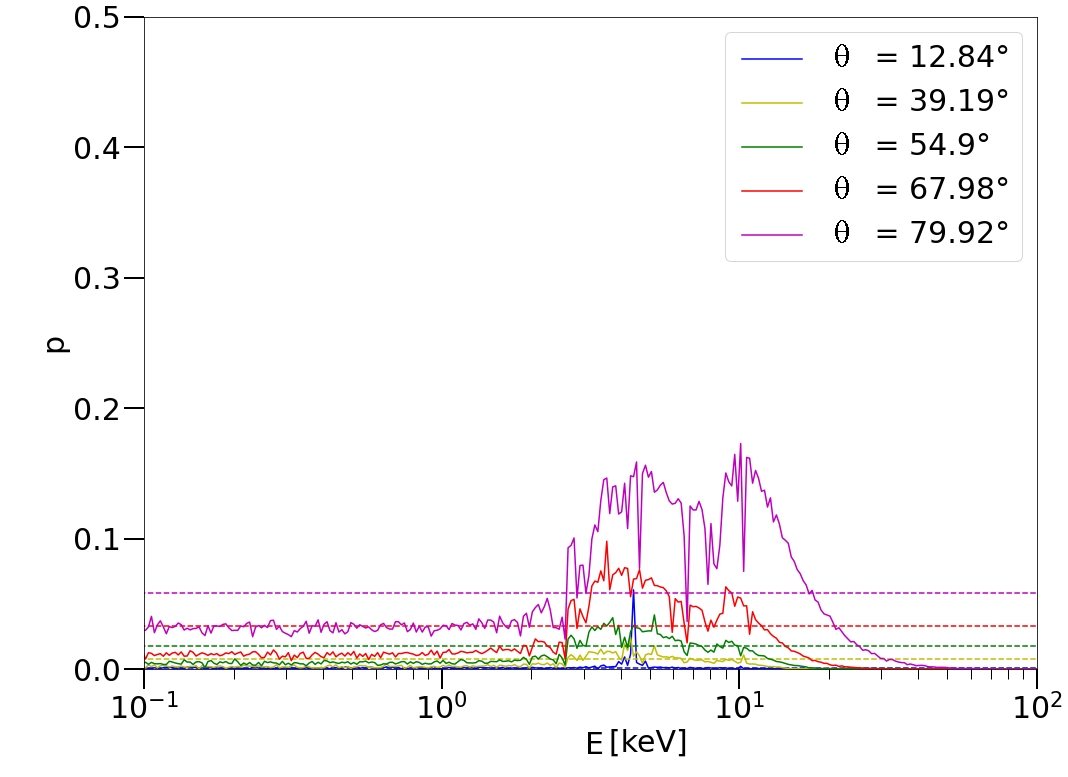}}
    \put(20,90){\scalebox{.55}{$\tau_\textrm{T,max} = 0.67$}}
    \put(20,80){\scalebox{.55}{$n_\textrm{H} = 10^{16} \ \textrm{cm}^{-3}$}}
    \end{picture}
    \begin{picture}(137,99)
    \put(0,0){\includegraphics[width=0.33\textwidth]{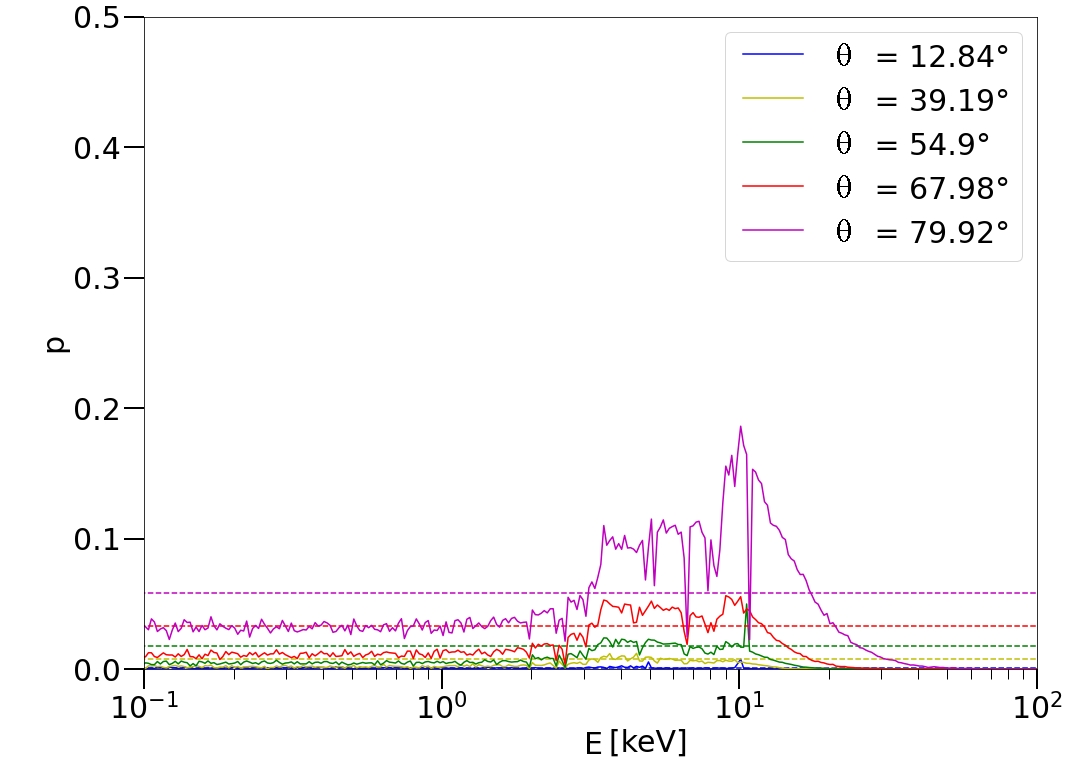}}
    \put(20,90){\scalebox{.55}{$\tau_\textrm{T,max} = 0.67$}}
    \put(20,80){\scalebox{.55}{$n_\textrm{H} = 10^{15} \ \textrm{cm}^{-3}$}}
    \end{picture}
    \begin{picture}(137,99)
    \put(0,0){\includegraphics[width=0.33\textwidth]{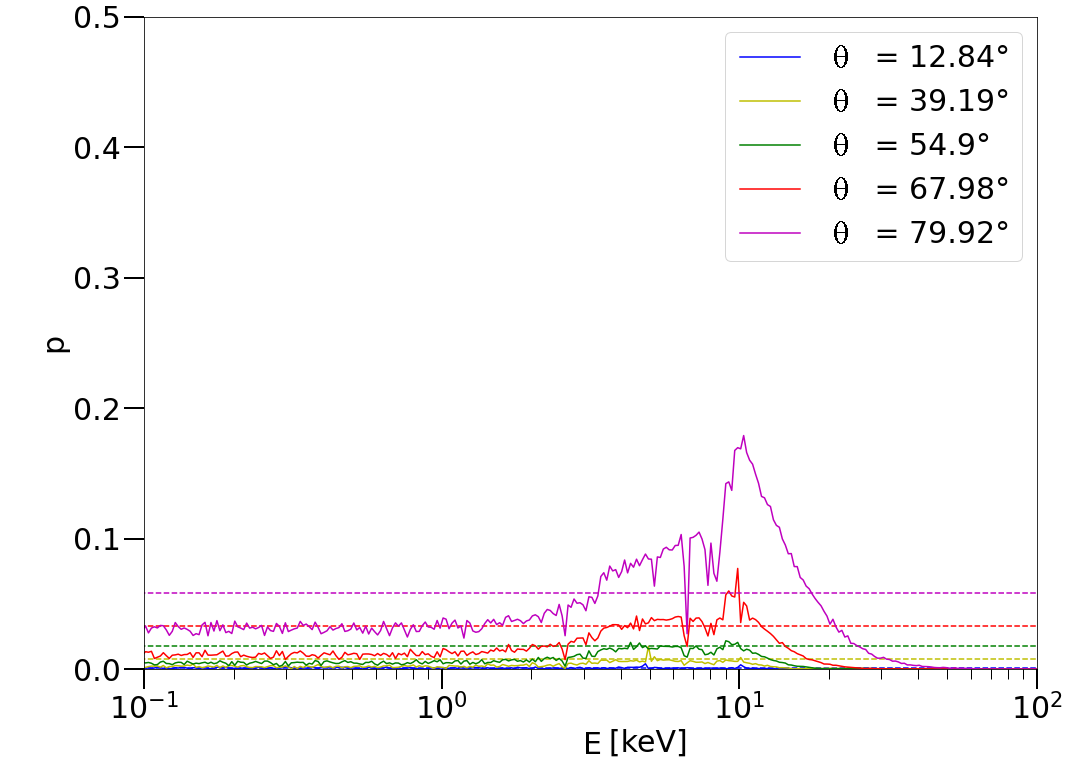}}
    \put(20,90){\scalebox{.55}{$\tau_\textrm{T,max} = 0.67$}}
    \put(20,80){\scalebox{.55}{$n_\textrm{H} = 10^{14} \ \textrm{cm}^{-3}$}}
    \end{picture}\\
    \begin{picture}(137,99)
    \put(0,0){\includegraphics[width=0.33\textwidth]{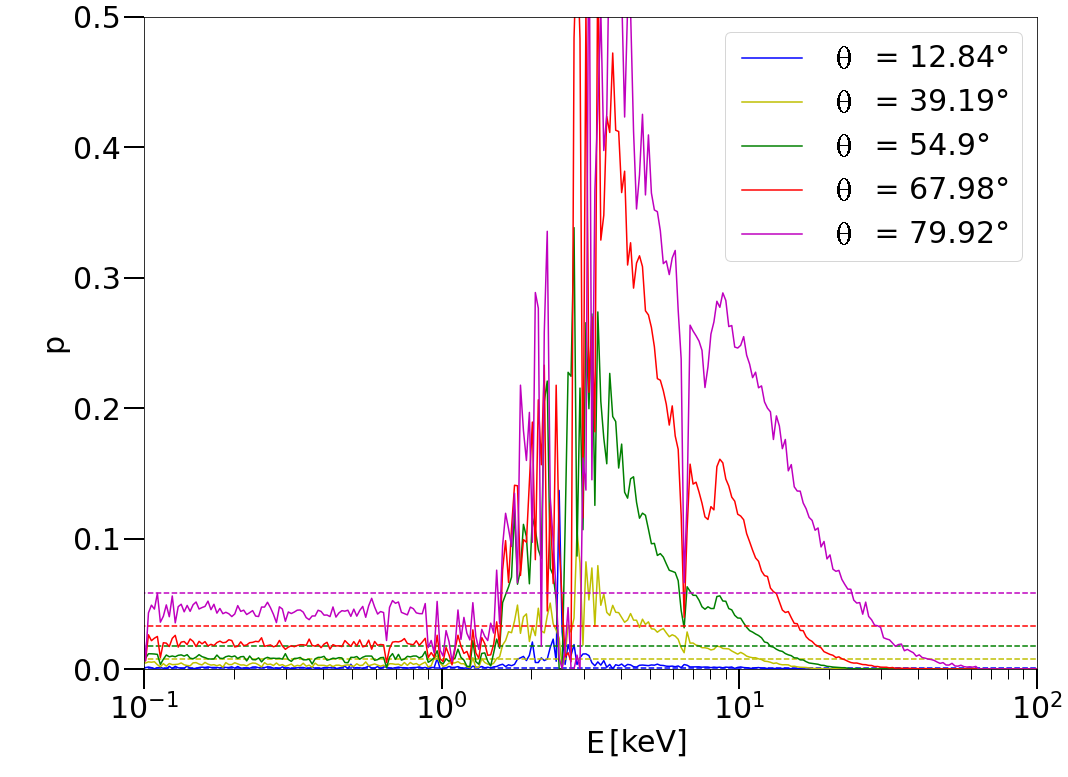}}
    \put(20,90){\scalebox{.55}{$\tau_\textrm{T,max} = 1$}}
    \put(20,80){\scalebox{.55}{$n_\textrm{H} = 10^{18} \ \textrm{cm}^{-3}$}}
    \end{picture}
    \begin{picture}(137,99)
    \put(0,0){\includegraphics[width=0.33\textwidth]{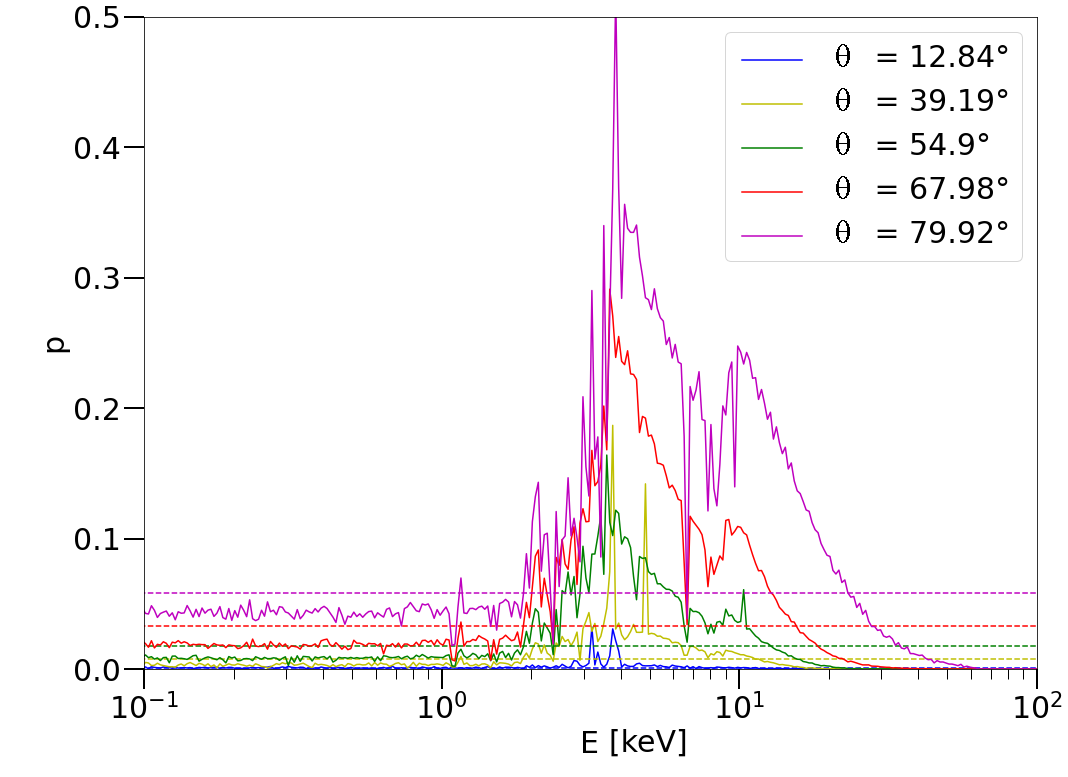}}
    \put(20,90){\scalebox{.55}{$\tau_\textrm{T,max} = 1$}}
    \put(20,80){\scalebox{.55}{$n_\textrm{H} = 10^{17} \ \textrm{cm}^{-3}$}}
    \end{picture}
    \begin{picture}(137,99)
    \put(0,0){\includegraphics[width=0.33\textwidth]{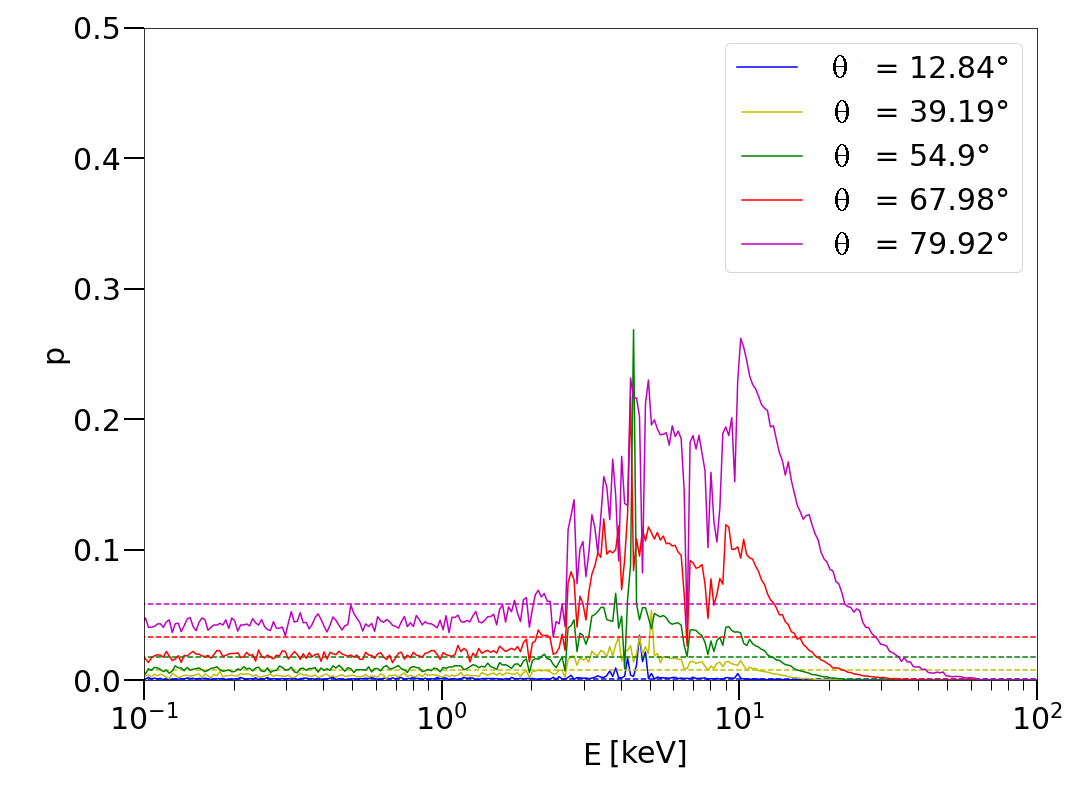}}
    \put(20,90){\scalebox{.55}{$\tau_\textrm{T,max} = 1$}}
    \put(20,80){\scalebox{.55}{$n_\textrm{H} = 10^{16} \ \textrm{cm}^{-3}$}}
    \end{picture}
    \begin{picture}(137,99)
    \put(0,0){\includegraphics[width=0.33\textwidth]{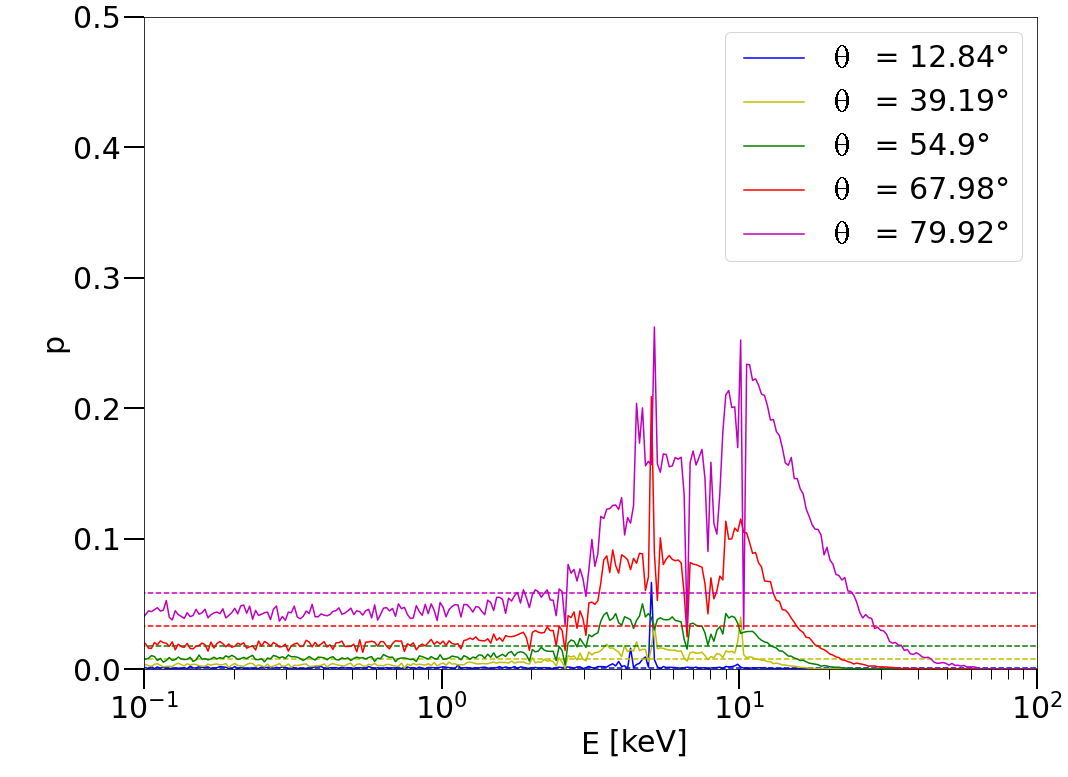}}
    \put(20,90){\scalebox{.55}{$\tau_\textrm{T,max} = 1$}}
    \put(20,80){\scalebox{.55}{$n_\textrm{H} = 10^{15} \ \textrm{cm}^{-3}$}}
    \end{picture}
    \begin{picture}(137,99)
    \put(0,0){\includegraphics[width=0.33\textwidth]{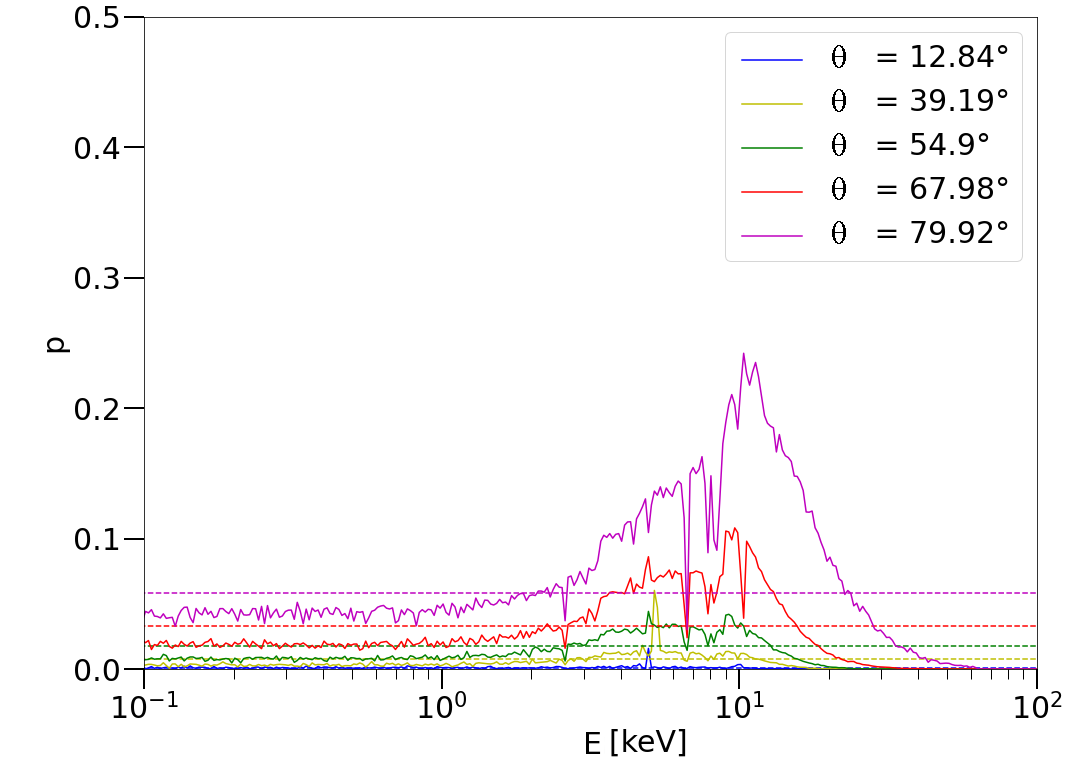}}
    \put(20,90){\scalebox{.55}{$\tau_\textrm{T,max} = 1$}}
    \put(20,80){\scalebox{.55}{$n_\textrm{H} = 10^{14} \ \textrm{cm}^{-3}$}}
    \end{picture}\\
    \begin{picture}(137,99)
    \put(0,0){\includegraphics[width=0.33\textwidth]{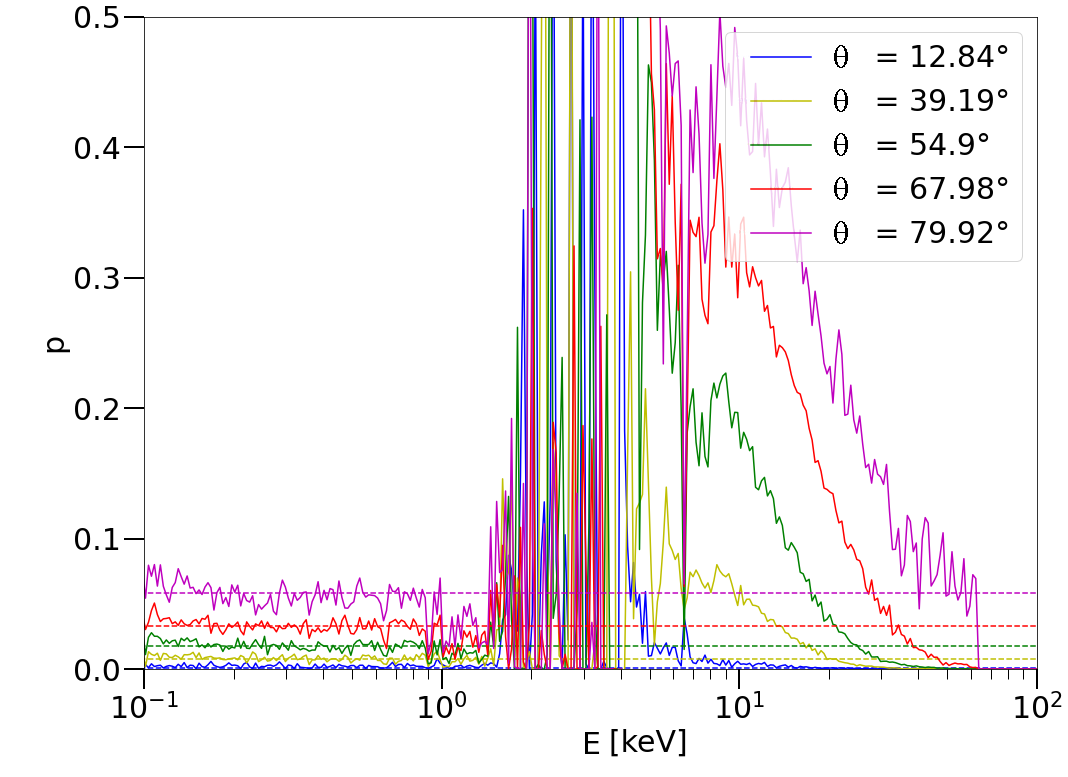}}
    \put(20,90){\scalebox{.55}{$\tau_\textrm{T,max} = 3$}}
    \put(20,80){\scalebox{.55}{$n_\textrm{H} = 10^{18} \ \textrm{cm}^{-3}$}}
    \end{picture}
    \begin{picture}(137,99)
    \put(0,0){\includegraphics[width=0.33\textwidth]{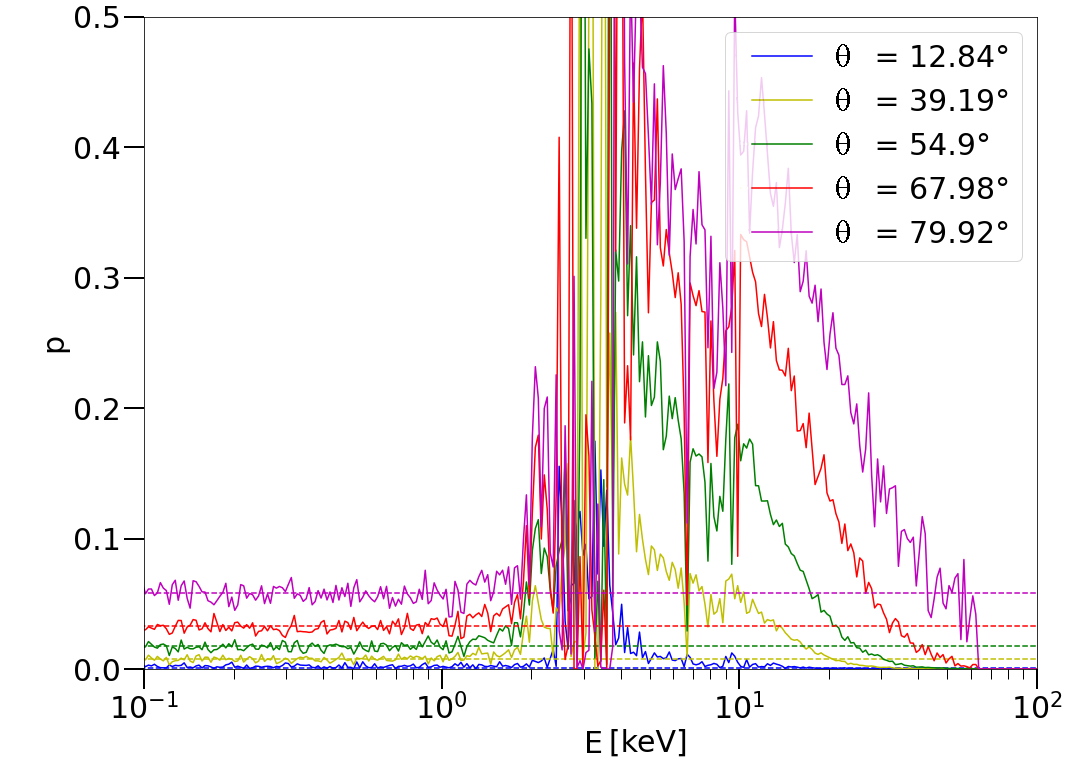}}
    \put(20,90){\scalebox{.55}{$\tau_\textrm{T,max} = 3$}}
    \put(20,80){\scalebox{.55}{$n_\textrm{H} = 10^{17} \ \textrm{cm}^{-3}$}}
    \end{picture}
    \begin{picture}(137,99)
    \put(0,0){\includegraphics[width=0.33\textwidth]{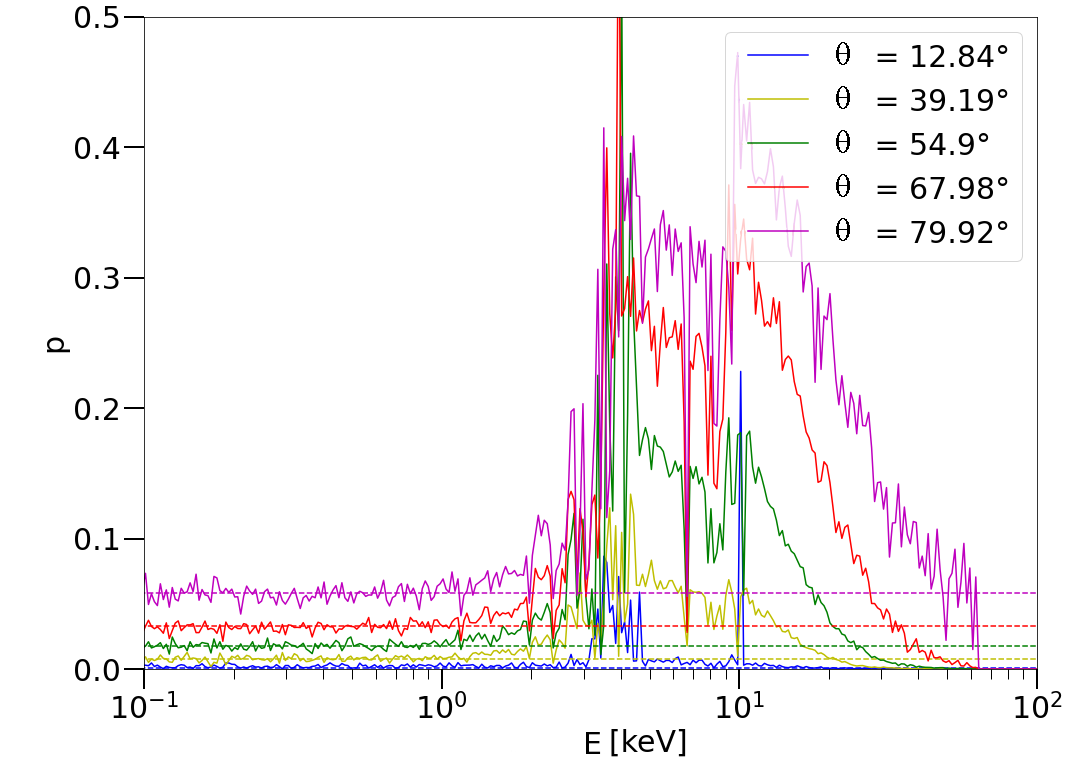}}
    \put(20,90){\scalebox{.55}{$\tau_\textrm{T,max} = 3$}}
    \put(20,80){\scalebox{.55}{$n_\textrm{H} = 10^{16} \ \textrm{cm}^{-3}$}}
    \end{picture}
    \begin{picture}(137,99)
    \put(0,0){\includegraphics[width=0.33\textwidth]{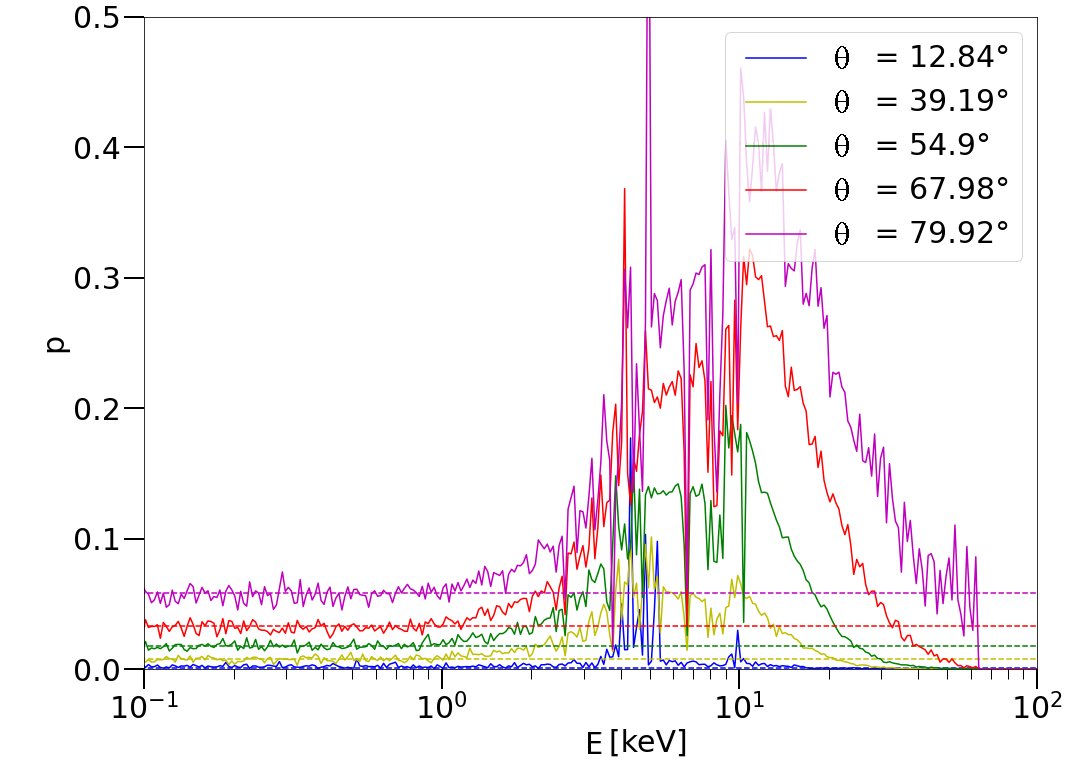}}
    \put(20,90){\scalebox{.55}{$\tau_\textrm{T,max} = 3$}}
    \put(20,80){\scalebox{.55}{$n_\textrm{H} = 10^{15} \ \textrm{cm}^{-3}$}}
    \end{picture}
    \begin{picture}(137,99)
    \put(0,0){\includegraphics[width=0.33\textwidth]{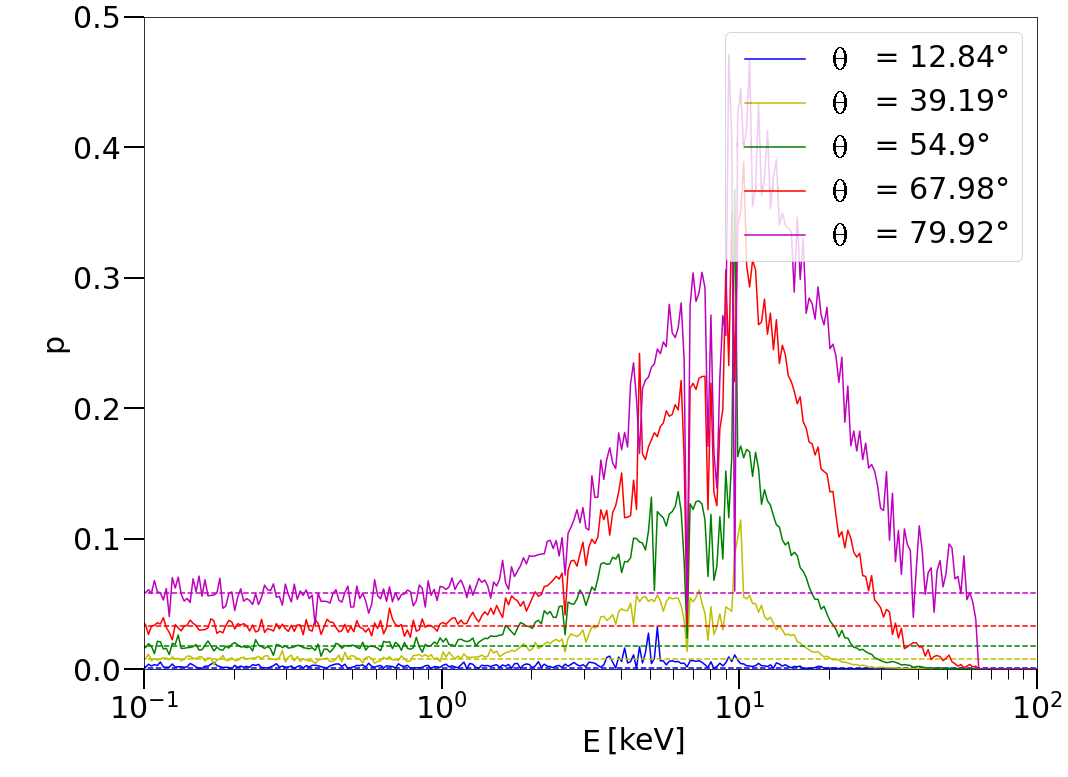}}
    \put(20,90){\scalebox{.55}{$\tau_\textrm{T,max} = 3$}}
    \put(20,80){\scalebox{.55}{$n_\textrm{H} = 10^{14} \ \textrm{cm}^{-3}$}}
    \end{picture}
    \caption{\footnotesize{The polarisation degree, $p$, versus energy obtained from {\tt STOKES} (solid lines) for various observer's inclinations $\theta$ in the color code, corresponding to the cases shown in Figure \ref{absorbed_spectra}. From left to right the density of the slab decreases ($n_\textrm{H} = 10^{18},10^{17},10^{16},10^{15},10^{14} \ \textrm{cm}^{-3}$, respectively) and the slab is more ionized for the same external illumination. Top row: the results for $\tau_\textrm{T,max} = 0.67$. Middle row: the results for $\tau_\textrm{T,max} = 1$ (corresponding directly to the spectra shown in bottom pannels of Figure \ref{absorbed_spectra}). Bottom row: the results for $\tau_\textrm{T,max} = 3$. We indicate the pure-scattering limit for semi-infinite atmosphere given by Chandrasekhar-Sobolev approximation for the same inclination values (dashed horizontal lines). The obtained polarisation angle from {\tt STOKES} is constant with energy and parallel with the slab.}}
    \label{absorbed_pd}
\end{figure}
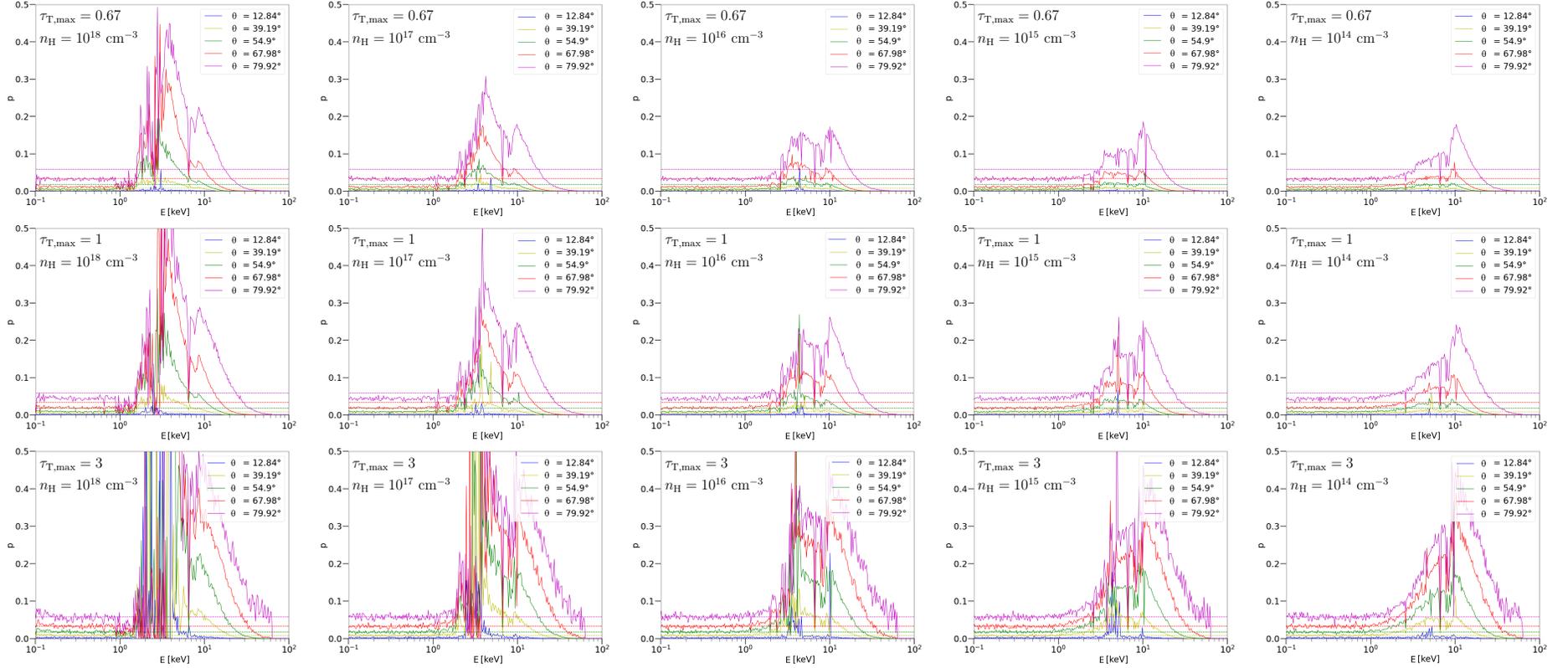
\end{landscape}

~\

At energies where the spectra are heavily absorbed, the polarisation degree rises steeply. If the photo-electric opacity is higher at a particular energy band, it induces polarisation through absorption combined with the underlying black-body distribution. An inclined observer sees photons that are arriving in vertical directions to the escaping surface region with more probability than photons that would travel diagonally through the slab without being absorbed. Hence, the~geometry is effectively reduced and the observer sees polarized photons roughly according to (\ref{thomson_polarization}). For this reason, absorption raises polarisation with increasing slab thickness $\tau_\textrm{T,max}$ and for a given thickness with $\theta$ at \textit{all} energies. It is just that at particular energies, the absorbtion effects may be more significant relative to elsewhere and the general presence of absorption effects is linked to the ionization state. The cases with high or moderate absorption reproduce the~results in \cite{Taverna2021} (see e.g. figure 3 or 8), the spectra are highly absorbed mostly around 3 keV and polarisation rather decreases with energy at~2--8~keV, because of the high gain of polarisation around 3 keV. If the density is low enough, so that the slab is almost fully ionized, the polarisation is steady or increases with energy at 2--8 keV, as at about 3 keV there is small absorption left and~only the~peak at about 10 keV remains in the broadband view. This peak is caused by absorption on top of scattering due to remaining ionization edges, particularly an iron edge at 9 keV, and due to Compton down-scattering effects. The latter causes higher polarisation of emergent photons at higher energies that scattered fewer times with respect to those emergent at lower energies, similarly to~the~effects in reflection described in Section \ref{sp_props_reflection}. The state of complete ionization that would result in~pure-scattering results and energy-independent polarisation degree for elastic scattering at $E \lessapprox 2$ keV and with even lower 2--8 keV average value, could not be reached in the studied conditions with the~radiative transfer codes, if not ad~hoc forced to neglect the absorption effects.

Although absorption in this model scenario can in principle raise polarisation to ever increasing polarisation, as the inclination dependent single-scattering limit (\ref{thomson_polarization}) allows, scattering-induced polarisation has a finite polarisation limit given by the Chandrasekhar-Sobolev result for semi-infinite atmosphere in Thomson regime, approximated by (\ref{viir}). This limit is also inclination dependent and drawn in dashed lines in~Figure \ref{absorbed_pd}. In \cite{Taverna2021}, it was proved that in pure-scattering regime the {\tt CLOUDY} and {\tt STOKES} simulations reach Chandrasekhar's limit for $\tau_\textrm{T,max} \gtrapprox 3$ for low enough energies (to ignore the down-scattering effects at higher energies). In our results that include absorption and Compton down-scattering, we typically observe a plateau in polarisation in between 0.2--1 keV that for $\tau_\textrm{T,max} \gtrapprox 3$ also precisely reaches the~Chandrasekhar's value. Therefore, at these energies we are at the pure-scattering Thomson regime with negligible absorption contribution. Elsewhere polarisation at high $\tau_\textrm{T,max}$ can exceed this limit, as absorption is added on top of the scattering results, because additional anisotropy in photon directions inside the medium is enforced and down-scattering energy redistribution effects on polarisation are scaled too. The~$\tau_\textrm{T,max}$ parameter does not have any theoretical constraint, because it serves as a free parameter of our model, which can approximate a more complex model or real observation with any value between zero and infinity. The effects of incident changing black-body temperature on the emergent polarisation will be discussed in the next section.

~\

If density is monotonically increased, the spectra are first heavily absorbed at~around 3 keV, which causes the additional increase of polarisation, forming a~large polarisation peak at 2--6 keV. The same patterns in spectra and polarisation with high absorption added by density occurs for higher $T_\textrm{BB}$, but generally higher densities are needed to absorb X-ray illumination with higher $T_\textrm{BB}$ [see (\ref{xibb})]. As soon as a particular high ionization threshold is reached by decreasing density (right side of the figures), the results of the~simulation look identical for~ever decreasing density and a state of full ionization cannot be reached in~the~numerical simulations. This could also justify the~constant-density assumption in such highly ionized limit, apart from comparisons to~hydrostatic-equilibrium models. It is important to note that high ionization ($\xi_\textrm{BB} \gtrapprox 10^5 \, \textrm{erg}\cdot \textrm{cm} \cdot \textrm{s}^{-1}$) is actually the~most natural state of such slabs in PIE under these conditions. For~illumination with temperature as low as $k_\textrm{B}T_\textrm{BB} = 0.5$ keV \citep[for e.g. 4U1630-47 the mean was $k_\textrm{B}T_\textrm{BB} \approx 1.15$ keV, ][]{Ratheesh2023}, we already require $n_\textrm{H} \gtrapprox 10^{20.5} \ \textrm{cm}^{-3}$ to see the strong absorption features near 3 keV. Such extreme astrophysical densities are already beyond the convergence limits and trustworthy computations with {\tt TITAN} and the atomic data used, as well as they are difficult to~be believed to~exist in \textit{upper} layers of~the~XRB disc atmospheres, although sometimes considered in~literature \citep[see e.g.][]{Liu2023}. Moreover, XRB sources like 4U1630-47 or LMC X-3 that we will focus on in Section \ref{chap05} do not show significant (orders-of-magnitude) flux absorption in~the~soft state, they are rather highly ionized with thermal peak well representing a multi-temperature black-body curve with a few ionization edges and spectral lines.

~\

We also tested that similar absorption behavior occurs by keeping the slab density fixed and decreasing the ionization parameter by decreasing flux normalization, i.e. as if the slab was a wind layer placed further and further from a diffuse thermally radiating body. The results are shown in Figures \ref{wind_absorbed_spectra} and~\ref{wind_absorbed_pd} in Appendix \ref{supplementary}. Again, even for a few orders of magnitude lower $\xi$ than $\xi_\textrm{BB} \approx 2.3\times 10^7 \, \textrm{erg} \cdot \textrm{cm} \cdot \textrm{s}^{-1}$ for this case, we are in the highly ionized limit where all spectropolarimetric results are identical up to some numerical noise. The~fact that polarisation is ionization dependent, but it does not matter in~what conditions the~gas achieved such ionization state, makes it difficult to~disentangle the origin(s) of emission and geometry of the studied medium within the~inner-accreting system, if spectroscopy of both continuum and lines also shows degeneracies in~this manner. A reduction of~complete photon isotropy of the source of~radiation would affect polarisation for lower optical depths.

~\

The same dependency on absorption with identical spectropolarimetric features was found by Lorenzo Marra in various regimes of PIE with {\tt CLOUDY}, as well as in CIE with {\tt CLOUDY}, if orders of magnitude higher slab temperatures ($T_\textrm{slab} \gtrapprox 10^8 \, K$) were assumed than in the CIE examined in \cite{Taverna2021}. Therefore, with these recent advances altogether, we enlarged the still narrow view of the problem that was provided in \cite{Taverna2020} not only in energy range, but in the ionization level. We showed how (with energy and amplitude) the absorption effects disappear and that there is some sort of saturation for high enough ionization, but not full ionization, as we still see some absorption effects mostly due to a remaining iron edge. The temperatures required to reach such high ionization limit in CIE would lead to strongly inverted temperature profiles with the top of the atmosphere being hotter than the emitting layer of~the~disc, which is not expected for realistic atmospheres \citep{Davis2005, Rozanska2011}. Thus, if high ionization is required, CIE is perhaps a less viable model than PIE, which produces a monotonically decreasing temperature profile counted from the side of external illumination.

\subsection{The limit of high ionization}\label{high_ionization_limit}

Because the absorbed cases produce polarisation profiles with energy that are highly degenerate with other parameters in play and because we argued that we do not model properly thick and absorbing slabs, we will now focus only on~the~highly ionized results. Not only that high ionization is expected, especially in inner parts of the disc, which tend to have higher contribution to the total flux after relativistic integration \citep[see e.g.][]{Dovciak2004b}, they are more supported by the first \textit{IXPE} results for XRBs in the soft state that tend to show increasing polarisation degree with energy in 2--8 keV (see Section \ref{chap05}). We will argue later that relativistic effects for a standard disc integration in summed polarisation tend to rather depolarize at higher energies and we have argued already that any increase of absorption due to $n_\textrm{H}$ change or $\xi$ change starts raising the polarisation first near 3 keV, causing at parts or in the whole 2--8 keV range locally a steady or decreasing polarisation with energy. Hence, an efficient mechanism for production of polarisation rising with energy is required, which the highly ionized slab may offer.\footnote{\,It is worth to mention that the solution presented here provides only one viable scenario for \textit{locally} increasing 2--8 keV polarisation with energy to explain some of the data. Others will be mentioned in Sections \ref{chap05} and \ref{conclusions}.}

In the high ionization limit, black-body temperature forms the underlying photon distribution, but the ionization structure is temperature and density independent, which is ideal for constructing radial profiles of emission that strongly depend on effective temperature and density in the Novikov-Thorne model. This would not be the case for absorbed polarisation profiles, as in Figure \ref{absorbed_pd}, where the $\xi(T_\textrm{BB},n_\textrm{H})$ fine-tuning of ionization in the local comoving frame with~the~disc would have to match in the net result for a particular disc temperature and density radial profile, including polarisation modification due to GR effects.

~\

Figures \ref{tr_ionized_spectra} and \ref{tr_ionized_pd} show again for a few selected temperatures and high densities (within the high ionization saturation of polarisation profiles): the ionization levels, continuum and line optical depths, spectra from {\tt STOKES} (only for~$\tau_\textrm{T,max} = 1$), and the resulting polarisation degree with energy. We refer to~the~detailed study of \citet[see their figures 1 and 4]{Davis2005} for estimates on~energy-dependent contributions of various processes to the total opacity of~XRB local disc emission. Photoelectric absorption cross-section generally declines as $\sim E^{-3}$ and is already less important compared to scattering above $\approx 0.2$ keV (compare the $\tau_\textrm{eff}/N_\textrm{H}$ values with $\sigma_\textrm{T}$ with polarisation shape with energy around 0.1 keV). The scattering cross-section is constant up to $\approx 0.5$ MeV, below which $\sigma_\textrm{KN} \approx \sigma_\textrm{T}$. However, in the \textit{IXPE} energy range, this trend of scattering importance relatively to absorption is reversed due to the presence of a strong iron ionization edge at 9 keV, even for the highly ionized slabs. It causes the~aforementioned stronger polarisation than the pure-scattering limit at~about 10 keV, which is also convolved with the underlying black-body photon distribution, as down-scattering effect on polarisation is in play. This remaining peak in polarisation is then shifted in energy, as the~underlying black-body peak is shifted. The polarisation peak is flattened towards higher energies, but we observe the black-body distribution effects combined with the~remaining sharp ionization edge at 9 keV.

We can conclude in the following way. High absorption and Compton down-scattering effects (as opposed to the Thomson elastic limit) along with high inclination generally results in high average polarisation degree, increasing with energy in 2--8 keV. The slope of polarisation with energy in 2--8 keV is given by $\tau_\textrm{T,max}$ and $\mu_\textrm{e}$ and to a lesser extent by $T_\textrm{BB}$. For highly ionized sources with high enough mean temperatures, the pure-scattering Thomson plateau at soft X-rays reaches 2 keV. At 2 keV, we are thus closer to pure Thomson scattering than at~the~upper part of~the~\textit{IXPE} range. If Compton up-scattering effects were considered in the simulation, they would presumably act against the down-scattering energy redistribution of photons, thus lifting this energy threshold where polarisation matches the Thomson scattering results to higher energies. This remains to be quantitatively tested with the latest version of {\tt STOKES} \textit{v2.34}.

~\

Hence, the value of polarisation fraction at 2 keV in some spectral states can constrain the lower limit on inclination according to (\ref{viir}) in optically thick regime (any higher inclination value can be achieved with finite $\tau$, which is an~ad~hoc parameter of our model) with some assumption on observed geometry. The \textit{IXPE} observed slope of polarisation degree with energy at 2--8 keV can in~addition serve as a first-order estimation tool for inclination, knowing the~neglect of local up-scattering and relativistic integration effects.

Typically, if considering a photosphere located on top of equatorial inner disc, a subsequent relativistic integration would result in depolarisation; hence, requiring even higher local slab inclinations with respect to the observed polarisation at 2 keV when compared with non-relativistic models. Near black hole, the local emission angle with respect to normal is rather decreased due to strong-gravity effects, plus more trajectories with various polarisation angles are combined in~GR \citep{Dovciak2004b}.
\begin{landscape}
\begin{figure}
    \centering
    \begin{picture}(137,99)
    \put(0,0){\includegraphics[width=0.33\textwidth]{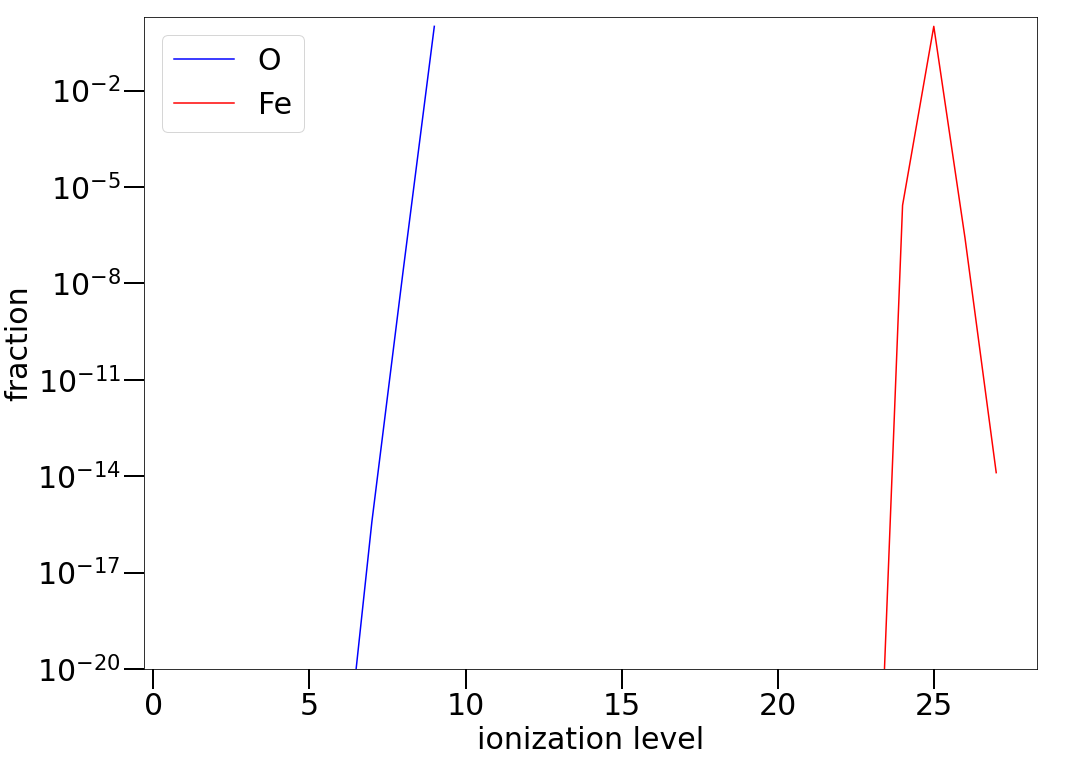}}
    \put(72,20){\scalebox{.55}{$k_\textrm{B}T_\textrm{BB} = 0.23 \ \textrm{keV}$}}
    \end{picture}
    \begin{picture}(137,99)
    \put(0,0){\includegraphics[width=0.33\textwidth]{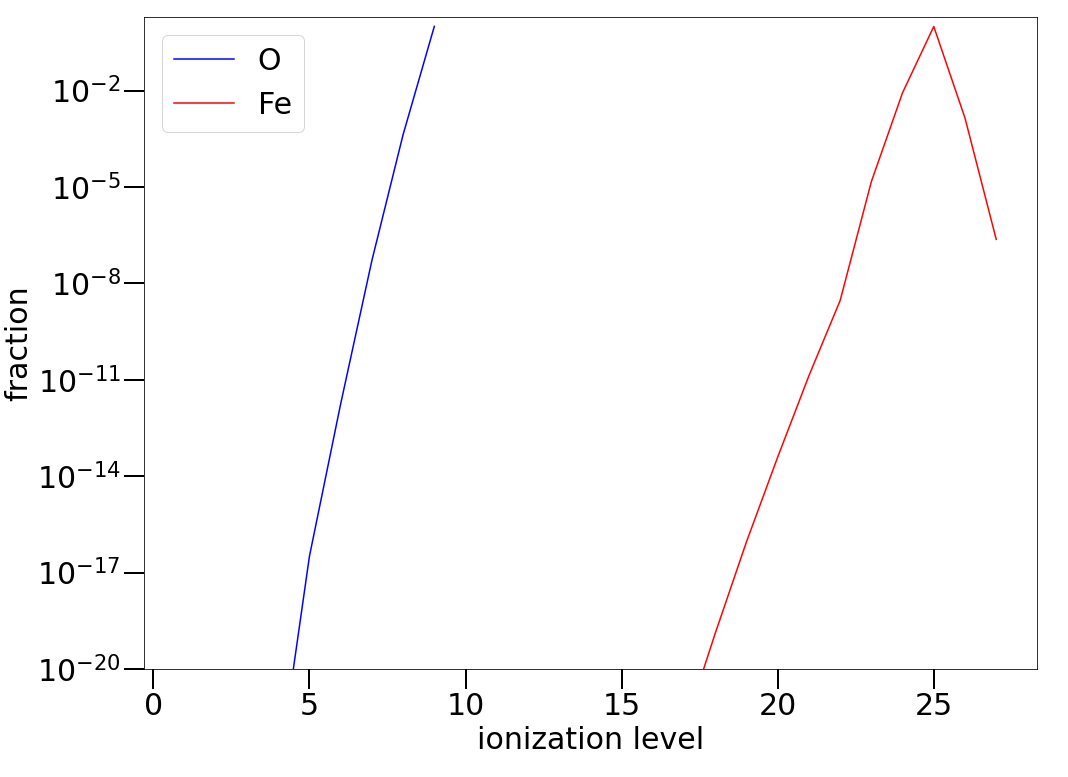}}
    \put(72,20){\scalebox{.55}{$k_\textrm{B}T_\textrm{BB} = 0.50 \ \textrm{keV}$}}
    \end{picture}
    \begin{picture}(137,99)
    \put(0,0){\includegraphics[width=0.33\textwidth]{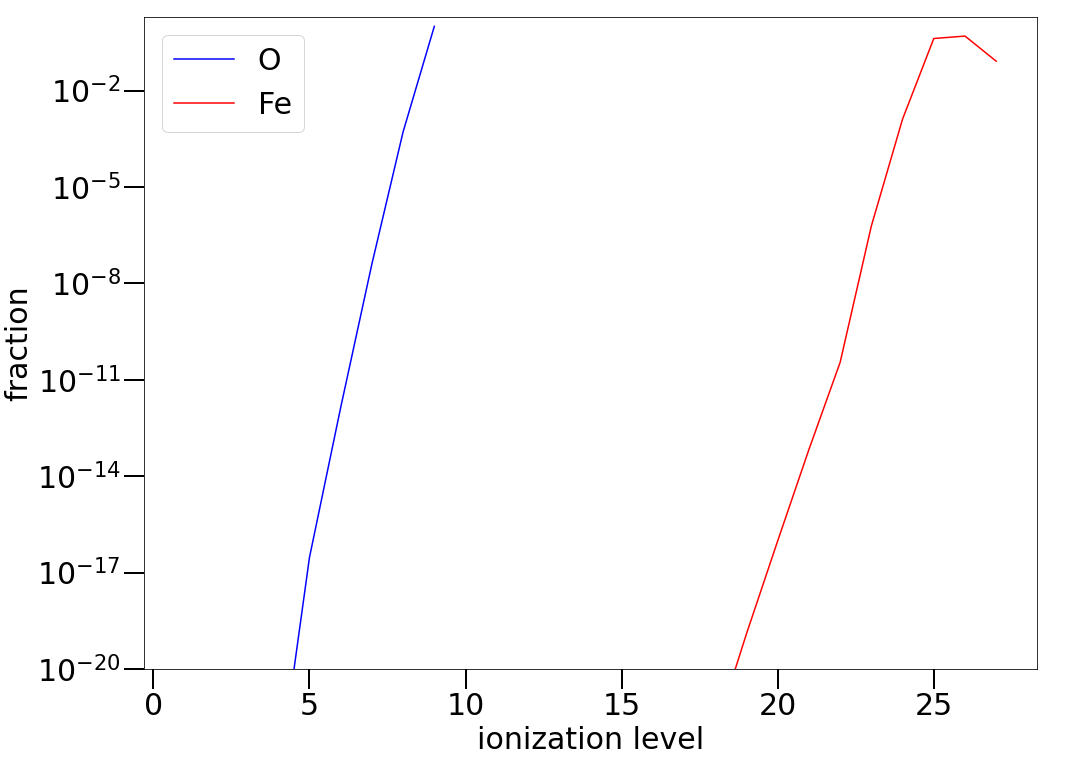}}
    \put(72,20){\scalebox{.55}{$k_\textrm{B}T_\textrm{BB} = 0.80 \ \textrm{keV}$}}
    \end{picture}
    \begin{picture}(137,99)
    \put(0,0){\includegraphics[width=0.33\textwidth]{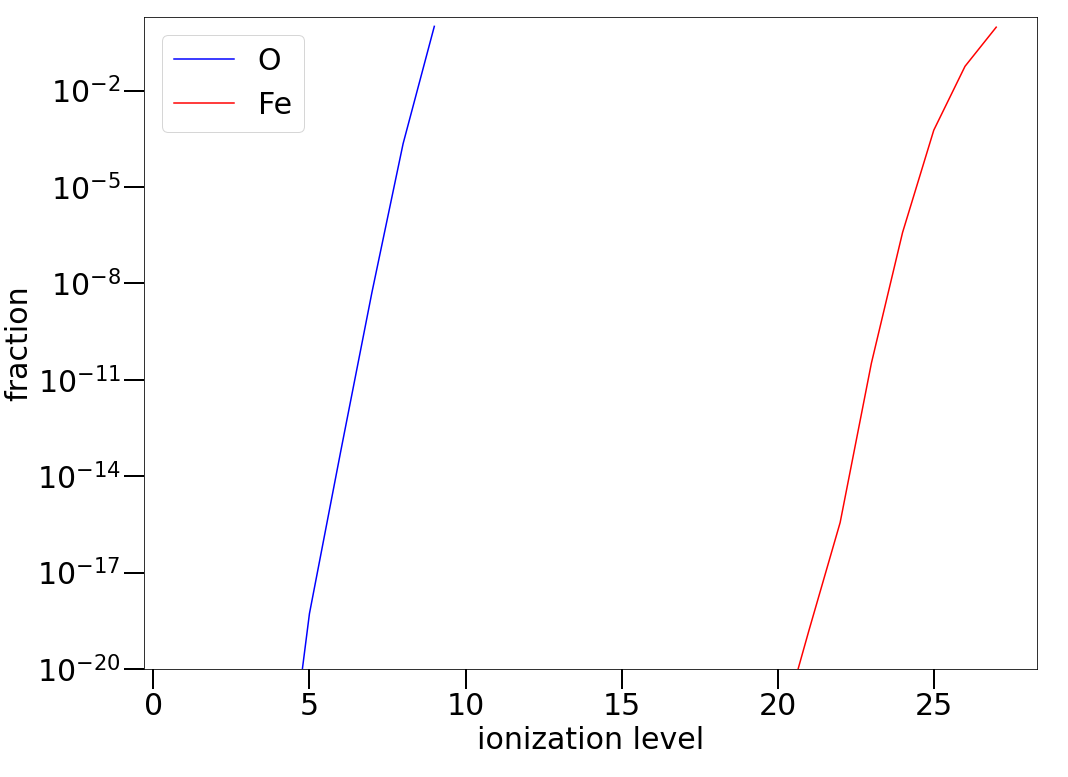}}
    \put(72,20){\scalebox{.55}{$k_\textrm{B}T_\textrm{BB} = 1.15 \ \textrm{keV}$}}
    \end{picture}
    \begin{picture}(137,99)
    \put(0,0){\includegraphics[width=0.33\textwidth]{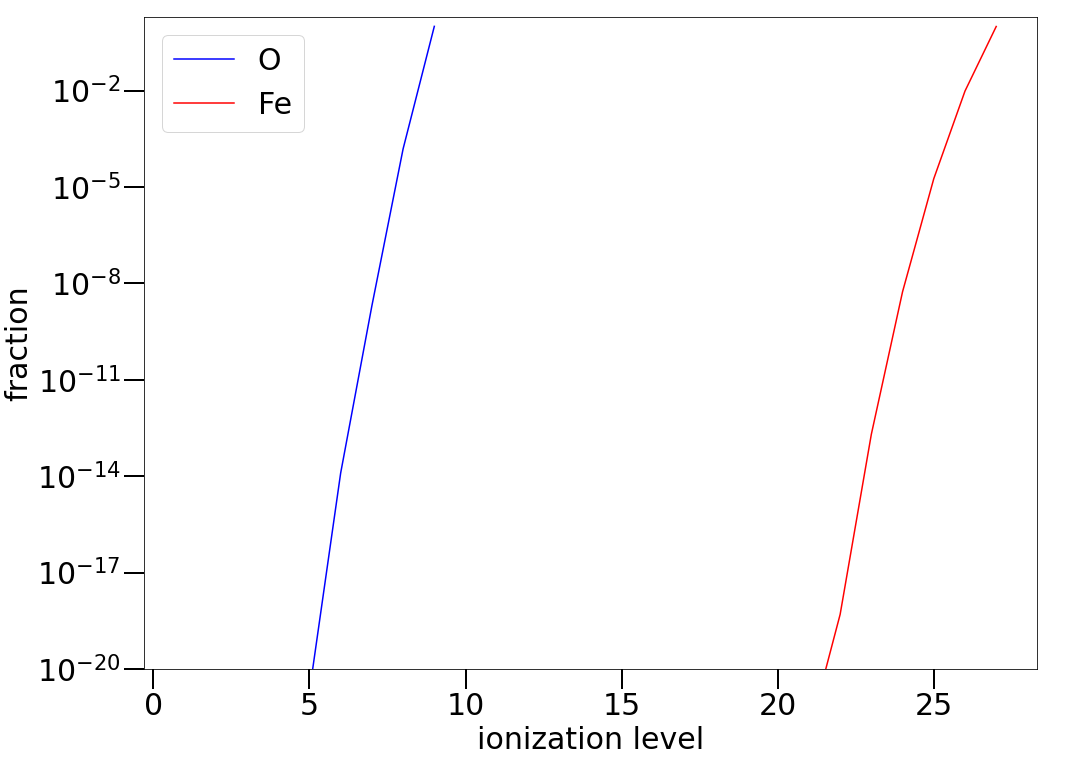}}
    \put(72,20){\scalebox{.55}{$k_\textrm{B}T_\textrm{BB} = 1.40 \ \textrm{keV}$}}
    \end{picture}\\
    \begin{picture}(137,99)
    \put(0,0){\includegraphics[width=0.33\textwidth]{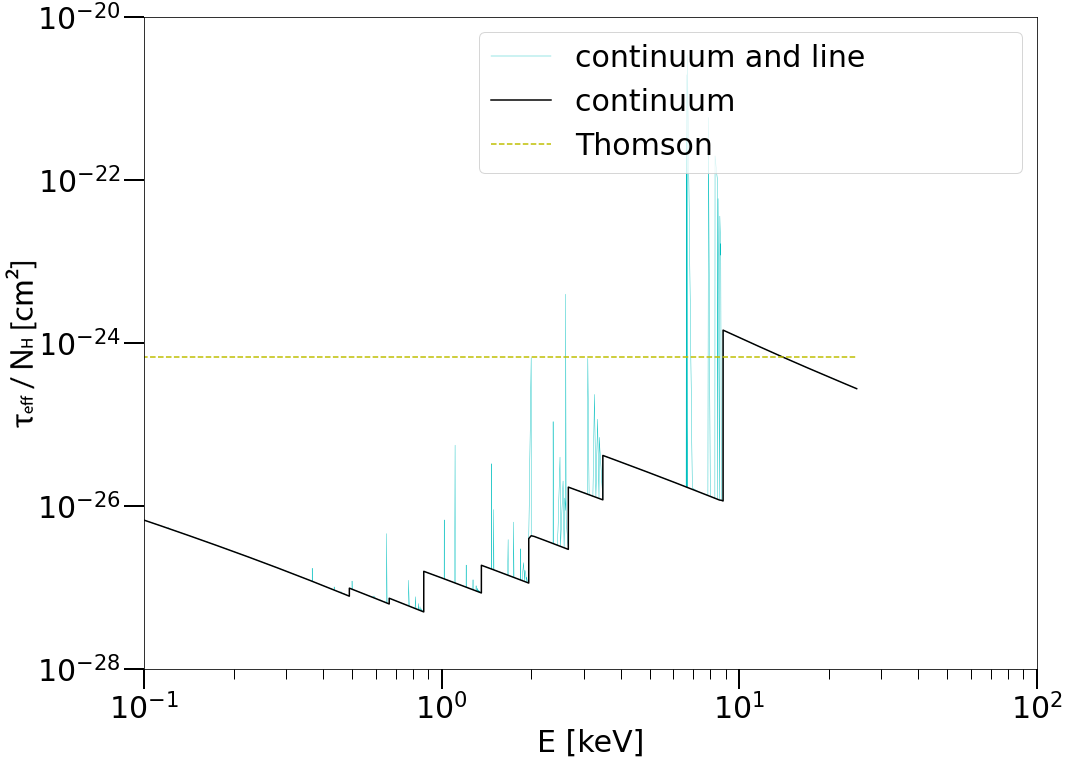}}
    \put(72,20){\scalebox{.55}{$k_\textrm{B}T_\textrm{BB} = 0.23 \ \textrm{keV}$}}
    \end{picture}
    \begin{picture}(137,99)
    \put(0,0){\includegraphics[width=0.33\textwidth]{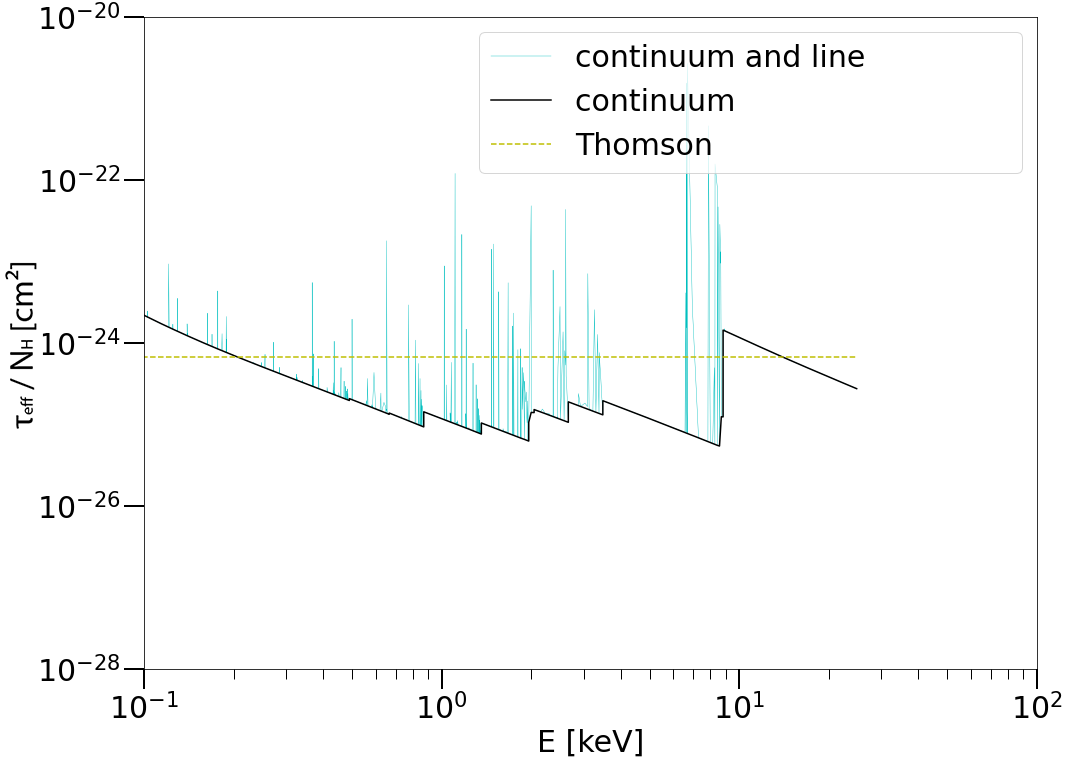}}
    \put(72,20){\scalebox{.55}{$k_\textrm{B}T_\textrm{BB} = 0.50 \ \textrm{keV}$}}
    \end{picture}
    \begin{picture}(137,99)
    \put(0,0){\includegraphics[width=0.33\textwidth]{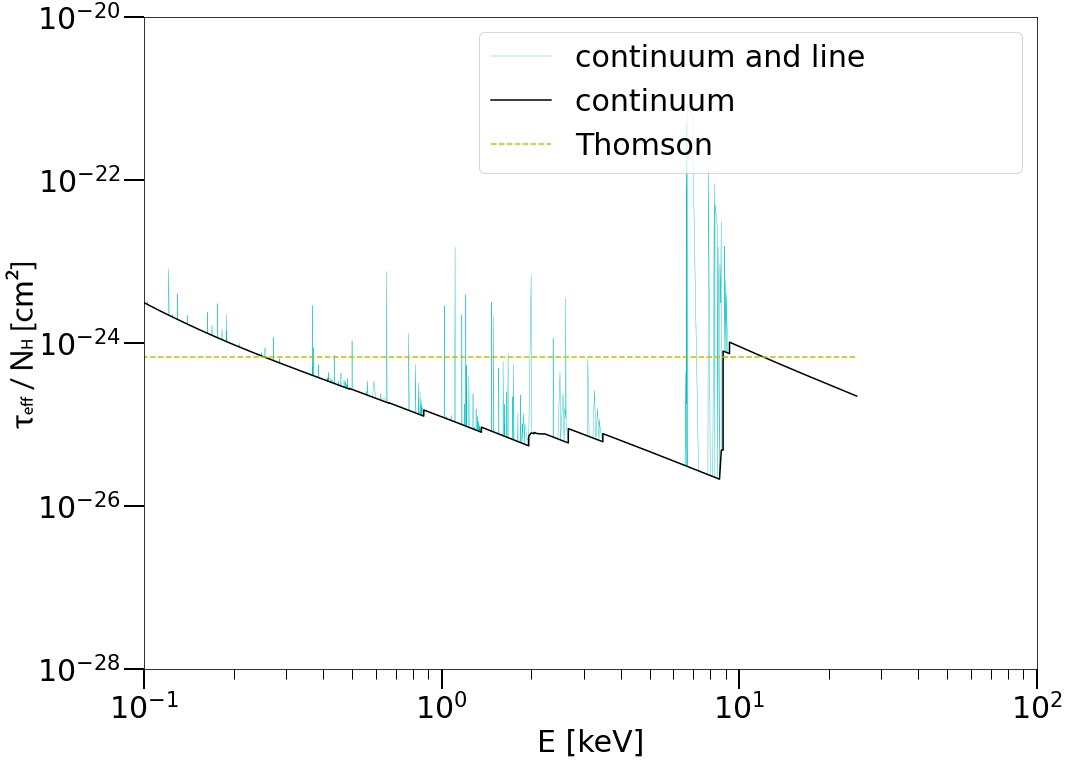}}
    \put(72,20){\scalebox{.55}{$k_\textrm{B}T_\textrm{BB} = 0.80 \ \textrm{keV}$}}
    \end{picture}
    \begin{picture}(137,99)
    \put(0,0){\includegraphics[width=0.33\textwidth]{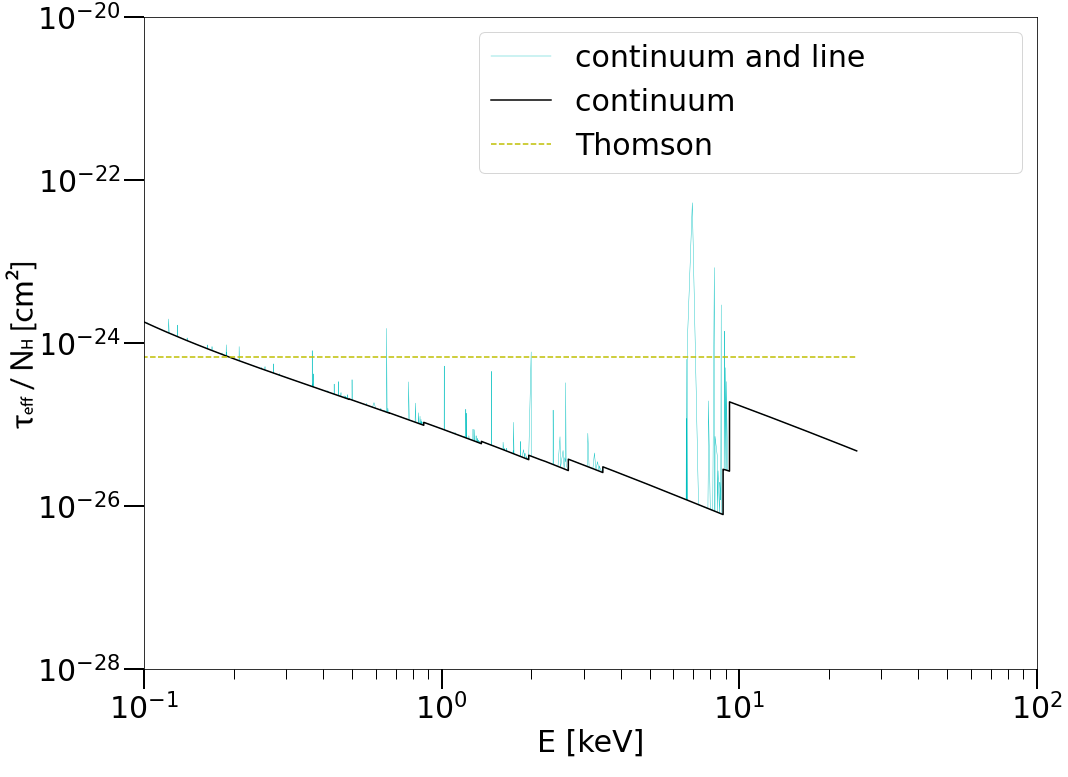}}
    \put(72,20){\scalebox{.55}{$k_\textrm{B}T_\textrm{BB} = 1.15 \ \textrm{keV}$}}
    \end{picture}
    \begin{picture}(137,99)
    \put(0,0){\includegraphics[width=0.33\textwidth]{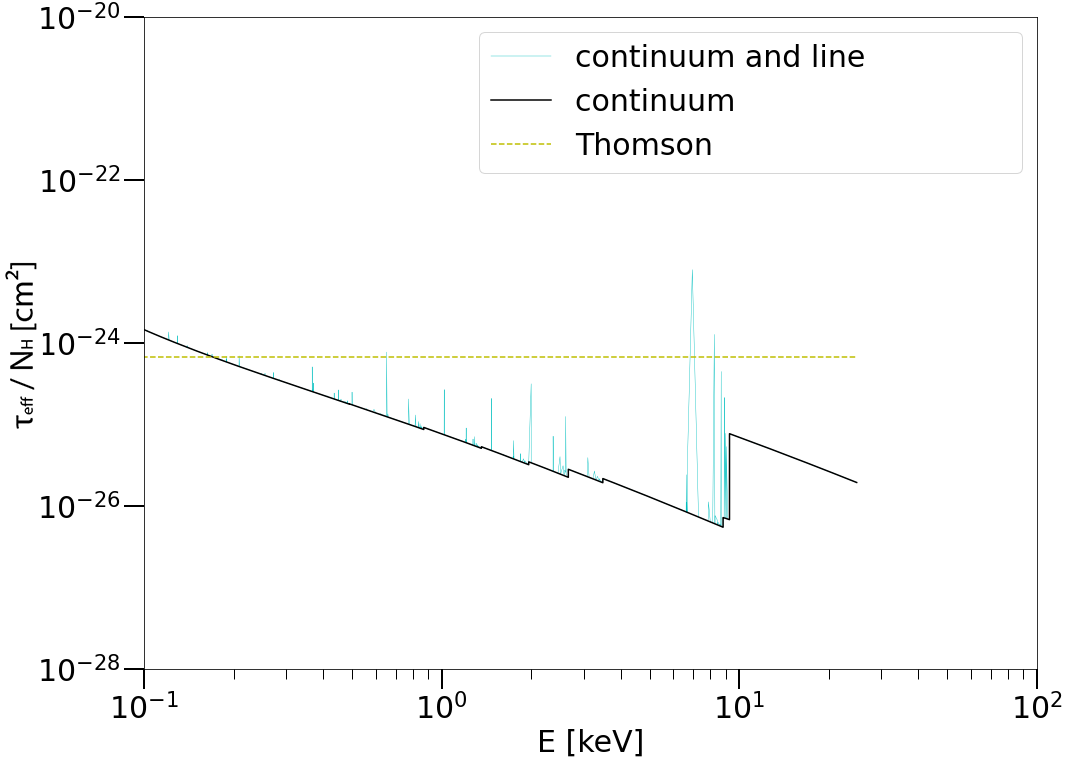}}
    \put(72,20){\scalebox{.55}{$k_\textrm{B}T_\textrm{BB} = 1.40 \ \textrm{keV}$}}
    \end{picture}\\
    \begin{picture}(137,99)
    \put(0,0){\includegraphics[width=0.33\textwidth]{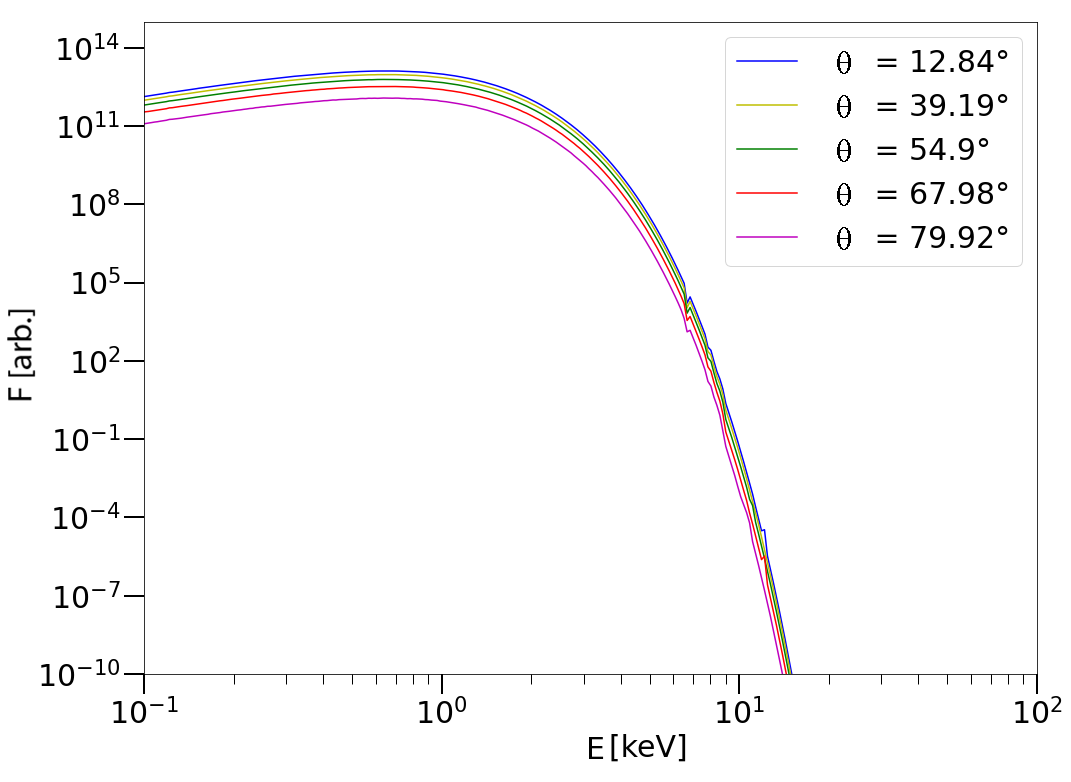}}
    \put(22,30){\scalebox{.55}{$\tau_\textrm{T,max} = 1$}}
    \put(22,20){\scalebox{.55}{$k_\textrm{B}T_\textrm{BB} = 0.23 \ \textrm{keV}$}}
    \end{picture}
    \begin{picture}(137,99)
    \put(0,0){\includegraphics[width=0.33\textwidth]{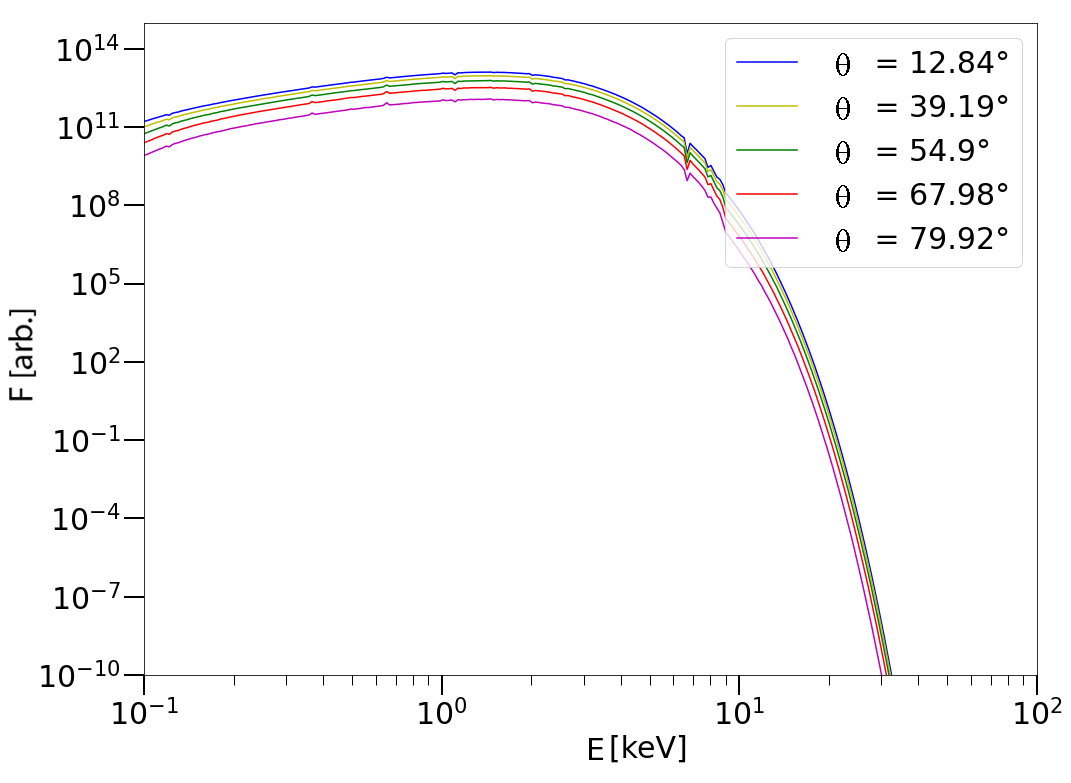}}
    \put(22,30){\scalebox{.55}{$\tau_\textrm{T,max} = 1$}}
    \put(22,20){\scalebox{.55}{$k_\textrm{B}T_\textrm{BB} = 0.50 \ \textrm{keV}$}}
    \end{picture}
    \begin{picture}(137,99)
    \put(0,0){\includegraphics[width=0.33\textwidth]{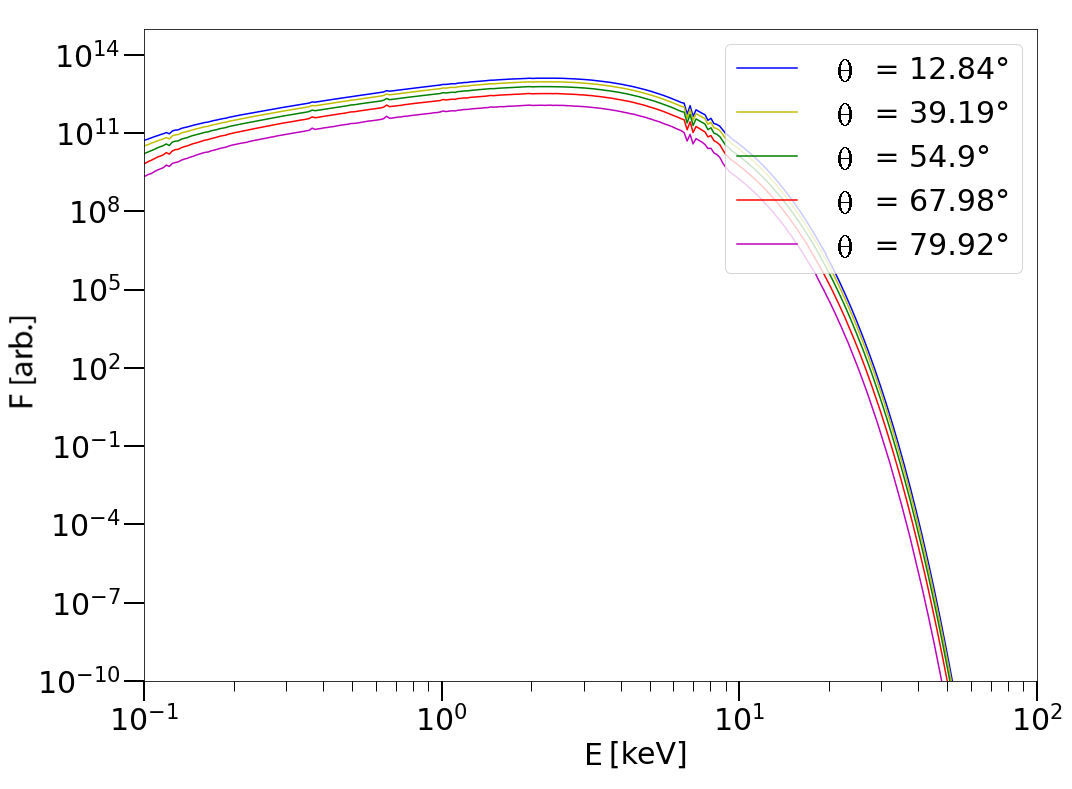}}
    \put(22,30){\scalebox{.55}{$\tau_\textrm{T,max} = 1$}}
    \put(22,20){\scalebox{.55}{$k_\textrm{B}T_\textrm{BB} = 0.80 \ \textrm{keV}$}}
    \end{picture}
    \begin{picture}(137,99)
    \put(0,0){\includegraphics[width=0.33\textwidth]{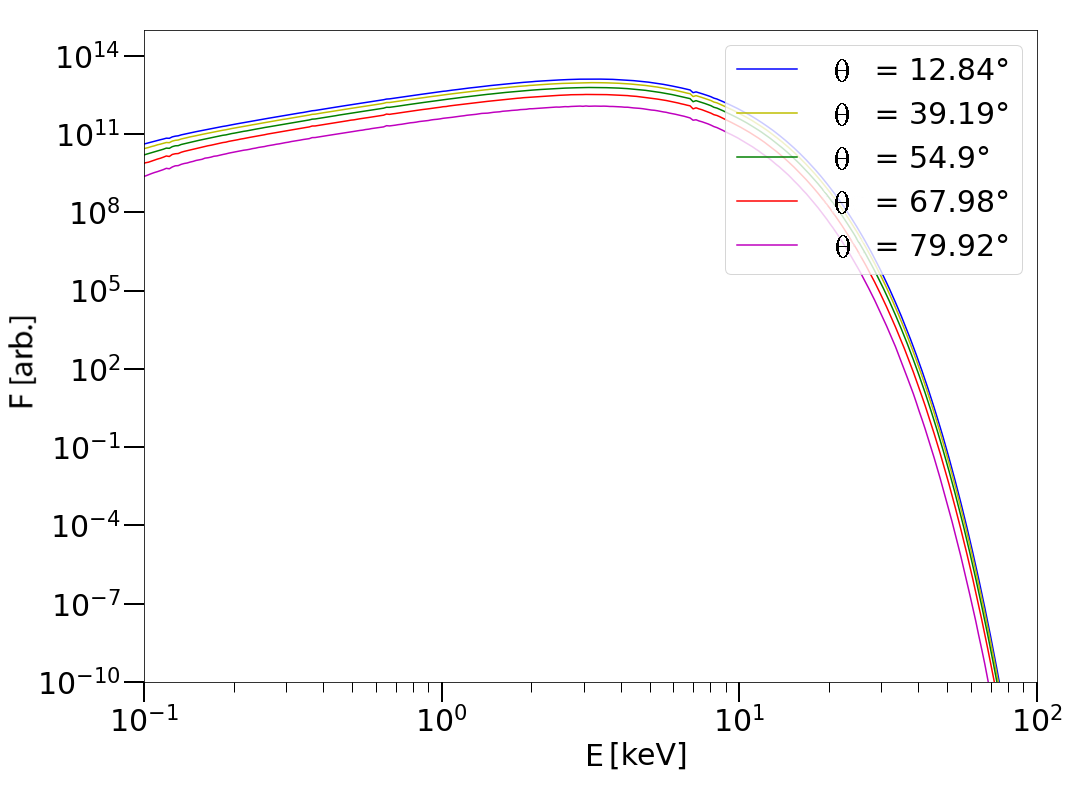}}
    \put(22,30){\scalebox{.55}{$\tau_\textrm{T,max} = 1$}}
    \put(22,20){\scalebox{.55}{$k_\textrm{B}T_\textrm{BB} = 1.15 \ \textrm{keV}$}}
    \end{picture}
    \begin{picture}(137,99)
    \put(0,0){\includegraphics[width=0.33\textwidth]{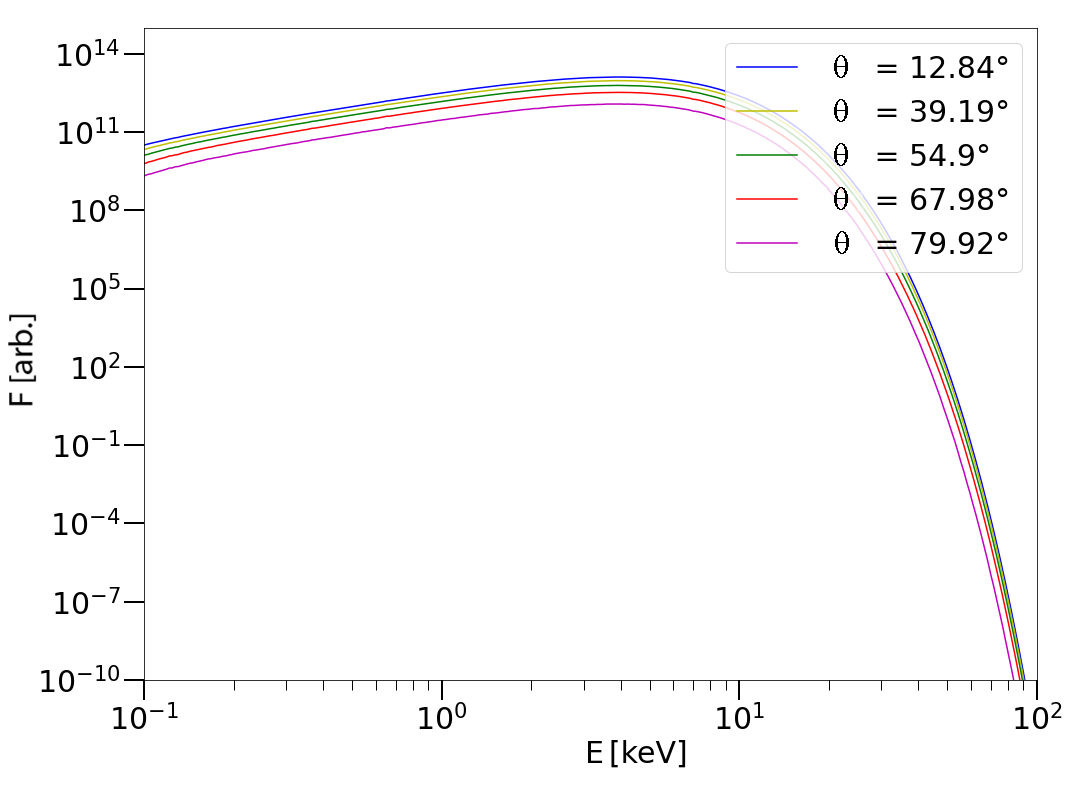}}
    \put(22,30){\scalebox{.55}{$\tau_\textrm{T,max} = 1$}}
    \put(22,20){\scalebox{.55}{$k_\textrm{B}T_\textrm{BB} = 1.40 \ \textrm{keV}$}}
    \end{picture}
    \caption{\footnotesize{Various highly ionized cases of transmitted single-temperature unpolarized black-body radiation through a constant density slab, as computed with {\tt TITAN} and {\tt STOKES} in PIE. From left to right the temperature of the illuminating black body increases ($k_\textrm{B}T_\textrm{BB} = 0.23,0.50,0.80,1.15,1.40$ keV, respectively), while the densities are high ($n_\textrm{H} = 10^{15},10^{20},10^{20.5},10^{20.5},10^{20.5} \ \textrm{cm}^{-3}$, respectively), but not high enough to depart from the saturation state and to cause significant absorption for a given $\xi = \xi_\textrm{BB}$. For any lower density we would see identical result. Top row: fractional abundance of ions of oxygen (blue) OI--OIX and iron (red) FeI--FeXXVII from {\tt TITAN} ionization structure precomputations. Middle row: the effective optical depth, $\tau_\textrm{eff}$, divided by column density, $N_\textrm{H}\, [\textrm{cm}^{-2}]$, versus energy for continuum processes (black) and including lines (cyan) from {\tt TITAN} ionization structure precomputations. Thomson cross-section $\sigma_\textrm{T}$ (yellow) represents pure scatterings. Bottom row: spectra, $F$, per energy bin for various observer's inclinations $\theta$ in the color code for $\tau_\textrm{T,max} = 1$, obtained from {\tt STOKES}.}}
    \label{tr_ionized_spectra}
\end{figure}
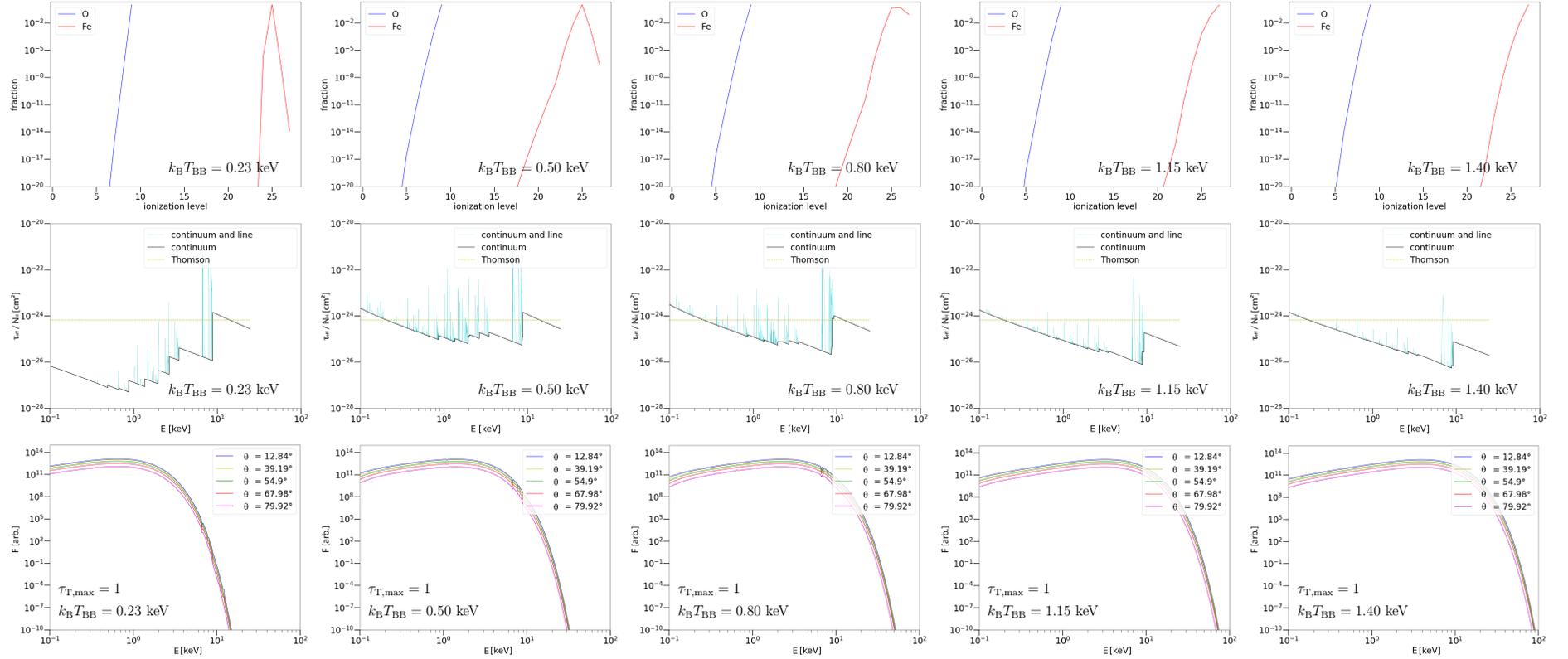
\end{landscape}
\begin{landscape}
\begin{figure}
    \centering
    \begin{picture}(137,99)
    \put(0,0){\includegraphics[width=0.33\textwidth]{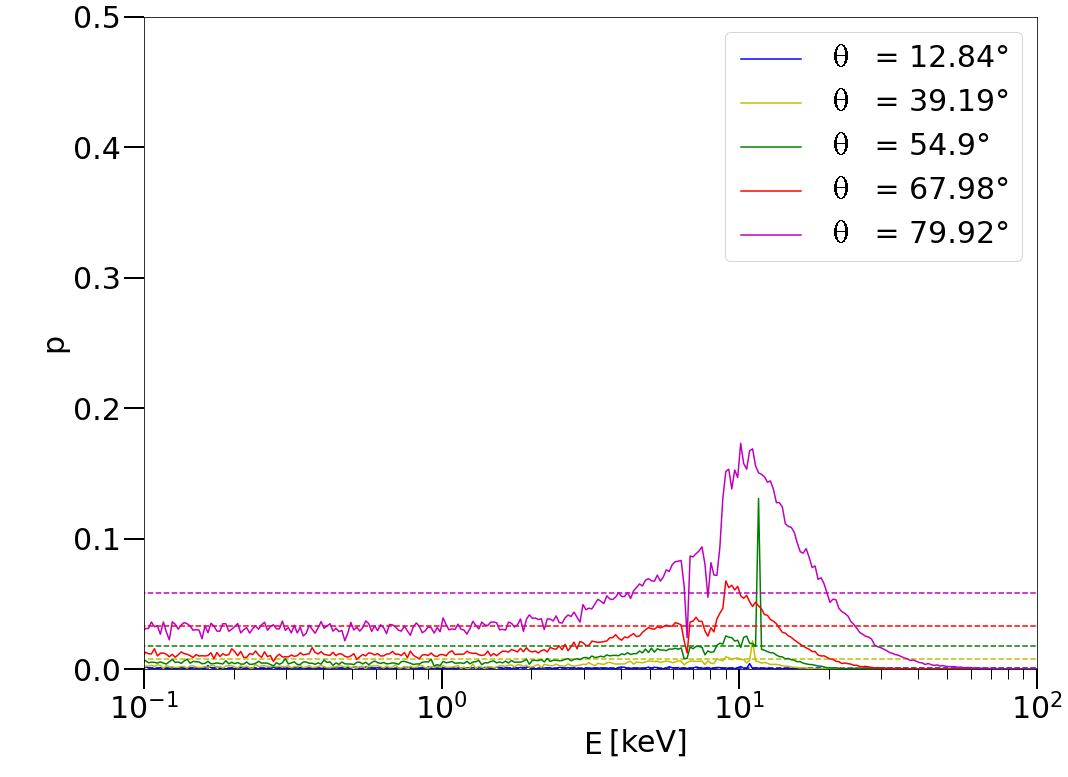}}
    \put(20,90){\scalebox{.55}{$\tau_\textrm{T,max} = 0.67$}}
    \put(20,80){\scalebox{.55}{$k_\textrm{B}T_\textrm{BB} = 0.23 \ \textrm{keV}$}}
    \end{picture}
    \begin{picture}(137,99)
    \put(0,0){\includegraphics[width=0.33\textwidth]{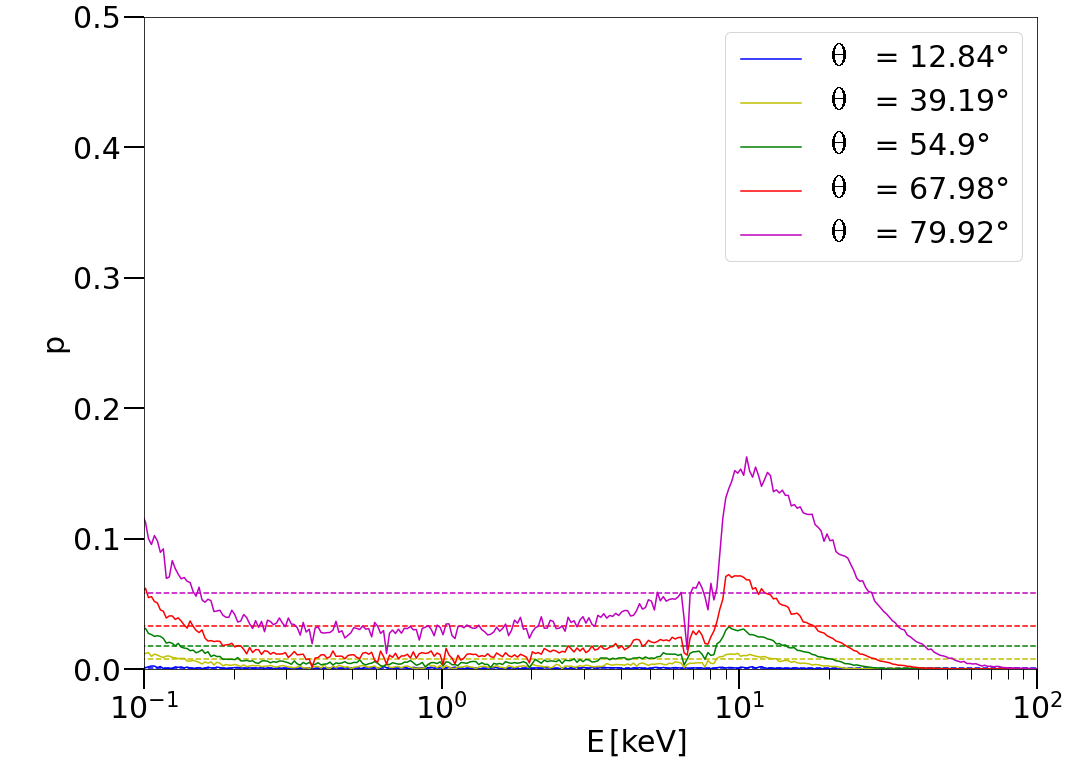}}
    \put(20,90){\scalebox{.55}{$\tau_\textrm{T,max} = 0.67$}}
    \put(20,80){\scalebox{.55}{$k_\textrm{B}T_\textrm{BB} = 0.50 \ \textrm{keV}$}}
    \end{picture}
    \begin{picture}(137,99)
    \put(0,0){\includegraphics[width=0.33\textwidth]{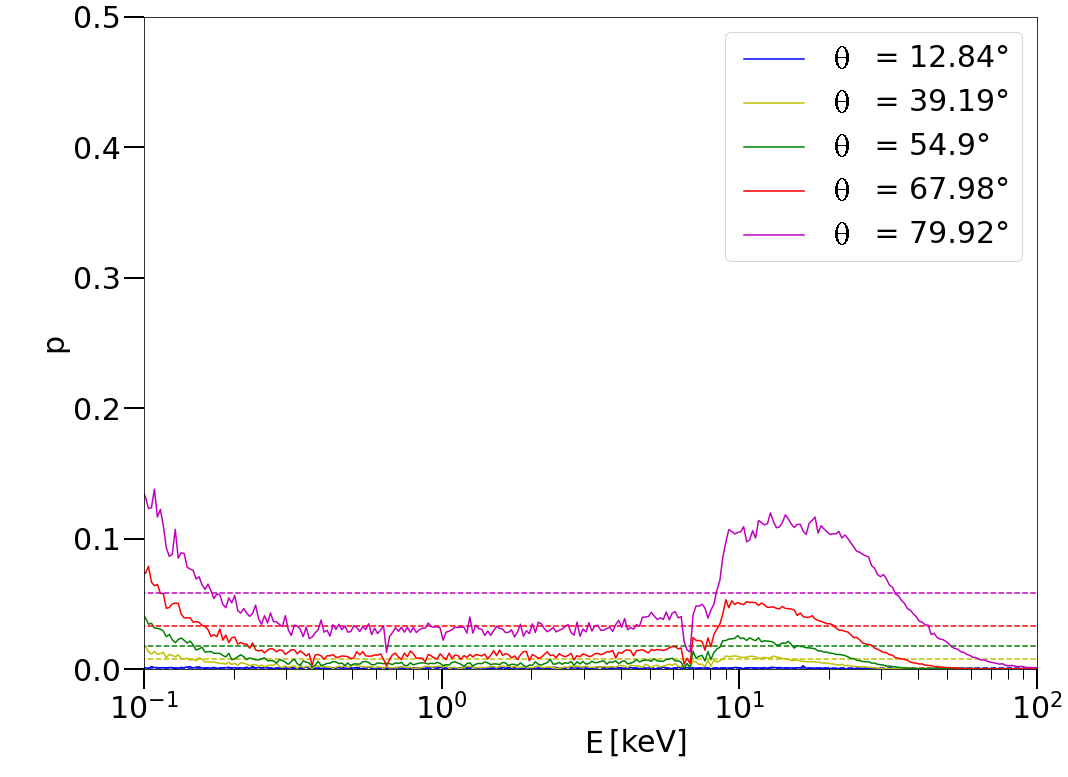}}
    \put(20,90){\scalebox{.55}{$\tau_\textrm{T,max} = 0.67$}}
    \put(20,80){\scalebox{.55}{$k_\textrm{B}T_\textrm{BB} = 0.80 \ \textrm{keV}$}}
    \end{picture}
    \begin{picture}(137,99)
    \put(0,0){\includegraphics[width=0.33\textwidth]{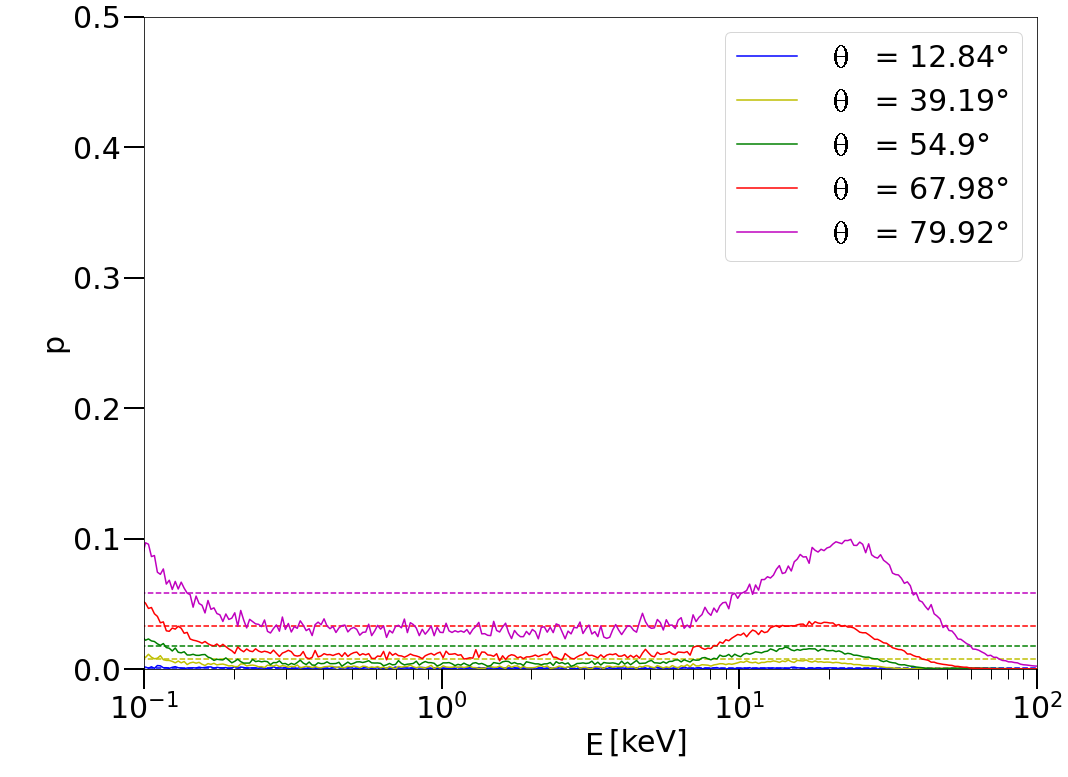}}
    \put(20,90){\scalebox{.55}{$\tau_\textrm{T,max} = 0.67$}}
    \put(20,80){\scalebox{.55}{$k_\textrm{B}T_\textrm{BB} = 1.15 \ \textrm{keV}$}}
    \end{picture}
    \begin{picture}(137,99)
    \put(0,0){\includegraphics[width=0.33\textwidth]{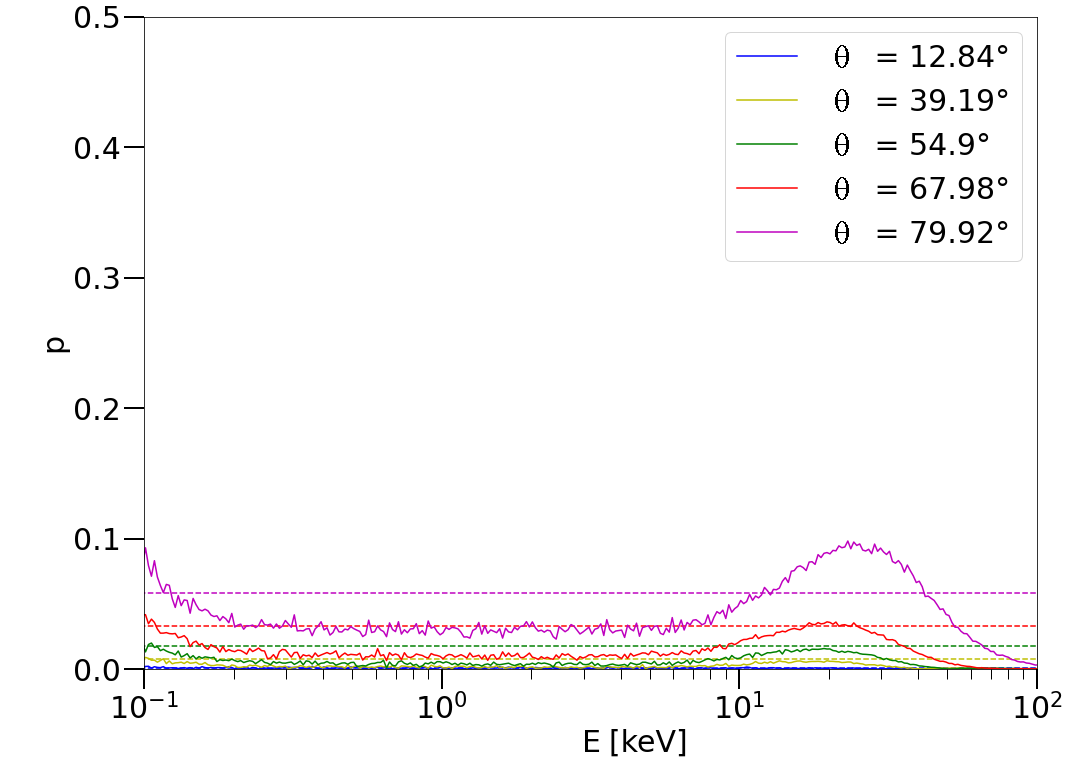}}
    \put(20,90){\scalebox{.55}{$\tau_\textrm{T,max} = 0.67$}}
    \put(20,80){\scalebox{.55}{$k_\textrm{B}T_\textrm{BB} = 1.40 \ \textrm{keV}$}}
    \end{picture}\\
    \begin{picture}(137,99)
    \put(0,0){\includegraphics[width=0.33\textwidth]{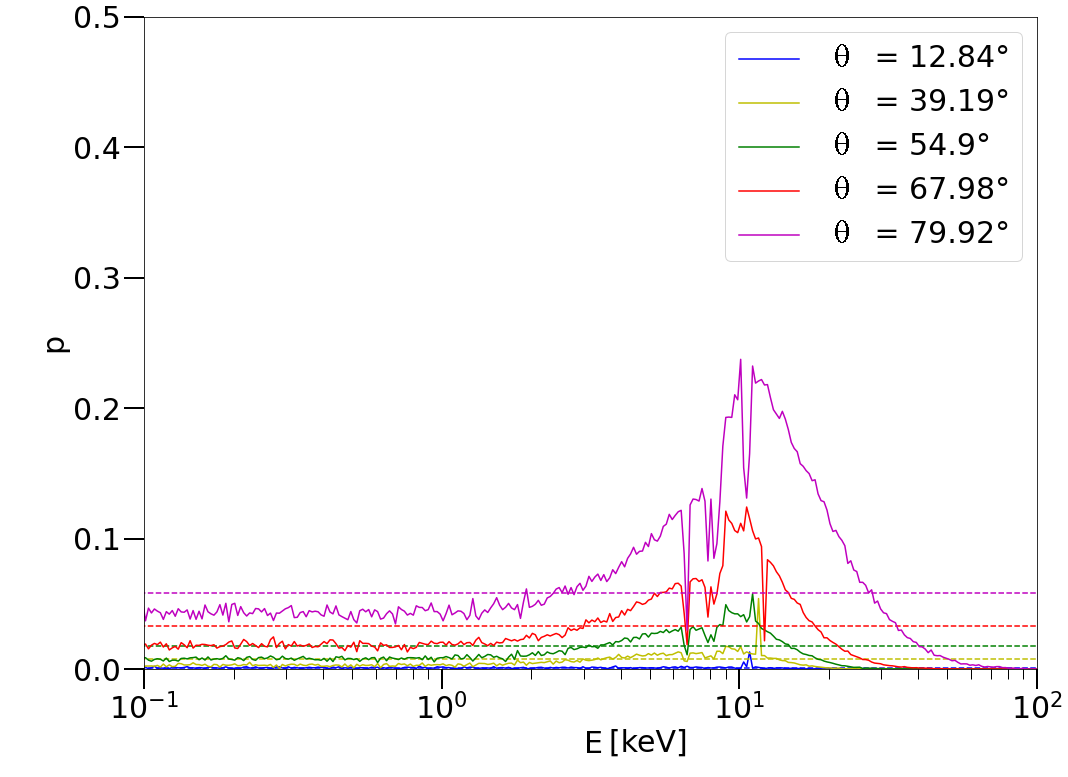}}
    \put(20,90){\scalebox{.55}{$\tau_\textrm{T,max} = 1$}}
    \put(20,80){\scalebox{.55}{$k_\textrm{B}T_\textrm{BB} = 0.23 \ \textrm{keV}$}}
    \end{picture}
    \begin{picture}(137,99)
    \put(0,0){\includegraphics[width=0.33\textwidth]{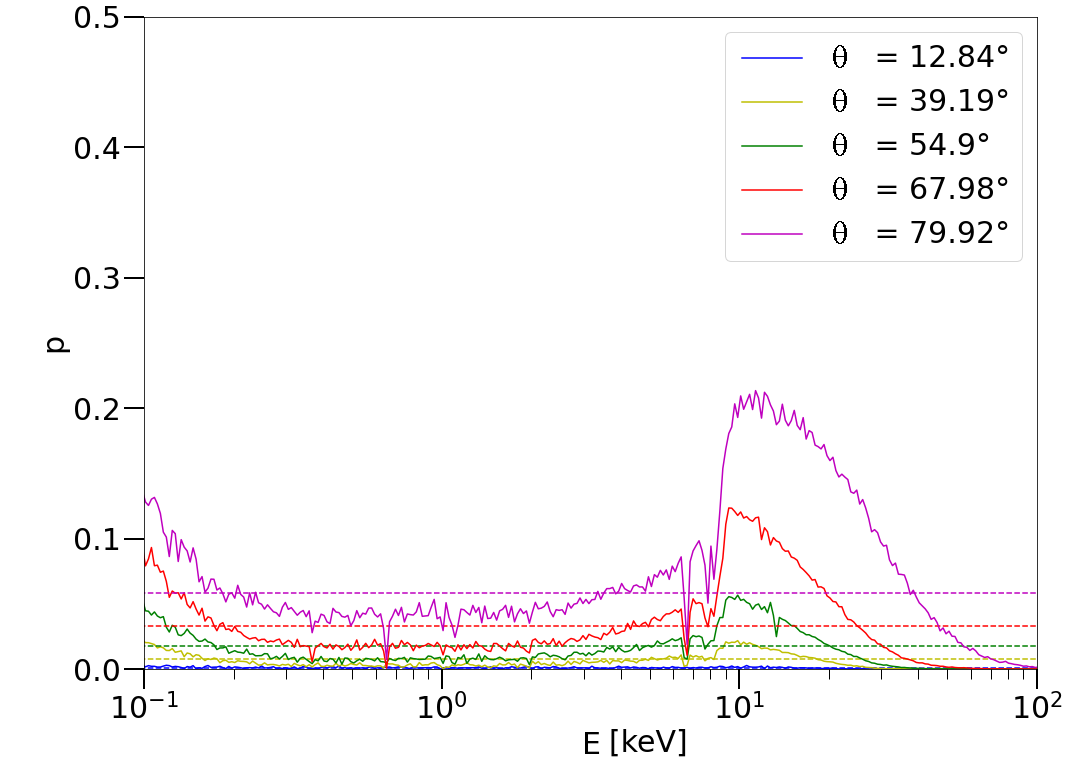}}
    \put(20,90){\scalebox{.55}{$\tau_\textrm{T,max} = 1$}}
    \put(20,80){\scalebox{.55}{$k_\textrm{B}T_\textrm{BB} = 0.50 \ \textrm{keV}$}}
    \end{picture}
    \begin{picture}(137,99)
    \put(0,0){\includegraphics[width=0.33\textwidth]{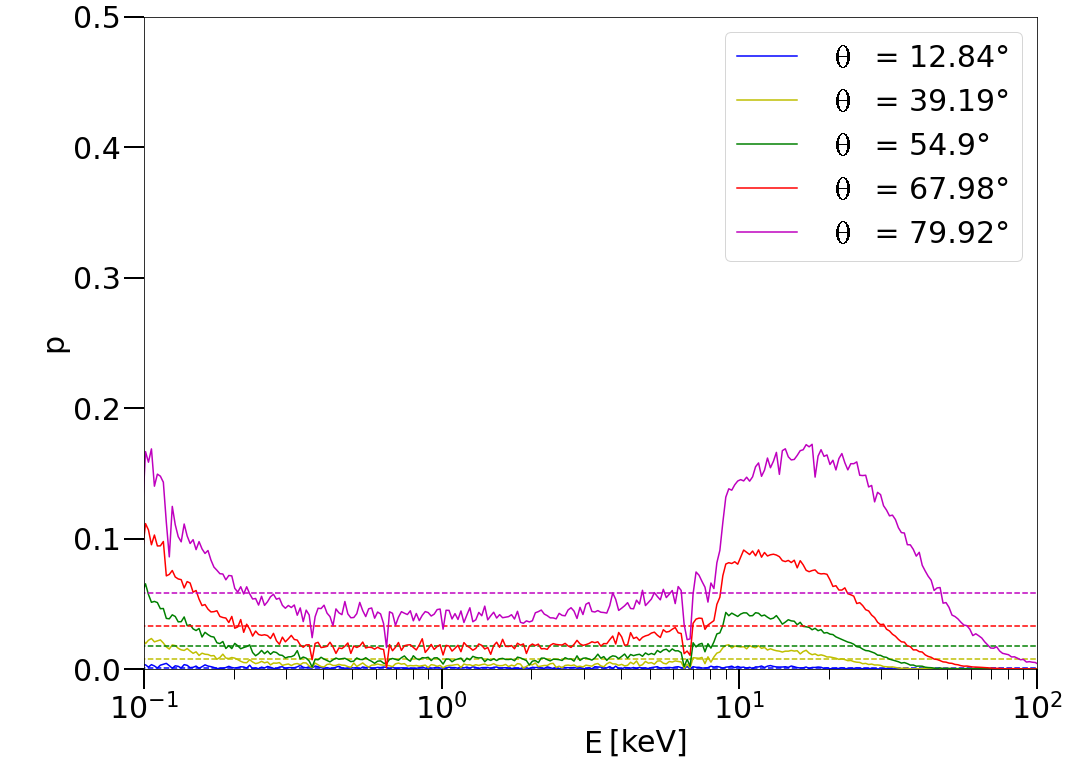}}
    \put(20,90){\scalebox{.55}{$\tau_\textrm{T,max} = 1$}}
    \put(20,80){\scalebox{.55}{$k_\textrm{B}T_\textrm{BB} = 0.80 \ \textrm{keV}$}}
    \end{picture}
    \begin{picture}(137,99)
    \put(0,0){\includegraphics[width=0.33\textwidth]{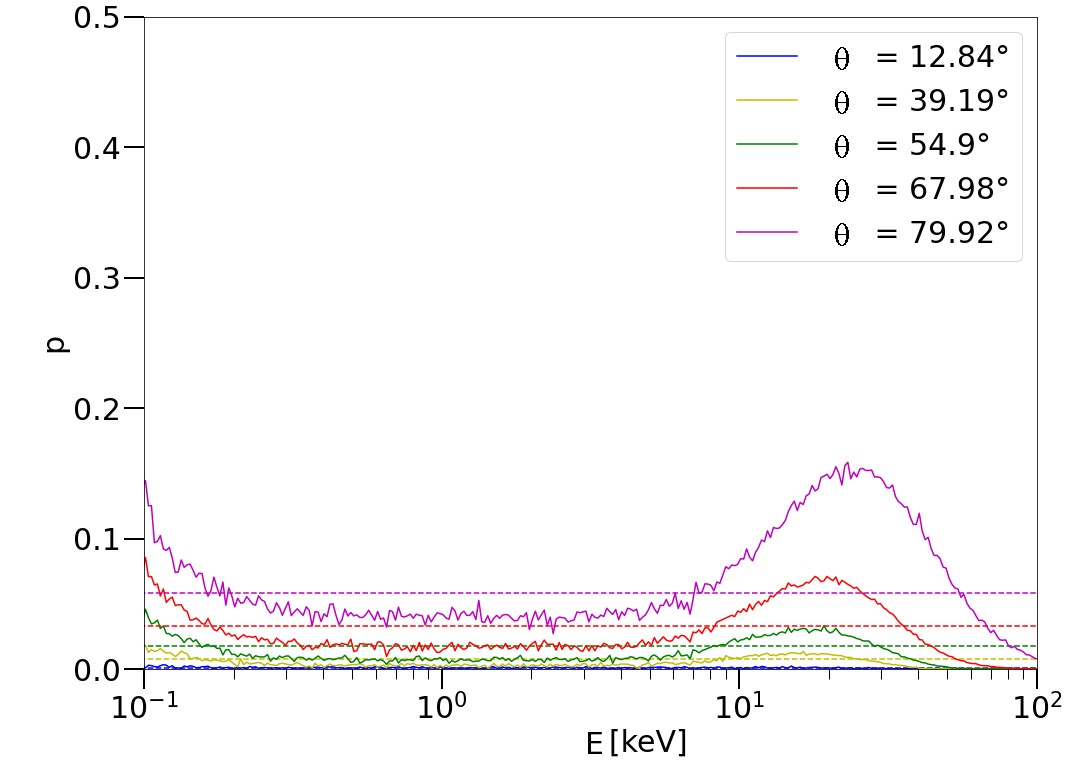}}
    \put(20,90){\scalebox{.55}{$\tau_\textrm{T,max} = 1$}}
    \put(20,80){\scalebox{.55}{$k_\textrm{B}T_\textrm{BB} = 1.15 \ \textrm{keV}$}}
    \end{picture}
    \begin{picture}(137,99)
    \put(0,0){\includegraphics[width=0.33\textwidth]{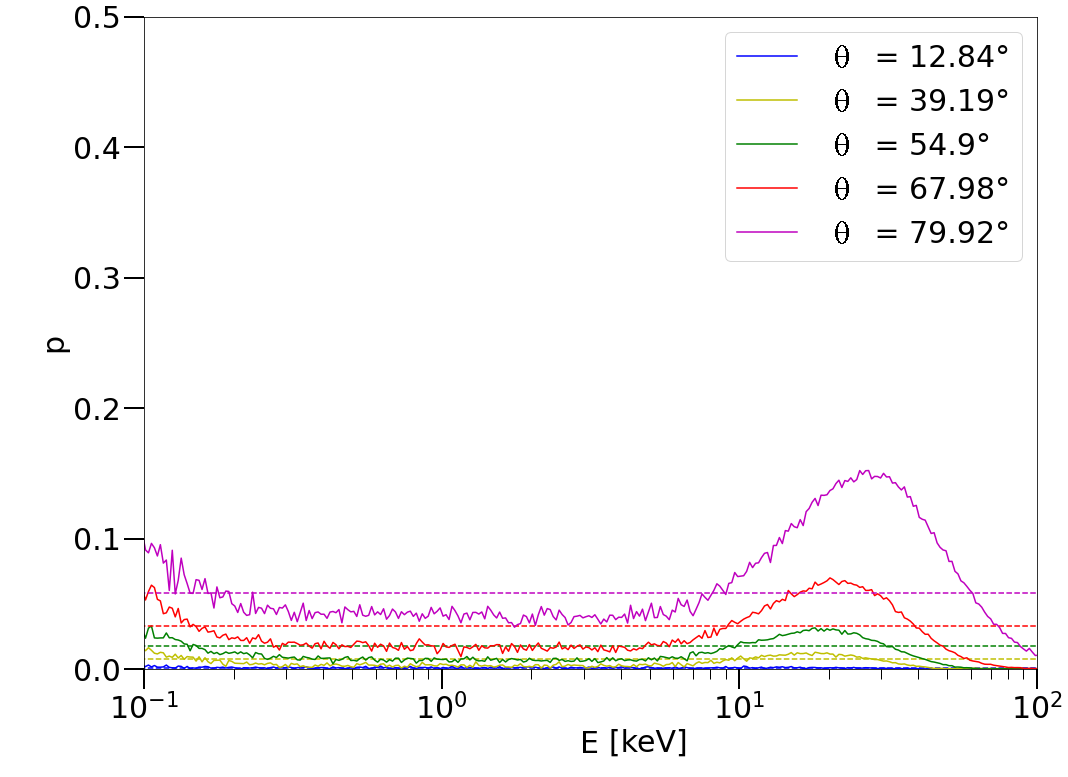}}
    \put(20,90){\scalebox{.55}{$\tau_\textrm{T,max} = 1$}}
    \put(20,80){\scalebox{.55}{$k_\textrm{B}T_\textrm{BB} = 1.40 \ \textrm{keV}$}}
    \end{picture}\\
    \begin{picture}(137,99)
    \put(0,0){\includegraphics[width=0.33\textwidth]{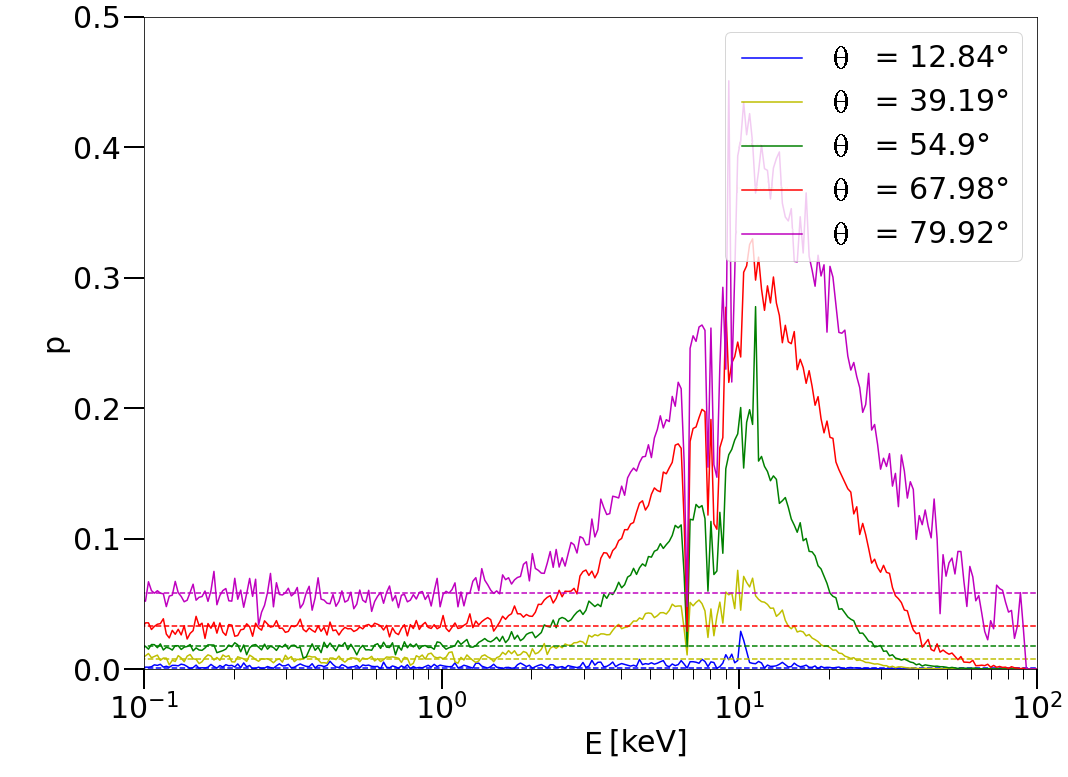}}
    \put(20,90){\scalebox{.55}{$\tau_\textrm{T,max} = 3$}}
    \put(20,80){\scalebox{.55}{$k_\textrm{B}T_\textrm{BB} = 0.23 \ \textrm{keV}$}}
    \end{picture}
    \begin{picture}(137,99)
    \put(0,0){\includegraphics[width=0.33\textwidth]{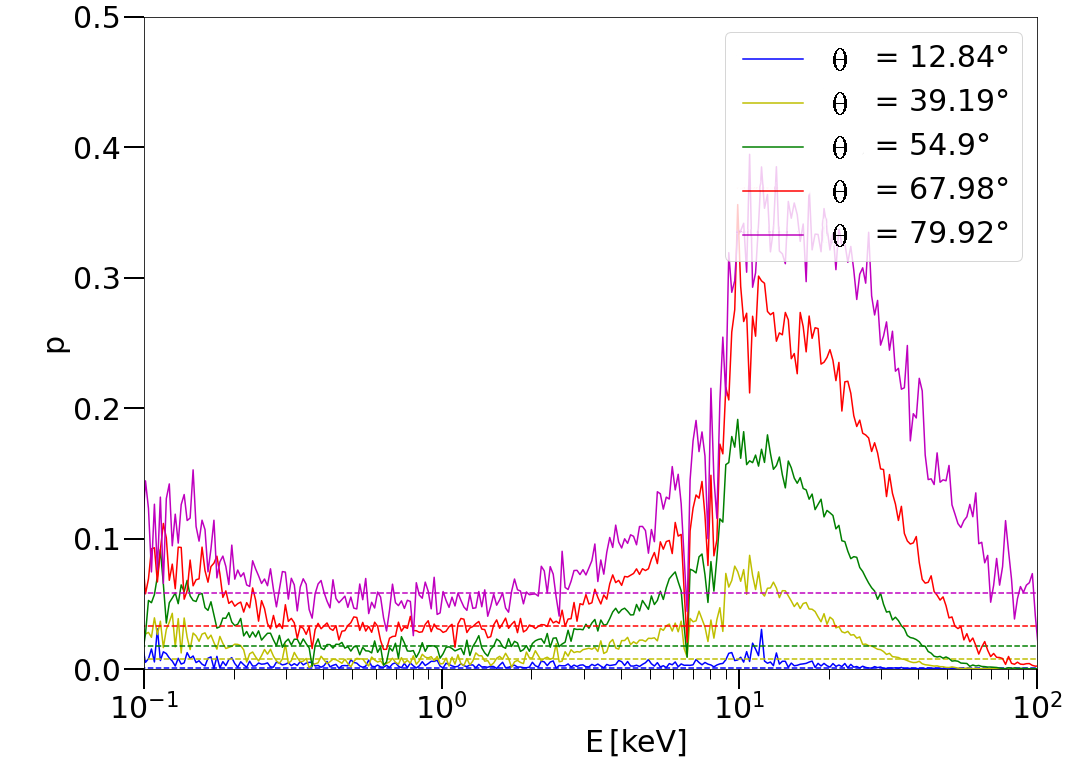}}
    \put(20,90){\scalebox{.55}{$\tau_\textrm{T,max} = 3$}}
    \put(20,80){\scalebox{.55}{$k_\textrm{B}T_\textrm{BB} = 0.50 \ \textrm{keV}$}}
    \end{picture}
    \begin{picture}(137,99)
    \put(0,0){\includegraphics[width=0.33\textwidth]{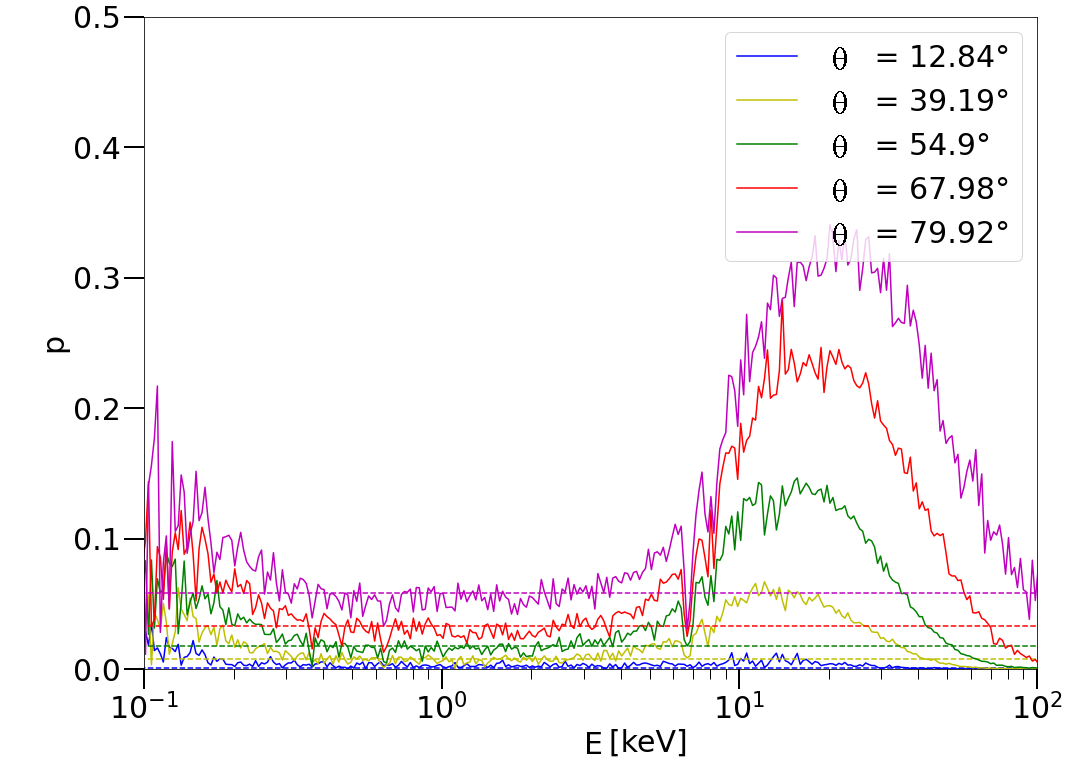}}
    \put(20,90){\scalebox{.55}{$\tau_\textrm{T,max} = 3$}}
    \put(20,80){\scalebox{.55}{$k_\textrm{B}T_\textrm{BB} = 0.80 \ \textrm{keV}$}}
    \end{picture}
    \begin{picture}(137,99)
    \put(0,0){\includegraphics[width=0.33\textwidth]{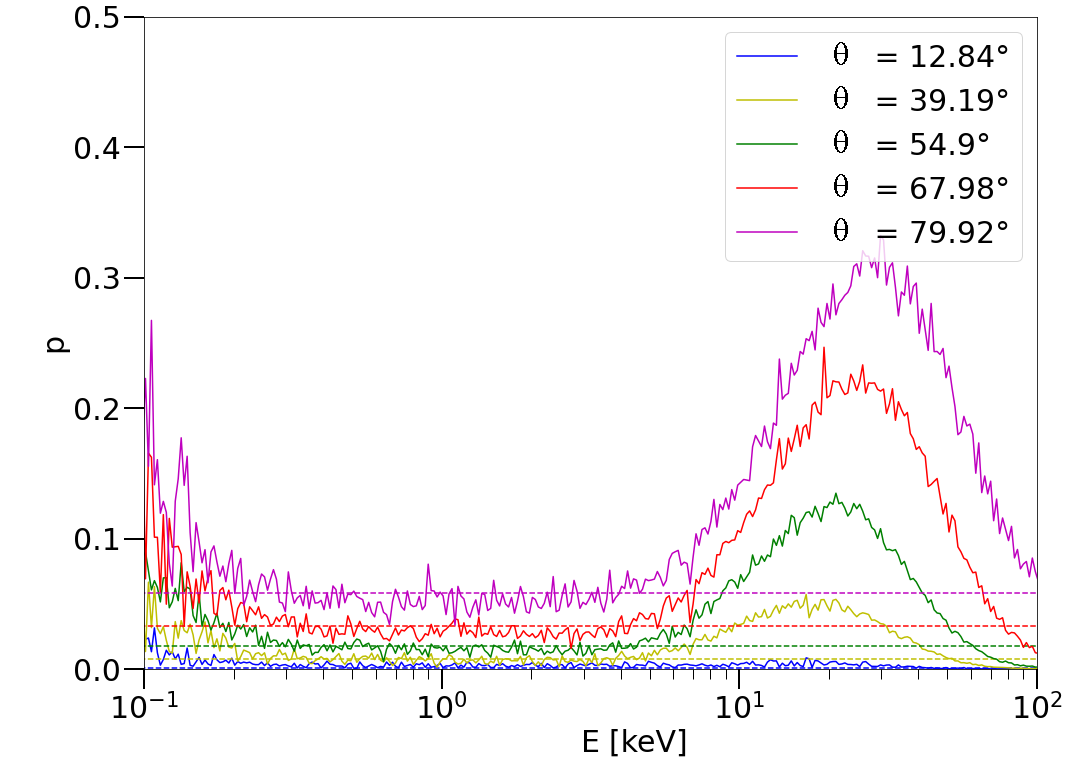}}
    \put(20,90){\scalebox{.55}{$\tau_\textrm{T,max} = 3$}}
    \put(20,80){\scalebox{.55}{$k_\textrm{B}T_\textrm{BB} = 1.15 \ \textrm{keV}$}}
    \end{picture}
    \begin{picture}(137,99)
    \put(0,0){\includegraphics[width=0.33\textwidth]{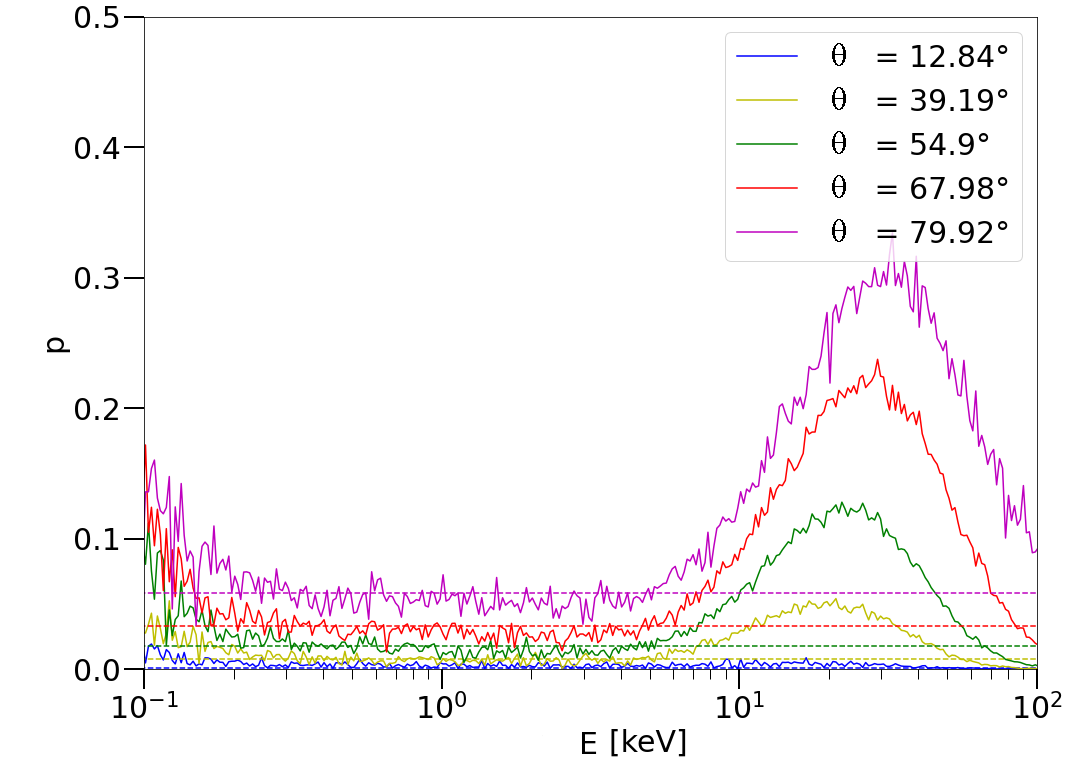}}
    \put(20,90){\scalebox{.55}{$\tau_\textrm{T,max} = 3$}}
    \put(20,80){\scalebox{.55}{$k_\textrm{B}T_\textrm{BB} = 1.40 \ \textrm{keV}$}}
    \end{picture}
    \caption{\footnotesize{The polarisation degree, $p$, versus energy obtained from {\tt STOKES} (solid lines) for various observer's inclinations $\theta$ in the color code, corresponding to~the~cases shown in Figure \ref{tr_ionized_spectra}. From left to right the temperature of the illuminating black body increases ($k_\textrm{B}T_\textrm{BB} = 0.23,0.50,0.80,1.15,1.40$ keV, respectively), while the densities are high ($n_\textrm{H} = 10^{15},10^{20},10^{20.5},10^{20.5},10^{20.5} \ \textrm{cm}^{-3}$, respectively), but not high enough to depart from the saturation state and to cause significant absorption for a given $\xi = \xi_\textrm{BB}$. For any lower density we would see identical result. Top row: the results for $\tau_\textrm{T,max} = 0.67$. Middle row: the~results for $\tau_\textrm{T,max} = 1$ (corresponding directly to the spectra shown in bottom pannels of Figure \ref{tr_ionized_spectra}). Bottom row: the results for $\tau_\textrm{T,max} = 3$. We indicate the pure-scattering limit for semi-infinite atmosphere given by Chandrasekhar-Sobolev approximation for the same inclination values (dashed horizontal lines). The~obtained polarisation angle from {\tt STOKES} is constant with energy and parallel with the slab.}}
    \label{tr_ionized_pd}
\end{figure}
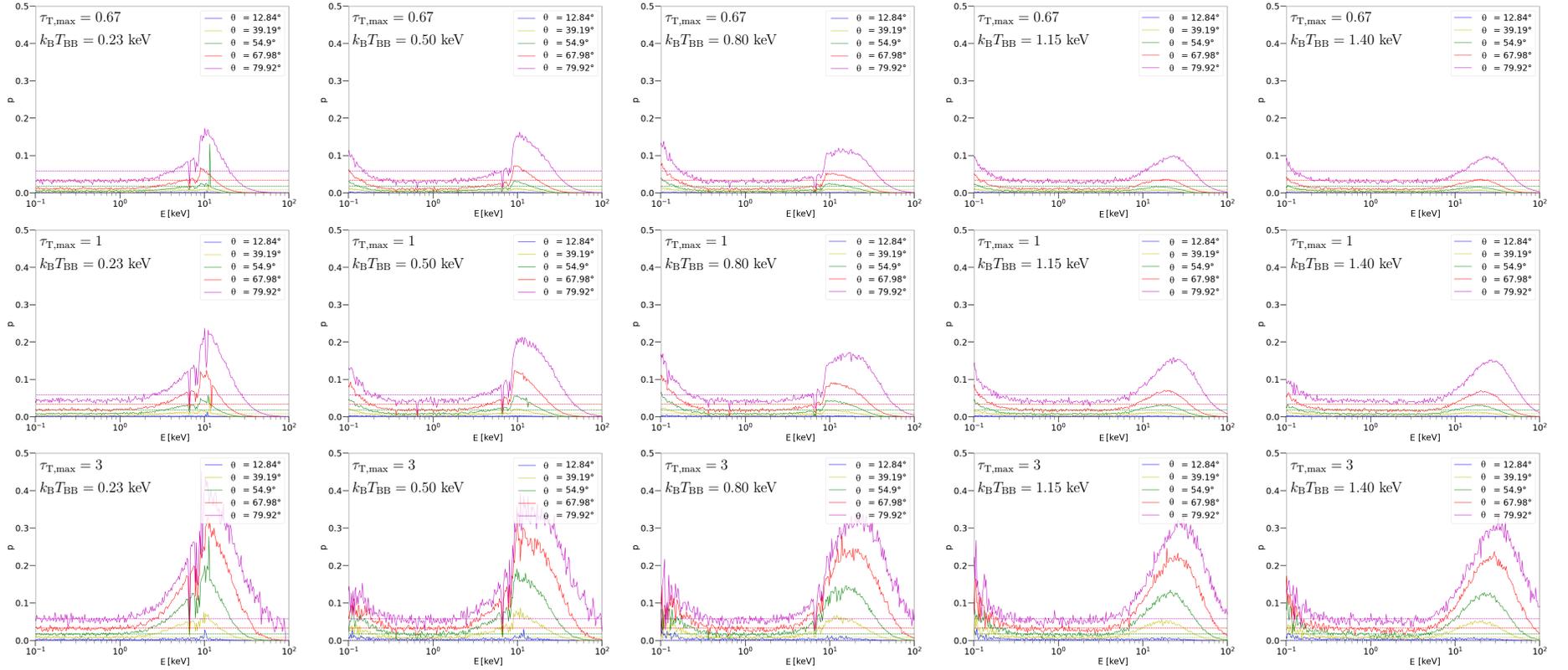
\end{landscape}

We constructed tables in the high ionization limit that allow for disc integration and more valuable fitting than directly in the local comoving frame, using the~relativistic {\tt KY} codes. The density independence of the highly ionized slab results allowed us to select one density $n_\textrm{H} = 10^{18} \ \textrm{cm}^{-3}$ and compute a~grid of~polarisation fractions with energy in the 1--20 keV range that is dependent only on~$\mu_\textrm{e} = \{ 0.025,0.075,...,0.925,0.975\}$, $k_\textrm{B}T_\textrm{BB} = \{0.5, 0.8, 1.15, 1.4\}$~keV and the~critical free parameter $\tau_\textrm{T,max} = \{0.2,0.4,0.67,1,1.3,1.5,2,3,5,7,10\}$ that controls the level of absorption and down-scattering effects on polarisation by scaling the slab height, but not to~the~detriment of ionization level, which is fixed. We manually checked that no strong absorption features near 3 keV emerged for the selected choice of values, i.e. that we stay in~the~desired highly ionized limit (for~$\xi = \xi_\textrm{BB}$) and that all has converged (an~MC effect of artificially high polarisation for low photon statistics was found). Because some stochastic numerical noise is unavoidable (but not artificially high polarisation due to~very low computational times) and because any disc integrator would appreciate to interpolate smooth functions, we opted for~smoothing the~results by performing linear regression of~the~model curves for~two polynomials of third order ($y = A'x^3 + B'x^2 +C'x + D'$), one applied below the 9 keV edge at 1--9~keV, and the other at 9--20 keV. Hence eight coefficients\footnote{\,Available at \url{https://doi.org/10.6084/m9.figshare.24018594.v1} on the day of submission, as part of the results of this dissertation.} were provided for the parameter space here above, which are meant to be implemented to relativistic integrators for fast data fitting of highly ionized sources. The 1--20 keV total energy range was chosen, because of anticipated relativistic energy shifts of~photons further out from the 2--8 keV \textit{IXPE} band.

~\

Last but not least, we ought to check in which conditions this fitting model is not valid anymore. One should worry any time when the fits are ending in corners of the parameter space. Here particularly low $T_\textrm{BB}$ and high $\tau_\textrm{T,max}$ could mean a departure from the highly ionized state. Because repeatedly high $\tau_\textrm{T,max}$ were needed to explain the \textit{IXPE} data (see Section \ref{chap05}), an interesting theoretical question emerged whether for ever increasing $\tau_\textrm{T,max}$ by changing height of~the~layer (and keeping the high ionization level) we would get some saturation in the increasing polarisation, or how to extrapolate the MC results for higher $\tau_\textrm{T,max}$. Although numerous nights were spent on this problem with blurry conclusions and a numerical experiment would require years on available institutional computational clusters, it was eventually abandoned even as a toy-model explanation of~extreme polarisation increasing with energy, because of the necessity of loosing the required ionization levels on the further (not illuminated) side of the slab for thick slabs that would lead to additional absorption features near 3 keV (see Figures \ref{absorbed_spectra} and \ref{absorbed_pd}). Also because of the neglect of internal sources in such model that would on the contrary depolarize the net output, providing sources of radiation with lower distance to~the~surface than the external source. We checked with additional vertically stratified calculations that given the lowest temperatures and highest densities, the further side of the slab looses the required level of ionization at $\tau_\textrm{T,max} \gtrapprox 10$ and the~models should generally not be used beyond this value. Higher $\tau_\textrm{T,max}$ may be in some cases explored, but the limits of validity are density and temperature dependent.

~\

The sensitivity of the results to the assumption of no internal sources can be evaluated by comparison with the model of Valery Suleimanov in \cite{Ratheesh2023}, which shows great similarities with the provided results for local emission of highly ionized disc rings \citep[figure B1, bottom left panel, ][]{Ratheesh2023}, despite using a completely different radiative transfer method. The~same energy-independent plateau in polarisation that we obtained below 2~keV is formed. And~equally to our model, the polarisation dependence on~inclination in this range follows exactly the pure-scattering law (\ref{viir}). The dependency of polarisation on ionization edge is more pronounced compared to~our case and~the~energy-independent plateau is extended up to 9 keV, where a~sharp increase in~polarisation occurs due to the remaining iron edge. The~result is then substantially different compared to ours for the outer disc rings \citep[figure B1, bottom right panel, ][]{Ratheesh2023}, where the highly ionized limit is not valid anymore. The~non-relativistic integration of this model \citep[figure B2, dotted lines, ][]{Ratheesh2023} preserves the pure-scattering result \mbox{at~soft~(0.1--1~keV)} \mbox{and~mid~(1--10~keV)} X-rays, suggesting higher weighting of the highly ionized disc rings.

Note that the relativistic integration \citep[figure B2, solid lines, ][]{Ratheesh2023}, apart from depolarisation with slight energy dependence, alters the energy dependence of the resulting polarisation angle. In~Section \ref{chap05} we will show results where our local model was relativistically integrated, taking this effect into account too for data fitting of the observed polarisation angle with energy.

One may also compare our results to the sophisticated MHD simulations done by \citet[][figures 7--9, 12, 15]{Davis2009}, which include the effects of~magnetic fields with carefully computed effects of Faraday rotation and inhomogeneities on~the~net X-ray polarisation. The predicted polarisation is parallel to~the~disc and constant according to the Chandrasekhar-Sobolev profile in between \mbox{0.2--4~keV} without the affects of Faraday rotation. The Faraday rotation in addition depolarizes the signal and causes increasing polarisation with energy at \textit{soft} \mbox{X-rays} and non-trivial polarisation orientation, in accordance with earlier study of \cite{Gnedin2006}. These works are departing from earlier estimates by \cite{Gnedin1978,Silantev1979,Laor1990,Agol1996,Blaes1996,Agol1998,Agol2000}.

\subsection{Summary}

Let us provide a quick summary of the above. Using a combined radiative transfer and MC method, we constructed simulations of a passive plane-parallel slab of~constant density that alters the polarisation state of the unpolarized illuminating isotropic single-temperature black-body source of radiation, located from the~opposite side of the slab with respect to a generally inclined observer. The~slab was held in PIE (a few comparisons with CIE approach showed similar results, but for higher temperatures needed inside the simulated top atmospheric layers) and the obtained vertical stratification was in general negligible for moderate optical depths. If the illuminating flux is low, or the density or optical depth is high, the polarisation may be on average high (even tens of \%) in 2--8 keV, decreasing with energy due to a strong peak at 3 keV. This peak gradually disappears by e.g. lowering the slab density and a certain polarisation degree profile with energy establishes for high enough ionization (a state of full ionization could not be reached in the studied PIE conditions) with lower polarisation in 2--8 keV and with a steady or increasing polarisation degree with energy. As long as the~slab optical thickness is higher than $\tau_\textrm{T,max} \gtrapprox 0.4$, the resulting polarisation angle is constant with energy and parallel with the slab. The polarisation degree in the highly ionized regime reaches Thomson scattering values at 2 keV given by the~Chandrasekhar-Sobolev analytical results, if $\tau_\textrm{T,max} \gtrapprox 3$. The model should not be used above $\tau_\textrm{T,max} \gtrapprox 10$, as beyond we depart from the high level of ionization in the PIE conditions studied. In general, the simulation is a toy model (introducing a free parameter of $\tau_\textrm{T,max}$ of the passive slab) of a real atmosphere on top of hot dissipative layers closer to the disc mid-plane. A more realistic models for comparison include internal sources of radiation, compute the entire semi-infinite problem (assuming e.g. a hydrostatic equilibrium) and include the effects of magnetic fields. However, due to simplicity, our model is suitable for~first-order estimates on e.g. inclination, if compared with {\it IXPE} observations of~XRBs with highly ionized thermally dominated spectra, especially if relativistically integrated across a disc with assumed geometry and temperature profile. This was facilitated by tabulating a smoothed function of the polarisation state with energy in 1--20 keV for a relevant part of the explored parameter space, dependent only on inclination, optical depth and local effective temperature.

\newpage
\ 
\vspace{4.5cm} 
\begin{figure}[h]
        \centering
	\includegraphics[trim={0cm 0cm 0cm 1cm},clip,width=0.6\textwidth]{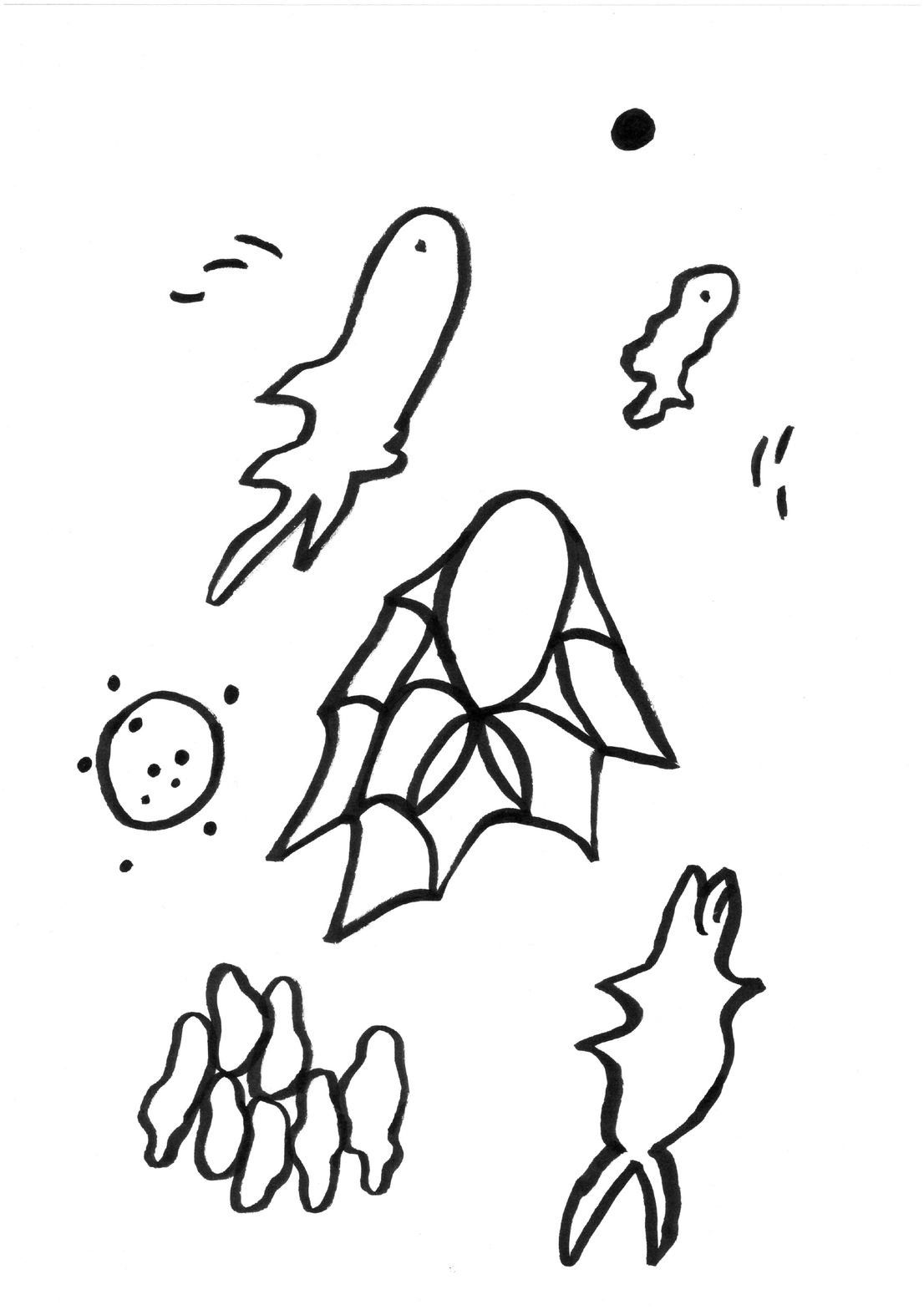}
\end{figure}
\chapter{Total emission from an inner-accreting region}\label{chap02}

In this chapter, we characterize emission from the entire inner disc-corona system in the lamp-post geometry and compare it to a toy model simulating a~sandwich geometry. We present the local reflection tables that were described in~Section~\ref{reflection_tables_TS}, integrated across the disc for a distant observer. All relativistic effects are included, apart from returning radiation after the first reflection from~the~disc. The~first reflection is accounted for and it will be of our prime interest. For~these computations, a new {\tt KY} routine, called {\tt KYNSTOKES}, was developed, based on~the~radiative transfer kernels of the previously published {\tt KY} codes. Thermal disc radiation is not included; hence, the results (also with respect to~the~assumed densities) are rather applicable as toy models of inner-accreting regions of AGNs than XRBs.

~\

Essentials of the {\tt KYNSTOKES} lamp-post model with implementation of~the~{\tt TITAN} and {\tt STOKES} local reflection tables were developed and described in~\cite{Podgorny2020}, which we will only briefly summarize. Over the scope of this dissertation, {\tt KYNSTOKES} in the lamp-post regime was tested and the results were discussed in more depth and context, as published in \cite{Podgorny2023}, but with the previous version of reflection tables from \cite{Podgorny2021}. Here we will list the findings from \cite{Podgorny2023} and~plot the corresponding and~additional figures with an updated {\tt KYNSTOKES} routine\footnote{\,Available at \url{https://projects.asu.cas.cz/dovciak/kynstokes} on the day of submission, including documentation and tabular dependencies.}, interpolating in~the~local reflection tables unified in $\xi$ that were described in Section \ref{reflection_tables_TS}.\footnote{\,We also note a small plotting error in the black curves (the lowest energy bin) of top-right panels in figures 6, 7, 9, 10, 15 and 16 of \cite{Podgorny2023}, which is corrected here.}

In addition, the slab coronal model was implemented to {\tt KYNSTOKES}\footnote{\,Available inside the same version at \url{https://projects.asu.cas.cz/dovciak/kynstokes} on the day of submission, including documentation and tabular dependencies.} based on earlier work of Michal Dovčiak and the results were discussed with respect to~the~lamp-post model in {\tt KYNSTOKES} over the scope of this dissertation. We provide the related key points at the end of this chapter, but we note that this part is still awaiting a peer review. The same holds for dependency of the emission on the inner disc radius, which was not covered in \cite{Podgorny2023} even for~the~lamp-post model (although the public version of the code allows to truncate the disc) and which we also add here for both studied geometrical archetypes of~the~corona.

~\

The {\tt KYNBBRR} routine,  suitable for fitting thermal radiation of XRBs, was recently updated thanks to Michal Dovčiak to its {\tt KYNEBBRR} variant, which implements the local thermal emission from highly ionized disc atmosphere computed with {\tt TITAN} and {\tt STOKES}, i.e. the one described at the end of the previous chapter. The relativistically integrated results obtained with this routine will be presented only briefly as part of the observational results in Chapter \ref{chap05} and not here, because the implementation, testing and discussion was done by colleagues that were involved in \cite{Dovciak2008,Taverna2020, Taverna2021, Mikusincova2023} and that developed the {\tt KYNBBRR} model.

~\

Aside to the main results obtained with {\tt KYNSTOKES}, we also propose in Appendix \ref{integration} an alternative way of \textit{integrating} the disc emission inside the {\tt KY} codes in general. It would, however, require rewriting large portions of the main scripts to provide thorough testing, or perhaps developing a new code. Because the outcomes in gained efficiency did not seem outstanding on the first sight, we decided not to continue further in this direction. Another reason is that the proposed alternative scenario for disc integration was intended to improve precision in detailed relativistic smearing of e.g. spectral lines, which is beyond the resolution of any current or forthcoming X-ray polarimeter, hence slightly out of the main focus of this dissertation.

\section{The lamp-post coronal model}\label{kynstokes_lamppost}

A motivation for this study was possible improvement of the model in \cite{Dovciak2011}, which considered the single-scattering Chandrasekhar's formulae (\ref{chandra1}) for reflection-induced polarisation and the {\tt NOAR} spectral computations of~reflecting neutral accretion disc. The ionization structure of the disc was not considered, nor any spectral lines. Only the unpolarized primary emission was studied in \cite{Dovciak2011} and although the code was updated for polarisation of the primary radiation for the AGN modeling described in \cite{Marin2018c,Marin2018b}, an error was found by Michal Dovčiak in the GR computations of~polarisation angle change used therein. Then, as part of this dissertation, the~computations of the change of polarisation angle due to GR effects between the lamp and the disc and between the lamp and the observer, originally \mbox{performed} by Michal Dovčiak, were revisited and verified. This is shown in Appendix \ref{change_PA}. \mbox{Independently,} we also discovered an error in \cite{Dovciak2011} in~the~\mbox{local} reflection prescriptions for polarisation \citep[fixed and described in~appendix B of][]{Podgorny2023}, which propagated into the global computations of~both polarisation angle and degree, i.e. figs 6 and 8–14 in \cite{Dovciak2011}. Thus, it is worth to revisit the entire discussion of the lamp-post results for a distant observer that appeared in \cite{Dovciak2011}, instead of doing a one-to-one comparative study that would show only the effects of modeling improvements.

~\

We refer to Figure \ref{corona}a for a sketch of the parametrization and to Sections \ref{relativistic} and \ref{codes_here} for the main physical setup of the model. In the plots below, we will assume three cases of disc ionization, given by the luminosity of the primary (increases the local $\xi$) and black-hole mass (scales the distance in physical units between the source and the disc): ``low ionization'' ($M = 1\times 10^8\,M_{\odot}$ and the observed 2--10 keV flux $L_{\textrm{X}}/L_{\textrm{Edd}} = 0.001$), ``moderate ionization'' ($M = 3\times 10^6\,M_{\odot}$ and the observed 2--10 keV flux $L_{\textrm{X}}/L_{\textrm{Edd}} = 0.01$), and ``high ionization'' ($M = 1\times 10^5\,M_{\odot}$ and the observed 2--10 keV flux $L_{\textrm{X}}/L_{\textrm{Edd}} = 0.1$). The disc is assumed to have a constant radial density profile, given by the local reflection tables with $n_\textrm{H} = 10^{15} \ \textrm{cm}^{-3}$, but in principle the code can scale density radially by means of local ionization parameter change. The disc inner radius $r_{\mathrm{in}}$ is fixed at~the~ISCO and $r_{\mathrm{out}} = 400 \, r_\textrm{g}$, if not stated otherwise. We will study the~case of reflected-only radiation with $N_\textrm{p}/N_\textrm{r} = 0$ and the case of added primary source with $N_\textrm{p}/N_\textrm{r} = 1$, i.e. a consistent one-to-one normalization of the primary and reflection contribution, according to the local reflection tables. The primary source is point-like and isotropic, which is one of the most critical simplifications of this work, as neither the flux nor the polarisation is expected to be isotropic in~the~lamp-post regime due to Comptonization of disc seed photons \citep[e.g.][]{Zhang2019b, Krawczynski2022a, Zhang2022, Zhang2023}. The local reflection tables are interpolated per each geodesic in the local reflection angles as well as in the three precomputed polarisation states of primary irradiation. The~arbitrary state of primary radiation for the locally emergent Stokes parameters $I$, $Q$ and $U$, commonly denoted as $S$, is obtained through
\begin{equation}\label{Sorig}
\begin{split}
    S(p_0, \Psi_0+\Delta\Psi) = & \, S(0,-) + p_0\Big\{ \Big[S(1,0) - S(0,-)\Big]\cos{2(\Psi_0+\Delta\Psi)} \\
    &\quad + \left[S\left(1,\frac{\pi}{4}\right) - S(0,-)\right]\sin{2(\Psi_0+\Delta\Psi)} \Big\}.
\end{split}
\end{equation}
thanks to the linearity\footnote{\,Any equation of this kind is only a representation of the Stokes vector in a particular basis in~3-dimensional vector space (omitting the Stokes $V$, as we do not solve for circular polarisation). Departing from the Müller formalism described e.g. in \cite{Chandrasekhar1960}, we can define Stokes parameters through $I_\textrm{l}$, $I_\textrm{r}$, $U$ in any basis, having $I = I_\textrm{l} + I_\textrm{r}$, \mbox{$Q = I_\textrm{l} - I_\textrm{r} = Ip\cos{2\Psi}$}, $U = U = Ip\sin{2\Psi}$. One can use $S^\textrm{L} = (I,0,0)$, $S^\textrm{R} = (0,I,0)$, $S^\textrm{U} = (0,0,I)$ for~a~basis representation, as the base vectors need not to be physical, i.e. representing a~polarisation state of~a~photon with non-zero intensity. Then
\begin{equation}
\begin{split}
    S = & \, (I_\textrm{l}, I_\textrm{r}, U) \\ = & \, \left(\frac{I}{2}\left[1+p\cos{2\Psi}\right], \frac{I}{2}\left[1-p\sin{2\Psi}\right], Ip\sin{2\Psi}\right)  \\
     = & \, \frac{1}{2}(S^\textrm{L}+S^\textrm{R}) + p\left[\frac{1}{2}\left( S^\textrm{L}-S^\textrm{R}\right) \cos{2\Psi} - S^\textrm{U}\sin{2\Psi}\right] \, .
\end{split}
\end{equation}
Since the reflection from the disc and follow-up transfer to infinity is a linear process,
we can compose the final result for arbitrary initial corona polarisation degree and angle,
\mbox{$S(p_0, \Psi_0+\Delta\Psi)$}, from the result for unpolarized corona, $S(0,-)$, corona fully polarized parallelly
with the~system axis, $S(1,0)$, and corona fully polarized 45\degr with respect to the axis, $S(1,\frac{\pi}{4})$. In this way we obtain (\ref{Sorig}).} of Stokes parameters.

\subsection{Spectropolarimetric properties}\label{lamppost_spproperties}

\begin{figure}[!htb]\centering
	\includegraphics[width=\textwidth]{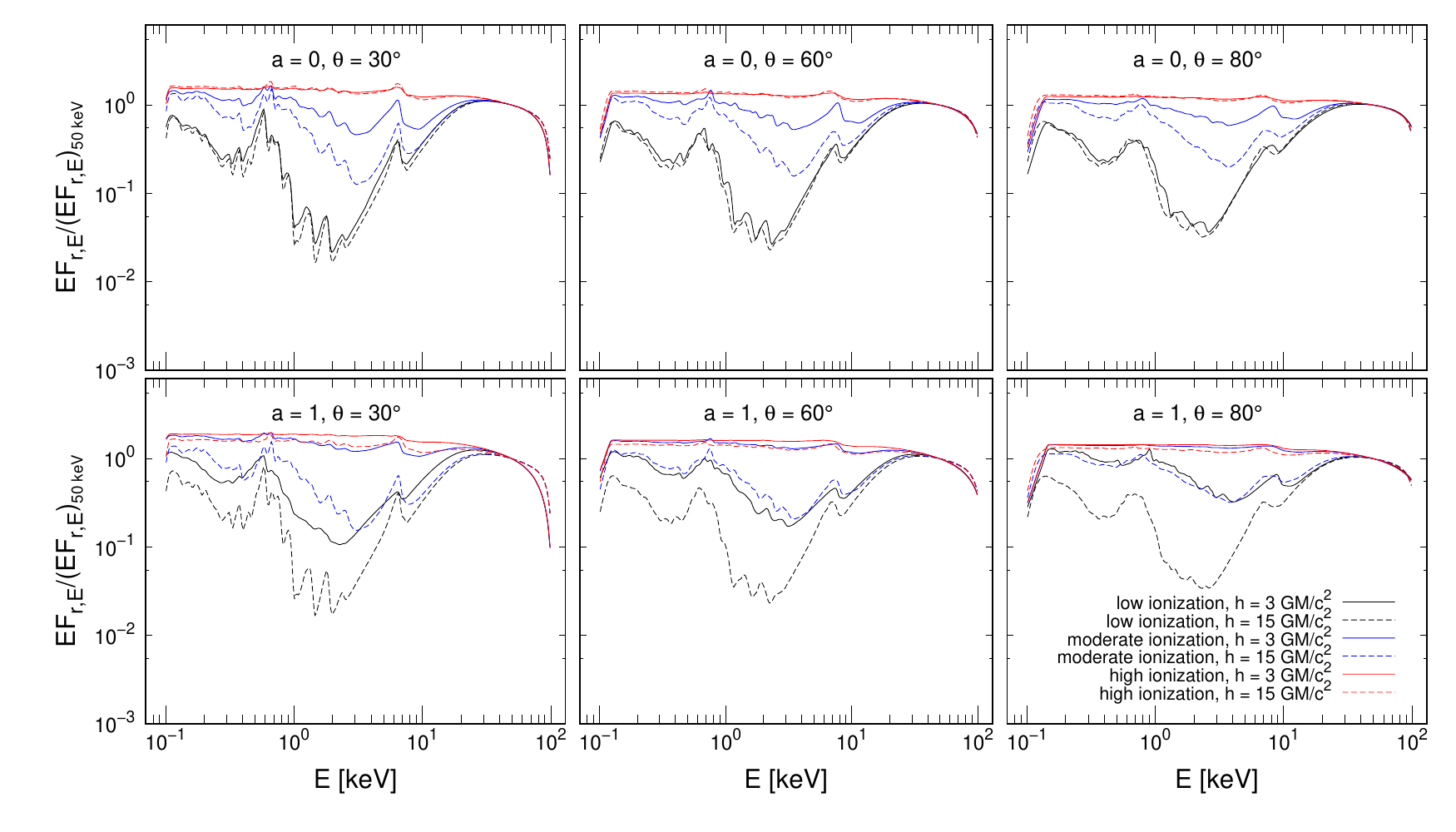}
	\caption{\footnotesize{The reflected-only spectra of the accretion disc for distant observer, $EF_\mathrm{r,E}$, normalized to value at 50 keV, obtained by {\tt KYNSTOKES} for black-hole spins $a = 0$ (top) and $a = 1$ (bottom), disc inclinations $\theta = 30^{\circ}$ (left), $\theta = 60^{\circ}$ (middle) and  $\theta = 80^{\circ}$ (right), $\Gamma = 2$ and unpolarized primary radiation, using the {\tt STOKES} local reflection model in the lamp-post scheme. We show cases of two different heights of the primary point-source above the disc $h = 3 \textrm{ } r_\textrm{g}$ (solid lines) and $h = 15 \textrm{ } r_\textrm{g}$ (dashed lines), and neutral disc (for $M = 1\times 10^8\,M_{\odot}$ and observed 2--10 keV flux $L_{\textrm{X}}/L_{\textrm{Edd}} = 0.001$, black lines), moderately ionized disc (for $M = 3\times 10^6\,M_{\odot}$ and observed 2--10 keV flux $L_{\textrm{X}}/L_{\textrm{Edd}} = 0.01$, blue lines) and highly ionized disc (for $M = 1\times 10^5\,M_{\odot}$ and observed 2--10 keV flux $L_{\textrm{X}}/L_{\textrm{Edd}} = 0.1$, red lines).}}
	\label{fig:K3_R_unpol_I.}
\end{figure}

\begin{figure}[!htb]\centering
	\includegraphics[width=\textwidth]{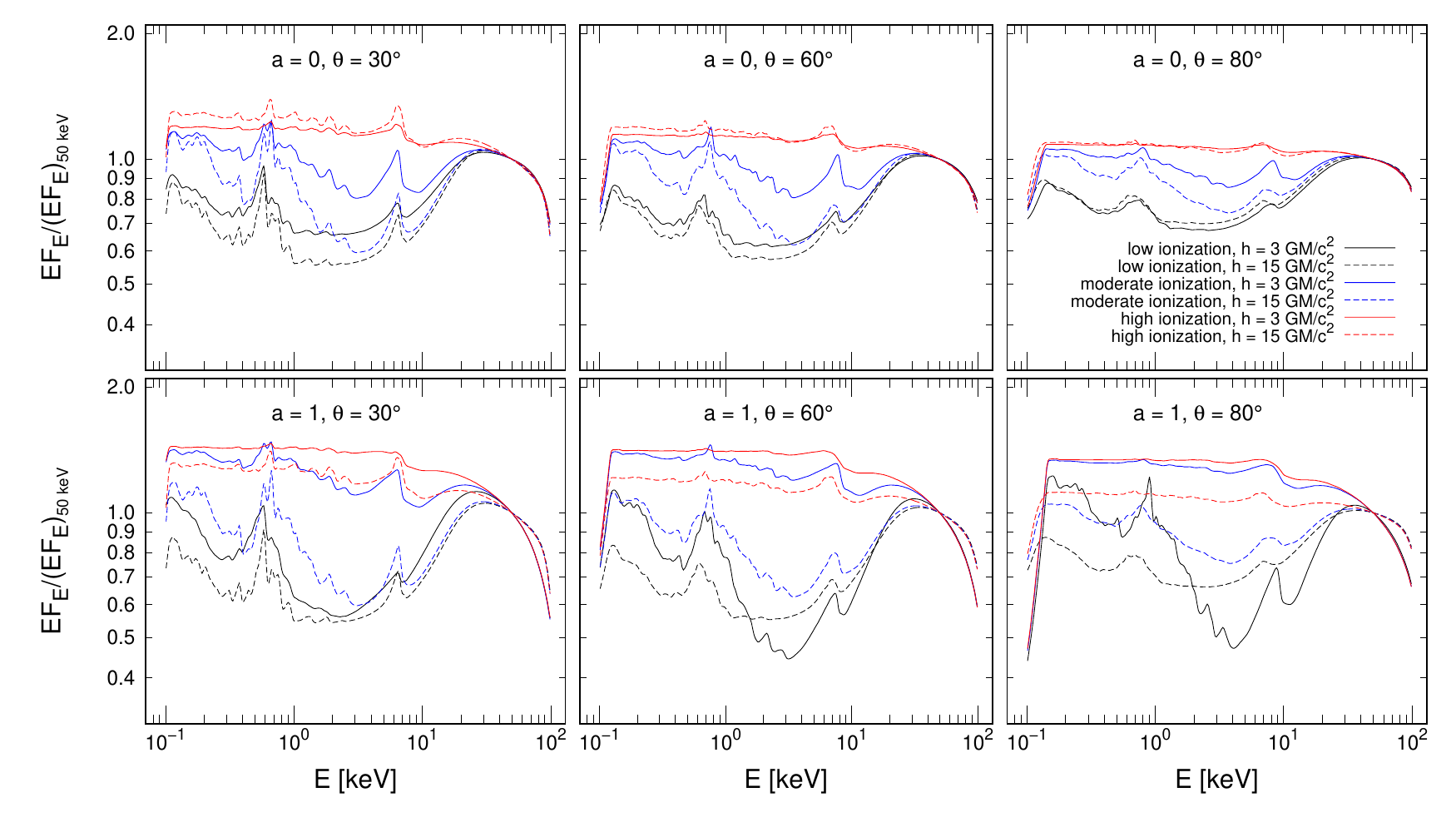}
	\caption{\footnotesize{The total spectra of the accretion disc for distant observer, $EF_\mathrm{E}$, normalized to~value at 50 keV, for the same parametric setup as in Figure \ref{fig:K3_R_unpol_I.}, displayed in the same manner.}}
	\label{fig:K3_RP_unpol_I.}
\end{figure}

We will first show the results for $\Gamma = 2$ and unpolarized primary radiation. The~obtained reflected-only and total spectra are shown in Figures \ref{fig:K3_R_unpol_I.} and~\ref{fig:K3_RP_unpol_I.} for~two extreme black-hole spins, three inclinations, two lamp-post heights and~three cases of disc ionization. The spectra are corrected for slope change and renormalized to value at 50 keV to see the distinct line and continuum features clearly. In all cases, the relativistic smearing occurs, compared to the local reprocessing profile. The relativistic effects are the strongest when most radiation reaches the inner disc (i.e. for low lamp-post heights and high black-hole spin) and for high inclinations. The added primary radiation suppresses the reflection features, and the differences from a plain power-law are minor (the $y$-axis in Figure \ref{fig:K3_RP_unpol_I.} is stretched compared to Figure \ref{fig:K3_R_unpol_I.}). The relativistic routines are well-tested \citep[see e.g.][]{Dovciak2004b} and the spectral results resemble other lamp-post reflection models \citep{Garcia2014}.

~\

\begin{figure}[!htb]\centering
	\includegraphics[width=\textwidth]{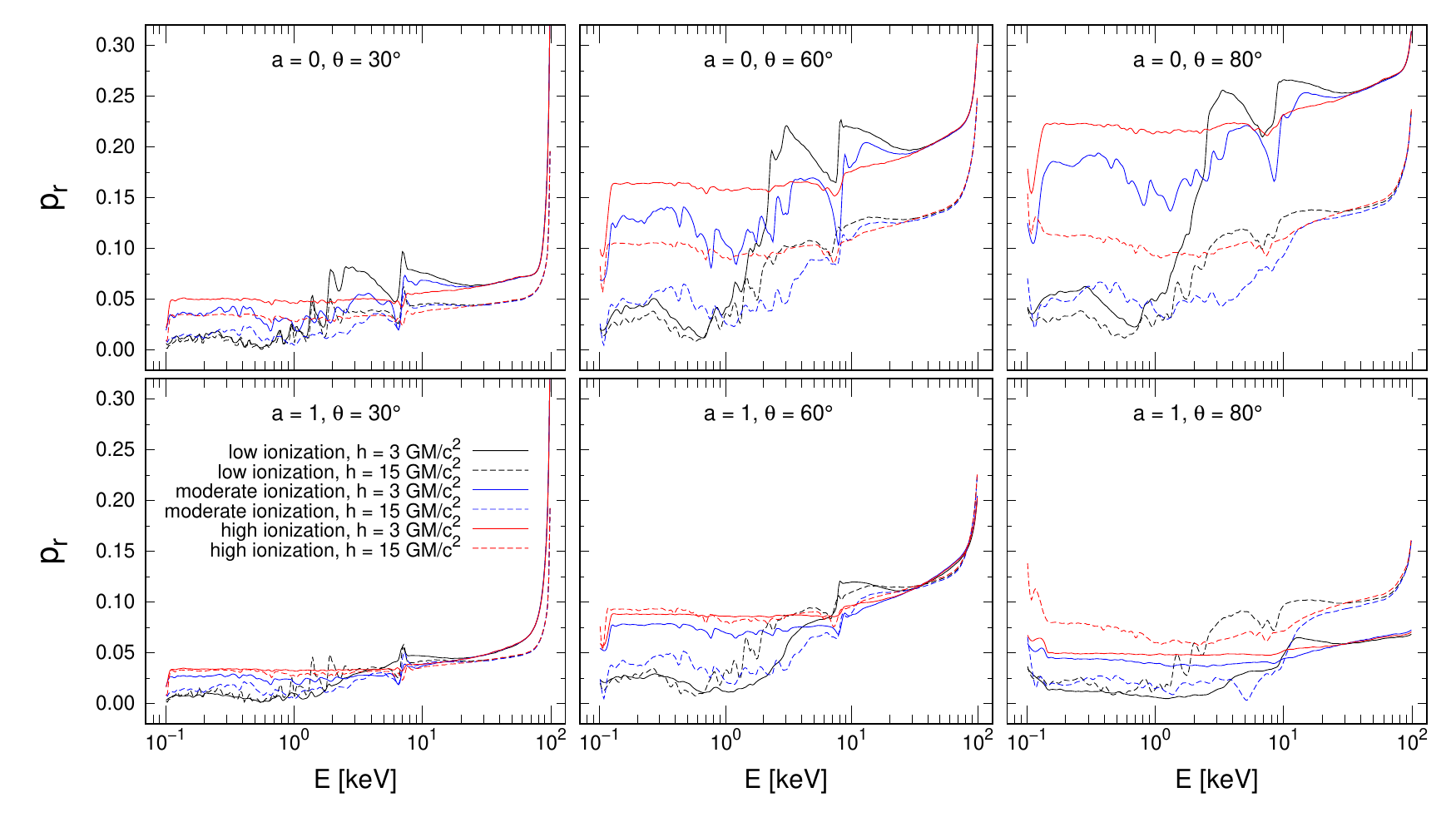}
	\caption{\footnotesize{The reflected-only polarisation degree, $p_\mathrm{r}$, versus energy for the same parametric setup as in Figure \ref{fig:K3_R_unpol_I.}, displayed in the same manner.}}
	\label{fig:K3_R_unpol_pdeg.}
\end{figure}

\begin{figure}[!htb]\centering
	\includegraphics[width=\textwidth]{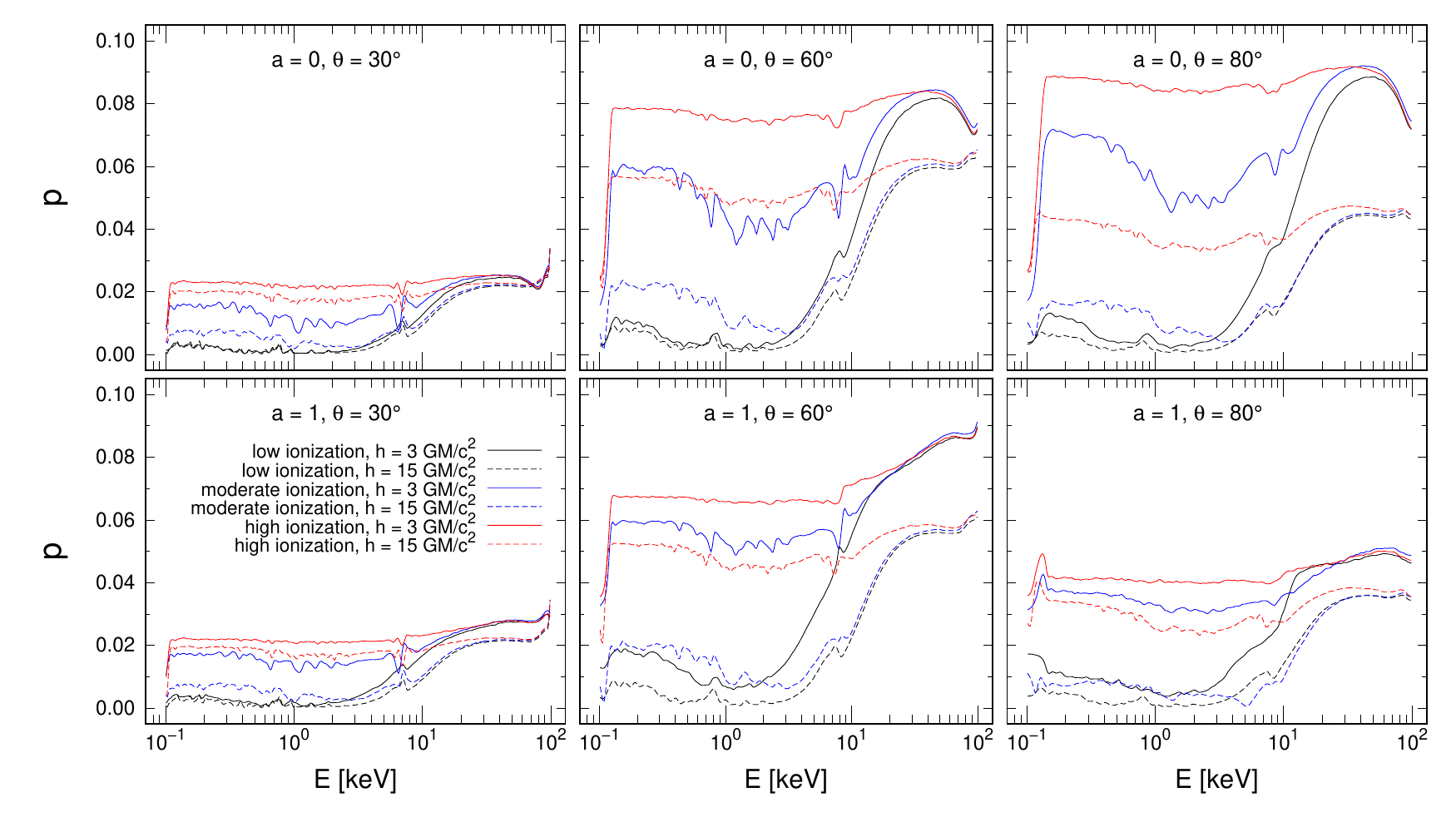}
	\caption{\footnotesize{The total polarisation degree, $p$, versus energy for the same parametric setup as in~Figure~\ref{fig:K3_R_unpol_I.}, displayed in the same manner.}}
	\label{fig:K3_RP_unpol_pdeg.}
\end{figure}

The corresponding reflected-only and total polarisation fractions with energy are shown in Figures \ref{fig:K3_R_unpol_pdeg.} and \ref{fig:K3_RP_unpol_pdeg.}, respectively. In the simplistic scenario presented, we estimate from the inner-accreting regions of AGNs the X-ray maximum of~$\approx 25\%$ polarisation fraction of the reflected component and the X-ray maximum of~$\approx 9\%$ polarisation fraction of the total signal for unpolarized primary in~the~most favorable lamp-post configurations. The relativistically integrated value of~reflected polarisation is slightly higher than in the uniformly integrated results presented in Figure \ref{local_not_in_mue}, because the transfer function intervenes as a~weighting factor and the light-bending effects apply close to the black hole. The~possibility of amplification of polarisation of the locally reflected emission due to relativistic effects was pointed out already in \cite{Dovciak2011}. Nonetheless, we see similar behavior of polarisation with global $\theta$ as with $\delta_\textrm{e}$ for~the~uniformly integrated local reflection tables. Apart from the highest $a$ and highest $\theta$ where depolarisation occurs due to the strong-gravity effects, i.e. superposition of geodesics with differently rotated polarisation angles (see Appendix \ref{change_PA}). The~relativistic effects on spectral lines are predicted also for polarisation, although no current or planned X-ray polarimeter would be able to observe AGNs (or XRBs) with good-enough energy resolution that would resolve individual lines. The polarisation for highly ionized discs preserves the near energy independence from local reflection tables, but compared to the local reflection tables the disc ionization in the global perspective also weights non-uniformly the flux contributions from different parts of the disc (combined with the transfer function effects).

\begin{figure}[!htb]\centering
	\includegraphics[width=\textwidth]{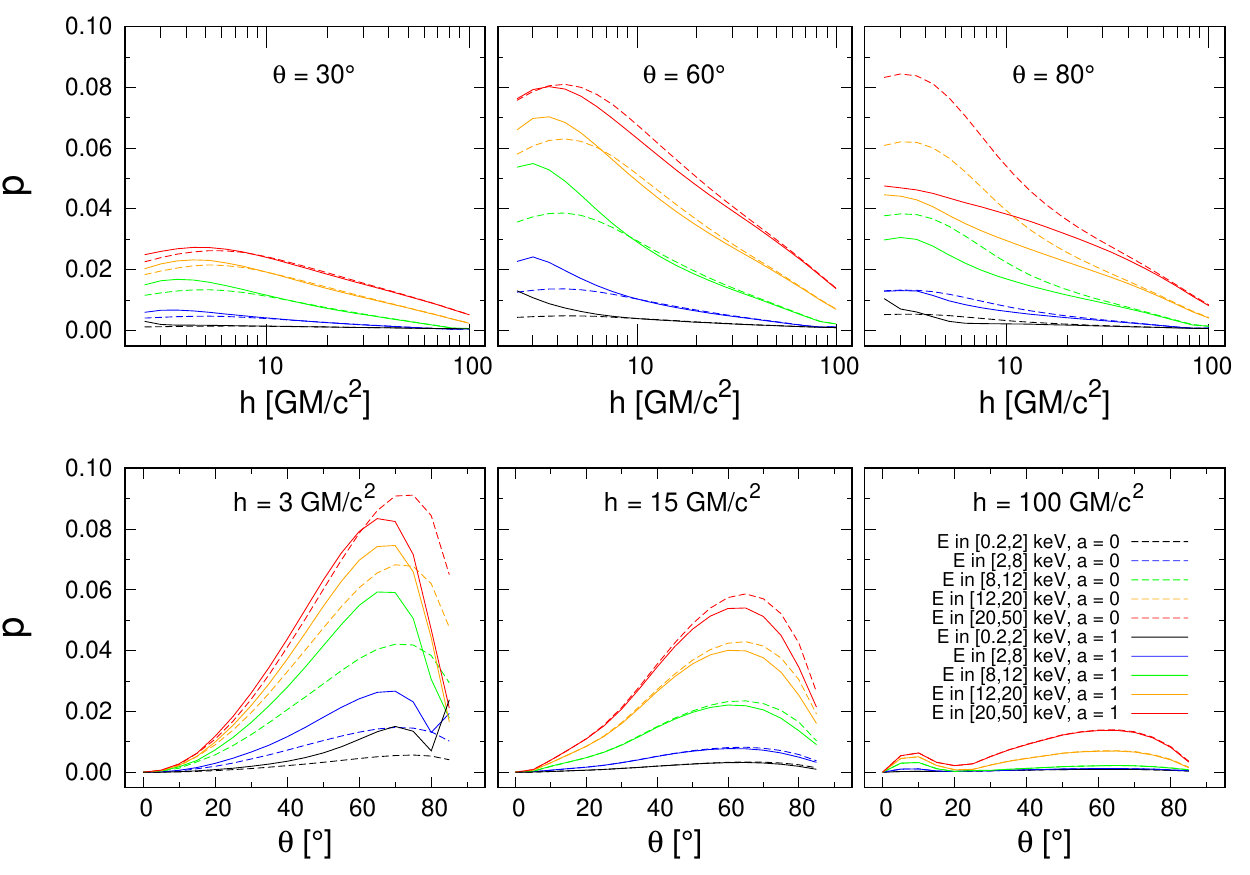}
	\caption{\footnotesize{Top panel: the total average polarisation degree, $p$, versus height of the primary point-source above the disc for disc inclinations $\theta = 30^{\circ}$ (left), $\theta = 60^{\circ}$ (middle) and  $\theta = 80^{\circ}$ (right), $\Gamma = 2$, unpolarized primary radiation, $M = 1\times 10^8\,M_{\odot}$ and observed 2--10 keV flux $L_{\textrm{X}}/L_{\textrm{Edd}} = 0.001$, i.e. neutral disc. We show cases of two black-hole spins $a = 0$ (dashed) and~$a = 1$ (solid) and for energy bands $E \in [0.2, 2]$ keV (black),  $E \in [2, 8]$ keV (blue), \mbox{$E \in [8, 12]$~keV} (green), $E \in [12, 20]$ keV (orange), and $E \in [20, 50]$ keV (red). Bottom panel: the total average polarisation degree, $p$, versus disc inclination for heights of the primary point-source above the~disc $h = 3 \textrm{ } r_\textrm{g}$ (left), $h = 15 \textrm{ } r_\textrm{g}$ (middle) and  $h = 100 \textrm{ } r_\textrm{g}$ (right), for the same configuration as on the top panel.}}
	\label{fig:fig9_tot_neutral.}
\end{figure}

\begin{figure}[!htb]\centering
	\includegraphics[width=\textwidth]{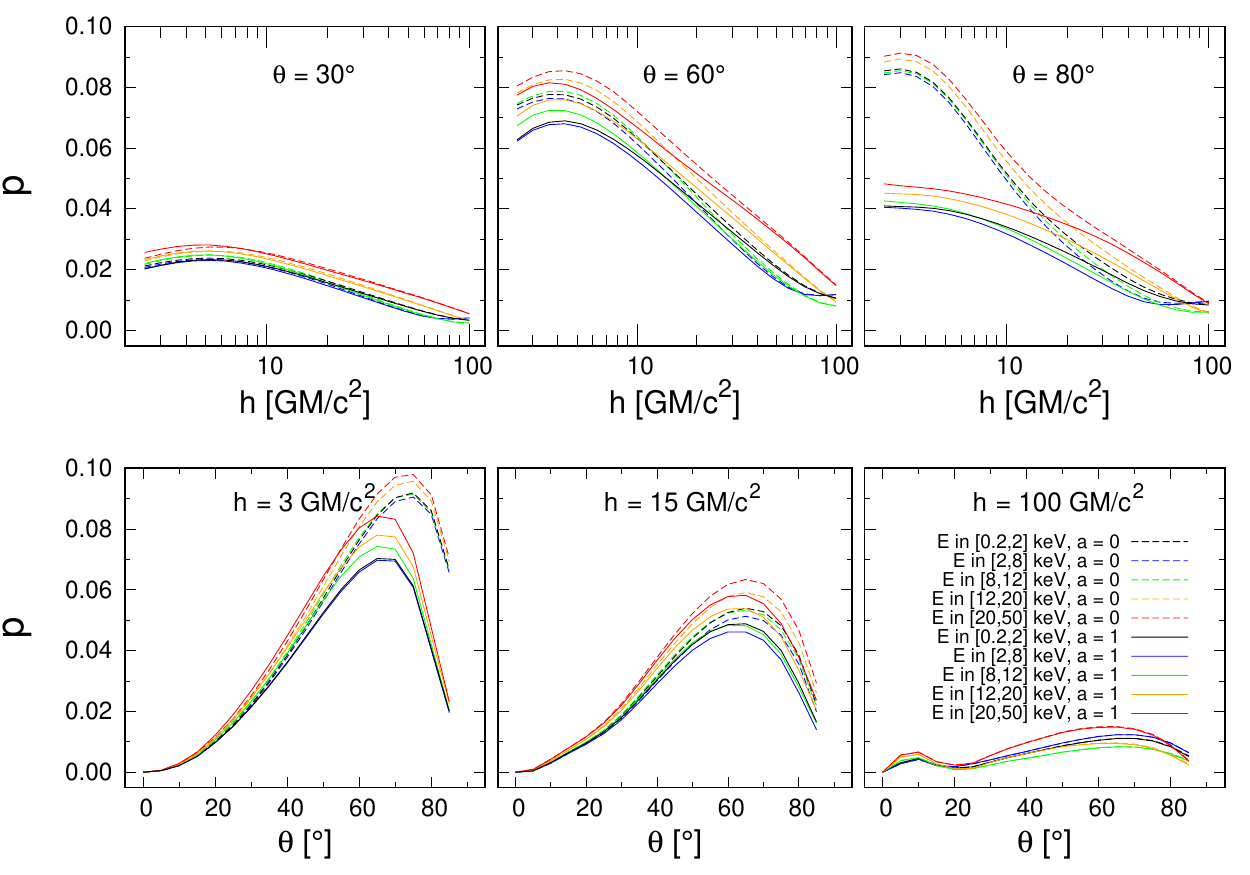}
	\caption{\footnotesize{The total average polarisation degree, $p$, versus energy for ionized disc \mbox{($M = 1\times 10^5\,M_{\odot}$} and observed 2--10 keV flux $L_{\textrm{X}}/L_{\textrm{Edd}} = 0.1$), otherwise for the same parametric setup as in Figure \ref{fig:fig9_tot_neutral.}, displayed in the same manner.}}
	\label{fig:fig9_tot_ionized.}
\end{figure}

The dependency of total polarisation degree on $h$ and $\theta$ is shown in more detail in Figures \ref{fig:fig9_tot_neutral.} and \ref{fig:fig9_tot_ionized.}, integrated in various energy bins and for the neutral and~ionized disc cases, respectively. Only if the lamp-post is located close to~the~event horizon, the relativistic effects depolarize significantly. Then the total polarisation departs from the otherwise monotonic dependency of $p(h)$, which is given by the~reflection fraction and by the change of the dominant incident inclination angle for the local reflection tables. It was checked that the differences in these figures would be negligible, if the accretion disc was considered up~to~$r_{\mathrm{out}} = 1000 \, r_\textrm{g}$ \citep{Podgorny2023}. The dependency of polarisation fraction on inclination is also nearly monotonic, apart from very high inclinations, where we obtained a drop in polarisation, thus a peak in polarisation is formed at~$\theta \approx 65^{\circ}$. This is caused by the lower reflection fraction with respect to~the~primary at high heights and by the relativistic light-bending effects. Although the~highly inclined views of the black hole's vicinity would be observationally blocked by the dusty torus in AGNs, we keep the modeling discussion, because in Chapter \ref{chap04} we will use {\tt KYNSTOKES} for high $\theta$ as a spectropolarimetric model input of~the~equatorial illumination of such tori. At low $h$ and high $\theta$, the effects of different spin on the resulting polarisation are the clearest, due to~the~largest relativistic distortion in such scenarios. Comparing the two figures, which vary in disc ionization level, we can clearly see the energy independence of~highly ionized reflection. The~fast rotating black holes allow for higher reflection contribution (when the disc is considered down to ISCO, whose location changes with $a$) and increase of polarisation at low lamp-post heights and low or moderate inclinations. But at~extreme inclinations, the black-hole spin rather acts as a~depolarizer.

\begin{figure}[!htb]\centering
	\includegraphics[width=\textwidth]{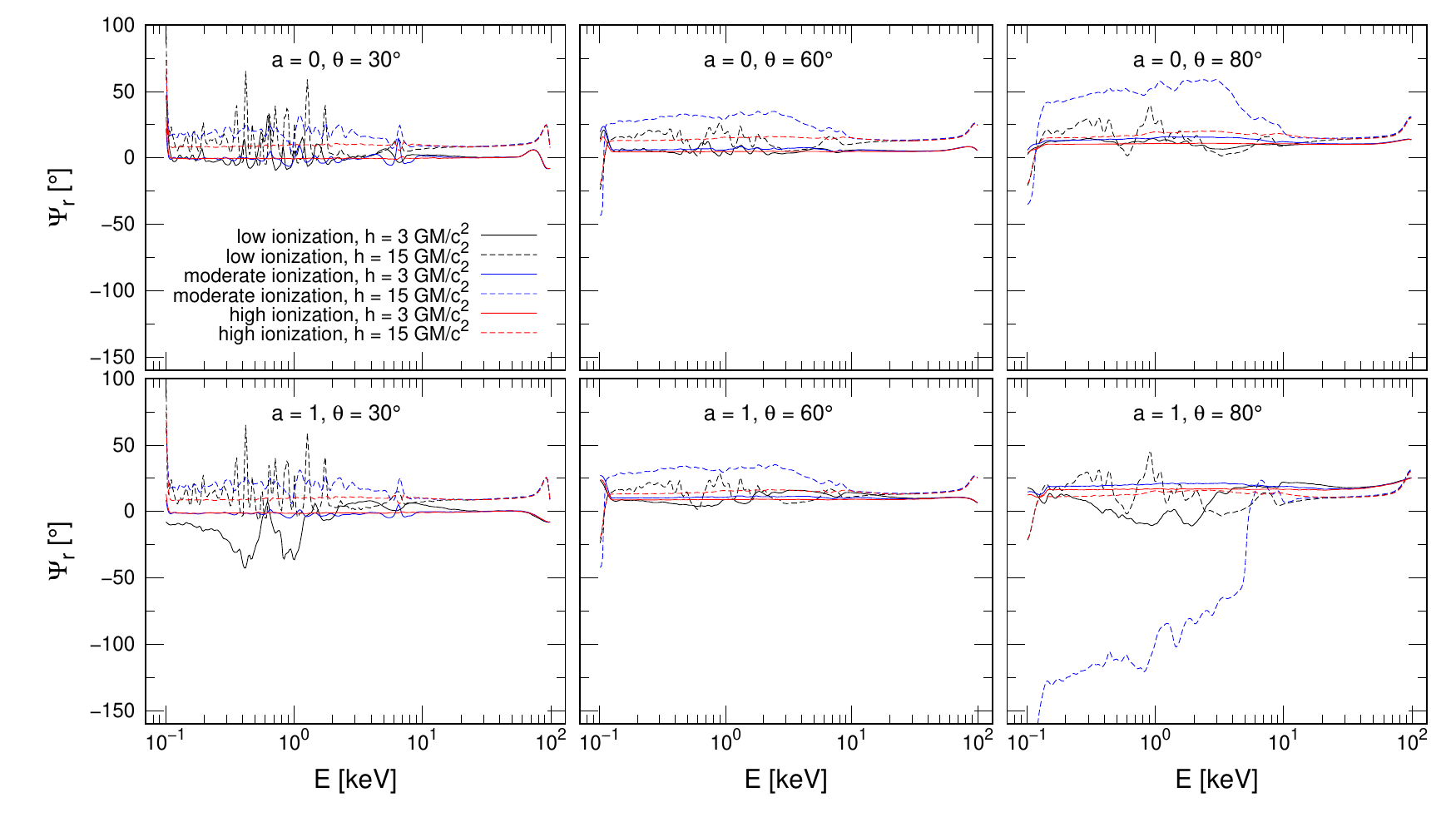}
	\caption{\footnotesize{The reflected-only polarisation angle, $\Psi_\mathrm{r}$, versus energy for the same parametric setup as in Figure \ref{fig:K3_R_unpol_I.}, displayed in the same manner.}}
	\label{fig:K3_R_unpol_pang.}
\end{figure}

Figure \ref{fig:K3_R_unpol_pang.} shows the corresponding polarisation angle versus energy (the primary is considered unpolarized, so this quantity is identical for the reflected-only and total signal). For the lamp-post model, we obtain polarisation nearly parallel with the axis of symmetry of the system. Only small deviations occur in~the~continuum due to relativistic effects and in the spectral lines due to undefined local $\Psi$ for unpolarized fluorescent emission. The resulting polarisation angle orientation after the relativistic corrections is not obvious for high lamp-post heights, but we conclude that even in these cases the dominant single-scattering polarisation genesis occurs on the left and right sides of the equatorial disc from observer's point of view, which symmetrically adds to net polarisation nearly parallel with the~black-hole rotation axis. We note that in detailed relativistic studies of coronal Comptonization of disc seed photons \citep{Ursini2022, Krawczynski2022a}, the resulting polarisation angle needs not to be either parallel or perpendicular to the axis of symmetry due to GR rotation effects. Figures \ref{fig:fig10_tot_neutral.} and \ref{fig:fig10_tot_ionized.} represent the same as in Figures \ref{fig:fig9_tot_neutral.} and \ref{fig:fig9_tot_ionized.}, respectively, but for~the~net polarisation angle, where the subtle differences are clearer when plotted integrated in different energy bins.

~\

\begin{figure}[!htb]
	\centering
        \begin{subfigure}{0.48\textwidth}
            \includegraphics[width=\textwidth]{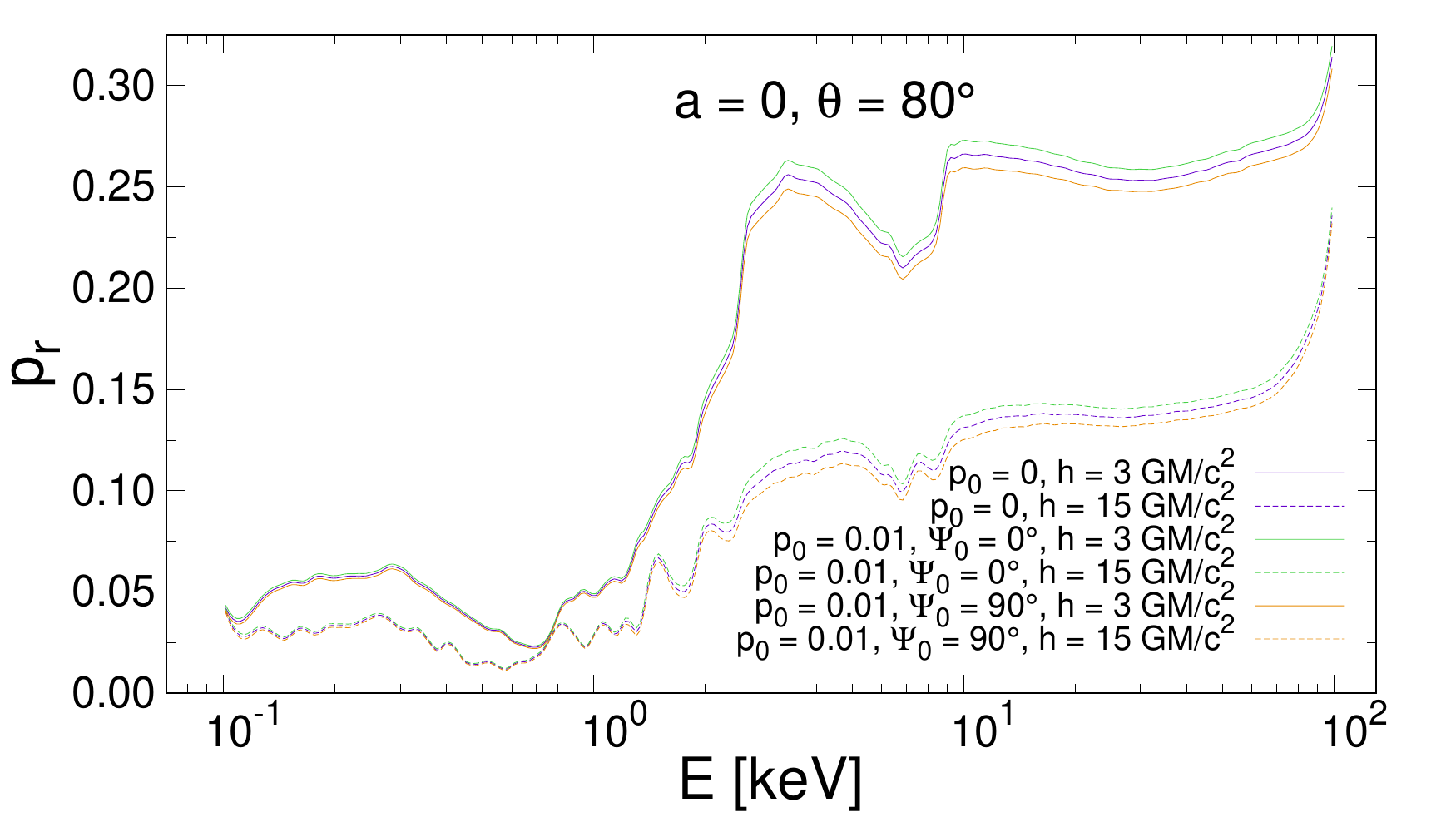}
        \end{subfigure}
        \begin{subfigure}{0.48\textwidth}
            \includegraphics[width=\textwidth]{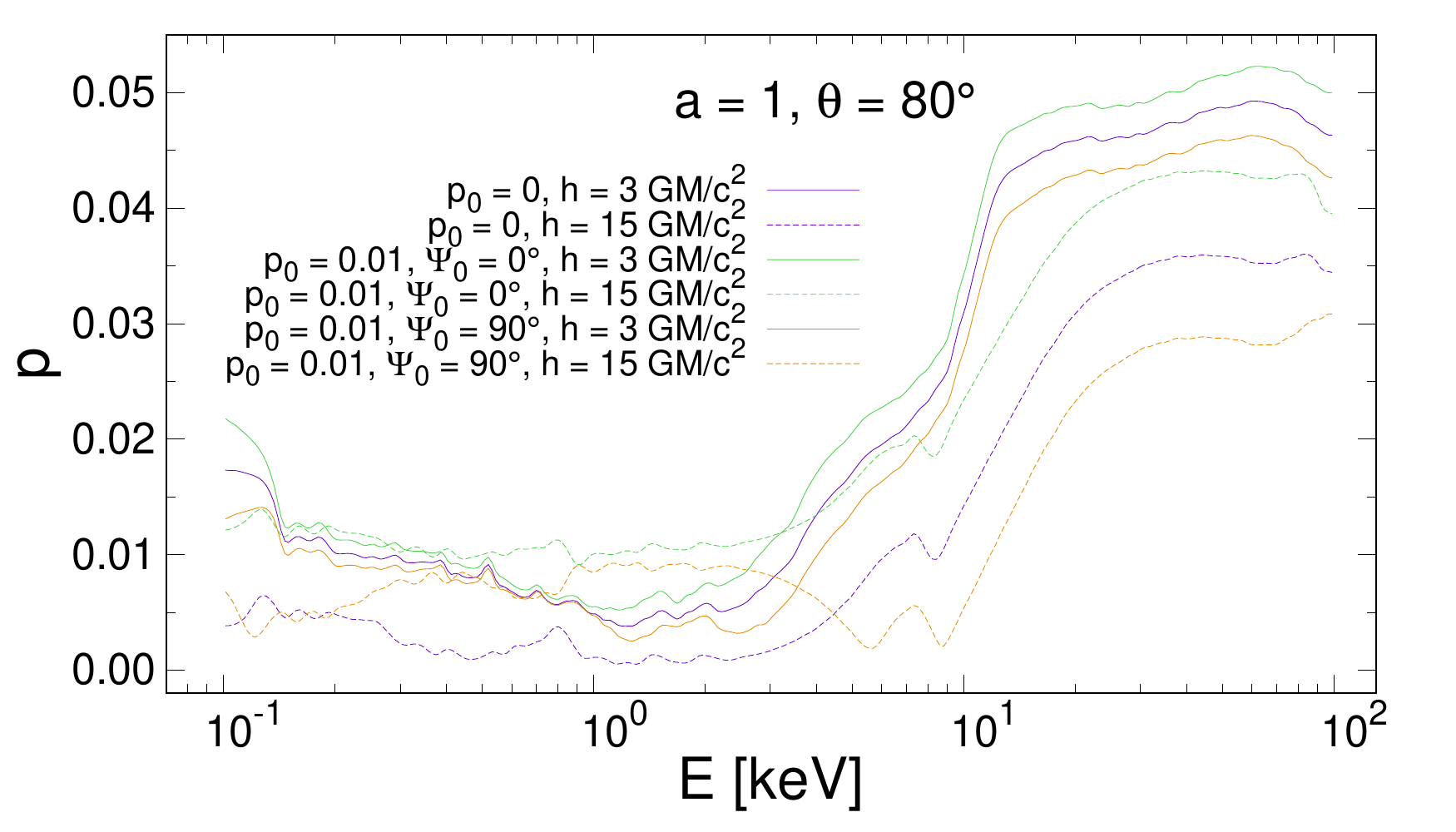}
        \end{subfigure}
        \begin{subfigure}{0.48\textwidth}
            \includegraphics[width=\textwidth]{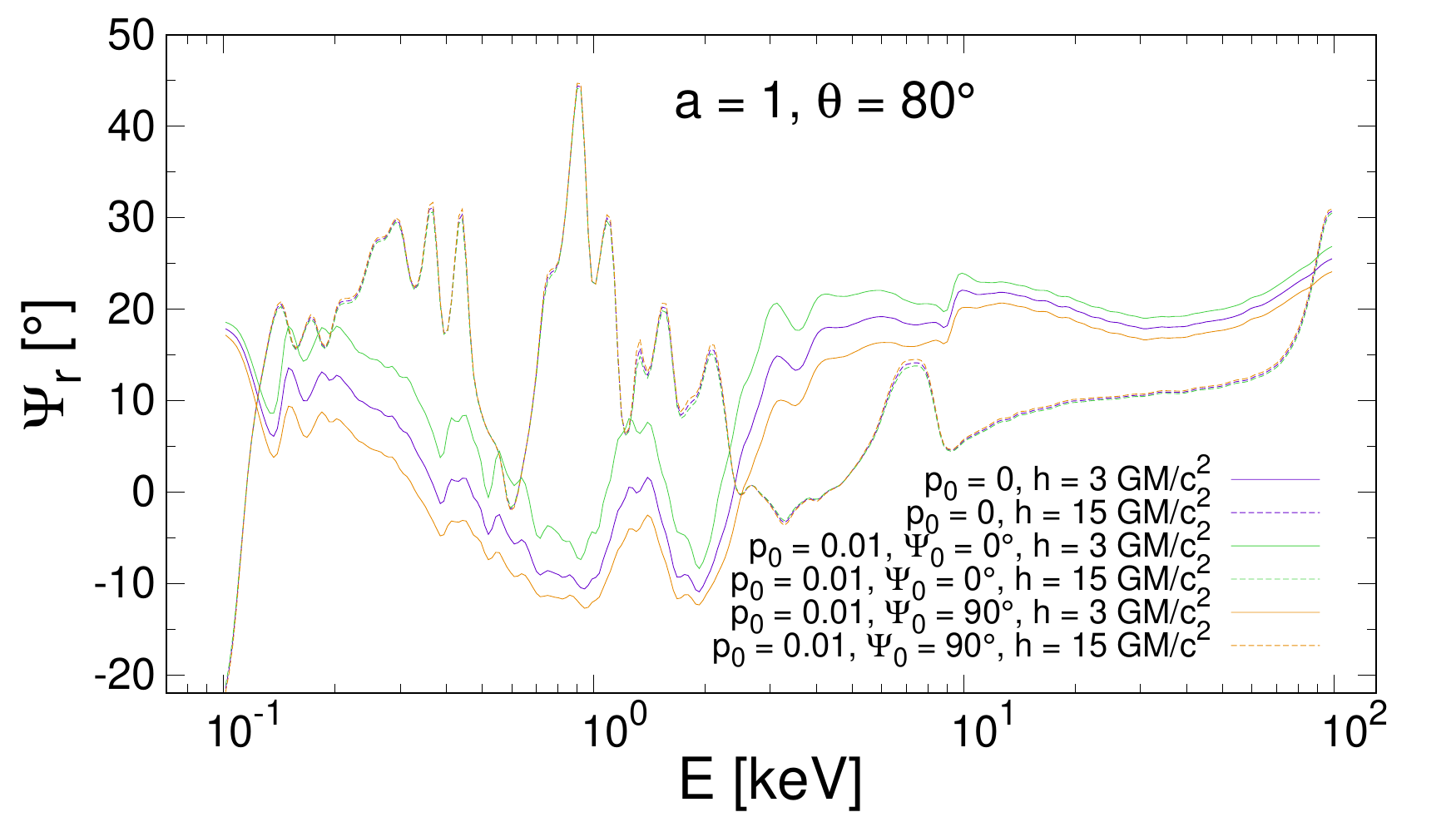}
        \end{subfigure}
 	\caption{\footnotesize{Top left: the reflected-only polarisation degree, $p_\mathrm{r}$, versus energy of the accretion disc for distant observer, obtained by {\tt KYNSTOKES} for black-hole spin $a = 0$, disc inclination $\theta = 80^{\circ}$, $\Gamma = 2$ and for the neutral disc ($M = 1\times 10^8\,M_{\odot}$ and observed 2--10 keV flux $L_{\textrm{X}}/L_{\textrm{Edd}} = 0.001$), using the {\tt STOKES} local reflection model in the lamp-post scheme. We show cases of two different heights of the primary point-source above the disc $h = 3 \textrm{ } r_\textrm{g}$ (solid lines) and $h = 15 \textrm{ } r_\textrm{g}$ (dashed lines), and three different polarisation states of the primary source: $p_0 = 0$ (purple), $p_0 = 0.01$ and $\Psi_0 = 0^{\circ}$ (green), $p_0 = 0.01$ and $\Psi_0 = 90^{\circ}$ (orange). Top right: the total polarisation degree, $p$, versus energy for black-hole spin $a = 1$, otherwise for the same parametric setup, displayed in the same manner. Bottom: the reflected-only polarisation angle, $\Psi_\mathrm{r}$, versus energy for black-hole spin $a = 1$, otherwise for the same parametric setup, displayed in the same manner.}}
    \label{K3A_unique}
\end{figure}

\begin{figure}[!htb]\centering
	\includegraphics[width=\textwidth]{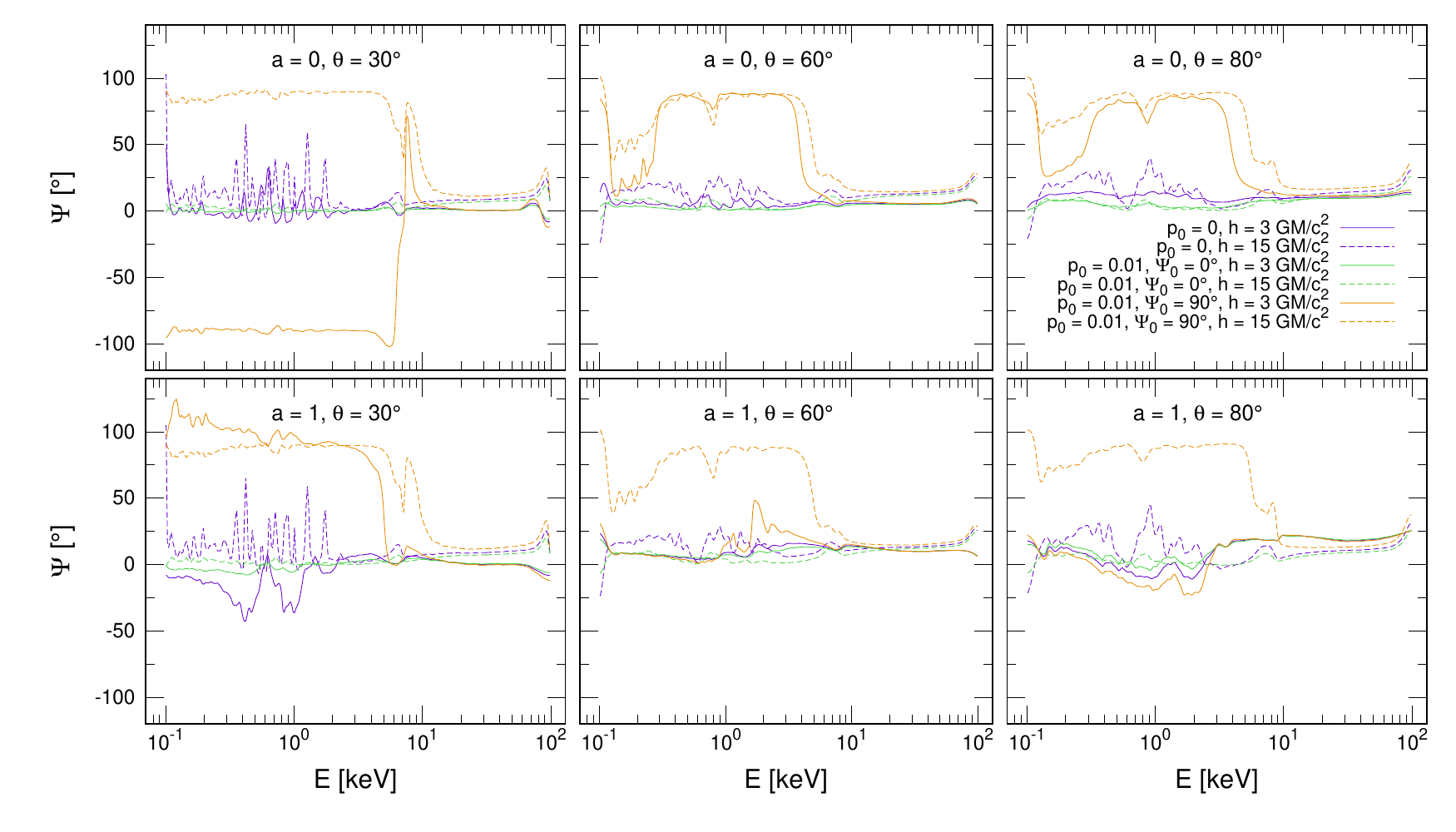}
	\caption{\footnotesize{The total polarisation angle, $\Psi$, versus energy of the accretion disc for distant observer, obtained by {\tt KYNSTOKES} for black-hole spins $a = 0$ (top) and $a = 1$ (bottom), disc inclinations $\theta = 30^{\circ}$ (left), $\theta = 60^{\circ}$ (middle) and  $\theta = 80^{\circ}$ (right), $\Gamma = 2$ and neutral disc ($M = 1\times 10^8\,M_{\odot}$ and observed 2--10 keV flux $L_{\textrm{X}}/L_{\textrm{Edd}} = 0.001$), using the {\tt STOKES} local reflection model in the lamp-post scheme. We show cases of two different heights of the primary point-source above the disc $h = 3 \textrm{ } r_\textrm{g}$ (solid lines) and $h = 15 \textrm{ } r_\textrm{g}$ (dashed lines), and three different polarisation states of the primary source: $p_0 = 0$ (purple), $p_0 = 0.01$ and $\Psi_0 = 0^{\circ}$ (green), $p_0 = 0.01$ and $\Psi_0 = 90^{\circ}$ (orange).}}
	\label{fig:K3A_RP_polar_n_pang_mainbody.}
\end{figure}

We will now release the assumption of unpolarized primary, while keeping $\Gamma = 2$, in order to discuss the effects of changing primary polarisation. The~effect on the resulting spectra is negligible, hence we will only study the resulting polarisation state. We plot the cases of $p_0 = 0$ (purple lines in figures), $p_0 = 0.01$ and $\Psi_0 = 0^{\circ}$ (green lines in figures), $p_0 = 0.01$ and $\Psi_0 = 90^{\circ}$ (orange lines in figures), which are well within conservative estimates of coronal polarisation \citep{Beheshtipour2017, BeheshtipourThesis, Tamborra2018, Ursini2022, Krawczynski2022a}. We omit the results for highly ionized discs, because they show identical effects of changing incident polarisation, only with a generally higher contribution of reflected component with respect to~the~primary. Figure \ref{K3A_unique} shows for selected cases of $a$ and $\theta$ the~reflected-only and total polarisation degree, and the reflected-only polarisation angle, for~a~neutral disc case and for two distinct lamp-post heights. The~same simulations for~different $a$ and~$\theta$ are shown in Figures \ref{fig:K3A_R_polar_n_pdeg_relative.}--\ref{fig:K3A_RP_polar_i_pang_relative.}, relatively to the case with unpolarized primary. We only provide here the usual six panels of different $a$ and~$\theta$ in Figure \ref{fig:K3A_RP_polar_n_pang_mainbody.} for the total value of polarisation angle for the neutral disc case, which is theoretically more interesting (see below).

The resulting polarisation fraction (reflected-only or total) preserves qualitatively the same response at hard X-rays (where mostly Compton scattering is involved) to incident polarisation state for all studied cases. The values differ from the case with unpolarized primary by $\lessapprox 1\%$, with lower emergent polarisation for~horizontally polarized incident photons (orange lines) and higher emergent polarisation for vertically polarized incident photons (green lines). If we test higher incident polarisation fraction, the resulting difference from the case with unpolarized primary (purple lines) scales linearly in the same qualitative direction. The~differences are symmetric with respect to the unpolarized primary case, which has roots in the way that local reflection is constructed via (\ref{Sorig}). The emergent reflected-only polarisation angle is also symmetric with respect to~the~unpolarized case with differences $\lessapprox 1^{\circ}$. The symmetry is broken when the~primary radiation is added to the total polarisation angle and degree, especially at soft X-rays. This is seen in Figure \ref{fig:K3A_RP_polar_n_pang_mainbody.}, where the orange lines drawn for~horizontal incident polarisation, deviate significantly at soft X-rays from the~parallel polarisation inherent to the reflected emission. At soft X-rays for a neutral disc, the reflected flux contribution is not significant with respect to the primary flux and the polarisation degree also tends to unite at $1\%$, if prescripted for the primary. Figure \ref{fig:fig10_tot_neutral_P2.} shows that the polarisation angle transition for $p_0 = 0.01$ and $\Psi_0 = 90^{\circ}$ needs not to always occur in the 2--8 keV important for \textit{IXPE} or \textit{eXTP}. It depends on~the~chosen $h$ and $\theta$.

~\

\begin{figure}[!htb]\centering
	\includegraphics[width=\textwidth]{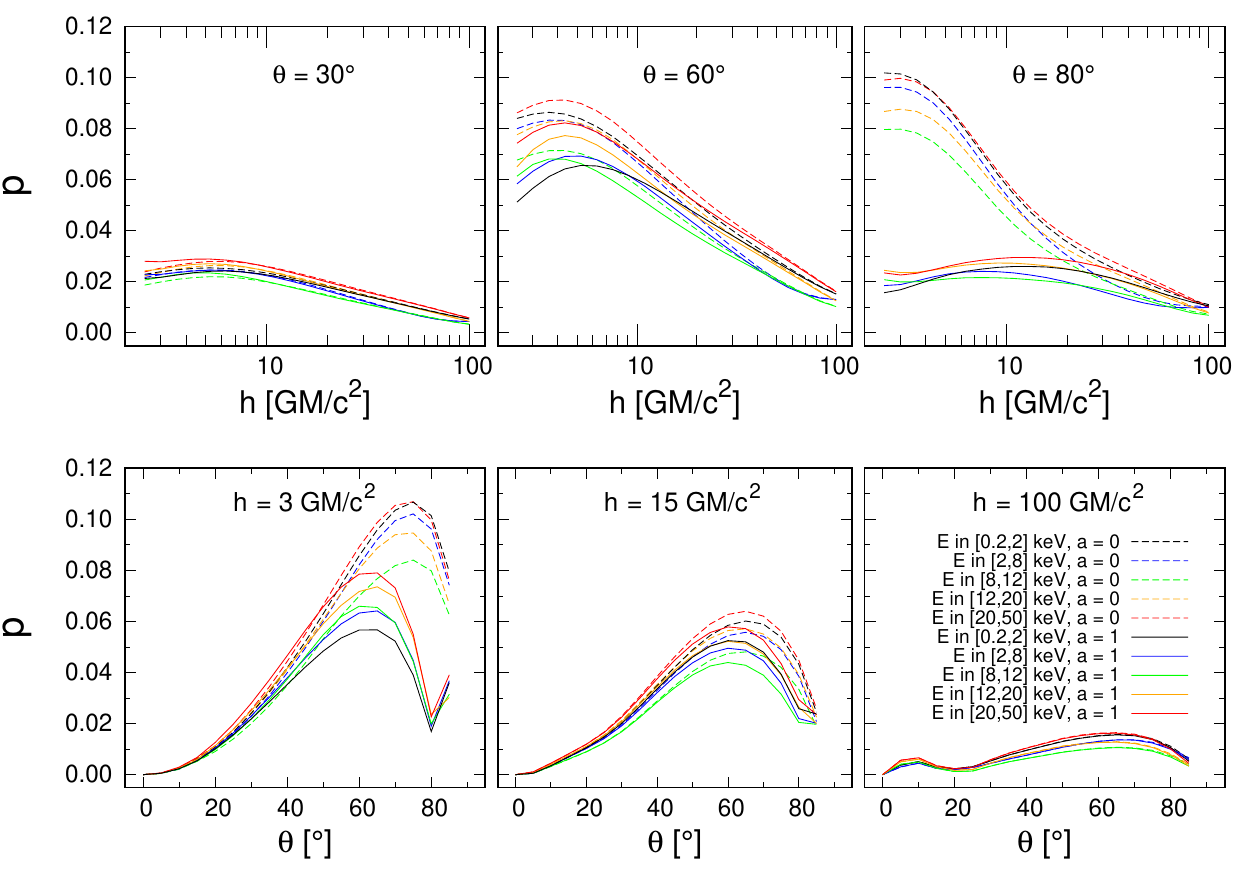}
	\caption{\footnotesize{The total average polarisation degree, $p$, versus energy for~$\Gamma = 3.0$, otherwise for~the~same parametric setup as in Figure \ref{fig:fig9_tot_ionized.}, displayed in the same manner.}}
	\label{fig:fig9_tot_ionized_30.}
\end{figure}

Let us briefly examine the role of $\Gamma$. Apart from the slope of primary radiation, which weights differently polarized photons at different energies, the disc ionization is affected. The local ionization parameter given by (\ref{xi}) is still changing with $\Gamma$ for a fixed normalization given by $L_{\textrm{X}}/L_{\textrm{Edd}}$ and~$M$ and~fixed $n_\textrm{H}$. Higher photon index makes the spectral lines and absorption in~the~reprocessed emission more pronounced (depending on the energy band), since less energetic photons ionize the disc. This imprints to the energy dependence of~polarisation degree at infinity. For higher $\Gamma$, we see a slight depolarisation in~between $0.8$--$3$ keV where the polarisation degree with energy had a~nearly constant curve for $\Gamma = 2$ for highly ionized (red curves) cases in Figures~\ref{fig:K3_R_unpol_pdeg.} and~\ref{fig:K3_RP_unpol_pdeg.}. The effect is, however, lower with the new code than reported in~\cite{Podgorny2023} due to~higher interpolation precision, using the newer version of the {\tt TITAN} and {\tt STOKES} tables. We notice an overall increase of polarisation for~$a = 0$ and decrease for $a = 1$ between $\Gamma = 2$ and $\Gamma = 3$, as relativistic effects play a~role. We provide the same total polarisation degree versus inclination and~lamp-post height as in Figure \ref{fig:fig9_tot_ionized.} (showing the results for $\Gamma = 2$) but for $\Gamma = 3$ in Figure \ref{fig:fig9_tot_ionized_30.}. The~$y$-axis is extended with respect to the previously shown case with $\Gamma = 2$.

~\

Finally, we will investigate the role of disc truncation for polarisation gain by reflection, which especially in the pure lamp-post model should introduce effectively a new geometry of emission. As described in Chapter \ref{introduction}, determination of~the~inner disc radius can be of vital importance for our understanding of~the~accretion mechanisms. This holds not only for~the~hard spectral state of~XRBs, where the disc is believed to be significantly truncated. The effects of~reducing spatial symmetry for reflection inside the system in this way are similar in polarisation induction across all X-ray energies, so we will only provide the integrated result in 2--8 keV band, important for the {\it IXPE} and {\it eXTP} missions. For~three different inclinations and various cases of incident polarisation, we will plot the reflected-only and total polarisation degree and angle versus $r_\textrm{in}$, in between ISCO and $100 \, r_\textrm{g}$. Since wide range of disc truncations is studied, we extend the outer disc radius to $r_{\mathrm{out}} = 1000 \, r_\textrm{g}$ specifically for these calculations. We also add the corresponding reflection fraction $R_\textrm{f}$, i.e. the observed energy-integrated reflected-only flux divided by the observed energy-integrated total flux in 2--8 keV. The dependencies are similar for different $a$, so we show only the results for $a = 1$, which is more interesting than a Schwarzschild black hole, because it allows to examine spatial regions closer to the event horizon, assuming a standard accretion disc with sharp truncation on the inner side.

\begin{figure}[!htb]\centering
	\includegraphics[width=\textwidth]{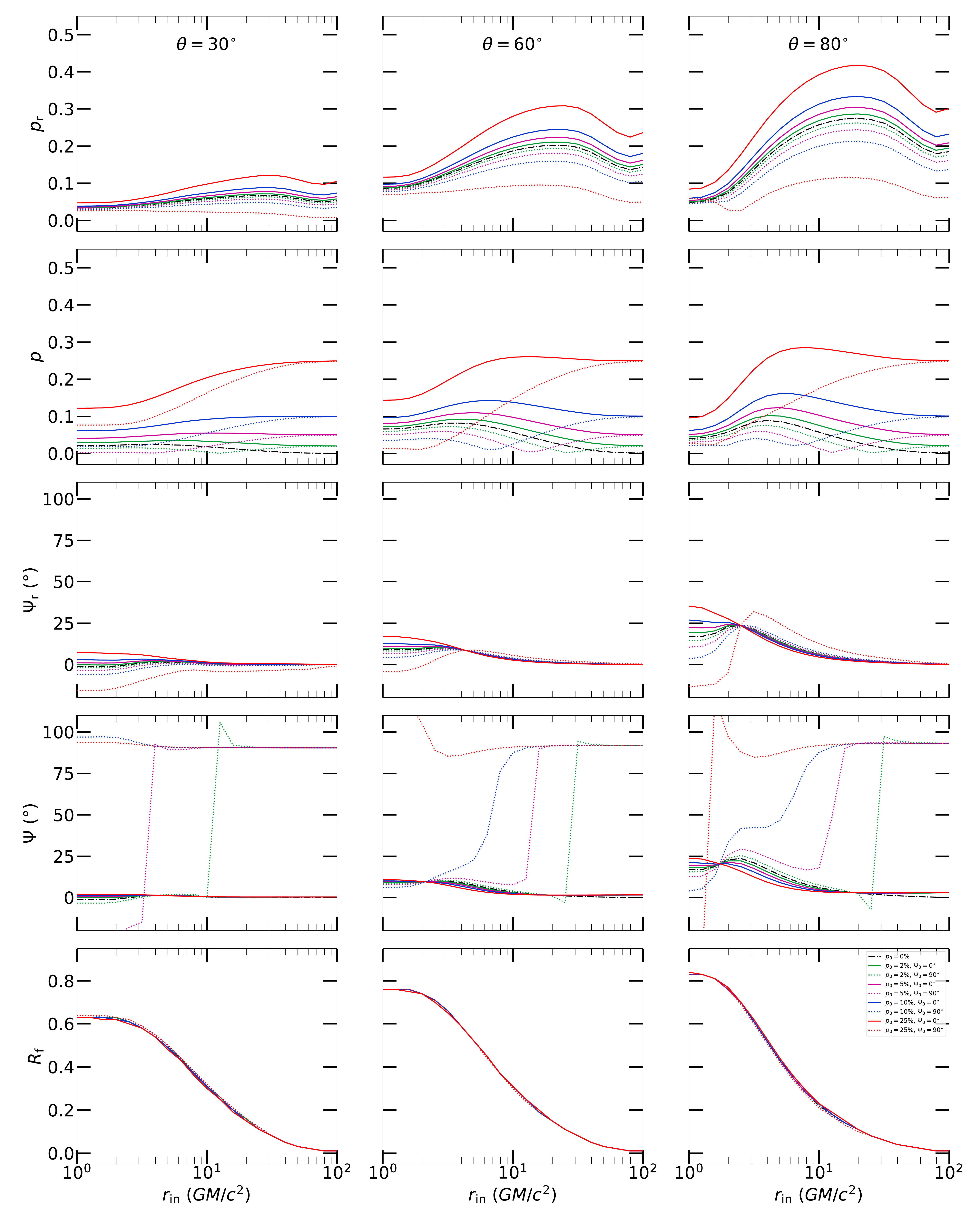}
	\caption{\footnotesize{The spectropolarimetric output of {\tt KYNSTOKES} in the lamp-post regime with respect to disc truncation, integrated in 2--8 keV band. From top to bottom we show the reflected-only polarisation degree, $p_\textrm{r}$, the total polarisation degree, $p$, the reflected-only polarisation angle $\Psi_\textrm{r}$, the total polarisation angle, $\Psi$, and the reflection fraction, $R_\textrm{f}$, all as a function of the inner disc radius $r_\textrm{in}$. Various cases of the primary polarisation state are shown in the color code and~line style. We show the results for $\Gamma = 2$, $a = 1$ and $\theta = 30^{\circ}$ (left column), $\theta = 60^{\circ}$ (middle column) and $\theta = 80^{\circ}$ (right column).}}
	\label{rin_ionized_lp_h3}
\end{figure}

Figures \ref{rin_ionized_lp_h3} and \ref{rin_ionized_lp_h15} show the results for $h = 3 \, r_\textrm{g}$ and $h = 15 \, r_\textrm{g}$, respectively, for highly ionized disc. The same for nearly neutral disc is shown in Figures \ref{rin_neutral_lp_h3} and \ref{rin_neutral_lp_h15}. The more we truncate the disc, the more we limit the allowed scattering space and the higher polarisation we obtain. The trade-off for~more detectable higher polarisation fraction is, however, the reflection fraction, which obviously decreases with truncation for lamp-post coronae. For highly ionized discs, the~increase of reflected-only polarisation degree with $r_\textrm{in}$ saturates and~at~low lamp-post heights, we observe again a decrease of polarisation degree for~$r_\textrm{in} \gtrapprox 30 \, r_\textrm{g}$. The~lower the $R_\textrm{f}$ is, the more the total polarisation result for~a~distant observer reaches the polarisation prescription at the lamp-post. The~polarisation change between the lamp-post and the observer is not significant (see Appendix \ref{change_PA}). All of~the~disc truncation effects are inclination dependent.

\subsection{Comparison to other models}

Let us compare the lamp-post results within the same {\tt KY} relativistic integration, but using different local reflection computations. We will now assume again the full disc without any truncation. For this we will use the {\tt KYNLPCR} routine (courtesy of Michal Dovčiak) that assumes reflection for a purely neutral disc, computed with {\tt NOAR} \citep{Rozanska2002}, and the {\tt KYNXILLVER} routine (courtesy of~Michal Dovčiak) that assumes the {\tt XILLVER} spectral reflection tables \citep{Garcia2013}, which we juxtaposed locally to the {\tt TITAN} and {\tt STOKES} tables in Section \ref{local_comparison}. Figure \ref{fig:K5_I_Lin.} shows the relativistically integrated spectra from {\tt KYNSTOKES}, {\tt KYNLPCR} and {\tt KYNXILLVER} routines for unpolarized primary with $\Gamma = 2$, $h = 3 \textrm{ } r_\textrm{g}$, two cases of disc ionization, two black-hole spins and~three inclinations. The largest differences occur, as anticipated from the local reflection tables, in the soft X-rays. {\tt KYNLPCR} underpredicts flux contribution for~the~more realistic nearly neutral disc below 1 keV for $a = 0$ and below 10 keV for~$a = 1$. The {\tt KYNSTOKES} spectral results are systematically higher normalized than the~{\tt KYNXILLVER} results for the low ionization cases, and lower normalized than the~{\tt KYNXILLVER} results for the high ionization cases. The difference may be also due to angular treatment of~the~local reflections. Three angles are considered in~the~{\tt TITAN} and {\tt STOKES} tables, compared to the isotropically irradiated and~azimuthally averaged reflection in {\tt XILLVER}.

~\

\begin{figure}[!htb]\centering
	\includegraphics[width=\textwidth]{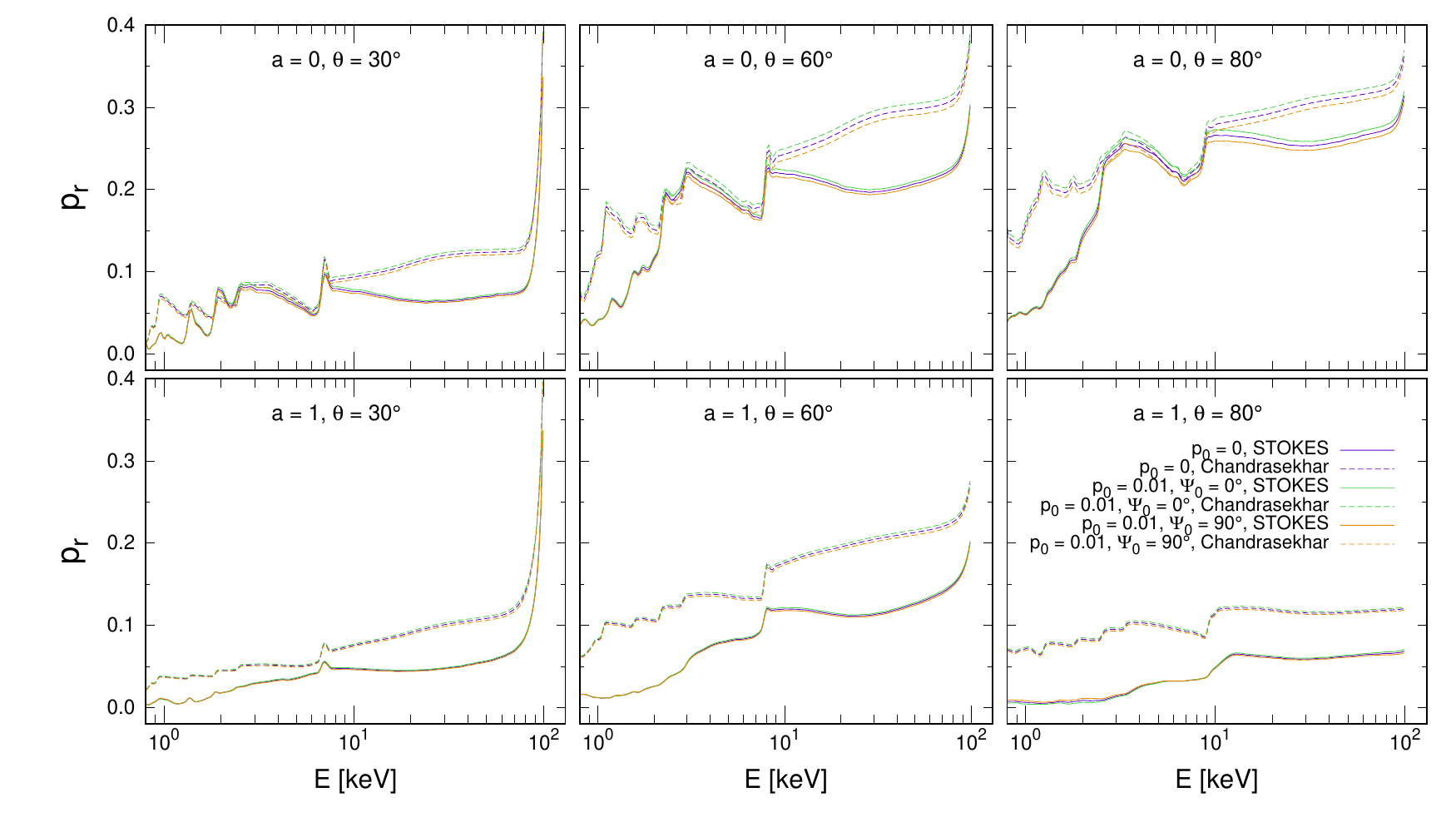}
	\caption{\footnotesize{The reflected-only polarisation degree versus energy, $p_\mathrm{r}$, of the accretion disc for~distant observer in the lamp-post scheme for black-hole spins $a = 0$ (top) and $a = 1$ (bottom), disc inclinations $\theta = 30^{\circ}$ (left), $\theta = 60^{\circ}$ (middle) and  $\theta = 80^{\circ}$ (right), $\Gamma = 2$, height of~the~primary point-source above the disc $h = 3 \textrm{ } r_\textrm{g}$ and neutral disc (for $M = 1\times 10^8\,M_{\odot}$ and~observed 2--10 keV flux $L_{\textrm{X}}/L_{\textrm{Edd}} = 0.001$). We compare cases of {\tt KY} disc integration with the~{\tt STOKES} local reflection tables (solid lines) and {\tt NOAR} local reflection tables with Chandrasekhar's analytical approximation for polarisation (dashed lines), and three different polarisation states of the~primary source: $p_0 = 0$ (purple), $p_0 = 0.01$ and $\Psi_0 = 0^{\circ}$ (green), $p_0 = 0.01$ and $\Psi_0 = 90^{\circ}$ (orange).}}
	\label{fig:K5_p_Ln.}
\end{figure}

\begin{figure}[!htb]\centering
	\includegraphics[width=\textwidth]{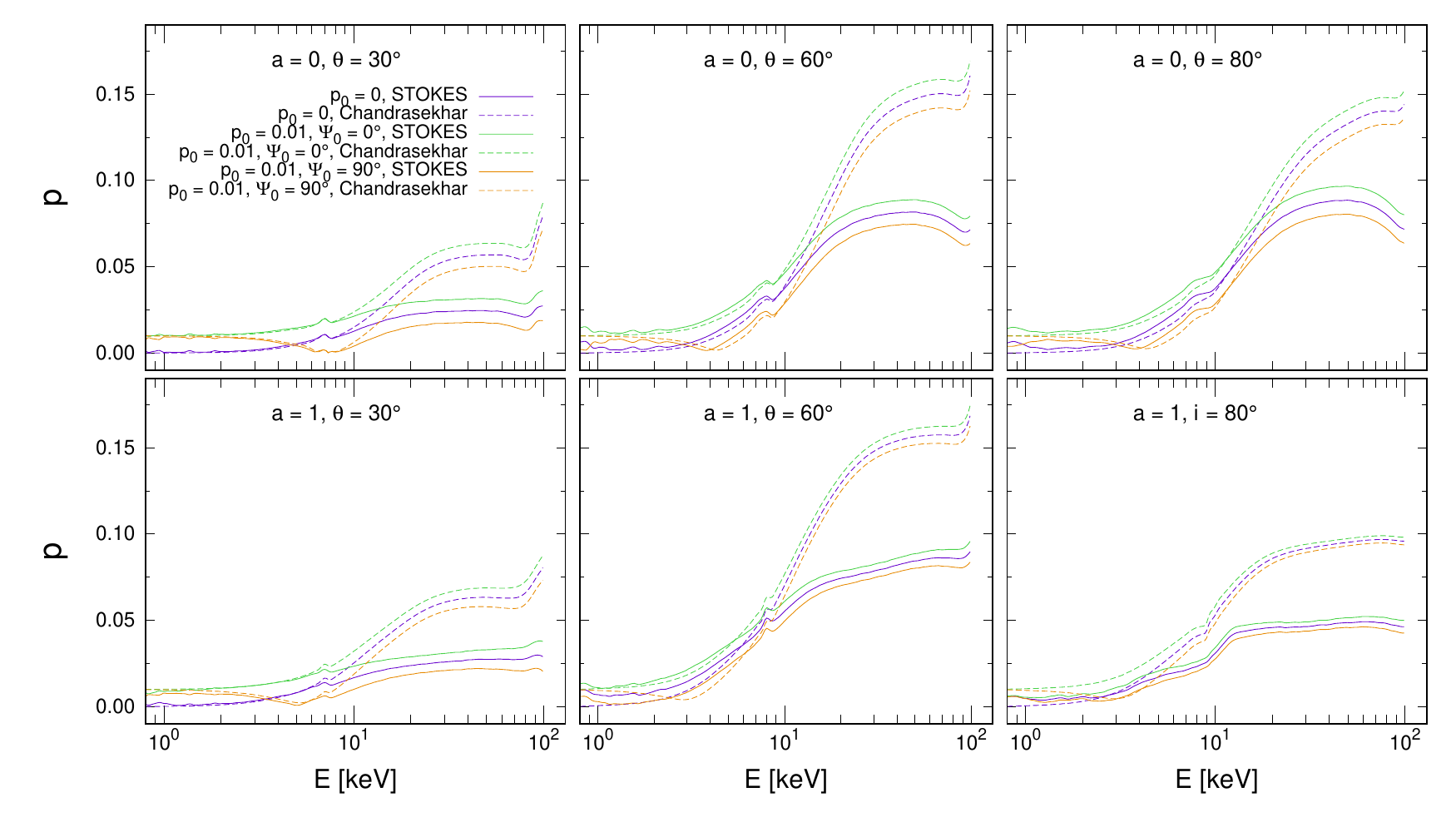}
	\caption{\footnotesize{The total polarisation degree, $p$, versus energy for distant observer for the same parametric setup as in Figure \ref{fig:K5_p_Ln.}, displayed in the same manner.}}
	\label{fig:K5_p_Ln_RP.}
\end{figure}

\begin{figure}[!htb]\centering
	\includegraphics[width=\textwidth]{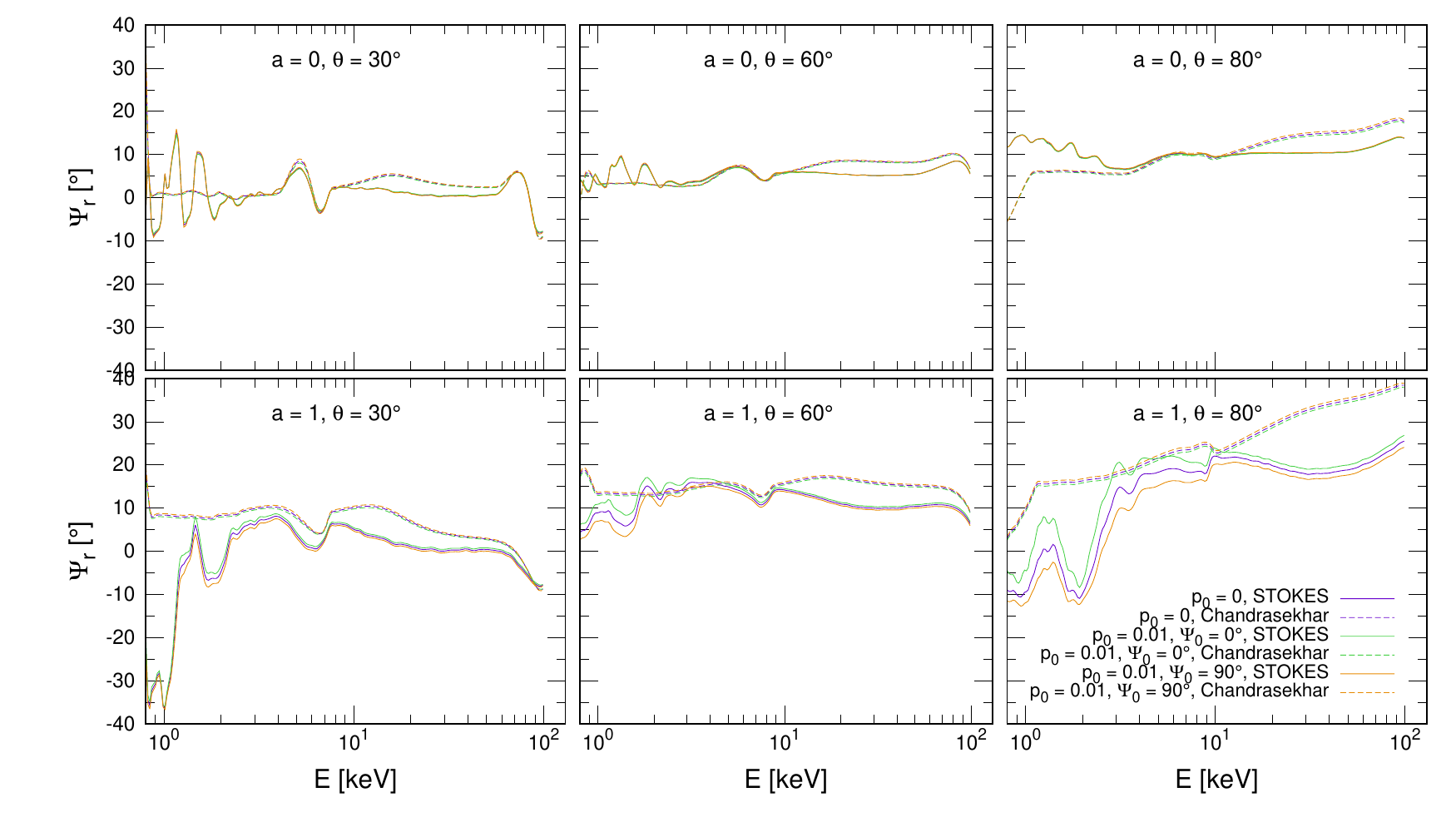}
	\caption{\footnotesize{The reflected-only polarisation angle, $\Psi_\mathrm{r}$, versus energy for distant observer for~the~same parametric setup as in Figure \ref{fig:K5_p_Ln.}, displayed in the same manner.}}
	\label{fig:K5_Psi_Ln.}
\end{figure}

\begin{figure}[!htb]\centering
	\includegraphics[width=\textwidth]{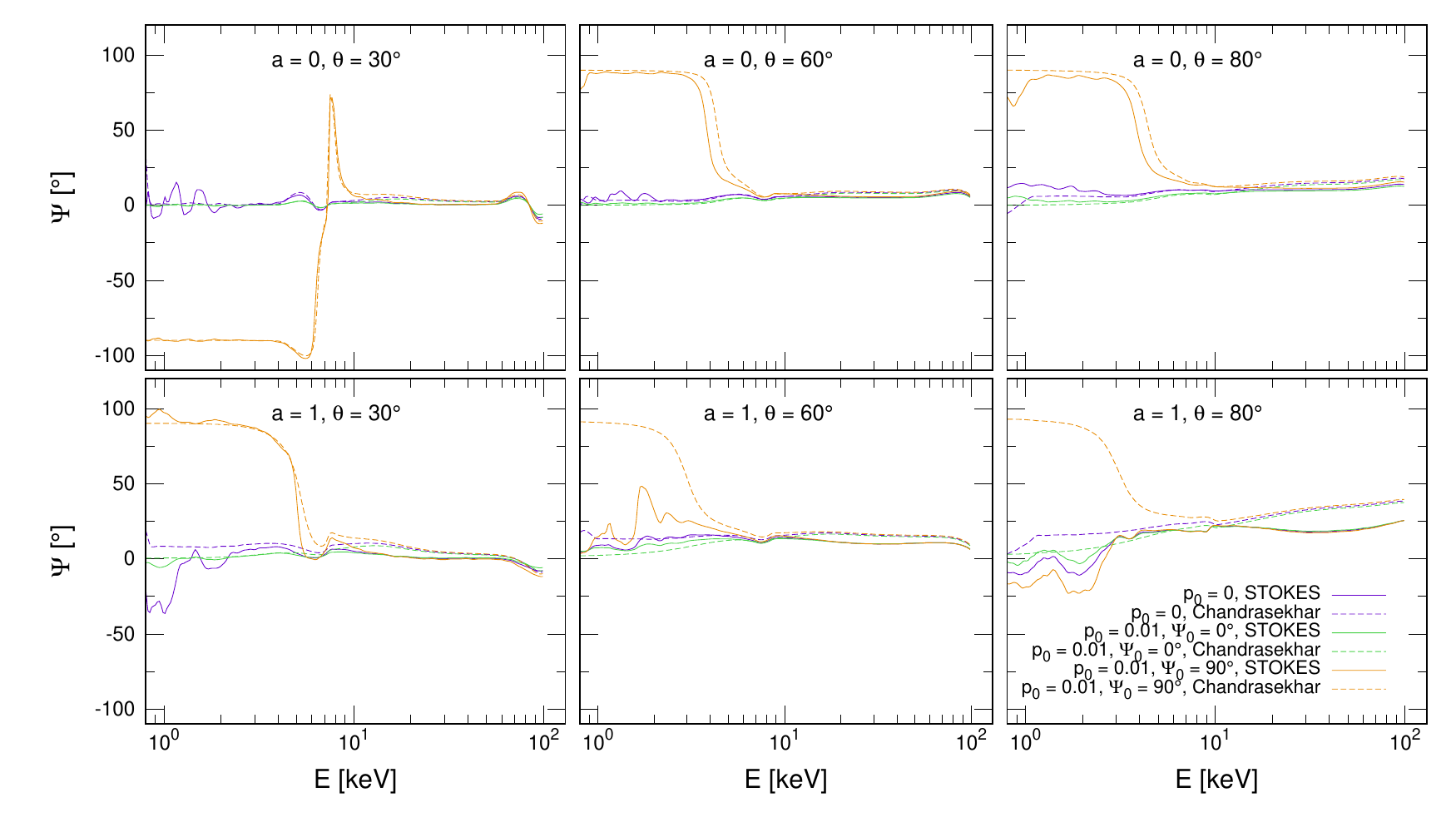}
	\caption{\footnotesize{The total polarisation angle, $\Psi$, versus energy for distant observer for the same parametric setup as in Figure \ref{fig:K5_p_Ln.}, displayed in the same manner.}}
	\label{fig:K5_Psi_Ln_RP.}
\end{figure}

The {\tt KYNLPCR} routine, with which the results in \cite{Dovciak2011} were computed, contains also local reflection polarisation calculations through the single-scattering Chandrasekhar's formulae \ref{chandra1}. Figures \ref{fig:K5_p_Ln.} and \ref{fig:K5_p_Ln_RP.} provide comparisons with {\tt KYNSTOKES} of the reflected-only and total polarisation degree versus energy, respectively, for changing incident polarisation, two black-hole spins, three inclinations, neutral accretion disc ($M = 1\times 10^8\,M_{\odot}$ and observed 2--10 keV flux $L_{\textrm{X}}/L_{\textrm{Edd}} = 0.001$ for {\tt KYNSTOKES}), unpolarized primary with $\Gamma = 2$ and $h = 3 \textrm{ } r_\textrm{g}$. The same for reflected-only and total polarisation angle is shown in Figures \ref{fig:K5_Psi_Ln.} and \ref{fig:K5_Psi_Ln_RP.}, respectively. The relativistic effects are the same for both codes. But as shown in Figure \ref{polcomp_chandra}, at X-ray energies $\gtrapprox 10$ keV the locally induced polarisation by single-scattering significantly overpredicts the resulting polarisation. We now see the full energy dependence, including relativistic effects. The small differences in local reflection polarisation angle result in nearly identical polarisation orientation predictions for the distant observer. The only exception being the initially horizontally polarized primary radiation, which in the case of {\tt KYNLPCR} is always dominating over the reflection in soft X-rays, as there the reflection component is practically absent.

~\

Direct comparisons with the {\tt RELXILL} relativistic lamp-post implementation of {\tt XILLVER} tables would require a detailed analysis of normalization in~the~first place. It would, however, be a valuable next step, because the original {\tt RELXILL} codes \citep{Garcia2014, Dauser2014} now include returning radiation \citep{Dauser2022}. Apart from the critical assumption of lamp-post isotropic emission and apart from many improvements already suggested for~the~local \text{reflection} tables in Section \ref{local_comparison}, it is also highly desirable to implement higher-order reflections in {\tt KYNSTOKES} and cross-validate the pioneering study of~\text{\cite{Dauser2022}} from the spectroscopic point of view. The effects of returning radiation on the resulting polarisation can be at least qualitatively predicted, based on~the~works of \cite{Schnittman2009, Schnittman2010, Taverna2020}, which were carried for returning radiation effects on polarisation of thermal radiation of~accretion discs in XRBs, assuming the diffuse Chandrasekhar's formulae for~reflection.

It is argued in \cite{Dauser2022} that the returning radiation provides the~largest difference for high black-hole spins ($a \gtrapprox 0.9$) and low lamp-post heights ($h \lessapprox 5 \, r_\textrm{g}$), due to contributions from the inner-most parts of the disc ($r \lessapprox 10 \, r_\textrm{g}$). Although azimuthally dependent, the net polarisation fraction given by the self-irradiation in \cite{Schnittman2009, Schnittman2010} (without and with an extended corona considered) was $\approx 10\%$ with polarisation angle parallel to~the~axis of symmetry. Both were constant in energy. In {\tt KYNSTOKES}, the resulting polarisation tends to be aligned with the principal axis and of the order of a few \%. Despite the different source of irradiation in \cite{Schnittman2009, Schnittman2010} than we would have for once reprocessed radiation returning to the disc, we could thus expect in our cases an additional polarisation effect, rather than depolarisation. Nonetheless, a detailed calculation would have to prove such predictions, because the local flux angular distribution could play a role, as well as the~relative flux contributions onto and from different disc regions, as well as the~spectral shape of~a~reprocessed power-law, compared to the thermally radiating discs. Moreover, in~Section~\ref{lamppost_spproperties} we proved the non-negligible impact of incident polarisation on~the~integrated reprocessed result.

\section{The sandwich coronal model}\label{kynstokes_extended}

Here we consider the extended corona with negligible height, sitting on top of~the~reflecting accretion disc, as described in Section \ref{codes_here}. As there is no photon travelling considered between the source of X-rays and the disc, the~total emission is given by the same geodesics from the equatorial plane, either from the~primary source, or from the~reprocessed radiation given by the {\tt TITAN} and {\tt STOKES} local reflection tables, described in Section \ref{reflection_tables_TS} and considered in~the~model at~the~same equatorial location, corotating with the source of~emission in~the~equatorial plane. Locally, we uniformly integrate the reflection tables in~the~incident angles $\mu_\textrm{i}$ and azimuthal emission angles $\Phi_\textrm{e}$, to account for the diffuse character of the~source. The disc is also here assumed to have a~constant radial density profile with $n_\textrm{H} = 10^{15} \ \textrm{cm}^{-3}$, according to the local reflection tables again. Equivalently to~the~previous section, the disc inner radius $r_{\mathrm{in}}$ is fixed at~the~ISCO and $r_{\mathrm{out}} = 400 \, r_\textrm{g}$, if not stated otherwise. We will again study the case of~reflected-only radiation with $N_\textrm{p}/N_\textrm{r} = 0$ and the case of added primary source with $N_\textrm{p}/N_\textrm{r} = 1$, normalized according to the local reflection tables and the radial emissivity profile. We stress that the model in this setup is not realistic mainly because the coronal photons re-processed inside the~disc would inevitably be re-processed again in~the~corona before reaching the~observer. Hence, the results of~this model should be viewed as an extreme case of an extended corona, to~which a moderately extended lamp-post model would be converging for ever larger (and~already unrealistic) sizes. In other words, as a direction of results (assuming some monotony) discussed in Section \ref{kynstokes_lamppost}, if we increased the size of~the~lamp-post corona, which is in practice difficult, as our computational approach relies on symmetry simplifications. We refer to~the~X-ray polarisation studies of~\cite{Poutanen1996,Schnittman2010,Krawczynski2022a} that deal with the subsequent re-processing in~the~corona in~sandwich geometries.

The radial emissivity profile of the primary source is given by the power-law index $q$. In order to not further complicate the discussion with changing the~physical scale units with $M$, we fix the black-hole mass at $M = 2 \times 10^6 M_\odot$. The user-defined observed 2--10 keV flux $L_{\textrm{X}}/L_{\textrm{Edd}}$ in {\tt KYNSTOKES} is considered as global, meaning this value irradiates each ring surface area of $2\pi r \textrm{d}r$ given by the chosen binning, multiplied by $r^{-q}$. The local reflection ionization parameter is scaled accordingly from $\xi_0$ defined at $r = 6 \, r_\textrm{g}$, which is available above ISCO for any black-hole spin. In order to reasonably compare our results to the lamp-post scenario where the radial irradiation profile is given, we chose the $q = 3$, $a = 0$ and $\theta = 30^{\circ}$ values as default and searched for such values of $L_{\textrm{X}}/L_{\textrm{Edd}}$ that produce similar levels of low, moderate and high disc ionizations, as obtained in Section \ref{lamppost_spproperties} for the {\tt KYNSTOKES} lamp-post variant. This turns out to be  $L_{\textrm{X}}/L_{\textrm{Edd}}\vert_{q = 3,\theta = 30^{\circ}} = 10^{-5}$, $L_{\textrm{X}}/L_{\textrm{Edd}}\vert_{q = 3,\theta = 30^{\circ}} = 10^{-2}$ and $L_{\textrm{X}}/L_{\textrm{Edd}}\vert_{q = 3,\theta = 30^{\circ}} = 10^{2}$ for the low, moderate and high ionization cases, respectively. Due to the chosen fixation of $\xi_0$ at $r = 6 \, r_\textrm{g}$ we then used $L_{\textrm{X}}/L_{\textrm{Edd}}$ for different $q$ and $\theta$ renormalized according to
\begin{equation}
     \left. L_{\textrm{X}}/L_{\textrm{Edd}}\right|_{q,\theta} = \frac{\left. L_{\textrm{X}}/L_{\textrm{Edd}}\right|_{q',\theta'}r^{q-q'}(2-q')\cos{(\theta)} (r_{\textrm{out}}^{2-q}-r_{\textrm{in}}^{2-q})}{(2-q)\cos{(\theta')}(r_{\textrm{out}}^{2-q'}-r_{\textrm{in}}^{2-q'})}\textrm{ ,}
\end{equation}
where $r = 6 \, r_\textrm{g}$, $r_{\textrm{in}} = 1 \, r_\textrm{g}$, $r_{\textrm{out}} = 400 \, r_\textrm{g}$, $q' = 3$ and $\theta' = 30^{\circ}$. It accounts for~the~difference in radial emissivity, hence similar ionization features are produced.

\begin{figure}[!htb]\centering
	\includegraphics[width=\textwidth]{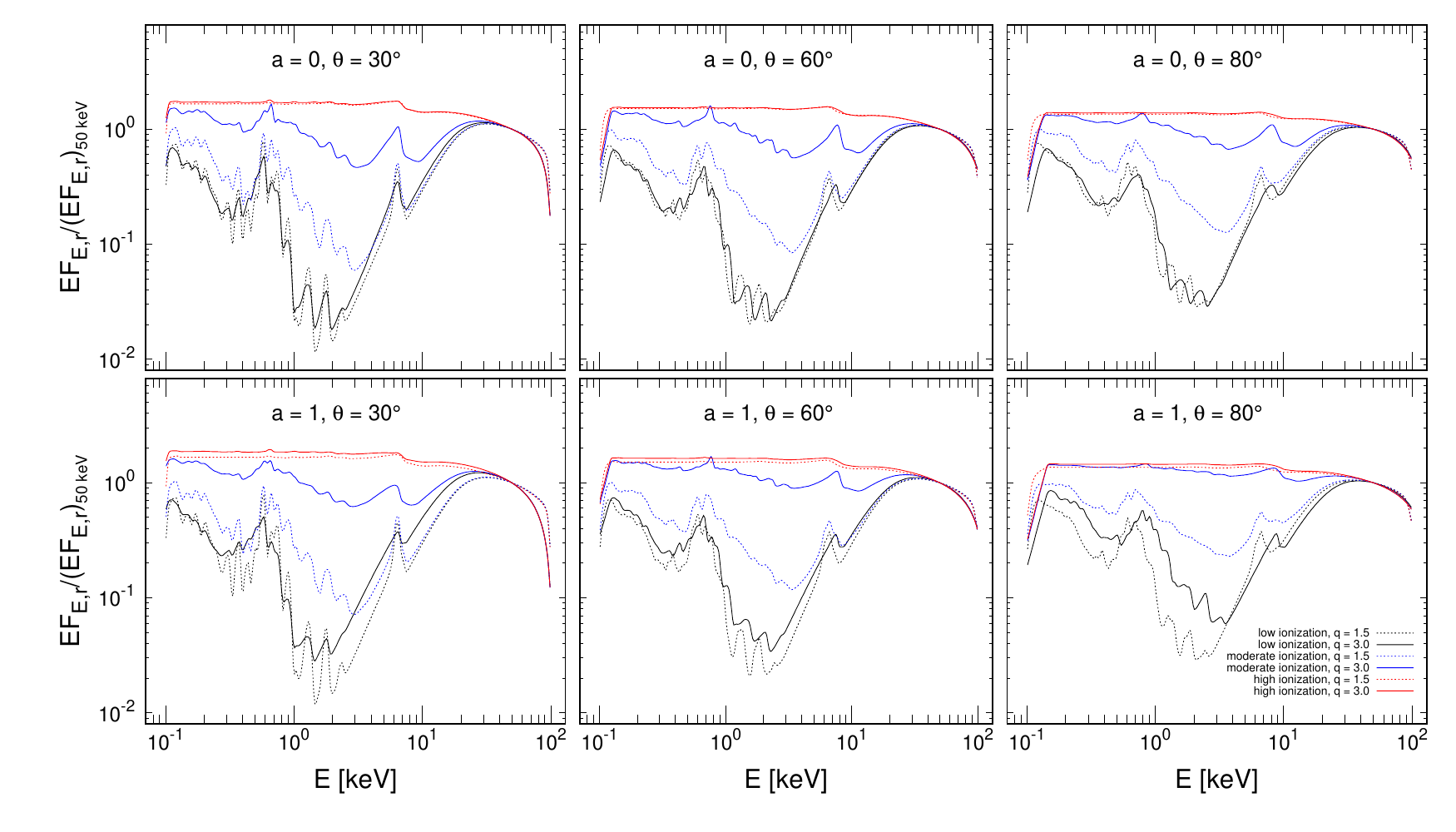}
	\caption{\footnotesize{The reflected-only spectra of the accretion disc for distant observer, $EF_\textrm{E,r}$, normalized to value at 50 keV, obtained by {\tt KYNSTOKES} for black-hole spins $a = 0$ (top) and $a = 1$ (bottom), disc inclinations $\theta = 30^{\circ}$ (left), $\theta = 60^{\circ}$ (midlle) and  $\theta = 80^{\circ}$ (right), $\Gamma = 2$, $M = 2\times 10^6\,M_{\odot}$, and unpolarized primary radiation, using the {\tt STOKES} local reflection model in the extended coronal scheme. We show cases of two different radial illumination power-law indices $q = 1.5$ (dotted lines) and $q = 3.0$ (solid lines), and neutral disc (observed 2--10 flux $L_{\textrm{X}}/L_{\textrm{Edd}} \approx 10^{-5}$, see text for details, black lines), moderately ionized disc (observed 2--10 flux $L_{\textrm{X}}/L_{\textrm{Edd}} \approx 10^{-2}$, see text for details, blue lines) and highly ionized disc (observed 2--10 flux $L_{\textrm{X}}/L_{\textrm{Edd}} \approx 10^{2}$, see text for details, red lines).}}
	\label{fig:K4_R_unpol_I.}
\end{figure}

\begin{figure}[!htb]\centering
	\includegraphics[width=\textwidth]{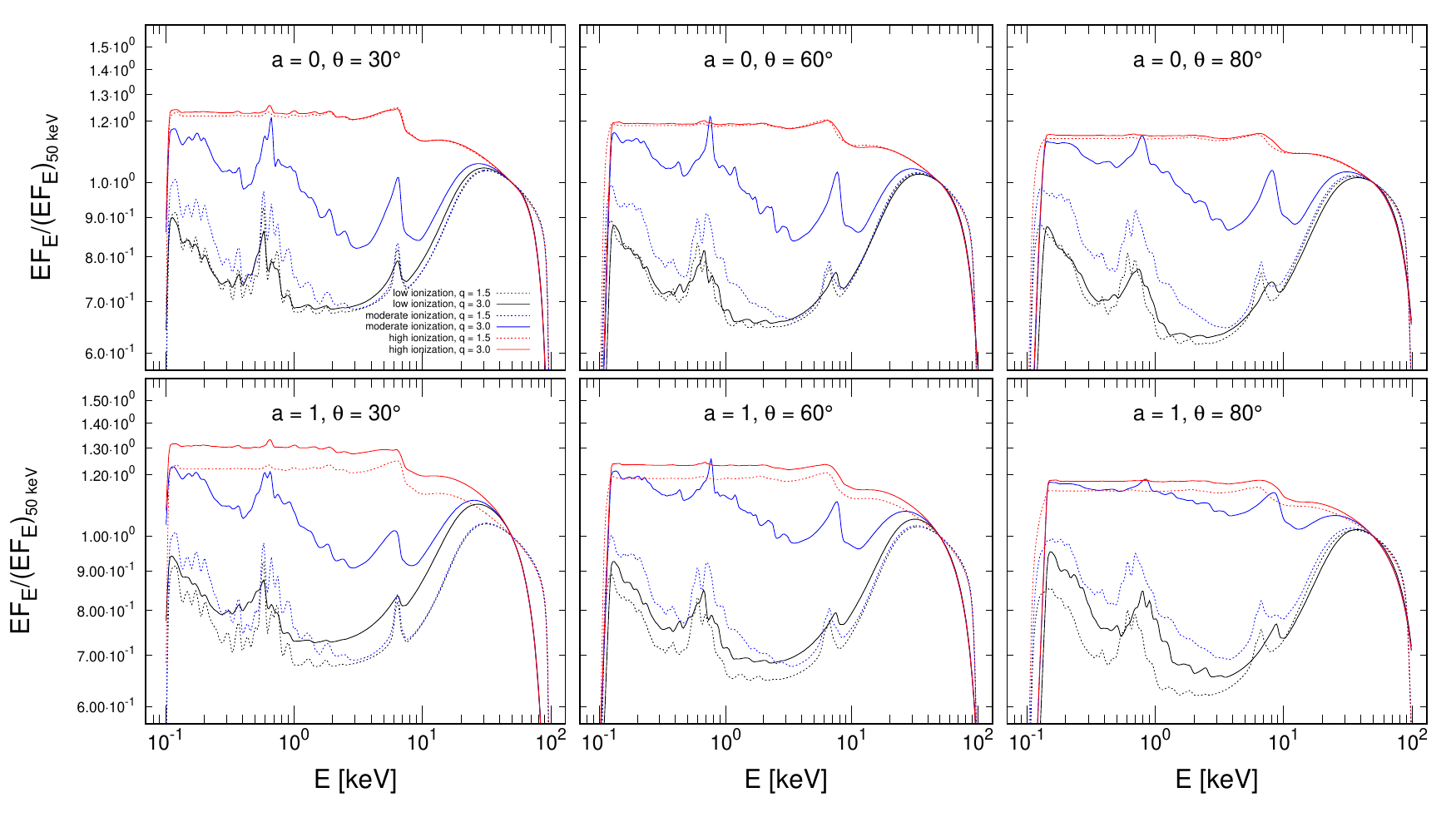}
	\caption{\footnotesize{The total spectra of the accretion disc for distant observer, $EF_\textrm{E}$, normalized to~value at 50 keV, for the same parametric setup as in Figure \ref{fig:K4_R_unpol_I.}, displayed in the same manner.}}
	\label{fig:K4_RP_unpol_I.}
\end{figure}

Figures \ref{fig:K4_R_unpol_I.} and \ref{fig:K4_RP_unpol_I.} show the reflected-only and total spectra, respectively, for the three cases of disc ionization, two values of $q$, two spins and three inclinations, all for unpolarized primary with $\Gamma = 2$. We again compensate for~the~primary power-law slope and renormalize to the value at 50 keV to see the spectral features clearly. Comparing to Figures \ref{fig:K3_R_unpol_I.} and \ref{fig:K3_RP_unpol_I.} for the lamp-post model, we see similar effects of relativistic integration across the disc. The radial emissivity index $q$ plays similar role for the sandwich corona as the height $h$ for lamp-post. The lower the height, the steeper irradiation profile we obtain for the lamp-post. At $r \gtrapprox 50 \, r_\textrm{g}$, the irradiation profile for lamp-post models approaches $q = 3$ for~any~$h$ \citep{Dovciak2014}. Only for high black-hole spin and high inclinations the relativistic effects on the primary radiation are significant enough that the~two geometries produce qualitatively different spectral profiles in our model.

~\

\begin{figure}[!htb]\centering
	\includegraphics[width=\textwidth]{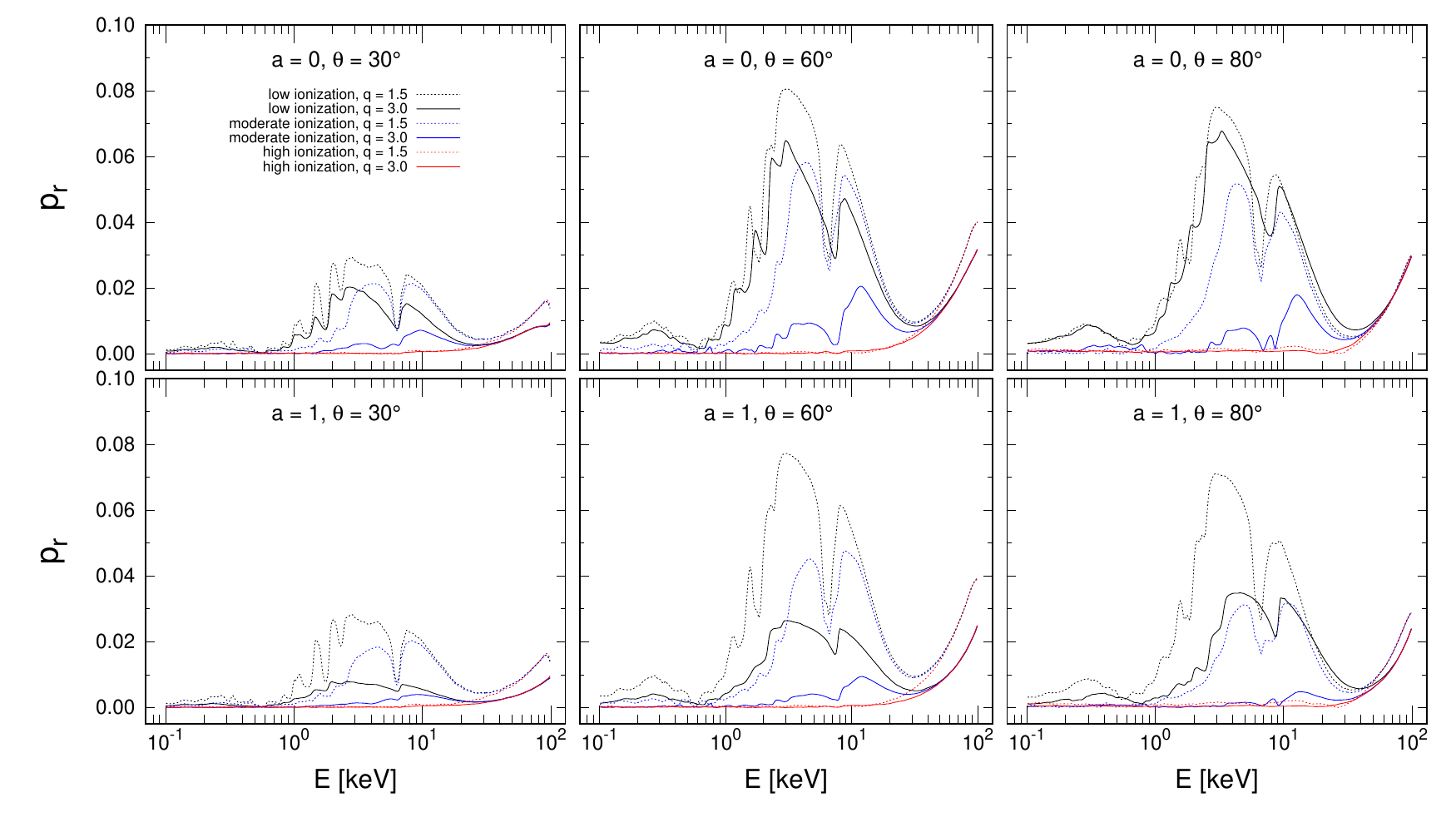}
	\caption{\footnotesize{The reflected-only polarisation degree, $p_\mathrm{r}$, versus energy for the same parametric setup as in Figure \ref{fig:K4_R_unpol_I.}, displayed in the same manner.}}
	\label{fig:K4_R_unpol_pdeg.}
\end{figure}

\begin{figure}[!htb]\centering
	\includegraphics[width=\textwidth]{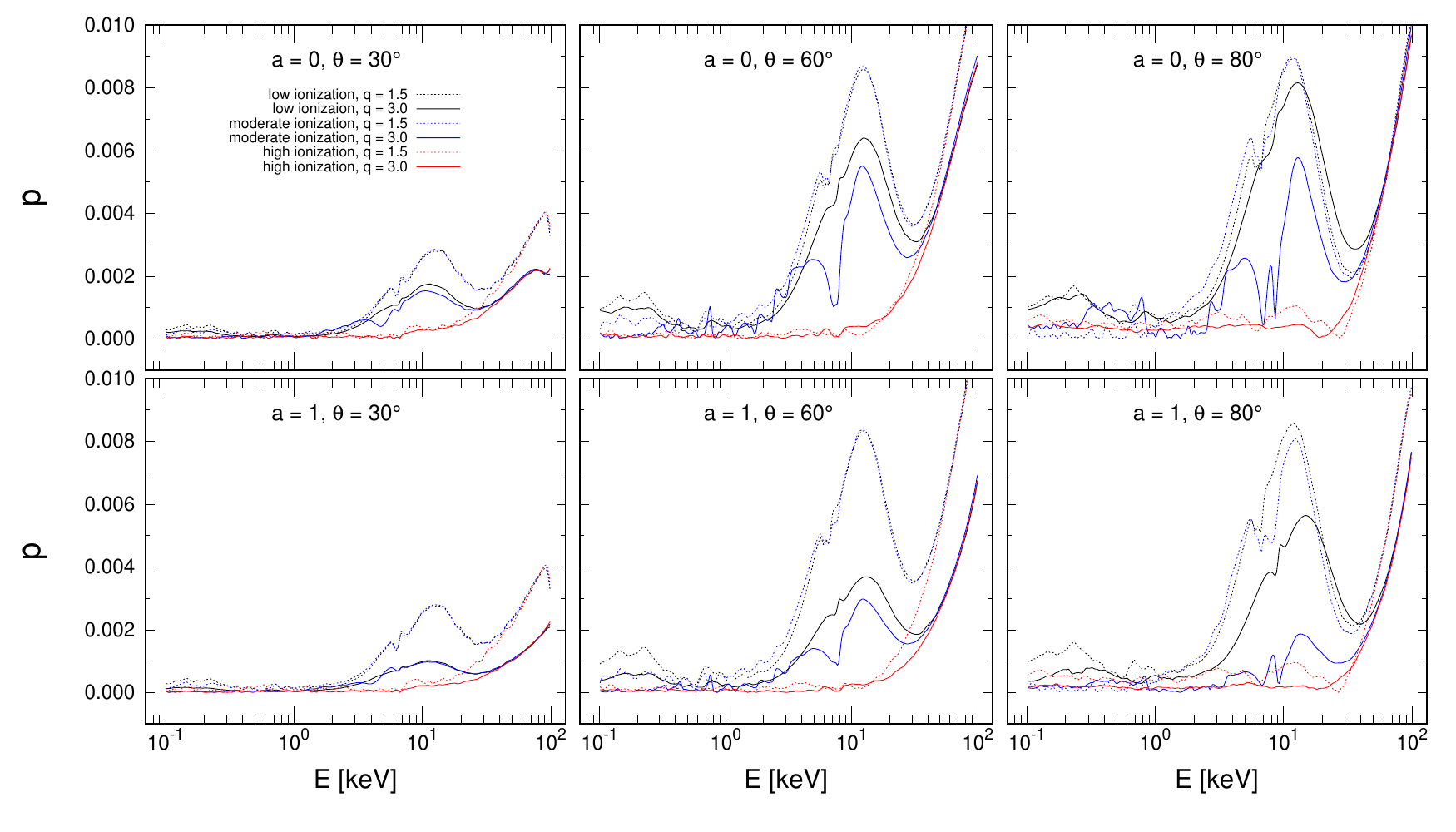}
	\caption{\footnotesize{The total polarisation degree, $p$, versus energy for the same parametric setup as in Figure \ref{fig:K4_R_unpol_I.}, displayed in the same manner.}}
	\label{fig:K4_RP_unpol_pdeg.}
\end{figure}

Polarisation signatures, on the contrary, vary significantly. Figures \ref{fig:K4_R_unpol_pdeg.} and~\ref{fig:K4_RP_unpol_pdeg.} provide the corresponding reflected-only and total polarisation profile with energy, respectively. We obtain almost full depolarisation in the soft X-rays. In~the 1--100 keV range, the reflected-only results follow the dependency of poalrization with energy of the local reflection tables (cf. Figure \ref{local_not_in_mue}), preserving the~dip of reduced polarisation in the iron line complex at 6--7 keV. The~absorption-induced peak in reflection that forms in 1--20 keV range is skewed towards higher energies when the primary radiation is added, because this coronal emission is unpolarized and with decreasing flux contribution with energy dilutes polarisation more at lower energies. If different $\Gamma$ is assumed, the reflected polarisation degree is diluted by the energy dependence of the unpolarized primary accordingly.

The maximum amplitude of reflected-only polarisation in the X-rays is only $\approx 8\%$ in the most favorable configurations, while the majority of the parameter space ends up below $\approx 2\%$. We note that the studied sandwich model is extremely simplistic and we would expect yet another reflected-only and total polarisation result, if a different coronal model was assumed. More sophisticated coronal studies \citep{Ursini2022, Krawczynski2022a, Krawczynski2022, Gianolli2023, Tagliacozzo2023} typically result in~higher polarisation of direct coronal emission in the slab models compared to~lamp-posts of non-negligible size, because with respect to the disc seed photons and the~observer, the slab model represents the more asymmetric situation. For~reflection of coronal radiation off the disc to the observer, the lamp-post scenario is in turn the more asymmetric variant for a distant observer.

~\

\begin{figure}[!htb]\centering
	\includegraphics[width=\textwidth]{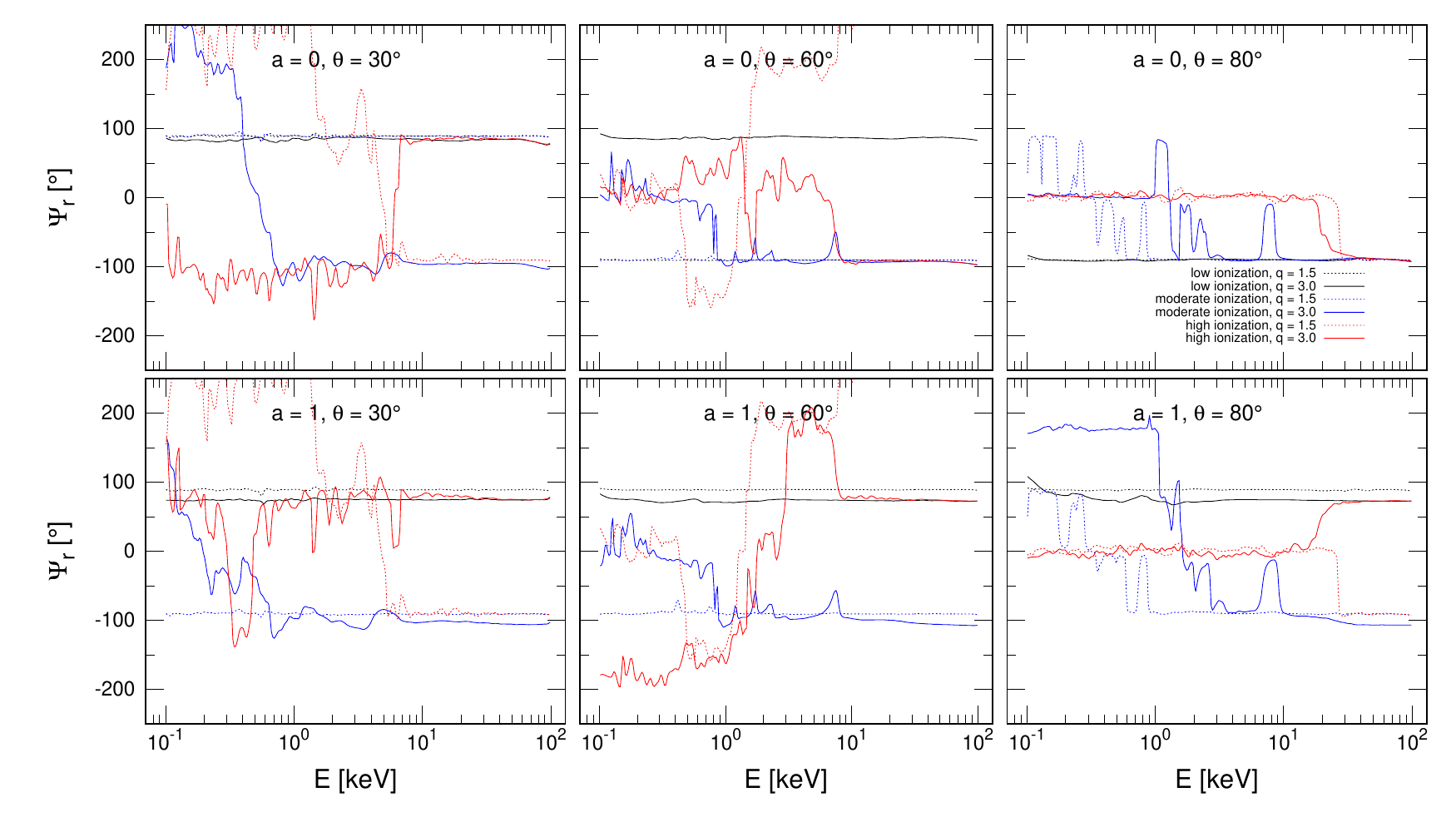}
	\caption{\footnotesize{The reflected-only polarisation angle, $\Psi_\mathrm{r}$, versus energy for the same parametric setup as in Figure \ref{fig:K4_R_unpol_I.}, displayed in the same manner.}}
	\label{fig:K4_R_unpol_pang.}
\end{figure}

Figure \ref{fig:K4_R_unpol_pang.} shows the corresponding polarisation angle with energy. Since we study the unpolarized primary case, the plot is the same with the primary radiation or without. For the sandwich model, we observe typically a dominant polarisation perpendicular to the axis of symmetry \textit{in reflection}. The reason for~this is the isotropic illumination of the disc, which locally produces polarisation rather parallel with the disc (see Sections \ref{modelling} and \ref{reflection_tables_TS}). Only for~the~highest inclinations the prevailing polarisation position angle for an observer at spatial infinity is rather parallel with the axis of symmetry for soft and mid X-rays and~for moderate or highly ionized cases. The value where the switch in~resulting polarisation occurs is generally also dependent on $\Gamma$. For the direct coronal emission, we expect, however, quite the opposite than from reflection off the disc: the lamp-post coronae typically produce polarisation perpendicular to the rotation axis, while the sandwich models typically produce polarisation parallel to the principal axis \citep{Ursini2022, Krawczynski2022a, Krawczynski2022, Gianolli2023, Tagliacozzo2023}.

~\

We studied the dependency on changing primary polarisation again for~the~cases of $1\%$ polarized primary with either parallel or perpendicular polarisation, juxtaposed to the unpolarized primary case. This is shown in Figures \ref{fig:K4A_R_polar_i_pdeg.}--\ref{fig:K4A_RP_polar_n_pang.}, for the reflected-only and total polarisation degree and~angle and~for~the~neutral and ionized disc. The effects of changing primary polarisation on spectra are again negligible. In most cases, we obtain the highest reflection-induced polarisation for~the~originally horizontally polarized primary. However, the role of the three cases in the net amplitude of reflected polarisation changes with energy. The added primary radiation does not always polarize or depolarize in the prescribed way at the equatorial plane, because each emission region is (i) weighted differently due to radial emissivity profile, (ii) weighted differently due to the relativistic transfer function, (iii) obtaining a different poalrization angle change from the original value in the equatorial plane due to GR effects between the disc and the observer, which results in net depolarisation for a distant observer.

~\

We briefly examined also the dependency on disc truncation by extending the~outer disc radius to $r_{\mathrm{out}} = 1000 \, r_\textrm{g}$ and plotting the polarisation quantities and $R_\textrm{f}$ with respect to $r_\textrm{in}$ similarly to the lamp-post results in Section \ref{lamppost_spproperties}. Figure \ref{rin_ionized_ext_q3} shows a representative case of $a = 1$, $q = 3$ and highly ionized disc reflection for three different inclinations. For the extended corona, the results depend strongly on the chosen primary emission model. For our choice, we do not obtain almost any change of $R_\textrm{f}$ with $r_\textrm{in}$. The polarisation fraction only slightly increases with $r_\textrm{in}$. The resulting polarisation angle is slightly rotated, depending on the inner disc extension below $10 \, r_\textrm{g}$ due to the impact of gravity on the polarisation angle along the photon's trajectory.

\newpage
\ 
\vspace{4.5cm} 
\begin{figure}[h]
        \centering
	\includegraphics[trim={0cm 0cm 0cm 0cm},clip,width=0.6\textwidth]{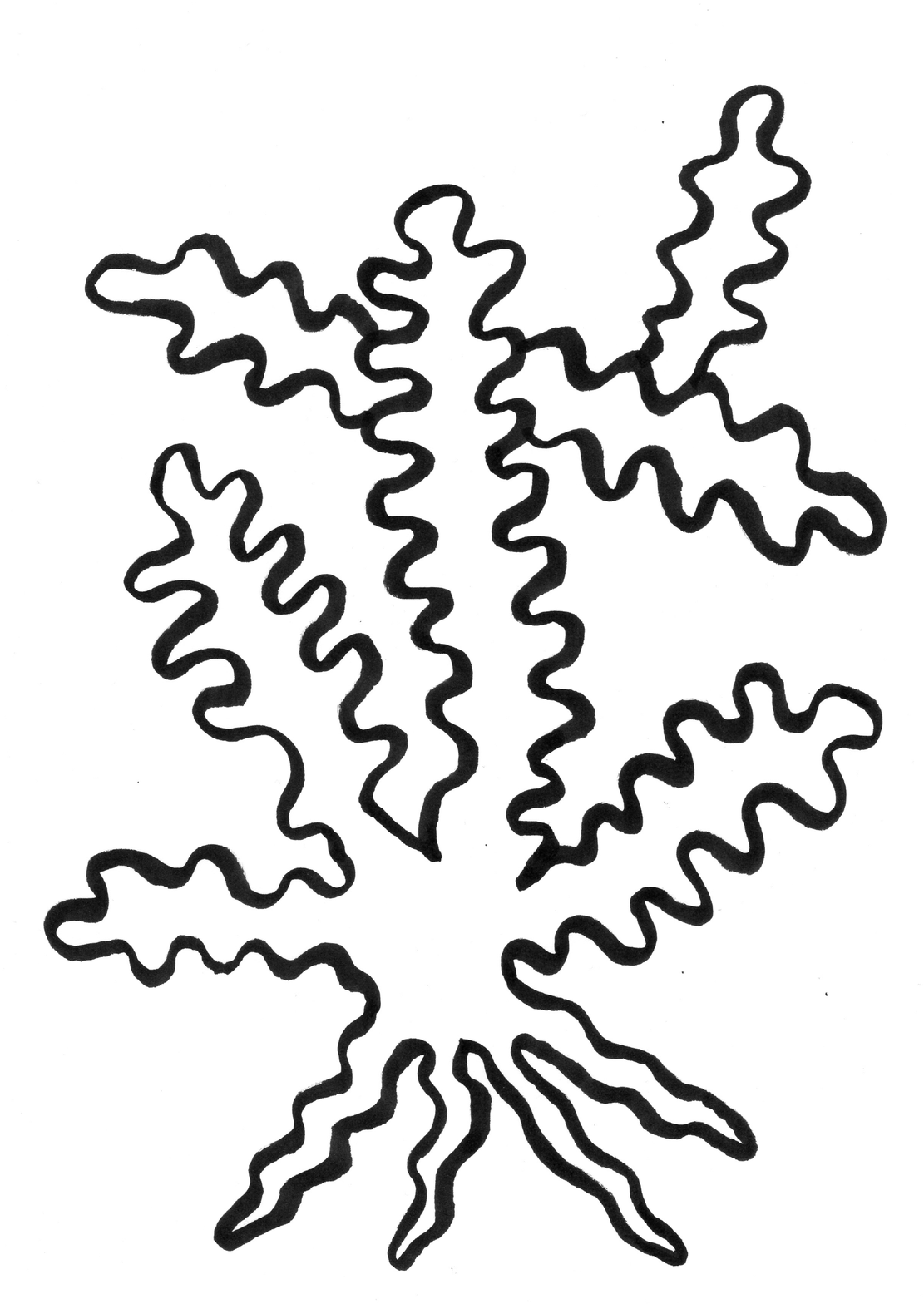}
\end{figure}
\chapter{Contribution of distant components of accreting systems}\label{chap04}

\begin{figure}[h]
	\centering
	\begin{minipage}[t]{0.49\textwidth}
		\includegraphics[width=\textwidth]{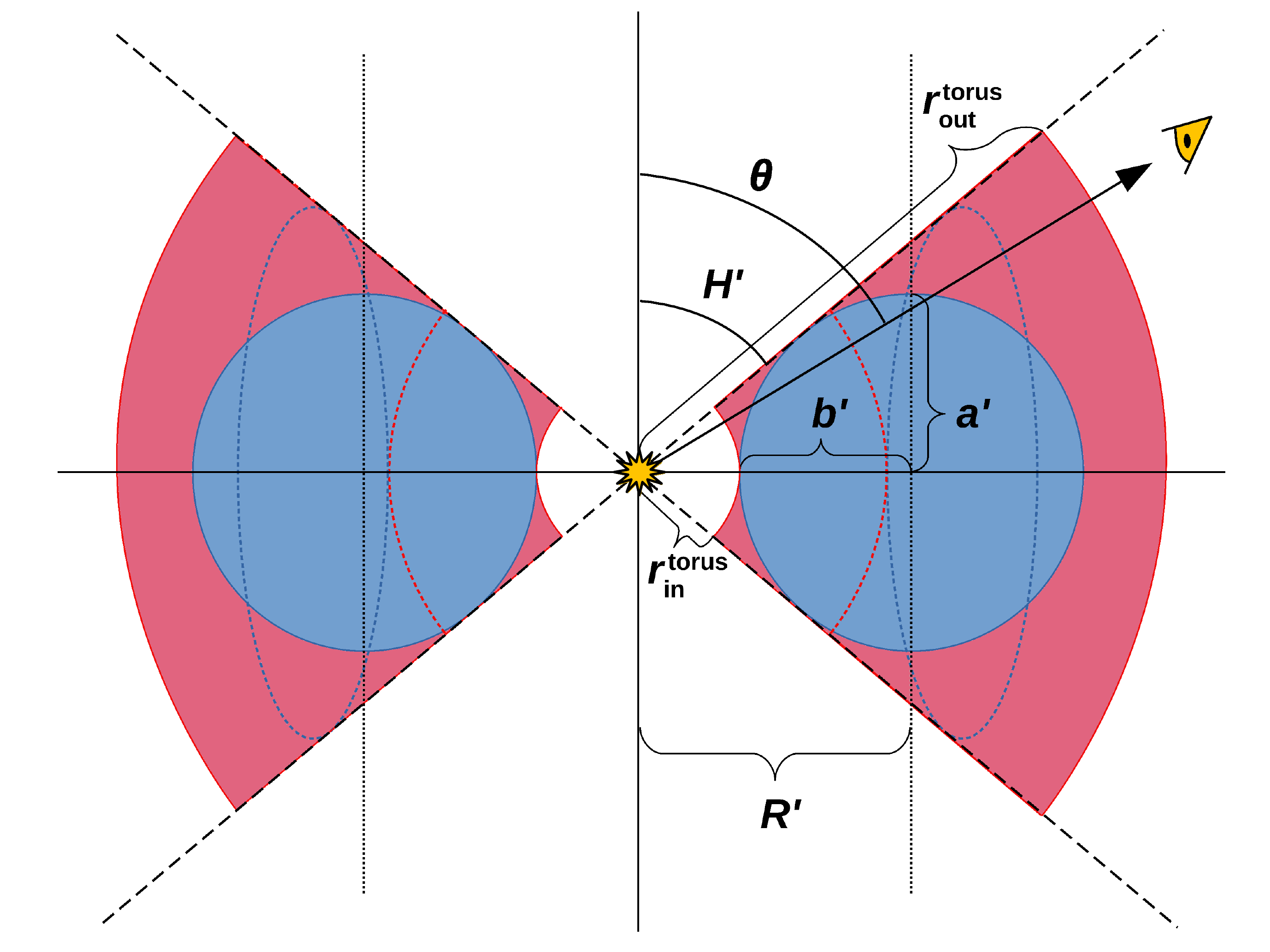}
		\caption{\footnotesize{Wedge-shaped, circular and elliptical toroidal structures, equatorially covering the central source of isotropic power-law emission. These homogeneous and static distant cold media are rotationally symmetric around the principal axis and mirror symmetric with respect to the equatorial plane. Image adapted from \cite{Podgorny2023c}.}}
		\label{equatorial}
	\end{minipage}
	\hfill
	\begin{minipage}[t]{0.49\textwidth}
		\includegraphics[width=\textwidth]{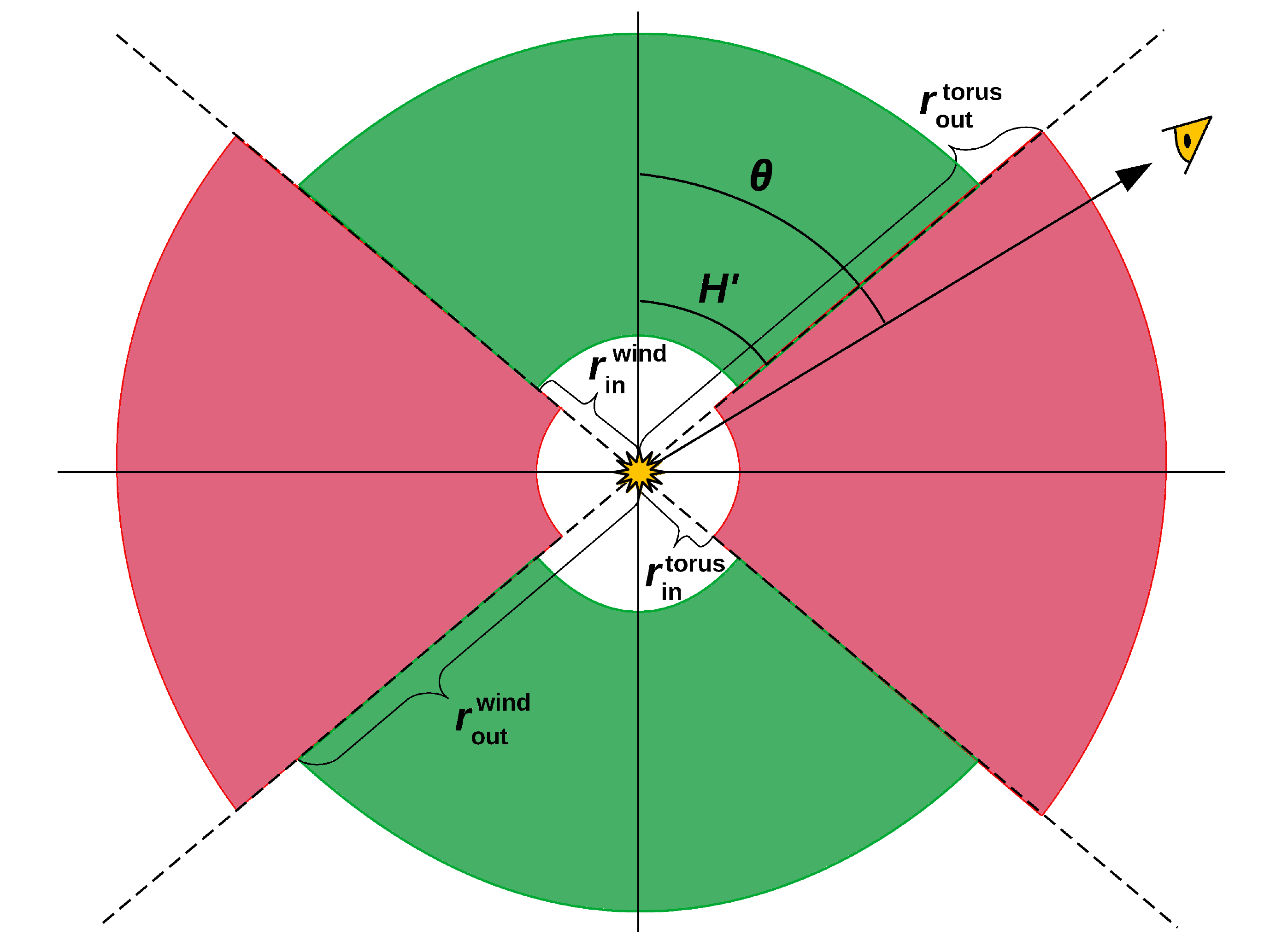}
		\caption{\footnotesize{The complete AGN model assuming polar winds in rotationally symmetric wedge-like cones around the principal axis. Wedge-like equatorial tori are kept glued to~the~polar scatterers. We assume a~disc-corona source of emission in the lamp-post geometry located in the center. Image adapted from \cite{Podgorny2023d}.}}
		\label{total_model}
	\end{minipage}
\end{figure}

In this chapter, we continue the trend of zooming out in physical size and we model the X-ray polarisation properties of distant circumnuclear components of~accreting compact objects. Wedge-like and toroidal shapes will be assumed for studies of equatorial reprocessing first, which are parametrized in Figure \ref{equatorial}. Wedge-like equatorial scatterers combined with conical polar outflows will also be studied at the end of this chapter, as in Figure \ref{total_model}. The details of the modeling will be given as part of the results below.

\section[Reprocessing in equatorial obscurers with Monte Carlo \newline simulations]{Reprocessing in equatorial obscurers \newline with Monte Carlo simulations}\label{equatorial_MC}

We begin with studying reprocessing in regions residing symmetrically around and near the equatorial plane. As suggested in Chapter \ref{introduction}, these can be either the~parsec-scale dusty tori of AGN, or more ionized outflows in funnel-like structures, originating from the outer accretion discs of stellar-mass compact objects. The~results presented in this section were produced with {\tt STOKES} \citep[\textit{v2.07}, identical with the version used for production of results in][]{Marin2018c, Marin2018b} and~published in \cite{Podgorny2023c}, only with a different notation. The~simplification with respect to the {\tt STOKES} results presented in Chapter \ref{chap01} is that partial ionization is treated only by means of uniform free-electron density. The free electrons are added on top of uniformly distributed atomic species with $A_\textrm{Fe} = 1$ and solar abundances from \cite{Asplund2005}. No clumpiness or ionization fronts are taken into account and no radiative transfer code is linked to the MC simulation to solve for true ionization structure. Hence the reflected continuum spectra show a typical $\sim -E^{-3}$ decay towards the soft X-rays, owing to photoelectric absorption in~cold media mixed with electrons of diverse origin (owing to e.g. streams following magnetic fields). With respect to a proper treatment of ionization we preserve in reflection the total flux amplification by means of scattering on free and bound electrons, but regarding energy dependence of the reflected spectra, we introduce artificially more absorption at soft X-rays. The same holds for transmission through a patch of matter: scattering effects follow the density of free electrons, but the energy dependence of absorption is biased towards the~soft X-rays. In any case, this leads to artificially enhanced polarisation towards the~soft X-rays. Comptonization, synchrotron and thermal radiation effects are neglected too, which is more justifiable for~the~distant regions than for~the~studies of~the~inner disc-corona system. We show MC results obtained only with {\tt STOKES} and no other code is used, in order to discuss the main principles and diversity of possible results.

We assume a simple isotropic static source of arbitrarily polarized power-law radiation in the center of coordinates, representing the disc-corona region shrunk to negligible size. In order to examine the effects of changing primary irradiation, we simulated unpolarized and $2\%$ polarized primary parallelly or perpendicularly to the axis of symmetry for a range of photon indices: \mbox{$1.2 \leq \Gamma \leq 2.8$}. We preserve axial symmetry, mirror symmetry with respect to the equatorial plane, and~homogeneous density along the scattering region, as it is already nearly impossible to cover such reduced parameter space in full scope within one discussion section. The scattering regions neither possess any bulk motion with respect to~the~source of radiation. No polar scattering regions are included for~results presented in~this section. The photons can change direction, or be truly absorbed anywhere in~the~equatorial scattering region. They can also escape the~reprocessing material and encounter it again at a different surface location. Thus, the~model takes into account the~effects of self-irradiation.

At first, we plan to examine the interaction of light and matter in the circumnuclear components theoretically. Thus, we forbid the registration of direct radiation for the following plots, in order to understand the main driving mechanisms inside the system for reflection-induced polarisation. This is important to~mention especially with respect to the type-1 viewing angles, where we aim to~examine the effects of only the reprocessed light. We will add direct radiation to these simulations in Chapter \ref{chap05} where we will apply the equatorial reprocessing MC model to the interpretation of {\it IXPE} observations of Cygnus X-3, an XRB caught in hard and intermediate spectral states. This will allow us to evaluate the~impact of direct radiation more clearly with respect to this section. It will also allow us to discuss the results from a different perspective and~to~assess the~model sensitivity to a few other assumptions.\footnote{\,Such as the impact of changing chemical composition by means of removal of neutral hydrogen. The winds in Cygnus X-3 are severely hydrogen-depleted due to the presence of~a~Wolf-Rayet companion star in a late stellar evolution phase \citep{Fender1999, Kallman2019}.}

Distinct geometrical variants of the equatorial scattering region are considered, as is shown in Figure \ref{equatorial}.  We study
\begin{itemize}
    \item Case A: a wedge-shaped torus defined by $r_\textrm{in}^\textrm{torus}$, $r_\textrm{out}^\textrm{torus}$ and its half-opening angle $H'$;
    \item Case B: a true geometrical torus with an elliptical profile given by the~parameters $R'$, $a'$ and $b'$;
    \item Case C: a subcase of Case B representing a torus with circular profile, i.e.~when $a' = b'$.
\end{itemize}
When there is a comparison between the wedge-shaped and circular or elliptical tori, we link $r_\textrm{in}^\textrm{torus}$ of the wedge to the $r_\textrm{in}^\textrm{torus} = R'-b'$ inner-most equatorial circle of the circular or elliptical tori, to examine the changing convexity or concavity effects of the inner toroidal wall located in the same place in equatorial plane. For Case B, we set $b'$ relative to $R'$ and study various eccentricities from prolate to oblate tori: $b' = R'/8$, $b' = R'/4$, $b' = R'/2$, $b' = 3R'/4$, where $R' = b' + r_\textrm{in}^\textrm{torus}$ for a fixed $r_\textrm{in}^\textrm{torus}$. Therefore, by changing the eccentricity, we also trace the effects of changing curvature of the inner reflecting surface. For a given $H'$ measured from the pole, we then define for Case B the parameter $a'$ self-consistently via
\begin{equation}\label{size_trafo}
    a' = \tan\left( \frac{\pi}{2} - H'\right) \sqrt{R'^2-b'^2} \textrm{ .}
\end{equation}
Case C is denoted as a special case of the elliptical torus, because it will be used frequently in the follow-up section as a referential MC case study for comparisons with a different model. In addition, for Case C we also compute $a' = b'$ in~a~different way than for the Case B. For Case C
\begin{equation}\label{opening_angle}
    a' = b' = r_\textrm{in}^\textrm{torus}\frac{\cos{H'}}{1-\cos{H'}} \textrm{ ,}
\end{equation}
where again $r_\textrm{in}^\textrm{torus}$ is given. In this way, for a fixed $H'$, when comparing Case C to Cases B with any $b'/R'$ ratio, the torus radii $R'$ are different, but the inner boundary remains the same between the two cases. Case B would equal Case C, if $b'/R'=\cos{H'}$. We always directly compare the geometrical shapes with the~same half-opening angle $H'$ and observer's inclination $\theta$.

The default parameter values are physically and observationally motivated \citep[see][and references therein]{Goosmann2007, Marin2018c, Marin2018b}. We choose $r_\textrm{in}^\textrm{torus} = 0.05$ in this section, representing a fraction of a parsec for~an~AGN torus inner boundary, or an outer edge of an accretion disc around supermassive black hole. Nevertheless, the units are arbitrary and the results can be used for~orders-of-magnitude lighter objects, as for polarisation genesis in this approach only the relative sizes inside the system matter. For Case A we set \mbox{$r_\textrm{out}^\textrm{torus} = 10$,} representing the outermost AGN torus boundary in parsecs. The relative distance of the outer edge of the reprocessing region to the center with respect to~characteristic sizes plays, however, little role compared to the inner-boundary location and shape, where the photons exhibit their first scatterings, which are determinative for the total geometry. This holds as long as the column densities of~the~entire scattering and absorption region are kept constant and high enough.

We refer to a ``type-1'' observer when $\theta < H'$. We refer to a ``type-2'' observer when $\theta > H'$. We will examine the output in 20 emission inclination bins linearly spaced in $0 \leq \cos{\theta} \leq 1$ and one averaged azimuthal bin. Due to high expected absorption at soft X-rays, we consider only 200 logarithmically spaced energy bins between 1 and 100 keV. We do not consider any relativistic effects, so the result is either parallelly or perpendicularly polarized with respect to the principal axis due to meridional mirror symmetry from the observer's point of view. Hence, we will adopt the notation of positive or negative polarisation degree, if the polarisation angle is parallel or perpendicular to the central axis of symmetry, respectively. 

We display the results with respect to equatorial column density of neutral hydrogen $N_\textrm{H}^\textrm{eq} = n_\textrm{H}L$, which is given by the uniform neutral hydrogen density $n_\textrm{H}$ and $L = r_\textrm{out}^\textrm{torus} - r_\textrm{in}^\textrm{torus}$ for Case A and $L = 2b'$ for Cases B and C. We also define the equatorial Thomson optical depth $\tau_\textrm{e}^\textrm{eq} = \sigma_\textrm{T}n_\textrm{e}L$, where $n_\textrm{e}$ is the uniform density of free electrons in the scattering region. We work with three cases of partial ionization, referring to them as ``low'', ``moderate'' and ``high'' ionization level ($\tau_\textrm{e}^\textrm{eq} = 0.07, 0.7, 7$, respectively, for $N_\textrm{H}^\textrm{eq} = 10^{25}  \textrm{ cm}^{-2}$ and $\tau_\textrm{e}^\textrm{eq} = 0.007, 0.07, 0.7$, respectively, for $N_\textrm{H}^\textrm{eq} = 10^{24}  \textrm{ cm}^{-2}$). In this way, we simplistically study a range of ionization levels from nearly fully ionized medium to nearly neutral medium, corrected for density change of the atomic species. 

~\

\begin{figure}[!htb]\centering
	\includegraphics[width=\textwidth]{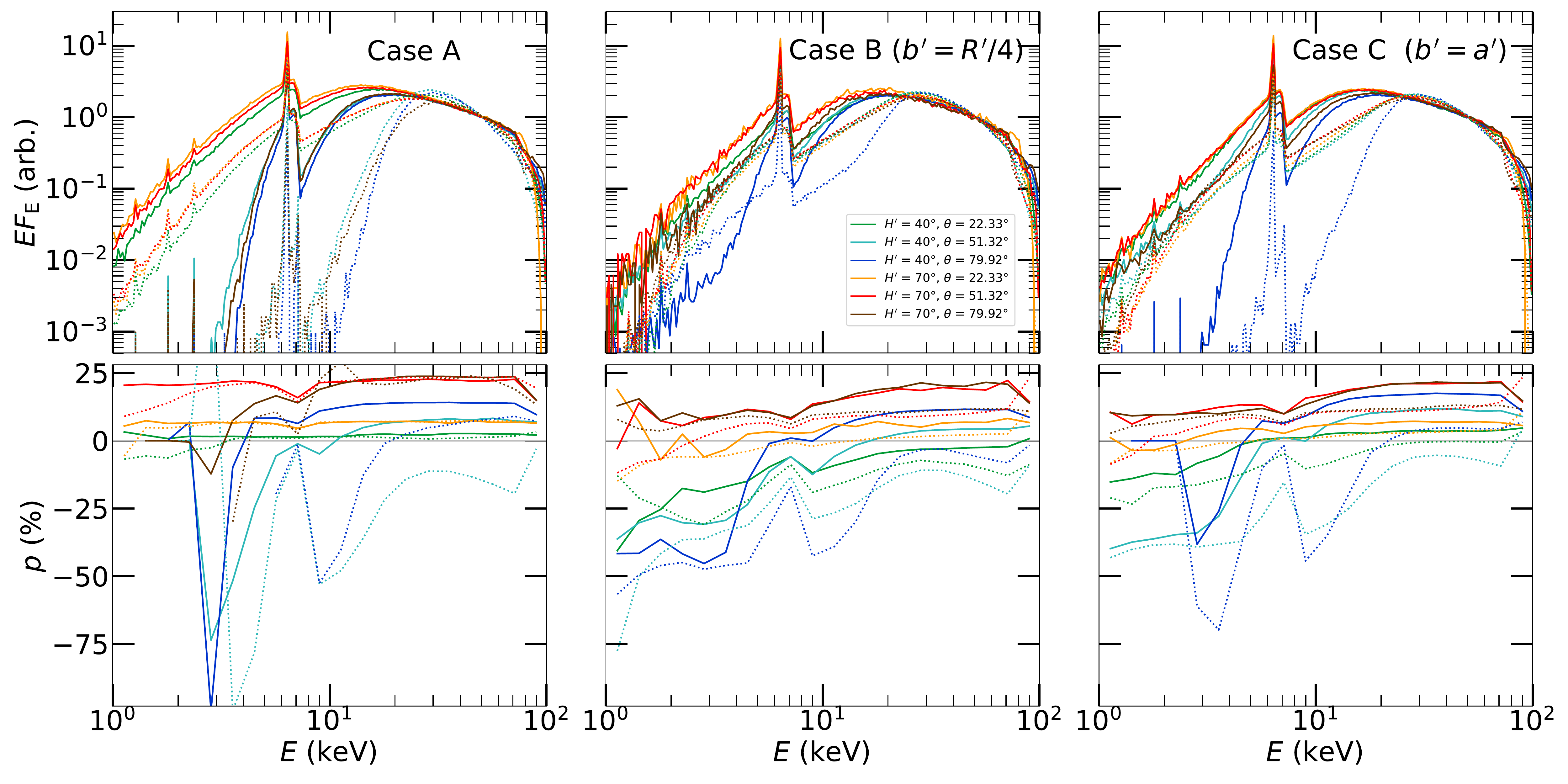}
	\caption{\footnotesize{Top: the spectra, $EF_\mathrm{E}$, normalized to value at 50 keV. Bottom: the corresponding polarisation degree, $p$, versus energy. We show the Cases A, B (for $b' = R'/4$) and C from left to right. Various inclinations $\theta$ and half-opening angles $H'$ for both type-1 and type-2 observers are shown in the color code. We show the torus column densities $N_\textrm{H}^\textrm{eq} = 10^{24}  \textrm{ cm}^{-2}$ and $\tau_\textrm{e}^\textrm{eq} = 0.7$ (solid lines) and $N_\textrm{H}^\textrm{eq} = 10^{25}  \textrm{ cm}^{-2}$ and $\tau_\textrm{e}^\textrm{eq} = 7$ (dotted lines), i.e. the highly ionized cases. The remaining parameters are $r_\textrm{in}^\textrm{torus} = 0.05$ (Case A), $r_\textrm{out}^\textrm{torus} = 10$ (Case A) and~the~primary input was set for $\Gamma = 2$ and $p_0 = 2\%$ for all displayed cases. The detailed parametric dependencies are shown in following figures when integrated in energy. Image adapted from \cite{Podgorny2023c}.}}
	\label{energy_dependent}
\end{figure}

We first show the results for $2\%$ parallelly polarized primary and $\Gamma = 2$. We will only complicate the discussion with changing primary radiation properties at~a~later stage, similarly to the previous chapter. Figure \ref{energy_dependent} shows the reprocessed spectra and corresponding polarisation state versus energy (binned in 10 neighboring energy bins for better photon statistics) for a selection of type-1 and~type-2 viewing angles, given by combinations of $\theta$ and $H'$. We display different geometrical archetypes of the reprocessing region and two different densities of~the~atomic species, assuming the highly ionized case of free electron density.

We obtain depolarisation in fluorescent spectral lines, although in the presented MC model we especially do not examine the spectral lines in detail compared e.g. to \cite{Dorodnitsyn2010, Dorodnitsyn2011} that studied the AGN warm absorbers, where surprisingly high X-ray polarisation in spectral lines was obtained due to resonant line scatterings. Moreover, in advanced geometries and with self-irradiation, compared to the previously studied slab-like problems, the~level of polarisation in emission lines is more dependent on subsequent reprocessings conditions of the line-emitted photons.

The obtained diversity in polarisation and the trends in polarisation angle switch with energy and other parameters can be explained in the following way. Although absorption in slab-like problems typically enhances polarisation, for~a~three dimensional toroidal structure this may not always be the~case. Relative flux contributions from different regions of the reflecting object need to~be considered, including the dominant polarisation angle obtained from each part. Depending on the ionization level, we see polarized reflection from the~inner side of the funnel, the flux contribution from each section is inclination and geometry dependent and the reprocessing occurs in various volumes, which depends on~the~uniform density of the medium in our model. In scattering-dominated steady-state media, the large-scale material elongation and orientation with respect to the source and with respect to the observer determines the dominant scattering plane, hence orientation of the resulting polarisation \citep[e.g.][]{Kartje1995, Goosmann2007, Schnittman2010, Krawczynski2022}. Here we obtain perpendicular net polarisation, if the reflection from the~inner side, which is the~furthest (or the closest) to the observer prevails. If we obtain the~highest polarized flux contribution from the left and right sides of~the~torus from observer's point of view, resulting polarisation parallel with the axis of symmetry occurs, such as in equatorially elongated coronae near accretion discs. Then higher overall flux absorption may cause actually decrease of~polarisation, typically in the hard X-rays, because essentially two orthogonally polarized components compete with each other and together produce the total polarisation state. Figure \ref{energy_dependent} shows rather perpendicularly polarized emission in~the~highly absorbed, softer X-rays, and stronger parallelly polarized component rather in~the~Compton reflected hard X-rays. The details are, however, much geometry dependent.

~\

\begin{figure}[!htb]\centering
	\includegraphics[width=\textwidth]{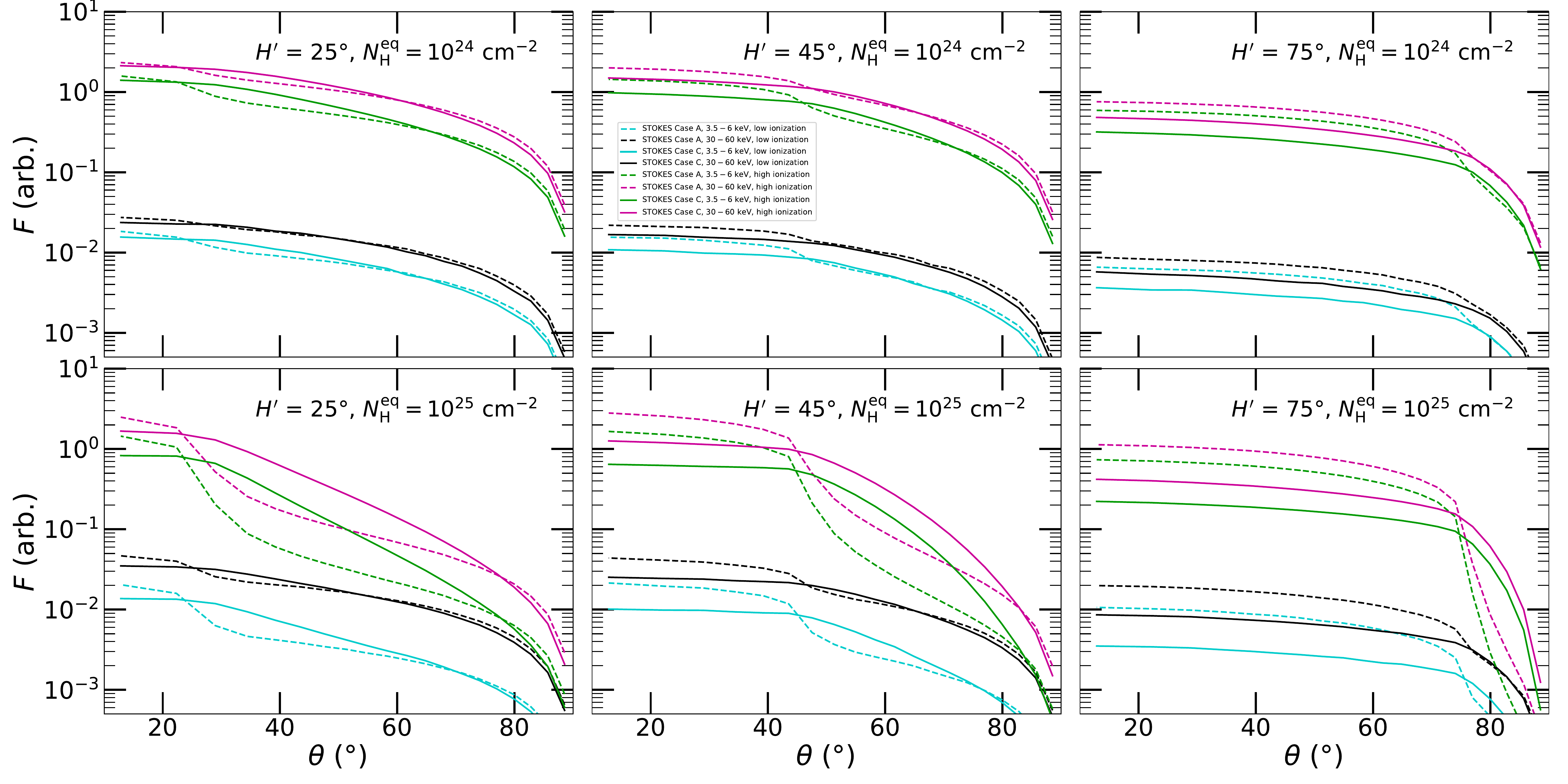}
	\caption{\footnotesize{Case A (dashed lines) compared to Case C (solid lines): we show the energy-integrated flux, $F$, versus inclination, $\theta$, for $H' = 25\degr$ (left), $H' = 45\degr$ (middle), and $H' = 75\degr$ (right) and for $N_\textrm{H}^\textrm{eq} = 10^{24}  \textrm{ cm}^{-2}$ (top) and $N_\textrm{H}^\textrm{eq} = 10^{25}  \textrm{ cm}^{-2}$ (bottom). We show the results for \textit{low} level of ionization, i.e. $\tau_\textrm{e}^\textrm{eq} = 0.007$ for \mbox{$N_\textrm{H}^\textrm{eq} = 10^{24}  \textrm{ cm}^{-2}$} and $\tau_\textrm{e}^\textrm{eq} = 0.07$ for~\mbox{$N_\textrm{H}^\textrm{eq} = 10^{25}  \textrm{ cm}^{-2}$,} integrated in 3.5--6 keV (blue) and 30--60 keV (black), as well as for~\textit{high} level of ionization, i.e. $\tau_\textrm{e}^\textrm{eq} = 0.7$ for $N_\textrm{H}^\textrm{eq} = 10^{24}  \textrm{ cm}^{-2}$ and $\tau_\textrm{e}^\textrm{eq} = 7$ for $N_\textrm{H}^\textrm{eq} = 10^{25}  \textrm{ cm}^{-2}$, integrated in 3.5--6 keV (green) and 30--60 keV (magenta). The remaining parameters are $r_\textrm{in}^\textrm{torus} = 0.05$ (Case A), $r_\textrm{out}^\textrm{torus} = 10$ (Case A) and the primary input was set to $\Gamma = 2$ and~$p_0 = 2\%$ for all displayed cases. Image adapted from \cite{Podgorny2023c}.}}
	\label{AvsC_Imue}
\end{figure}

In order to discuss the spectral continuum without any spectral line contamination, we integrate the collected photons in 3.5--6 keV and 30--60 keV in~the~following. If partial ionization was properly precomputed, we expect it would affect the 3.5--6 keV band only, causing less absorption towards the soft X-rays. This means higher flux and reduced polarisation fraction (for any sign, as both cases of~net polarisation orientations are a result of competing scattering geometries only). Figure \ref{AvsC_Imue} shows the relative flux drop with inclination for Case A versus Case C for various half-opening angles, equatorial column densities and for~the~low and high ionization examples. For the wedge-shaped torus, the line-of-sight column density with respect to the source is a step function of inclination, i.e. from zero to the same value as the equatorial column density below the grazing inclination angle. For elliptical tori, the torus becomes gradually opaque in the line of sight when viewed more and more inclined. We see this change of received flux with inclination below the~grazing angle for the two geometrical cases, more prominently for higher densities and regardless of the energy band or ionization. Higher ionization causes higher reflectivity of the torus, hence higher total received flux for the same irradiating number of photons.

\begin{figure}[!htb]\centering
	\includegraphics[width=\textwidth]{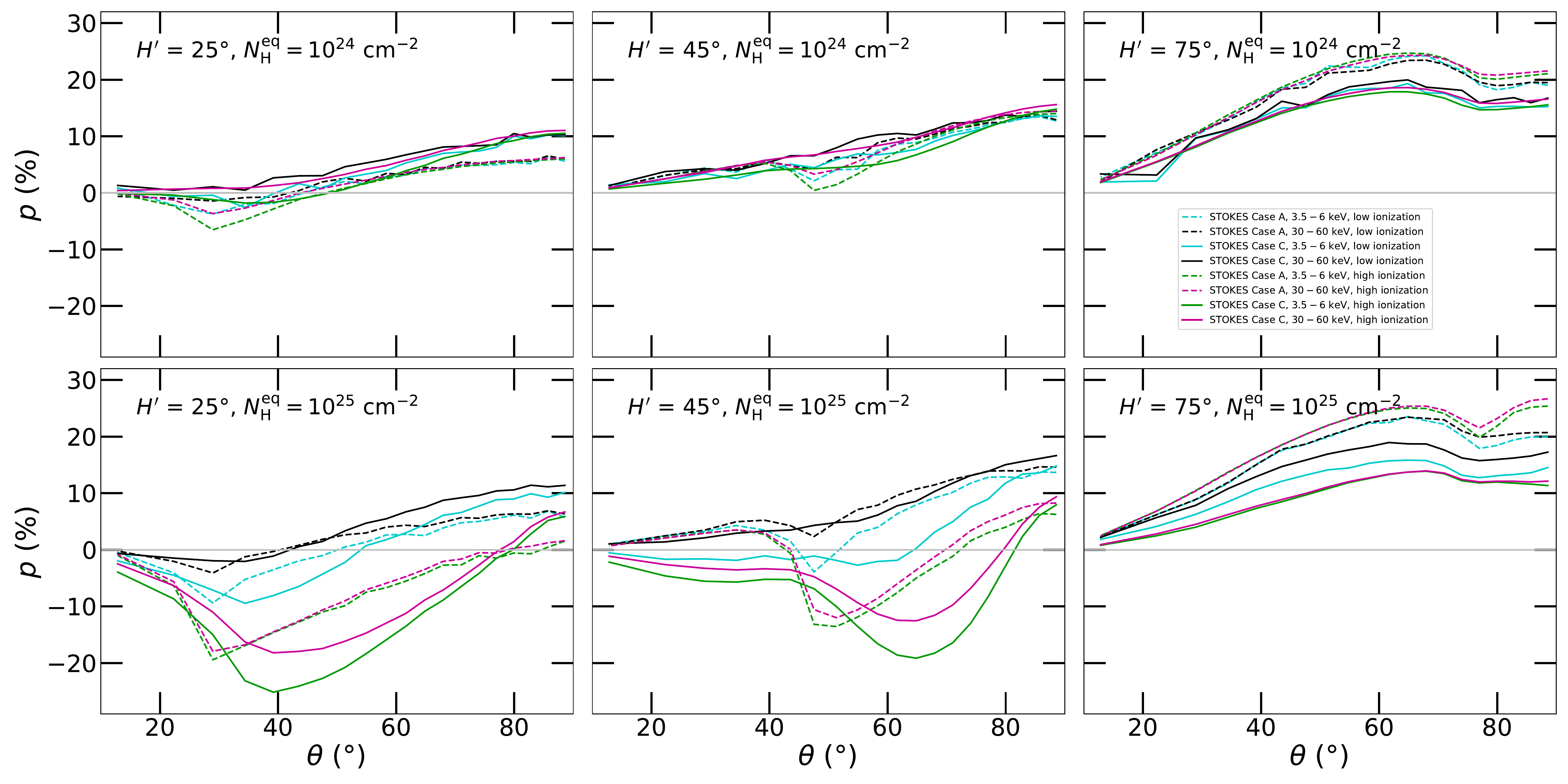}
	\caption{\footnotesize{The same as in Figure \ref{AvsC_Imue}, but energy-averaged polarisation degree, $p$, versus inclination $\theta$ is shown. Image adapted from \cite{Podgorny2023c}.}}
	\label{AvsC_pmue}
\end{figure}

Figure \ref{AvsC_pmue} shows the corresponding polarisation state versus inclination for~various half-opening angles. The same from a different perspective, i.e. for~polarisation versus the half-opening angle for various inclinations is shown in Figure \ref{AvsC_ptheta}. From these figures, we see with more clarity how the density (causing reprocessing in different volumes) drives the competition between the two orthogonal configurations linked to the mirror symmetry and to the dominant scattering events that occur rather in the equatorial plane, or rather in the meridional plane. The tested lower densities provide almost exclusively parallel polarisation, reaching $\approx 20\%$ in the most suitable geometries, which are typically high inclinations and high half-opening angles. At extremely high $H'$, we observe a~peak in inclination and a truly equatorial observer would see again slightly less polarized emission than a slightly lower inclined one. For low half-opening angles, the perpendicular polarisation can reach $\approx 25\%$, again non-monotonically with inclination and depending on other parameters. In this regime, there is depolarisation with lower densities, as the photons scatter in large volumes of the equatorial medium, thus on average through higher diversity of directions. For high enough densities, the ``periscope'' polarisation induction by reflection on the inner walls is stronger for higher ionization of the medium, as it then strongly reflects in the meridional plane for rather low half-opening angles compared to~contributions from other scattering directions. Some dependency of polarisation on~energy is observed due to the energy-dependent contribution of absorption and the aforementioned two dominant regimes. A wedge-shaped torus is more likely to produce higher parallel oriented polarisation (or less polarized perpendicular polarisation) than a circular torus. The equatorial inner toroidal radius is held the same for~both configurations, so the difference is the convexity or concavity of the inner wall in~this case, which is critical for first-order scatterings.

The switch in polarisation angle and non-monotonic polarisation degree with inclination were also reported for similar investigations in the optical and UV bands \citep{Goosmann2007,Marin2012}, as these are mostly geometrical effects, rather independent of the energy band. For a type-2 observer, we obtained almost no difference in the results, if zero-scattered photons were added (see Chapter \ref{chap05}). The results are then similar, although not equivalent, to~the~X-ray polarisation estimates provided for obscuration models in \cite{Ghisellini1994, Goosmann2011, Ratheesh2021,Ursini2023, Veledina2023, Tanimoto2023,Tomaru2023}. In~Chapter \ref{chap05}, we will provide perhaps a more comparable view to some of~these studies. In~particular, the work of \cite{Goosmann2011} used the~same code as us, but the~exact modeling conditions were different and the~study was targeting particularly the AGN in NGC 1068. Here we aim to provide a much broader overview of the configuration space and to show the striking diversity of~possible results for~a~general source.

~\

\begin{figure}[!htb]\centering
	\includegraphics[width=\textwidth]{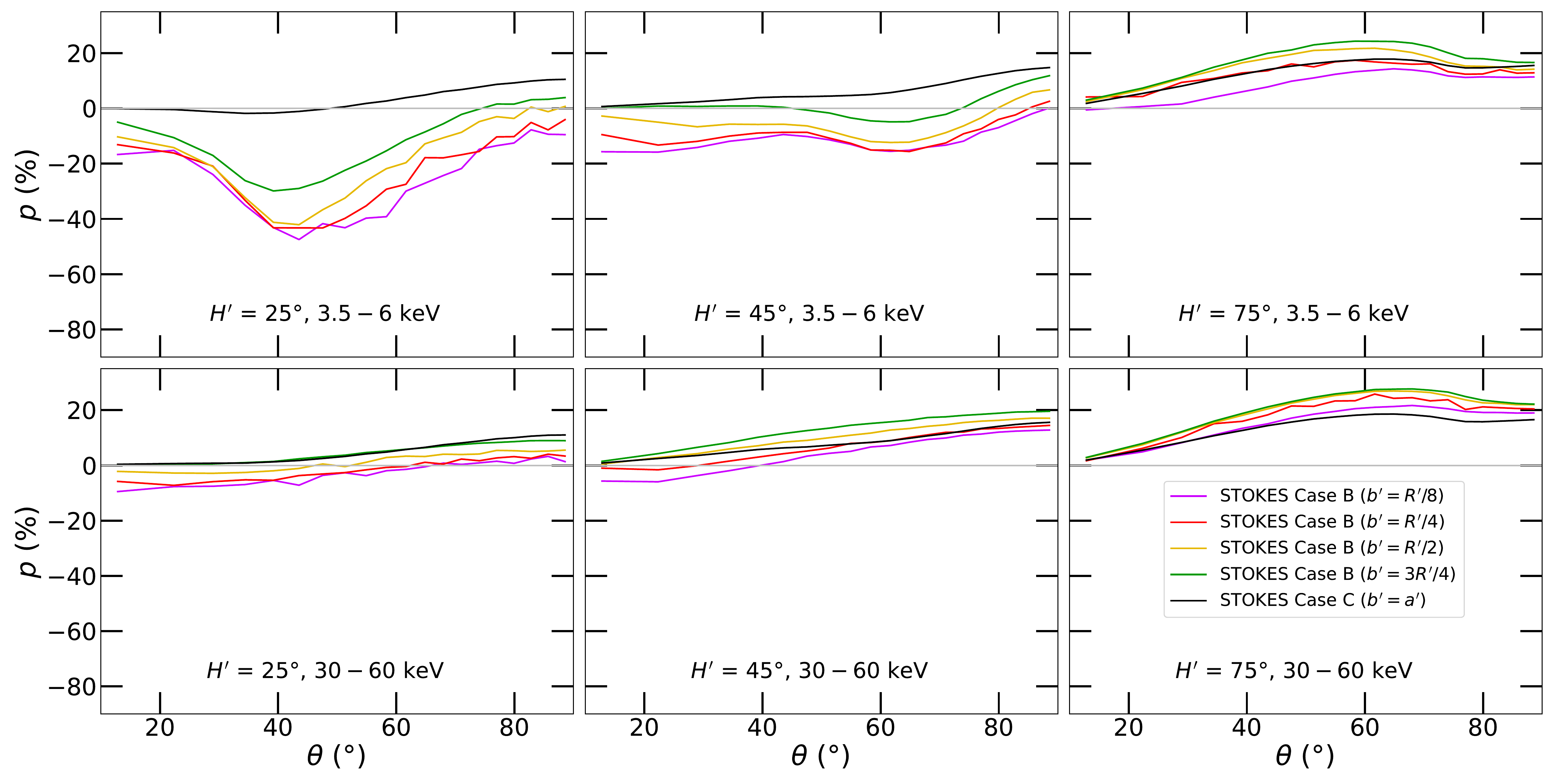}
	\caption{\footnotesize{Case B for various ellipse eccentricities compared to the circular Case C in energy-averaged polarisation degree, $p$, versus inclination $\theta$ for $H' = 25\degr$ (left), $H' = 45\degr$ (middle), and~$H' = 75\degr$ (right). We show the results integrated in 3.5--6 keV (top) and 30--60 keV (bottom) and for $N_\mathrm{H}^\textrm{eq} = 10^{24} \textrm{ cm}^{-2}$. The black curve corresponds to the Case C ($b' = a'$). The~magenta, red, yellow, green curves correspond to the Case B for $b' = R'/8$, $b' = R'/4$, $b' = R'/2$, $b' = 3R'/4$, respectively. The results are shown for \textit{high} level of ionization, i.e.~$\tau_\textrm{e}^\textrm{eq} = 0.7$, see corresponding Figure \ref{BvsC_pmue24_low_io} for low ionization. The primary input was set to~$\Gamma = 2$ and $p_0 = 2\%$ for all displayed cases. Image adapted from \cite{Podgorny2023c}.}}
	\label{BvsC_pmue24}
\end{figure}

\begin{figure}[!htb]\centering
	\includegraphics[width=\textwidth]{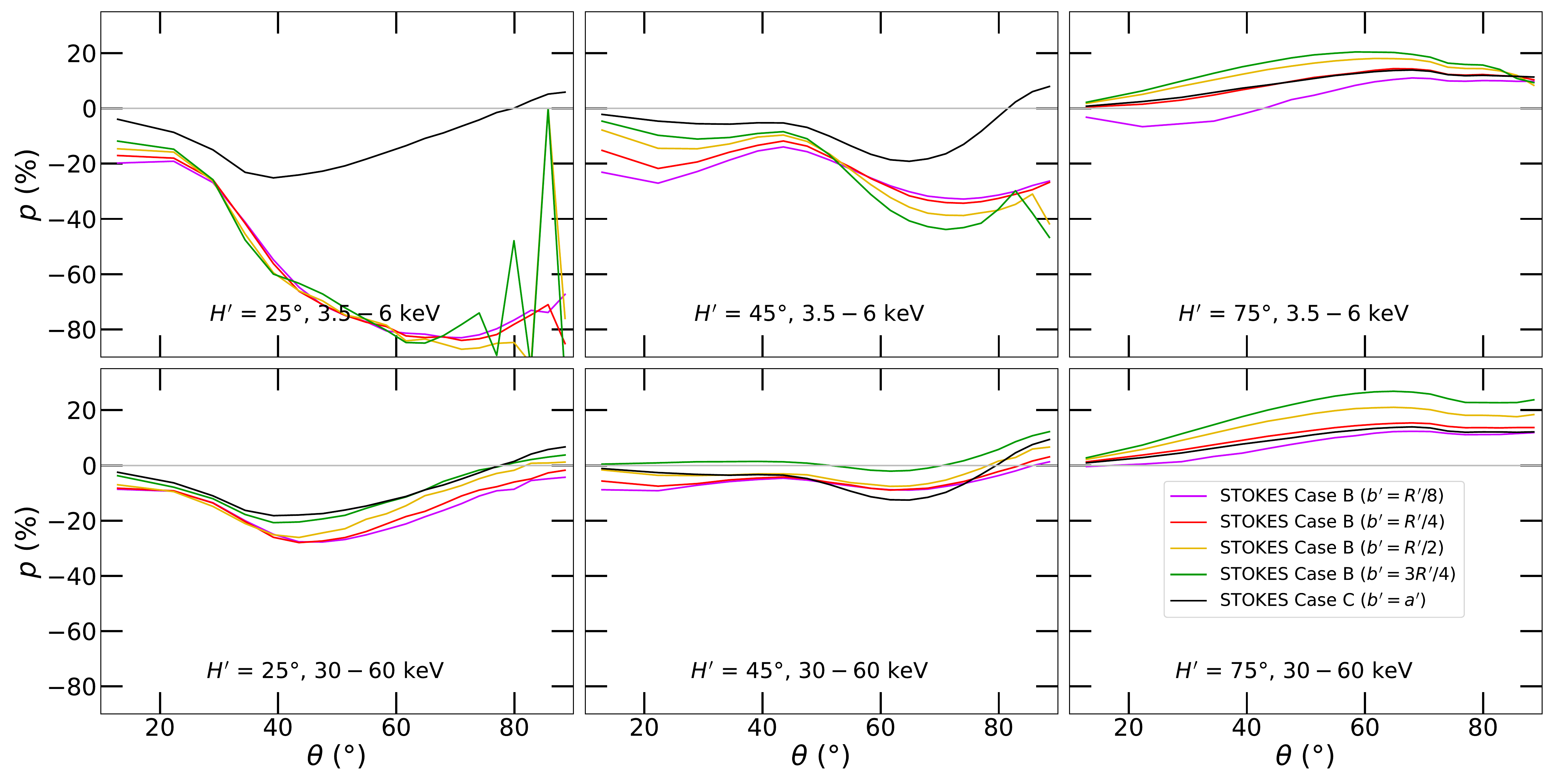}
	\caption{\footnotesize{The same as in Figure \ref{BvsC_pmue24}, but for $N_\mathrm{H}^\textrm{eq} = 10^{25} \textrm{ cm}^{-2}$. The results are shown for~\textit{high} level of ionization, i.e. $\tau_\textrm{e}^\textrm{eq} = 7$, see corresponding Figure \ref{BvsC_pmue25_low_io} for low ionization. Image adapted from \cite{Podgorny2023c}.}}
	\label{BvsC_pmue25}
\end{figure}

We will now turn attention to the effects of changing eccentricity of the elliptical tori. In the same plotting style, we compare in Figures \ref{BvsC_pmue24}, \ref{BvsC_pmue25}, \ref{BvsC_ptheta24}, \ref{BvsC_ptheta25} the~polarisation versus inclination and half-opening angle for the same two different densities, the same two energy bands and the high level of ionization. We fix $r_\textrm{in}^\textrm{torus}$ and vary $R'$, thus we track the effects of compactness of the entire reprocessing structure with fixed distance to the source. Case C is also plotted over a range of prolate and oblate Cases B, but it does not always represent an!intermediate case due to the adopted definitions (see above). The more oblate the torus is, i.e. equatorially elongated, the more parallel oriented polarisation it produces. In this sense, the most oblate Cases B with $b' = 3R'/4$ provide results closest to the previously shown wedge-shaped Case A, as there are high column density conditions for light rays that pass under inclinations near the half-opening angle. The dependency on eccentricity is reversed only for low half-opening angles, particularly for soft energies, high densities and high inclinations, i.e. when the~absorption effects cause preferentially perpendicular polarisation orientation for more oblate tori under extreme obscuration (with low numerical noise seen e.g. in top-middle panel of Figure \ref{BvsC_pmue25} and in comparison with Case A in dashed green curve of bottom-middle panel of Figure \ref{AvsC_pmue}). In this part of the configuration space, the curvature sign (whether convex or concave with respect to~the~central emission point) of the inner reflecting surface plays larger role. The studies of~\cite{Ghisellini1994, Ratheesh2021, Ursini2023, Veledina2023} used a conical shape of the obscurer. Although detailed comparisons are beyond the scope of our work, we anticipate the the geometry adopted therein with essentially no inner wall would vary from the elliptical tori presented here most clearly in this corner of the parameter space, perhaps even more from the wedge-shaped Case A torus.

The same dependencies of polarisation on inclination are also provided in Figures \ref{BvsC_pmue24_low_io} and \ref{BvsC_pmue25_low_io}, but for~low ionization. Despite more numerical noise being present, because more radiation is absorbed for the same number of simulated photons\footnote{\,We used $N_\textrm{tot} = 10^{10}$ simulated photons, which meant roughly 2 months to compute all section results on the available institutional clusters.}, we obtained qualitatively the same polarisation response to changing geometry. The results are even quantitatively similar for lower densities, but for~higher densities the dispersion in polarisation with respect to the range of eccentricities is smaller in the parallel oriented regime. The reason is the absorption on atoms under these density conditions, which is large enough even for the most prolate cases, compared to pure reflection on free electrons. The furthermost side of the torus with respect to the observer is then unable to produce enough perpendicularly polarized photons to significantly depolarize the parallelly polarized component. In the follow-up section, we will provide a comparison of different ionization cases of the presented MC model within one figure when compared to~a~different model.

~\

\begin{figure}[!htb]\centering
	\includegraphics[width=\textwidth]{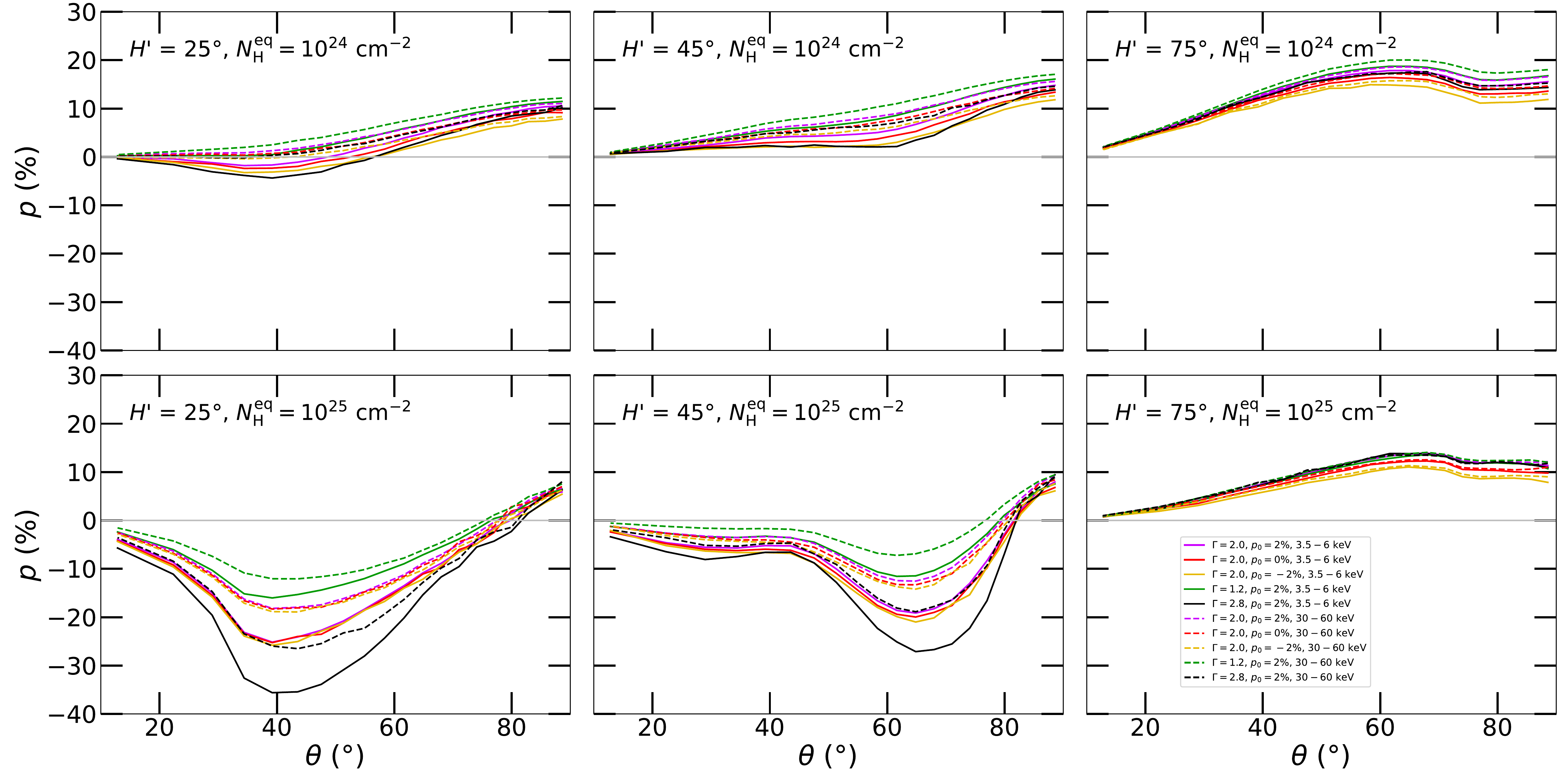}
	\caption{\footnotesize{Case C only for different cases of primary radiation: energy-averaged polarisation degree, $p$, versus inclination $\theta$ for \mbox{$H' = 25\degr$} (left), $H' = 45\degr$ (middle), and \mbox{$H' = 75\degr$} (right). We show the results integrated in 3.5--6 keV (solid) and 30--60 keV (dashed) \mbox{and~for~$N_\mathrm{H}^\textrm{eq} = 10^{24} \textrm{ cm}^{-2}$} and $\tau_\textrm{e}^\textrm{eq} = 0.7$ (top) and $N_\mathrm{H}^\textrm{eq} = 10^{25} \textrm{ cm}^{-2}$ and $\tau_\textrm{e}^\textrm{eq} = 7$ (bottom). The~magenta curves correspond to the $2\%$ parallelly polarized primary with $\Gamma = 2.0$. The~red and~yellow correspond to the unpolarized and $2\%$ perpendicularly polarized primary with $\Gamma = 2.0$, respectively. The green and black curves correspond to the $2\%$ parallelly polarized primary with $\Gamma = 1.2$ and $\Gamma = 2.8$, respectively. The results are shown for \textit{high} ionization, although other ionization levels show similar dispersion in polarisation with respect to~the~changing primary input. This dispersion is different with respect to~the~chosen geometry, while the~displayed Case C represents an intermediate result from all tested cases. Image adapted from \cite{Podgorny2023c}.}}
	\label{C_primary}
\end{figure}

Lastly, we discuss the effects of changing primary radiation. Figure \ref{C_primary} shows the polarisation results in the same plotting style for a range in $\Gamma$ and different primary polarisation state, all for Case C and high ionization of the equatorial reprocessing region. Because we do not include direct radiation, the incident polarisation state changing by $2\%$ in polarisation fraction impacts the resulting polarisation by $\lessapprox 2\%$ for a generally inclined observer. The~result is depolarized or additionally polarized with respect to the case with unpolarized primary, following the incident polarisation orientation with respect to the polarisation angle for unpolarized primary. The change in photon index $\Gamma$ plays a much larger role, also depolarizing or additionally polarizing with respect to the intermediate $\Gamma = 2$ example. The change in net polarisation with respect to the~unpolarized primary or $\Gamma = 2$ case is qualitatively in the~same direction in the plots, when the polarisation fraction passes the zero point and the polarisation angle switches, which again suggests two competing directions of polarisation generation. We omit the discussion of spectral lines, where the equivalent width would be also $\Gamma$-dependent. The energy dependence of polarisation, which is geometry and density dependent (see Figure \ref{energy_dependent}), plays a role while $\Gamma$ influences the relative importance of soft and hard X-rays. We simply weight more over higher or lower polarised energy bins, when changing $\Gamma$ and plotting the energy-integrated results. The~difference can be even as large as $\approx 20\%$ in 3.5--6 keV polarisation for~the~demonstrated large change in photon index for high column densities. In Chapter \ref{chap05}, we will evaluate the~degeneracy in~the~results for a single 2--8~keV polarimetric observation by {\it IXPE}. When testing other ionization levels in the obscuring media, the~dispersion in~polarisation with respect to the changing properties of primary radiation appears similar. It varies, however, with chosen geometry. The~displayed Case C represents an intermediate result. The more prolate elliptical tori are more sensitive to the change of $\Gamma$ (up to $30\%$ difference in polarisation in~the~most critical configurations in~3.5--6~keV), while the more oblate elliptical tori and wedge-shaped tori show minor changes (maximum of~$10\%$ difference in~polarisation in~the~most sensitive configurations in 3.5--6 keV).

We do not aim to study the imprints of extended or unisotropic sources. Any such consideration would require a separate model design. Perhaps even more useful configuration that we do not consider is thermal emission (or combined thermal and power-law emission) from the central disc-corona region, appearing in the soft state of XRBs, which may be also physically covered in highly obscuring equatorial outflows. Because we obtained different results for different slope of~the~primary power-law, we expect that adding a strong thermal radiation would affect the results in a way that increasing $\Gamma$ does. The change in~spectral shape should play a larger role than any changed primary polarisation of such spectral state \citep[in fact, the first {\it IXPE} results do not reveal a significant change in polarisation for XRB spectral changes,][]{Cavero2023, Ingram2023b, Steiner2024}. We refer to the study of \cite{Tomaru2023} that obtained parallelly polarized results from scattering of thermal emission in~optically thin winds.

\section[Simple models of axially symmetric reflecting structures]{Simple models of axially symmetric \newline reflecting structures}\label{simple_models}

\begin{figure}[!htb]
	\centering
	\begin{minipage}[b]{0.49\textwidth}
		\includegraphics[width=\textwidth]{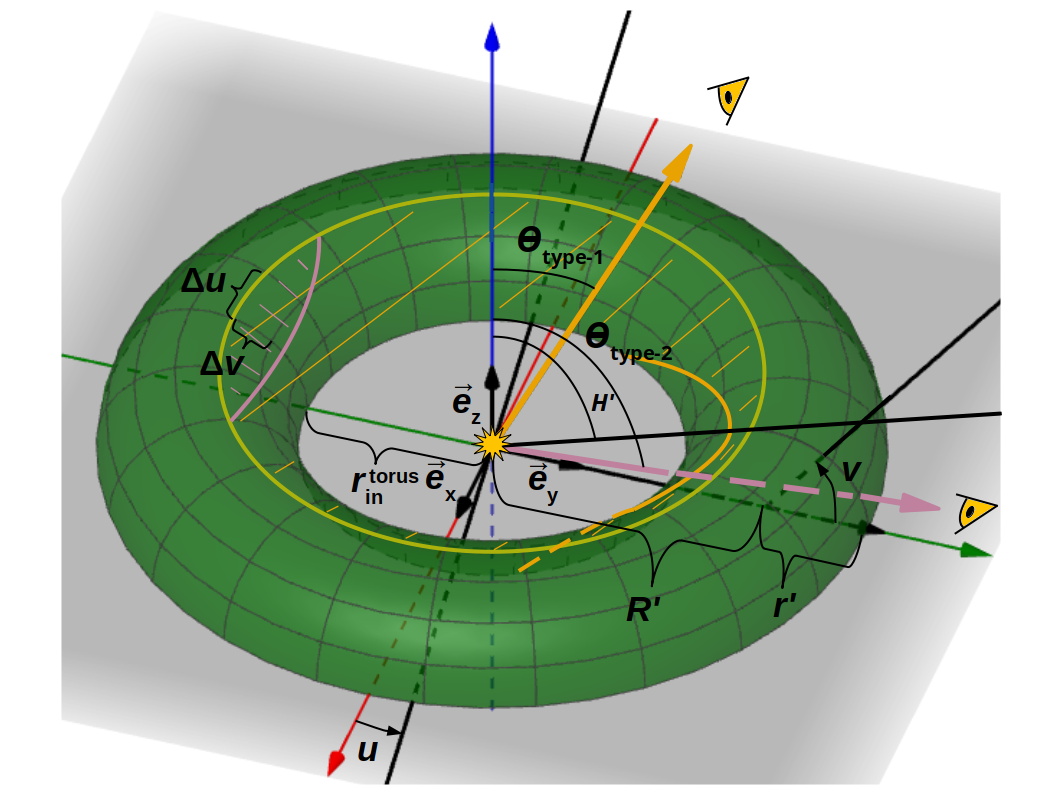}
		\caption{\footnotesize{A circular toroidal reflector that is considered for the results presented in Section \ref{simple_torus_reflection}. Being located in~the~center of~the~coordinate system, the X-ray power-law of arbitrary incident polarisation and power-law index $\Gamma$ is isotropically illuminating the inner side of~the~torus surface. We neglect the half-space below the equator, as if there was fully absorbing material in the equatorial plane. The~observer, generally inclined at~\mbox{$0\degr < i < 90\degr$}, is viewing the source \mbox{in~the~$yz$-plane}. The material only reflects once by means of local reprocessings, i.e.~the~photons emerge from the same macroscopic location, as they hit. The surface regions not directly illuminated by the~central source are not accounted for, i.e. the $v$ values lower than reaching the~tangent points of~the~torus with the half-opening angle cone that are represented by the~yellow shadow boundary. The~illuminated part cannot be in addition self-obscured by the~torus in~the~observer's line of sight to~be accounted for~in~the~integration. The~lower boundaries of~the~areas that are obscured by the~opaque torus for an observer~in~two generic $\theta_\textrm{type-1}$ and~$\theta_\textrm{type-2}$ inclinations are schematically drawn in~orange and~pink colors, respectively. The visible reflecting areas for~a~generic type-1 and type-2 observer are then shaded in~orange and pink, respectively. Image adapted from \cite{Podgorny2023e}.}}
		\label{xsstokes_mo}
	\end{minipage}
	\hfill
	\begin{minipage}[b]{0.49\textwidth}
		\includegraphics[width=\textwidth]{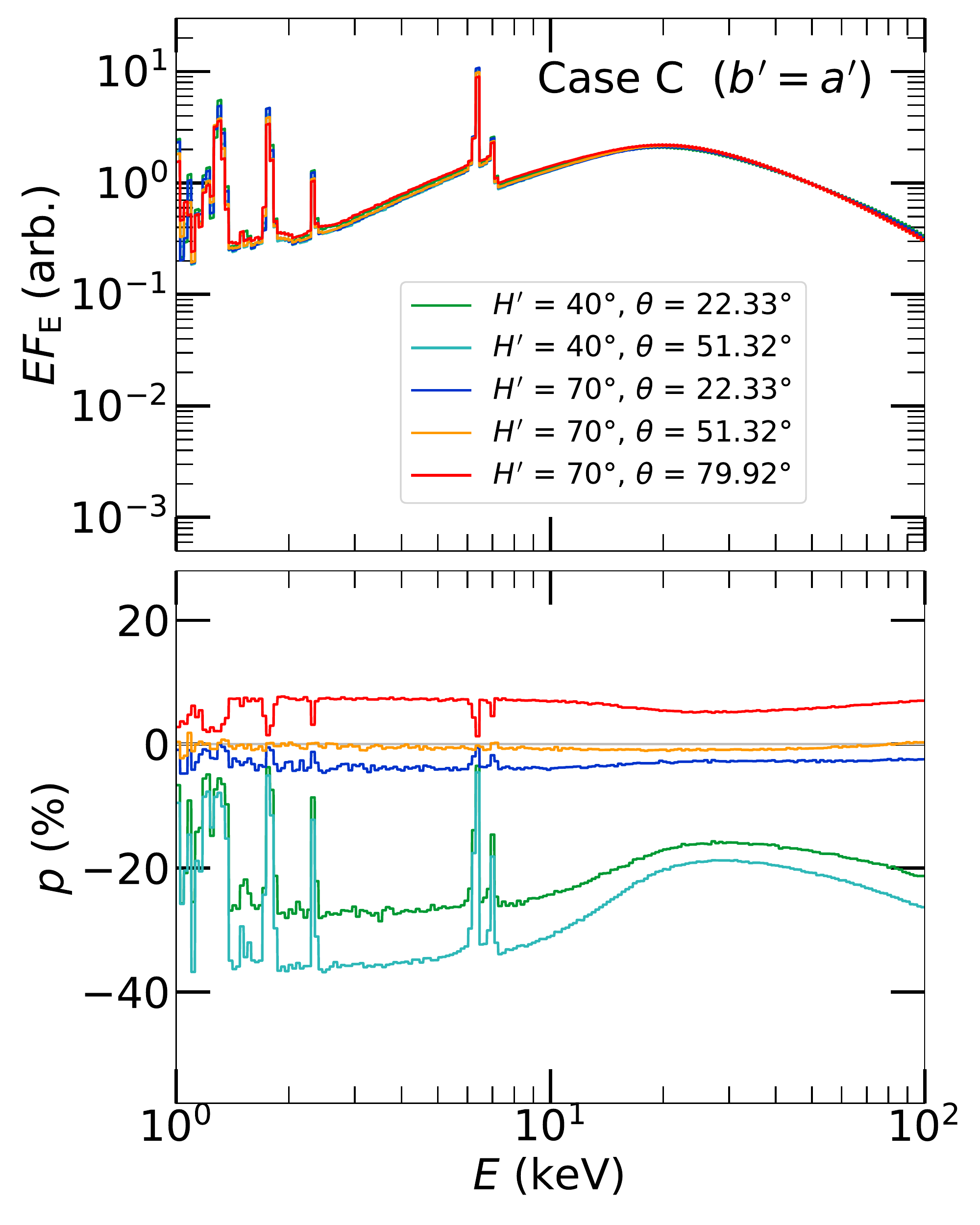}
		\caption{\footnotesize{The energy-dependent results from the {\tt xsstokes\_torus} model. Top: the~spectra, $EF_\mathrm{E}$, normalized to value at 50 keV. Bottom: the corresponding polarisation degree, $p$, versus energy. For easy comparisons, we show the {\tt xsstokes\_torus} results in the same way as the results from pure MC simulations are given in Figure \ref{energy_dependent}. Various inclinations $\theta$ and half-opening angles $H'$ for both type-1 and type-2 observers are shown in the color code. The primary input was set to $\Gamma = 2$ and $p_0 = 2\%$ for all displayed cases. The highest discrepancies between the pure MC model and the simplistic scenario occurs for high inclinations and high half-opening angles. The curvature and relative size effects of the torus surface are eliminated from such comparisons, because the torus in both models (Case C) has identical profile. The detailed parametric dependencies are shown in the following figures when integrated in energy. Image adapted from \cite{Podgorny2023e}.}}
		\label{energy_dependent_simple}
	\end{minipage}
\end{figure}

In this section, we first explore a simplified model of reflection off the inner walls of a geometrical torus, using the same isotropically emitted power-law, originating in the center of coordinates with a prescribed incident polarisation state. We still do not account for any direct radiation. By ``reflection'' we mean reprocessing, as usual, because we will use the spectropolarimetric local reflection tables, described in Section \ref{reflection_tables_TS} \citep[in the version published in ][]{Podgorny2021}, integrated over the torus surface. However, only a ``single-reprocessing'' mode is allowed, meaning that the photons can enter and escape the scattering region only once. Because the local reflection model was developed for accretion discs's atmospheres, we will use only the most neutral part of the tables, i.e. the lowest $\xi$ available, to represent the cold parsec-scale dusty tori of AGNs. The torus has a circular profile (Case C in Section \ref{equatorial_MC}), is fully opaque, and we consider only illumination of the upper half-space, as if there was an absorbing optically thick material extending from the outer disc to the inner torus boundary. Figure \ref{xsstokes_mo} shows the model parametrization. An observer can be inclined at any angle $0\degr < \theta < 90\degr$ and the torus is assumed to have again some half-opening angle $H'$. The inner surface of the torus is illuminated from the equatorial plane up~to~its shadow boundary, given by the half-opening angle. The observer's inclination then additionally constrains an area that is the intersection of the visible surface and the illuminated surface that we account for. These are marked in Figure \ref{xsstokes_mo} as orange-shaded and pink-shaded regions for a generic type-1 ($\theta < H'$) and~type-2 ($\theta > H'$) observer, respectively. In this setup, we may directly compare the~results to the MC modeling for Case C, described in Section \ref{equatorial_MC}. We can then assess to~what extent the X-ray polarisation is given solely by the geometrical parameters $\theta$ and $H'$ and by single reprocessing, in contrast to self-irradiation and partial transparency effects allowed in the MC modeling.

The numerical implementation is in detail described in Appendix \ref{torus_model}, as the~calculation was divided in three steps, each comprising of a newly developed routine that can be of wider use. The final model, called {\tt xsstokes}, is compatible with {\tt XSPEC} and efficiently interpolates for arbitrary state of primary polarisation, using equation 
\begin{equation}\label{interpolate_Stokes}
\begin{split}
    S(p_0, \Psi_0) = & \, S(0,-) + p_0\Big\{ \left[S(0,-)-S\left(1,\frac{\pi}{2}\right)\right]\cos{2(\Psi_0)} \\
    &\quad + \left[S\left( 1,\frac{\pi}{4}\right) - S(0,-)\right]\sin{2(\Psi_0)} \Big\} \, ,
\end{split}
\end{equation}
which is just a variant\footnote{\,In order to get (\ref{interpolate_Stokes}) from (\ref{Sorig}), one inserts $S(1,\frac{\pi}{2})$ to the left side, to obtain
\begin{equation}
    S\left(1,\frac{\pi}{2}\right) = 2S(0,-) - S(1,0)\textrm{ ,}
\end{equation}
and then replaces $S(1,0) = 2S(0,-) - S(1,\frac{\pi}{2})$ in (\ref{Sorig}).} of (\ref{Sorig}) that was used for the same problem in Chapter \ref{chap02}, but could not be used here, as we work with the older version of local reflection tables, pre-computed for unpolarized, $100\%$ diagonally polarized and $100\%$ horizontally polarized emission. The same interpolation scheme works for any axially symmetric reflector. Thus, in this variant of an equatorial toroidal obscuring medium, we use a subversion name {\tt xsstokes\_torus}\footnote{\,Available at \url{https://github.com/jpodgorny/xsstokes_torus} on the day of~submission, including documentation and tabular dependencies. The work is still in progress on~the~day of~submission.}. But we also modelled one other reflection scenario imitating a faraway disc in the equatorial plane, which in the exact same way interpolates for arbitrary incident polarisation.

This effort resulted in another {\tt XSPEC} compatible model, a subvariant of~{\tt xsstokes}, called {\tt xsstokes\_disc}\footnote{\,Available at \url{https://github.com/jpodgorny/xsstokes_disc} on the day of~submission, including documentation and tabular dependencies. The work is still in progress on the day of~submission.}. It represents the same nearly neutral local reflection tables from \cite{Podgorny2021} integrated uniformly, i.e. with no weighting applied, in all azimuthal emission angles $\Phi_\textrm{e}$ and in a range of high incident inclination angles: $0 \leq \mu_\textrm{i} \leq M_\mathrm{i}$, where $M_\mathrm{i} = 0.3$. Such model illustrates a distant region of~the~accretion disc, illuminated by a central vertically extended hot inner-accretion flow of Comptonized emission and/or a reflecting outer accretion disc with increasing geometrical thickness. We note, however, that the~latter geometrical imagination results in a range of surface emission inclination angles for~a~particular distant observer, if the reflecting medium deviates from the equatorial plane non-linearly, while we keep a particular emission angle $\theta$ (adopted as $\delta_\textrm{e}$ from the original tables) as a free parameter, which drives the~polarisation results to a significant extent.\footnote{\,Such thick disc scenarios should be rather approximated by the reflecting torus model for~high half-opening angles.} We assume the disc to be truncated (or supported in {\tt XSPEC} by a different reflection model in the inner hot parts\footnote{\,Direct coronal emission would also here have to be added in {\tt XSPEC} to the presented reflection model, when fitting real data.}) and the inner-accreting system to be unobscured even for large inclination angles. The~results of the {\tt xsstokes\_disc} model are presented here below, in comparison to~the~corresponding limiting case of a torus with high half-opening angle.

~\

All results presented in this section and in Appendix \ref{torus_model} are part of~\cite{Podgorny2023e}, which is, however, still under review process. Therefore, we present these results as preliminary and note that the model will still undergo many changes in the near future. For example, the parameter $M_\mathrm{i}$ is being currently implemented as a model parameter, representing the coronal physical size (large $M_\mathrm{i}$ is equivalent to the {\tt KYNSTOKES} results in sandwich geometry without kinetic and relativistic effects). For the reflection on a torus, we plan to add a new parameter $B = \{0,1\}$, representing a switch between taking the reflecting area below the torus equator into account or not. Another pending issue is the fact that for every half-opening angle there is an inclination above which the observer does not see any illuminated surface area on the torus. This is important for small half-opening angles and the $H'$ parameter should be thus transformed to~range between 0 and 1, effectively spanning all allowed values per inclination $\theta$, in order to not be able to fit in {\tt XSPEC} at higher inclinations.

For all results presented in this section, we keep the notation of a positive and~negative polarisation degree for parallel or perpendicular polarisation orientation of the incident or emergent radiation, respectively. No other polarisation angle cases are considered on the input and no relativistic effects are considered inside the simulations that would result in symmetry breaking.

\subsection{Reflection from inner walls of a geometrical torus}\label{simple_torus_reflection}

Let us first discuss the results of the {\tt xsstokes\_torus} model. Figure \ref{energy_dependent_simple} shows the~energy-dependent spectra and polarisation for a selection of $\theta$ and $H'$, displayed in the same way as in Figure \ref{energy_dependent}. Compared to the MC simulations presented in~Section \ref{equatorial_MC}, we obtain much lower numerical noise and generally higher predicted polarisation fraction in the perpendicular orientation, while qualitative behavior of~the~polarisation degree and angle with $\theta$ and $H'$ is similar. The~depolarisation in~spectral lines is preserved. From the spectra it is clear that even the~most neutral part of the local reflection tables does not allow strong absorption towards the soft X-rays and the associated continuum polarisation fraction is rather constant in energy in 1--4 keV. We obtain closer results to the elliptical tori class (Cases B and C) presented in Section \ref{equatorial_MC} than to the wedge-shaped tori (Case A) from Section \ref{equatorial_MC}. The case of the highest inclination and the smallest half-opening angle scenario from Figure \ref{energy_dependent} is not shown, because in~{\tt xsstokes\_torus} we have the zero-transparency condition and~the~reflecting area is already eclipsed in such case.

\begin{figure}[!htb]\centering
	\includegraphics[width=\textwidth]{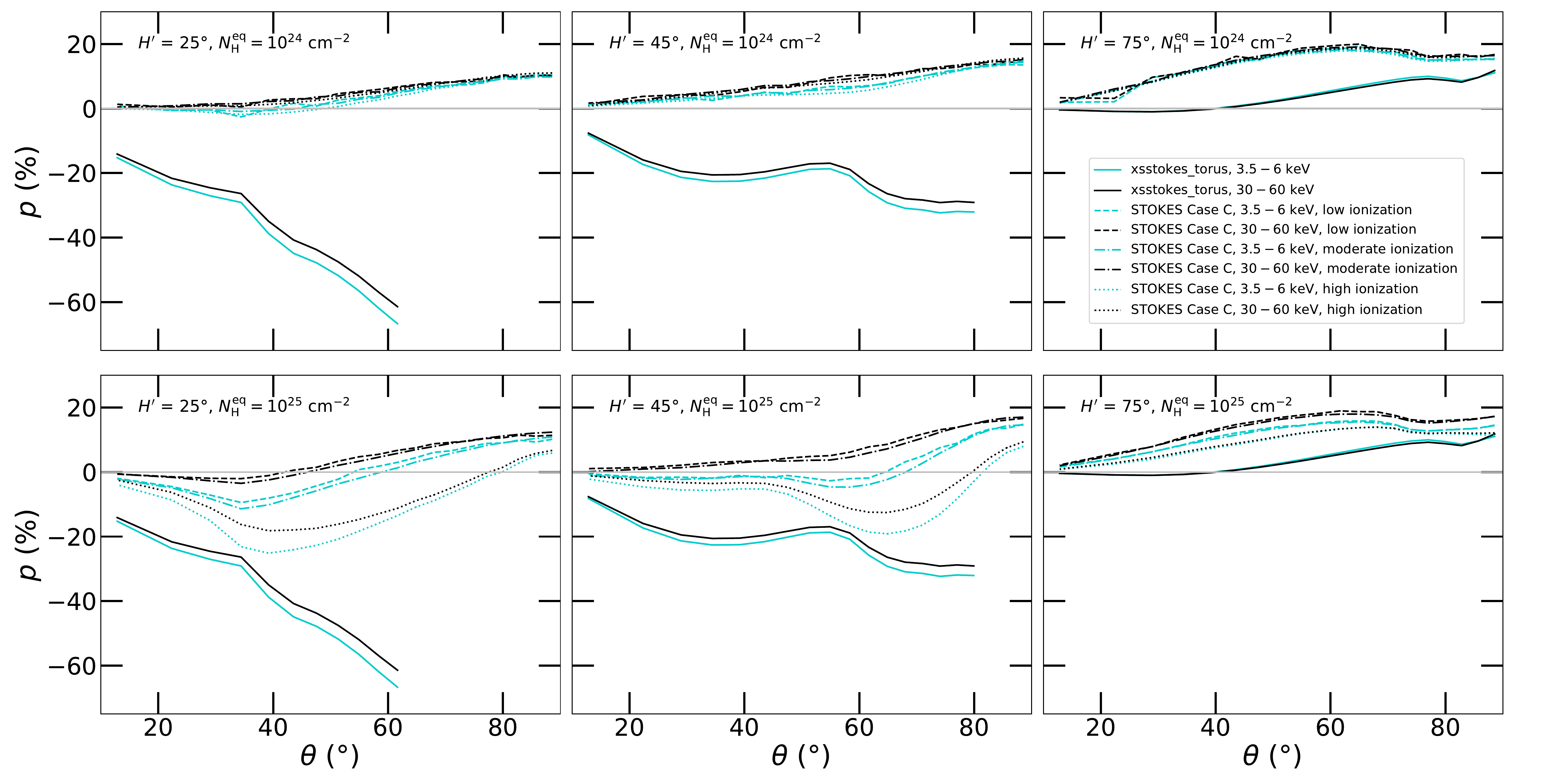}
	\caption{\footnotesize{Comparison of the pure {\tt STOKES} simulation of Case C from Section \ref{equatorial_MC} (with different line styles assigned to different ionization levels) with the {\tt xsstokes\_torus} results (solid). We plot the polarisation degree, $p$, versus inclination $\theta$ for $H' = 25\degr$ (left), $H' = 45\degr$ (middle), and~$H' = 75\degr$ (right). We show the results integrated in 3.5--6 keV (blue) and 30--60 keV (black) and for $N_\textrm{H}^\textrm{eq} = 10^{24}  \textrm{ cm}^{-2}$ in {\tt STOKES} Case C (top) and $N_\textrm{H}^\textrm{eq} = 10^{25}  \textrm{ cm}^{-2}$ in {\tt STOKES} Case C (bottom). The primary input was set to $\Gamma = 2$ and $p_0 = 2\%$ for all displayed cases. Image adapted from \cite{Podgorny2023e}.}}
	\label{CvsD_pmue}
\end{figure}

Figure \ref{CvsD_pmue} and \ref{CvsD_ptheta} show the polarisation results versus inclination and half-opening angle, respectively, integrated in the 3.5--6 keV and 30--60 keV bands that are not contaminated by spectral lines. On these figures, we compare directly the~X-ray polarisation from {\tt xsstokes\_torus} with the MC modeling results from Section \ref{equatorial_MC} for various ionization levels and equatorial column densities. Albeit they both show the same data from a different perspective, these two figures are additionally illustrative, because we did not compare different ionization levels of~the~MC model in Section \ref{equatorial_MC} in the same panels. We cut the {\tt xsstokes\_torus} parameter space for low half-opening angles and high inclinations, where the illuminated area is fully obscured. It is clear from these figures that the highest polarisation contribution from the perpendicularly polarized component arises from the torus with highest free electron density contribution alongside high neutral hydrogen density (low transparency conditions). This is also the case that resembles most the {\tt xsstokes\_torus} model, which takes into account only the highly reflective and perpendicularly polarized component originating on the furthermost side of~the~inner walls of the torus from the observer. The {\tt xsstokes\_torus} model results in about twice higher polarisation fraction for low and moderate half-opening angles. It also predicts parallelly oriented net polarisation for high half-opening angles, but then the resulting polarisation fraction is on the contrary about twice lower than for the pure MC model due to~the~lack of strong flux contribution of the parallelly polarized component, originating in reflection from left and right sides of the torus inner walls from the observer's perspective. In the MC simulations, the parallelly polarized component is strengthened by means of~partial transparency.

\subsection{Reflection from a faraway disc}\label{simple_disc_reflection}

\begin{figure}[!htb]\centering
	\includegraphics[width=\textwidth]{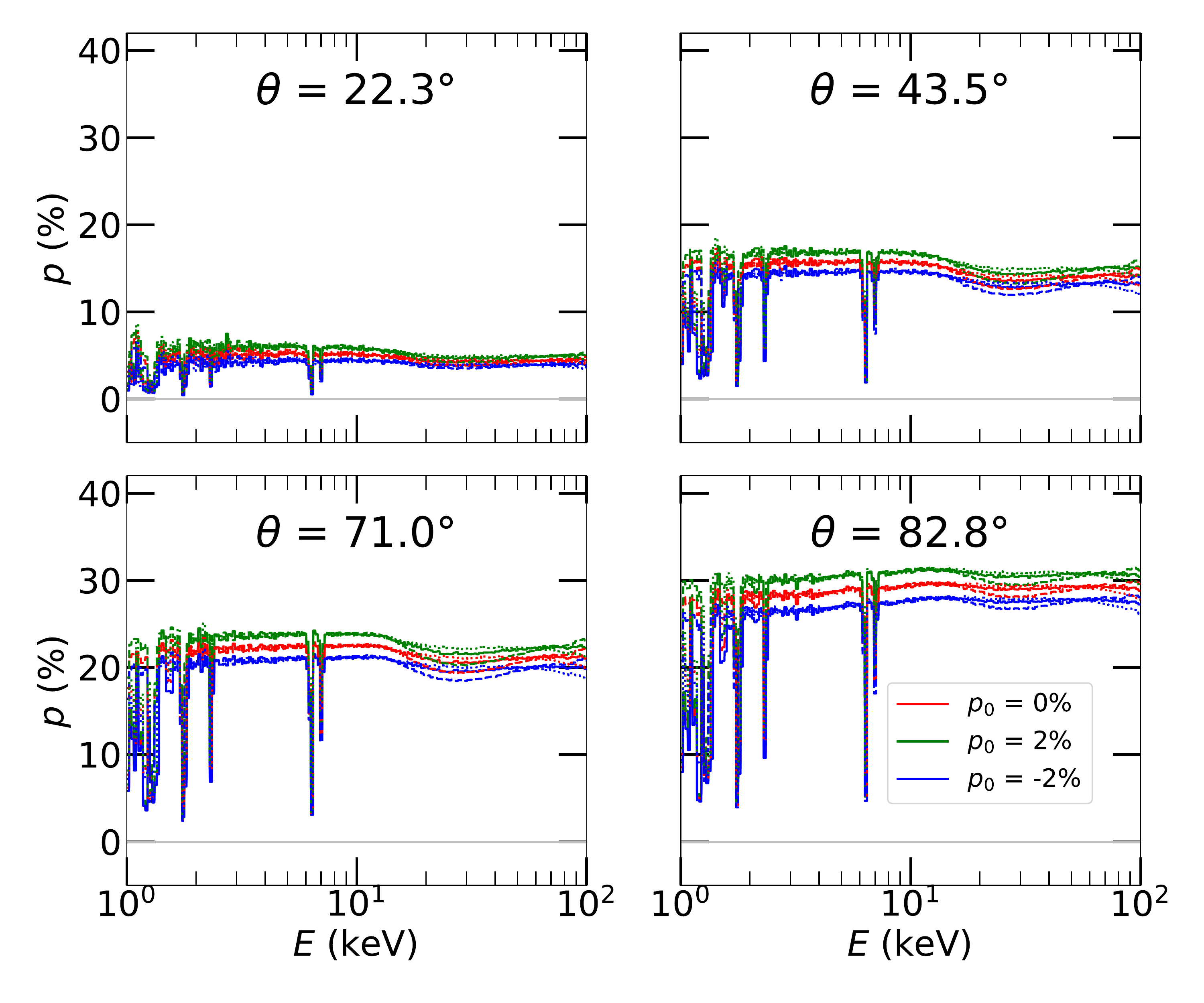}
	\caption{\footnotesize{The polarisation degree, $p$, versus energy for distant disc reflection computed by {\tt xsstokes\_disc}. We show the results for various cosines of the observer's inclination: $\mu_\textrm{e} = 0.925$ (top left panel), $\mu_\textrm{e} = 0.725$ (top right panel), $\mu_\textrm{e} = 0.525$ (bottom left panel), $\mu_\textrm{e} = 0.125$ (bottom right panel). The color code represents different incident polarisations of the primary power-law: unpolarized (red), $2\%$ parallelly polarized (green), and $2\%$ perpendicularly polarized (blue). The solid lines correspond to $\Gamma = 2.0$, dashed lines to $\Gamma = 1.2$, dotted lines to $\Gamma = 2.8$. Image adapted from \cite{Podgorny2023e}.}}
	\label{disc_energy_dependent}
\end{figure}

We now return to the problem of disc reflection with no SR or GR effects applied, as computed by the {\tt xsstokes\_disc} model. Figure \ref{disc_energy_dependent} shows the polarisation dependence on energy for various observer's inclinations, primary power-law indices and primary polarisation states. Unlike the result for reflection from a~toroidal structure that has non-negligible vertical extension from the equatorial plane, here we obtain only parallel polarisation angle on the output. This is natural regarding polarisation arising from Compton scattering (see Section \ref{processes}) in no distortion of space-time geometry, when the source of illumination is located in~the~center of~the~system and the scattering plane is rather aligned with the matter distribution. Nearly the same orientation of polarisation by disc reflection was obtained also in~the~lamp-post geometries with {\tt KYNSTOKES} in Section \ref{lamppost_spproperties}, where the disc was reaching to the ISCO and GR effects were included. The received polarisation fraction by disc reflection in {\tt xsstokes\_disc} can be any value between $0\%$ up to $\approx 30\%$, strongly depending on the inclination of~the~observer, which is the main driver of polarisation here. The spectral lines are again depolarized, as evident from the local disc reflection tables. The energy profile of polarisation degree agrees with that of the {\tt xsstokes\_torus} model, seen in Figure \ref{energy_dependent_simple}, that uses the same local reflection tables.

~\

What varies from the torus reflection is the impact of changing primary power-law index on the emergent polarisation, which is negligible in the majority of~the~parameter space covered by {\tt xsstokes\_disc} (apart from $\approx 2\%$ difference in polarisation around 30 keV for high inclinations). The response of~the~reflected polarisation degree to the incident polarisation is ``linear'' in the same sense as for the {\tt KYNSTOKES} results presented in Section \ref{lamppost_spproperties} or as for the reflection from~a~geometrical torus described above. The received emission has enhanced polarisation compared to unpolarized primary case by less or the same percentage of the incident polarisation fraction, if the incident polarisation is oriented parallelly, and is depolarized compared to unpolarized primary case by less or the~same percentage, if the incident polarisation is oriented perpendicularly. These effects are nearly energy-independent.

~\

\begin{figure}[!htb]\centering
    \begin{minipage}[b]{0.49\textwidth}
        \includegraphics[width=\textwidth]{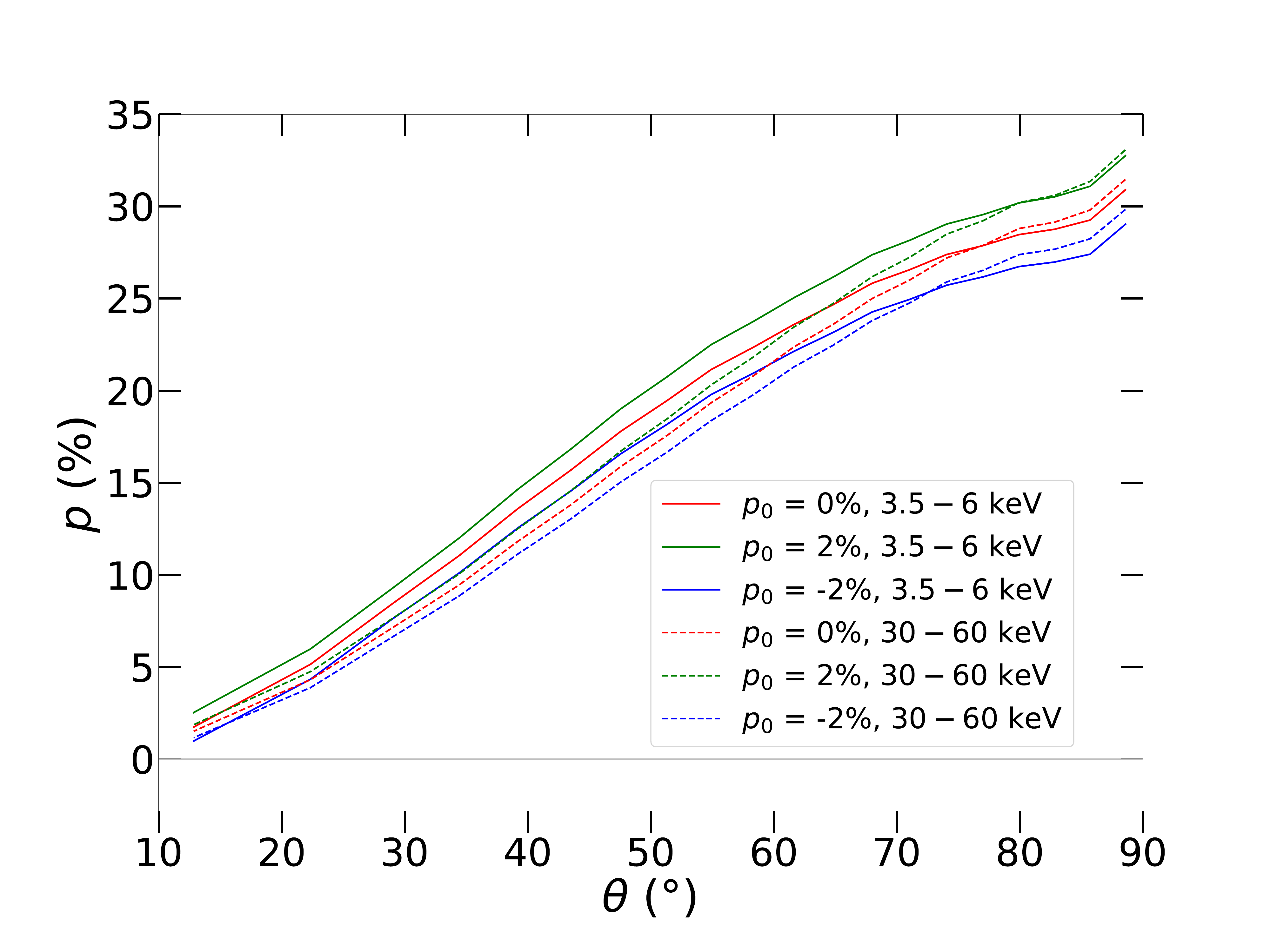}
        \caption{\footnotesize{The computed energy-averaged polarisation fraction, $p$, with {\tt xsstokes\_disc} in 3.5--6 keV (solid lines) and 30--60 keV (dashed lines) versus observer's inclination $\theta$ for $\Gamma = 2$. The color code represents different incident polarisations of the primary power-law: unpolarized (red), $2\%$ parallelly polarized (green), and $2\%$ perpendicularly polarized (blue). Image adapted from \cite{Podgorny2023e}.}}
	\label{disc_mue_dependent}
    \end{minipage}
    \hfill
    \begin{minipage}[b]{0.49\textwidth}
        \includegraphics[width=\textwidth]{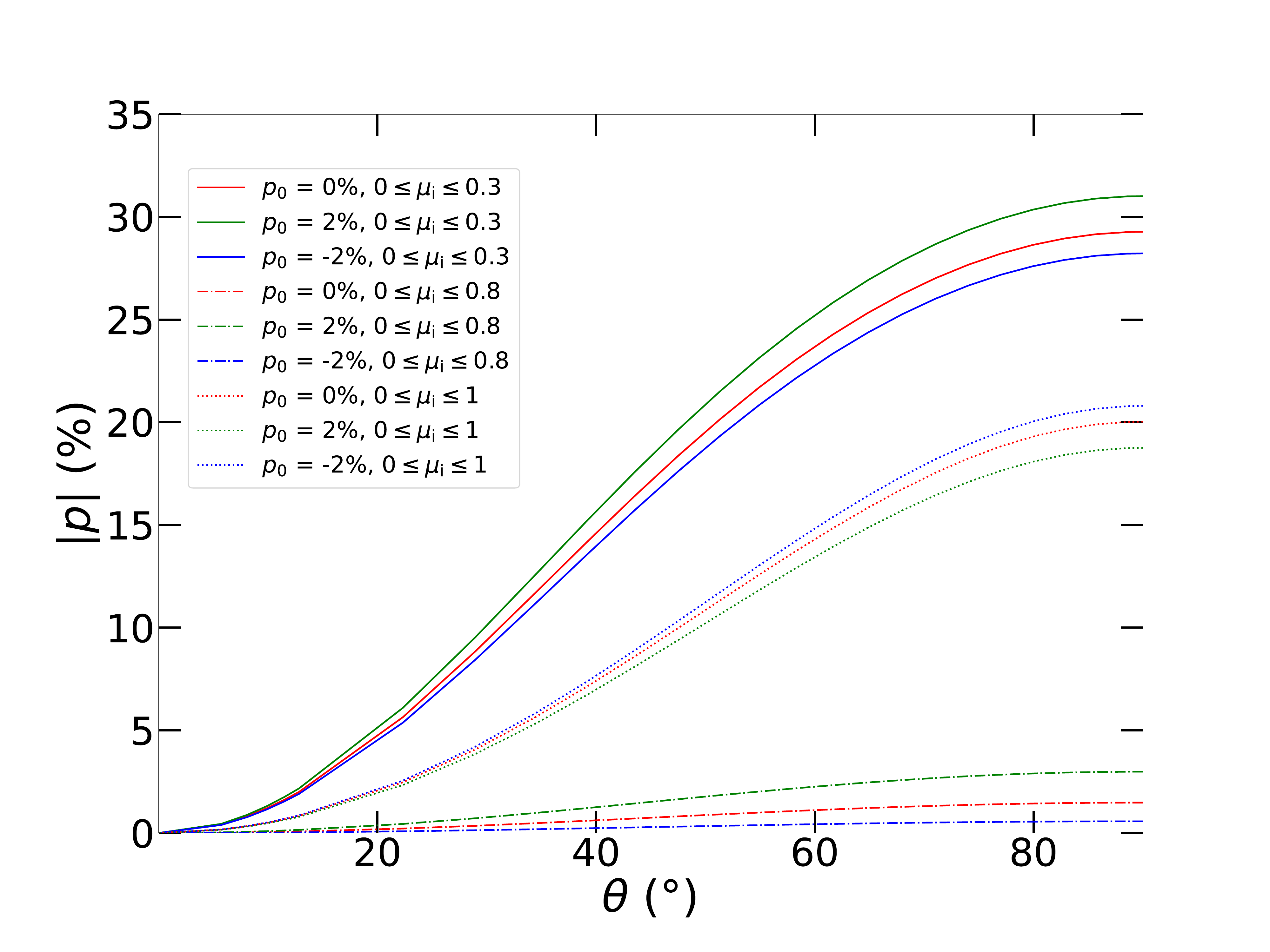}
	\caption{The absolute value of polarisation, $\lvert p\rvert$,  versus observer's inclination $\theta$. The results are from Chandrasekhar's single-scattering approximation (\ref{chandra1}) integrated uniformly in~all azimuthal emission angles $\Phi_\textrm{e}$ and~in~selected ranges of cosines of incident angles \mbox{$0 \leq \mu_\textrm{i} \leq 0.3$} (solid lines), \mbox{$0 \leq \mu_\textrm{i} \leq 0.8$} (dashed lines) and \mbox{$0 \leq \mu_\textrm{i} \leq 1$} (dotted lines). The color code represents different incident polarisation states of~the~primary, unpolarized and $2\%$ parallelly and perpendicularly polarized. Image adapted from \cite{Podgorny2023e}.}
	\label{chandra_comparison}
    \end{minipage}
\end{figure}

To study the inclination dependence in more detail, we again integrate the~polarisation results in the 3.5--6 keV and 30--60 keV bands, which is shown in Figure \ref{disc_mue_dependent} for one chosen $\Gamma$. 
We obtain here a monotonic and roughly linear dependency of polarisation degree on observer's inclination. This is a consistent limit of the torus reflection models for high half-opening angles (see e.g. Figure \ref{CvsD_pmue}, right panels), similarly to the net polarisation angle results. Due to effectively higher asymmetry in the geometry of scattering for~the~truncated disc reflection, we obtain higher polarisation fraction than with the torus reflection models where we operate with more competing scattering directions with respect to the observer. The {\tt xsstokes\_disc} results displayed in~Figure~\ref{disc_mue_dependent} are in line with figure 11 from \cite{Ratheesh2021}, which solved for a similar equatorial disc reflection problem with codes from \cite{Matt1989, Matt1991}. Although the dependency on inclination is consistent between the~two figures, \cite{Ratheesh2021} obtained about twice lower resulting polarisation, likely because they solved for more isotropic illumination of the disc. Our mitigation of~the~incident angles to $0 \leq \mu_\textrm{i} \leq 0.3$ then again reduces the~allowed scattering plane directions. The inclination dependence of polarisation obtained in our model is also consistent with the results predicted by single-scattering Chandrasekhar's approximation for reflection (\ref{chandra1}), integrated in $\mu_\textrm{i}$ and $\Phi_\textrm{e}$ accordingly, which is shown in Figure \ref{chandra_comparison}. The fact that the absolute values of~polarisation degree and angle are matching too is natural, as for nearly neutral media we expect results closer to the single-scattering results, while highly ionized reflection should rather be approximated by the diffuse Chandrasekhar's approximation.

We add two other cases of integration ranges of the incident angles for Chandrasekhar's single-scattering approximation in Figure \ref{chandra_comparison}. The unpolarized result is not achieved for fully isotropic irradiation, but rather for $0 \leq \mu_\textrm{i} \lessapprox 0.8$. If we continue to increase the range of illumination angles beyond this range up~to~$0 \leq \mu_\textrm{i} \leq 1$, the resulting polarisation switches orientation to become parallel with the disc, as the rays arriving at the disc in vertical direction begin to~dominate the polarisation output. We thus display the absolute value of~polarisation, $\lvert p \rvert$, to~directly compare the obtained polarisation fraction between the~three cases. The~fully isotropic irradiation case reaches maximum polarisation fraction at $\approx 20\%$ for edge-on observers, as we mentioned already in~Section \ref{modelling}. The~orthogonal net polarisation then causes inverted response of~the~emergent polarisation degree to the incident polarisation angle. The resulting orientation means that elementary right-angle principles are in play and that the same explanations are behind the dominant polarisation orientation emerging from the relativistic lamp-post and sandwich coronal reflection models discussed in Chapter \ref{chap02}. While the lamp-post, at least for low heights $h$, is rather causing the~projected photon paths to be equatorial and results in disc-reflected polarisation angles nearly parallel with the~principal axis, the extended slab coronal model presented imposes higher relative contribution of photons emitted vertically towards the~disc (the local reflection tables in {\tt KYNSTOKES} for extended coronae are indeed integrated in the entire range of~$0 \leq \mu_\textrm{i} \leq 1$), which in most cases results in prevailing disc-reflected polarisation angles parallel to the disc.

\section[Polarized X-ray emission from an entire active galactic \newline nucleus]{Polarized X-ray emission from an entire \newline active galactic nucleus}\label{full_AGN_model_MC}

Last but not least, we attempted for a toy model of an entire AGN that is schematically shown and parametrized in Figure \ref{total_model}. The modeling is described in~\cite{Podgorny2023d} and summarized here below. We return to the MC simulations of the reprocessings in parsec-scale components of AGNs performed entirely with {\tt STOKES} \textit{v2.07}, as in Section \ref{equatorial_MC}. We now account also for direct emission. We assume a 3-component model consisting of (i) a point source of~emission located in~the~center of coordinates, representing the inner disc-corona system, (ii)~a~homogeneous wedge-shaped equatorial torus (Case A) discussed in Section \ref{equatorial_MC}, which is glued at fixed half-opening angle $H' = 60^\circ$ to (iii) two axially symmetric homogeneous polar scattering regions of the same wedge-shaped meridional profile, representing polar winds in form of NLRs. The same model was adopted for type-1 ($\theta = 20^\circ$) and type-2 ($\theta = 70^\circ$) AGN studies in \cite{Marin2018c,Marin2018b}. Here we replace the central emission lamp-post model used in~\cite{Marin2018c,Marin2018b} with the {\tt KYNSTOKES} model from \cite{Podgorny2023} in~the~lamp-post regime. The~difference from \cite{Marin2018c,Marin2018b} in the central illumination is the newly computed local disc reflection tables that allow to~study partially ionized disc effects and a correction in the polarized GR radiative transfer between the lamp and the disc. We will argue that even a simple 3-component model with circumnuclear components of uniform density and one geometry can produce diverse X-ray polarisation outcome, similarly to the discussion in Section \ref{equatorial_MC}. General observational prospects for this model with respect to the \textit{IXPE} and \textit{eXTP} missions, i.e. without a focus on a particular source, will be given in~Chapter \ref{chap05}.

As in \cite{Marin2018c,Marin2018b}, regarding the central emission region we tested the~limiting examples of black-hole spins $a = 0$ and $a = 1$, while examining a~nearly neutral disc \cite[$L_\textrm{X}/L_\textrm{Edd} =  0.001$ and $M/M_\odot =  10^8 $  in {\tt KYNSTOKES}, comparable to the neutral disc computations in][]{Marin2018b,Marin2018c} and highly ionized disc reflection ($L_\textrm{X}/L_\textrm{Edd} = 0.1 $ and $M/M_\odot = 10^5 $ in {\tt KYNSTOKES}). We evaluated the cases of unpolarized and $2\%$ parallelly and perpendicularly polarized primary power-law (with $\Gamma = 2$), prescribed at the lamp-post, which are denoted by $p_0 = 0\%$, $p_0 = 2\%$ and \mbox{$p_0 = -2\%$.}\footnote{\,For the plots with resulting polarisation state with energy we will, however, show the~polarisation angle next to a positively defined polarisation degree, as GR effects might result in~different than parallelly or perpendicularly oriented polarisation.} We stress that the lamp-post coronal geometry is a) simplistic and b)~not supported by the recent \textit{IXPE} discoveries (see Chapter \ref{chap05}). However, it is in line with one-to-one comparisons to \cite{Marin2018c,Marin2018b} and by assessing different polarisation angles, we may also mimic different coronal geometries.\footnote{\,For the primary emission in lamp-post geometries, 
we expect rather orthogonal polarisation to the principal axis. The opposite parallel orientation is typically the theoretical prediction for slab coronal geometries, elongated in the equatorial plane \citep{Ursini2022, Krawczynski2022a, Krawczynski2022}.} Investigation of the impact of coronae with non-negligible size with respect to~the~distant components or of any sort of thermal radiation in the X-rays is exceeding the extent of this work.

In the same way as in \cite{Marin2018c,Marin2018b}, we consider semi-isotropic illumination of the parsec-scale scattering regions, i.e. the polar winds are illuminated by the~{\tt KYNSTOKES} output for $\theta = 20^{\circ}$ and the equatorial regions by the~{\tt KYNSTOKES} output for $\theta = 70^{\circ}$. Relativistic effects are neglected in the subsequent MC simulation that models only the reprocessing in the distant equatorial and polar components, which happens simultaneously (one photon can escape and reach any scattering region at any time during its journey). The faraway scatterers are considered cold, as they do not thermally radiate, but we again treat the~partial ionization of these regions simplistically by means of free uniform density of electrons $n_\textrm{e}$. We consider different values neutral hydrogen column densities of the torus and polar winds, denoted by $N_\textrm{H}^\textrm{eq} = n_\textrm{H}(r_\textrm{out}^\textrm{torus} - r_\textrm{in}^\textrm{torus})$ and~$N_\textrm{H}^\textrm{wind} = n_\textrm{H}(r_\textrm{out}^\textrm{wind} - r_\textrm{in}^\textrm{wind})$, respectively. We use again the solar abundance from \cite{Asplund2005} with $A_\mathrm{Fe} = 1.0$. For type-2 viewing angles of AGNs, in~addition to~the~absorbing polar winds, the cases of ionized polar winds\footnote{\,For ionized winds we opted for $\tau_\textrm{e} = \sigma_\textrm{T}(r_\textrm{out}^\textrm{wind}-r_\textrm{in}^\textrm{wind})n_\textrm{e} = 0.03$, as further increase of free electron density introduces overly ionized component that dominates the X-ray polarisation output.} and~no polar winds are considered, as in \cite{Marin2018b}. Direct comparisons with previous studies are the main reason for the configurations adopted. The neutral polar winds are less physical, but could be an illustrative example in the spirit of aspiring to understand what inside the system causes what on the output and which processes are negligible compared to others. We refer to Tables \ref{table_model_grid_type1} and~\ref{table_model_grid_type2} that summarize the model grid for type-1 and type-2 AGNs, respectively, and introduce the 2--8 keV average polarisation results (in positive and negative polarisation convention), as well as observational prospects that will be elaborated on~in~Chapter \ref{chap05}. We refer to \cite{Marin2018c,Marin2018b} for details on~the~parameter space explored that is physically and observationally motivated and~that we fully adopt here, because the principal aim is to track to what extent the~change in~inner emission region plays a role. We will show, however, that other (unexpected) discrepancies from \cite{Marin2018c,Marin2018b} were also identified with the~latest simulations.
\begin{table*}
\centering
	\caption{{\footnotesize The main configuration space for type-1 AGNs. In addition to those parameters studied in \citet{Marin2018c,Marin2018b}, the effect of partial disc ionization was tested with {\tt KYNSTOKES}. The column densities $N_\textrm{H}^\textrm{eq}$ and $N_\textrm{H}^\textrm{wind}$ are provided in $\textrm{cm}^{-2}$. Top value in~each slot (black bold) is the model unfolded 2--8 keV average polarisation degree, $p$, for a given configuration. The rest of the table will be discussed in Chapter \ref{chap05}. The three values in the left column in each slot (blue) are the estimated observational times, $T_\textrm{obs} \, [\textrm{Ms}]$, that are required for the model $|p|$ to exceed the simulated $M\!D\!P_{99\%}$ for \textit{IXPE} (evaluates whether the polarisation is to be detected at a $99\%$ confidence level, although here we compare only the unfolded model), using the unweighted approach in {\tt IXPEOBSSIM}, if the observed X-ray flux is \mbox{$F_\mathrm{X, 2-10} = 1 \times 10^{-10}, 5 \times 10^{-11}, 1 \times 10^{-11} \, \textrm{erg cm}^{-2} \, \textrm{s}^{-1}$} from top to bottom, respectively, and that were linearly interpolated in the computed $\{T_\textrm{obs},F_\textrm{X,2-10}\}$ space. The three values in the right column in each slot (red) are the estimated observed X-ray fluxes, $F_\mathrm{X, 2-10} \, [10^{-10} \, \textrm{erg cm}^{-2} \, \textrm{s}^{-1}]$, that are required for the model $|p|$ to exceed the simulated $M\!D\!P_{99\%}$, using the unweighted approach in {\tt IXPEOBSSIM}, if the observational time is $T_\mathrm{obs} = 0.5, 1, 1.5 \, \textrm{Ms}$ from top to bottom, respectively, and that were linearly interpolated in~the~computed $\{T_\textrm{obs},F_\textrm{X,2-10}\}$ space. The results are serving as first-order estimates only, see Chapter \ref{chap05} for details. For a particular source, more specific information on its composition may be obtained, which enables detectability predictions with higher accuracy. The $M\!D\!P_{99\%}$ values can be also slightly reduced by means of weighted approach in {\tt IXPEOBSSIM}, which is available for real data analysis with \textit{IXPE} \citep{DiMarco2022}. The observational times needed for~\textit{IXPE} may be reduced by about a factor of 4 for \textit{eXTP} due to its larger effective mirror area and other instrumental differences.}}
 \resizebox{0.75\textwidth}{!}{%
\begin{tabular}{ccc|llllll|ll|ll|}
\cline{4-13}
\multicolumn{1}{l}{}            & \multicolumn{1}{l}{}    & \multicolumn{1}{l}{} & \multicolumn{10}{c}{{\color[HTML]{000000} type-1 AGNs (absorbing winds only)}}                                                                                                                                                                                                                                                                                                                                                                                                                            \\ \cline{4-13} 
\multicolumn{1}{l}{}            & \multicolumn{1}{l}{}    & \multicolumn{1}{l}{} & \multicolumn{6}{c}{{\color[HTML]{000000} $N_\textrm{H}^\textrm{wind} =10^{21} $}}                                                                                                                     & \multicolumn{2}{c}{{\color[HTML]{000000} $N_\textrm{H}^\textrm{wind} =10^{22} $}}                                                                                                                     &  \multicolumn{2}{c}{{\color[HTML]{000000}  $N_\textrm{H}^\textrm{wind} =10^{23}$}}                                                                                                                                                                                   \\ \cline{4-13} 
\multicolumn{1}{l}{}            & \multicolumn{1}{l}{}    & \multicolumn{1}{l}{} & \multicolumn{2}{l}{{\color[HTML]{000000} $N_\textrm{H}^\textrm{eq}=10^{23}$}}       & \multicolumn{2}{l}{{\color[HTML]{000000} $N_\textrm{H}^\textrm{eq}=10^{24}$}}       & \multicolumn{2}{l}{{\color[HTML]{000000} $N_\textrm{H}^\textrm{eq}=10^{25} $}}       & \multicolumn{2}{l}{{\color[HTML]{000000} $N_\textrm{H}^\textrm{eq}=10^{24} $}}       & \multicolumn{2}{l}{{\color[HTML]{000000} $N_\textrm{H}^\textrm{eq}=10^{24}$}}           \\
\hline \hline
\end{tabular}}
\label{table_model_grid_type1}
\end{table*}

\begin{table*}
\centering
	\caption{{\footnotesize The same as in Table \ref{table_model_grid_type1}, but for the studied type-2 AGNs cases. The tested flux values for interpolation of observational times (in blue from top to bottom in each configuration) are $F_\mathrm{X, 2-10} = 1 \times 10^{-11}, 9 \times 10^{-12}, 8 \times 10^{-12} \, \textrm{erg cm}^{-2} \, \textrm{s}^{-1}$.}}
 \resizebox{\textwidth}{!}{%
\begin{tabular}{ccc|llllll|llllll|llllll|}
\cline{4-21}
\multicolumn{1}{l}{}            & \multicolumn{1}{l}{}    & \multicolumn{1}{l}{}                                                                                                        & \multicolumn{18}{c}{{\color[HTML]{000000} type-2 AGNs}}                                                                                                                                                                                                                                                                                                                                                                                                                                                           \\ \cline{4-21} 
\multicolumn{1}{l}{}            & \multicolumn{1}{l}{}    & \multicolumn{1}{l}{}                                                                                   & \multicolumn{6}{c}{{\color[HTML]{000000} absorbing winds with $N_\textrm{H}^\textrm{wind} =10^{21}$}}                                                                                                                     & \multicolumn{6}{c}{{\color[HTML]{000000} ionized winds}}                                                                                                                     & \multicolumn{6}{c}{{\color[HTML]{000000} no winds}}                                                                                                                     \\
\hline \hline
\end{tabular}}
\label{table_model_grid_type2}
\end{table*}

~\

\begin{figure}
	\includegraphics[width=0.49\columnwidth]{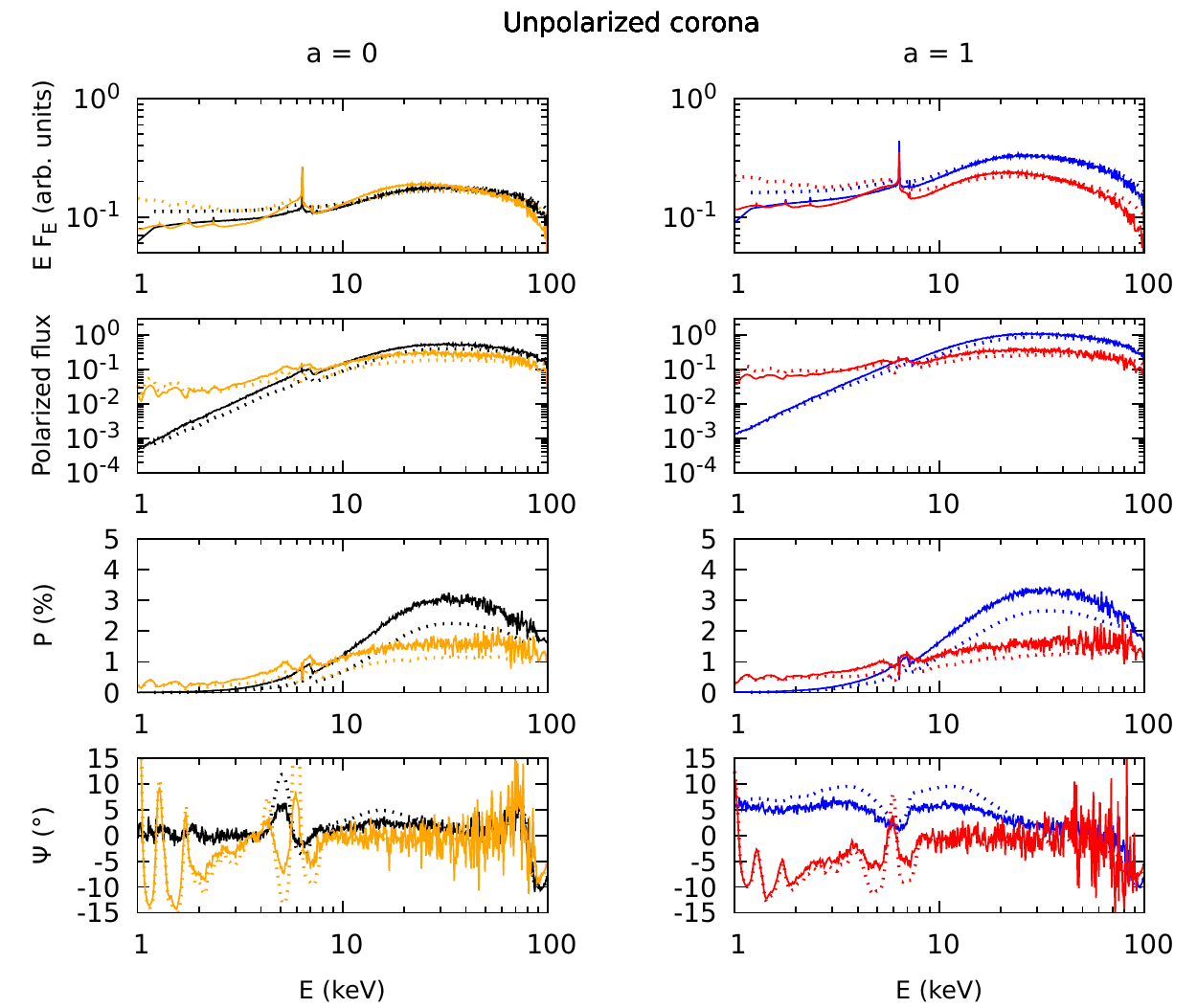}
        \includegraphics[width=0.49\columnwidth]{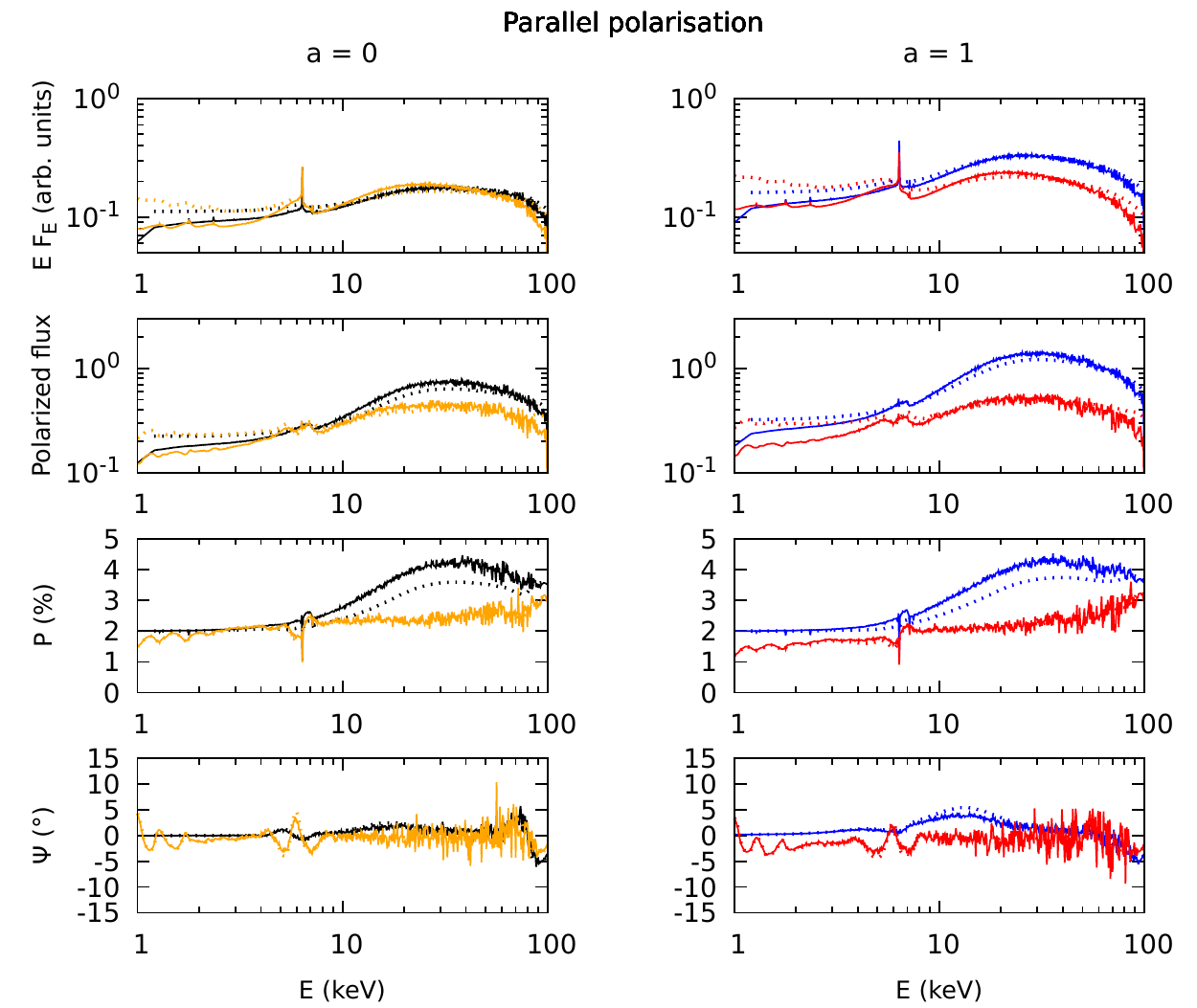}
	\caption{{\footnotesize The total type-1 AGN emission in the $\theta = 20\degr$ direction in the case of unpolarized primary radiation (solid lines) and the incident disc-corona emission in the polar direction (dotted lines). We display from top to bottom the energy-dependent flux (in arbitrary units, corrected for the primary power-law slope) and the corresponding polarized flux, polarisation degree and polarisation angle. From left to right: the case of unpolarized primary for \mbox{$a = 0$} and~$a = 1$ and the case of $2\%$ parallelly polarized primary for $a = 0$ and $a = 1$. The computations from \citet{Marin2018c} are displayed in black and blue. The new computations for~\textit{ionized} disc are displayed in yellow and red. The winds are neutral with the column density \mbox{$N_\textrm{H}^\textrm{wind} = 10^{21} \, \textrm{cm}^{-2}$.} The column density of the torus is $N_\textrm{H}^\textrm{eq} = 10^{24} \, \textrm{cm}^{-2}$. See \citet{Marin2018c} for the remaining parameters. Image adapted from \cite{Podgorny2023d}.}}
	\label{ionizedKY_UnpolRG_Results_nh24_absorbing_winds}
\end{figure}

\begin{figure}
	\includegraphics[width=0.49\columnwidth] {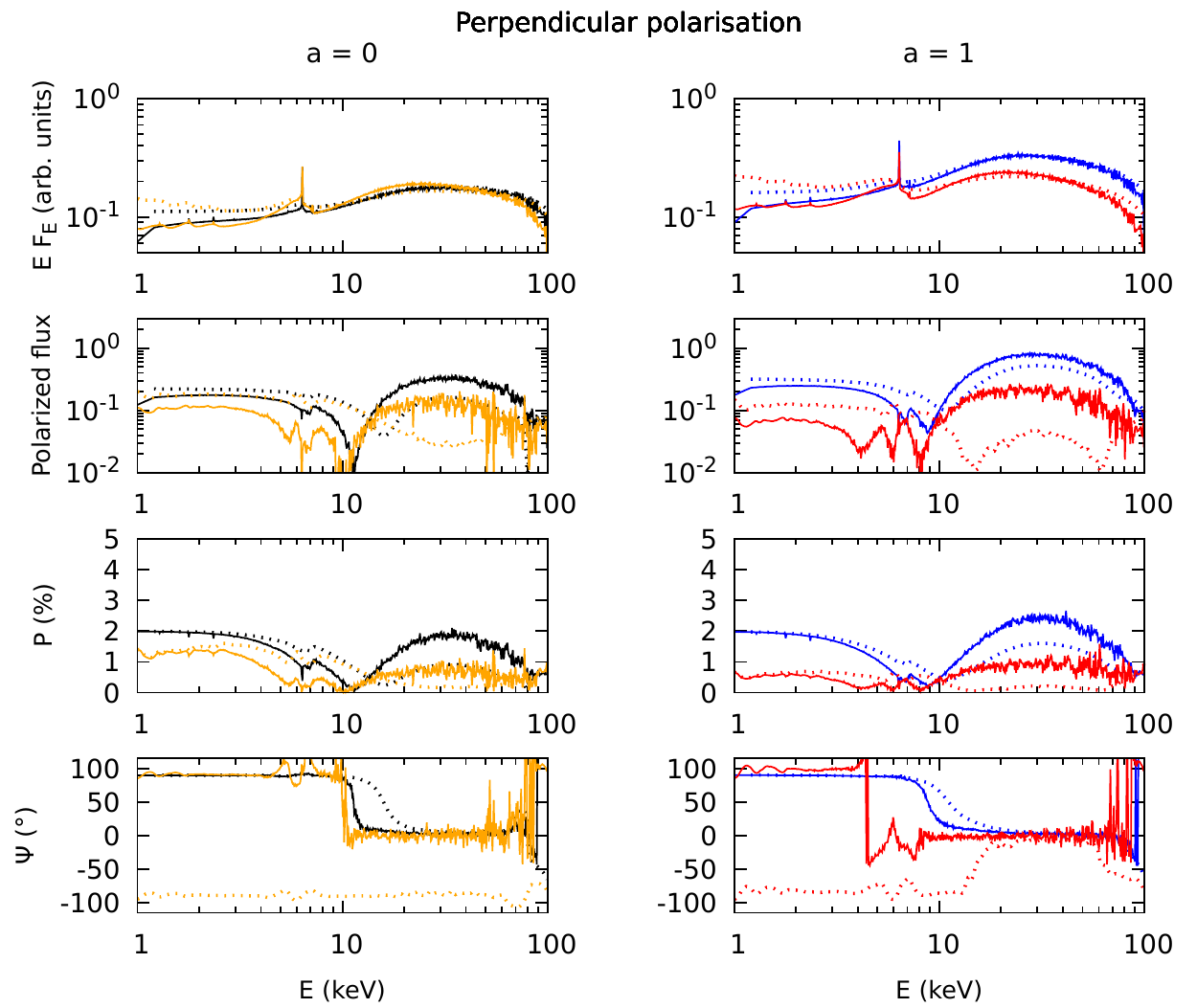}
        \includegraphics[width=0.49\columnwidth]{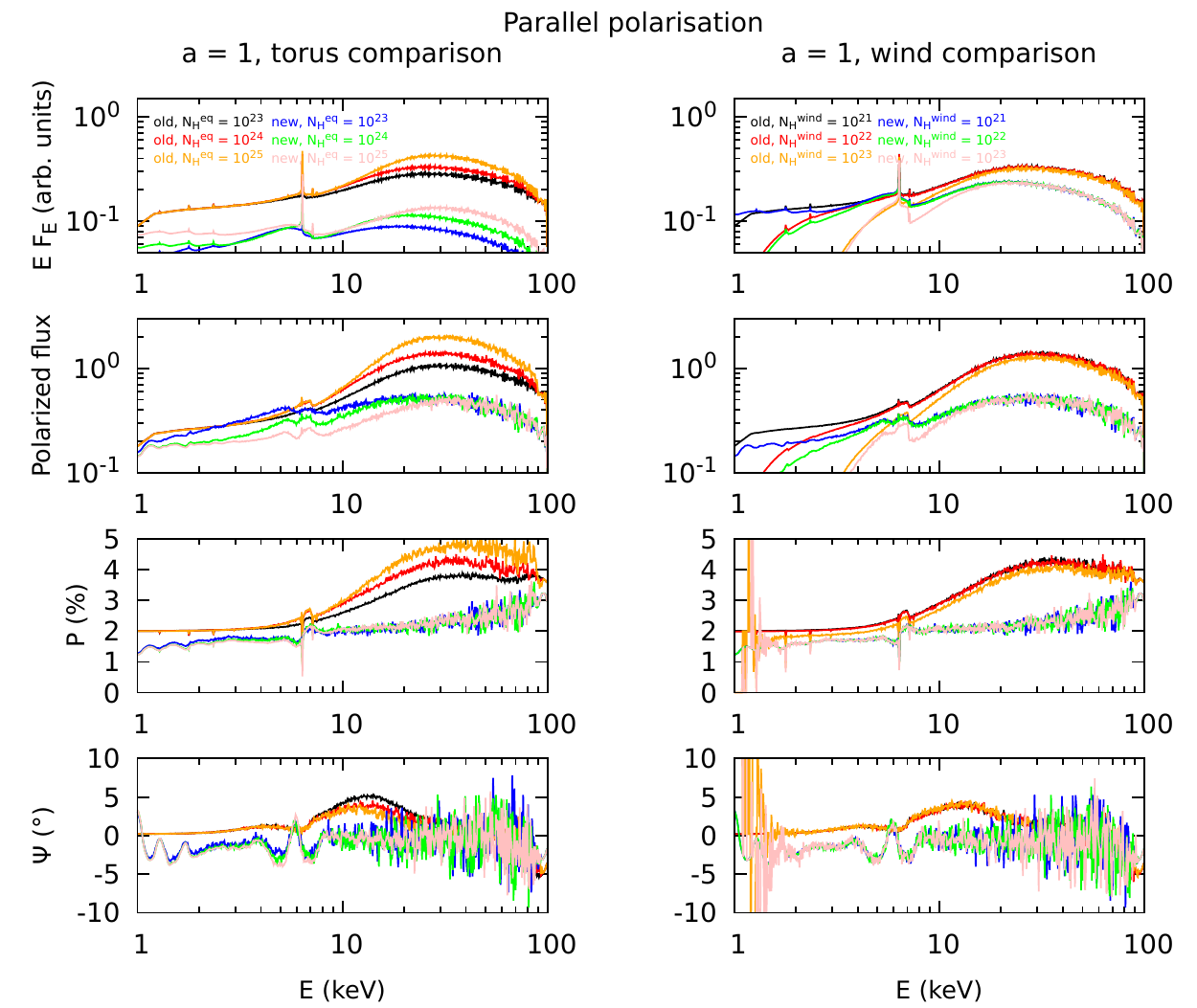}
	\caption{{\footnotesize From top to bottom the energy-dependent flux (in arbitrary units, corrected for~the~primary power-law slope) and~the~corresponding polarized flux, polarisation degree and~polarisation angle. Left: the same as in Figure \ref{ionizedKY_UnpolRG_Results_nh24_absorbing_winds}, but for $2\%$ perpendicularly polarized coronal radiation for $a = 0$ and $a = 1$. Right: comparison of torus densities and polar wind densities for type-1 AGNs. The total AGN emission for the comparisons are in the $\theta = 20\degr$ direction and for the case of $2\%$ parallelly polarized primary radiation and $a = 1$.  For the comparisons of different tori ($3^\textrm{rd}$ column from the left) we display the results for torus column densities: $N_\textrm{H}^\textrm{eq} = 10^{23} \, \textrm{cm}^{-2}$ (black), $N_\textrm{H}^\textrm{eq} = 10^{24} \, \textrm{cm}^{-2}$ (red), $N_\textrm{H}^\textrm{eq} = 10^{25} \, \textrm{cm}^{-2}$ (orange) for~the~computations from \citet{Marin2018c} and the same torus column densities $N_\textrm{H}^\textrm{eq} = 10^{23} \, \textrm{cm}^{-2}$ (blue), $N_\textrm{H}^\textrm{eq} = 10^{24} \, \textrm{cm}^{-2}$ (green), $N_\textrm{H}^\textrm{eq} = 10^{25} \, \textrm{cm}^{-2}$ (pink) for the new computations for \textit{ionized} disc. The winds are neutral with the column density $N_\textrm{H}^\textrm{wind} = 10^{21} \, \textrm{cm}^{-2}$. For the comparisons of different winds ($4^\textrm{th}$ column from the left) we display the results for wind column densities: $N_\textrm{H}^\textrm{wind} = 10^{21} \, \textrm{cm}^{-2}$ (black), $N_\textrm{H}^\textrm{wind} = 10^{22} \, \textrm{cm}^{-2}$ (red), $N_\textrm{H}^\textrm{wind} = 10^{23} \, \textrm{cm}^{-2}$ (orange) for the computations from \citet{Marin2018c} and the same wind column densities $N_\textrm{H}^\textrm{wind} = 10^{21} \, \textrm{cm}^{-2}$ (blue), $N_\textrm{H}^\textrm{wind} = 10^{22} \, \textrm{cm}^{-2}$ (green), $N_\textrm{H}^\textrm{wind} = 10^{23} \, \textrm{cm}^{-2}$ (pink) for~the~new computations for~\textit{ionized} disc. See \citet{Marin2018c} for~the~remaining parameters. The torus has the column density of $N_\textrm{H}^\textrm{eq} = 10^{24} \, \textrm{cm}^{-2}$. Image adapted from \cite{Podgorny2023d}.}}
	\label{ionizedKY_PolGRPerp_Results_nh24_absorbing_winds}
\end{figure}

\begin{figure}
	\includegraphics[width=0.49\columnwidth]{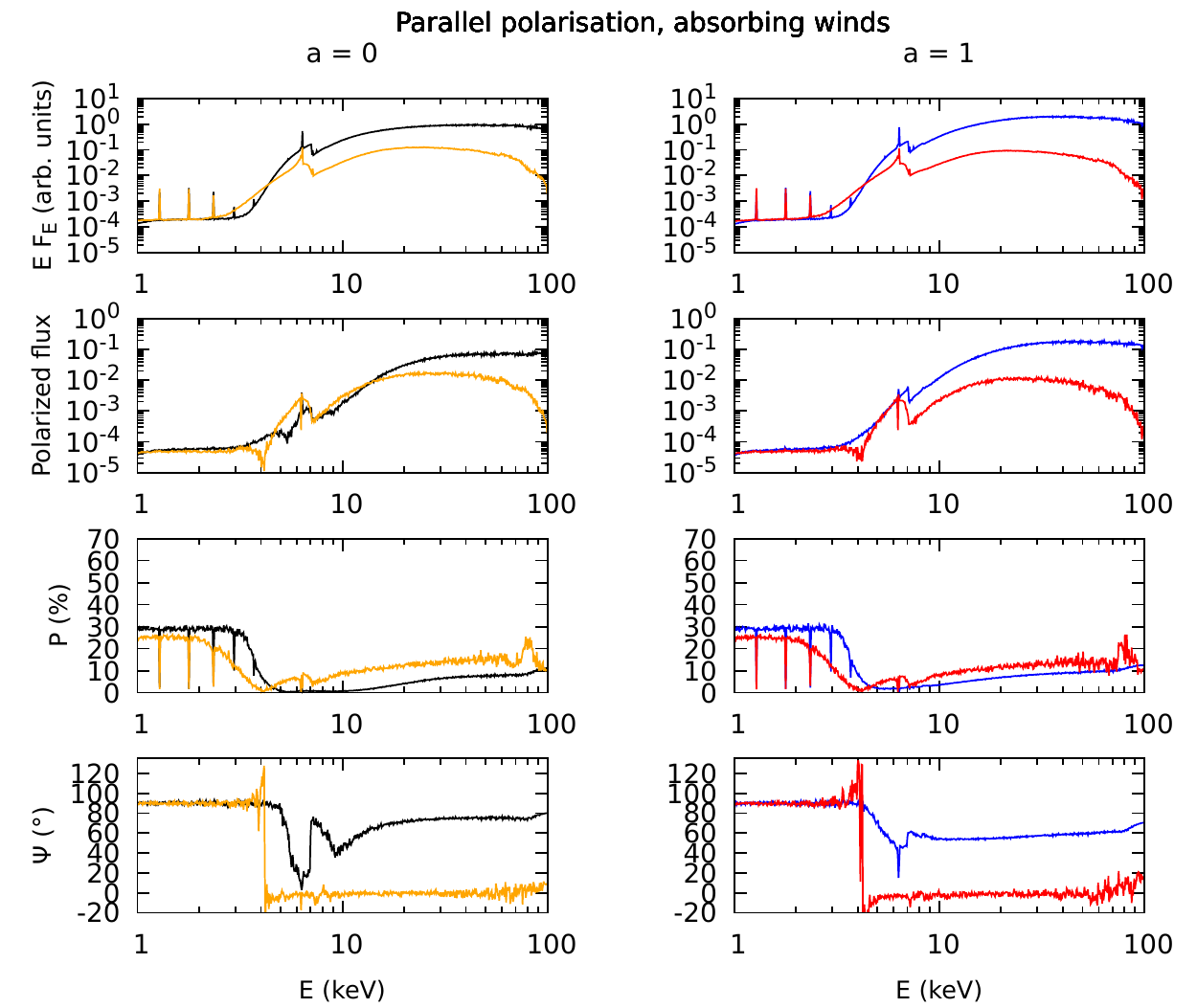}
        \includegraphics[width=0.49\columnwidth]{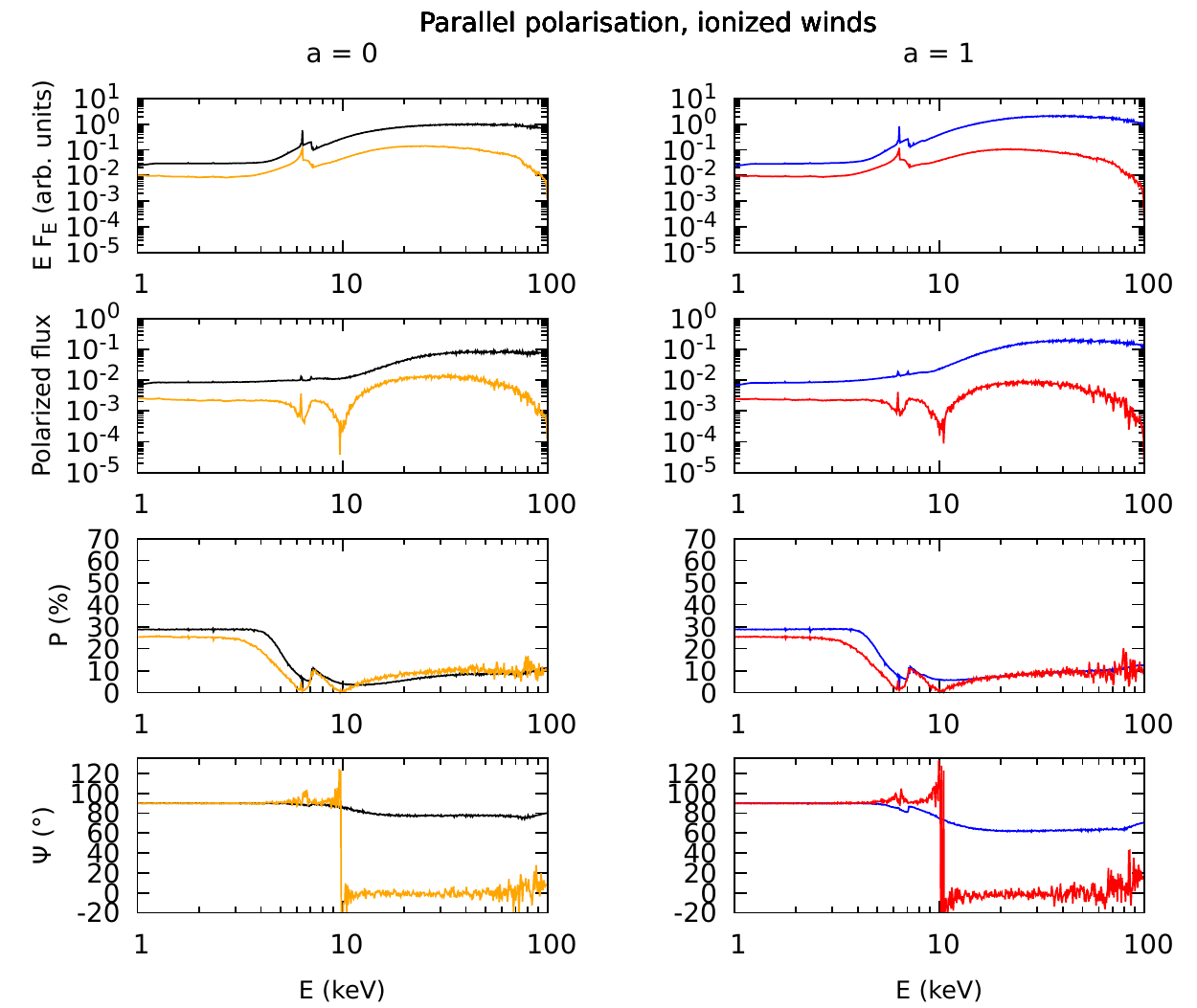}
	\caption{{\footnotesize The total type-2 AGN emission in the $\theta = 70\degr$ direction in the case of $2\%$ parallelly polarized coronal radiation. We display from top to bottom the energy-dependent flux (in~arbitrary units, corrected for the primary power-law slope) and the corresponding polarized flux, polarisation degree and polarisation angle. From left to right: the case of absorbing winds with $N_\textrm{H}^\textrm{wind} = 10^{21} \, \textrm{cm}^{-2}$ for $a = 0$ and $a = 1$ and the case of ionized winds for $a = 0$ and $a = 1$. The computations from \citet{Marin2018b} are displayed in black and blue. The new computations for \textit{ionized} disc are displayed in yellow and red. The torus column density is set to $N_\textrm{H}^\textrm{eq} = 10^{24} \, \textrm{cm}^{-2}$. See \citet{Marin2018b} for the remaining parameters. Image adapted from \cite{Podgorny2023d}.}}
	\label{type2_ionizedKY_PolGRPara_Results_nh24_absorbing_winds}
\end{figure}

Although the entire grid from \cite{Marin2018c, Marin2018b} was reproduced, we show here only the most representative spectral and polarisation results versus energy in the full 1--100 keV studied range. We will compare our latest synthetic spectra and polarisation directly in the same panels to the older simulations from \cite{Marin2018c, Marin2018b}, which are differentiated by the color code. Figures \ref{ionizedKY_UnpolRG_Results_nh24_absorbing_winds} and \ref{ionizedKY_PolGRPerp_Results_nh24_absorbing_winds} show type-1 AGNs comparing different cases of primary radiation, black-hole spins, torus densities and polar wind densities \citep[all for ionized accretion discs, which is the case that should differ more from ][]{Marin2018c,Marin2018b}. Comparisons of type-2 AGNs for the absorbing winds and ionized winds cases are shown in~Figure~\ref{type2_ionizedKY_PolGRPara_Results_nh24_absorbing_winds}. The 2--8 keV average results, which were obtained by adding counts in $I$, $Q$ and $U$ in the same bins, are provided for the entire computational grid in Tables \ref{table_model_grid_type1} and \ref{table_model_grid_type2}. To complete the picture, we show the incident radiation properties in the $\theta = 20^{\circ}$ and $\theta = 70^{\circ}$ directions for type-1 and type-2 AGNs, respectively, in Figures \ref{ionizedKY_TF_PO_PA_inputs}--\ref{type2_neutralKY_TF_PO_PA_inputs}. These figures then again allow for a direct comparison with the incident radiation used for the simulations in \cite{Marin2018c, Marin2018b} in the same directions. We note that the orientation convention for $\Psi$ was different in \cite{Marin2018c, Marin2018b} by $90^\circ$ from the one adopted here; thus, we rotate the results from the old simulations to stick to the current conventions.

~\

The results can already serve for first-order estimates of X-ray polarisation of radio-quiet AGNs. All in all, we discovered more discrepancy between the~old and~new simulations for type-2 AGNs. This might sound surprising, because the~spectropolarimetric properties of the central source should play a role mostly for unobscured sources. We discovered, however, some errors in the computations of type-2 AGNs in \cite{Marin2018b}, where the total output of the simulation was, in an unknown way, wrongly rotated to resemble the central emission. This caused the type-2 AGNs to preserve a strong dependency on the central emission properties in \cite{Marin2018b}, which we no longer see in the latest results. Our latest results for type-1 AGNs rather agree, but in the 10--100 keV band we obtained about half of the polarisation fraction that was previously estimated in~\cite{Marin2018c}. Regarding energy dependence for both type-1 and~type-2 AGNs, we obtained most dissimilarities from the older publications rather in~the~hard X-rays (10--100 keV) than in~the~2--8~keV band important for~{\it IXPE}.

~\

Let us first discuss in more depth the studied type-1 AGNs. We showed in~Section \ref{equatorial_MC} that even for the type-1 observers the composition and geometry of~the~equatorial scattering region greatly affects the X-ray polarisation properties of the \textit{reprocessed} radiation. With the \textit{total} emission results presented in this section we, however, estimate that the properties of the parsec-scale components may alter the X-ray polarisation degree output only up to a few \% for type-1 AGNs with respect to the polarisation fraction of the bare nuclear component. Typically the parallelly polarized primary cases result in unaltered polarisation degree. The unpolarized primary cases result in $\approx 1\%$ additional polarisation with either parallel or perpendicular orientation (for meridional symmetry reasons), depending on the properties of the parsec-scale reprocessing components and observer's inclination (see Section \ref{equatorial_MC} for details). The perpendicularly polarized incident emission tends to be gradually depolarized on the output with energy: at soft X-rays preserving its original polarisation state and then reaching the unpolarized state at around 10 keV, continuing at even higher energies in parallel polarisation direction with increasing polarisation degree with energy. Similarly to the arguments for equatorial reprocessings that we used in Section \ref{equatorial_MC}, we see a competition between a dominantly perpendicularly polarized primary emission and dominantly parallelly polarized reprocessed radiation in~the~last case mentioned. As we showed in Section \ref{lamppost_spproperties} with the {\tt KYNSTOKES} code in lamp-post regime, especially for the neutral disc with low black-hole spin values, the reflection component is negligible compared to the primary in the soft X-rays. Hence, the primary radiation properties are rather preserved, at least towards the poles. In the equatorial directions, the nuclear emission may be parallelly polarized at all studied energies (see Figure \ref{type2_ionizedKY_TF_PO_PA_inputs}). The~reprocessing in distant components then adds a parallelly polarized contribution to the total signal especially in the hard X-rays, where the photons are less absorbed. The energy transition for~the~cases with perpendicularly polarized corona appears sharper and at slightly lower energies than in \cite{Marin2018c} due to~the~model differences and due to choice of a particular inclination bin (see below).

In \cite{Marin2018c}, a comparison for different torus and wind densities was provided for type-1 AGNs, which we also compare to our latest results in Figure \ref{ionizedKY_PolGRPerp_Results_nh24_absorbing_winds} (right). Performing the same analysis, we reached a reasonable agreement for comparison of different wind densities, but we note again some differences of unknown origin for the torus comparisons. Part of the dissimilarities can be attributed to the choice of a particular inclination bin (see below). In general, we also observe a slight swing in the polarisation angle around the iron line at 6.4 keV with respect to the nearly energy-independent polarisation angle seen in \cite{Marin2018c}. This may be due to the lower predicted polarisation in the line itself (after the reprocessing, which generally adds polarisation to an unpolarized fluorescent emission at the photon creation location), as the polarisation angle is undefined for unpolarized light.

~\

As in \cite{Marin2018b}, for type-2 AGNs we obtain the same prediction for~$\gtrapprox 25\%$ perpendicularly polarized emission in the soft X-rays that arises from high absorption in the dusty torus and high contribution to polarisation from the polar winds, which reflect dominantly in the right scattering angles in~the~meridional plane. The trade-off for detectability by polarimeters at these energies is obviously the low X-ray flux caused by Compton thickness. As advertised before, we conclude that the GR effects do not play any role for the type-2 X-ray polarimetric observers. More generally speaking, the spectropolarimetric properties of~the~primary radiation and the inner-accreting region characteristics, now also probed for partial ionization of the disc atmospheres thanks to the latest {\tt KYNSTOKES} code, are expected to be to a large extent washed out for Compton-thick AGNs. In~the~hard X-rays, the resulting emission is parallelly polarized with a slight increase with energy up to $\approx 10\%$ near 100 keV.

~\

We discovered that in general the results are highly dependent on the inclination bin chosen when proceeding with 20 inclination bins linearly spaced in~cosine of $\theta$, as we did. We opted for $\cos{\theta} = 0.375$ and $\cos{\theta} = 0.975$ in~the~simulations for type-2 and type-1 AGNs, respectively, being close to the stated $\theta = 70\degr$ and~$\theta = 20\degr$ viewing angles stated in \cite{Marin2018c,Marin2018b}. But if, for example, the~neighboring simulation bins $\mu_\textrm{e} = 0.325$ and $\mu_\textrm{e} = 0.925$ are selected, $\Psi$ may change in the order of a few degrees and $p$ in the order of $1\%$. The energy transition in $p$ and $\Psi$ that served as a diagnostic tool in \cite{Marin2018c,Marin2018b} is shifted more by the inclination bin chosen than by any claimed effects therein. Results from Section \ref{equatorial_MC} indicate the the choice of the half-opening angle (fixed to $H' = 60^\circ$ here) will also affect the X-ray polarisation properties at least for the type-2 AGNs. The $\theta$ and $H'$ parameters, alongside composition, shape and possible misalignment of the parsec-scale components thus appear to be determinative for the result, especially for the type-2 AGNs, and should be estimated before any more detailed interpretation from the X-ray polarimetric data is drawn. Constraints on such parameters can be obtained by additional modeling and from X-ray spectroscopic observations or polarimetry in other wavelengths \textit{for a particular} AGN \citep[e.g.][]{Antonucci1985, Smith2004, Goosmann2011, Gaskell2012, Marin2018d, Hutsemekers2019, Marin2020, Stalevski2023}.

We envision that the simulation results presented here would be also more diverse, if non-isotropic coronal emission was implemented to {\tt KYNSTOKES}, but the~true sensitivity to this assumption cannot be tested in the present scope. We also note that we used the semi-isotropic illumination of the parsec-scale components in the subsequent MC simulation, which is another simplification that could be tested for model sensitivity perhaps more easily, just by binning in higher angular resolution in the {\tt STOKES} input and by relaunching the simulations. It would require, however, some additional technical effort due to the {\tt STOKES} {\it v2.07} code's architecture. Any more sophisticated inner-accreting model is also possible to test within {\tt STOKES} (more easily if assumed to be of negligible size, as in the results above), if the expected central source emission is provided in a tabular format. We expect here an effect on the studied cases of type-1 AGNs, because even with a handful of first {\it IXPE} results, coronal properties of some AGNs were indeed already constrained through these observations \citep[][see Chapter \ref{chap05} for detailed discussion]{Marinucci2022, Gianolli2023, Tagliacozzo2023, Ingram2023}.

\newpage
\ 
\vspace{4.5cm} 
\begin{figure}[h]
        \centering
	\includegraphics[trim={0cm 0cm 0cm 0cm},clip,width=0.6\textwidth]{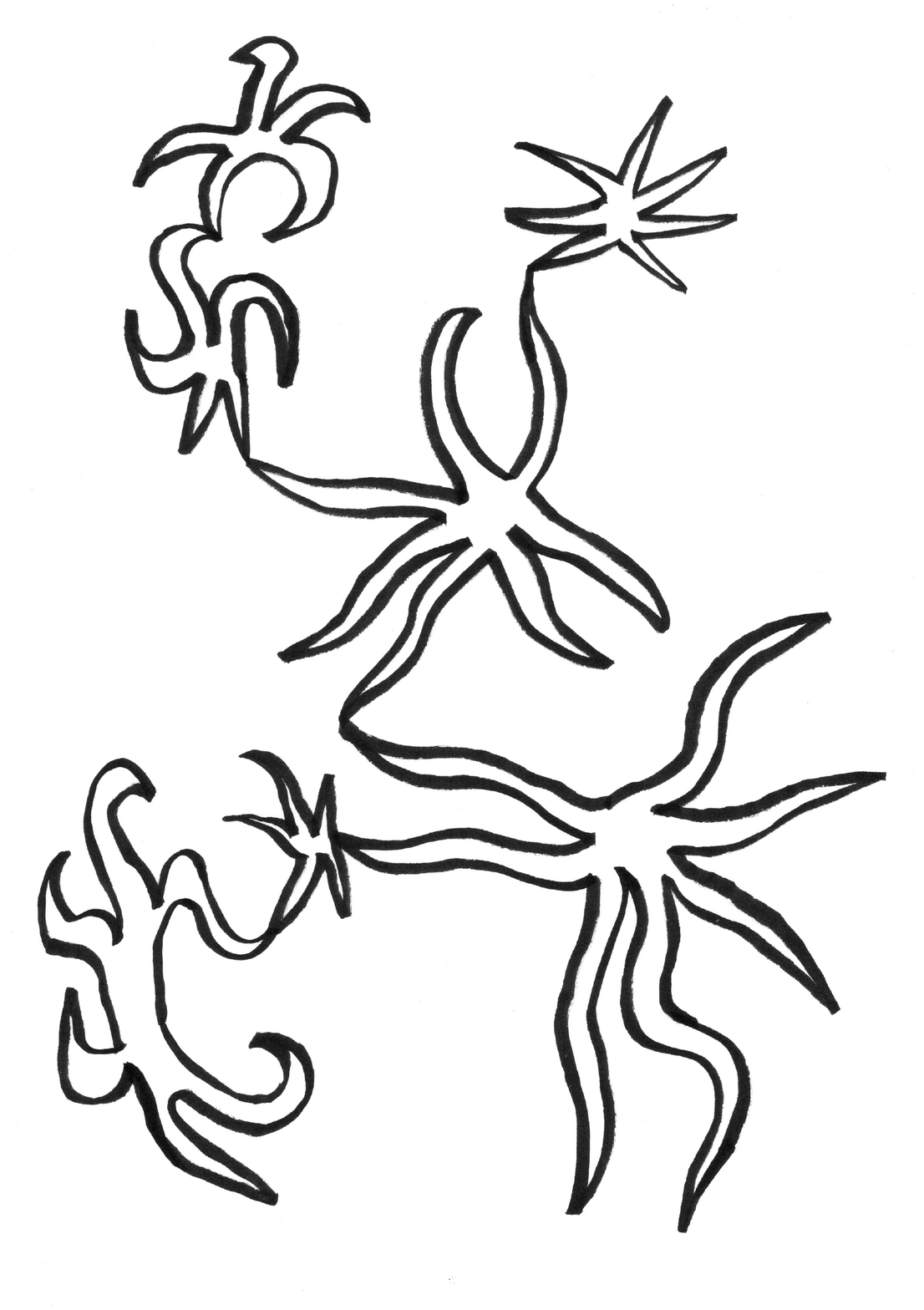}
\end{figure}
\chapter{X-ray polarimetric observations of accreting black holes}\label{chap05}

In this chapter, we provide the observational analysis and perspective with the~latest operating or planned X-ray polarimeters. Since X-ray polarimetry is currently experiencing a boom thanks to the {\it IXPE}, the results presented in this chapter and their context are highly dynamic and may be in some cases outdated shortly after the submission of the manuscript.

~\

During the first year and a half of operations, {\it IXPE} observed a number of~accreting black holes, complementarily to the observations of other interesting X-ray sources that were outlined in Section \ref{cases}. A specific category contains blazars, which show very distinct properties due to their orientation towards us \citep{DiGesu2022,Ehlert2022,Liodakis2022,DiGesu2023,Ehlert2022b,Kim2023,Marshall2023,Middei2023,Peirson2023}. In terms of observational traits, inclined accreting black holes exhibit properties rather closer to the previously mentioned weakly magnetized accreting neutron stars, which in addition to the disc, corona and~more distant components possess a boundary layer of the compact object.

Radio-quiet accreting black holes that we primarily study in this dissertation are expected to be X-ray polarized mostly due to scattering and absorption mechanisms, because ordered magnetic fields are not strongly present. The first {\it IXPE} results confirm the long-standing theoretical expectations \citep[e.g.][]{Soffita2017, Weisskopf2022} that sources possessing strong and ordered magnetic fields will be generally higher X-ray polarized than those that do not. Thus, it was rather a pleasant surprise that many accreting black holes still revealed a few percent polarized X-rays in 2--8 keV. We will now introduce these {\it IXPE} results in more detail.

~\

According to the expectations, the observed bright Galactic XRBs provided so far more positive polarisation detections with {\it IXPE} than the generally fainter extragalactic AGNs. However, we have argued enough with all the numerical models presented that even an observed polarisation upper limit can still be theoretically highly informative.

We will thus begin with presenting the recent X-ray polarimetric discoveries in accreting stellar-mass black holes and then continue with the more mysterious (and arguably more complex) AGN analogs. As part of this dissertation, we contributed to the discussions in the {\it IXPE} science analysis and simulations working group -- in the topical working groups of accreting stellar-mass black holes and radio-quiet AGNs. We participated in the observation analyses of the {\it IXPE} team campaigns listed below and loosely affected the large collaboration efforts by the ongoing theoretical work presented in this dissertation. If more detailed investigations were performed for a particular source as part of this dissertation, we devote to it a specific section below. Simultaneously, we carried simulations that should assist with future observational planning for accreting black holes with {\it IXPE} or other X-ray polarimeters mentioned in Section \ref{space_observatiories}. These are also provided below.

\section{Accreting stellar-mass black holes}\label{IXPE_results_XRB}

The first observed XRB by {\it IXPE} was Cygnus X-1 \citep{Krawczynski2022}, caught in its hard spectral state in May 2022 ($4.01\% \pm 0.20\%$ in 2--8 keV at~a~$68\%$ confidence level) and June 2022 ($3.84\% \pm 0.31\%$ in 2--8 keV at~a~$68\%$ confidence level). The obtained polarisation is perfectly aligned with the observed projected direction of the radio jet in the source \citep{MillerJones2021} in both observations. Most importantly, this confirms that a) the regions many orders of~magnitude\footnote{\,The leading figure 2 in \cite{Krawczynski2022} compares radio jet emission from \cite{MillerJones2021}, which is billion-km spatially extended against $\approx 2000$-km diameter of~the~inner accreting region, which is X-ray dominated, surrounding a $60$-km wide black-hole event horizon \citep{Krawczynski2022}.} smaller in physical size that {\it IXPE} covered are inherently linked to~the~large scale jet emission and b) that the corona is more likely equatorially elongated \citep[but note the results of][which are in contradiction to that]{Dexter2023, Moscibrodzka2023}. Number of explanations occurred for the surprisingly high polarisation degree, the most promising being higher inclination of the accretion disc or coronal medium than the orbital inclination \citep[$\theta = 27.5^{\circ} \pm 0.8^{\circ}$,][]{MillerJones2021} \citep{Krawczynski2022, Zdziarski2023}, or an outflowing Comptonization component at relativistic speeds \citep{Poutanen2023, Dexter2023}. In May 2023, Cygnus X-1 was observed again by {\it IXPE}, but in~its soft state \citep[$1.99\% \pm 0.17\%$ in 2--8 keV at a $68\%$ confidence level,][]{Dovciak2023, Jana2023, Steiner2024}. Although the polarisation degree decreased, the polarisation angle remained very similar, parallel with the~jet, despite strong presence of~the~thermal component in 2--8 keV. This raises questions on the origin of polarisation in the soft state, as usually the thermal disc emission models predict emergent polarisation parallelly to the accretion disc (see Section \ref{modelling}). But it could still be the case that the polarisation of thermal component was just diluting the still dominant polarisation from Comptonization component, orthogonally oriented.

Another example of a hard-state XRB with observed high X-ray polarisation by \textit{IXPE} was Swift J1727.8-1613 \citep{Dovciak2023c,Dovciak2023b,Veledina2023b,Ingram2023b}, a Galactic black-hole candidate discovered in August 2023 through a strong X-ray outburst \citep{Kennea2023, Negoro2023}. Although not much information about the object is known yet, it was caught first in the hard state and subsequently in transition to the soft state. The {\it IXPE} detection of $4.71\% \pm 0.27\%$ 2--8 keV polarisation (at a $68\%$ confidence level) in the hard state, aligned with the observed submillimeter polarisation \citep[which is about twice lower,][]{Vrtilek2023}, promises similar polarisation emergence conditions to the hard state of Cygnus X-1. During the transition to the soft state, the polarisation properties seem again not to be changed significantly \citep{Ingram2023b}, similarly to the Cygnus X-1 data \citep{Steiner2024}, which is much surprising, as the spectral energy distribution changes significantly. QPOs were already detected in the \textit{IXPE} flux data of~Swift J1727.8-1613 \citep{Ingram2023b,Veledina2023b}, so additional detailed analysis may reveal the expected corresponding swing in X-ray polarisation angle. Thus altogether, this newly discovered source promises gold mines for our understanding of accreting black holes, not only through the eyes of X-ray polarimetry, as many multiwavelength campaigns are ongoing.

~\

{\it IXPE} observed a number of XRB sources in the soft spectral state, which may be classified as more typical high/soft states (cf. Section \ref{all_XRBs}) than the~state-transited two sources mentioned above. The first one being 4U1630-47 ($8.3\% \pm 0.3\%$ in 2--8 keV at a $68\%$ confidence level), observed in September 2022 and providing surprisingly high polarisation degree, increasing with energy \citep{Rawat2023, Kushawa2023, Ratheesh2023}. The polarisation angle was nearly constant with energy, however, the orientation of the source on the sky from other data is unknown. We will return to the intricate possibilities of interpretation of such polarisation state in the soft state below, but what is perhaps equally puzzling is little change in polarisation when transitioning into a different state, similarly to the cases of Cygnus X-1 and Swift J1727.8-1613. The source was observed again in March of 2023 in the steep power-law state \citep{Cavero2023} and although the polarisation has decreased ($6.8\% \pm 0.2\%$ in 2--8 keV at a $68\%$ confidence level), the polarisation angle remained almost unchanged. Albeit neither here, nor in the cases of Cygnus X-1 and Swift J1727.8-1613, the second set of observations does not represent an XRB in a ``clear'' spectral state, we note that the state transitions occurred in two different directions (departing from the high/soft state in 4U1630-47 and departing from the low/hard state in Cygnus X-1 and Swift J1727.8-1613). If the thermal radiation was polarized perpendicularly to the disc, it would mean significant theoretical reconsiderations.

So far, no XRB was caught by {\it IXPE} in a clear high/soft state that would have both, statistically significant polarisation detection and known orientation on~the~sky. {\it IXPE} observed the prototypical soft-state XRB, LMC X-1, in October 2022 \citep{Podgorny2023b}. It meant only a hint of $1.0\% \pm 0.4\%$ polarisation (in 2--8 keV at a $68\%$ confidence level, but more securely stating that the observed polarisation is below the $M\!D\!P_{99\%} = 1.1\%$ in 2--8 keV, hence statistically insignificant at a $99\%$ confidence level) in the projected ionization cone direction observed in optical band \citep{Cooke2008}, i.e. most likely perpendicularly to~the~projected accretion disc plane. Low orbital inclination, if linked to the inner X-ray source inclination, is in line with the low observed polarisation. We also note that we observed a slight dilution of the thermal radiation by hard \mbox{X-ray} power-law, which may be another reason for low polarisation due to mixture of~two perpendicularly polarized components. Thus, the LMC X-1 observation does not exclude the Chandrasekhar-Sobolev predictions for thermally radiating discs, if the~power-law component was polarized perpendicularly to~the~thermal one and was driving the resulting polarisation (i.e. if it was diluted by the thermal emission in polarisation, not vice versa). As this source is remarkably stable, additional \textit{IXPE} observations of LMC X-1 may, however, reveal the true orientation of polarisation of thermal radiation as well as prove a variability with orbital phase, which was detected with low significance during the first \textit{IXPE} observation \citep{Podgorny2023b}. The latest X-ray spectroscopic analysis suggests a~presence of~a~warm and optically thick surface skin of the disc \citep{Zdziarski2023b}, which is in~accordance with the inverse disc temperature-luminosity relation for LMC X-1 \citep{Gierlinski2004} and~which could affect the resulting polarisation. Due to low statistics, it was not possible to constrain the~black-hole spin, nor disc inclination from X-ray polarimetry in this source, using the~{\tt KYNBBRR} code. Especially the determination of~spin in (not only) LMC X-1 is of~paramount importance for gravitational-wave, stellar-evolution and cosmological studies \citep[e.g.][]{Qin2019, Belczynski2021, Mehta2021, Fishbach2022, Shao2022}.

The last two XRBs caught in true soft state were 4U1957+115 (\mbox{$1.88\% \pm 0.37\%$} in 2--8 keV at~a~$68\%$ confidence level) and LMC X-3 ($3.2\% \pm 0.6\%$ in 2--8 keV at a $68\%$ confidence level), observed in May 2023 and July 2023, respectively \citep{Akash2023, Majumder2023, Svoboda2023, Kushwaha2023b, Marra2023}. They both show increasing polarisation with energy and constant polarisation angle with unknown relation to the system projected orientation. 4U1957+115 is possible to fit with conventional models of~relativistic thermally radiating standard discs, confirming a rather highly spinning black hole \citep{Marra2023}. However, the case of LMC X-3 is already slightly better fitted with thick disc atmospheres alongside lower spin values \citep{Svoboda2023}, similarly to the 4U1630-47 case, where even relativistic outflowing velocities are needed to explain the high polarisation degree to~emerge in such scenarios \citep{Ratheesh2023}. Thus, it suggests itself that the low spin of~4U1957+115 fitted with conventional models may not be of high confidence, if the~same polarisation models will fail at other sources with similar spectral properties.

~\

Last but not least, two XRBs were observed in highly obscured conditions. The ULX object SS433 was observed in its eastern lobe \citep[$17.3\% \pm 4.3\%$ in 2--5 keV at~a~$68\%$ confidence level,][]{Kaaret2023}. The extremely high polarisation may be attributed to the magnetic fields accompanying the jet, being oriented parallel to the observed projected jet direction \citep{Stirling2004} and producing synchrotron emission polarized in the perpendicular direction \citep{Kaaret2023}. Some reflection of the X-rays from inner side of the obscuring funnel may play a role too, because the polarisation is in the direction perpendicular to~the~projected jets. The latter polarisation generation mechanism is certainly applicable to~the~case of Cygnus X-3, which shows only a reflected X-ray power-law in spectra and was caught in the hard and intermediate (minor flaring) states in October and November 2022 ($20.6\% \pm 0.3\%$ in 2--8 keV at~a~$68\%$ confidence level) and December 2022 ($10.4\% \pm 0.3 \%$ in 2--8 keV at a $68\%$ confidence level), respectively \citep{Veledina2023}. The polarisation angle is orthogonal to~the~observed direction of polar ejecta \citep{Fender1999} within the measurement uncertainties. We will elaborate more below on this source, as models developed in this dissertation were used for the data interpretation. As described in Section \ref{equatorial_MC}, it is no surprise that highly obscured accreting black holes may also provide large polarisation by means of scattering and absorption only, because the~isotropy of~emission inside the system is significantly reduced.

\subsection{Analysis of the Cygnus X-1 {\it IXPE} observations}

Only a small fraction of {\it IXPE} observations of compact accreting objects showed significant disc reflection, where the {\tt KYNSTOKES} model developed in this dissertation would be directly applicable. Apart from the weakly magnetized neutron star Cygnus X-2 \citep{Farinelli2023}, where the analysis with {\tt KYNSTOKES} is ongoing, we used it for the first {\it IXPE} observation of Cygnus X-1 \citep{Krawczynski2022}. In particular we were curious if the results in the lamp-post regime could explain the observations. Mostly the observed polarisation angle disfavored the lamp-post coronal types, as the estimates from {\tt KerrC}, {\tt COMPPS} and {\tt MONK} codes (that are focused on the Comptonization component) suggested, although a possibility for the observed polarisation produced by relativistically outflowing lamp-post corona was recently reexamined by \cite{Dexter2023, Moscibrodzka2023}. In any such lamp-post assumptions, a first-order estimate on the reflection fraction versus polarisation degree gained by reflection can be made with the~{\tt KYNSTOKES} model in~lamp-post regime, which was described in Section \ref{kynstokes_lamppost} and which provides almost exclusively polarisation orientation parallel with the~rotation axis and polarisation degree increasing with energy, supporting the observed trends in Cygnus X-1 \citep{Krawczynski2022}. The polarisation by reflection is treated through Chandrasekhar's diffuse formulae in {\tt KerrC} and is neglected in {\tt MONK}. Therefore we still tested possibilities that {\tt KYNSTOKES} offers, especially regarding the truncated accretion disc and extreme incident polarisation, because if the $\approx 10\%$ of~reflection fraction in~2--8~keV \citep{Krawczynski2022} could be in any scenario $40\%$ polarized, it could explain the $4\%$ total polarisation fully for an unpolarized corona. Figures similar to Figure \ref{rin_ionized_lp_h3} were drawn for~a~large parametric space. It was concluded that with the observed 2--8 keV disc reflection fraction of $\approx 10\%$, for any disc truncation the polarisation gained by reflection in~the~equatorial plane cannot be significant enough for lamp-posts to~lead the~observed total polarisation values.

~\

Another numerical experiment, in collaboration with Wenda Zhang and~Michal Dovčiak and later on also with Sudeb Ranjan Datta, was made to trace the disc illumination with {\tt MONK} and to obtain the radius and energy-dependent spectropolarimetric values in the equatorial plane with {\tt MONK}. All available parameter constraints on Cygnus X-1 in the literature were considered. The inner disc inclination was assumed as the orbital one \citep{MillerJones2021}. The~black-hole spin was assumed as $a = 0.998$.  For simplicity we also assumed the inner disc radius to extend to the ISCO in this experiment. The lamp-post corona in {\tt MONK} was static, spherical, had a radius of $10 \, r_\textrm{g}$ and Thomson optical depth $\tau_\textrm{T} = 0.6$ and was located at height $12 \, r_\textrm{g}$. Then we used {\tt KYNSTOKES} in the lamp-post regime with the same polarisation state and primary power-law ($\Gamma = 2$) prescribed at the corona and studied the reflected-only output. This was done at~each energy band and per each disc ring, subsequently superposed.

\begin{figure}[!htb]\centering
	\includegraphics[width=\textwidth]{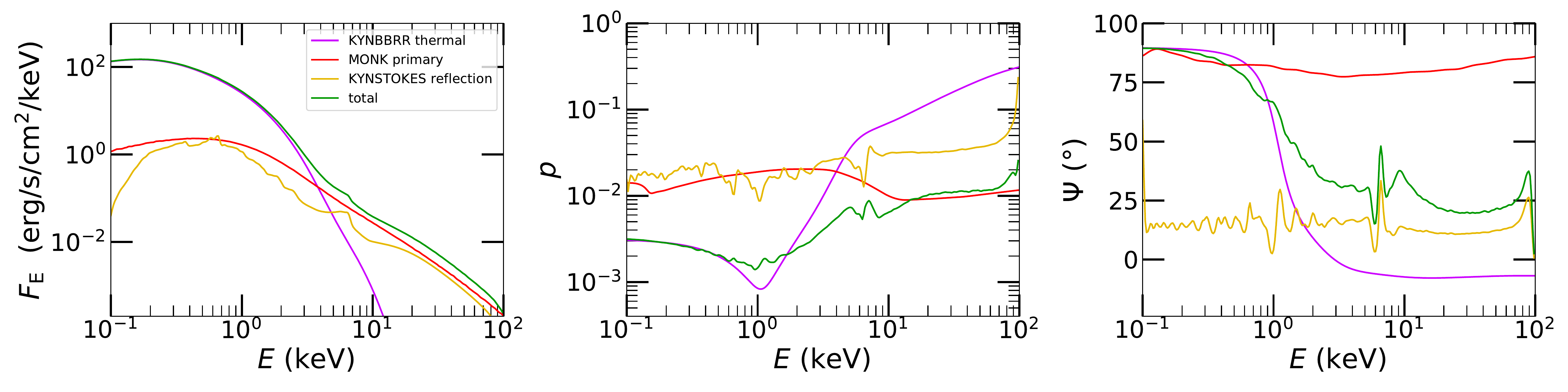}
	\caption{\footnotesize{Combined unfolded and unabsorbed simulation of the main spectral components observed in the hard state of Cygnus X-1 in the {\it IXPE} related observational campaign in May of 2022 \citep{Krawczynski2022}. The thermal component was simulated by {\tt KYNBBRR} model, the lamp-post primary emission component was simulated by {\tt MONK} code and the relativistic disc reflection component was simulated by {\tt KYNSTOKES}. We display the spectrum (left) and~the~corrsponding polarisation degree (middle) and angle (right).}}
	\label{cygx1_combined_simulation}
\end{figure}

Although the superposed {\tt KYNSTOKES} results were similar to the already presented results with initial conditions independent of energy and disc radius (see Section \ref{kynstokes_lamppost}), we concluded that a more realistic estimate for direct interpretation of X-ray polarisation of XRBs in the hard state would be possible only after either a) implementing non-isotropic coronal emission into {\tt KYNSTOKES} or b) linking only the disc reprocessing and relativistic radiative transfer from the~disc to~the~distant observer from {\tt KYNSTOKES} directly to the {\tt MONK} equatorial output, as a~subroutine of {\tt MONK}, which currently neglects coronal photons reflected from the disc. The~option b) is currently being investigated, being limited by required MC photon counts needed for obtaining the polarisation input in~the~equatorial plane in sufficient resolution within reasonable computational times. In the above mentioned attempt with {\tt KYNSTOKES}, when combining the obtained reflected-only signal with the direct primary from {\tt MONK} and with the thermal radiation from {\tt KYNBBRR}, all unabsorbed and normalized relative to each other according to~the~spectral analysis of Cygnus X-1 \citep{Krawczynski2022}, we could not obtain polarisation higher than $\approx 1\%$ on average in the {\it IXPE} band, although an increase with energy was obtained. The polarisation angle in 2--8 keV was also misaligned by $\approx 30^{\circ}$ with the principal axis. The resulting total simulation and component decomposition is shown in Figure \ref{cygx1_combined_simulation}.

\subsection[Analysis of the 4U1630-47 and LMC X-3 {\it IXPE} \newline observations]{Analysis of the 4U1630-47 and LMC X-3 {\it IXPE} \newline observations}

The {\it IXPE} observations of XRBs 4U1630-47 and LMC X-3 in their soft states, which were difficult to explain using standard models, were a motivation for~more theoretical work during this dissertation. The model for thermal radiation passing through a passive comoving atmosphere that was developed based on~the~{\tt TITAN}, {\tt CLOUDY} and {\tt STOKES} codes, was fully described in Section \ref{local_transmission}. Here we will add information on how this model was used for direct comparison with the~data in~\cite{Ratheesh2023} and \cite{Svoboda2023}. Although such model is in~many senses non-standard, it is currently the only modeling direction, alongside the~bremsstrahlung models of \cite{Krawczynski2023}, which allows for~high enough polarisation degree of X-ray disc emission with polarisation increase with energy in 2--8 keV. The produced polarisation angle is constant with energy, which is as well in line with the data. However, if any clear evidence for~2--8 keV polarisation of thermal disc emission from XRBs came in~the~future alongside known projected primary disc axis orientation of the source, which would be parallel to~the~detected polarisation, then the presented model would not be applicable to explain such data, because it produces polarisation direction parallel with the~disc. The same validity limit holds for the assumption of~high ionization, which we use according to the X-ray spectral characteristics of~4U1630-47 and~LMC X-3, based on simultaneous spectral observational campaigns to~the~{\it IXPE} exposures \citep{Ratheesh2023,Svoboda2023}.

~\

Because the first {\it IXPE} observation of 4U1630-47 in the soft state meant a~particularly demanding data interpretation due to the measured extreme mean polarisation degree, we first made some preliminary estimates with the local absorbing atmosphere model in high ionization limit (see Section \ref{high_ionization_limit}), whether it can roughly explain such polarisation without any disc integration and relativistic effects. Figure \ref{analytical_semiinfinite} showed already the most probable 2--8 keV polarisation degree value and the most probable polarisation degree value in~the~lowest energy bin from \cite{Ratheesh2023}, plotted over the inclination dependence of~classical analytical X-ray polarisation estimates. As we argued in~Section~\ref{high_ionization_limit}, the polarisation value at 2 keV should already be close to the pure-scattering energy-independent plateau expected at soft X-rays, which is supported by both our model and the model of Valery Suleimanov \citep{Ratheesh2023}, and~which could be also altered by Faraday rotation towards lower energies \citep{Davis2009}. Hence, at 2 keV the result can be roughly approximated by the~Chandrasekhar-Sobolev profile for semi-infinite atmosphere with scattering-only induced polarisation properties. For the 2--8 keV average polarisation degree value, we have to add the absorption effects that increase polarisation in such scenarios, which can be done through the Loskutov-Sobolev model \citep{LoskutovSobolev1981}. Both analytical approximations match with the data near 2 keV and at 2--8 keV roughly at local inclinations $\delta_\textrm{e} \approx 80^{\circ}$ (see Figure \ref{analytical_semiinfinite}), which forms a first-order theoretical lower limit on inclination, as the relativistic disc integration will require even higher values.

\begin{figure}[!htb]
    \centering
    \includegraphics[height=7cm]{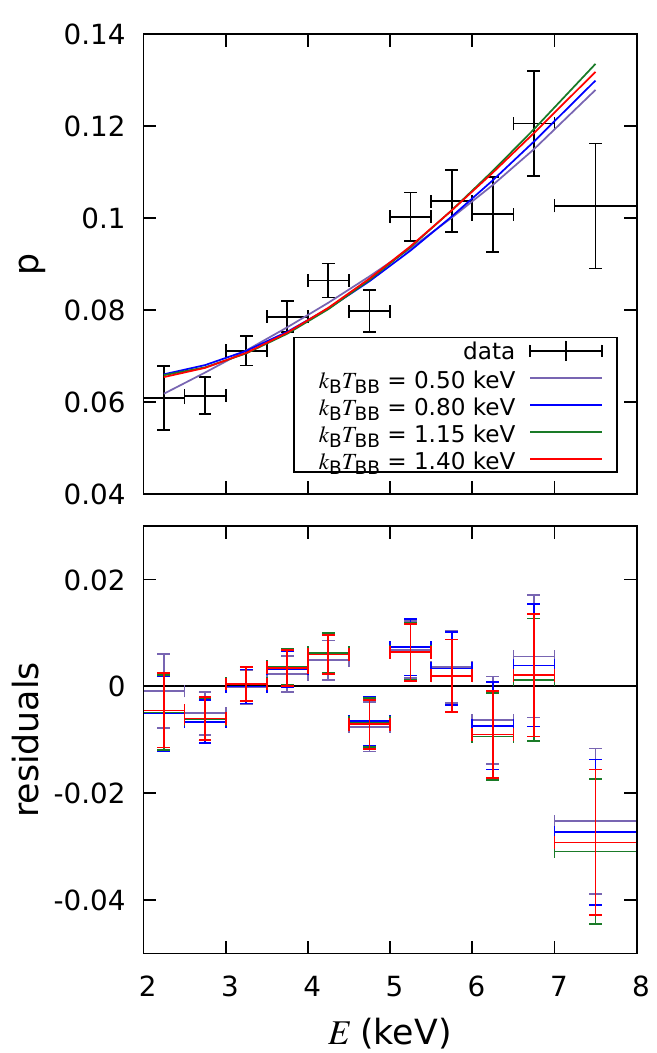}
    \includegraphics[height=7cm]{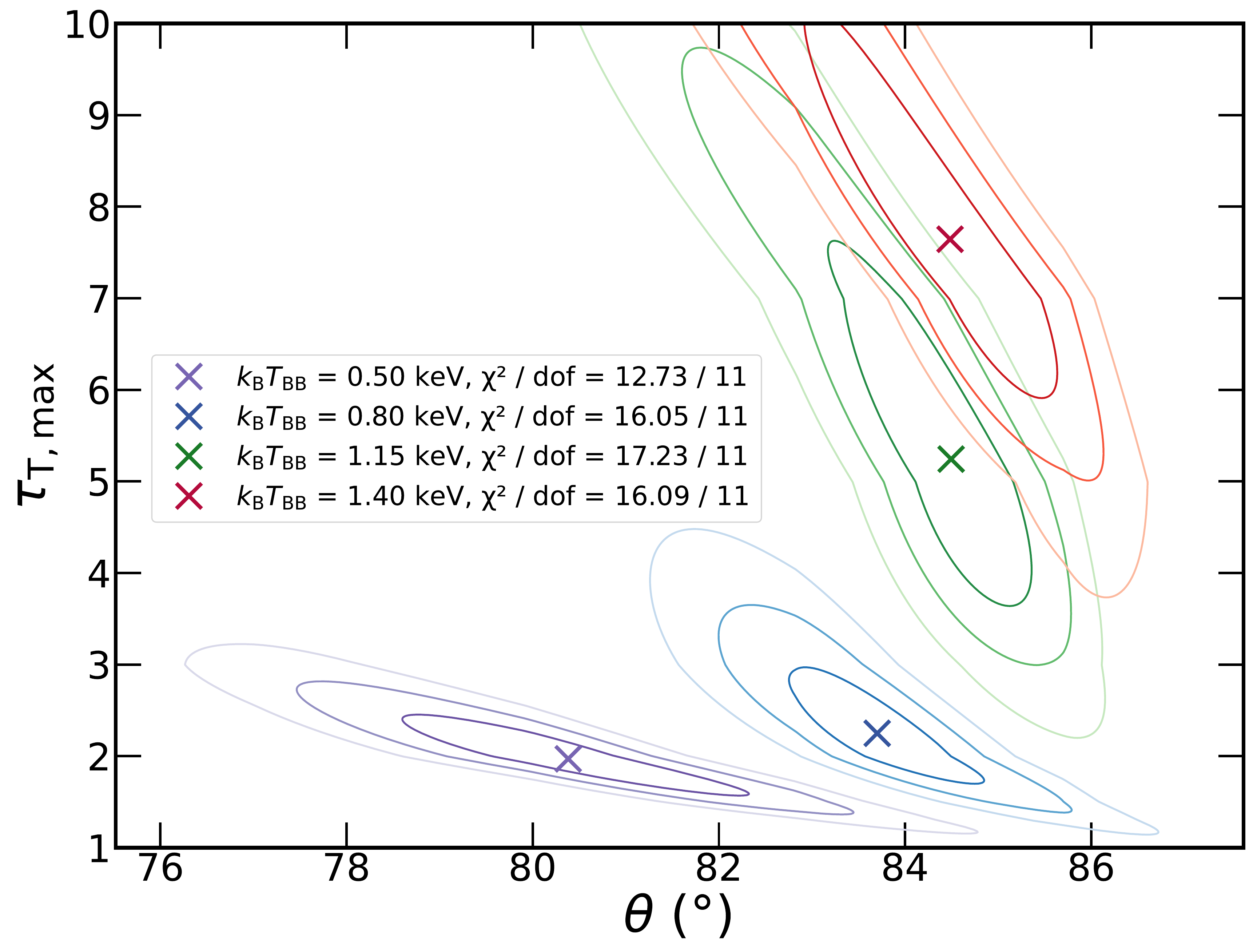}
    \caption{Best-fit local atmosphere solutions of polarisation degree, $p$, versus energy (left) and 2D contours at $1\sigma$, $2\sigma$, $3\sigma$ levels around their value in the \{$\theta$,$\tau_\textrm{T,max}$\} space (right) obtained with XSPEC by fitting the energy-dependent polarisation degree data with smoothed {\tt TITAN} and {\tt STOKES} models, fully described in Section \ref{high_ionization_limit}, for $kT_\textrm{BB} = 0.50$, $0.80$, $1.15$ and $1.40$ keV. The $\chi^2/\textrm{dof}$ values are indicated. The underlying data are from the first {\it IXPE} observation of the XRB 4U1630-47 in the soft spectral state \citep{Ratheesh2023}. Image adapted from \cite{Ratheesh2023}.}
    \label{fig:local_best_fits}
\end{figure}

Using the {\tt XSPEC table model generator} described in Appendix \ref{torus_model}, we created a tabular {\tt FITS} spectropolarimetric model suitable for {\tt XSPEC} that contained the smoothed polynomial approximation of polarisation degree with energy \mbox{in~1--20~keV,} which was described in Section \ref{high_ionization_limit}. Fitting this absorbing atmosphere model directly to the 4U1630-47 data without any relativistic integration delivers another simple estimate on the required inclination, alongside the~model parameter $\tau_\textrm{T,max}$. Figure \ref{fig:local_best_fits} shows the resulting fit of polarisation degree with energy in {\tt XSPEC}. Despite the dependency on $T_\textrm{BB}$ value, we conclude that, for~the~mean $kT_\textrm{BB} = 1.15$ keV obtained from simultaneous campaigns that were described in \cite{Ratheesh2023}, the results match well with $\tau_\textrm{T,max} \approx 5$ and~$\theta \approx 84^{\circ}$. This is already extreme compared to the usual $\theta \approx 78^{\circ}$ eclipsing upper limits on~inclination obtained for XRBs in e.g. \cite{deJong1996}.

As described in Section \ref{high_ionization_limit} and Chapter \ref{chap02}, the smoothed local model was then implemented to the {\tt KYNBBRR} code, forming a new variant {\tt KYNEBBRR}. The~implementation and fitting with {\tt KYNEBBRR} were performed by Michal Dovčiak and~Jiří Svoboda and were not part of this dissertation. A proper relativistic integration using this model for the first {\it IXPE} 4U1630-47 dataset required a relativistically outflowing atmosphere (with $v \approx 0.5c$), alongside high optical thickness ($\tau_\textrm{T,max} \approx 7$) and high inclination ($70^{\circ} \lessapprox \theta \lessapprox 75^{\circ}$) \citep{Ratheesh2023}. The~aberration-induced polarisation was not necessary for the data fit of~the~{\it IXPE} LMC X-3 observation with {\tt KYNEBBRR}, but the optically thick absorbing and Compton-scattering atmosphere provided also here a lower reduced $\chi^2$ than using the relativistic thermal emission model variant {\tt KYNBBRR} with pure-scattering Chandrasekhar's prescription for locally emergent polarisation, but still assuming the Novikov-Thorne accretion disc \citep{Svoboda2023}. For the case of LMC X-3, a slim disc instead of standard geometrically thin accretion disc would perhaps suit better also in polarisation, as it does already in pure spectral analysis \citep{Svoboda2023}. Nonetheless, the standard accretion disc theory is significantly challenged through such discoveries by {\it IXPE}.

\subsection{General observational prospects for equatorially obscured sources and analysis of the Cygnus X-3 {\it IXPE} observations}\label{cygx3_results}

We will be more detailed in description of the analysis in Cygnus X-3 than with the other sources, because additional modeling was performed as part of this dissertation to support the interpretations of the {\it IXPE} discoveries published in~\cite{Veledina2023}. The two aforementioned observations of Cygnus X-3 by {\it IXPE} were separated by about 2 months. Similarly to the discovery paper of~\cite{Veledina2023}, let us call the first observation ``main'' and the second target-of-opportunity observation ``ToO''. We will work only with the time-averaged values, although a small variability of polarisation with orbital phase was detected. The study of \cite{Veledina2023} contains a leading analytical model of single-scattering reflection on conical equatorial outflows, which is compared with the energy-independent 3.5--6 keV part of the X-ray polarisation data. The observations are interpreted as reflection from the inner walls of such funnel, which is assigned a ULX-like narrow opening angle. Apart from our theoretical investigations of reprocessings in distant equatorial obscurers that were already described in Section \ref{equatorial_MC}, we performed 2--8 keV simulations that include direct radiation and remove neutral hydrogen from the otherwise same list of absorbers, as in Section \ref{equatorial_MC}, to be closer to~the~assumed hydrogen-poor environment of outflows in Cygnus X-3 \citep{Fender1999, Kallman2019}. The results were analyzed and plotted in \cite{Veledina2023} in the same way and in direct relation to the analytical models studied therein. Soon after, we decided to elaborate in more depth on the MC analysis and we provided a full range of examples and an extended discussion as part of~\cite{Podgorny2023c}. The extended discussion, including visual presentation of the full atlas of results \citep[exceeding those that fitted in][]{Podgorny2023c}, is presented in this section and Appendix \ref{supplementary}.

The models described in Section \ref{equatorial_MC} are particularly useful for observational analysis of Cygnus X-3, because unlike the analytical models, they allow to assess the role of partial transparency and partial ionization, although the latter only simplistically by means of adjusting free electron density (thus introducing more polarisation at soft X-rays with respect to a proper treatment of ionization). The~partial transparency has been in fact proposed in \cite{Veledina2023} to~play a role in decrease of~the~observed polarisation degree from~the~main observation to the ToO observation. Apart from changing the chemical composition and adding direct radiation in~this section, we also fix the mid torus distance to $R'=0.5$ (again the true units are unimportant for polarisation production in~our model design) and unfreeze $r_\textrm{in}^\textrm{torus}$.\footnote{\,Then the effects of changing distance of the inner radius of the torus to the center can be studied for Case B. The parameter $b'$ is given by the ratios $b' = R'/8$, $b' = R'/4$, $b' = 3R'/4$ and~$a'$ is consistently computed from (\ref{size_trafo}). The wedge-shaped Case A has $r_\textrm{in}^\textrm{torus}$ computed from $r_\textrm{in}^\textrm{torus} = R' - b'$, where $b' = a'$ is from a circular torus with $b' = R'\cos{H'}$.} In this way, such additional models are an~opportunity to~test the sensitivity of~results to previously adopted assumptions. Otherwise the model setup is unchanged from Section \ref{equatorial_MC}. We do not add any polar scatters, we do not introduce misalignment, nor clumpiness or non-uniform densities. We keep the~convention of a positive and negative polarisation degree, corresponding to~the~parallel and perpendicular polarisation angle with respect to the axis of~symmetry, respectively, because again no relativistic effects are assumed and~the~result cannot be any other than orthogonal or parallel with respect to~the~leading directions.

In fact, the orientation of Cygnus X-3 on the sky is known not only thanks to observed radio-loud ejecta \citep{Fender1999}, but also to the corresponding IR and submillimeter polarisation \citep{Marti2000, MillerJones2004, Jones1994}. The X-ray polarisation obtained with {\it IXPE} is perpendicular to~the~projected jet directions within the measurement uncertainty \citep{Veledina2023}. This means that when comparing to the Cygnus X-3 data, we will focus on those parts of the parameter space where negative polarisation is expected and where type-2 viewing angles apply, as the spectra confirm a highly obscured source. However, we keep below the presentation of the full parameter space, as the presented models may be useful for interpretation of~other {\it IXPE} observations in the future and they are distinct from the models described in~Section~\ref{equatorial_MC} that served for more theoretical understanding of the reprocessing mechanisms. Similar obscuration conditions have been discussed in the context of~{\it IXPE} analysis of the type-2 AGN Circinus Galaxy \citep{Ursini2023}, the~ULX object SS433 \citep{Kaaret2023} and the accreting Be-transient X-ray pulsar LS V +44 17/RX J0440.9+4431 \citep{Doroshenko2023}. More observations of equatorially obscured accreting compact objects are likely to be performed by {\it IXPE} or other forthcoming X-ray polarimeters. Reprocessing in~distant equatorial covering regions was also considered in the {\it IXPE} analysis of~the~(unobscured) AGN NGC 4151 \citep{Gianolli2023}.

Only the cases of $\Gamma = 2$, $\Gamma = 3$ and unpolarized and $4\%$ parallelly polarized emission \citep[inspired by the observed value in the hard state of Cygnus X-1][]{Krawczynski2022} were computed. We considered 40 bins in~inclination in~the~same linear spacing and range as with the 20 bins treated in Section \ref{equatorial_MC}. This is because we will plot the results in form of polarisation heatmaps in~the~$\{\theta, H'\}$ space, to be able to directly compare the MC model to the analytical results provided in figure 4 of \cite{Veledina2023}, or to figure 6 of \cite{Ursini2023} that used the \cite{Ghisellini1994} MC code for the interpretation of the type-2 AGN in Circinus Galaxy. The new simulations were performed with $N_\textrm{tot} = 10^{10}$ simulated photons, which meant roughly 2 months to compute all on the available institutional clusters. As it will be shown below, even though the same energy binning as in Section \ref{equatorial_MC} was adopted, the photon statistics are barely enough for energy-dependent analysis in 3 bins across 2--8 keV. Because neutral hydrogen was removed, we describe the results in terms of equatorial density $N_\textrm{He}^\textrm{eq}$ of the remaining most abundant element, which is neutral helium. We believe that such values are then more informative to an experienced reader than if we had recalculated to the usual $N_\textrm{H}^\textrm{eq}$. Both $N_\textrm{He}^\textrm{eq}$ and $\tau_\textrm{e}^\textrm{eq}$ values are scaled accordingly, to~represent similar conditions to the two density cases and three partial ionization cases that were studied in Section \ref{equatorial_MC}.

~\

\begin{figure}[!htb]\centering
	\includegraphics[width=\textwidth]{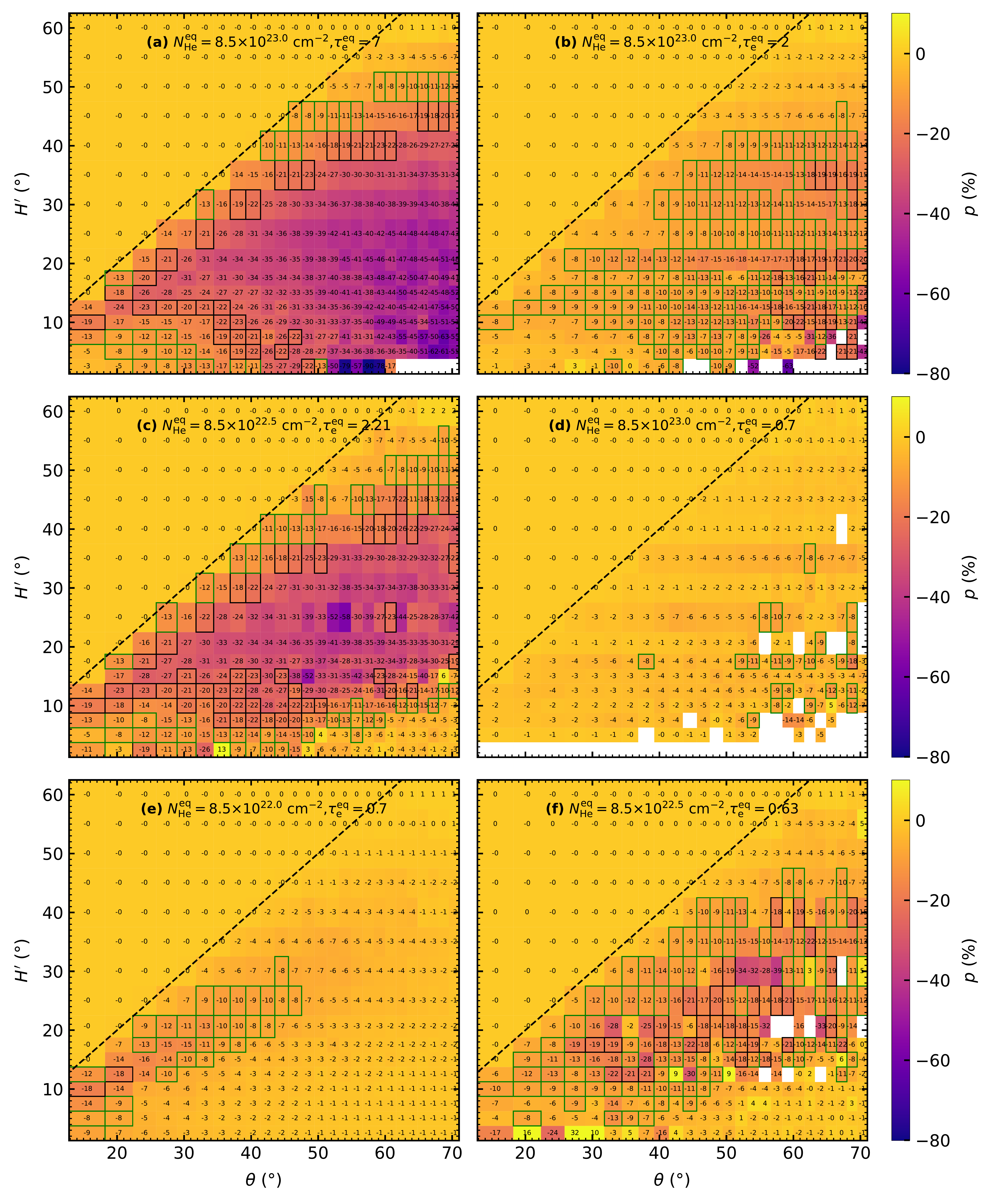}
	\caption{\footnotesize{In the color code we show the {\tt STOKES} model polarisation degree, $p$, integrated in~2--8~keV for the Case B elliptical equatorial obscurer with $b' = R'/4$ with respect to various inclinations $\theta$ and half-opening angles $H'$. All photons are registered and the black dashed lines mark the grazing inclination angles. The black and green rectangles represent the simulation results roughly matching the main and ToO \textit{IXPE} observations, respectively, of~an~X-ray binary system Cygnus X-3, extensively described in \citet{Veledina2023}. We show the~case of unpolarized primary radiation with $\Gamma = 2$. The left column shows the results for \textit{high} ionizations, how they differ when lowering the column densities, i.e. from top to bottom we show (a) $N_\textrm{He}^\textrm{eq} = 8.5 \times 10^{23.0} \ \textrm{cm}^{-2}$ and $\tau_\textrm{e}^\textrm{eq} = 7$, (c) $N_\textrm{He}^\textrm{eq} = 8.5 \times 10^{22.5} \ \textrm{cm}^{-2}$ and $\tau_\textrm{e}^\textrm{eq} = 2.21$, (e)~$N_\textrm{He}^\textrm{eq} = 8.5 \times 10^{22.0} \ \textrm{cm}^{-2}$ and $\tau_\textrm{e}^\textrm{eq} = 0.7$. In top right and center right we show the highest density case for \textit{lower} ionization levels, i.e. (b) $N_\textrm{He}^\textrm{eq} = 8.5 \times 10^{23.0} \ \textrm{cm}^{-2}$ and $\tau_\textrm{e}^\textrm{eq} = 2$ and (d) $N_\textrm{He}^\textrm{eq} = 8.5 \times 10^{23.0} \ \textrm{cm}^{-2}$ and $\tau_\textrm{e}^\textrm{eq} = 0.7$, respectively. The (f) bottom right panel shows the~moderate density case for \textit{lower} ionization, i.e. $N_\textrm{He}^\textrm{eq} = 8.5 \times 10^{22.5} \ \textrm{cm}^{-2}$ and $\tau_\textrm{e}^\textrm{eq} = 0.63$. Image adapted from \cite{Podgorny2023c}.}}
	\label{all_torus_0.25}
\end{figure}

We begin with analyzing the results in the integrated 2--8 keV band and~only turn in later stages to the advantages of energy-dependent analysis. Figure \ref{all_torus_0.25} shows the heatmaps of polarisation degree in the $\{\theta, H'\}$ space for selected cases in~$N_\textrm{He}^\textrm{eq}$ and $\tau_\textrm{e}^\textrm{eq}$ that show the expected depolarisation by lowering density and/or~ionization. When testing for particular examples of even lower densities and lower ionization levels (not shown), we obtained the limit of complete depolarisation. We obtained the general pattern of two distinct branches of solutions for type-2 observers, consistently with the analytical models in \cite{Veledina2023} and MC simulations in \cite{Ursini2023} (only the upper branch is observable in~the~range of figure 6 therein). The dashed black lines mark the~grazing inclination angles. The main observation is marked in black rectangles, considered as $p = 21\% \pm 3\%$ in 2--8 keV within simulation uncertainties. The~ToO observation is marked in green rectangles, considered as $p = 10\% \pm 3\%$ in \mbox{2--8~keV} within simulation uncertainties. The white regions represent empty bins, where the results did not converge within reasonable computational times to provide MC results with reasonable numerical noise. The panels are shown for~a~general Case B obscurer with $b' = R'/4$, unpolarized primary and $\Gamma = 2$. Figures \ref{all_torus_0.125}--\ref{all_torus_0.25_p4G3} provide examples of other geometries and primary radiation properties plotted in~the~exact same way. The lower branch of solutions is preferred for~the~interpretation of Cygnus X-3, as the steep increase in polarisation between the~grazing angle line and the upper contour of observed values suggests high time variability of polarisation, which is not observed \citep{Veledina2023}.

\begin{figure}[!htb]\centering
	\includegraphics[width=\textwidth]{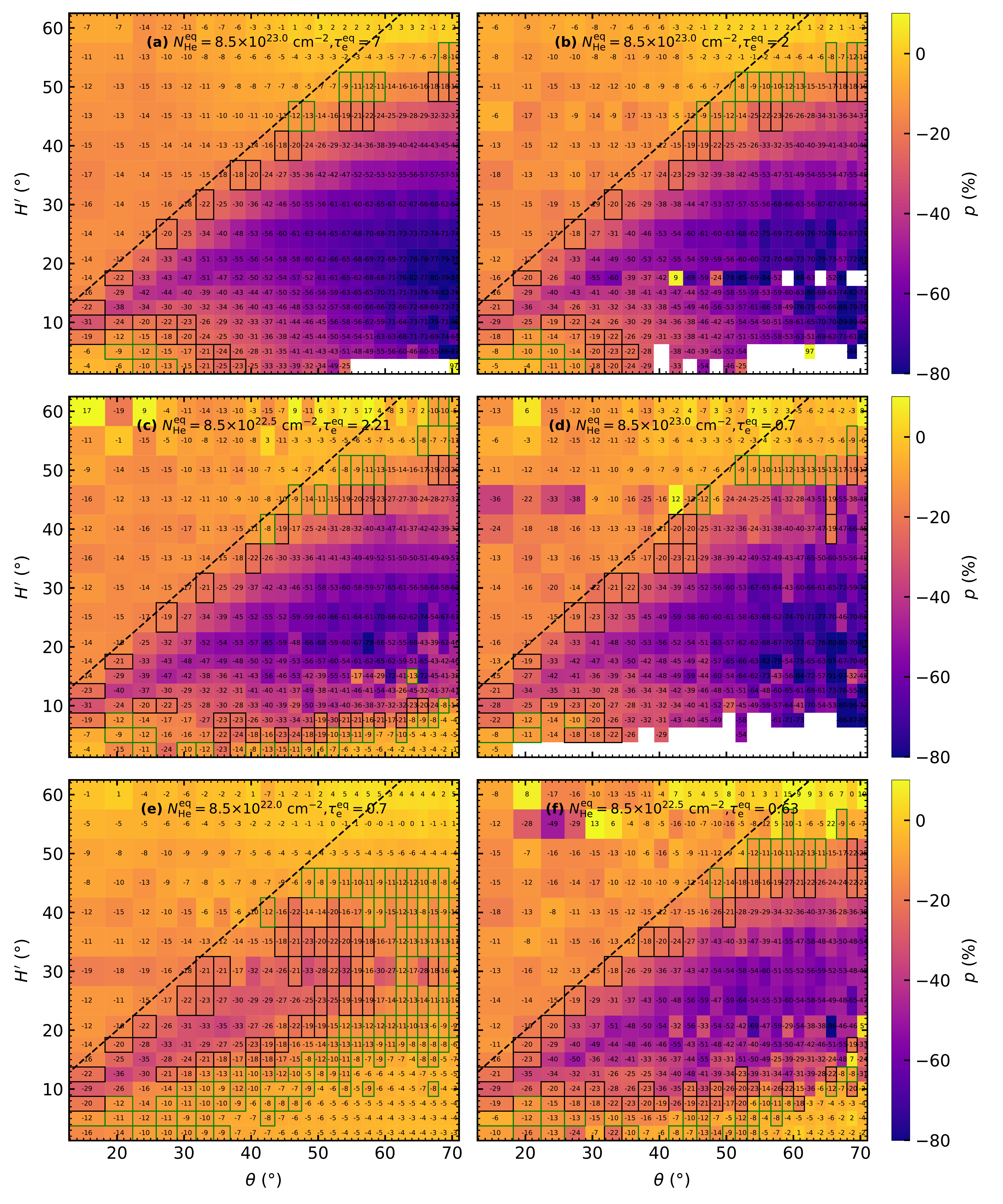}
	\caption{\footnotesize{The same as in Figure \ref{all_torus_0.25}, but for registered single-scattered photons only. Image adapted from \cite{Podgorny2023c}.}}
	\label{single_torus_0.25}
\end{figure}

To be able to evaluate the impact of single-scattered photons, we repeated all simulations with registering only the photons that scattered once before reaching the detectors in the MC simulation. In Figure \ref{single_torus_0.25}, we plot the corresponding results in the same way for the cases considered in Figure \ref{all_torus_0.25}. In~the~single-scattering regime, we obtain even closer solutions to the analytical single-scattering model adopted as leading in the 3.5--6 keV band in \cite{Veledina2023}. Similar comparison conclusions apply for the single-scattering regimes in~other geometries and for other primary radiation cases tested by us. There is still a principal difference between the analytical single-scattering model in \cite{Veledina2023} from our ``single-scattering regime'', because we still consider the absorption of photons. The leading model in \cite{Ursini2023} (figure 6) examines a truly neutral absorber, thus providing perhaps more dissimilar results to ours in the full regime, although a detailed one-to-one comparison would be required for further commentary on possible modeling discrepancies of~this model with respect to ours.

Despite the modeling differences of the {\tt STOKES} simulations presented here in~the~full regime, with respect to the results presented in Section \ref{equatorial_MC}, and despite the inclusion of spectral lines in the studied full 2--8 keV range, we obtained very similar behavior of polarisation with geometrical parameters for the type-2 viewing angles to that already described in Section \ref{equatorial_MC}. This means that for~the cases where reprocessing effects are critical for the emergent polarisation, the addition of direct radiation and the~chemical composition changes do not play a major role, although the latter will have to be reassessed once a more proper treatment of partial ionization is considered. Majority of the obscured parameter space shows perpendicular polarisation, as the dominant source of polarisation are photons scattered once on the inner walls of the reflecting funnel. The polarisation is indeed diluted for~lower densities by scattering in larger volumes and by providing less collimated photon trajectories inside the system. The ionization fraction also determines the~reflectivity and effectively the degree of asymmetry with respect to the type-2 observer, by means of selective directional absorption inside the system. For~type-1 viewing angles, the simulated results are corresponding to the incident radiation with up to $\approx 1\%$ difference in polarisation degree.

When comparing the results for different $b'$ by changing the value of $r_\textrm{in}^\textrm{torus}$, we identify qualitatively similar effects of changing compactness to those discussed around figure 7 in \cite{Goosmann2007} that were studying similar configurations in the optical and UV bands. The geometrically extended tori possess enlarged surface area eligible for first-order scatterings. Also in the~X-rays, it translates into less reduced geometry of scatterings, hence larger dispersion in~polarisation of individual rays, hence depolarisation. The two cases of~the~more prolate elliptical tori show rather similar polarisation predictions (up~to~$\approx 10^{\circ}$ difference in the model $\theta$ and $H'$ in the full 2--8 keV band between the $b' = R'/8$ and $b' = R'/4$ cases). The most oblate Case B ($b' = 3R'/4$) shows polarisation properties closer to the wedge-shaped torus. This is because for the~oblate elliptical structures the observer views the source of radiation through large line-of-sight column densities shortly below the grazing inclination angle, which results into very steeply increasing polarisation fraction with $H'$ or $\theta$ just below the dashed lines in figures, similarly to the Case A. The prolate elliptical or circular tori then perhaps better approximate real outflows arising from the equatorial plane that must have a gradual change in density and ionization before mixing with the ISM.

The expected depolarisation via density and ionization may be up to some extent constrained from simultaneous X-ray spectroscopy, despite the possibility of non-homogeneity and clumpiness in real equatorial winds, which we do not cover with our simple polarisation model. The inclination, half-opening angle, density and partial ionization effects seem to play much larger role than the~effects of changing geometrical profile of the obscurer \citep[cf. also the conical scattering geometry used in][]{Ursini2023, Veledina2023} or changing primary radiation (in the tested examples we obtain up to $\approx 10^{\circ}$ difference in~the~model $\theta$ and $H'$ in the full 2--8 keV band). The primary power-law slope can be obtained from simultaneous X-ray spectroscopy as well. Although detailed information on~the~X-ray spectral lines is typically available too (not only for the case Cygnus X-3), we do not have ambitions to model the effects of depolarisation in spectral lines in detail to assess the X-ray polarisation sensitivity to line related processess and possible parametrization, as here the situation is rather complex. Chemical composition intervenes combined with the physical conditions at the origin and~in~the~reprocessing matter that subsequently alter the polarisation state, according to the inner geometry of the system. The predicted high degeneracy of X-ray polarisation results can be also lifted by means of polarimetry in other wavelengths that can constrain not only the orientation of the system \citep[e.g.][]{Unger1987,Jones1994,Marti2000,Stirling2001,MillerJones2004,Cooke2008, Krawczynski2022, Kravtsov2023}, but also the fundamental $\theta$ and $H'$ parameters. This is frequently used within the scope of AGN unification scenario in IR, optical and UV polarimetry that is in addition able to provide hints on the geometry of the BLRs or polar scatterers with respect to the inner subparsec accretion engine \citep[e.g.][]{Antonucci1985, Oliva1998, Smith2004, Gaskell2012, Marin2018d, Hutsemekers2019, Marin2020, Stalevski2023}.

~\

\begin{figure}[!htb]\centering
	\includegraphics[width=\textwidth]{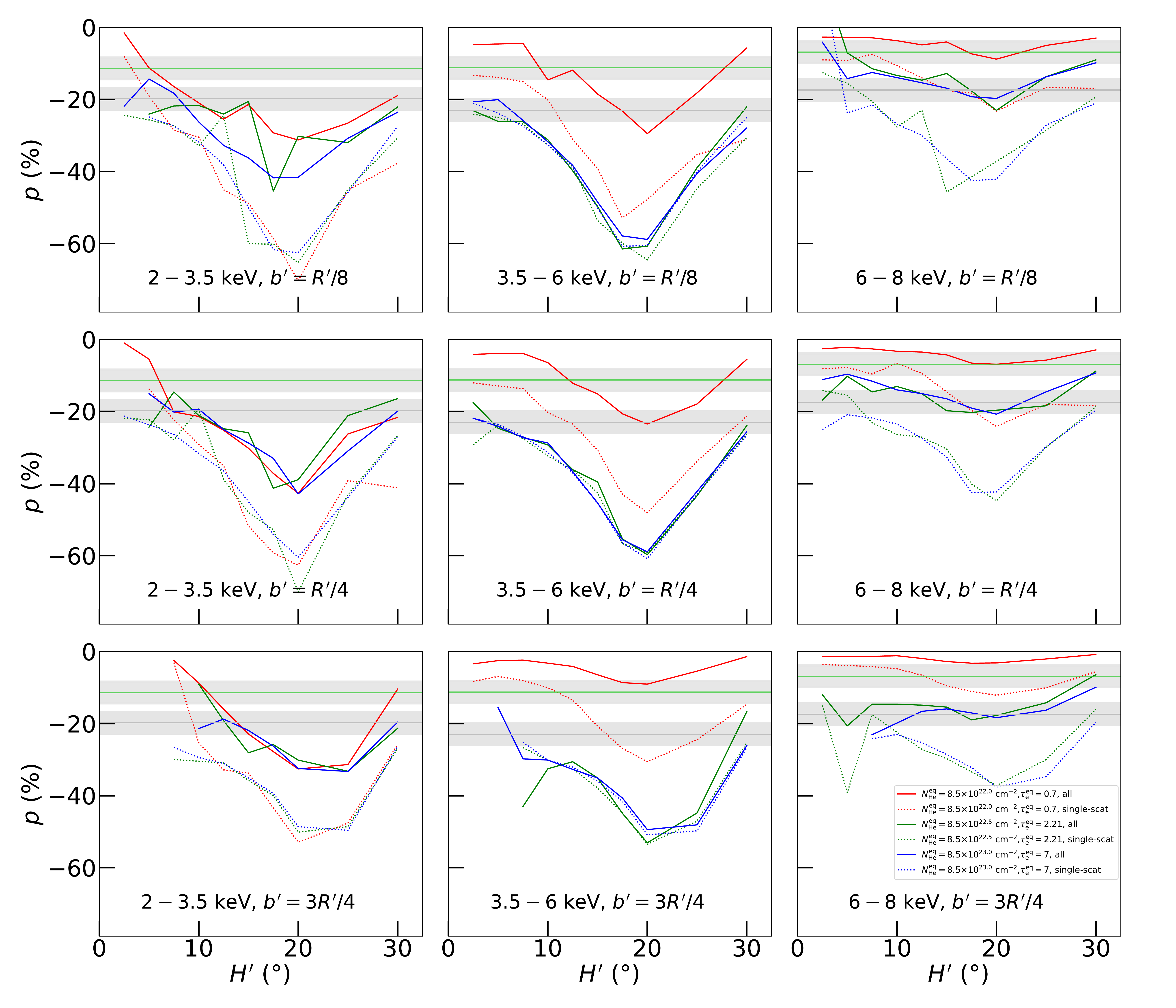}
	\caption{\footnotesize{The {\tt STOKES} model polarisation degree, $p$, averaged in 2--3.5 keV (left), 3.5--6 keV (middle) and 6--8 keV (right), versus half-opening angle $H'$ (in the obscured type-2 configurations only) for a selected $\theta = 30\degr$ inclination angle, corresponding to the orbital inclination in Cygnus X-3 \citet{Vilhu2009, Antokhin2022}. The corresponding energy-averaged polarisation data from the main and ToO observations are marked with gray and green horizontal lines, respectively, with shadowed areas corresponding to simulation uncertainties. The~top, middle and bottom panels correspond to the Case B simulations of elliptical tori with eccentricities given by $b' = R'/8$, $b' = R'/4$ and $b' = 3R'/4$, respectively. We show the~results for all registered photons (solid lines) and only those photons that scattered once (dotted lines). The different colors correspond to various column densities for \textit{high} ionization: $N_\textrm{He}^\textrm{eq} = 8.5 \times 10^{23.0} \ \textrm{cm}^{-2}$ and $\tau_\textrm{e}^\textrm{eq} = 7$ (blue), $N_\textrm{He}^\textrm{eq} = 8.5 \times 10^{22.5} \ \textrm{cm}^{-2}$ and $\tau_\textrm{e}^\textrm{eq} = 2.21$ (green), $N_\textrm{He}^\textrm{eq} = 8.5 \times 10^{22.0} \ \textrm{cm}^{-2}$ and $\tau_\textrm{e}^\textrm{eq} = 0.7$ (red). Image adapted from \cite{Podgorny2023c}.}}
	\label{slices_1_torus_0.25}
\end{figure}

\begin{figure}[!htb]\centering
	\includegraphics[width=\textwidth]{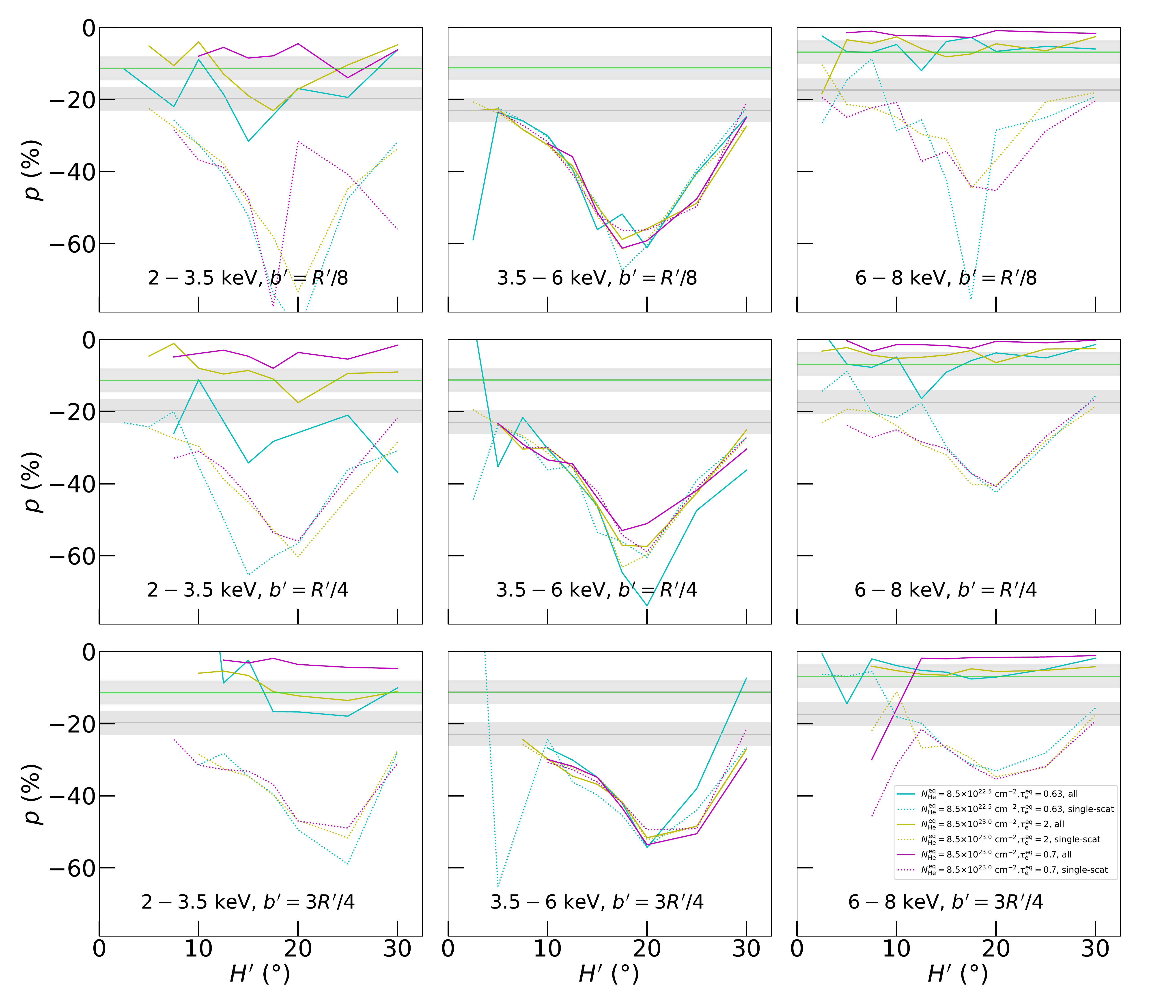}
	\caption{\footnotesize{The same as in Figure \ref{slices_1_torus_0.25}, but for \textit{lower} ionization levels. For~the~highest density case $N_\textrm{He}^\textrm{eq} = 8.5 \times 10^{23.0} \ \textrm{cm}^{-2}$ we show $\tau_\textrm{e}^\textrm{eq} = 2$ (yellow) and $\tau_\textrm{e}^\textrm{eq} = 0.7$ (magenta). For~the~moderate density case $N_\textrm{He}^\textrm{eq} = 8.5 \times 10^{22.5} \ \textrm{cm}^{-2}$ we show $\tau_\textrm{e}^\textrm{eq} = 0.63$ (cyan). Image adapted from \cite{Podgorny2023c}.}}
	\label{slices_2_torus_0.25}
\end{figure}

Within mid X-ray polarimetry, the last obvious way of reducing the available parametric space is obtaining \textit{energy-resolved} upper limit or polarisation detection (or combination of both with respect to energy). Cygnus X-3, towards which we will now turn again our attention, showed energy dependence of polarisation in~2--8~keV with remarkably small uncertainties thanks to its X-ray brightness, its overall high polarisation and the sensitivity of {\it IXPE}. Our Case A (wedge-shaped) MC model results in poor photon statistics for the obscured scenarios and only a noticeable depolarisation in the iron line residing in 6--8 keV was identified. However, apart from the Case A, we were able to acquire rough information on~the~energy dependence produced by our model, which is shown in Figures \ref{slices_1_torus_0.25} and \ref{slices_2_torus_0.25} for both: the full regime when all photons are registered and~the~single-scattering regime. We investigate the result in three energy bands: 2--3.5 keV, 3.5--6 keV and 6--8 keV, for the previously studied three shapes of~the~scattering region and six combinations of densities and ionization levels. We plot the~resulting polarisation versus the half-opening angle, assuming the system inclination of~$\theta = 30^{\circ}$ \citep{Vilhu2009, Antokhin2022}. The {\it IXPE} data for~Cygnus X-3 for~the~main and ToO observations are indicated by gray and~green horizontal lines, respectively, but the numerical noise is too high to provide acceptable geometry and composition constraints, as the two observations should match within one branch of solutions.

We prove that the single-scattering regime well represents the full model in~3.5--6~keV where no spectral lines are present, which supports the choice of~the~leading model in \cite{Veledina2023}. Although the conditions to determine the exact level of depolarisation in spectral lines with respect to the continuum can be complex, we obtain a noticeable depolarisation in the 2--3.5~keV and 6--8~keV bands with respect to 3.5--6 keV. The stronger depolarisation by fluorescent lines is more likely for lower ionization fraction.

Apart from the lines, also the energy-dependence of bound-free and free-free absorption affects the energy-dependence of resulting polarisation, which is inherently linked to the geometrical parameters of the system. The only case where the single-scattering approximation does not match the full result in 3.5--6 keV is when low densities and low ionization are in the game (red lines in Figure \ref{slices_1_torus_0.25}). In this case, the depolarisation in continuum energies is too strong due to~extended angular distribution of individual scattering events. The depolarisation through changing $N_\textrm{He}^\textrm{eq}$ is sensitive with respect to the Compton-thickness energy transition, which occurs in or near the 2--8 keV band.

A smaller decrease (increase actually in some low $N_\textrm{He}^\textrm{eq}$ examples) of polarisation fraction from 3.5--6 keV to 2--3.5 keV is obtained than from 3.5--6 keV to~6--8~keV. The reason is the obscuring medium being indeed more transparent at~higher energies, which causes monotonic decrease of polarisation with energy (see Figure \ref{energy_dependent}), unlike the depolarisation due to spectral lines, which occurs in~the~upper and lower energy bin simultaneously. This could possibly explain the leveling of polarisation in the ToO observation detected in 2--6 keV band \citep{Veledina2023}, compared to the main observation, if the interpretation of decreased density of the obscurer between the two observational periods was correct. We remind, however, that we still derive conclusions from polarisation arising from uniform density of the outflows and a simplistic partial ionization treatment affecting the soft X-rays, which may be critical assumptions difficult to~test without further modeling efforts.

\section{Accreting supermassive black holes}\label{IXPE_results_AGN}

The state-of-the-art scientific advancements in X-ray polarimetry of XRBs slightly turned the topical direction of this dissertation. But since AGNs were originally the primary assignment, forming the title of this work, let us turn our attention back to these objects from observational perspective, as of late 2023. We contributed to the core {\it IXPE} team analyses that are mentioned below with source-specific simulations, knowing the parameters of individual sources from literature, similarly to the general investigations described in Chapter \ref{chap04}. Apart from that, we also attempted for a broad observational perspective of AGNs with {\it IXPE} and~{\it eXTP}, as of 2023, to which we devote two specific sections here in the end.

~\

The first observed AGN by {\it IXPE} was the type-1 MCG 05-23-16 \citep[$< 3.2\%$ \mbox{in~2--8~keV} at a $99\%$ confidence level,][]{Marinucci2022, Tagliacozzo2023}, observed in two operational windows in May 2022 and November 2022. X-ray stability of the source allowed to add the two sets of observations. The~obtained upper limit can still constrain a lot on the inner coronal geometries and many cases of e.g. lamp-post or conical on-axis coronae were disfavored in~case of MCG 05-23-16. Similar unobscured and very bright AGN is IC 4329A \citep[$3.3\% \pm 1.9\%$ in 2--8 keV at a $90\%$ confidence level,][]{Pal2023, Ingram2023}, observed in March 2023, which allowed to evaluate the polarisation angle as aligned with the colimated large-scale radio ejecta \citep{Unger1987}. It is argued in \cite{Ingram2023} that the data favor more asymmetric coronae (perhaps outflowing) with elongated direction in the disc plane. Similar geometrical conclusions for the corona were derived for the December 2022 observations of~NGC 4151 \citep[$4.9\% \pm 1.1\%$ in 2--8 keV at a $68\%$ confidence level,][]{Gianolli2023}, where the reflection from distant parsec-scale components, however, complicated the spectropolarimetric disentangling. The detected X-ray polarisation in NGC 4151 was aligned with the direction of extended radio emission \citep{Harrison1986,Ulvestad1998}.

~\

Last but not least, the brightest type-2 AGN, the Circinus Galaxy \citep[\mbox{$20.0\% \pm 3.8\%$} in 2--6 keV at a $68\%$ confidence level,][]{Ursini2023}, was observed in July 2022, showing remarkably high X-ray polarisation. As we already mentioned, similar polarisation interpretation to the Cygnus X-3 Galactic black hole was put forward in \cite{Ursini2023} due to the pure reflection appearance of its spectra and polarisation orientation perpendicular to the projected radio jet \citep{Elmouttie1998,Curran2008}. We will elaborate below on the limits of such conclusions.

~\

Section \ref{full_AGN_model_MC} was devoted to a 3-component toy-model of a full AGN emission that was developed based on the {\tt KYNSTOKES} central emission model introduced in Section \ref{kynstokes_lamppost} and based on previous modeling efforts of parsec-scale components discussed in \cite{Marin2018c, Marin2018b} and Section \ref{equatorial_MC}. Although the parameter space covered by this model in Section \ref{full_AGN_model_MC} was certainly not complete, we will finish this chapter by providing \textit{general} polarisation observational prospects in~mid \mbox{X-rays} (2--8~keV) for all of the simulation cases studied in Section \ref{full_AGN_model_MC}. Such analysis is meant to provide rough ideas on how sensitive the contemporary instruments are with respect to a typical sample of AGNs. Contrary to the handful of sources that were already observed by {\it IXPE} and that were selected within the mission's targeting strategy, because they had the highest likelihood for X-ray polarisation detection, according to the literature.

We will use the observation simulation software {\tt IXPEOBSSIM} \citep[version \textit{30.2.1},][]{Baldini2022} created for the {\it IXPE} mission, which allows to insert unfolded models (such as those from Section \ref{full_AGN_model_MC}) and provide simulated data and~the~$M\!D\!P_{99\%}$ in a user-selected energy binning, using the latest {\it IXPE} instrumental response matrices (version 12). Given the similarity of instruments on board of~the~planned {\it eXTP} mission, the results for approximately four times lower observational times may be used for observational planning with {\it eXTP}, given its larger planned effective mirror area and higher planned sensitivity compared to~{\it IXPE} \citep{Zhang2019}. But even for {\it IXPE}, the obtained numbers should serve as first-order predictions only, because we will only compare the~unfolded model averaged in 2--8 keV to the simulated $M\!D\!P_{99\%}$. And because in~the~current state, {\tt IXPEOBSSIM} does not allow to use the weighted analysis, which could additionally lower the $M\!D\!P_{99\%}$ by even $10\%$ \citep{Baldini2022,DiMarco2022}.

We will begin with the type-1 AGNs studied in Section \ref{full_AGN_model_MC}, alongside briefly mentioning results for a bare AGN (with no equatorial and no polar scatterers), which the plain {\tt KYNSTOKES} model described in Chapter \ref{chap02} can predict. We will finish with general predictions for the type-2 AGNs studied in Section \ref{full_AGN_model_MC} with some remarks to the already performed {\it IXPE} observation of the Circinus Galaxy. We will keep the positive and negative polarisation degree notation for the rest of this chapter, as our full AGN modeling results (from Section \ref{full_AGN_model_MC}) that we will primarily work with produce either parallel or perpendicular net polarisation, respectively.

\subsection{General observational prospects for type-1 active galactic nuclei}

\begin{figure}[!htb]
	\centering
	\begin{minipage}[t]{0.49\textwidth}
		\includegraphics[width=\textwidth]{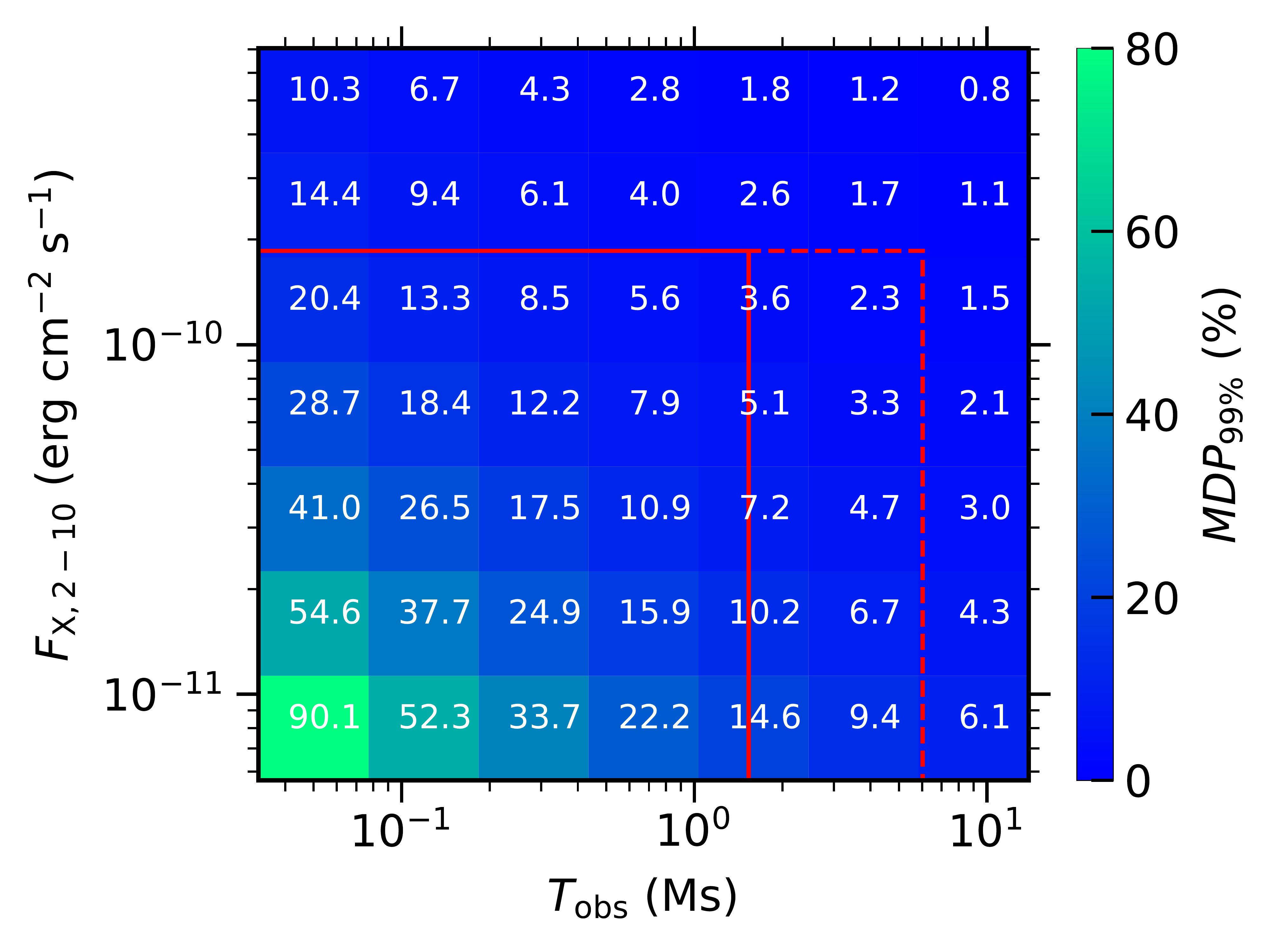}
		\caption{\footnotesize{An example of the $M\!D\!P_{99\%}$ obtained with {\tt IXPEOBSSIM} in one \text{2--8 keV} energy bin for various observational times $T_\textrm{obs}$ and observed X-ray fluxes $F_\mathrm{X, 2-10}$ of a type-1 AGN model input with $2\%$ parallelly polarized lamp-post emission, neutral accretion disc extending to the ISCO, black-hole spin $a = 1$, absorbing winds of $N_\textrm{H}^\textrm{wind} = 10^{23}  \textrm{ cm}^{-2}$ and equatorial torus of $N_\textrm{H}^\textrm{eq} = 10^{24}  \textrm{ cm}^{-2}$. The~solid red rectangle suggests a somewhat realistic window for~an~\textit{IXPE} detection that is on one hand given by the mission's observational strategy in~the~point-and-stare regime and on the other hand by the brightest AGNs on the sky. The~dashed line shows how the~window would approximately enlarge for \textit{eXTP} in~the~same energy band, given its planned larger effective mirror area compared to \textit{IXPE}. Image adapted from \cite{Podgorny2023d}.}}
		\label{mdp_example}
	\end{minipage}
	\hfill
	\begin{minipage}[t]{0.49\textwidth}
		\includegraphics[width=\textwidth]{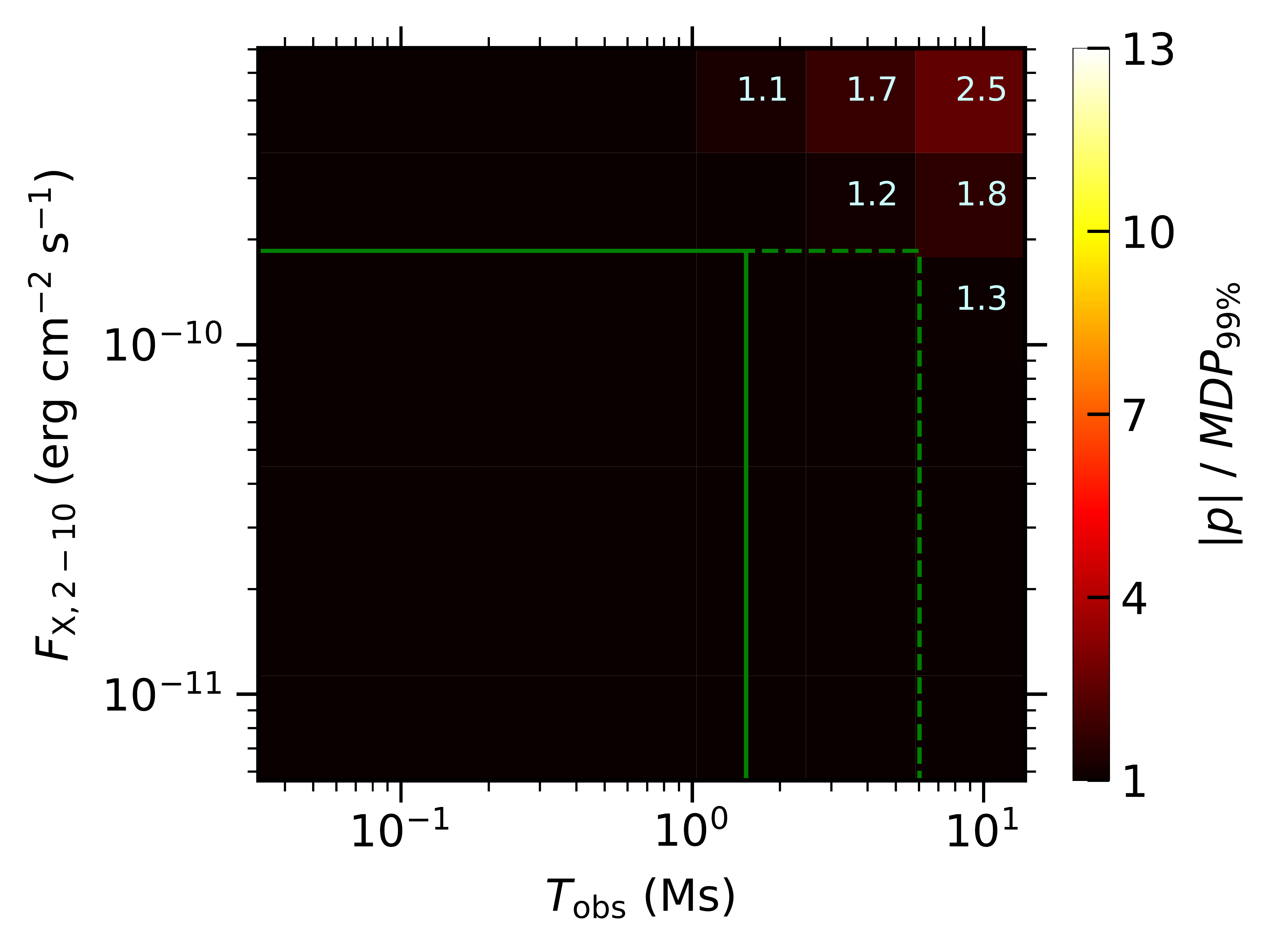}
		\caption{\footnotesize{The model polarisation degree, $p$, divided by the $M\!D\!P_{99\%}$ obtained with {\tt IXPEOBSSIM} in one 2--8 keV energy bin for~various observational times $T_\textrm{obs}$ and observed \mbox{X-ray} fluxes $F_\mathrm{X, 2-10}$. The polarisation degree is positive in this case. If the ratio of~$|p|/M\!D\!P_{99\%}$ is below 1, the polarisation is not detected at~the~$99\%$ confidence level and~we do not display the value and mark the corresponding region in black. The input model is a~type-1 AGN with $2\%$ parallelly polarized lamp-post emission, neutral accretion disc extending to~the~ISCO, black-hole spin $a = 1$, absorbing winds of~$N_\textrm{H}^\textrm{wind} = 10^{23}  \textrm{ cm}^{-2}$ and equatorial torus of~$N_\textrm{H}^\textrm{eq} = 10^{24}  \textrm{ cm}^{-2}$, i.e.~the~same as in Figure \ref{mdp_example}. The boundaries in green lines are the same as red lines in Figure \ref{mdp_example}, suggesting a somewhat realistic observational window. Image adapted from \cite{Podgorny2023d}.}}
		\label{f1}
	\end{minipage}
\end{figure}

Figure \ref{mdp_example} shows an example of~simulated $M\!D\!P_{99\%}$ values in 2--8 keV for one case of~type-1 AGN models as input and for a range of single-exposure observation times $T_\mathrm{obs}$ and X-ray source fluxes $F_\mathrm{X, 2-10}$ between 2 and 10 keV. Our simulated 2--8 keV $M\!D\!P_{99\%}$ results are consistent with the values provided in~the~aforementioned {\it IXPE} discovery papers, if the values from unweighted analysis were provided. We display an observational window for {\it IXPE} by marking a)~the~brightest known stable type-1 AGN sources as an upper limit on the \mbox{$y$-axis} \citep[$F_\mathrm{X, 2-10} \approx 1.8 \times 10^{-10} \, \textrm{erg cm}^{-2} \, \textrm{s}^{-1}$, ][]{Beckmann2006, Ingram2023} and b) the realistic maximum observational time focused on one source provided by \textit{IXPE} as an upper limit on the $x$-axis ($T_\textrm{obs} \approx 1.5 \, \textrm{Ms} \approx 17.4 \, \textrm{days}$).\footnote{\,Information is from private communication with the \textit{IXPE} team. It is the maximum observational time reserved in the mission's observing history per one source per one observing window in multiple subsequent exposures according to \textit{IXPE} telemetry limits. Note that for~stable sources \textit{IXPE} enables additive observations in multiple observational epochs, such as in~\cite{Tagliacozzo2023}.} If the~energy range is further reduced, $M\!D\!P_{99\%}$ typically increases due to the lower number of photons, but see Section \ref{all_polarimetry} for details. {\tt IXPEOBSSIM} allows to~adjust the uniformly defined Galactic column density, $N_\textrm{H}^\textrm{gal}$, which we hold fixed at~a~generic value of \mbox{$N_\textrm{H}^\textrm{gal} = 5 \times 10^{20} \, \textrm{cm}^{-2}$}.

~\

We then plotted for all simulated type-1 AGNs in Section \ref{full_AGN_model_MC} the absolute value of the model polarisation degree over the minimum detectable polarisation, $|p|/M\!D\!P_{99\%}$, in the color code (below 1 as black, which means not detectable at~a~$99\%$ confidence level) versus the same values of $F_\mathrm{X, 2-10}$ and $T_\mathrm{obs}$, as in~Figure~\ref{mdp_example}. Figure \ref{f1} shows one such example, which represents one of the most favorable configurations for detection. Rest of the simulated type-1 AGNs required roughly equivalent or higher exposure times or fluxes. We additionally tried linear interpolation of $F_\mathrm{X, 2-10}$ for three realistic $T_\mathrm{obs}$ values and vice versa where $|p| \approx M\!D\!P_{99\%}$, in order to provide rough polarisation detectability clues for a generic set of type-1 sources. The results are provided in Table \ref{table_model_grid_type1}, alongside the 2--8 keV polarisation state averages of the underlying model. Although the detectability prospects are not very optimistic on the first look, we note that for some sources multiple observations can be added and, apart from the above mentioned limits of our analysis, X-ray polarisation in \textit{certain} type-1 AGNs has already been discovered by {\it IXPE}, including clues on the properties of their inner-accreting regions \citep{Gianolli2023, Ingram2023}. And we note that even an upper limit on polarisation can constrain a non-negligible part of~the~parameter space \citep{Marinucci2022, Tagliacozzo2023}. Also in some cases \citep[e.g.][]{Ursini2023}, {\it IXPE} already proved to detect polarisation with greater statistical confidence in a \textit{reduced} energy range, rather than in the full 2--8 keV band that we examine here only.

~\

From the point of view of highest probability of detection with respect to~the~studied parameters of the inner-accreting regions, such as the inclination of~the~observer with respect to the accretion disc, we also tried to examine the~case of no equatorial and no polar scatterers. Table \ref{bare_nucleus} provides the average total polarisation in 2--8 keV (\textit{IXPE} and \textit{eXTP}) and 15--80 keV (\textit{XL-Calibur}) obtained with the {\tt KYNSTOKES} model presented in Chapter \ref{chap02} in the lamp-post and~extended sandwich corona regimes for a range of parameter values. When running the same {\tt IXPEOBSSIM} analysis for the raw {\tt KYNSTOKES} outputs in~the~lamp-post regime, only an extremely limited part of the tested parameter region produced detectable polarisation at the $99\%$ confidence level in the same estimated window for type-1 AGNs. The inclination of the accretion disc would have to be close to~$\theta = 60^{\circ}$, i.e. near the grazing angles for AGN obscuration by dusty tori, and~it would have to hold either a highly ionized disc (i.e. with high coronal luminosity and/or low black-hole mass) or a highly polarized primary (with $p_0 \gtrapprox 3 \%$). This holds for~nearly any realistic central black-hole spins and lamp-post heights, which are unlikely to be constrained through present-day X-ray polarimetry even in~the~most favorable scenarios inside the lamp-post geometry concept. Given the~low total polarisation degree results for the extended corona with {\tt KYNSTOKES} presented in Section \ref{kynstokes_extended} and the 2--8 keV averaged results compared to the lamp-post regime (note that the provided values in Table \ref{bare_nucleus}~are~for unpolarized primary only), we have not even tried the {\tt IXPEOBSSIM} simulations for the extended coronal version. We note, however, that our central disc-corona model is extremely simplistic and more realistic slab coronal models provide high (a few \%) polarisation predictions, especially for high inclinations \citep{Poutanen1996, Zhang2019b, Krawczynski2022a, Krawczynski2022}. It would be thus worth to test those coronal models in~the~same manner, which was outside of scope of this work. Moreover, the first sample of~{\it IXPE} observations of accreting black holes already resulted in a preference for a corona extending in the accretion disc plane, or left the question of coronal geometry unanswered.
\begin{table}
    \centering
	\caption{{\footnotesize The unfolded average total polarisation degree, $p \,[\%]$, between 2--8 keV (the \textit{IXPE} or~\textit{eXTP} mission range) and 15--80 keV (the \textit{XL-Calibur} balloon experiment range) for unpolarized primary radiation and $\Gamma = 2$, as obtained with the {\tt KYNSTOKES} code described in~Chapter \ref{chap02}. We show cases of neutral disc ($M = 1\times 10^8\,M_{\odot}$ and observed 2--10 keV flux $L_{\textrm{X}}/L_{\textrm{Edd}} = 0.001$, left cells in black) and highly ionized disc ($M = 1\times 10^5\,M_{\odot}$ and observed 2--10 keV flux $L_{\textrm{X}}/L_{\textrm{Edd}} = 0.1$, right cells in red) and various combinations of lamp-post heights $h$ (in~the~lamp-post regime) or radial emissivity power-law indices $q$ (in the extended sandwich corona regime), black-hole spins $a$ and disc inclinations $\theta$. We preserve the notation of~the~positive and negative polarisation degree, if the corresponding angle is parallel or perpendicular to~the~projected rotation axis, respectively. We note, however, that due to GR effects the resulting polarisation angle can be off the true main orthogonal directions by even $\approx 20^\circ$ (see Chapter \ref{chap02}).}}
	\resizebox{\textwidth}{!}{%
    \begin{tabular}{ccc|llllll|llllll}
    \cline{4-15} \multicolumn{1}{c}{}  
     &                \multicolumn{1}{c}{}            &  \multicolumn{1}{c}{}     & \multicolumn{6}{c}{\textit{IXPE} and \textit{eXTP} (2--8 keV) } & \multicolumn{6}{c}{\textit{XL-Calibur} (15--80 keV) }  
    
    \\ \cline{4-15} \multicolumn{1}{c}{}  
                                                     &     \multicolumn{1}{c}{}                       &     \multicolumn{1}{c}{}    & \multicolumn{2}{c}{$\theta = 30^{\circ}$}                        & \multicolumn{2}{c}{$\theta = 60^{\circ}$}                        & \multicolumn{2}{c}{$\theta = 80^{\circ}$} & \multicolumn{2}{c}{$\theta = 30^{\circ}$}                        & \multicolumn{2}{c}{$\theta = 60^{\circ}$}                        & \multicolumn{2}{c}{$\theta = 80^{\circ}$}                       \\ \hline \hline
                                                     &                          & $a = 0$ & {\color[HTML]{333333} 0.34} & {\color[HTML]{CB0000} 2.18} & {\color[HTML]{000000} 1.04} & {\color[HTML]{CB0000} 7.51} & {\color[HTML]{000000} 1.04} & {\color[HTML]{CB0000} 8.46} &{\color[HTML]{333333} 2.29} & {\color[HTML]{CB0000} 2.47} & {\color[HTML]{000000} 7.48} & {\color[HTML]{CB0000} 8.24} & {\color[HTML]{000000} 7.95} & {\color[HTML]{CB0000} 9.04} \\ \cline{3-15} 
                                                     & $h = 3 \textrm{ } r_\textrm{g}$  & $a = 1$ & {\color[HTML]{333333} 0.54} & {\color[HTML]{CB0000} 2.13} & {\color[HTML]{000000} 2.07} & {\color[HTML]{CB0000} 6.58} & {\color[HTML]{000000} 1.03} & {\color[HTML]{CB0000} 4.02} &{\color[HTML]{333333} 2.50} & {\color[HTML]{CB0000} 2.61} & {\color[HTML]{000000} 7.77} & {\color[HTML]{CB0000} 7.87} & {\color[HTML]{000000} 4.66} & {\color[HTML]{CB0000} 4.68} \\ \cline{2-15} 
                                                     &                          & $a = 0$ & {\color[HTML]{333333} 0.24} & {\color[HTML]{CB0000} 1.77} & {\color[HTML]{000000} 0.62} & {\color[HTML]{CB0000} 4.97} & {\color[HTML]{000000} 0.44} & {\color[HTML]{CB0000} 3.51} &{\color[HTML]{333333} 2.03} & {\color[HTML]{CB0000} 2.22} & {\color[HTML]{000000} 5.45} & {\color[HTML]{CB0000} 6.07} & {\color[HTML]{000000} 4.00} & {\color[HTML]{CB0000} 4.58} \\ \cline{3-15} 
    lamp-post corona      & $h = 15 \textrm{ } r_\textrm{g}$ & $a = 1$ & {\color[HTML]{333333} 0.24} & {\color[HTML]{CB0000} 1.72} & {\color[HTML]{000000} 0.59} & {\color[HTML]{CB0000} 4.58} & {\color[HTML]{000000} 0.37} & {\color[HTML]{CB0000} 2.56} &{\color[HTML]{333333} 1.98} & {\color[HTML]{CB0000} 2.17} & {\color[HTML]{000000} 5.12} & {\color[HTML]{CB0000} 5.68} & {\color[HTML]{000000} 3.26} & {\color[HTML]{CB0000} 3.68} \\ \hline \hline
                                                     &                          & $a = 0$ & {\color[HTML]{333333} -0.06} & {\color[HTML]{CB0000} -0.01} & {\color[HTML]{000000} -0.19} & {\color[HTML]{CB0000} 0.01} & {\color[HTML]{000000} -0.25} & {\color[HTML]{CB0000} 0.04} &{\color[HTML]{333333} -0.13} & {\color[HTML]{CB0000} -0.08} & {\color[HTML]{000000} -0.44} & {\color[HTML]{CB0000} -0.20} & {\color[HTML]{000000} -0.49} & {\color[HTML]{CB0000} -0.11} \\ \cline{3-15} 
                                                     & $q = 3.0$  & $a = 1$ & {\color[HTML]{333333} -0.04} & {\color[HTML]{CB0000} -0.00} & {\color[HTML]{000000} -0.11} & {\color[HTML]{CB0000} 0.00} & {\color[HTML]{000000} -0.14} & {\color[HTML]{CB0000} 0.02} &{\color[HTML]{333333} -0.08} & {\color[HTML]{CB0000} -0.05} & {\color[HTML]{000000} -0.27} & {\color[HTML]{CB0000} -0.12} & {\color[HTML]{000000} -0.38} & {\color[HTML]{CB0000} -0.08} \\ \cline{2-15} 
                                                     &                          & $a = 0$ & {\color[HTML]{333333} -0.07} & {\color[HTML]{CB0000} -0.01} & {\color[HTML]{000000} -0.24} & {\color[HTML]{CB0000} 0.02} & {\color[HTML]{000000} -0.30} & {\color[HTML]{CB0000} 0.06} &{\color[HTML]{333333} -0.21} & {\color[HTML]{CB0000} -0.02} & {\color[HTML]{000000} -0.55} & {\color[HTML]{CB0000} -0.24} & {\color[HTML]{000000} -0.42} & {\color[HTML]{CB0000} -0.07} \\ \cline{3-15} 
    sandwich corona      & $q = 1.5$ & $a = 1$ & {\color[HTML]{333333} -0.07} & {\color[HTML]{CB0000} -0.01} & {\color[HTML]{000000} -0.24} & {\color[HTML]{CB0000} 0.02} & {\color[HTML]{000000} -0.28} & {\color[HTML]{CB0000} 0.02} &{\color[HTML]{333333} -0.21} & {\color[HTML]{CB0000} -0.12} & {\color[HTML]{000000} -0.53} & {\color[HTML]{CB0000} -0.24} & {\color[HTML]{000000} -0.42} & {\color[HTML]{CB0000} -0.08} \\ \hline \hline
    \end{tabular}}
    \label{bare_nucleus}
\end{table}

\subsection[General observational prospects for type-2 active galactic nuclei and analysis of the Circinus Galaxy {\it IXPE} \newline observations]{General observational prospects for type-2 active galactic nuclei and analysis of the Circinus Galaxy {\it IXPE} observations}

\begin{figure}[!htb]\centering
	\includegraphics[width=\textwidth]{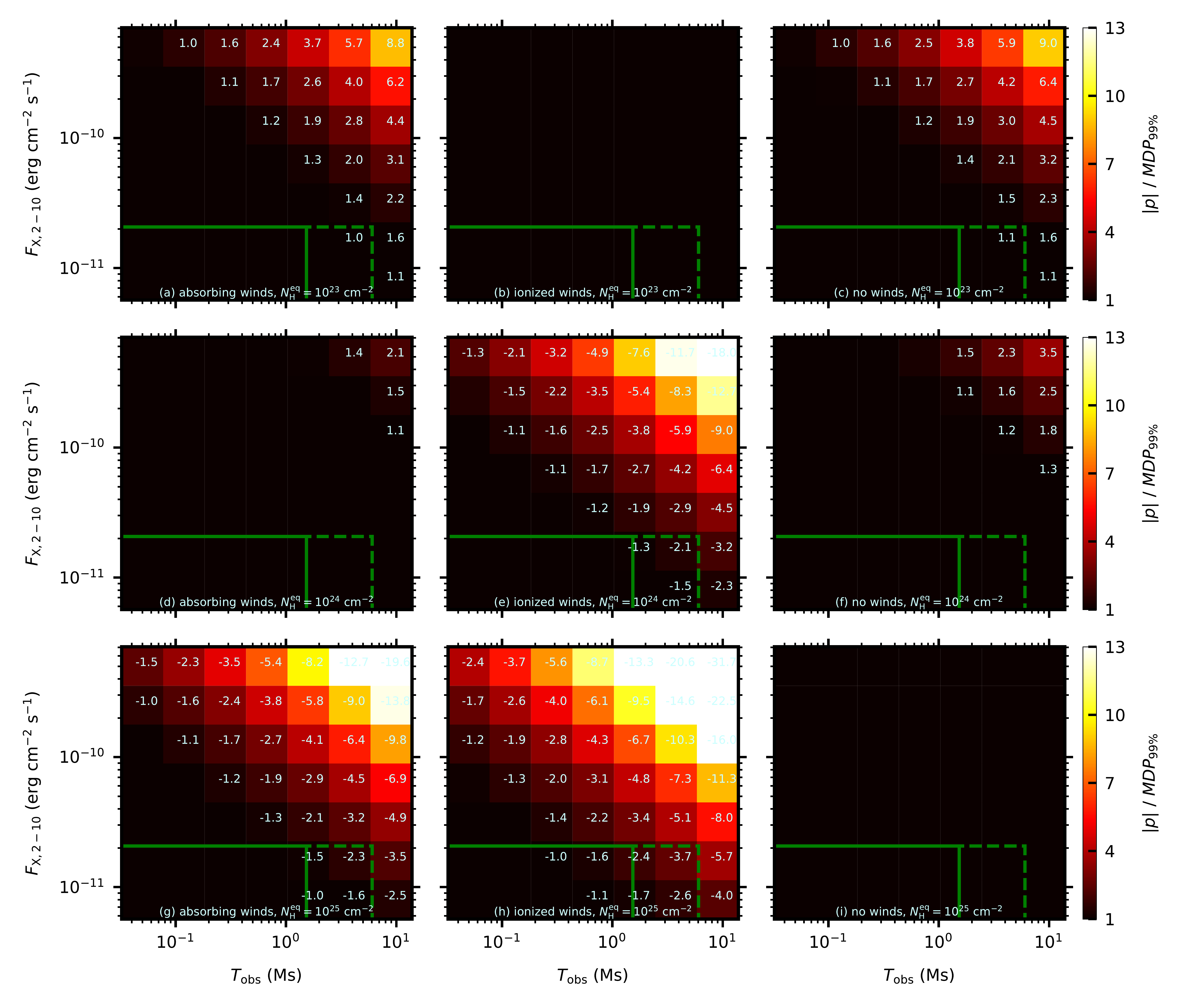}
	\caption{\footnotesize{The same as in Figure \ref{f1}, but the input model is a type-2 AGN with $2\%$ parallelly polarized lamp-post emission, neutral accretion disc extending to the ISCO, black-hole spin $a = 1$. We show nine different configurations of the parsec-scale components: (a) absorbing winds of $N_\textrm{H}^\textrm{wind} = 10^{21}  \textrm{ cm}^{-2}$ and equatorial torus of $N_\textrm{H}^\textrm{eq} = 10^{23}  \textrm{ cm}^{-2}$ (top-left), (b) ionized winds and equatorial torus of $N_\textrm{H}^\textrm{eq} = 10^{23}  \textrm{ cm}^{-2}$ (top-middle), (c) no winds and equatorial torus of $N_\textrm{H}^\textrm{eq} = 10^{23}  \textrm{ cm}^{-2}$ (top-right), (d) absorbing winds of $N_\textrm{H}^\textrm{wind} = 10^{21}  \textrm{ cm}^{-2}$ and equatorial torus of $N_\textrm{H}^\textrm{eq} = 10^{24}  \textrm{ cm}^{-2}$ (middle-left), (e) ionized winds and equatorial torus of $N_\textrm{H}^\textrm{eq} = 10^{24}  \textrm{ cm}^{-2}$ (center), (f) no winds and equatorial torus of $N_\textrm{H}^\textrm{eq} = 10^{24}  \textrm{ cm}^{-2}$ (middle-right), (g) absorbing winds of $N_\textrm{H}^\textrm{wind} = 10^{21}  \textrm{ cm}^{-2}$ and equatorial torus of $N_\textrm{H}^\textrm{eq} = 10^{25}  \textrm{ cm}^{-2}$ (bottom-left), (h) ionized winds and equatorial torus of $N_\textrm{H}^\textrm{eq} = 10^{25}  \textrm{ cm}^{-2}$ (bottom-middle), (i) no winds and~equatorial torus of $N_\textrm{H}^\textrm{eq} = 10^{25}  \textrm{ cm}^{-2}$ (bottom-right). The negative sign in front of the written values in~some panels indicates that in this case, the net model polarisation angle in 2--8 keV is orthogonal to the projected system axis of symmetry. In the color code, we keep the absolute value of polarisation degree for consistency of polarisation detectability estimates with other cases. The solid green rectangle again suggests a somewhat realistic window for an \textit{IXPE} detection that is on one hand given by the mission's observational strategy in the point-and-stare regime and on the other hand by the brightest type-2 AGNs on the sky. The~dashed line shows how the window would enlarge for \textit{eXTP} in the same energy band, given its planned larger effective mirror area compared to \textit{IXPE}. Image adapted from \cite{Podgorny2023d}.}}
	\label{all_heatmaps_type2}
\end{figure}

We performed the same {\tt IXPEOBSSIM} simulations with the type-2 AGN synthetic spectra and polarisation from Section \ref{full_AGN_model_MC} as inputs. Comparing to type-1 AGNs, here the situation is more favorable in high polarisation degree expected in~the~{\it IXPE} band. But as usual, high polarisation through more specific geometry of scattering (a reduced symmetry) is to the detriment of average expected photon count rate. The brightest type-2 AGNs are detected in X-ray fluxes at~least an~order of magnitude lower \citep[$F_\mathrm{X, 2-10} \approx 1.5 \times 10^{-11} \, \textrm{erg cm}^{-2} \, \textrm{s}^{-1}$, ][]{Bianchi2002, Tanimoto2022}. We mark this value along with the same maximum $T_\mathrm{obs}$ in~Figure~\ref{all_heatmaps_type2} that shows the same heatmaps of $|p|/M\!D\!P_{99\%}$ versus $F_\mathrm{X, 2-10}$ and $T_\mathrm{obs}$ for the type-2 cases with absorbing winds, ionized winds and~no winds for three different torus column densities and one set of parameter values related to the nucleus. In the provided numbers on each heatmap, we keep the sign of~$p$, i.e. we note the values of $p/M\!D\!P_{99\%}$, because for type-2 AGNs the~resulting polarisation angle can be either parallel or orthogonal to the principal axis. The~underlying average 2--8 keV unfolded model polarisation values were provided in Table \ref{table_model_grid_type2} already, alongside the same linear interpolation results for~three values of $F_\mathrm{X, 2-10}$ and $T_\mathrm{obs}$ where $|p| \approx M\!D\!P_{99\%}$, as in Table \ref{table_model_grid_type1}. The wide conclusions are similar to the type-1 AGNs: although the first AGN observations by \textit{IXPE} proved that for the \textit{brightest} and \textit{well-known} sources the~X-ray polarisation data can improve our knowledge significantly, the complexity and diversity of AGNs (especially of the parsec-scale components with respect to type-2 AGNs) is preventing us from providing a clear diagnostic tool or a qualitative guidance for~a~\textit{general} set of~sources, given the contemporary sensitivity of X-ray polarimetric instruments. Such goals are too ambitious for the near future. Regarding the~prospects for~plain detectability by {\it IXPE} or {\it eXTP} with respect to~faintness of~AGNs, the situation is perhaps more favorable towards type-2 sources rather than towards unobscured AGNs.

~\

In the case of \textit{absorbing winds}, the higher the equatorial obscuration is, the~higher the contribution from the perpendicularly polarized component we obtain in the 2--8 keV band over the parallel polarisation that is more likely arising from the equatorial scatterings. However, the flux has a reverse dependency on the torus thickness due to absorption. The dependency of such trade-off in~X-ray polarisation detectability on the torus density is interesting with regards to 2--8 keV $M\!D\!P_{99\%}$ for {\it IXPE}. We obtain higher probability of 2--8 keV detection for tori with $N_\textrm{H}^\textrm{eq} = 10^{23}  \textrm{ cm}^{-2}$ than for tori with $N_\textrm{H}^\textrm{eq} = 10^{24}  \textrm{ cm}^{-2}$. However, the~cases with $N_\textrm{H}^\textrm{eq} = 10^{25}  \textrm{ cm}^{-2}$ show again better observational prospects than for~$N_\textrm{H}^\textrm{eq} = 10^{24}  \textrm{ cm}^{-2}$, in fact the most promising out of the three examples. In~order to~explain this behavior, we will first analyze the other cases of polar winds, because their compositions plays an important role.

In the case of \textit{ionized winds}, we get low polarisation detection probabilities for low torus column density of $N_\textrm{H}^\textrm{eq} = 10^{23}  \textrm{ cm}^{-2}$. The situation is significantly more promising towards detections for $N_\textrm{H}^\textrm{eq} = 10^{24}  \textrm{ cm}^{-2}$ and even more for~$N_\textrm{H}^\textrm{eq} = 10^{25}  \textrm{ cm}^{-2}$. For the case with the lowest torus transparency, we obtain higher detectability for ionized winds compared to the absorbing winds. This is because for the ionized winds, we place an ideal polar reflector (causing high $\gtrapprox 25\%$ perpendicularly oriented polarisation) for photons emerging upwards and reaching the observer through the ``periscope''. In the case of \textit{no polar winds}, we get a reverse dependency of detectability with equatorial torus thickness. We obtain better observational conditions, if the torus is more transparent, which in addition to higher flux produces higher average 2--8 keV polarisation with parallel orientation (see Table \ref{all_heatmaps_type2}). Hence, the case of absorbing case produces an intermediate case, mixing these two effects. This is clear from the expected 2--8 keV polarisation angle that changes by $90^\circ$ between $N_\textrm{H}^\textrm{eq} = 10^{23}  \textrm{ cm}^{-2}$ and~$N_\textrm{H}^\textrm{eq} = 10^{25}  \textrm{ cm}^{-2}$. For the former case the delicate energy transition between the two polarisation states seen in the simulations (discussed in Section \ref{full_AGN_model_MC}) occurs at energies rather lower from the {\it IXPE} band. Although in the analysis presented here, the 2--8 keV detectability results are in addition related to~the~energy-dependent flux.

~\

Such discussion is worth applying to the so-far only observed type-2 AGN by {\it IXPE}: the Circinus Galaxy \citep[with Compton thickness $N_\textrm{H} > 10^{24}  \textrm{ cm}^{-2}$,][]{Arevalo2014, Kayal2023}, described and interpreted in \cite{Ursini2023}. The study provides a detailed spectropolarimetric analysis that suggests that the~observed polarisation of $17.6\% \pm 3.2\%$ in the 2--8 keV band (at $68\%$ confidence level) can be largely attributed to the equatorial scatterings, while the~polarisation from the polar reflection is claimed unconstrained. They assumed two distinct power-law indices $\Gamma = 1.6$ and $\Gamma = 3.0$ for the cold (equatorial) and~warm (polar) reflectors, respectively, that were kept fixed in the spectropolarimetric fit, which resulted in low total flux contribution of the warm reflector, thus low contribution to the net polarisation. If different assumptions were taken more in~favor of~the~warm reflector, it could affect the result of~the~polarisation component analysis.\footnote{\,In the appendix of \cite{Ursini2023}, it is discussed that such attempts did not improve the spectral fit that was executed before the spectropolarimetric fit and that the presented results are consistent with the spectral analysis of \cite{Marinucci2013}.}

Thus, given our models that in addition include the polar reflector \citep[compared to][where only the equatorial scattering region is considered]{Ursini2023}, we humbly propose reconsideration of the contribution to the polarisation detected in the Circinus Galaxy by {\it IXPE} that could be on the other hand almost entirely attributed to the partly ionized polar scattering region. This is supported furthermore by the fact that our simulations from Section \ref{full_AGN_model_MC} and those in \cite{Marin2018b} clearly lead to two contributions to the net polarisation state that are distinguished by a $90\degr$ switch in the polarisation angle. The observed polarisation in the Circinus Galaxy is orthogonal to the principal axis of the entire accreting system \citep{Ursini2023}. Hence, this orientation is consistent with our interpretation for polar scatterings, on average symmetric in~the~meridional plane. When studying the equatorial parsec-scale scattering regions in~detail in~Section~\ref{equatorial_MC}, we argued that the parallelly oriented summed polarisation vector is more likely to arise from equatorial reprocessings, \textit{if the half-opening and the observer's inclination are high enough}, which is the case of Circinus Galaxy \citep{Kayal2023}. In such cases, the equatorial component may rather serve as a depolarizer and dilutes the orthogonally and highly polarized warm reflection from the poles.

When considering the expected and observed energy dependence of polarisation in Circinus Galaxy, the above conclusions are also supported. Although it is beyond the scope of this work to examine combinations of different energy binnings in 2--8 keV range in detail, we clearly see that the simulations result in high (tens of \%) polarisation with perpendicular orientation at soft X-rays, thanks to the polar scatterer. This component then prevails more inside (or even above) the 2--8 keV band for higher column densities of the torus. In \cite{Ursini2023}, it is proposed that the non-detection of polarisation in 6--8 keV is due to the presence of unpolarized iron line, which is certainly true. However, it suggests itself that an additional effect of depolarisation in 6--8 keV can be due to the competing polarisation contribution in parallel direction with respect to the principal axis arising from the equatorial reprocessings. For high enough Compton thickness of the dusty torus that the source is likely possessing \citep{Arevalo2014, Kayal2023} this would not mean a switch of the polarisation angle to parallel orientation at higher energies within the 2--8 keV band, which is indeed not observed \citep{Ursini2023}, but only a dilution of polarisation with the same orientation. The switch would occur at even higher energies due to the reduced flux contribution from light passing directly through the opaque torus \citep[see Section \ref{full_AGN_model_MC} and ][]{Marin2018b}.

\newpage
\ 
\vspace{4.5cm} 
\begin{figure}[h]
        \centering
	\includegraphics[trim={0cm 0cm 0cm 0cm},clip,width=0.6\textwidth]{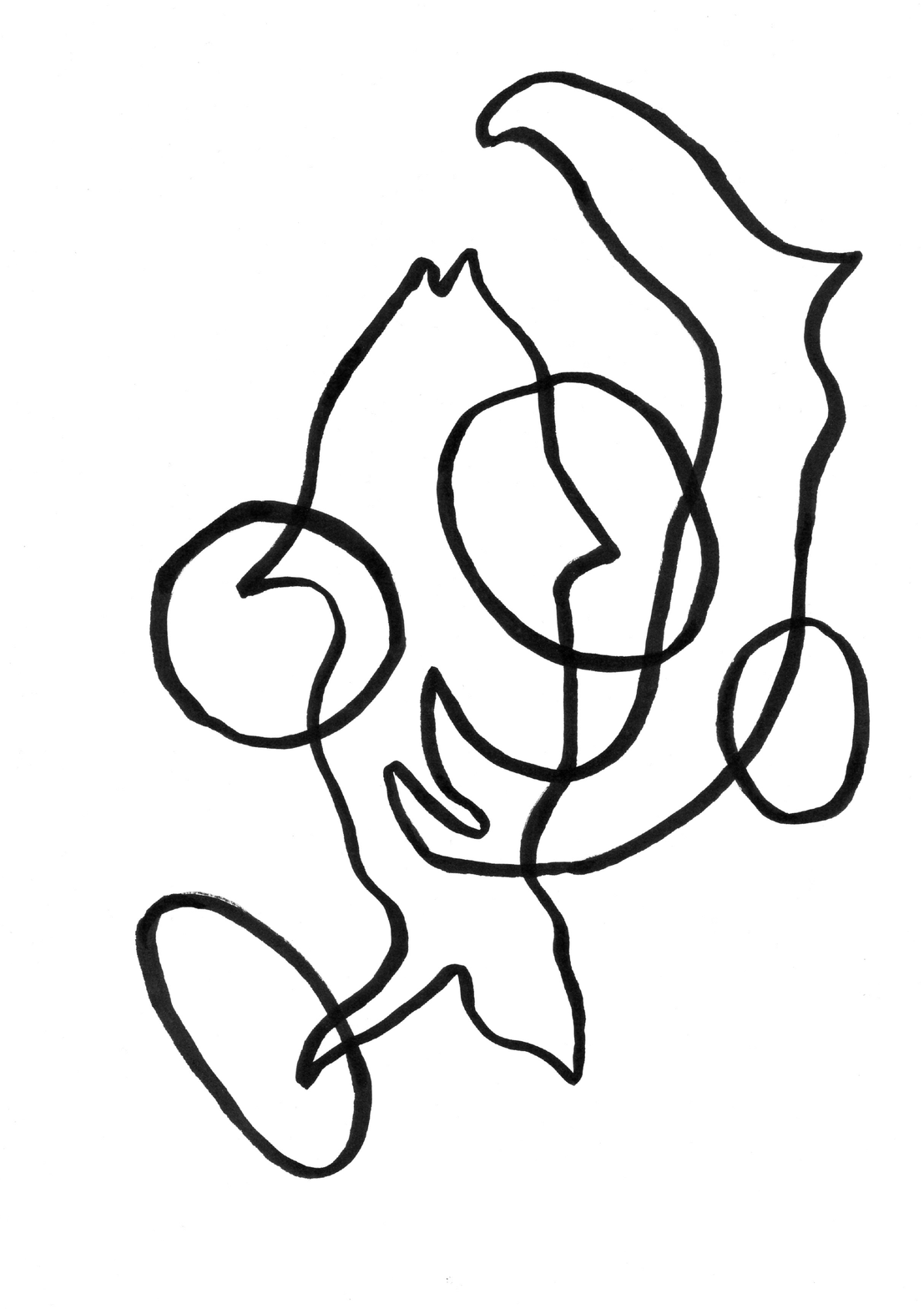}
\end{figure}
\chapter{Conclusions and future prospects}\label{conclusions}

Let us conclude and provide directions of useful continuation of this work. In~the~beginning, we promised that we will show how elementary geometrical principles may be used to derive information on the outer space. We argued that in~the~case of AGNs and XRBs, the (a)symmetry in the origin of X-ray emission yields polarisation signatures in the observed electromagnetic radiation, as~the~preferred orientation of the electric vector position angle is determined by~the~emission and reprocessing mechanisms, as well as superposition effects. In~this~way, any broken spatial symmetry in the interaction of light and~matter inside the system, from quantum Planck scales all the way to multiples of~black-hole radii where gravity matters, may lead to enhanced net polarisation in~a~beam of photons on~behalf of~the~naturally randomly polarized (unpolarized) light. In~the~X-rays near accreting black holes, this is associated mostly with magnetic fields, scattering and absorption mechanisms, as the main physical drivers \mbox{of~X-ray} polarisation modifications. We restricted ourselves to configurations where magnetic fields do not intervene significantly in the X-ray emission. We assumed a given background gravity field and matter configuration according to~known information on AGNs and XRBs, which comes from the interplay of~observations and logical deduction based on elementary conservation principles and known fundamental interactions.

Devoting the first half of our results to theory, we developed a series of X-ray spectropolarimetric numerical models from local radiative transfer computations applied to photospheres of accretion discs, to estimators of strong-gravity effects on the emergent light near the black-hole event horizon, to models of reprocessing in distant circumnuclear components. By constructing toy models of reasonable simplifications and assumptions -- to the detriment of computational expensiveness and hopefully not much to the detriment of accurate description of physical reality -- we provided predictions for X-ray polarimetric observations of AGNs and XRBs. These may serve for analysis of the recent discoveries by the leading X-ray polarimetric mission {\it IXPE}, as well as for observational prospects with the current and forthcoming X-ray polarimeters (such as {\it IXPE}, {\it XL-Calibur} -- already launched --, or {\it eXTP} -- planned to be launched in 2027). Despite being largely degenerated in the considered parameter space with respect to predicted X-ray polarisation, the models may result in significantly more informative potential, if combined with X-ray spectroscopic and timing techniques, or polarimetry of~AGNs and~XRBs at other wavelengths. The power of observational X-ray polarimetry in multi-wavelength and multi-instrument campaigns was demonstrated in the last part of the results where many recent {\it IXPE} results were presented, alongside direct applications of the numerical models developed in the first part of this work.

~\

In Chapter \ref{chap01} we presented a calculation for arbitrarily polarized X-ray power-laws reprocessed in constant-density disc atmospheres, comoving with the source of irradiation. We used an iterative radiative transfer equation solver {\tt TITAN} and~a~MC simulator {\tt STOKES}, aiming at polarisation estimation with high accuracy and developing models, primarily for AGNs. Both X-ray spectroscopy and~X-ray polarimetry are powerful independent methods to derive the spin of~the~central accreting black hole (not only) by means of reflected coronal radiation from inner regions of the accretion discs. High-fidelity reflection modeling is, however, necessary to infer information not only on the black-hole spin, but also on~the~system inclination and orientation, as well as on the accretion disc's properties and ionization structure. We showed that careful treatment of partial ionization of~the~disc's upper layers is critical for the resulting X-ray polarisation profiles with energy.

We may improve the presented model in several ways. In the future, it would be beneficial to release the assumption of constant density, as this is expected to~have an impact on both the resulting spectra and polarisation \citep[e.g.][]{Rozanska2002, Ross2007, Rozanska2011, Podgorny2021}. This would, however, require designing a completely new setup. Minor improvements and checks for model dependencies of the results are achievable with less effort. Apart from addressing different cut-off selections or cosmic matter abundances, adding more spectral lines, or attempting for pencil-like illumination in~{\tt TITAN}, perhaps the easiest first step towards generalization would be attempting for the same computations, just for different slab densities. High-density reflection tables, suitable for accretion discs of XRBs, can be compared in their spectral part much in the same way to the high-density {\tt XILLVER} \citep{Garcia2016} and~{\tt REFLIONX} \citep[the {\tt REFHIDEN} version of the standard tables][]{Ross2005, Ross2007} tables. Cross-validation at a deeper level in flux is necessary for assessing polarisation properties of this flux in reflecting high-density disc models.

We argued that with contemporary numerical facilities it is difficult to derive information on the inner-accreting structures in the vicinity of black holes from X-ray polarimetry of thermally emitting disc atmospheres. Due to the expected accretion disc X-ray properties, here the main applications are for XRBs, rather than AGNs. With the same codes and major assumptions as for the reflection studies, we constructed models of passive scattering and absorbing layers in~transmission, illuminated from the opposite side by isotropic single-temperature black-body radiation, representing the radiative power input from a significantly heated optically thick accretion disc below. The absorbing medium may cause substantial flux depletion in the transmitted radiation at about 3~keV, which results in rather high and decreasing polarisation degree with energy between 2--8~keV where {\it IXPE} operates. However, if the plane-parallel slab is ionized enough, which is a natural state of the tested plane-parallel slabs in photo-ionization equilibrium, the X-ray spectropolarimetric properties of the emergent radiation do not change significantly with further ionization by decrease of medium density or increase in irradiating intensity. Alongside typical X-ray signatures of~high ionization in~XRBs in the thermal state, the obtained model polarisation degree with energy, which is constant or increasing in 2--8 keV, may be used to explain some puzzling {\it IXPE} observations of XRBs in the high/soft state \citep[namely 4U1630-47 and LMC X-3,][]{Ratheesh2023, Svoboda2023}, by means of enhanced absorption and Compton down-scattering effects and possibly by adding outflowing velocities to the absorbing and
scattering medium that cause additional polarisation increase due to aberration. In such high ionization limit, however, the disc structure and~the~absorption and~scattering geometry is largely unconstrained by our model.

It would be useful to put more effort in constructing full semi-infinite atmosphere models in X-ray polarisation gained by transmission of thermal radiation, which so far consist of only few examples in the literature. A few more individual and brave attempts in detailed radiative transfer modeling -- without the ambitions for constructing a data-fitting tool -- can tell a lot in favoring or disfavoring of an entire class of less sophisticated theoretical works, which are typically more useful for direct data fitting. Although the model that was presented in~this dissertation fails at high slab thickness, there are possibilities of~adaptation. The {\tt TITAN} code allows for boundary conditions that could be used for thick atmospheres and using the information from {\tt TITAN} on the source function through intermediate layers in the non-LTE conditions (after convergence of the iterative ALI computations), we could compensate for the flux loss, which is inevitable in the MC {\tt STOKES} results for thick atmospheres. In a broader view, different illuminating spectra or combined illumination (most importantly the same slab seen in transmission for a black body and in reflection for a power-law: with the~vertical ionization structure \textit{consistently} precomputed) are also accessible, since we have well tested the two \textit{detached} illumination examples already, using the same method with {\tt TITAN} and {\tt STOKES}. Another critical update of this part of work on local spectropolarimetric radiative transfer would be to use the latest {\tt STOKES} \textit{v2.34} version that includes Comptonization effects, which should be tested first with the already performed models to track changes in the results, as we neglected inverse Compton scattering in this work.

Apart from the ongoing {\it IXPE} campaigns, much progress in our understanding of XRBs in the soft state is also expected from future (non-)detections of~polarisation in other X-ray energy bands. Such as in the 15--80 keV window that is covered by the \textit{XL-Calibur} experiment and that can reveal more on the continuation of the rise of polarisation with energy in 2--8 keV, which is currently detected in many XRBs. Observing new sources with {\it IXPE} with known projected orientation on the sky would also help to validate the above presented modeling directions, because our model of thermal emission passing through an~absorbing atmosphere is valid only if the resulting polarisation angle is parallel to~the~prevailing slab plane, similarly to most of the decades old analytical results. If the polarisation directions orthogonal to the disc plane become observationally preferred for thermal emission in the near future, either a delicate combination of~optically thin covering medium with particular distribution of internal sources, or geometries, or bulk motions would have to be assumed, or perhaps this would be exactly a small example in the history of science where a new gate opens and we might be missing something fundamental. Observing the same sources again and observing new sources with {\it IXPE} would both help in this sense.

~\

In Chapter \ref{chap02}, we provided two relativistic models of the entire disc-corona system: assuming the lamp-post geometry and the sandwich slab geometry. They incorporate the local reflection tables for partially ionized AGN accretion discs discussed in Chapter \ref{chap01} and provide first-order estimates of major quantities describing the inner-accreting regions via spectropolarimetric fitting in {\tt XSPEC} or by means of unfolded model comparison. Focusing on X-ray polarisation arising from reflection, we discussed the major spectropolarimetric properties of these regions. The predicted X-ray polarisation fraction gained by reflection from geometrically thin accretion discs extending to the ISCO in the lamp-post models is typically increasing with energy, of the order of a few \%, and with a polarisation angle oriented parallel to the principal axis. In~most of the tested scenarios, the X-ray polarisation obtained from reflection in accretion discs in the sandwich slab coronal models is less than $1\%$, peaking in the {\it IXPE} band and with polarisation angle typically parallel to the equatorial plane. In addition, {\tt KYNSTOKES} allows to assess the inner disc radius location through X-ray polarisation by reflection. In the lamp-post geometries, the polarisation gain by geometrical reduction is to~the~detriment of flux in the reflected component. Such diagnostics may be useful again towards the upcoming bright times of observational X-ray polarimetry and towards our understandings of the accretion mechanisms.

Regarding future development of the relativistic codes presented here, it is highly desirable to reimplement any newer version of the local reflection tables, especially towards any models more suitable for XRBs. Then, we need to release the isotropic emission assumption of the lamp-post corona or function with more realistic coronal emission by implementing the standard disc reflection from {\tt KYNSTOKES} for arbitrary input into {\tt MONK} or {\tt KerrC}. Studying different coronal properties in this way provides much room for new discoveries.

Departures from the Kerr metric are not possible to test with our models, but could be critical.  Returning radiation effects in Kerr spacetime for X-ray polarisation by disc reflection need to be thoroughly investigated already in the scope current models, as the recent spectral analysis of returning radiation in~\cite{Dauser2022} suggests that they could play a role.

Apart from {\tt KYNSTOKES} that studies reflection from a geometrically thin disc, we may also try to implement the new local reflection tables to the {\tt SLIMBH} model, built for slim discs, as these are in many theoretical and observational arguments preferred over the standard Novikov-Thorne accretion. The work of \cite{West2023} is currently also taking lead in assessing the critical assumption of the disc thickness, including full GR treatment.

~\

In Chapter \ref{chap04}, we provided a set of computations assessing X-ray polarisation properties of distant circumnuclear components. These are far enough from the~core, to be separated from GR effects that affect the source radiation near its origin and complicate the modeling. However, the reprocessed polarisation is still highly dependent on the composition, shape and size of the scattering regions. In~case of both AGNs and XRBs, this is more important for X-ray polarimetry of obscured sources, viewed rather edge-on. Rather than attempting for providing clear guidelines for observational analysis, which is too ambitous in the current state of knowledge, we argued that even with static and homogeneous toroidal equatorial obscurers, the diversity in possible X-ray polarisation production of~the~reprocessed signal is striking, reaching tens of \% in parallel or perpendicular orientation with respect to the axis of symmetry.

By means of comparing the detailed MC simulations with {\tt STOKES} to a simplified model of reflection from the inner toroidal surface, we showed that large part of the diversity in X-ray polarisation is due to geometry of the reprocessing medium only. The newly built simple model and its associated routines open up place for easy-to-implement modifications and for a new class of {\tt XSPEC} spectropolarimetric models suitable for studying reflection from axially symmetric surfaces without any relativistic effects. However, the remaining differences between the~simple approach and a full MC simulation (that accounted for e.g. partial transparency and self-irradiation effects) still proved to be important for~the~polarisation predictions, modifying the results by at least few \% in~the~testing \mbox{3.5--6~keV} and \mbox{30--60~keV} bands towards higher perpendicular polarisation fraction with respect to the axis of symmetry.

We envision that the additional modeling of non-homogeneous distant scattering regions, which we did not do here, is necessary to assess the sensitivity to this assumption. At least the AGN tori are largely believed to be clumpy \citep{Nenkova2002, Risaliti2002, Matt2003, Tristram2007} and~a~detailed X-ray polarisation model could be conversely used for examination of~the~clumpiness characteristics \citep{Marin2012b, Marin2013, Marin2015b}. We also completely ignored possible misalignment of the scattering regions, which would be of scope of another study of the same length, as releasing this assumption opens at least one new dimension in the parameter region \citep{Goosmann2011}. Another useful way of continuation of this work would be the study of different irradiating spectral distributions, or non-isotropic or spatially extended sources.

Since the presented modeling represents one of the first broad-parametric attempts for X-ray polarisation estimates from geometrically thick equatorial obscurers, it would be highly beneficial to carry out a detailed comparative study with respect to the simultaneously emerging analytical models \citep{Veledina2023} and MC codes \citep{Ghisellini1994,Ratheesh2021,Ursini2023,Meulen2023, Tomaru2023, Tanimoto2023} in~the~near future.

~\

In Chapter \ref{chap05}, we provided a few examples of usage of the new numerical models for the interpretation of the first {\it IXPE} observations. In addition, we provided general observational prospects for accreting black holes. Not only {\it IXPE}, operating in the mid X-rays, and {\it XL-Calibur}, operating in the hard X-rays are believed to bring interesting X-ray polarimetric discoveries in the near future. We also mentioned the {\it eXTP} mission, which should be a successor of {\it IXPE}, operating in~the~same energy band with about four times larger effective mirror area. The~soft X-ray sounding rocket experiment \textit{REDSoX}, planned to operate in \mbox{0.2--0.8~keV}, could also reveal a lot of X-ray polarisation mysteries around accreting black holes, given our models. Although in AGNs, the soft X-rays are usually a matter of debates already from the spectroscopic point of view due to~the~presence of~the~soft excess in flux, which we have not modelled thoroughly.

~\

Staying humble with the scientific progress that this individual work allowed, we stress the general utility of X-ray polarimetry in the context of high-energy astrophysics that was theoretically envisioned already in the 1970s and now is observationally becoming confirmed. For the case of accreting black holes either by strongly favoring locations of the hot emitting plasma, which are critically unknown inside the accretion flow, much like a microscope providing physical clues to unresolved structures. Or by showing surprises in high X-ray polarisation of~unknown origin in the reprocessings of thermal emission, arising from the~core. Independent clues to black-hole spin values, which are typically extremely high from X-ray spectroscopy (a surprising outcome for a dynamical evolution), or to~the~accretion disc's inclination, size and structure, which are for many sources completely unknown, are only a matter of time when larger samples of X-ray polarimetric observations will be collected. The same holds for characteristics of~equatorial and polar outflows and tests of the unification scenario for AGNs, which is dependent on the positioning and composition of parsec-scale components of various kinds that X-ray polarimetry can unravel.

If more observational surprises are yet to come and theoretical efforts lack behind, in such examples, it suggests itself to learn from other branches of astrophysics, such as the solar physics, where observational data as well as numerical simulations are typically more developed than for black-hole environments \citep[e.g.][]{Stenflo1994, DeglInnocenti2004,Heinzel2020}. One can also learn from laboratory X-ray polarimetry \citep[e.g.][]{Giles1995,Huard1997,Stohr2006,Grizolli2016,Shan2016,Yang2021,Stohr2023} that solves many quantum phenomena in greater detail than the astrophysical sources allow, despite being unique in the extreme physical conditions that are occurring elsewhere in our Universe.

\newpage
\ 
\vspace{4.5cm} 
\begin{figure}[h]
       \centering
	\includegraphics[trim={0cm 0cm 0cm 0cm},clip,width=0.6\textwidth]{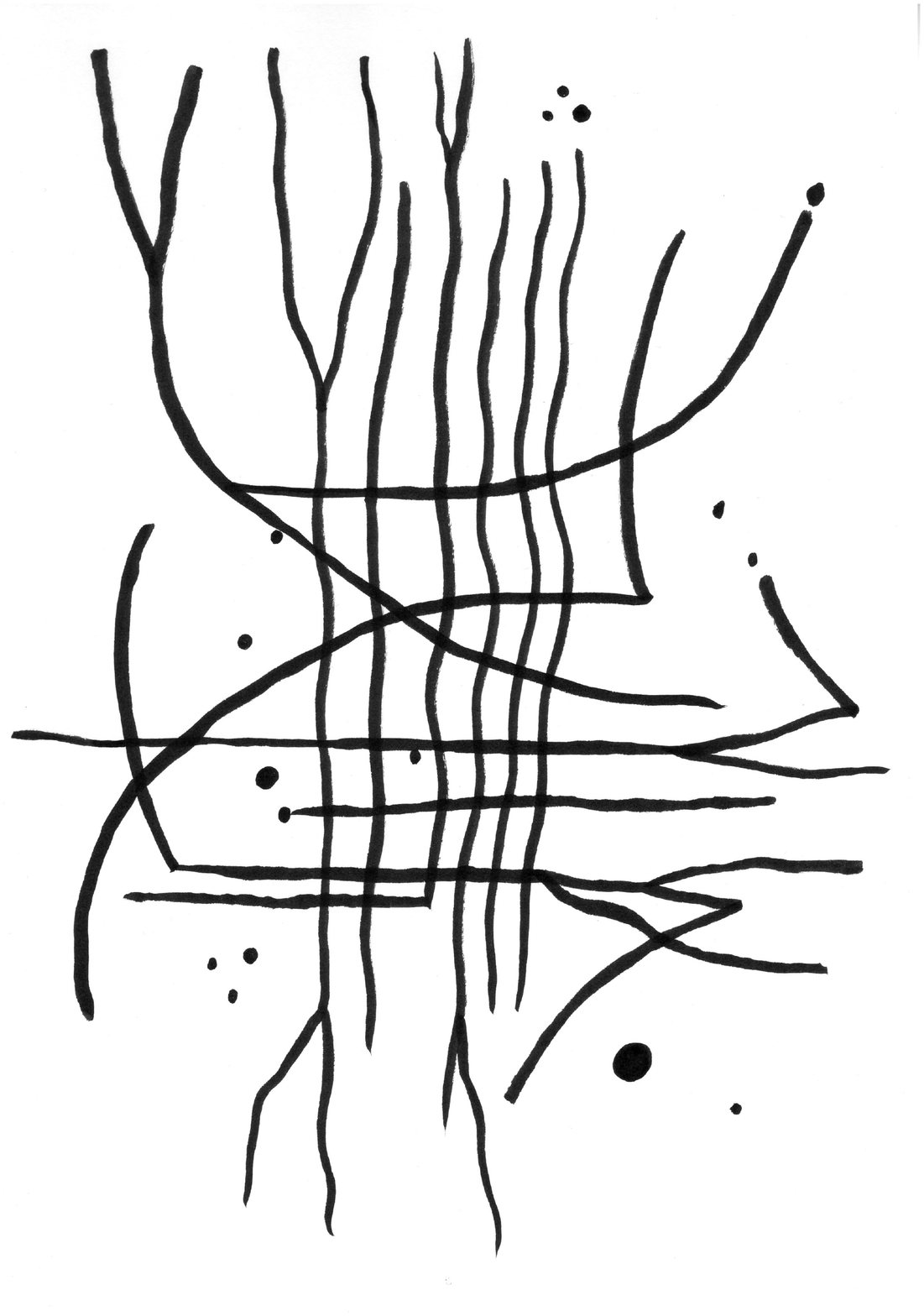}
\end{figure}

\bibliographystyle{plainnat}    

\renewcommand{\bibname}{Bibliography}


\bibliography{thesis}


%
%




\appendix

\chapter{Supplementary material}\label{supplementary}

\section*{Supplementary figures to Chapter \ref{chap01}}

\begin{figure}[!htb]\centering
	\includegraphics[width=\textwidth]{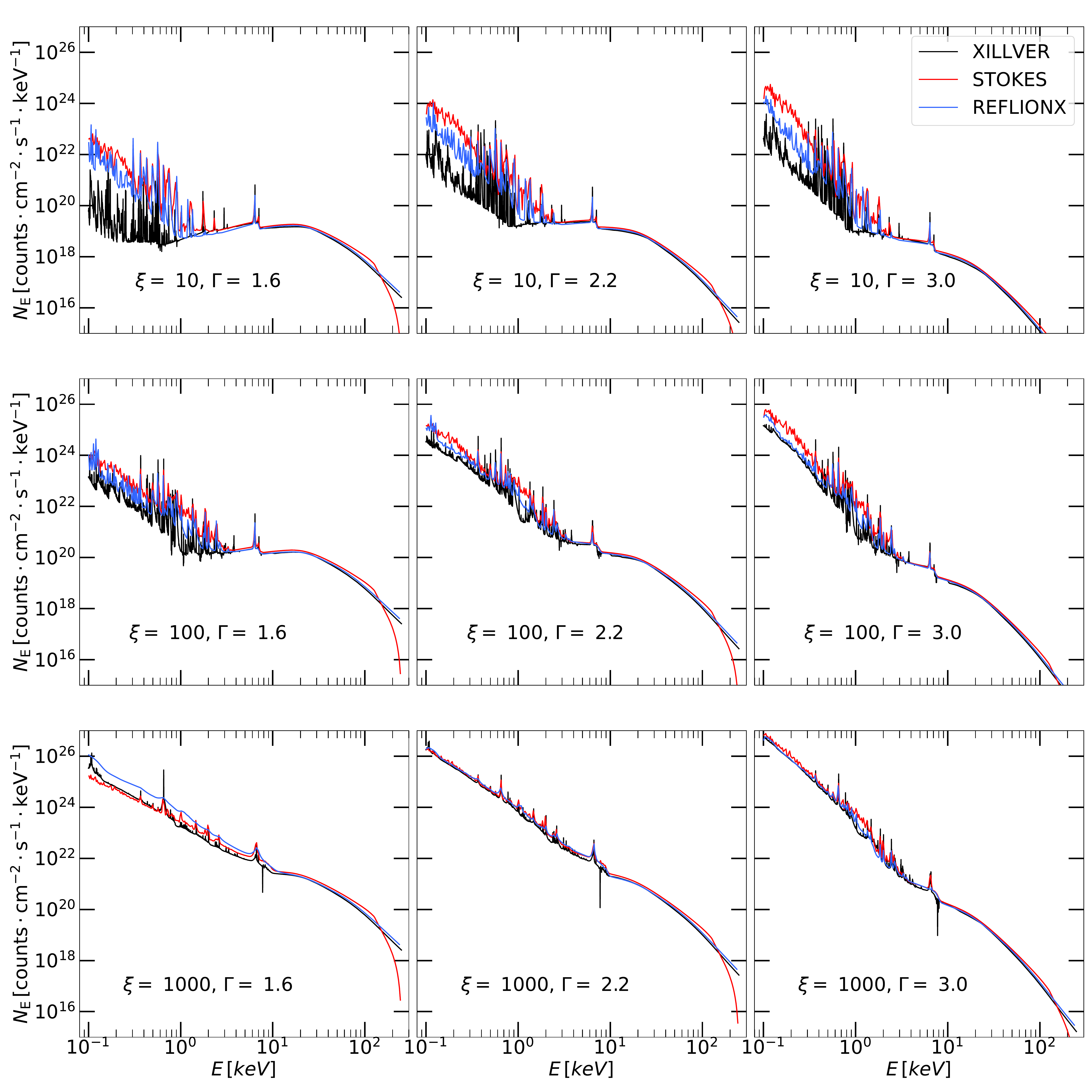}
	\caption{\footnotesize{The same as in Figure \ref{xillver_reflionx_compare}, but for various $\xi$ in $[\textrm{erg} \cdot \textrm{cm} \cdot \textrm{s}^{-1}]$ and $\Gamma$.}}
	\label{compare_all_xst}
\end{figure}

\begin{landscape}
\begin{figure}[h]
    \centering
    \begin{picture}(137,99)
    \put(0,0){\includegraphics[width=0.33\textwidth]{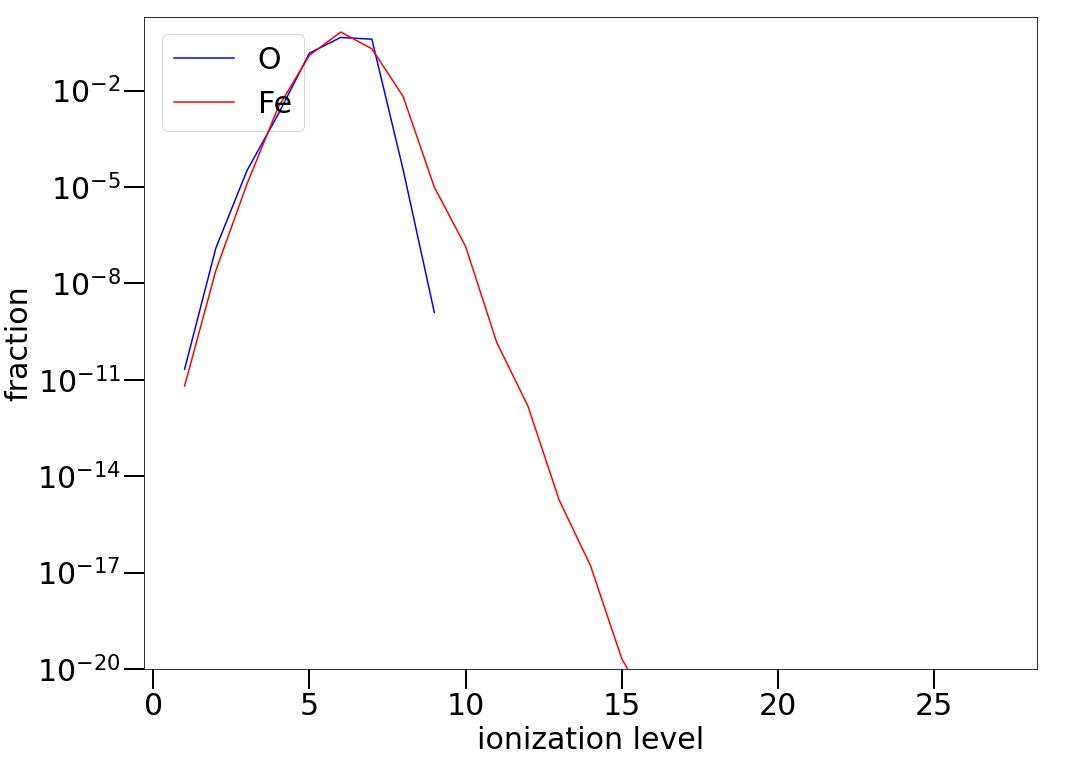}}
    \put(42,20){\scalebox{.55}{$\xi = 3\times 10^0 \, \, \textrm{erg} \cdot \textrm{cm} \cdot \textrm{s}^{-1}$}}
    \end{picture}
    \begin{picture}(137,99)
    \put(0,0){\includegraphics[width=0.33\textwidth]{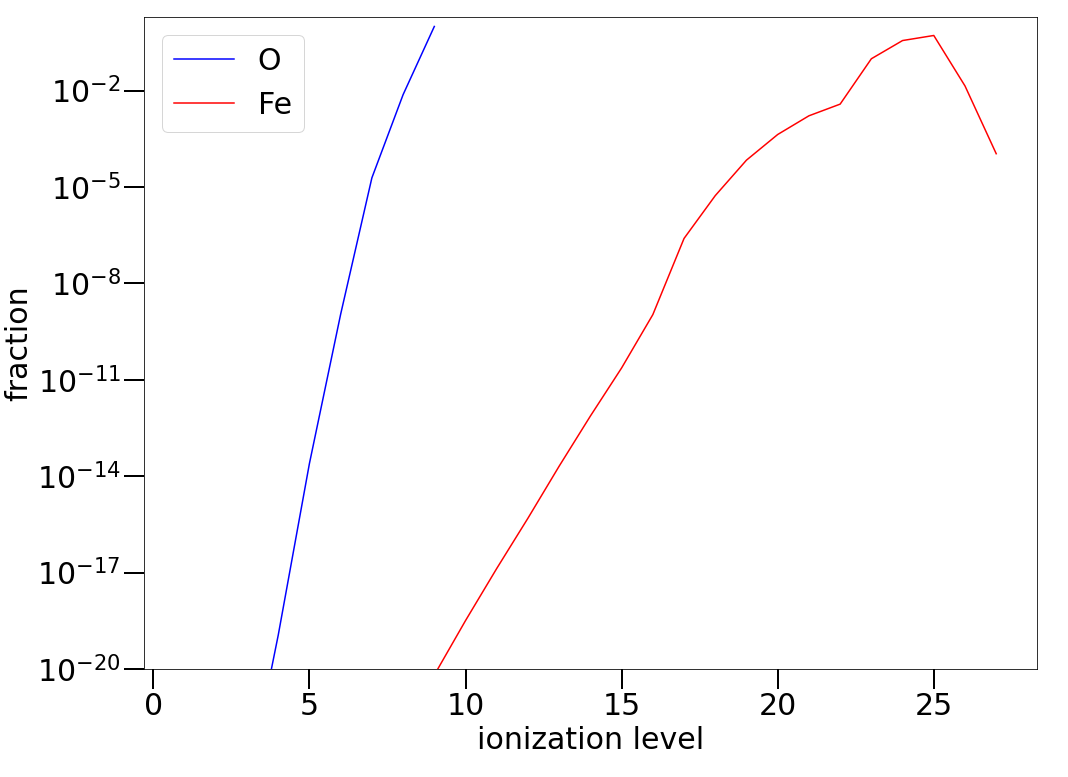}}
    \put(42,20){\scalebox{.55}{$\xi = 3\times 10^2 \, \, \textrm{erg} \cdot \textrm{cm} \cdot \textrm{s}^{-1}$}}
    \end{picture}
    \begin{picture}(137,99)
    \put(0,0){\includegraphics[width=0.33\textwidth]{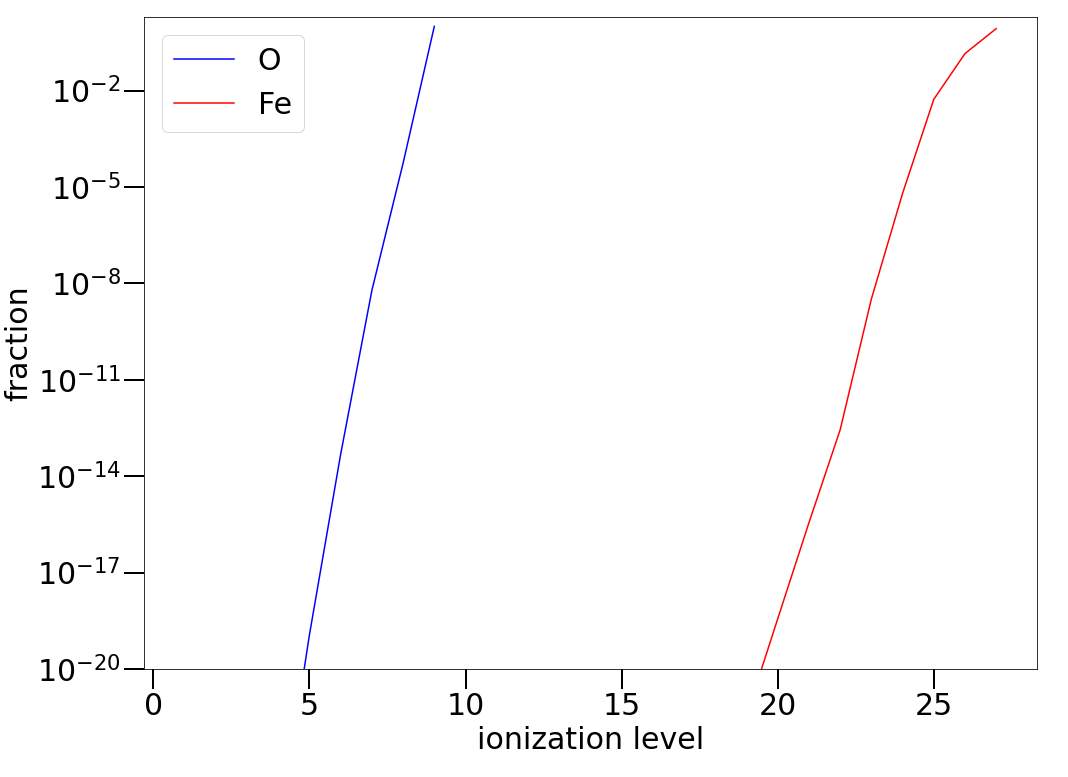}}
    \put(42,20){\scalebox{.55}{$\xi = 3\times 10^4 \, \, \textrm{erg} \cdot \textrm{cm} \cdot \textrm{s}^{-1}$}}
    \end{picture}
    \begin{picture}(137,99)
    \put(0,0){\includegraphics[width=0.33\textwidth]{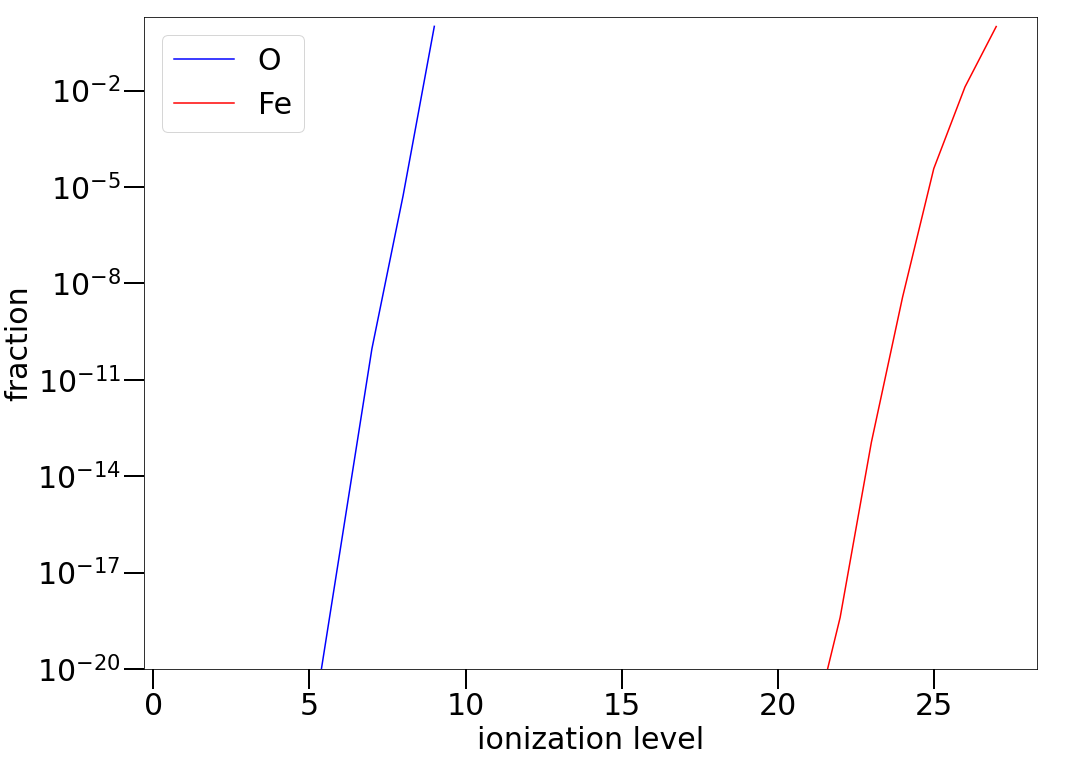}}
    \put(42,20){\scalebox{.55}{$\xi = 3\times 10^5 \, \, \textrm{erg} \cdot \textrm{cm} \cdot \textrm{s}^{-1}$}}
    \end{picture}
    \begin{picture}(137,99)
    \put(0,0){\includegraphics[width=0.33\textwidth]{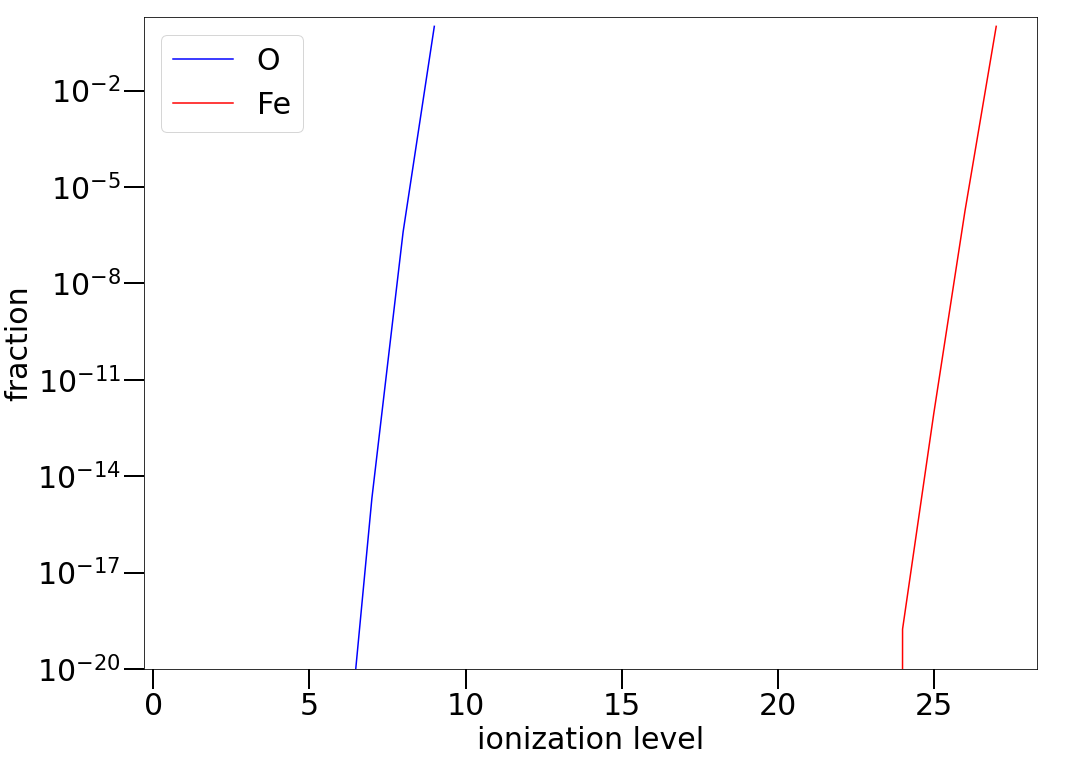}}
    \put(42,20){\scalebox{.55}{$\xi = 3\times 10^8 \, \, \textrm{erg} \cdot \textrm{cm} \cdot \textrm{s}^{-1}$}}
    \end{picture}\\
    \begin{picture}(137,99)
    \put(0,0){\includegraphics[width=0.33\textwidth]{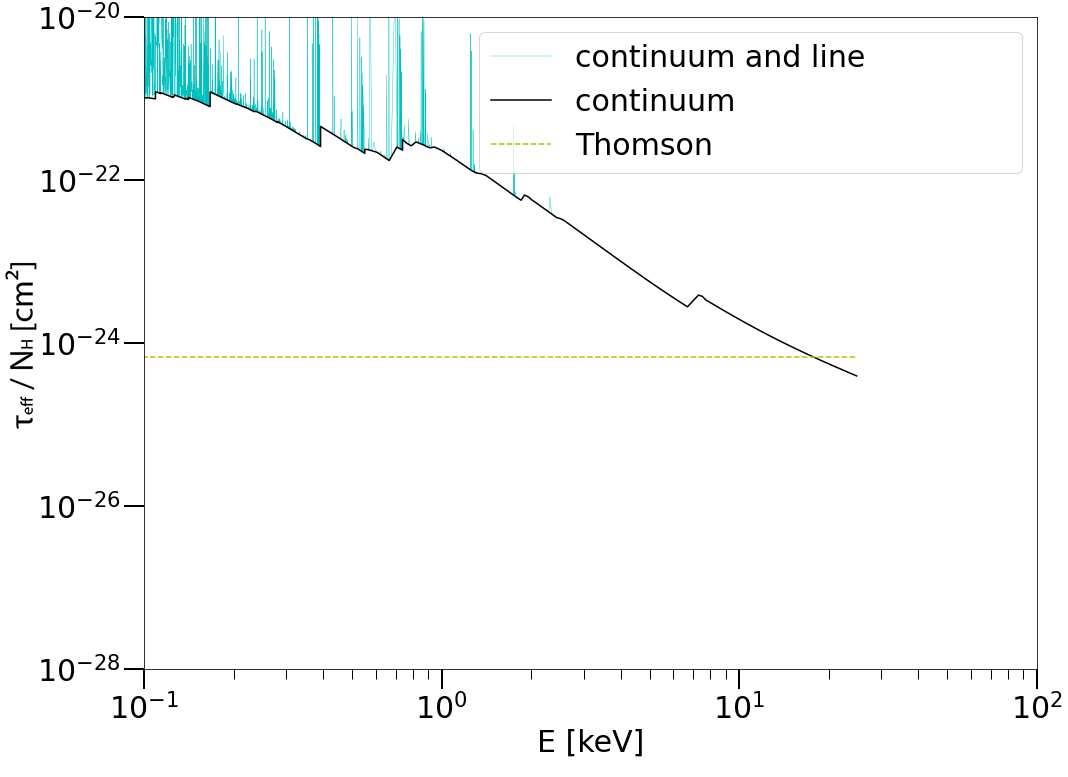}}
    \put(42,20){\scalebox{.55}{$\xi = 3\times 10^0 \, \, \textrm{erg} \cdot \textrm{cm} \cdot \textrm{s}^{-1}$}}
    \end{picture}
    \begin{picture}(137,99)
    \put(0,0){\includegraphics[width=0.33\textwidth]{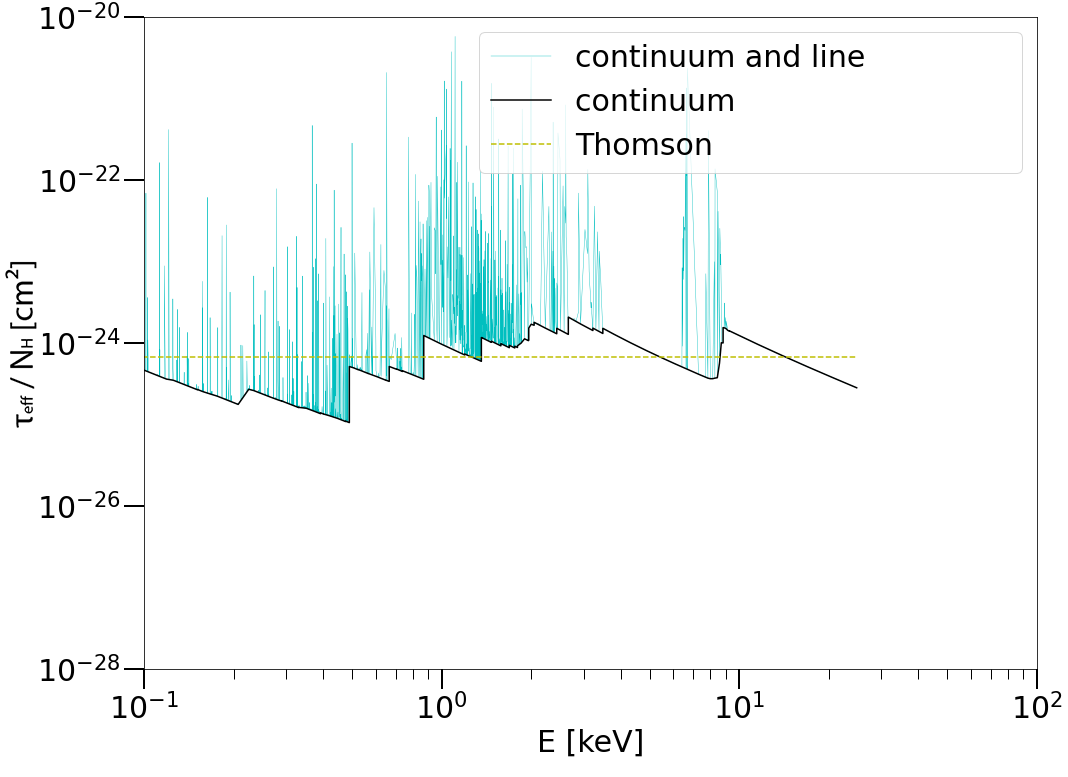}}
    \put(42,20){\scalebox{.55}{$\xi = 3\times 10^2 \, \, \textrm{erg} \cdot \textrm{cm} \cdot \textrm{s}^{-1}$}}
    \end{picture}
    \begin{picture}(137,99)
    \put(0,0){\includegraphics[width=0.33\textwidth]{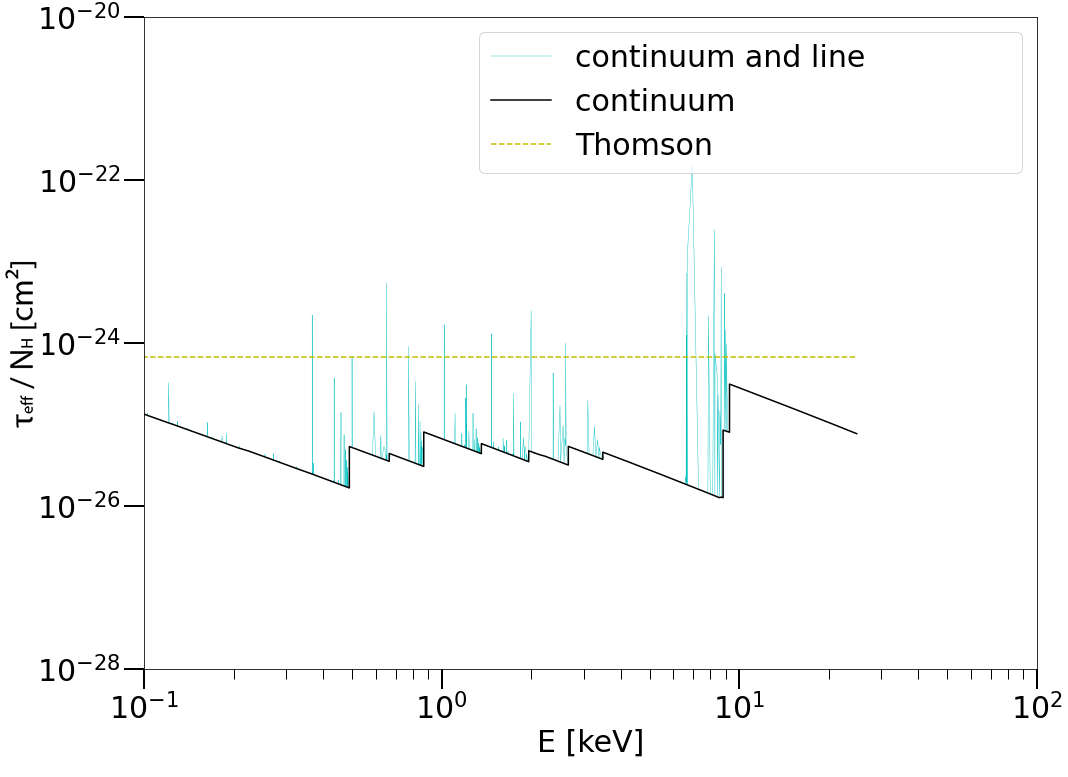}}
    \put(42,20){\scalebox{.55}{$\xi = 3\times 10^4 \, \, \textrm{erg} \cdot \textrm{cm} \cdot \textrm{s}^{-1}$}}
    \end{picture}
    \begin{picture}(137,99)
    \put(0,0){\includegraphics[width=0.33\textwidth]{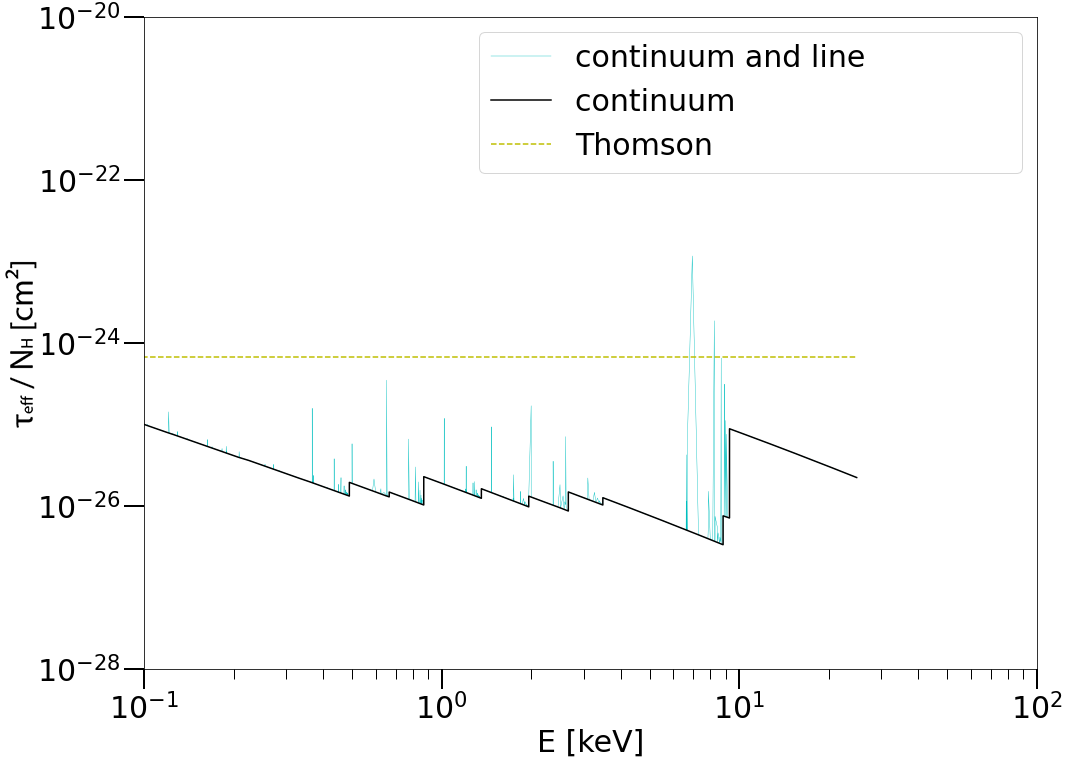}}
    \put(42,20){\scalebox{.55}{$\xi = 3\times 10^5 \, \, \textrm{erg} \cdot \textrm{cm} \cdot \textrm{s}^{-1}$}}
    \end{picture}
    \begin{picture}(137,99)
    \put(0,0){\includegraphics[width=0.33\textwidth]{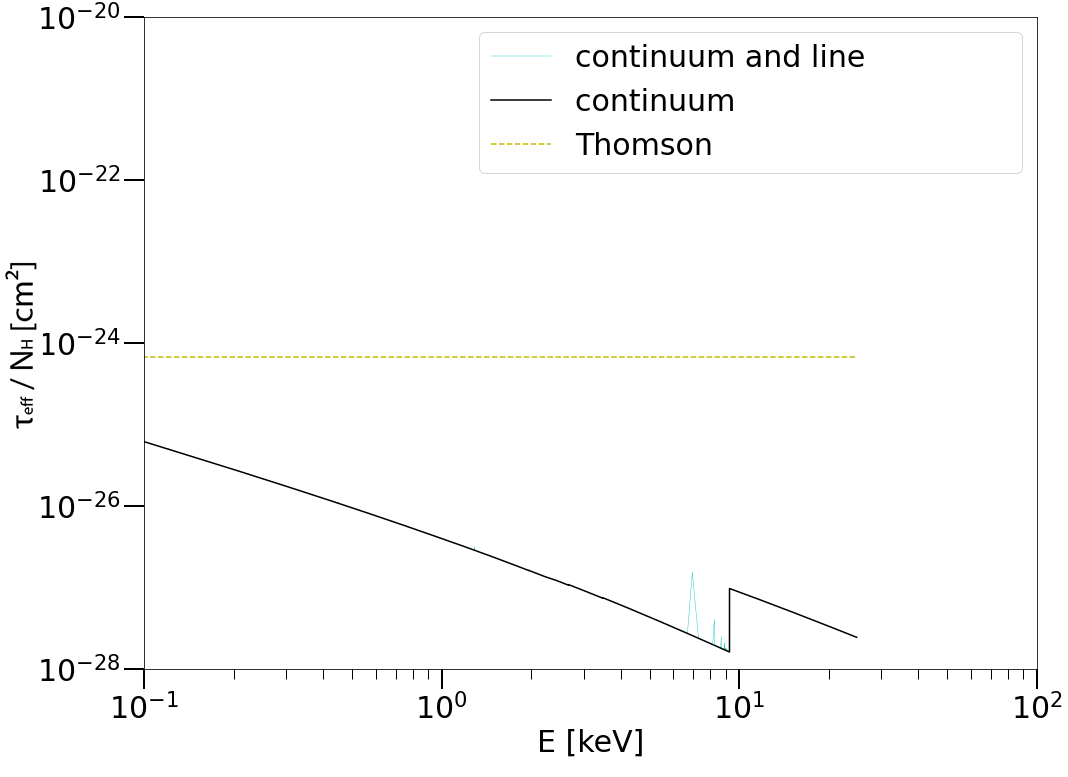}}
    \put(42,20){\scalebox{.55}{$\xi = 3\times 10^8 \, \, \textrm{erg} \cdot \textrm{cm} \cdot \textrm{s}^{-1}$}}
    \end{picture}
    \caption{\footnotesize{Various cases of transmitted single-temperature unpolarized black-body radiation with $k_\textrm{B}T_\textrm{BB} = 1.15$ keV through a constant density slab, as computed with {\tt TITAN} and {\tt STOKES} in PIE. From left to right the flux of the irradiation increases ($\xi = 3\times 10^0,3\times 10^2,3\times 10^4,3\times 10^5,3\times 10^8 \, \, \textrm{erg} \cdot \textrm{cm} \cdot \textrm{s}^{-1}$, respectively) and~the~slab is more ionized for the same density $n_\textrm{H} = 10^{18} \ \textrm{cm}^{-3}$. Top row: fractional abundance of ions of oxygen (blue) OI--OIX and iron (red) FeI--FeXXVII from {\tt TITAN} ionization structure precomputations. Bottom row: the effective optical depth, $\tau_\textrm{eff}$, divided by column density, $N_\textrm{H}\, [\textrm{cm}^{-2}]$, versus energy for continuum processes (black) and including lines (cyan) from {\tt TITAN} ionization structure precomputations. Thomson cross-section $\sigma_\textrm{T}$ (yellow) represents pure scatterings.}}
    \label{wind_absorbed_spectra}
\end{figure}
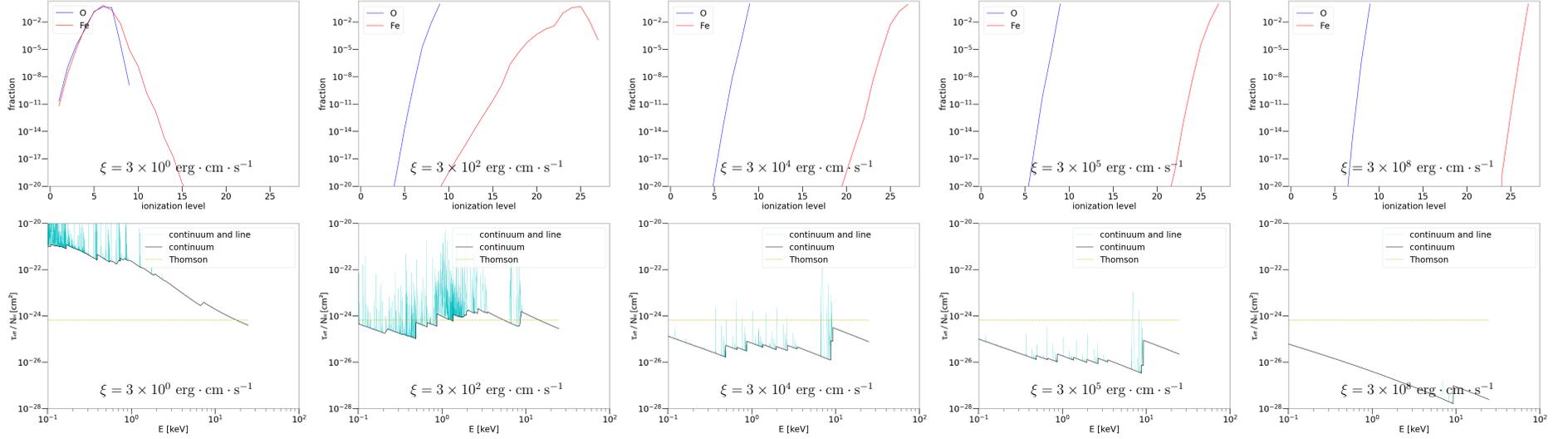
\end{landscape}
\begin{landscape}
\begin{figure}[h]
    \centering
    \begin{picture}(137,99)
    \put(0,0){\includegraphics[width=0.33\textwidth]{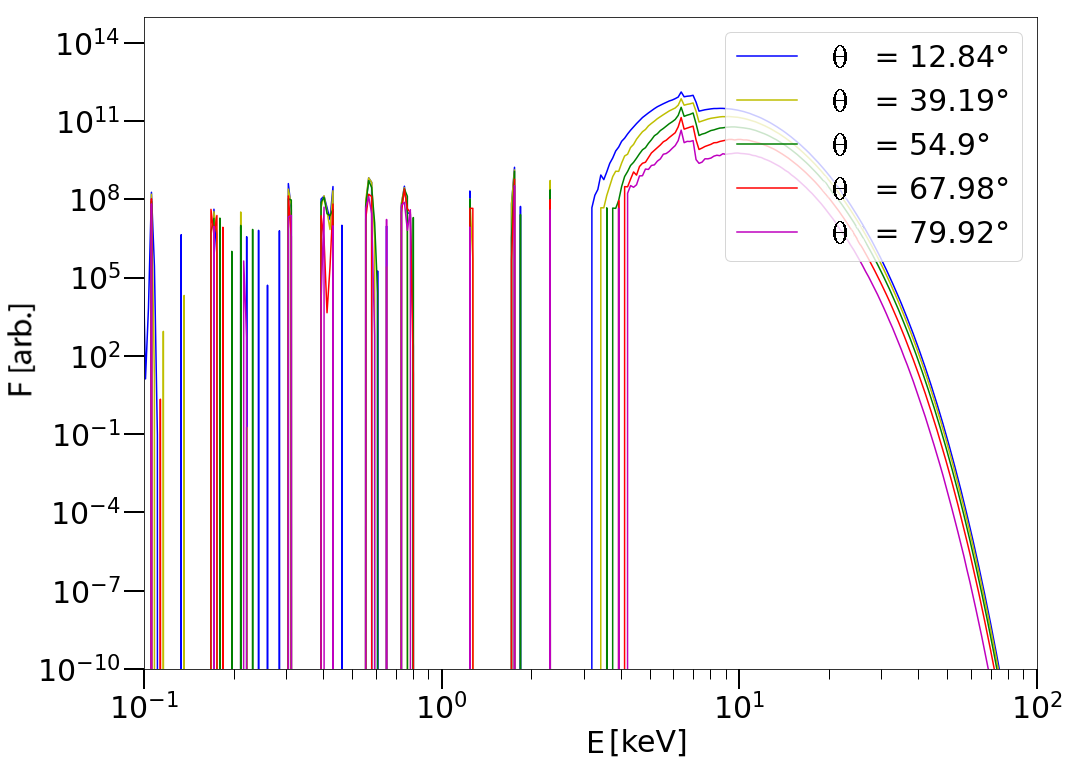}}
    \put(22,20){\scalebox{.55}{$\xi = 3\times 10^0 \, \, \textrm{erg} \cdot \textrm{cm} \cdot \textrm{s}^{-1}$}}
    \end{picture}
    \begin{picture}(137,99)
    \put(0,0){\includegraphics[width=0.33\textwidth]{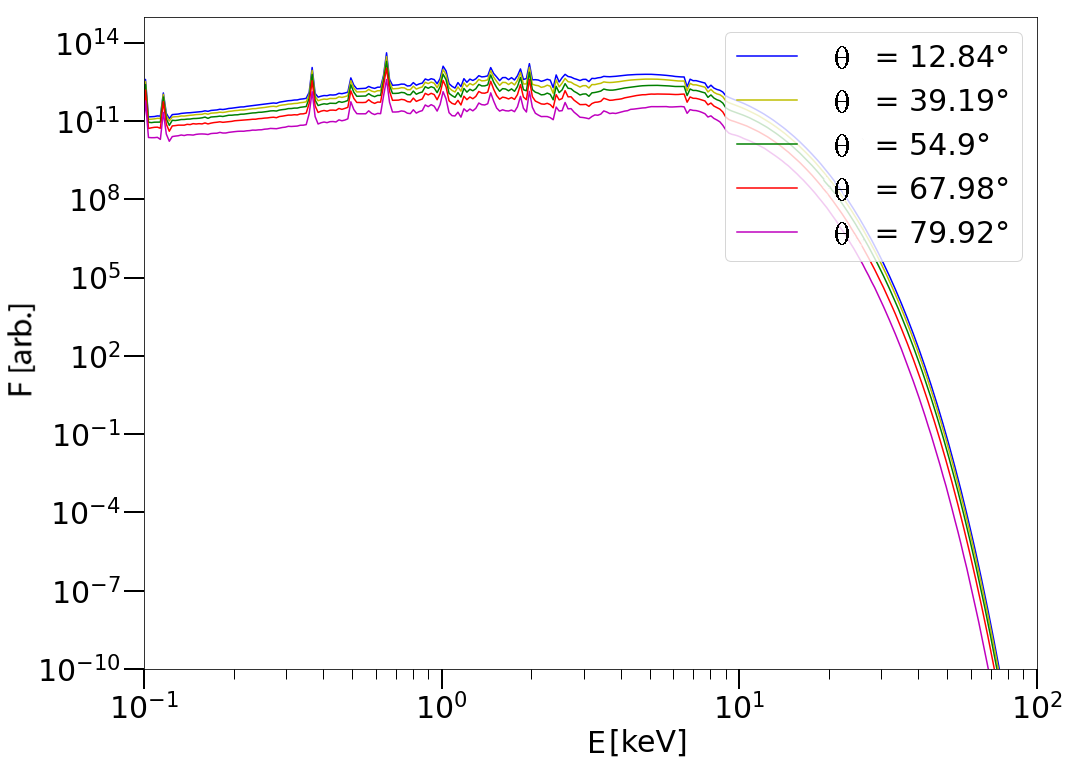}}
    \put(22,20){\scalebox{.55}{$\xi = 3\times 10^2 \, \, \textrm{erg} \cdot \textrm{cm} \cdot \textrm{s}^{-1}$}}
    \end{picture}
    \begin{picture}(137,99)
    \put(0,0){\includegraphics[width=0.33\textwidth]{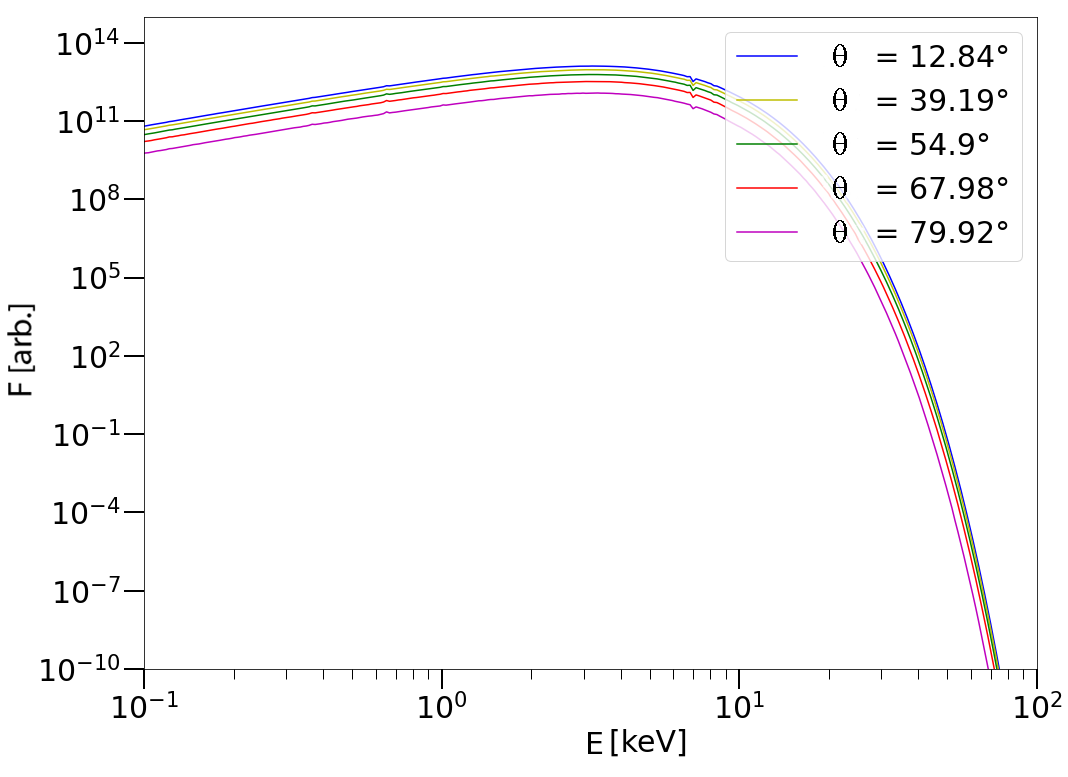}}
    \put(22,20){\scalebox{.55}{$\xi = 3\times 10^4 \, \, \textrm{erg} \cdot \textrm{cm} \cdot \textrm{s}^{-1}$}}
    \end{picture}
    \begin{picture}(137,99)
    \put(0,0){\includegraphics[width=0.33\textwidth]{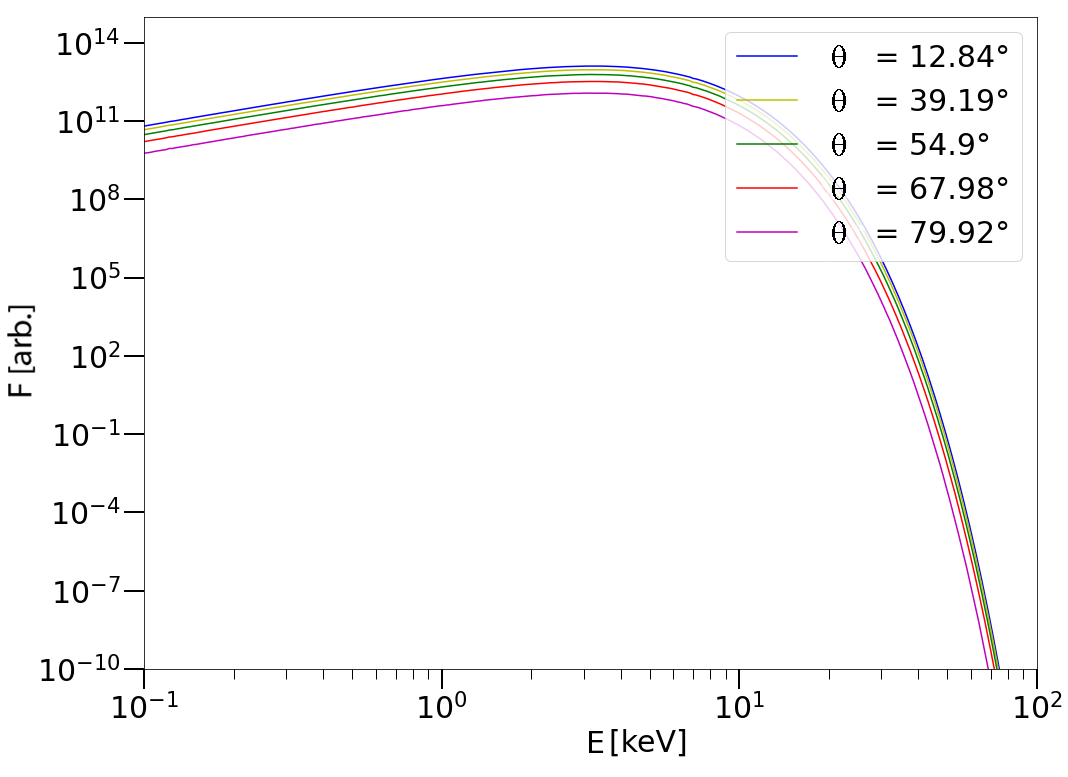}}
    \put(22,20){\scalebox{.55}{$\xi = 3\times 10^5 \, \, \textrm{erg} \cdot \textrm{cm} \cdot \textrm{s}^{-1}$}}
    \end{picture}
    \begin{picture}(137,99)
    \put(0,0){\includegraphics[width=0.33\textwidth]{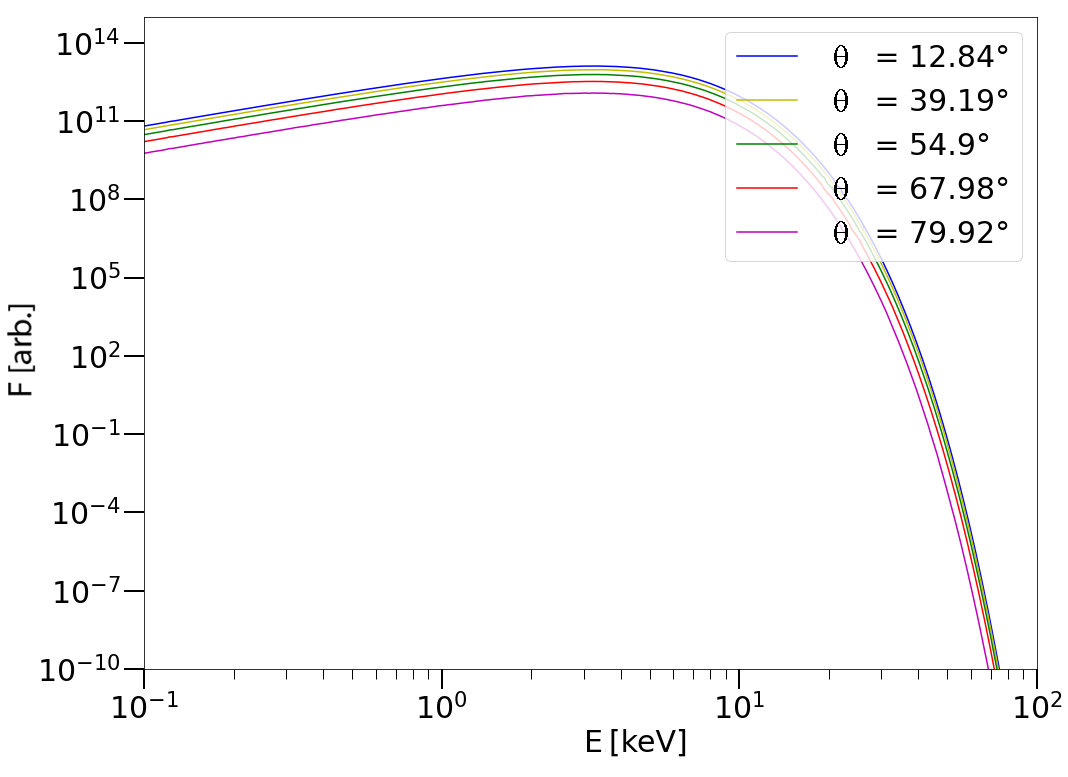}}
    \put(22,20){\scalebox{.55}{$\xi = 3\times 10^8 \, \, \textrm{erg} \cdot \textrm{cm} \cdot \textrm{s}^{-1}$}}
    \end{picture}\\
    \begin{picture}(137,99)
    \put(0,0){\includegraphics[width=0.33\textwidth]{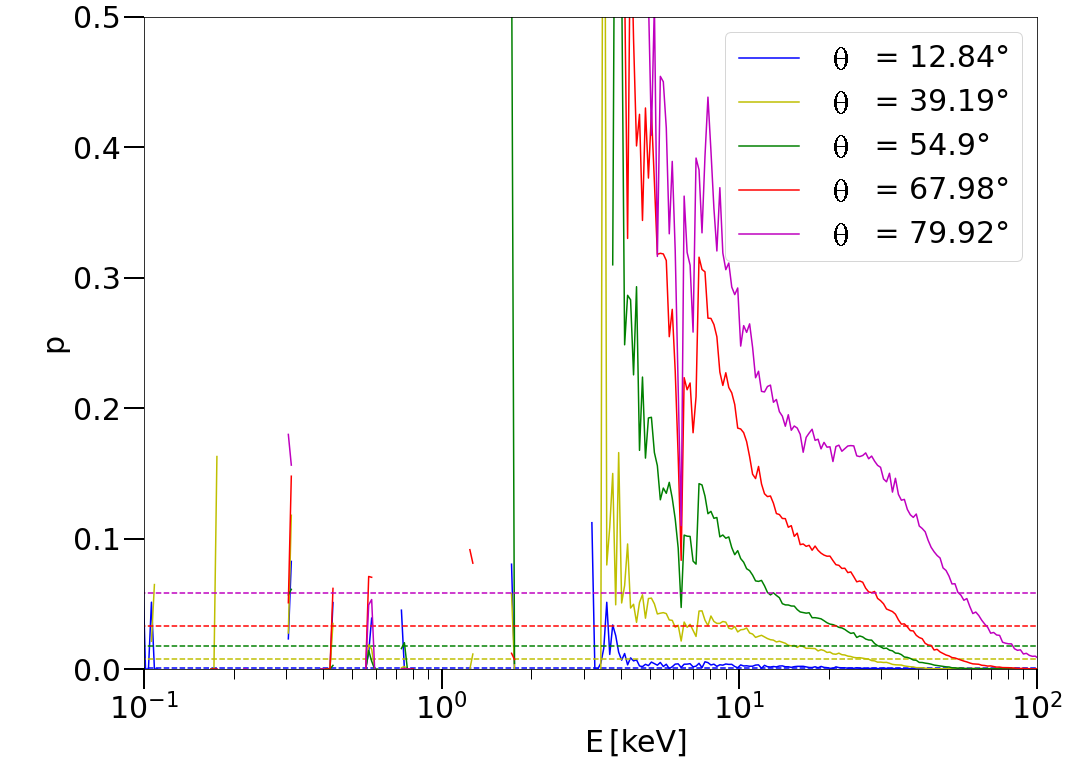}}
    \put(22,80){\scalebox{.55}{$\xi = 3\times 10^0 \, \, \textrm{erg} \cdot \textrm{cm} \cdot \textrm{s}^{-1}$}}
    \end{picture}
    \begin{picture}(137,99)
    \put(0,0){\includegraphics[width=0.33\textwidth]{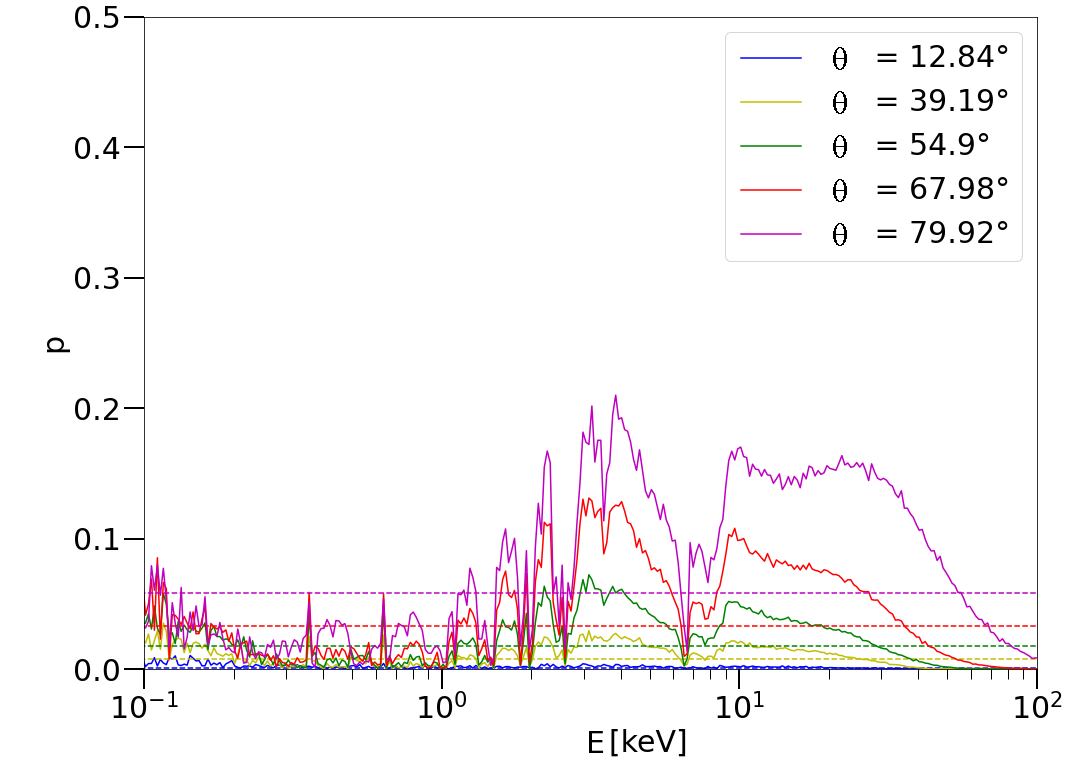}}
    \put(22,80){\scalebox{.55}{$\xi = 3\times 10^2 \, \, \textrm{erg} \cdot \textrm{cm} \cdot \textrm{s}^{-1}$}}
    \end{picture}
    \begin{picture}(137,99)
    \put(0,0){\includegraphics[width=0.33\textwidth]{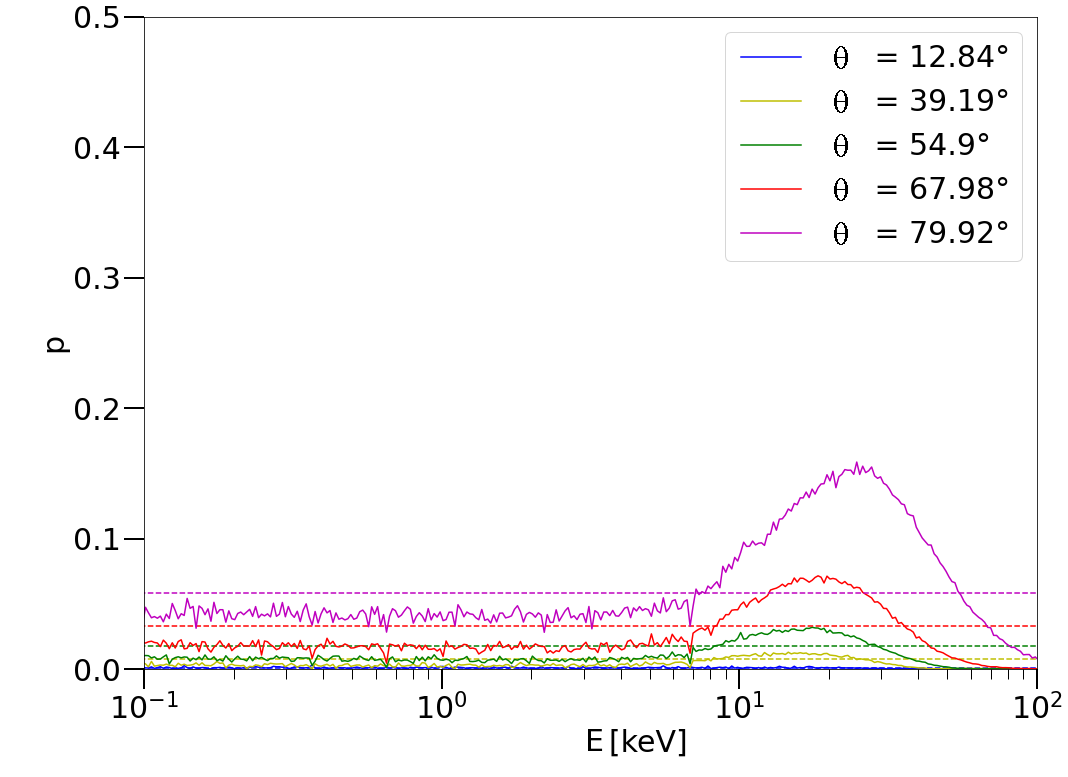}}
    \put(22,80){\scalebox{.55}{$\xi = 3\times 10^4 \, \, \textrm{erg} \cdot \textrm{cm} \cdot \textrm{s}^{-1}$}}
    \end{picture}
    \begin{picture}(137,99)
    \put(0,0){\includegraphics[width=0.33\textwidth]{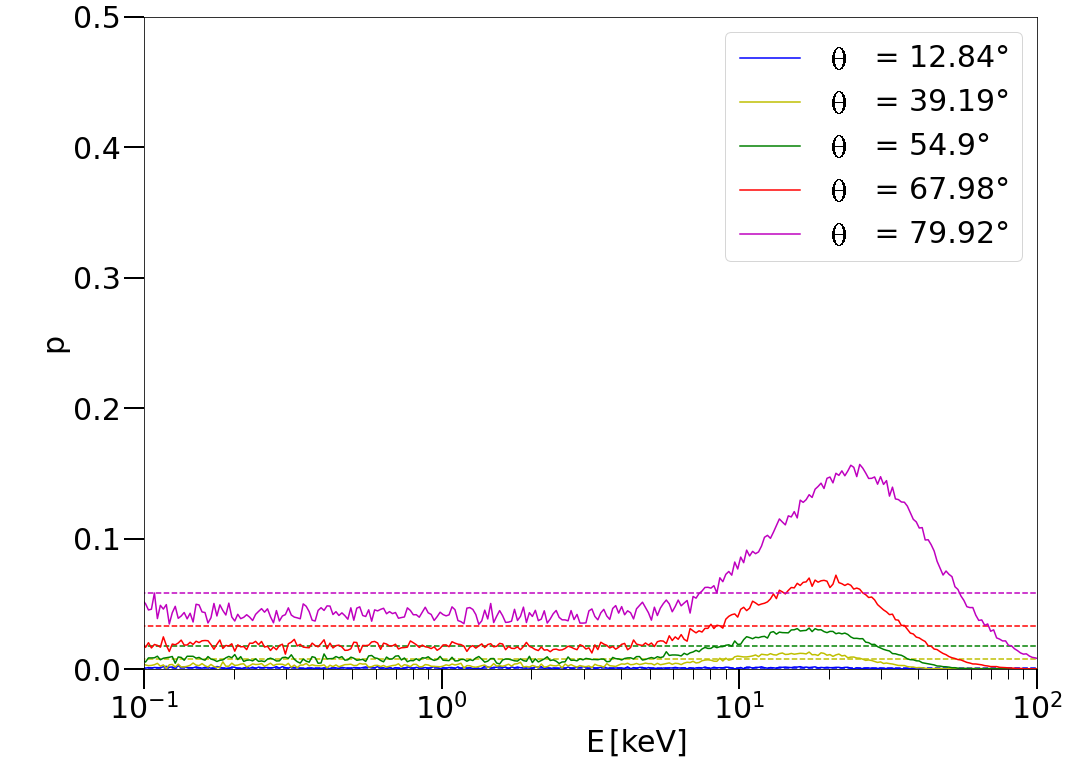}}
    \put(22,80){\scalebox{.55}{$\xi = 3\times 10^5 \, \, \textrm{erg} \cdot \textrm{cm} \cdot \textrm{s}^{-1}$}}
    \end{picture}
    \begin{picture}(137,99)
    \put(0,0){\includegraphics[width=0.33\textwidth]{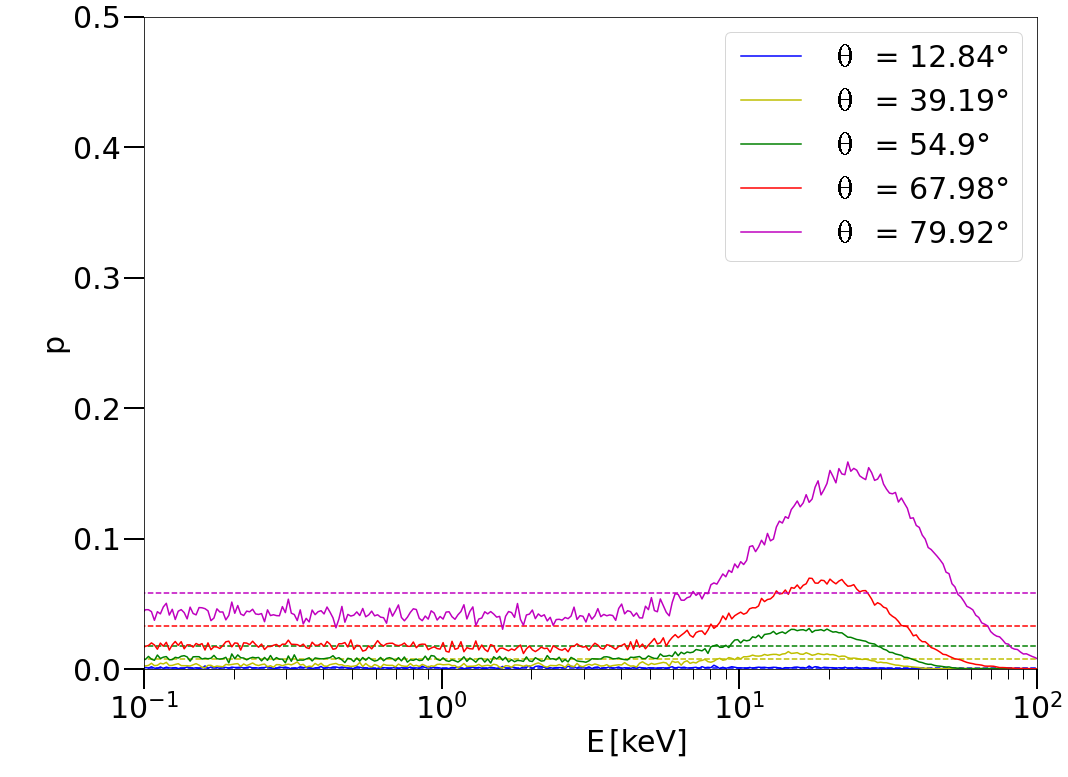}}
    \put(22,80){\scalebox{.55}{$\xi = 3\times 10^8 \, \, \textrm{erg} \cdot \textrm{cm} \cdot \textrm{s}^{-1}$}}
    \end{picture}
    \caption{\footnotesize{The {\tt STOKES} results for $\tau_\textrm{T,max} = 1$, corresponding to the cases shown in Figure \ref{wind_absorbed_spectra}. From left to right the flux of the irradiation increases ($\xi = 3\times 10^0,3\times 10^2,3\times 10^4,3\times 10^5,3\times 10^8 \, \, \textrm{erg} \cdot \textrm{cm} \cdot \textrm{s}^{-1}$, respectively) and the slab is more ionized for the same density $n_\textrm{H} = 10^{18} \ \textrm{cm}^{-3}$. Top row: spectra, $F$, per energy bin for various observer's inclinations $\theta$ in the color code. Bottom row: the corresponding polarisation degree, $p$, versus energy (solid lines) for various observer's inclinations $\theta$ in the color code. We indicate the pure-scattering limit for semi-infinite atmosphere given by Chandrasekhar-Sobolev approximation for~the~same inclination values (dashed horizontal lines). The obtained polarisation angle from {\tt STOKES} is constant with energy and parallel with the slab.}}
    \label{wind_absorbed_pd}
\end{figure}
\end{landscape}

\section*{Supplementary figures to Chapter \ref{chap02}}

\begin{figure}[!htb]\centering
	\includegraphics[width=\textwidth]{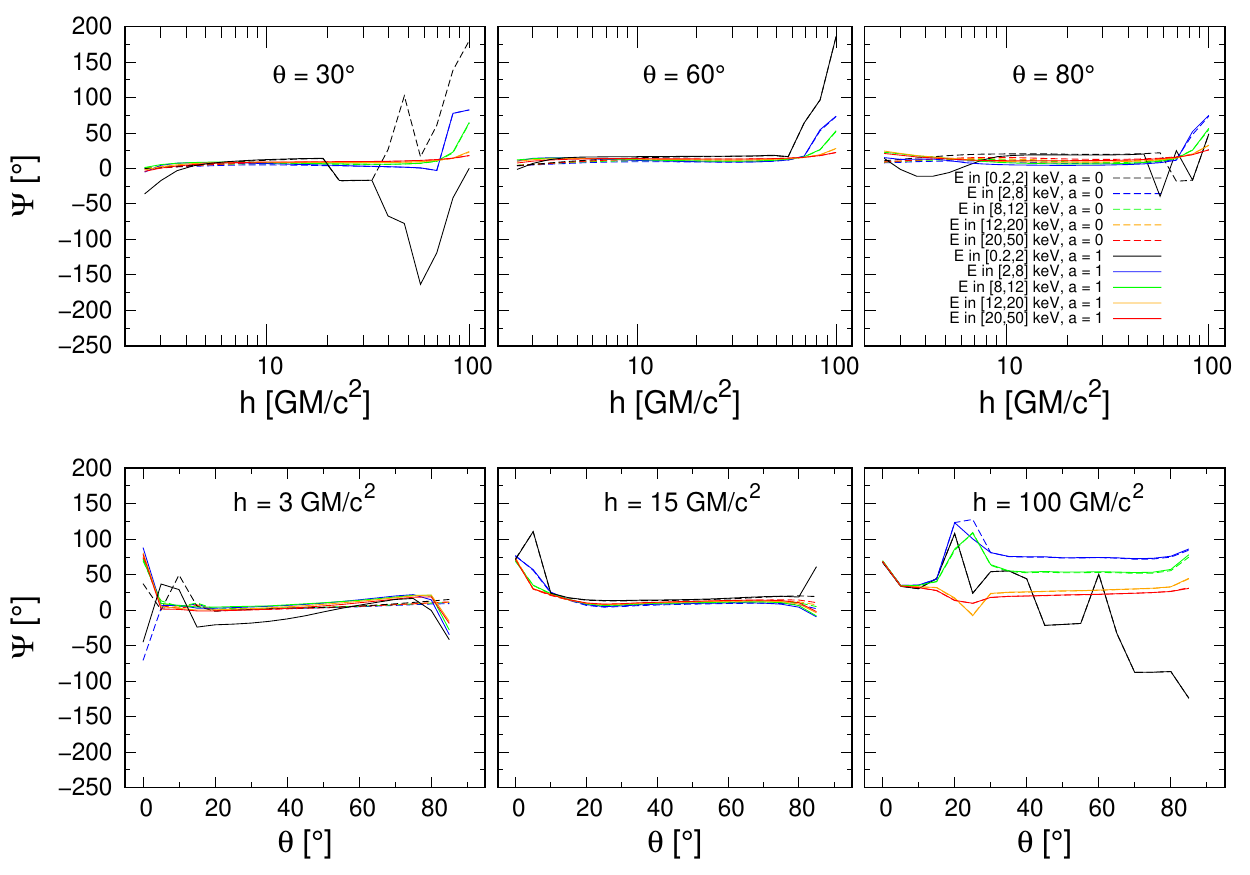}
	\caption{\footnotesize{Top panel: the total average polarisation angle, $\Psi$, versus height of the primary point-source above the disc for disc inclinations $\theta = 30^{\circ}$ (left), $\theta = 60^{\circ}$ (middle) and~$\theta = 80^{\circ}$ (right), $\Gamma = 2$, unpolarized primary radiation, $M = 1\times 10^8\,M_{\odot}$ and observed 2--10 keV flux $L_{\textrm{X}}/L_{\textrm{Edd}} = 0.001$, i.e. neutral disc. We show cases of two black-hole spins $a = 0$ (dashed) and~$a = 1$ (solid) and for energy bands $E \in [0.2, 2]$ keV (black),  $E \in [2, 8]$ keV (blue), \mbox{$E \in [8, 12]$~keV} (green), $E \in [12, 20]$ keV (orange), and $E \in [20, 50]$ keV (red). Bottom panel: the total average polarisation angle, $\Psi$, versus disc inclination for heights of the primary point-source above the~disc $h = 3 \textrm{ } r_\textrm{g}$ (left), $h = 15 \textrm{ } r_\textrm{g}$ (middle) and  $h = 100 \textrm{ } r_\textrm{g}$ (right), for the same configuration as on the top panel.}}
	\label{fig:fig10_tot_neutral.}
\end{figure}
\begin{figure}[!htb]\centering
	\includegraphics[width=\textwidth]{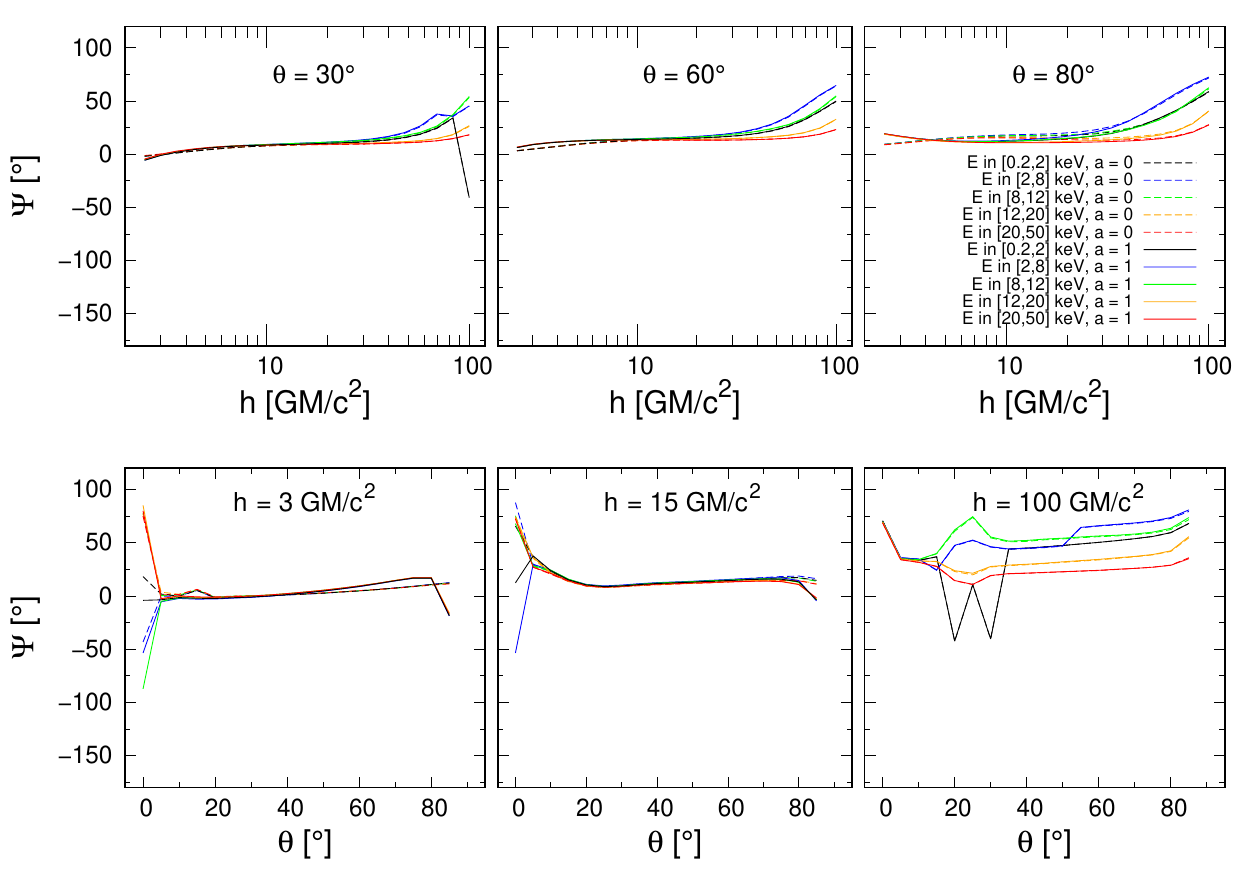}
	\caption{\footnotesize{The total average polarisation angle, $\Psi$, versus energy for ionized disc (\mbox{$M = 1\times 10^5\,M_{\odot}$} and observed 2--10 keV flux $L_{\textrm{X}}/L_{\textrm{Edd}} = 0.1$), otherwise for the same parametric setup as in Figure \ref{fig:fig10_tot_neutral.}, displayed in the same manner.}}
	\label{fig:fig10_tot_ionized.}
\end{figure}

\begin{figure}[!htb]\centering
	\includegraphics[width=\textwidth]{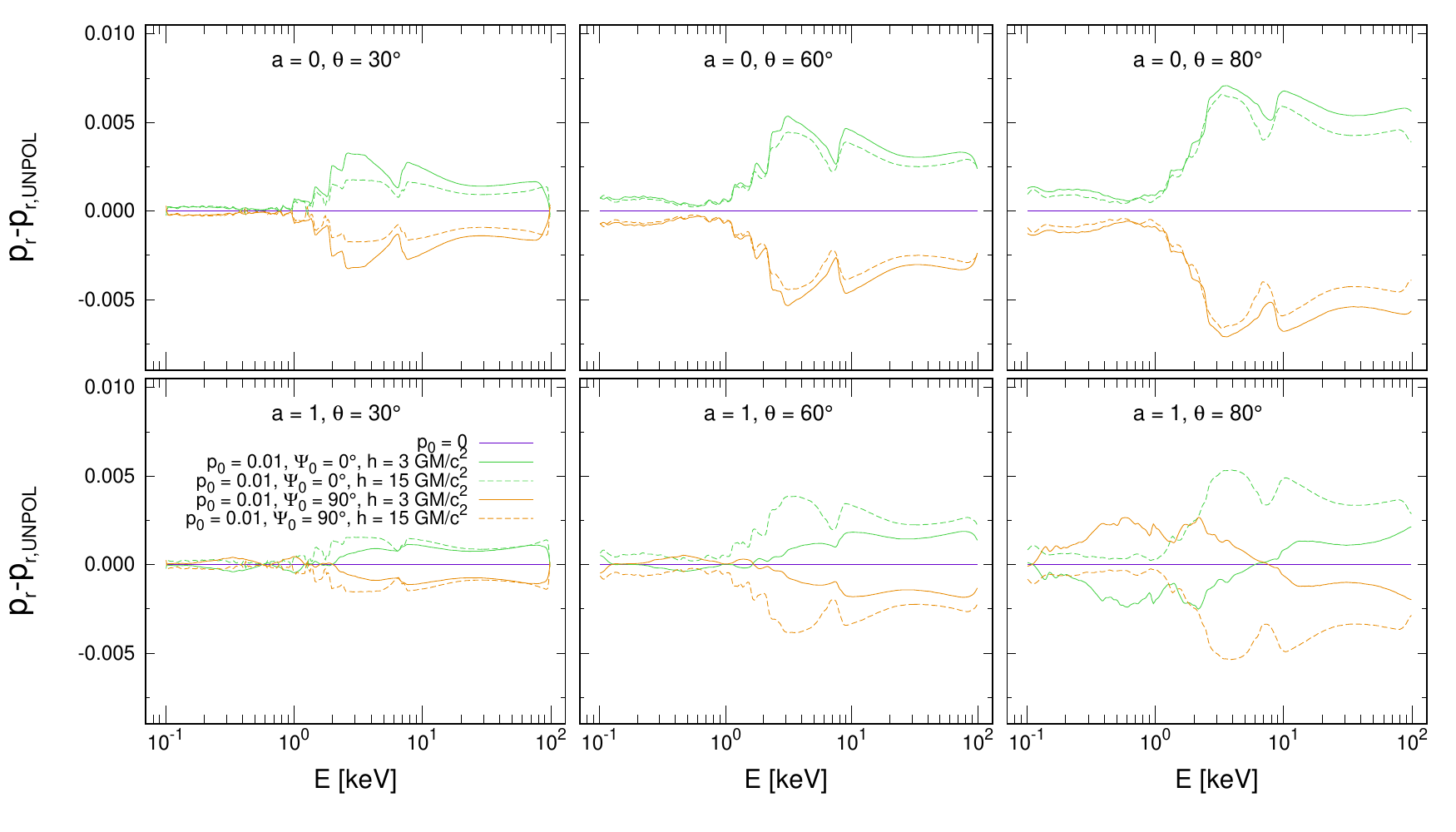}
	\caption{\footnotesize{The relative difference of the reflected-only polarisation degree versus energy of~the~accretion disc for distant observer from a case with unpolarized primary radiation: $p_\mathrm{r}-p_\mathrm{r,UNPOL}$, obtained by {\tt KYNSTOKES} for black-hole spins $a = 0$ (top) and $a = 1$ (bottom), disc inclinations $\theta = 30^{\circ}$ (left), $\theta = 60^{\circ}$ (middle) and  $\theta = 80^{\circ}$ (right), $\Gamma = 2$ and neutral disc ($M = 1\times 10^8\,M_{\odot}$ and observed 2--10 keV flux $L_{\textrm{X}}/L_{\textrm{Edd}} = 0.001$), using the {\tt STOKES} local reflection model in~the~lamp-post scheme. We show cases of two different heights of the primary point-source above the~disc $h = 3 \textrm{ } r_\textrm{g}$ (solid lines) and $h = 15 \textrm{ } r_\textrm{g}$ (dashed lines), and three different polarisation states of the primary source for which $p_\mathrm{r}$ was computed: $p_0 = 0$ (purple), $p_0 = 0.01$ and $\Psi_0 = 0^{\circ}$ (green), $p_0 = 0.01$ and $\Psi_0 = 90^{\circ}$ (orange).}}
	\label{fig:K3A_R_polar_n_pdeg_relative.}
\end{figure}
\begin{figure}[!htb]\centering
	\includegraphics[width=\textwidth]{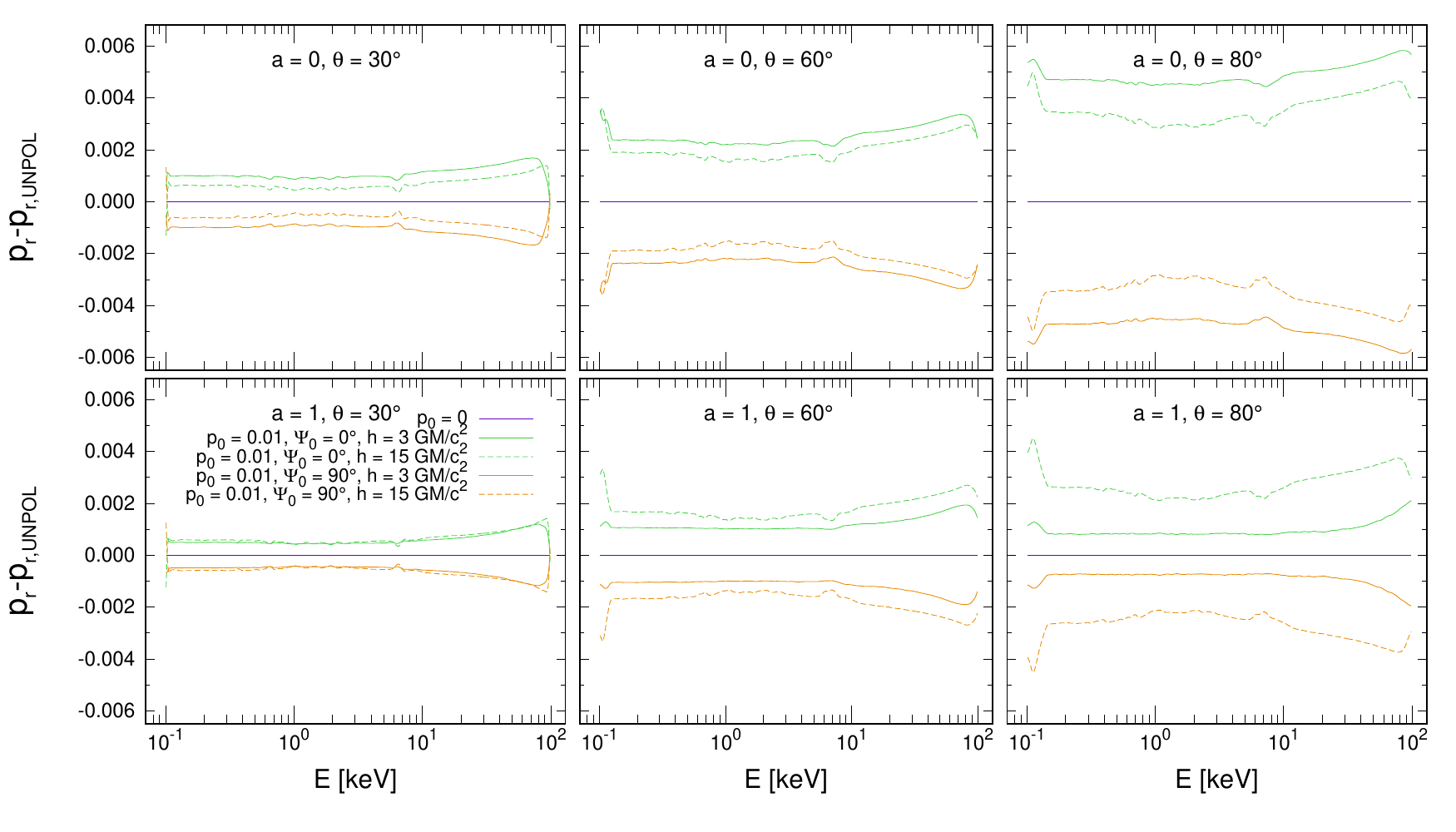}
	\caption{\footnotesize{The reflected-only polarisation degree, $p_\mathrm{r}-p_\mathrm{r,UNPOL}$, versus energy for highly ionized disc ($M = 1\times 10^5\,M_{\odot}$ and observed 2--10 keV flux $L_{\textrm{X}}/L_{\textrm{Edd}} = 0.1$), otherwise for~the~same parametric setup as in Figure \ref{fig:K3A_R_polar_n_pdeg_relative.}, displayed in the same manner.}}
	\label{fig:K3A_R_polar_i_pdeg_relative.}
\end{figure}
\begin{figure}[!htb]\centering
	\includegraphics[width=\textwidth]{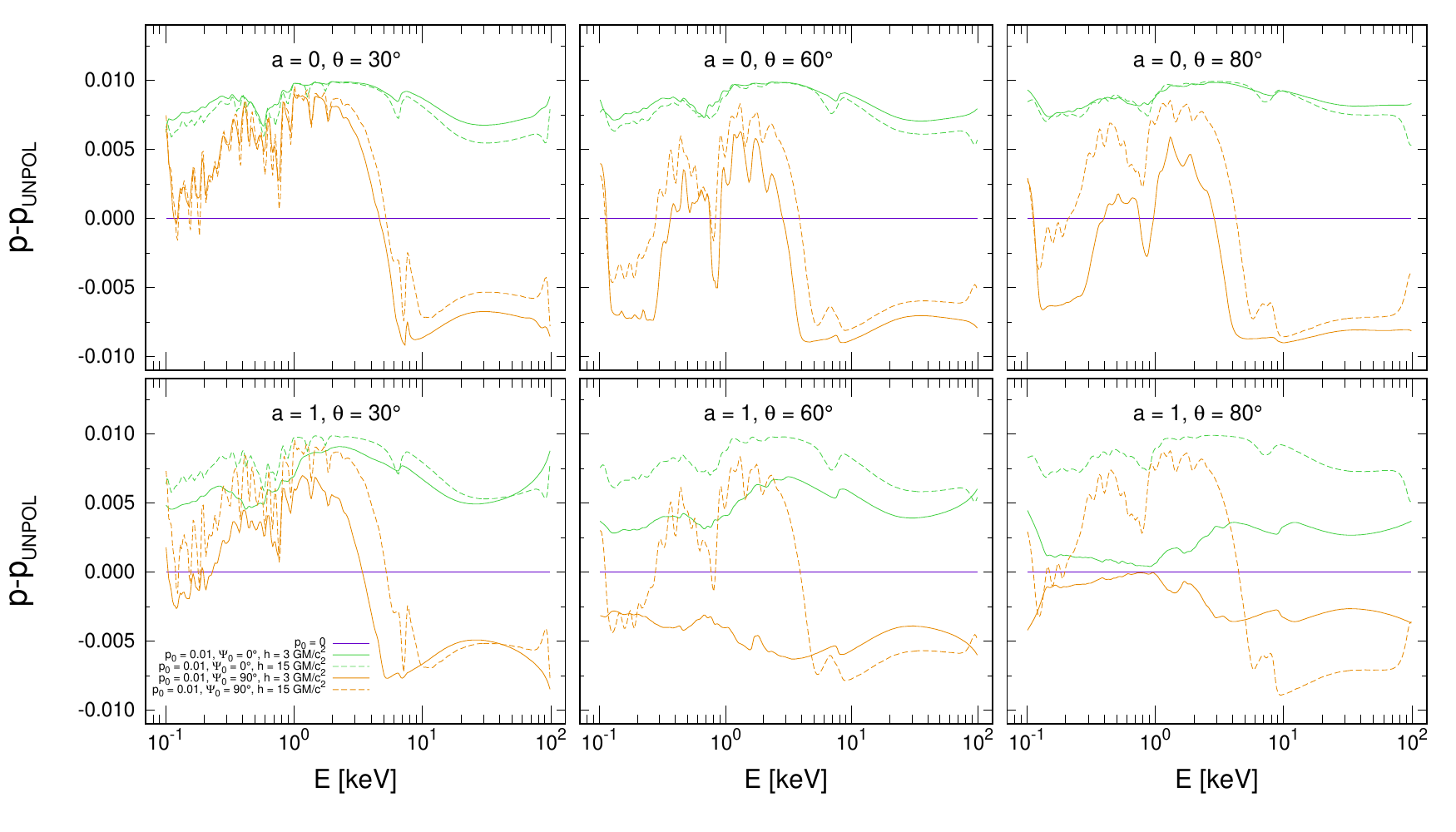}
	\caption{\footnotesize{The total polarisation degree, $p-p_\mathrm{UNPOL}$, versus energy for the same parametric setup as in Figure \ref{fig:K3A_R_polar_n_pdeg_relative.}, displayed in the same manner.}}
	\label{fig:K3A_RP_polar_n_pdeg_relative.}
\end{figure}
\begin{figure}[!htb]\centering
	\includegraphics[width=\textwidth]{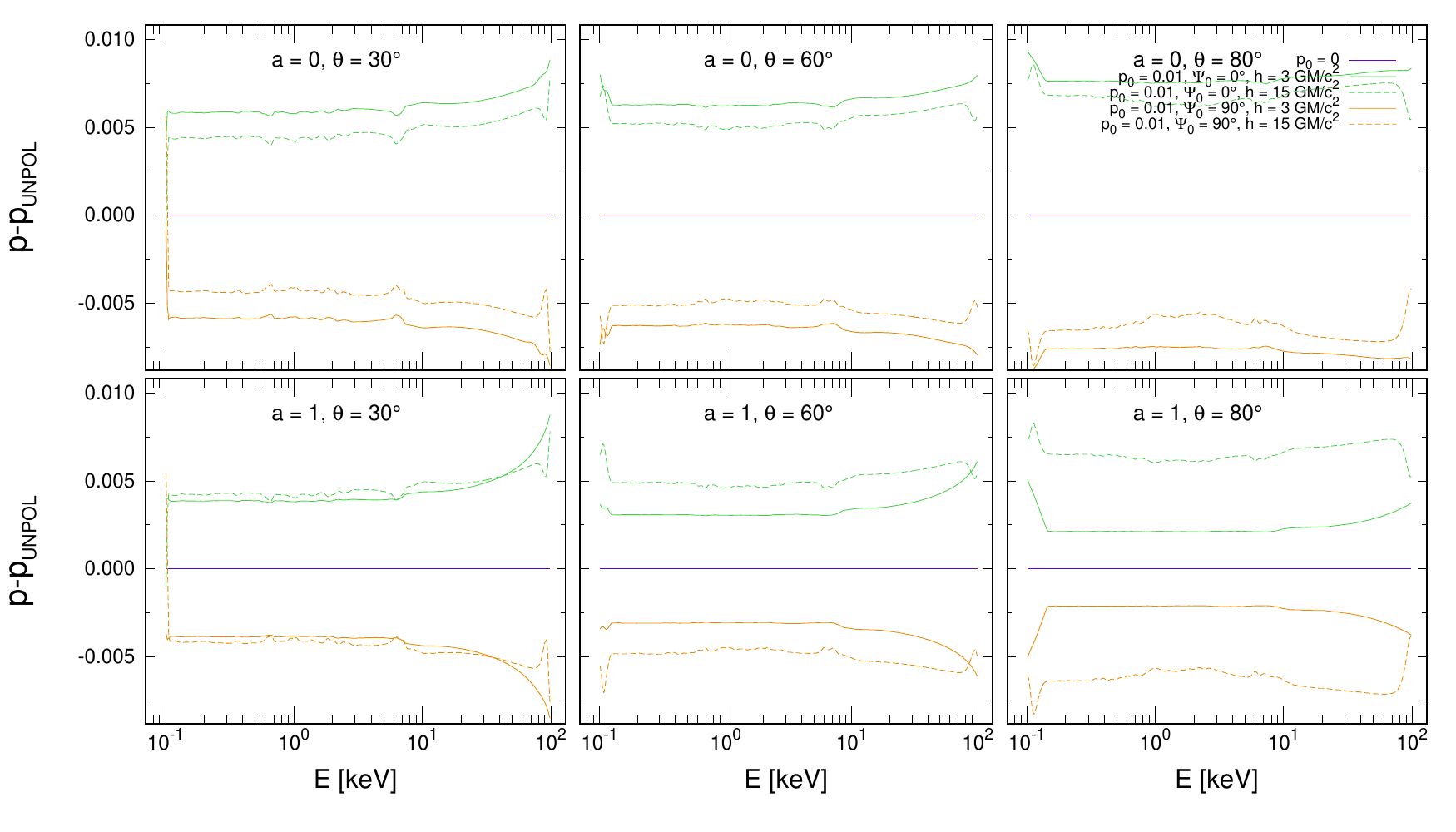}
	\caption{\footnotesize{The total polarisation degree, $p-p_\mathrm{UNPOL}$, versus energy for the same parametric setup as in Figure \ref{fig:K3A_R_polar_i_pdeg_relative.}, displayed in the same manner.}}
	\label{fig:K3A_RP_polar_i_pdeg_relative.}
\end{figure}
\begin{figure}[!htb]\centering
	\includegraphics[width=\textwidth]{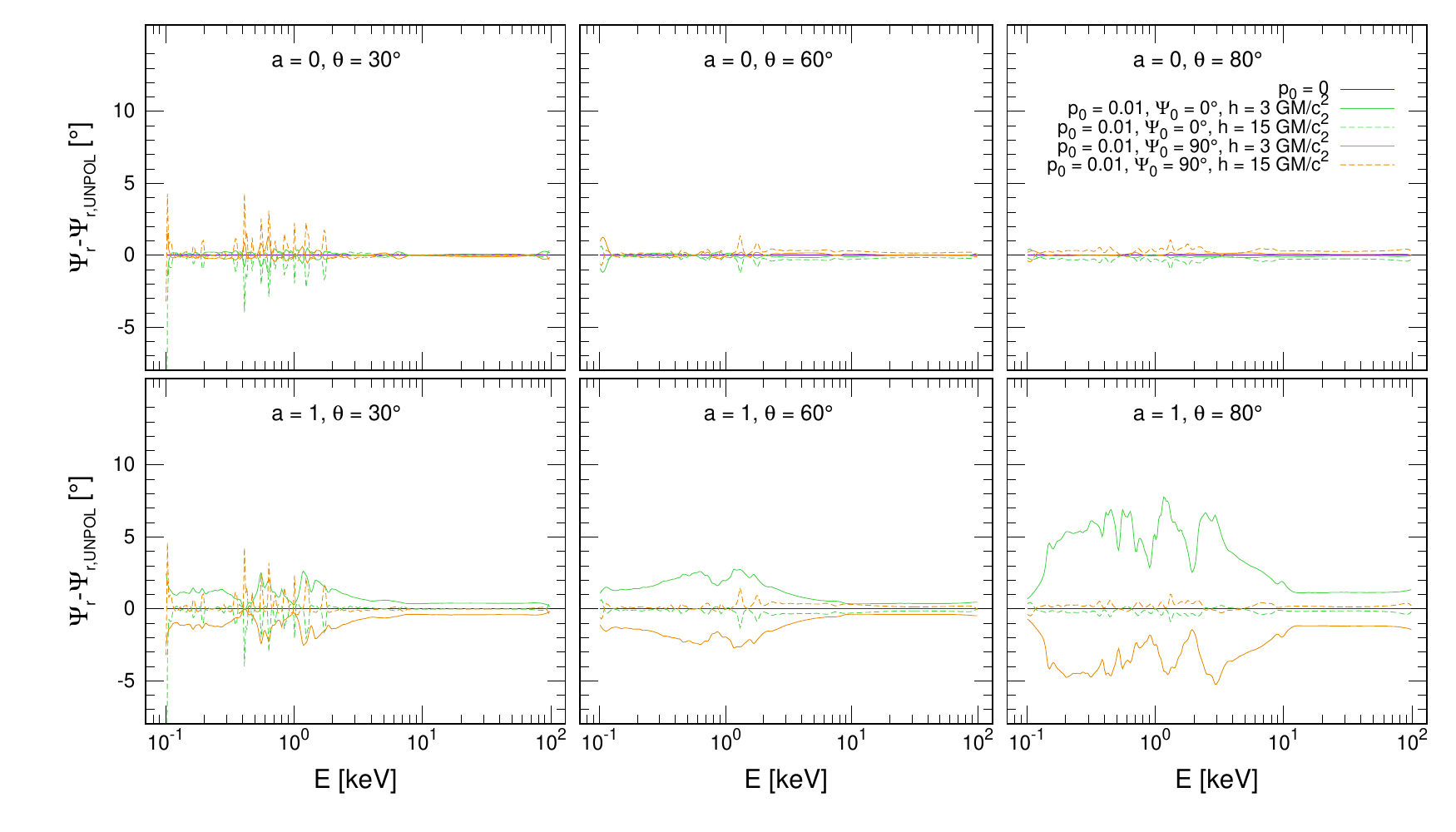}
	\caption{\footnotesize{The reflected-only polarisation angle, $\Psi_r-\Psi_\mathrm{r,UNPOL}$, versus energy for the same parametric setup as in Figure \ref{fig:K3A_R_polar_n_pdeg_relative.}, displayed in the same manner.}}
	\label{fig:K3A_R_polar_n_pang_relative.}
\end{figure}
\begin{figure}[!htb]\centering
	\includegraphics[width=\textwidth]{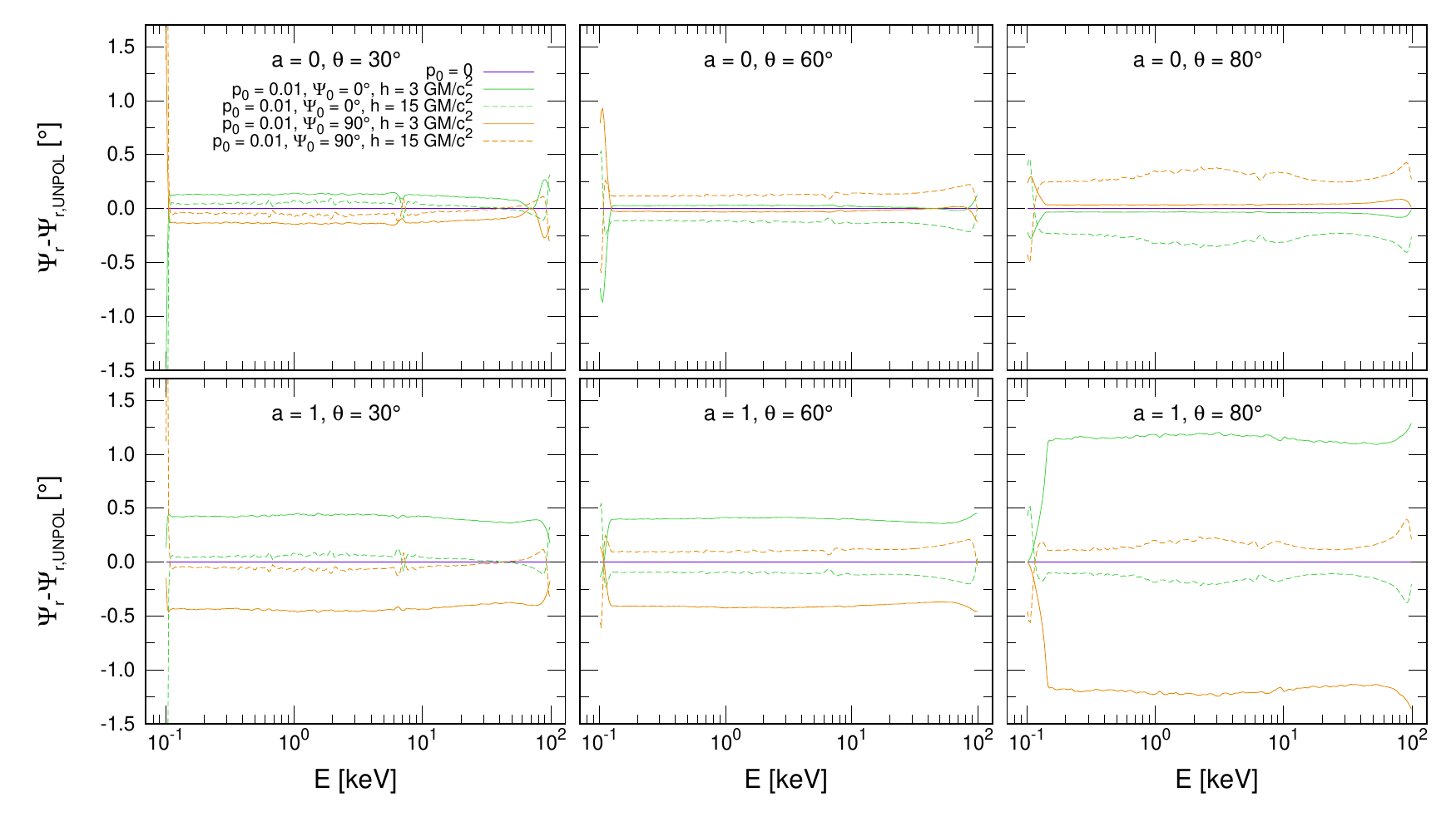}
	\caption{\footnotesize{The reflected-only polarisation angle, $\Psi_\mathrm{r}-\Psi_\mathrm{r,UNPOL}$, versus energy for highly ionized disc ($M = 1\times 10^5\,M_{\odot}$ and observed 2--10 keV flux $L_{\textrm{X}}/L_{\textrm{Edd}} = 0.1$), otherwise for~the~same parametric setup as in Figure \ref{fig:K3A_R_polar_n_pdeg_relative.}, displayed in the same manner.}}
	\label{fig:K3A_R_polar_i_pang_relative.}
\end{figure}
\begin{figure}[!htb]\centering
	\includegraphics[width=\textwidth]{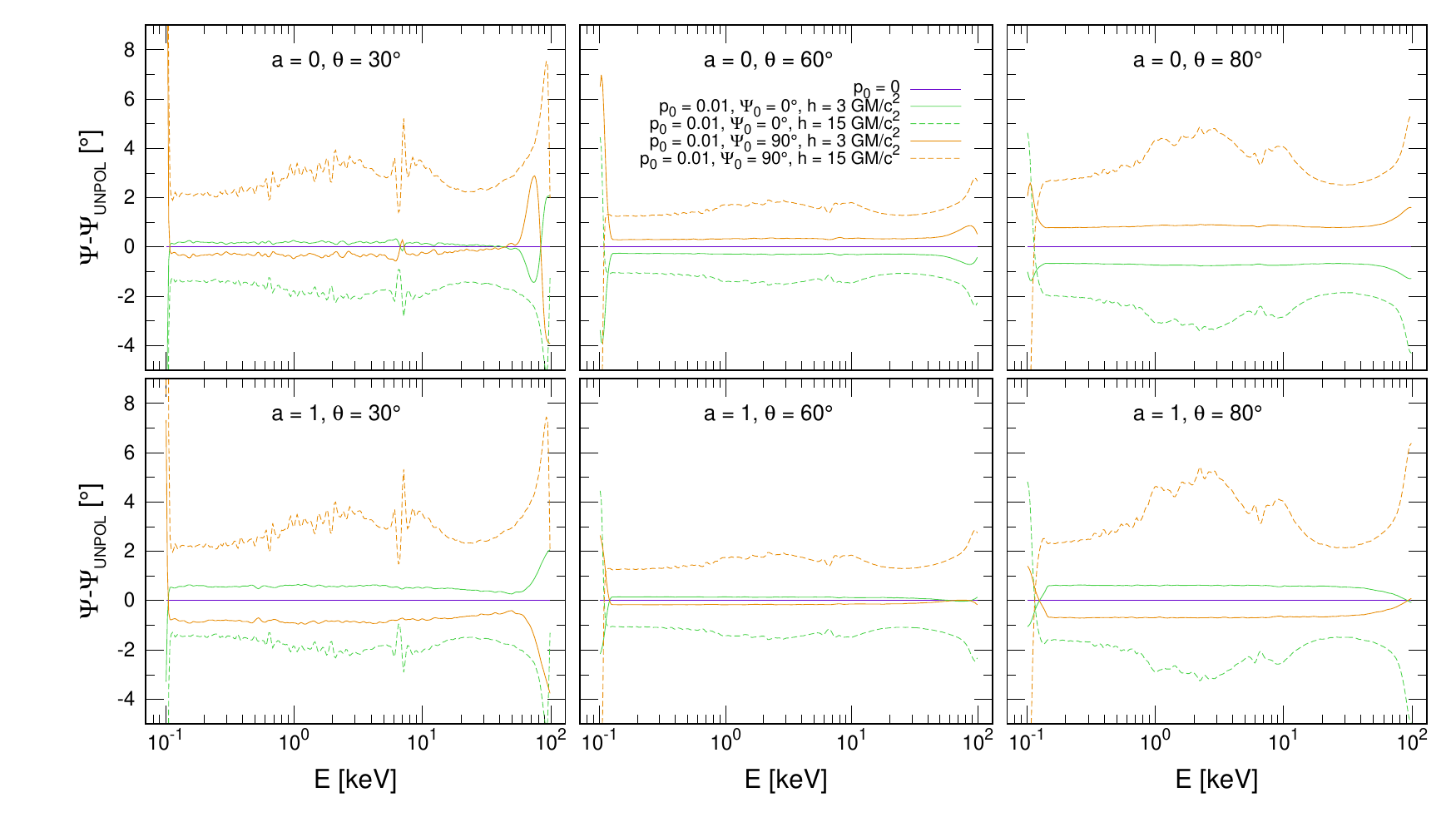}
	\caption{\footnotesize{The total polarisation angle, $\Psi-\Psi_\mathrm{UNPOL}$, versus energy for the same parametric setup as in Figure \ref{fig:K3A_R_polar_i_pang_relative.}, displayed in the same manner.}}
	\label{fig:K3A_RP_polar_i_pang_relative.}
\end{figure}

\begin{figure}[!htb]\centering
	\includegraphics[width=\textwidth]{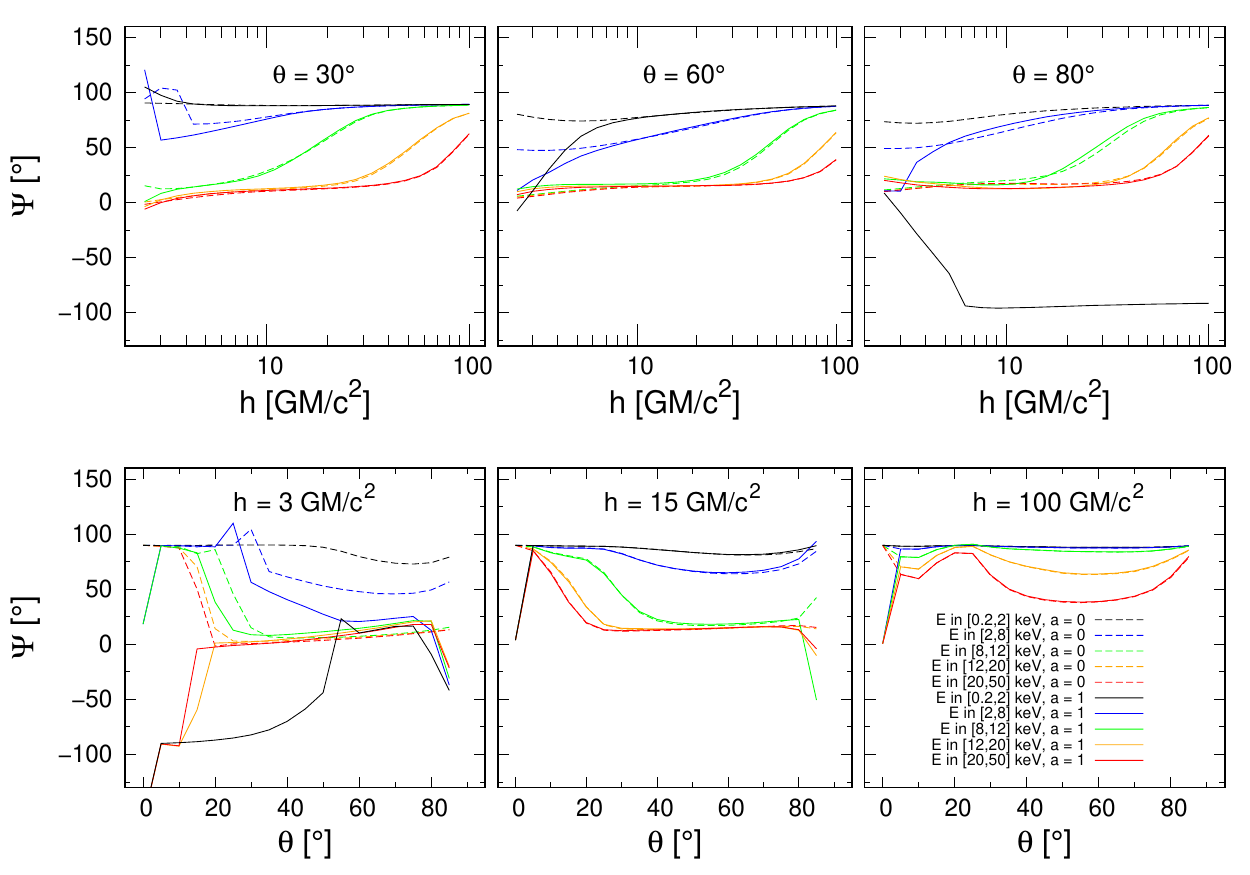}
	\caption{\footnotesize{The total average polarisation angle, $\Psi$, versus energy for primary polarisation of~$p_0 = 0.01$ and $\Psi_0 = 90^{\circ}$, otherwise for the same parametric setup as in Figure \ref{fig:fig10_tot_neutral.}, displayed in the same manner.}}
	\label{fig:fig10_tot_neutral_P2.}
\end{figure}

\begin{figure}[!htb]\centering
	\includegraphics[width=\textwidth]{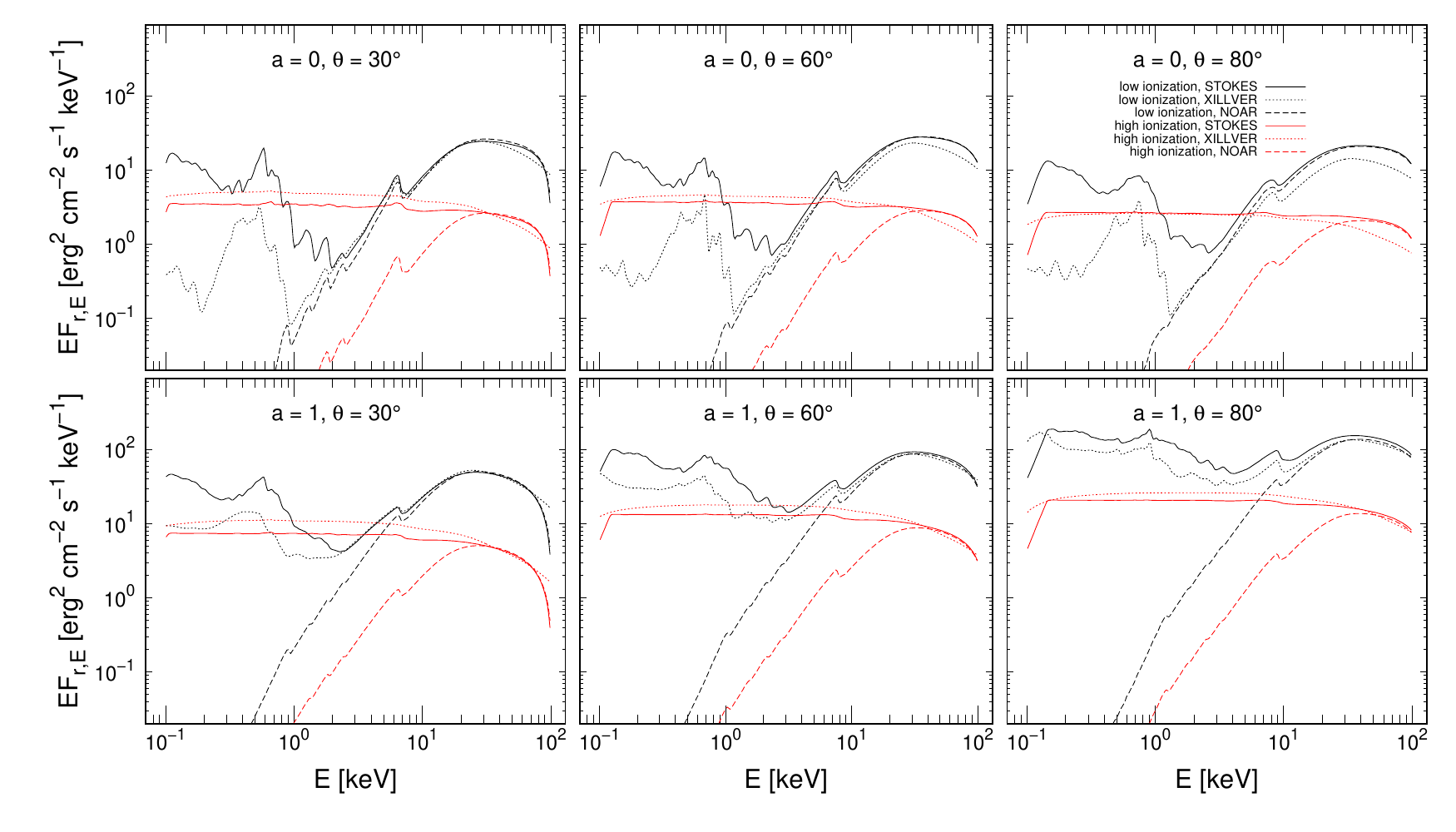}
	\caption{\footnotesize{The reflected-only spectra of the accretion disc for distant observer in the lamp-post scheme, $EF_\textrm{r,E}$, for black-hole spins $a = 0$ (top) and $a = 1$ (bottom), $\theta = 30^{\circ}$ (left), $\theta = 60^{\circ}$ (middle) and  $\theta = 80^{\circ}$ (right), $\Gamma = 2$, height of the primary point-source above the disc $h = 3 \textrm{ } r_\textrm{g}$ and unpolarized primary radiation. We compare cases of {\tt KY} disc integration with the {\tt STOKES} local reflection tables (solid lines), {\tt XILLVER} local reflection tables (dotted lines) and {\tt NOAR} local reflection tables (dashed lines), and neutral disc ($M = 1\times 10^8\,M_{\odot}$ and observed 2--10 keV flux $L_{\textrm{X}}/L_{\textrm{Edd}} = 0.001$, black lines) and highly ionized disc ($M = 1\times 10^5\,M_{\odot}$ and observed \mbox{2--10~keV} flux $L_{\textrm{X}}/L_{\textrm{Edd}} = 0.1$, red lines).}}
	\label{fig:K5_I_Lin.}
\end{figure}

\begin{figure}[!htb]\centering
	\includegraphics[width=\textwidth]{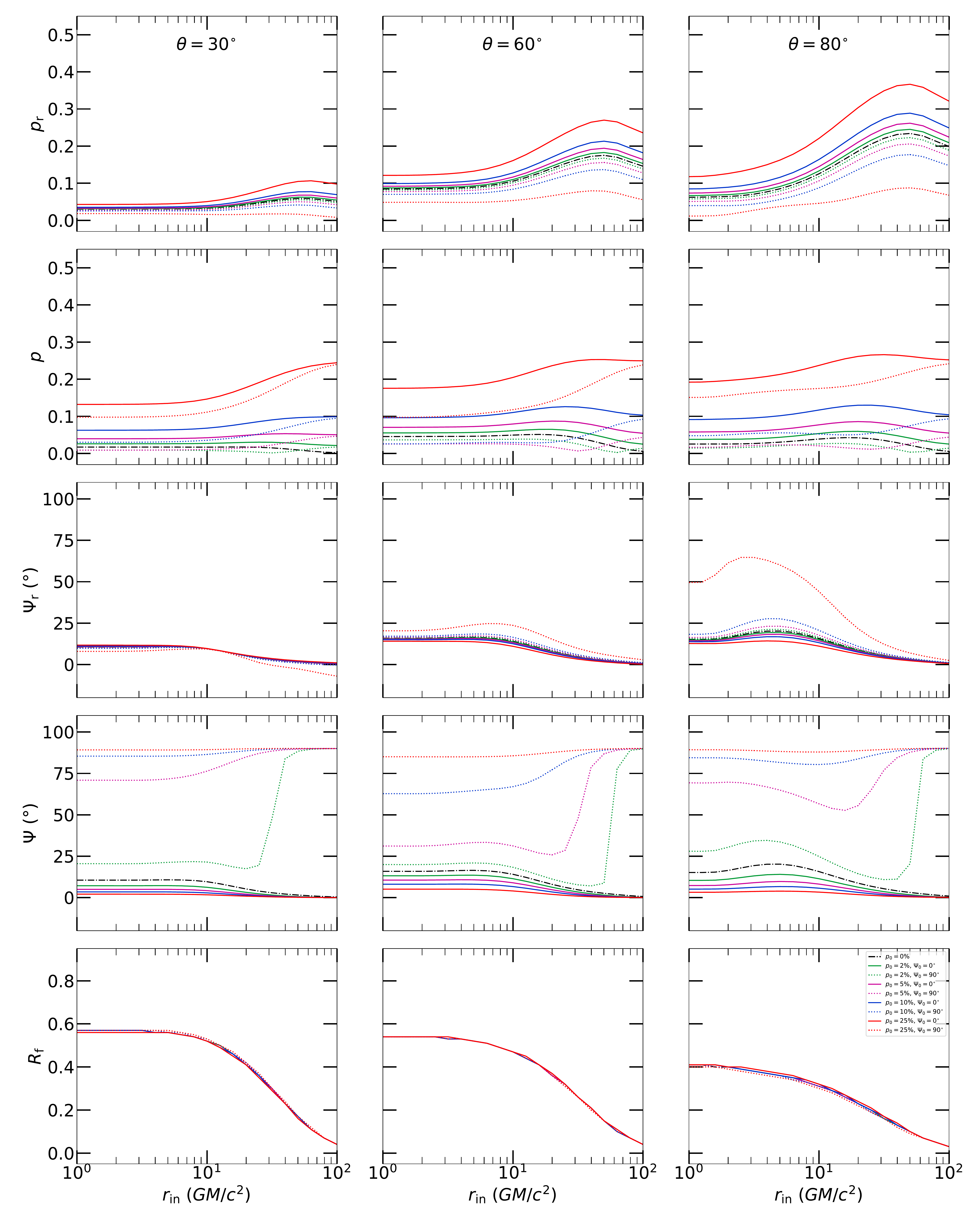}
	\caption{\footnotesize{The same as in Figure \ref{rin_ionized_lp_h3}, but for $h = 15 \textrm{ } r_\textrm{g}$.}}
	\label{rin_ionized_lp_h15}
\end{figure}
\begin{figure}[!htb]\centering
	\includegraphics[width=\textwidth]{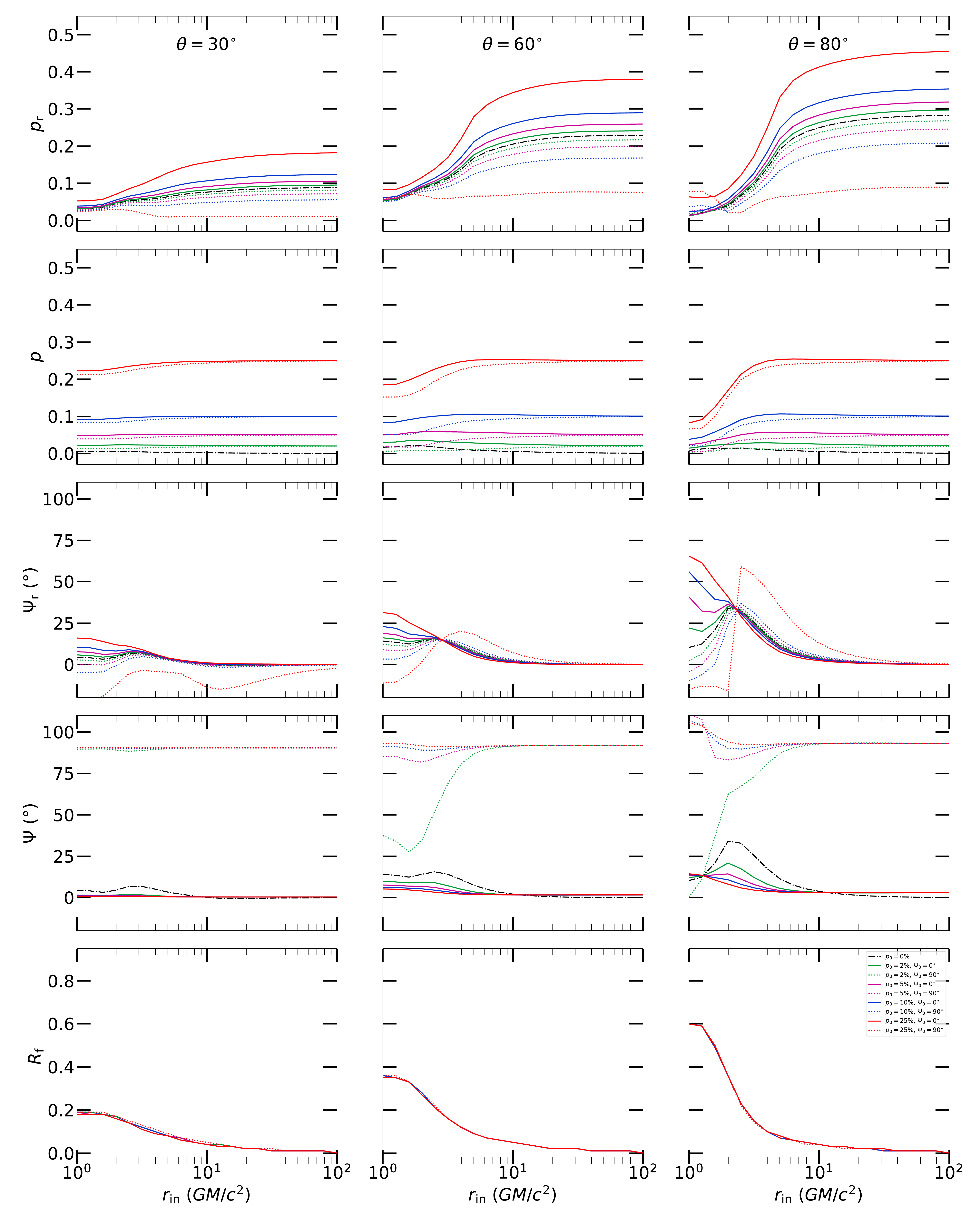}
	\caption{\footnotesize{The same as in Figure \ref{rin_ionized_lp_h3}, but for neutral disc.}}
	\label{rin_neutral_lp_h3}
\end{figure}
\begin{figure}[!htb]\centering
	\includegraphics[width=\textwidth]{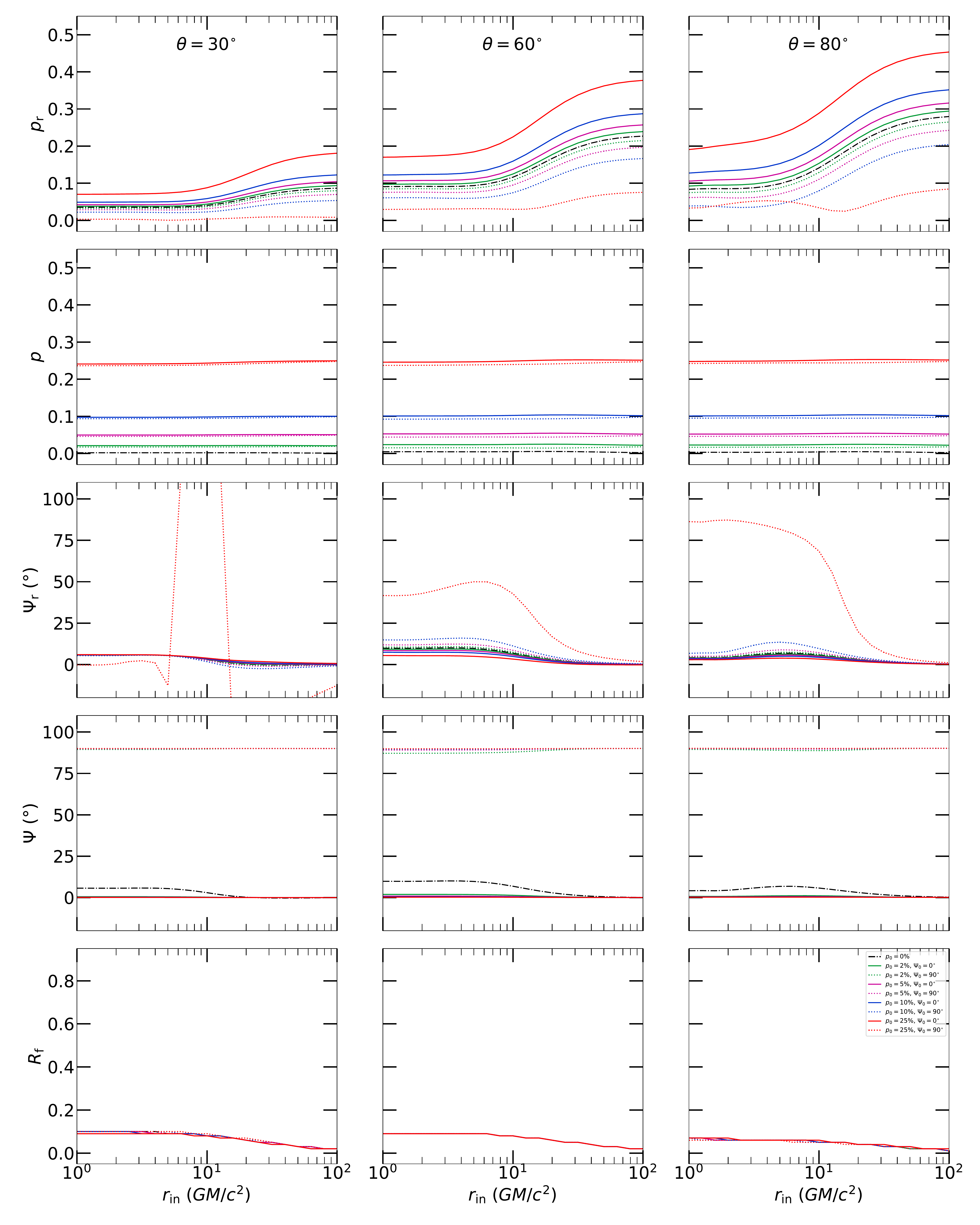}
	\caption{\footnotesize{The same as in Figure \ref{rin_ionized_lp_h3}, but for $h = 15 \textrm{ } r_\textrm{g}$ and neutral disc.}}
	\label{rin_neutral_lp_h15}
\end{figure}

\begin{figure}[!htb]\centering
	\includegraphics[width=\textwidth]{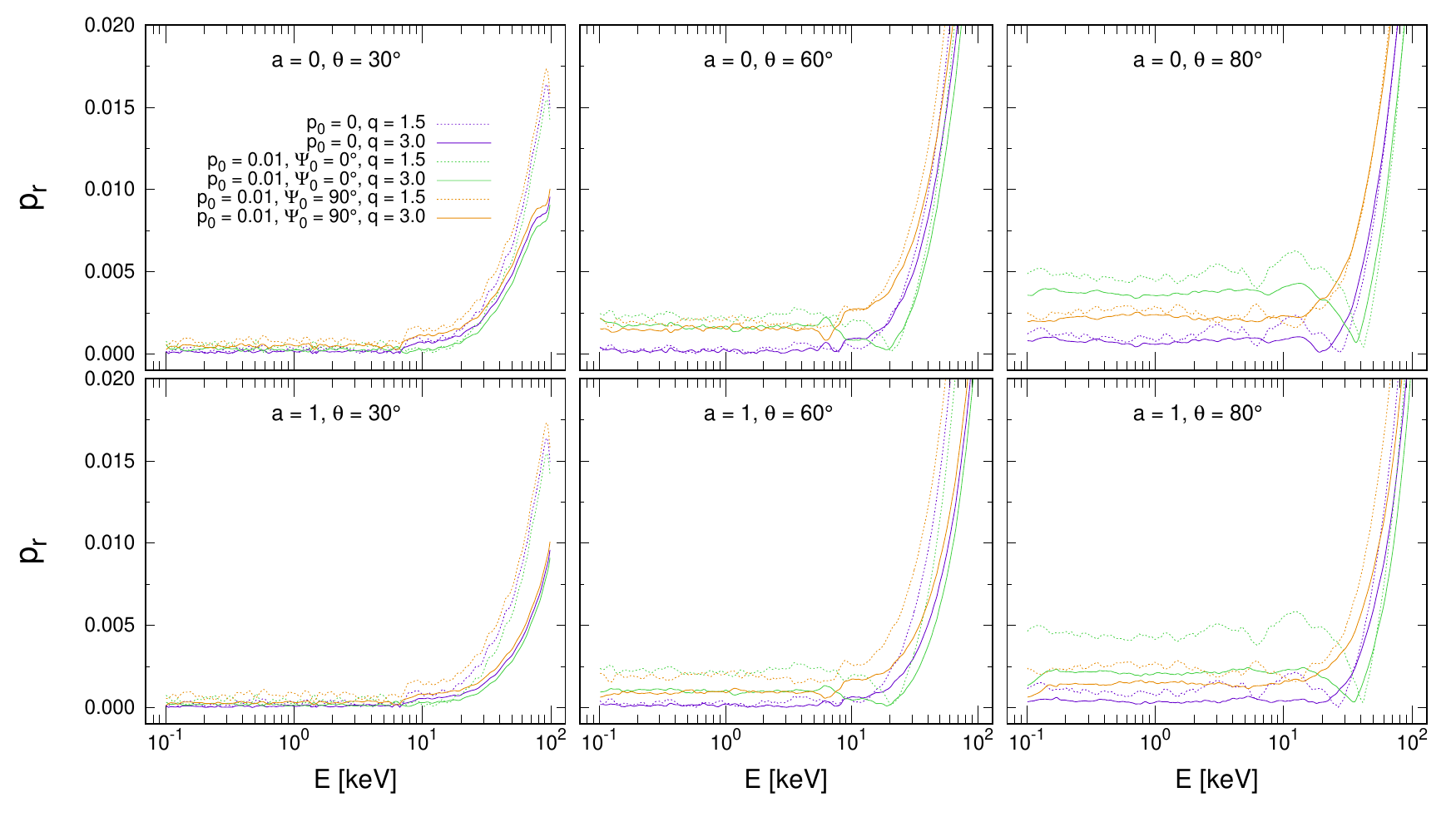}
	\caption{\footnotesize{The reflected-only polarisation degree, $p_\mathrm{r}$, versus energy of the accretion disc for~a~distant observer, obtained by {\tt KYNSTOKES} for black-hole spins $a = 0$ (top) and $a = 1$ (bottom), disc inclinations $\theta = 30^{\circ}$ (left), $\theta = 60^{\circ}$ (midlle) and  $\theta = 80^{\circ}$ (right), $\Gamma = 2$, \mbox{$M = 2\times 10^6\,M_{\odot}$,} and highly ionized disc (observed 2--10 flux $L_{\textrm{X}}/L_{\textrm{Edd}} \approx 10^{2}$, see text for~details), using the~{\tt STOKES} local reflection model in the extended coronal scheme. We show cases of two different radial illumination power-law indices $q = 1.5$ (dotted lines) and $q = 3.0$ (solid lines), and three different polarisation states of the primary source: $p_0 = 0$ (purple), $p_0 = 0.01$ and~$\Psi_0 = 0^{\circ}$ (green), $p_0 = 0.01$ and $\Psi_0 = 90^{\circ}$ (orange).}
	\label{fig:K4A_R_polar_i_pdeg.}}
\end{figure}
\begin{figure}[!htb]\centering
	\includegraphics[width=\textwidth]{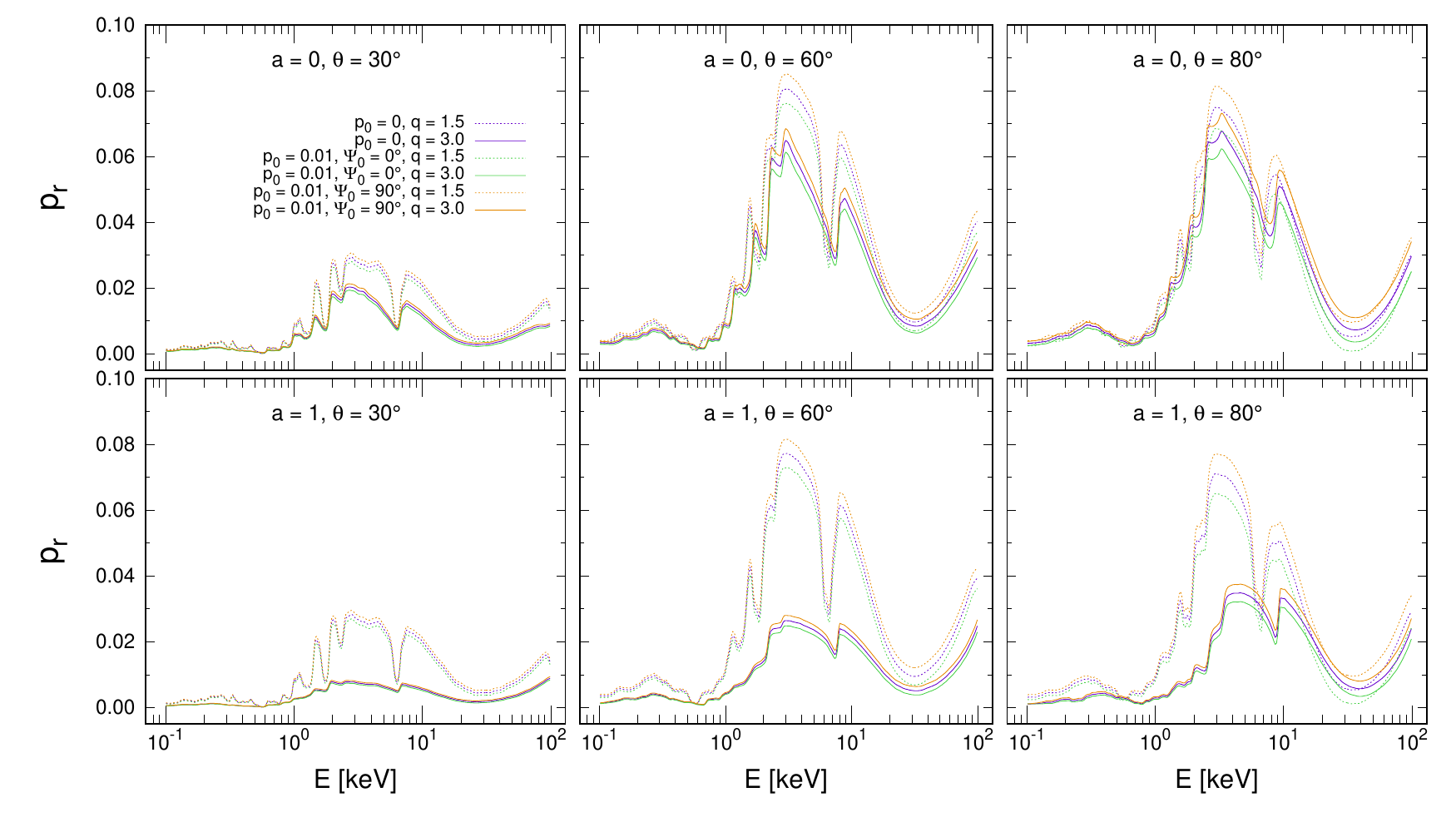}
	\caption{\footnotesize{The reflected-only polarisation degree, $p_\mathrm{r}$, versus energy for neutral disc (observed 2--10 flux $L_{\textrm{X}}/L_{\textrm{Edd}} \approx 10^{-5}$, see text for details), otherwise for the same parametric setup as in~Figure \ref{fig:K4A_R_polar_i_pdeg.}, displayed in the same manner.}}
	\label{fig:K4A_R_polar_n_pdeg.}
\end{figure}
\begin{figure}[!htb]\centering
	\includegraphics[width=\textwidth]{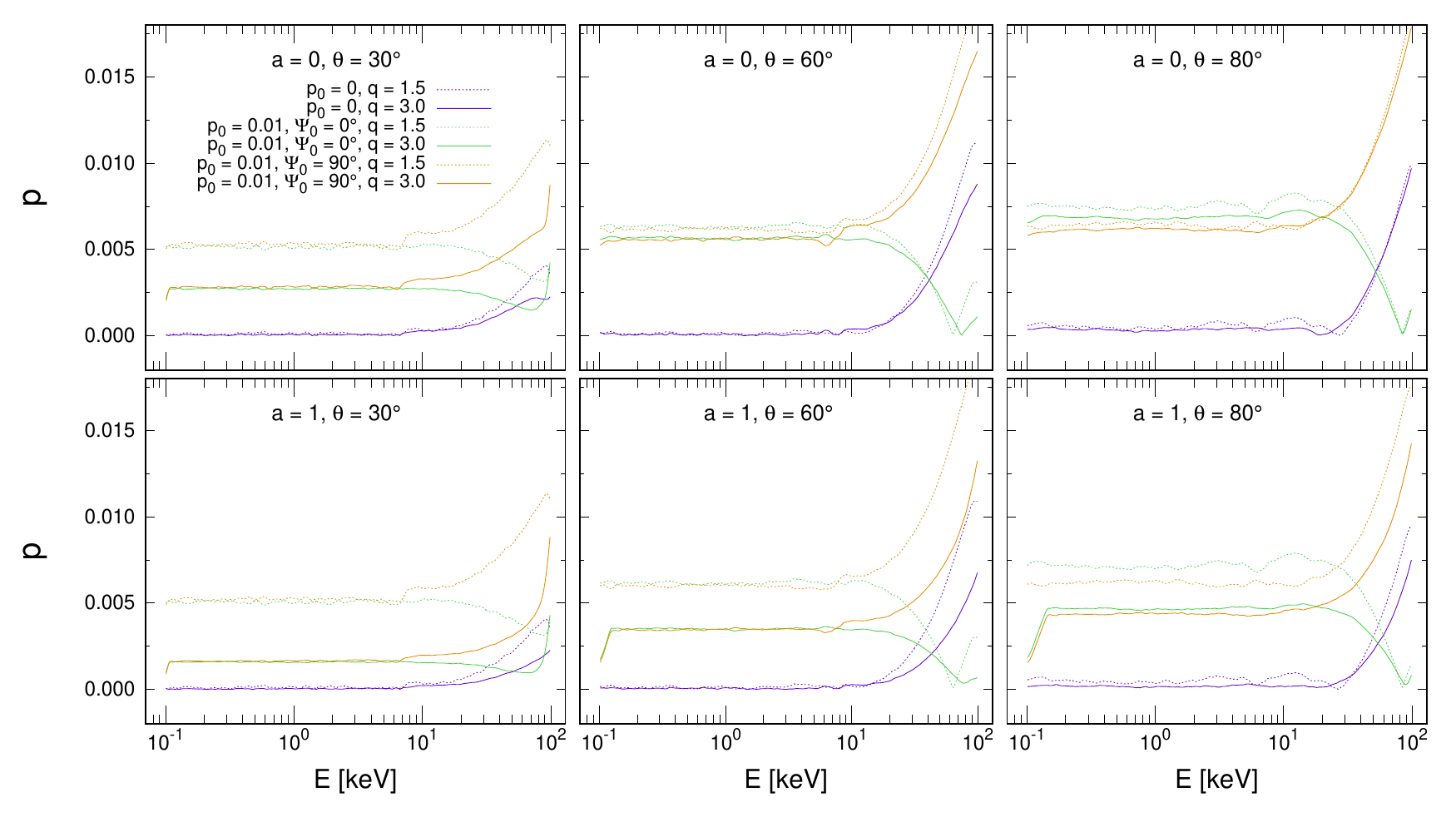}
	\caption{\footnotesize{The total polarisation degree, $p$, versus energy for the same parametric setup as in Figure \ref{fig:K4A_R_polar_i_pdeg.}, displayed in the same manner.}}
	\label{fig:K4A_RP_polar_i_pdeg.}
\end{figure}
\begin{figure}[!htb]\centering
	\includegraphics[width=\textwidth]{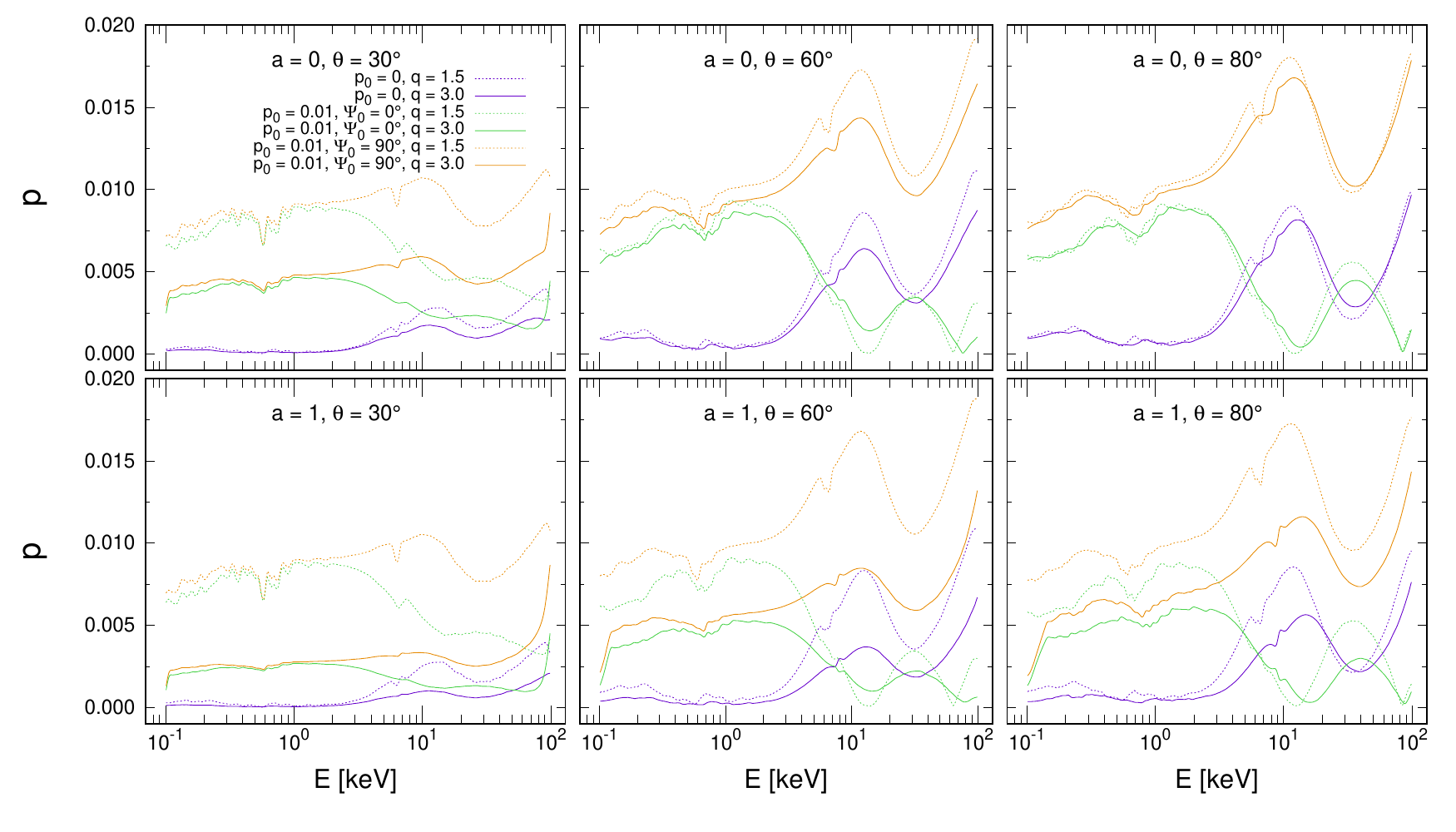}
	\caption{\footnotesize{The total polarisation degree, $p$, versus energy for the same parametric setup as in Figure \ref{fig:K4A_R_polar_n_pdeg.}, displayed in the same manner.}}
	\label{fig:K4A_RP_polar_n_pdeg.}
\end{figure}
\begin{figure}[!htb]\centering
	\includegraphics[width=\textwidth]{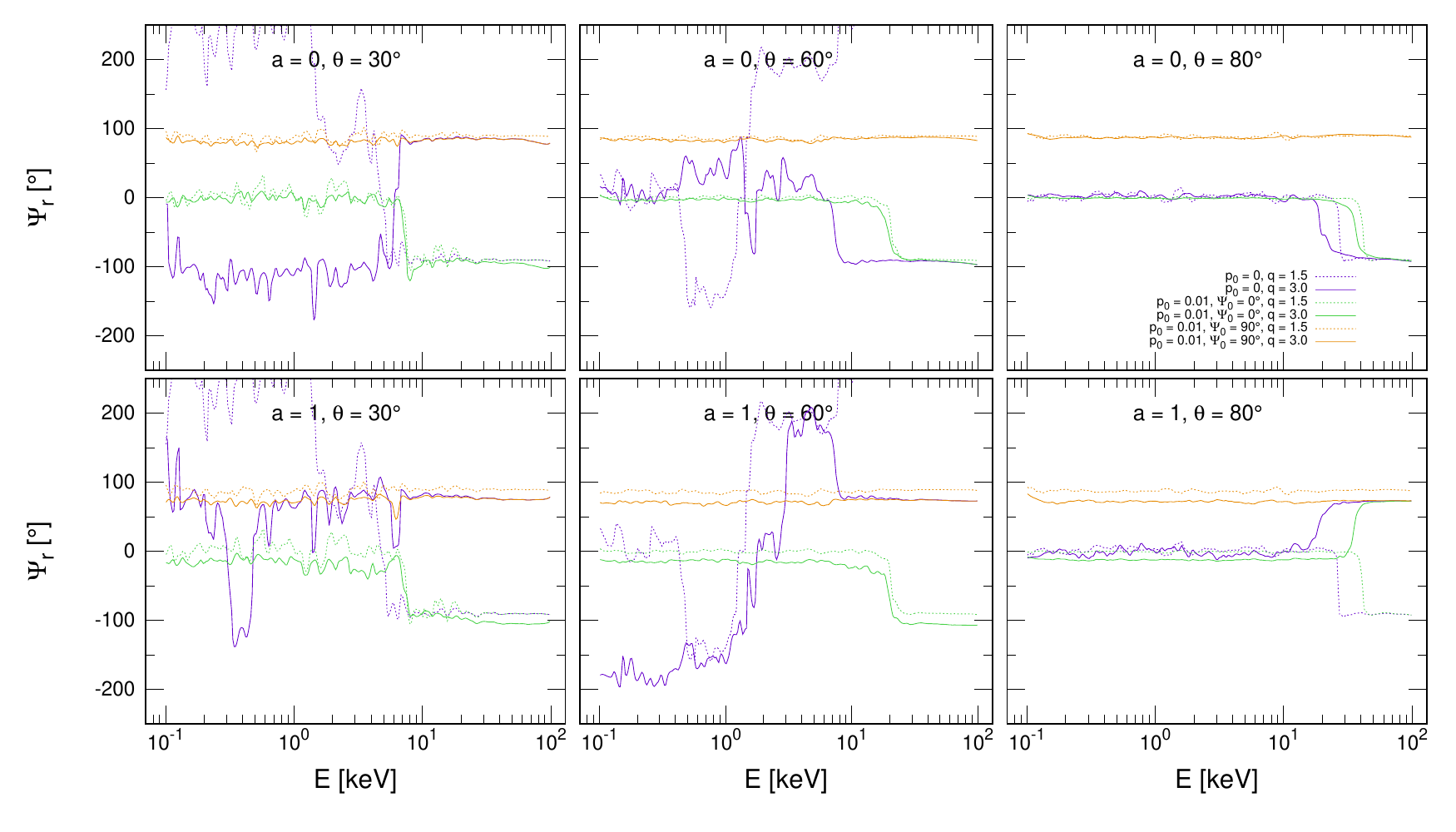}
	\caption{\footnotesize{The reflected-only polarisation angle, $\Psi_\mathrm{r}$,  versus energy for the same parametric setup as in Figure \ref{fig:K4A_R_polar_i_pdeg.}, displayed in the same manner.}}
	\label{fig:K4A_R_polar_i_pang.}
\end{figure}
\begin{figure}[!htb]\centering
	\includegraphics[width=\textwidth]{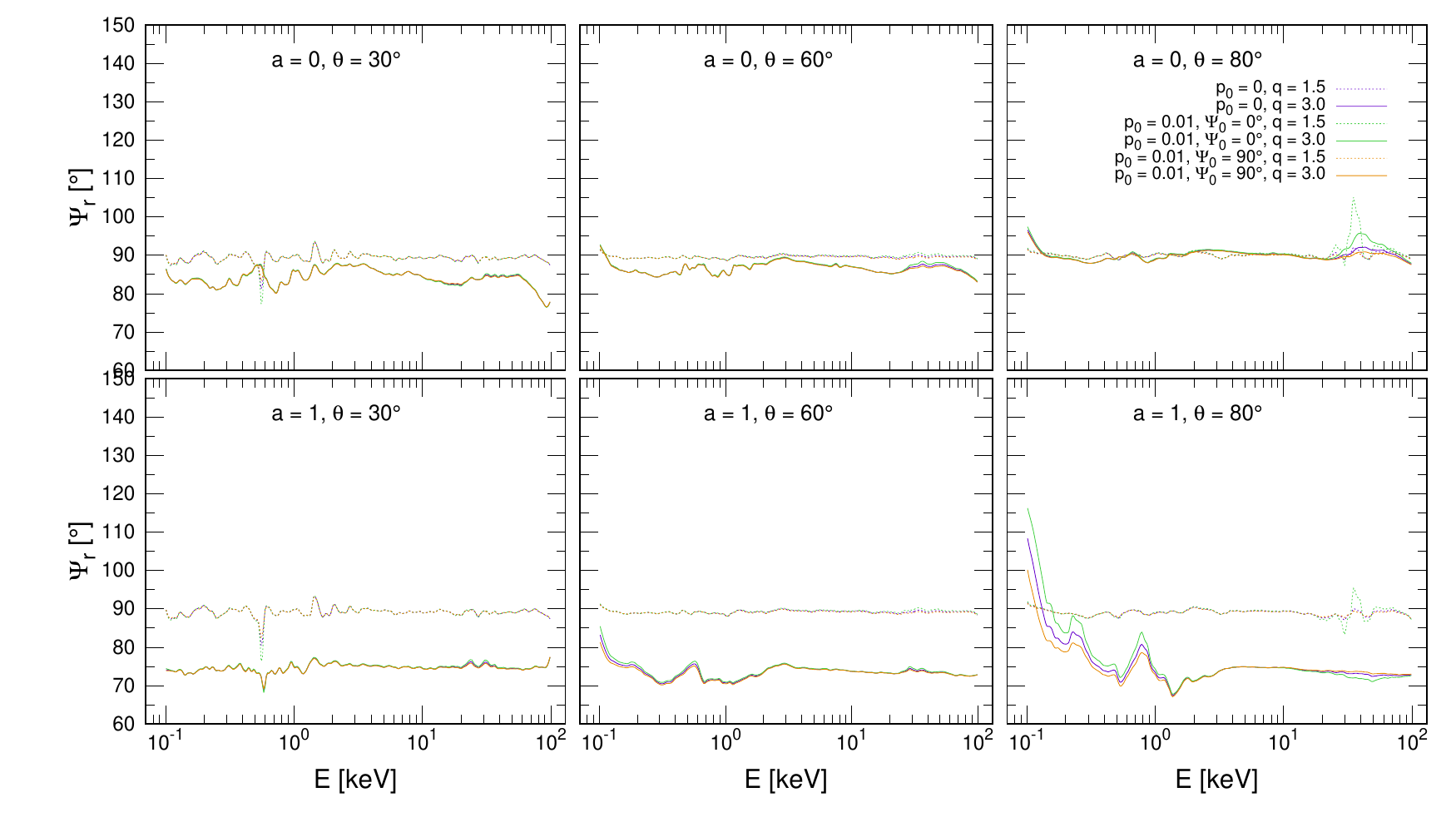}
	\caption{\footnotesize{The reflected-only polarisation angle, $\Psi_\mathrm{r}$,  versus energy for neutral disc (observed 2--10 flux $L_{\textrm{X}}/L_{\textrm{Edd}} \approx 10^{-5}$, see text for details), otherwise for the same parametric setup as in~Figure \ref{fig:K4A_R_polar_i_pdeg.}, displayed in the same manner.}}
	\label{fig:K4A_R_polar_n_pang.}
\end{figure}

\begin{figure}[!htb]\centering
	\includegraphics[width=\textwidth]{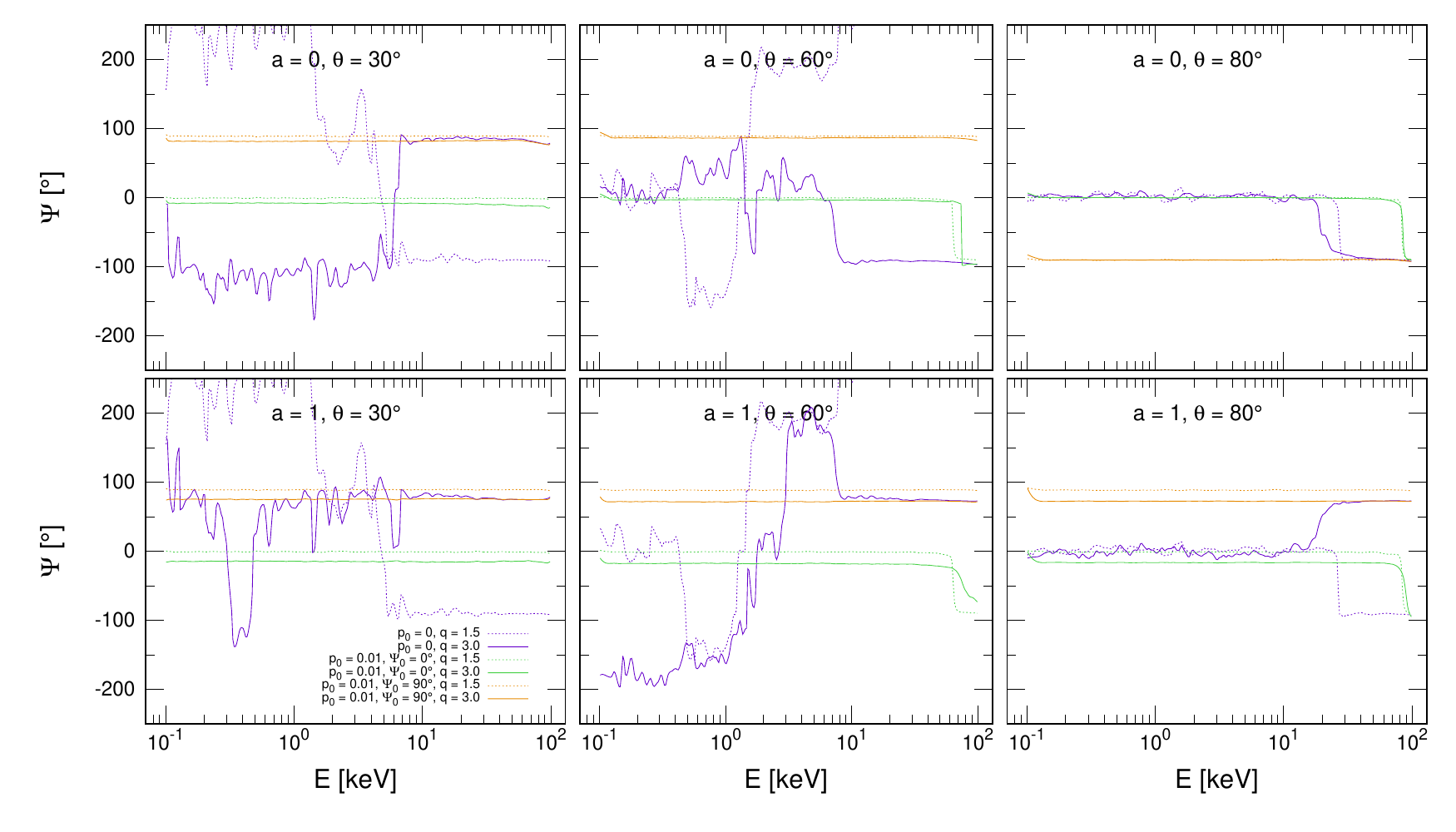}
	\caption{\footnotesize{The total polarisation angle, $\Psi$, versus energy for the same parametric setup as in Figure \ref{fig:K4A_R_polar_i_pang.}, displayed in the same manner.}}
	\label{fig:K4A_RP_polar_i_pang.}
\end{figure}
\begin{figure}[!htb]\centering
	\includegraphics[width=\textwidth]{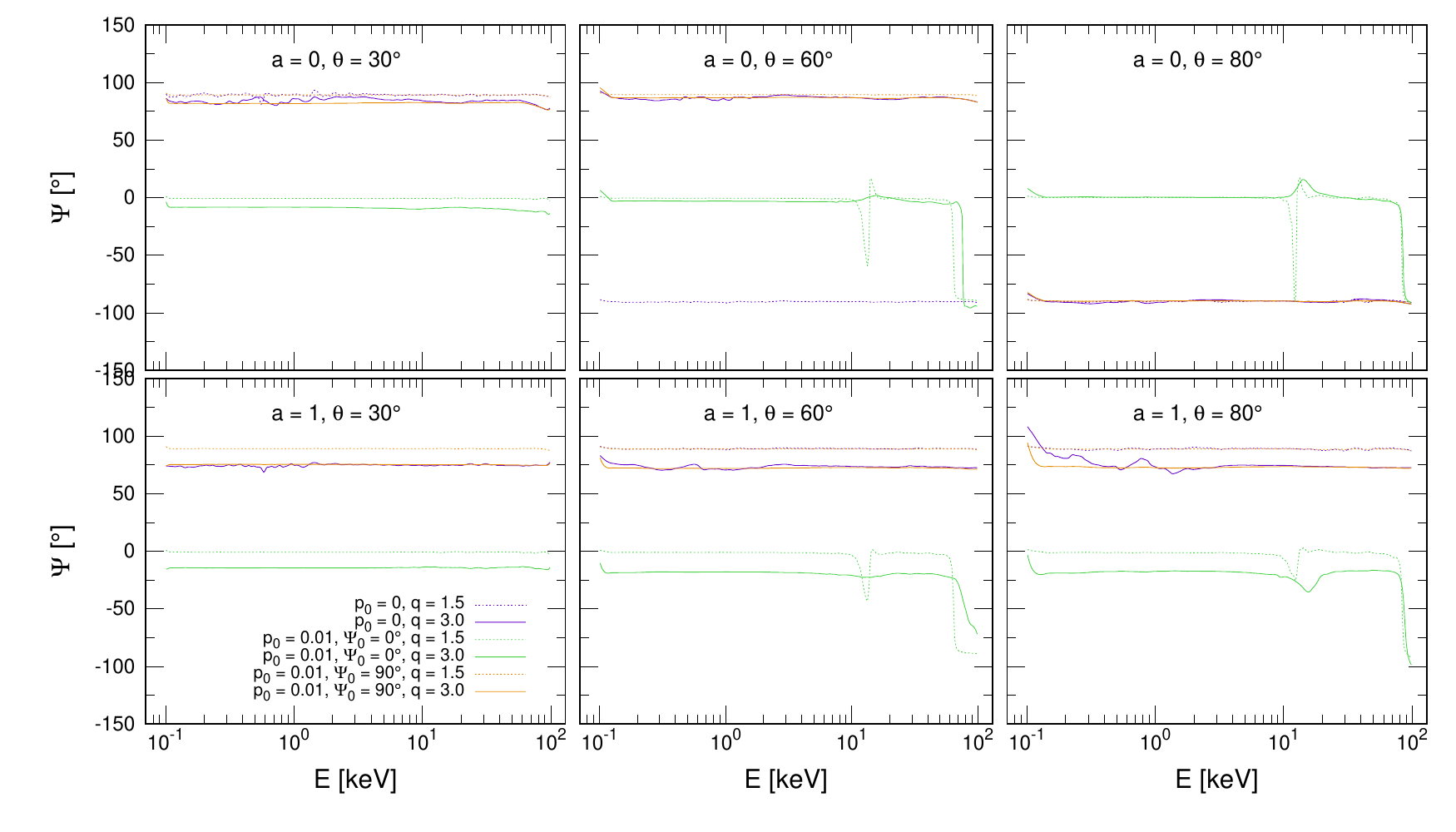}
	\caption{\footnotesize{The total polarisation angle, $\Psi$, versus energy for the same parametric setup as in Figure \ref{fig:K4A_R_polar_n_pang.}, displayed in the same manner.}}
	\label{fig:K4A_RP_polar_n_pang.}
\end{figure}

\begin{figure}[!htb]\centering
	\includegraphics[width=\textwidth]{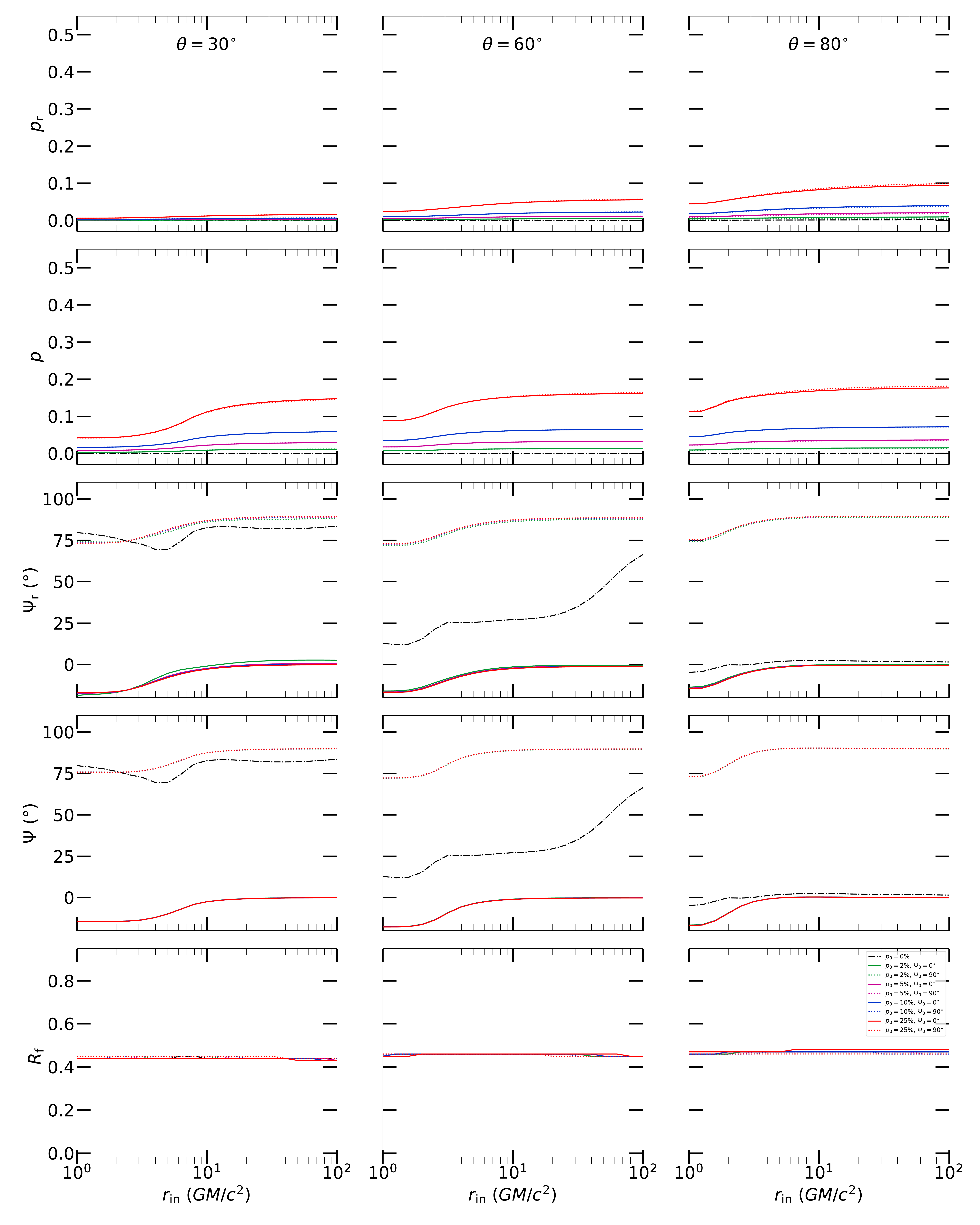}
	\caption{\footnotesize{The same as in Figure \ref{rin_ionized_lp_h3}, but for the extended sandwich corona with $q = 3$.}}
	\label{rin_ionized_ext_q3}
\end{figure}

\begin{landscape}
\end{landscape}

\section*{Supplementary figures to Chapter \ref{chap04}}

\begin{figure}[!htb]\centering
	\includegraphics[width=\textwidth]{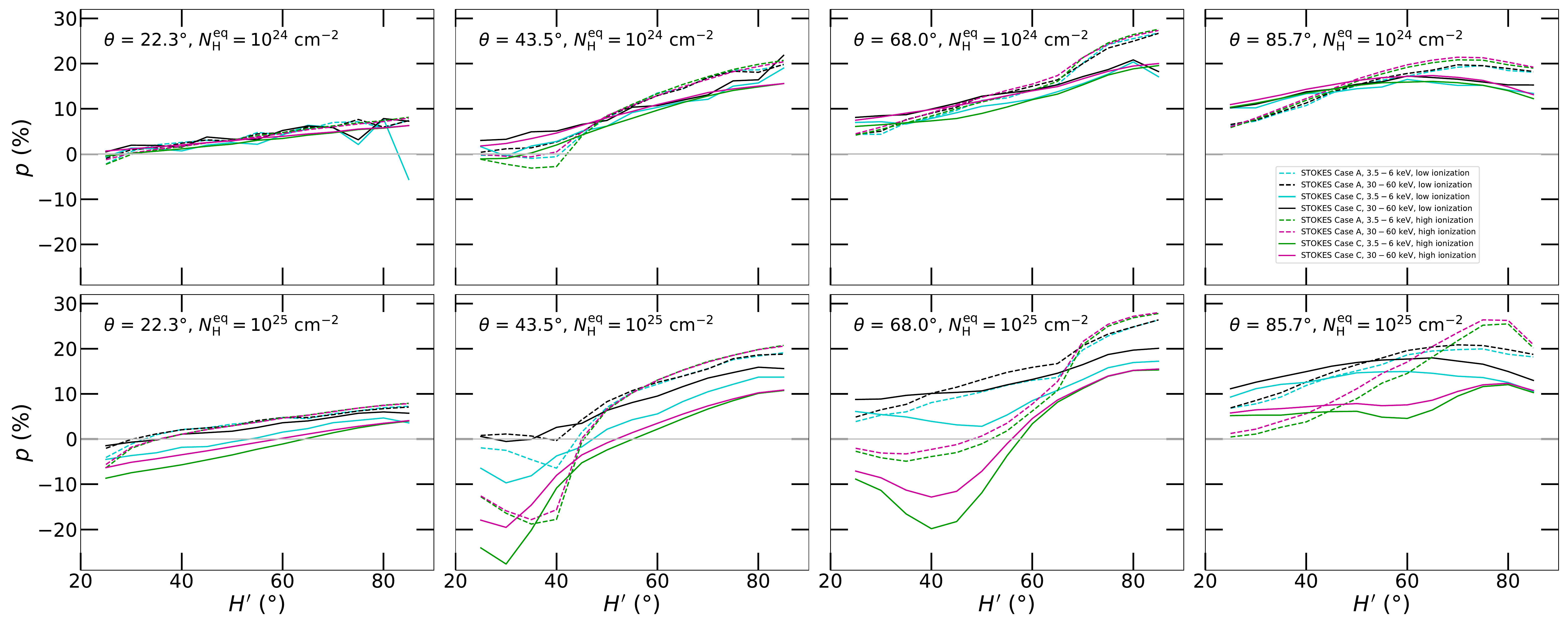}
	\caption{\footnotesize{The same as in Figure \ref{AvsC_pmue}, but energy-averaged polarisation degree, $p$, versus half-opening angle $H'$ is shown for $\theta = 22.3\degr$, $\theta = 43.5\degr$, $\theta = 68.0\degr$, and $\theta = 85.7\degr$ (left to right). Image adapted from \cite{Podgorny2023c}.}}
	\label{AvsC_ptheta}
\end{figure}

\begin{figure}[!htb]\centering
	\includegraphics[width=\textwidth]{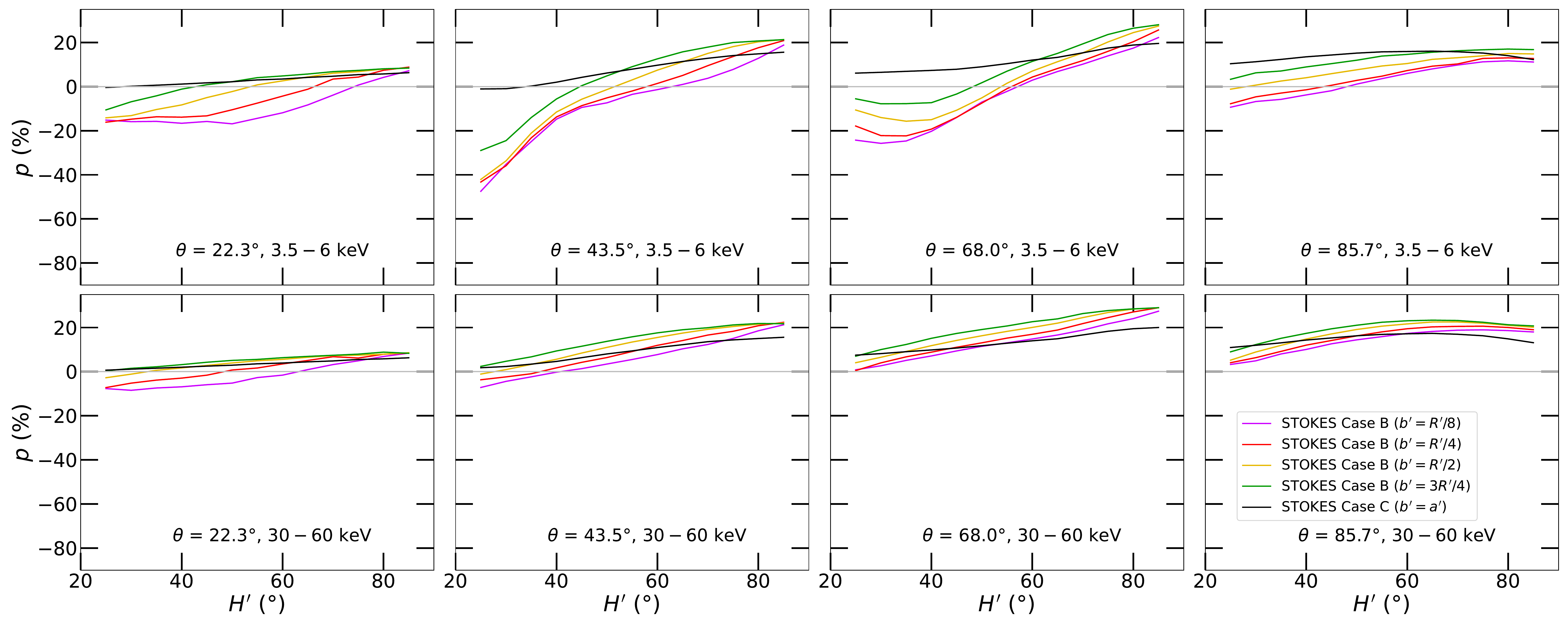}
	\caption{\footnotesize{The same as in Figure \ref{BvsC_pmue24}, but energy-averaged polarisation degree, $p$, versus half-opening angle $H'$ is shown for $\theta = 22.3\degr$, $\theta = 43.5\degr$, $\theta = 68.0\degr$, and $\theta = 85.7\degr$ (left to right). Image adapted from \cite{Podgorny2023c}.}}
	\label{BvsC_ptheta24}
\end{figure}
\begin{figure}[!htb]\centering
	\includegraphics[width=\textwidth]{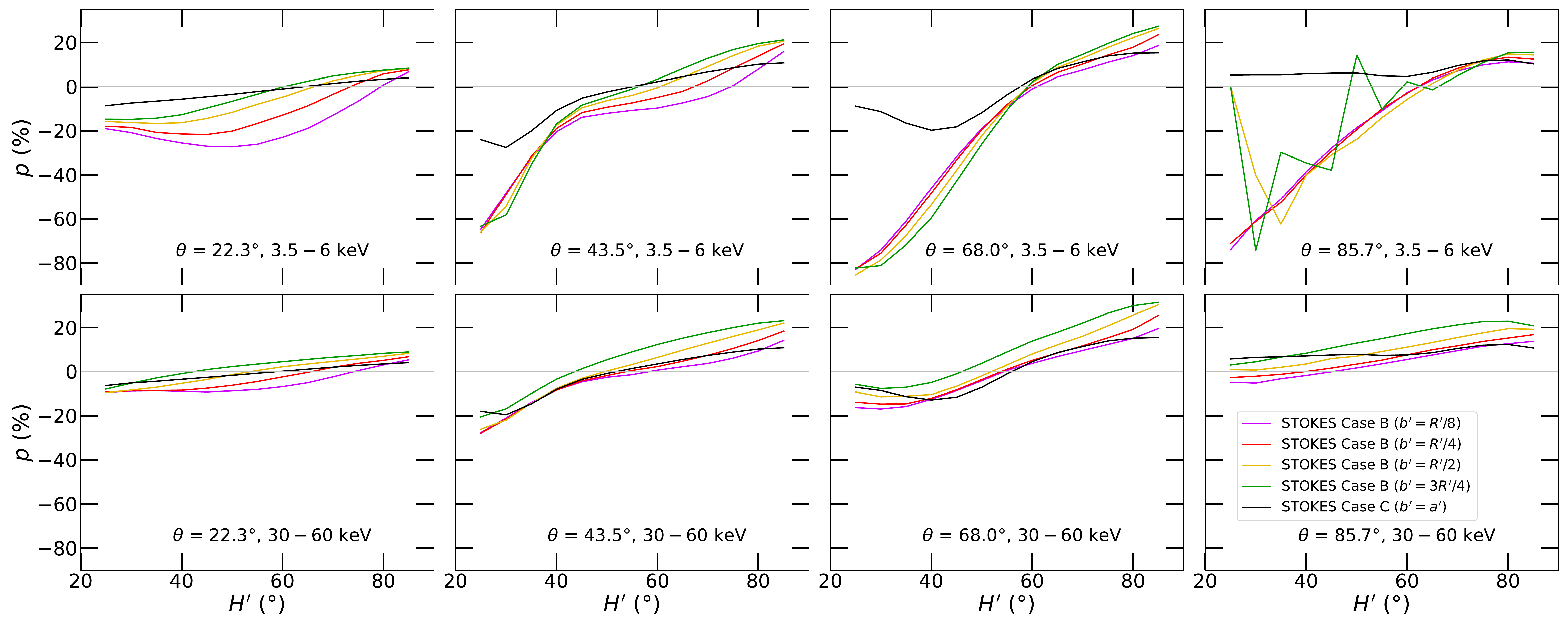}
	\caption{\footnotesize{The same as in Figure \ref{BvsC_ptheta24}, but for $N_\mathrm{H}^\textrm{eq} = 10^{25} \textrm{ cm}^{-2}$. Image adapted from \cite{Podgorny2023c}.}}
	\label{BvsC_ptheta25}
\end{figure}

\begin{figure}[!htb]\centering
	\includegraphics[width=\textwidth]{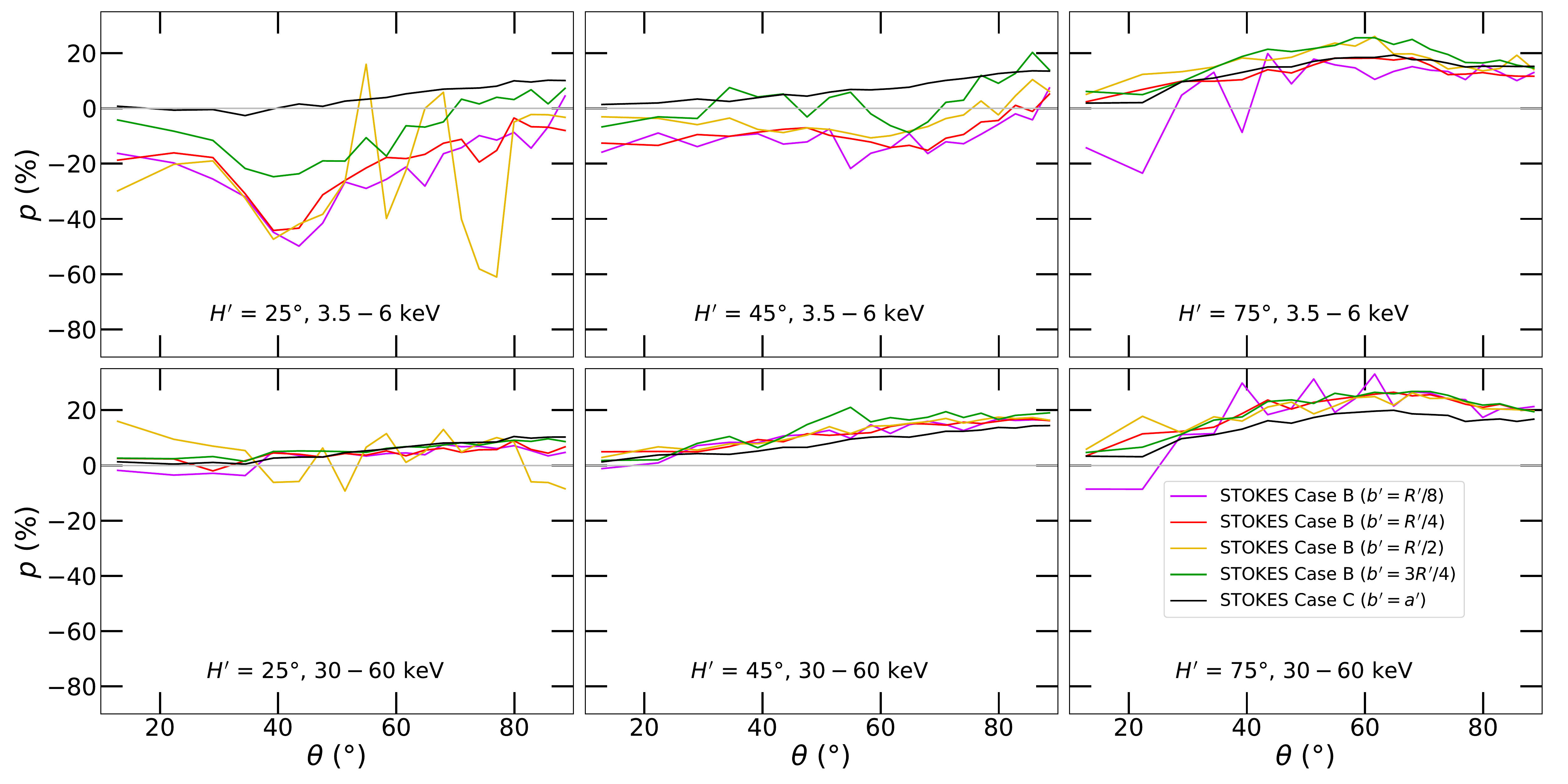}
	\caption{\footnotesize{The same as in Figure \ref{BvsC_pmue24}, but for \textit{low} level of ionization, i.e. $\tau_\textrm{e}^\textrm{eq} = 0.007$. Image adapted from \cite{Podgorny2023c}.}}
	\label{BvsC_pmue24_low_io}
\end{figure}
\begin{figure}[!htb]\centering
	\includegraphics[width=\textwidth]{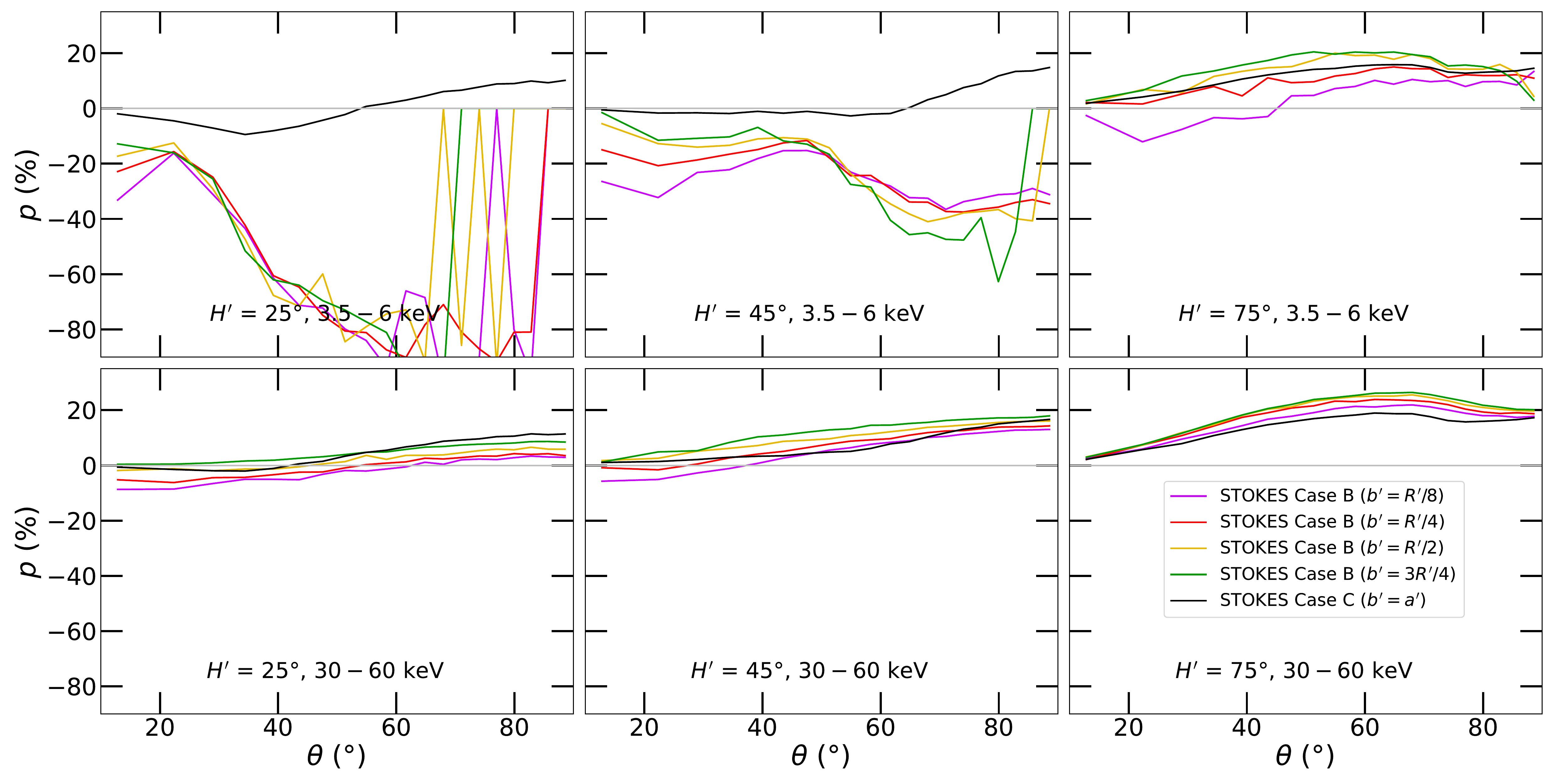}
	\caption{\footnotesize{The same as in Figure \ref{BvsC_pmue25}, but for \textit{low} level of ionization, i.e. $\tau_\textrm{e}^\textrm{eq} = 0.07$. Image adapted from \cite{Podgorny2023c}.}}
	\label{BvsC_pmue25_low_io}
\end{figure}

\begin{figure}[!htb]\centering
	\includegraphics[width=\textwidth]{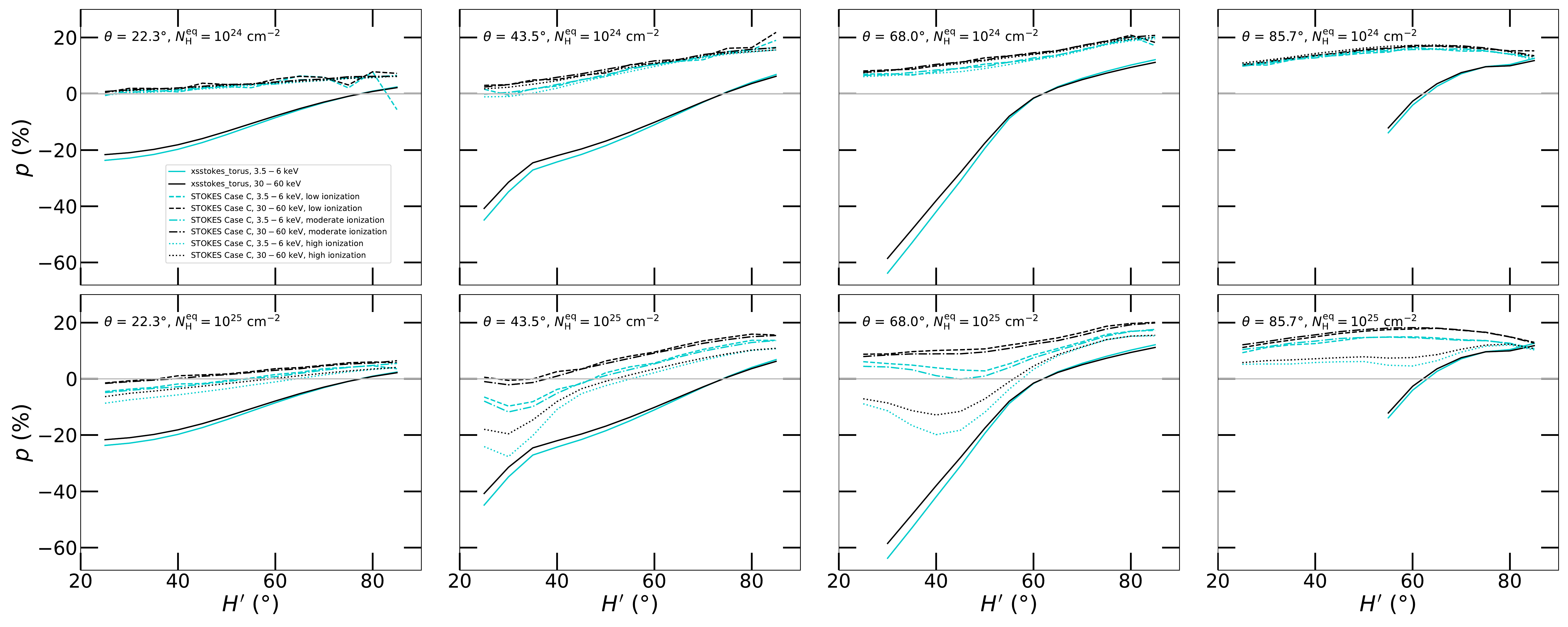}
	\caption{\footnotesize{The same as in Figure \ref{CvsD_pmue}, but energy-averaged polarisation degree, $p$, versus half-opening angle $H'$ is shown for $\theta = 22.3\degr$, $\theta = 43.5\degr$, $\theta = 68.0\degr$, and $\theta = 85.7\degr$ (left to right). Image adapted from \cite{Podgorny2023e}.}}
	\label{CvsD_ptheta}
\end{figure}

\begin{figure}
	\includegraphics[width=0.32\columnwidth]{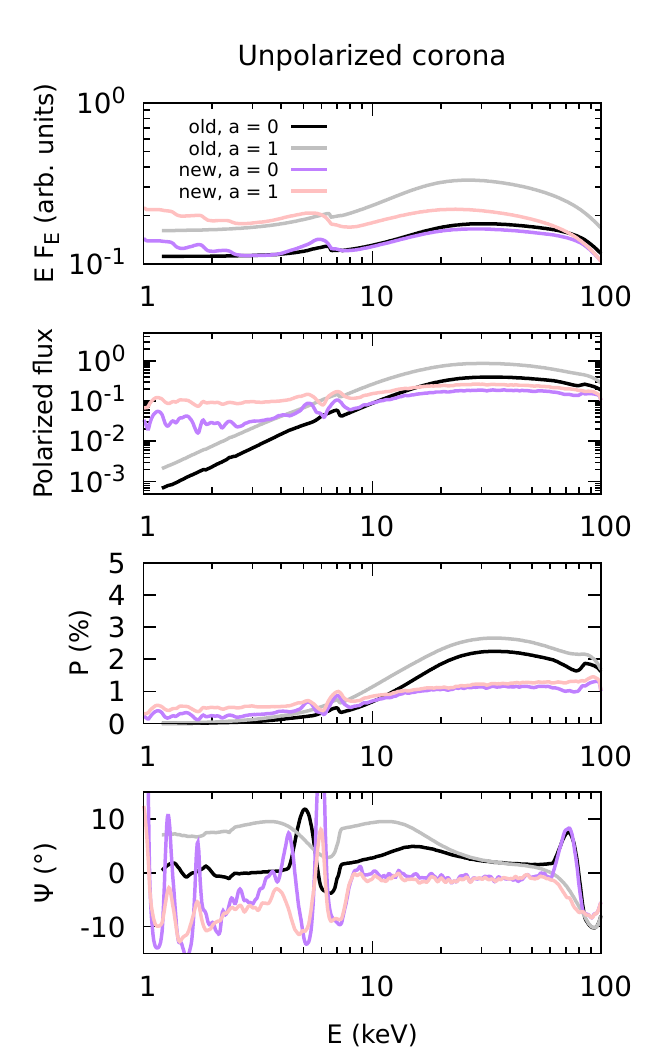}
        \includegraphics[width=0.32\columnwidth]
        {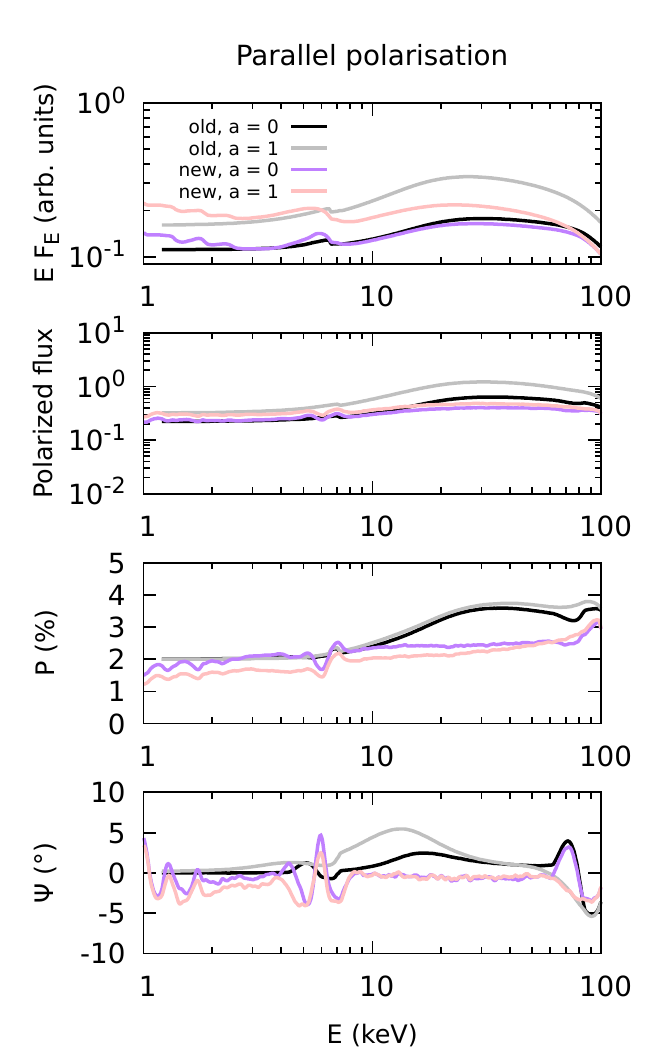}
        \includegraphics[width=0.32\columnwidth]{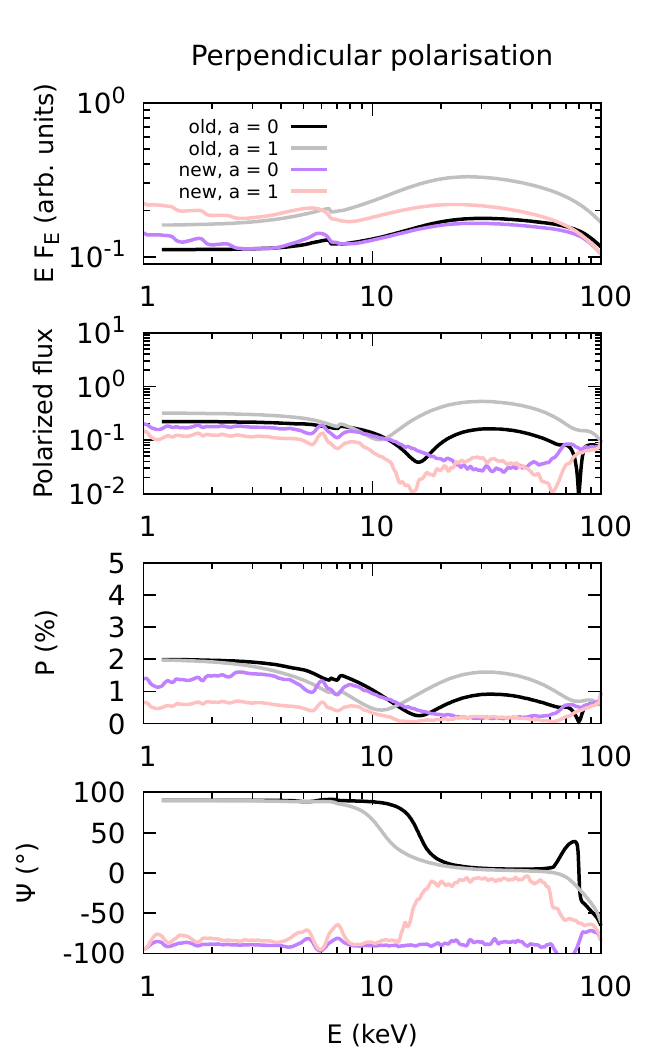}
	\caption{{\footnotesize The incident radiation in the polar directions for AGNs modelled in Section~\ref{full_AGN_model_MC}: in the case of unpolarized coronal radiation (left) and $2\%$ parallelly polarized coronal radiation (middle) and $2\%$ perpendicularly polarized coronal radiation (right). We display from top to bottom the energy-dependent flux (in arbitrary units, corrected for the primary power-law slope) and the corresponding polarized flux, polarisation degree and polarisation angle. The~computations from \citet{Marin2018c} are displayed in black and gray for $a = 0$ and~$a = 1$, respectively. The new computations with {\tt KYNSTOKES} for \textit{ionized} disc are displayed in~purple and pink for $a = 0$ and $a = 1$, respectively. Image adapted from \cite{Podgorny2023d}.}}
	\label{ionizedKY_TF_PO_PA_inputs}
\end{figure}
\begin{figure}
	\includegraphics[width=0.32\columnwidth]{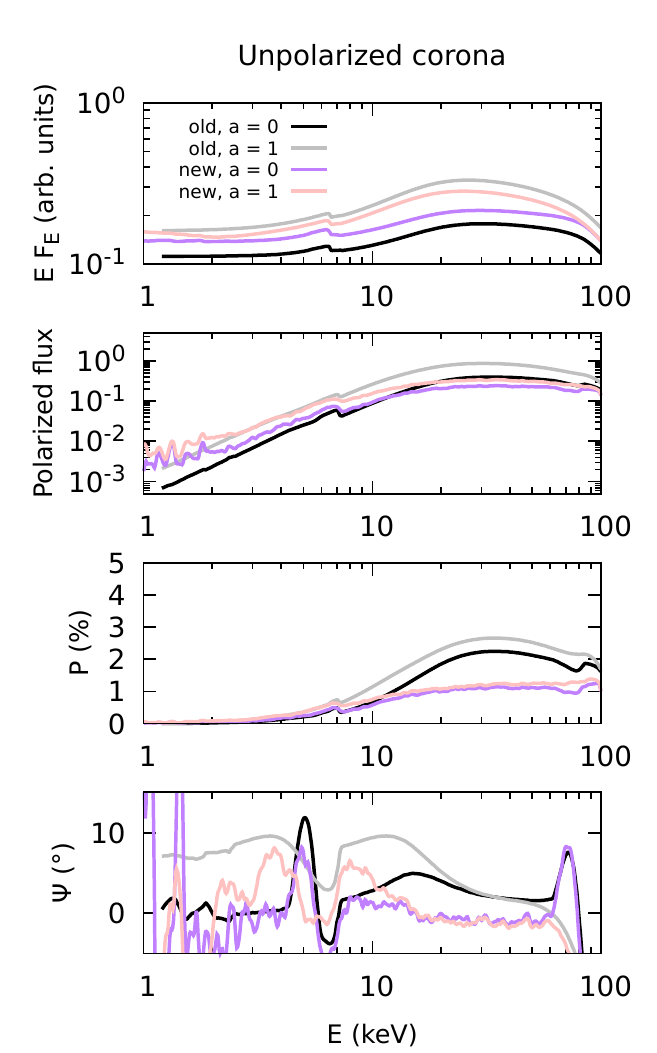}
        \includegraphics[width=0.32\columnwidth]
        {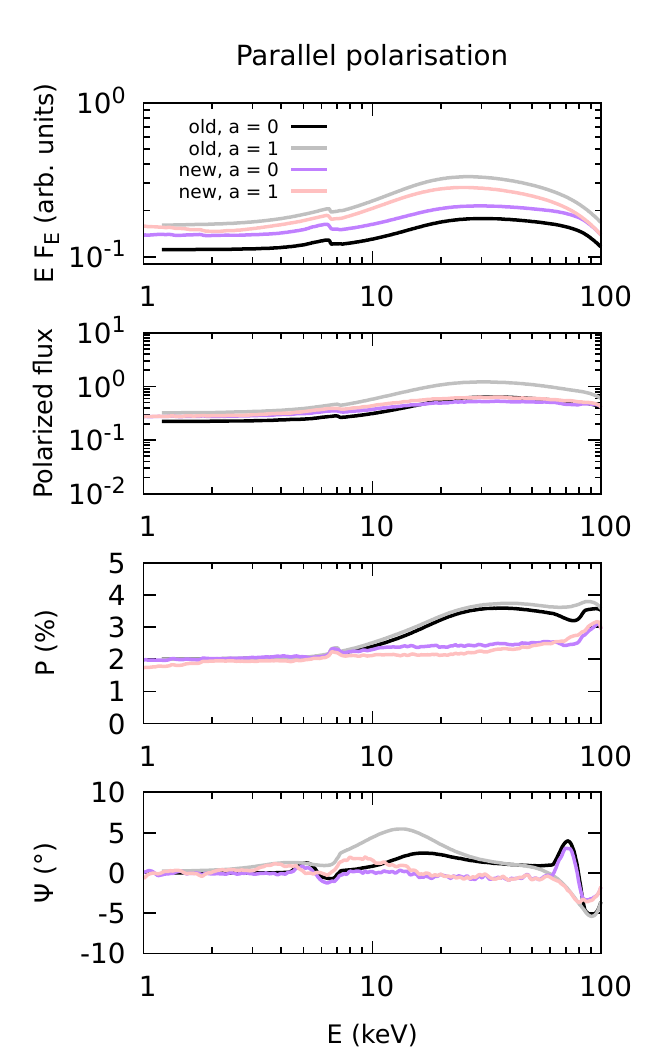}
        \includegraphics[width=0.32\columnwidth]{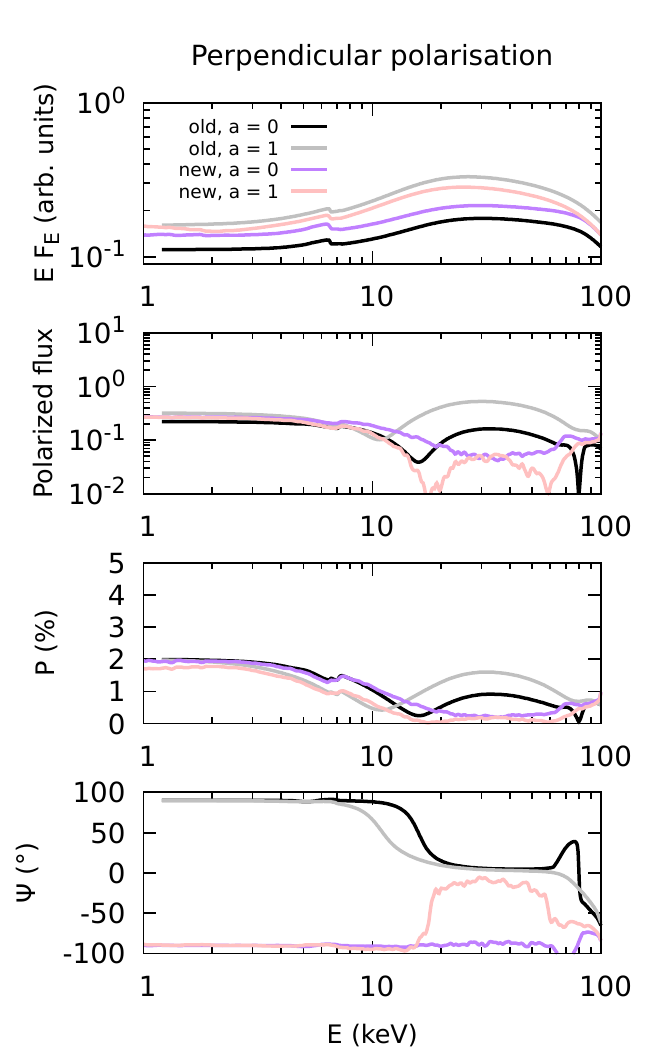}
	\caption{{\footnotesize The same as Figure \ref{ionizedKY_TF_PO_PA_inputs}, but for \textit{neutral} disc. Image adapted from \cite{Podgorny2023d}.}}
	\label{neutralKY_TF_PO_PA_inputs}
\end{figure}
\begin{figure}
	\includegraphics[width=0.32\columnwidth]{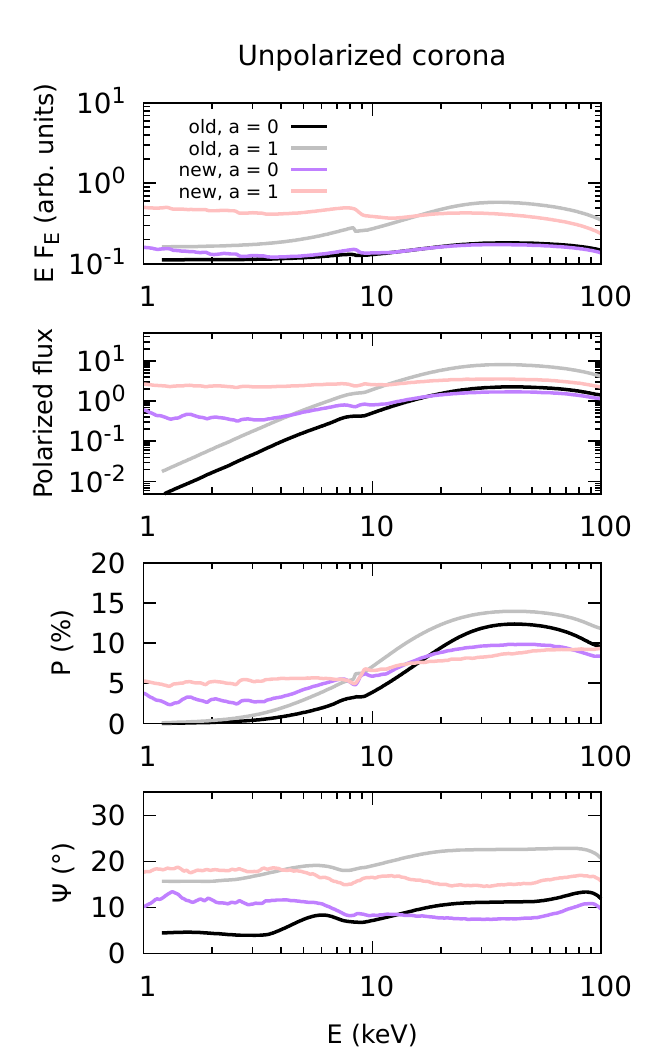}
        \includegraphics[width=0.32\columnwidth]
        {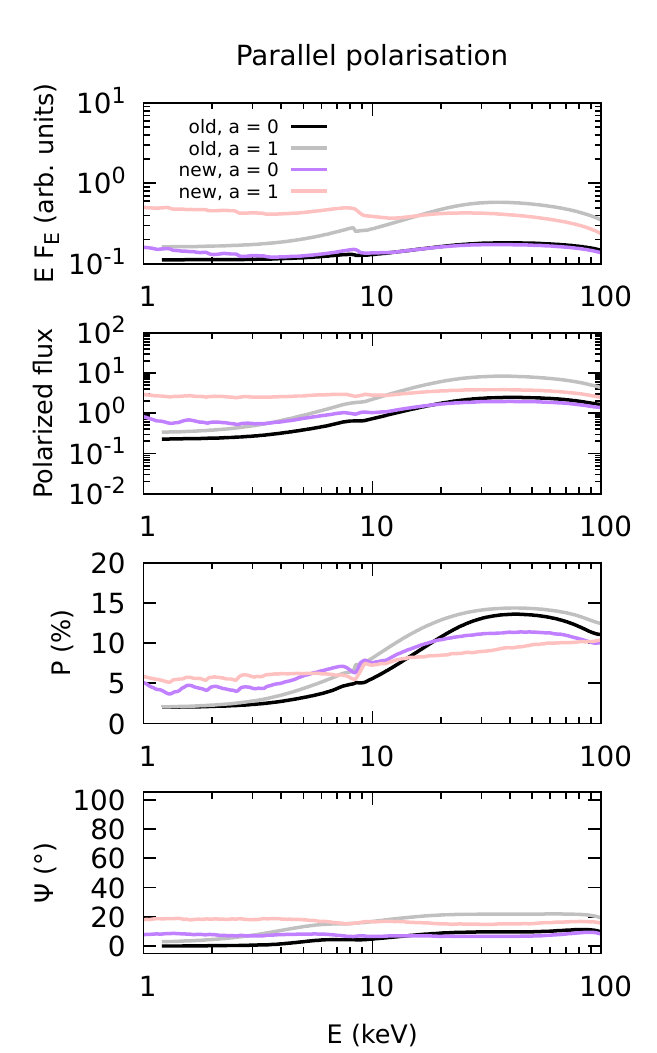}
        \includegraphics[width=0.32\columnwidth]{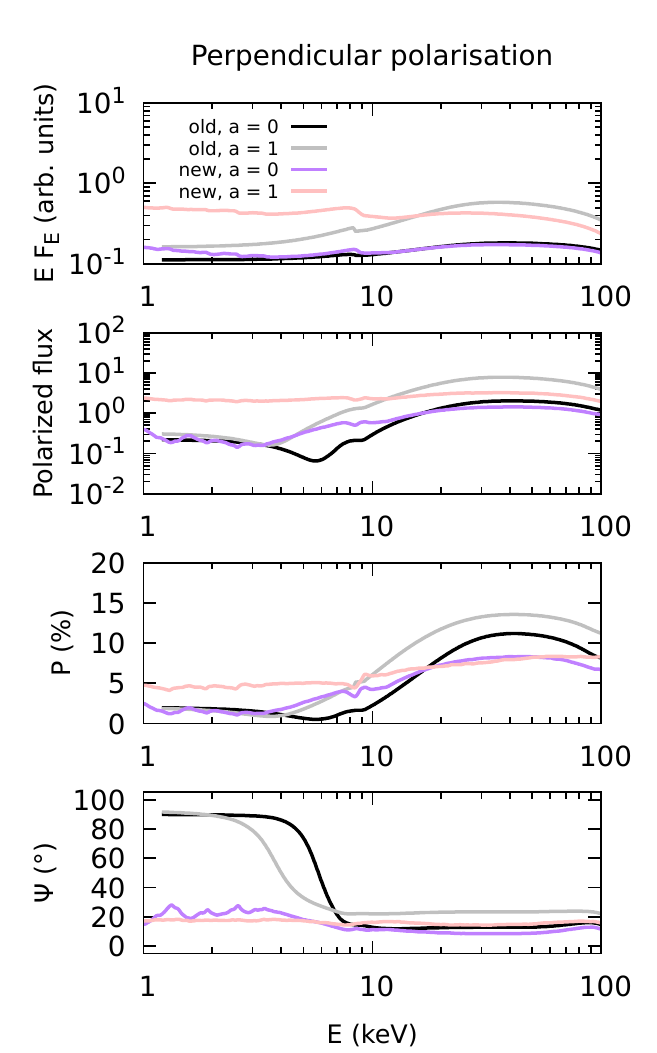}
	\caption{{\footnotesize The same as Figure \ref{ionizedKY_TF_PO_PA_inputs}, but in the equatorial directions for AGNs modelled in~Section \ref{full_AGN_model_MC}. Image adapted from \cite{Podgorny2023d}.}}
	\label{type2_ionizedKY_TF_PO_PA_inputs}
\end{figure}
\begin{figure}
	\includegraphics[width=0.32\columnwidth]{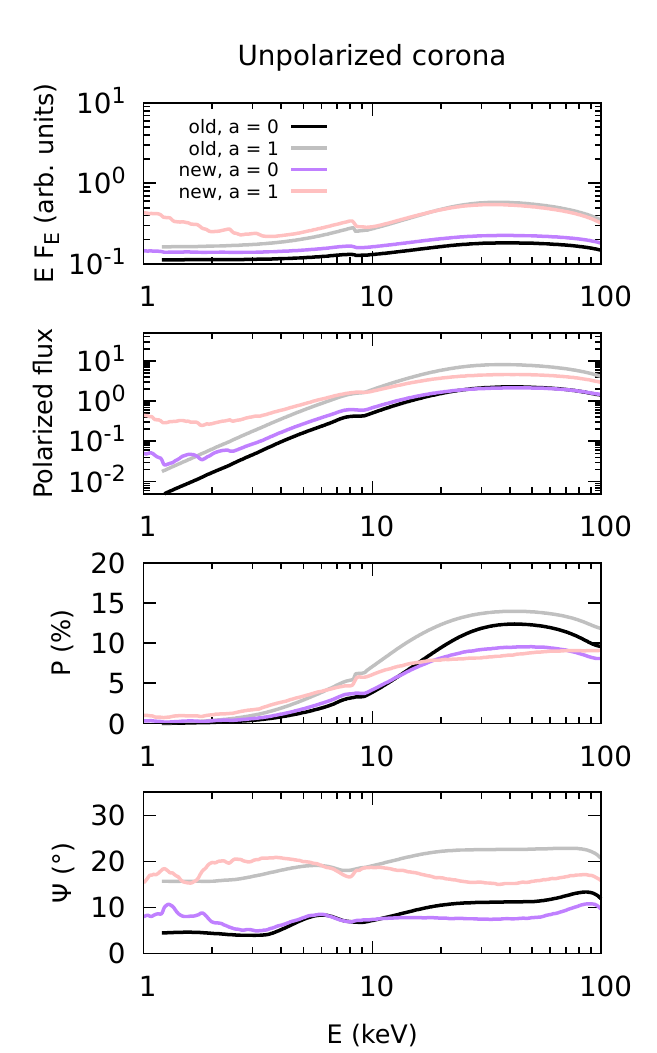}
        \includegraphics[width=0.32\columnwidth]
        {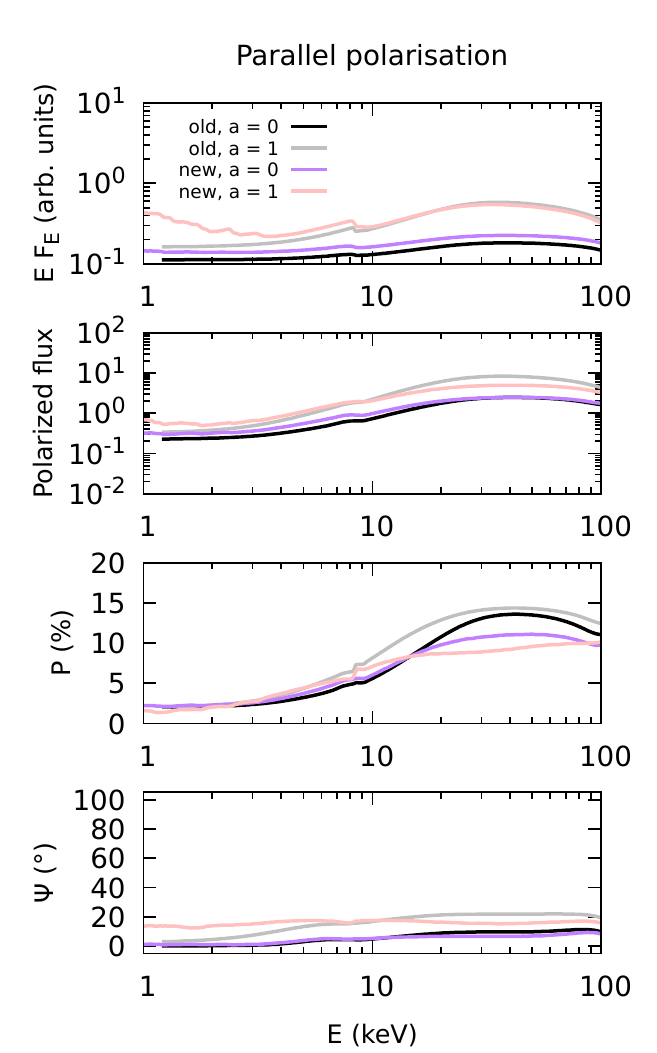}
        \includegraphics[width=0.32\columnwidth]{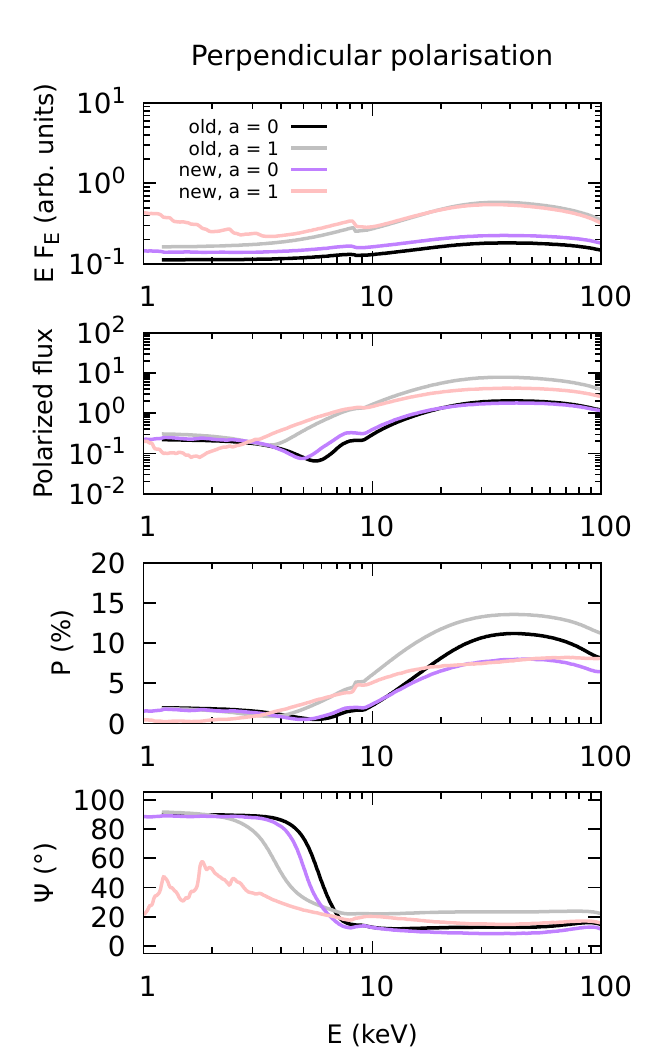}
	\caption{{\footnotesize The same as Figure \ref{type2_ionizedKY_TF_PO_PA_inputs}, but for \textit{neutral} disc. Image adapted from \cite{Podgorny2023d}.}}
	\label{type2_neutralKY_TF_PO_PA_inputs}
\end{figure}

\begin{landscape}
\end{landscape}

\section*{Supplementary figures to Chapter \ref{chap05}}

\begin{figure}[!htb]\centering
	\includegraphics[width=\textwidth]{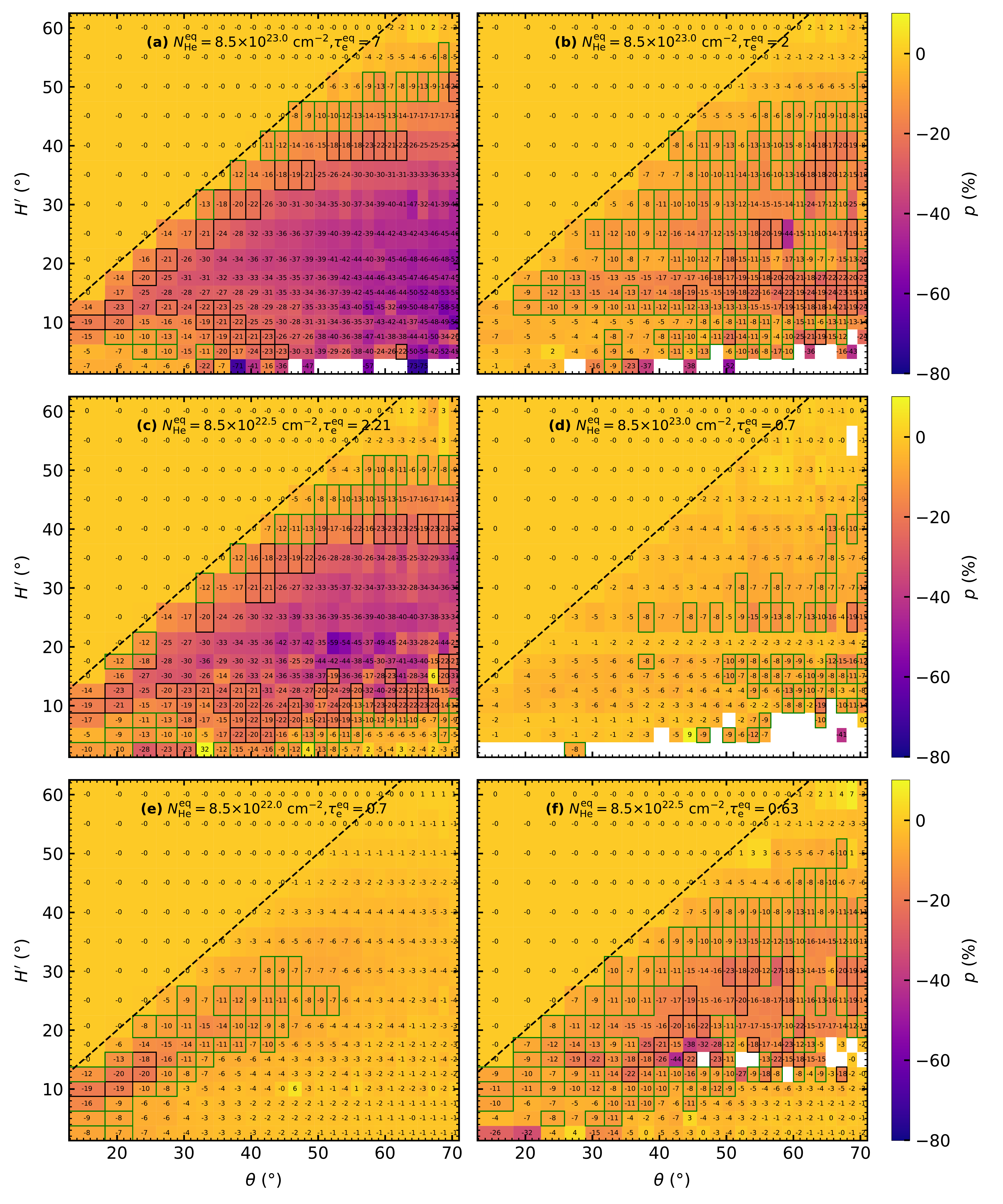}
	\caption{\footnotesize{The same as in Figure \ref{all_torus_0.25}, but for a Case B elliptical obscurer with $b' = R'/8$. Image adapted from \cite{Podgorny2023c}.}}
	\label{all_torus_0.125}
\end{figure}
\begin{figure}[!htb]\centering
	\includegraphics[width=\textwidth]{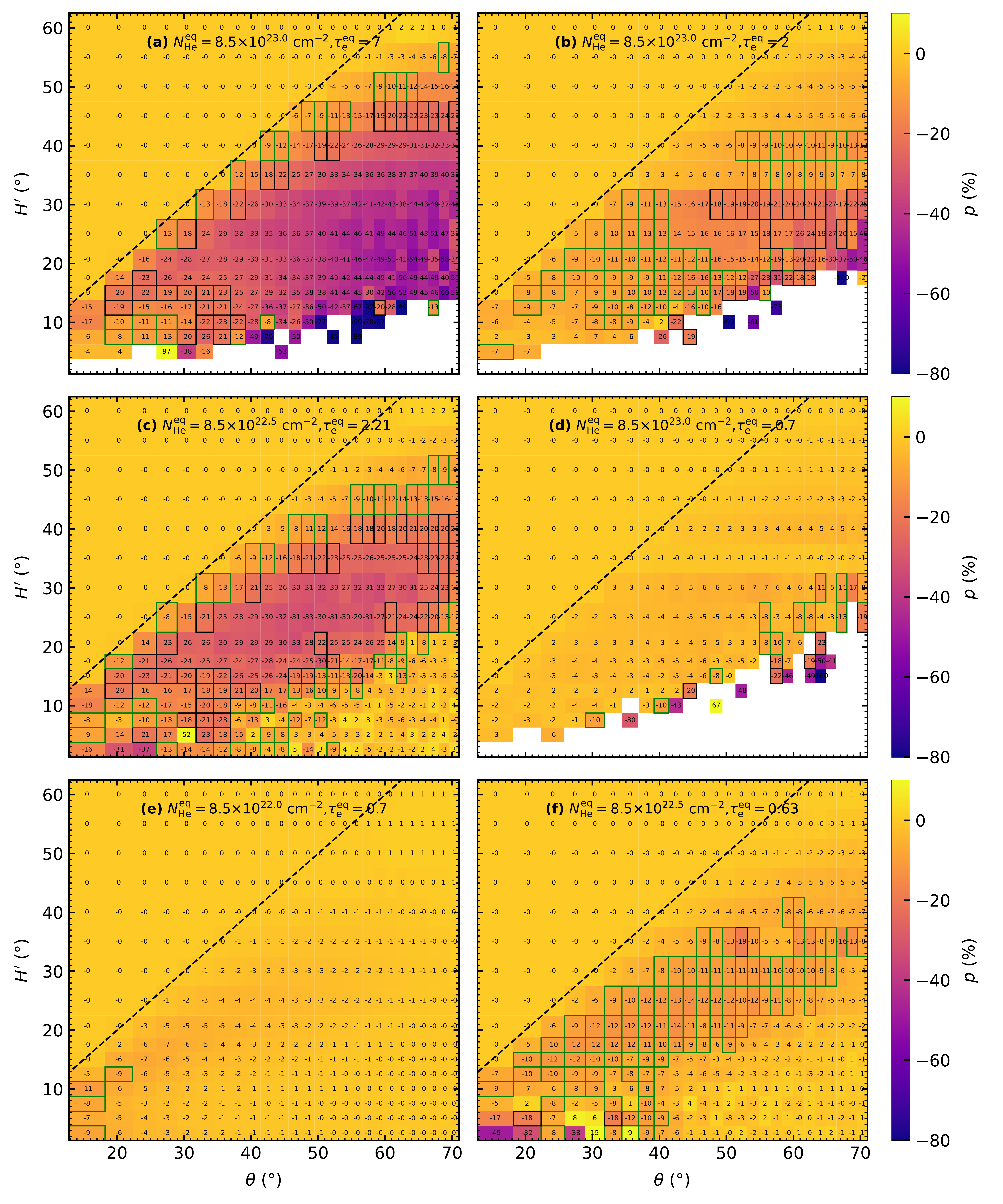}
	\caption{\footnotesize{The same as in Figure \ref{all_torus_0.25}, but for a Case B elliptical obscurer with $b' = 3R'/4$. Image adapted from \cite{Podgorny2023c}.}}
	\label{all_torus_0.75}
\end{figure}
\begin{figure}[!htb]\centering
	\includegraphics[width=\textwidth]{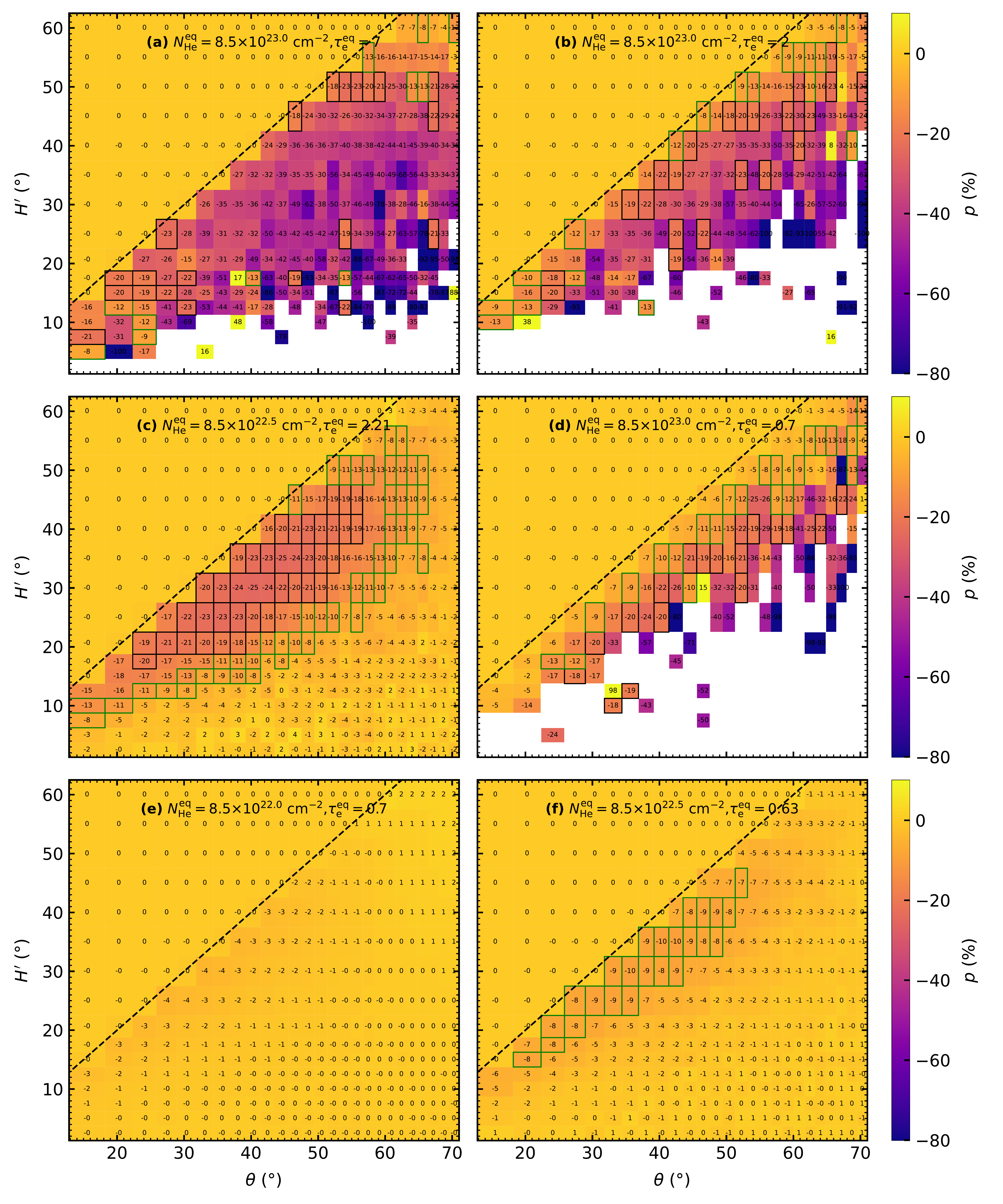}
	\caption{\footnotesize{The same as in Figure \ref{all_torus_0.25}, but for a Case A wedge-shaped obscurer with $r_\mathrm{out}^\textrm{torus} = 10$ and $r_\textrm{in}^\textrm{torus} = R' - b'$, where $b' = a'$ is consistently computed through (\ref{opening_angle}) for any circular torus, depending on inclination $\theta$ and half-opening angle $H'$. Image adapted from \cite{Podgorny2023c}.}}
	\label{all_flared}
\end{figure}

\begin{figure}[!htb]\centering
	\includegraphics[width=\textwidth]{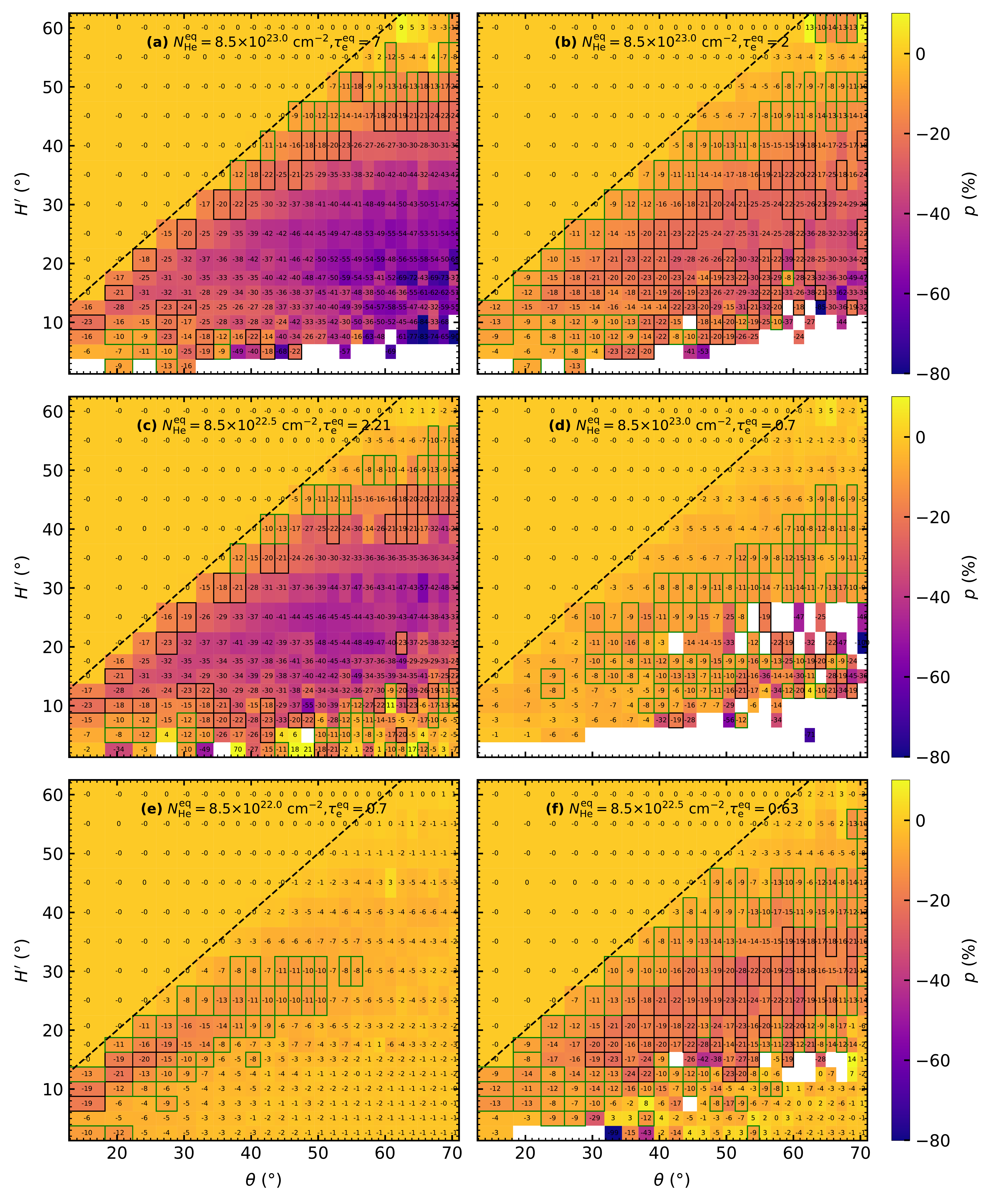}
	\caption{\footnotesize{The same as in Figure \ref{all_torus_0.25}, but for $\Gamma = 3$.}}
	\label{all_torus_0.25_G3}
\end{figure}
\begin{figure}[!htb]\centering
	\includegraphics[width=\textwidth]{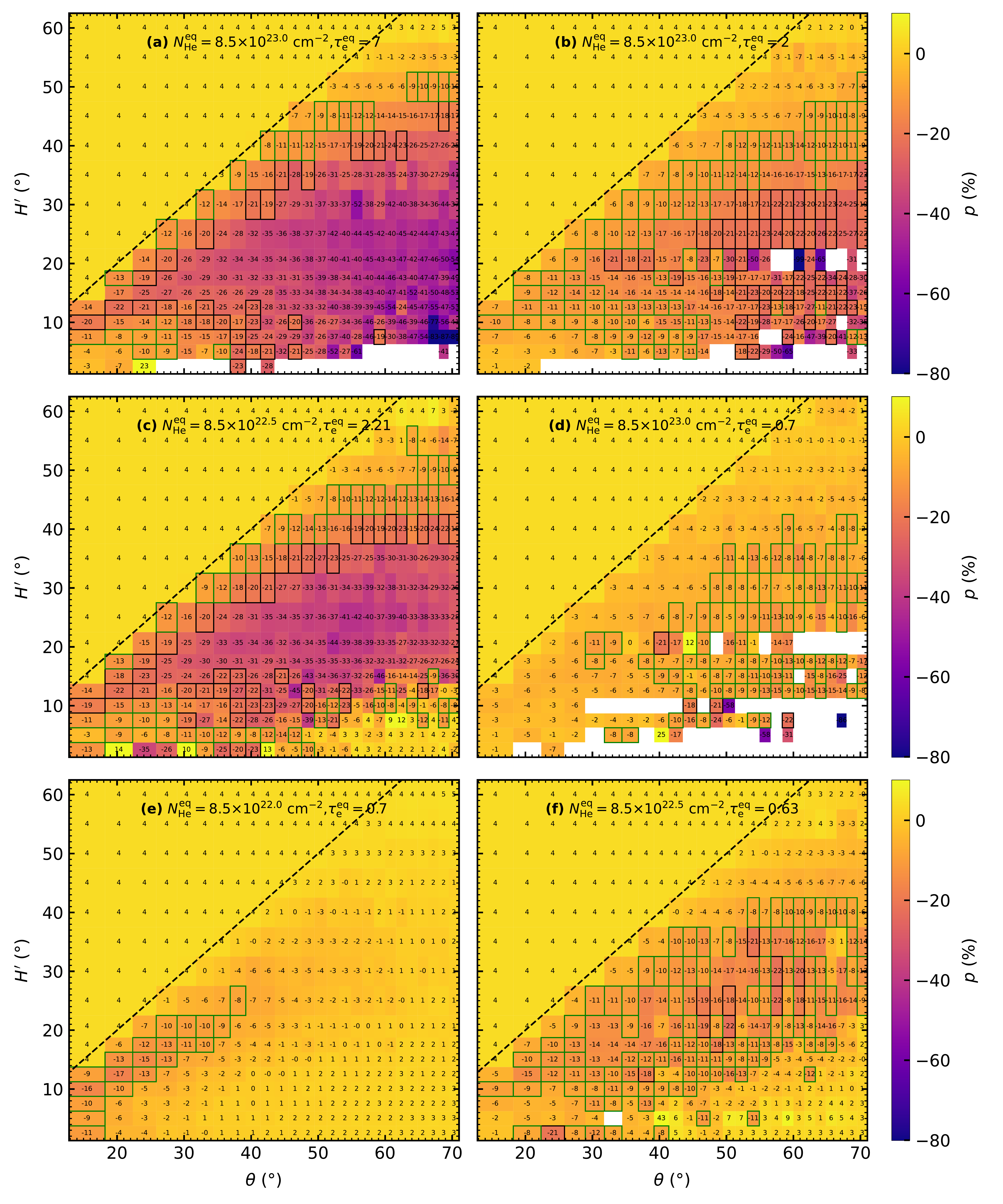}
	\caption{\footnotesize{The same as in Figure \ref{all_torus_0.25}, but for $p_0 = 4\%$.}}
	\label{all_torus_0.25_p4G2}
\end{figure}
\begin{figure}[!htb]\centering
	\includegraphics[width=\textwidth]{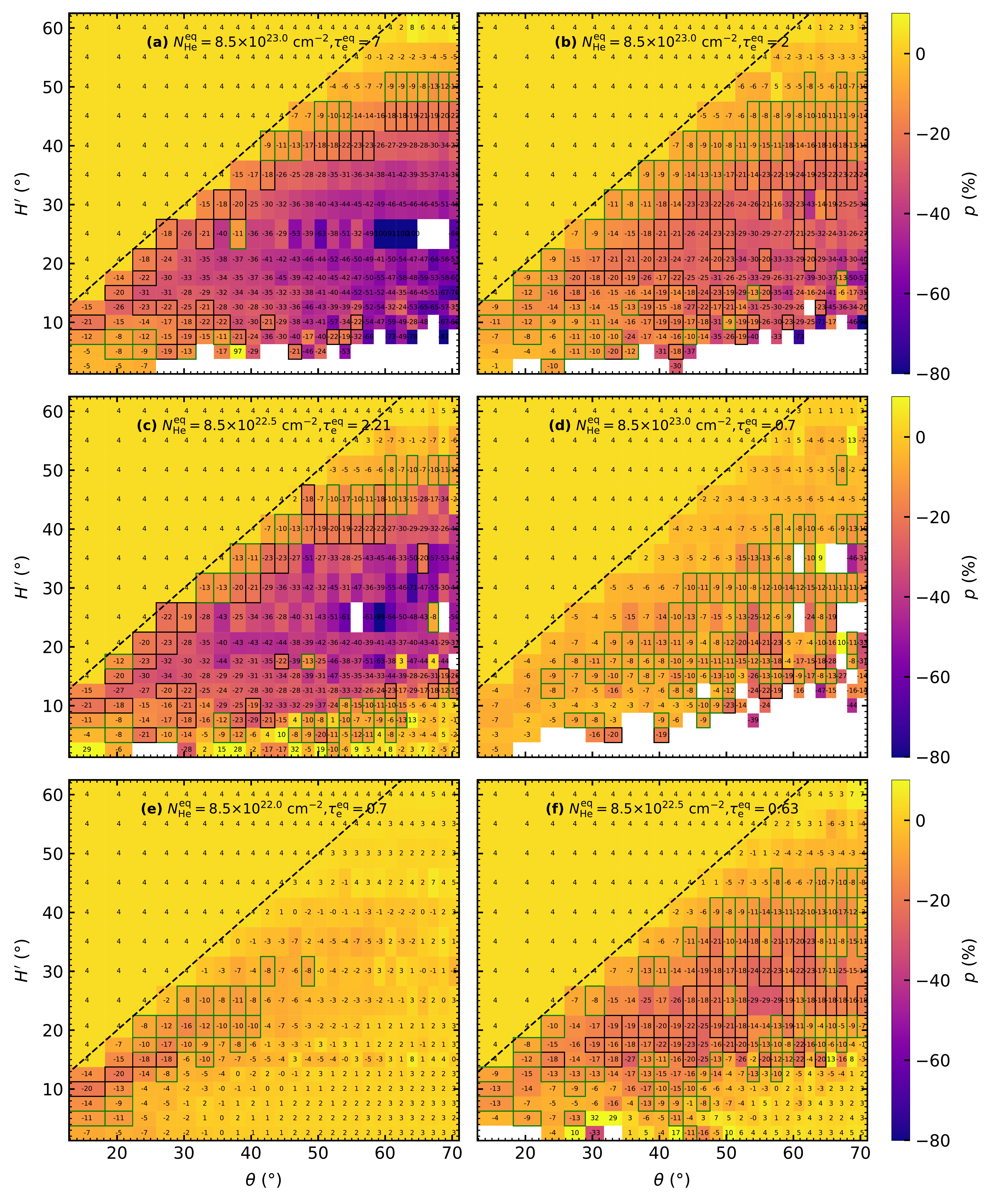}
	\caption{\footnotesize{The same as in Figure \ref{all_torus_0.25}, but for $p_0 = 4\%$ and $\Gamma = 3$.}}
	\label{all_torus_0.25_p4G3}
\end{figure}

\chapter{Singularities in general-relativistic surface integrals}\label{integration}

The disc emission in equatorial plane can be two-dimensionally integrated in multiple ways: using the $\alpha$ and $\beta$ impact parameters, the related Carter's constants of motion (see Appendix \ref{change_PA}), the Boyer-Lindquist coordinates $r$ and $\phi$, or using the $g$ factor instead of azimuthal $\phi$, and in forward or reversed integration order. Exact treatment of integrals of the kind (\ref{totalpol}) typically involves elliptical integrals of the first kind \citep{Cunningham1975} and some simplifications, e.g. in radial emissivity profile, in directionality of the local emission, or in axial symmetry. On~the~contrary, numerical integration and tabular precomputation of geodesics, the transfer function, or other intermediate results, using discretization techniques, may lead to lack of precision and may not always be the most efficient option. In~the~latter approach, detailed computations of relativistic smearing of e.g. spectral lines tend to be in particular computationally expensive due to singularities in the integrand of various, but known kinds. Realizing quickly that an exploration and comparison of all possibilities would be of scope of another PhD thesis, we at least made an attempt of substituting the singularities in $g$ and $r$ coordinates with Gaussian quadratures \citep{Gauss1814, Jacobi1826}, as this is quickly possible, provided the expressions in \cite{Cunningham1975} with singularity prescriptions. If~the~polynomial substitution proves to be significantly faster on~a~few examples, it may be worth of full implementation in the {\tt KY} codes, where the current {\tt ide} integrating subroutine works in the $r$ and $\phi$ coordinates and does not treat efficiently the singularities in $\phi$.

~\

We will depart from \cite{Cunningham1975}, equation (7), which is basically (\ref{cunningham}) written with
\begin{equation}
    g^{*} = \frac{g-g_\textrm{min}(r)}{g_\textrm{max}(r)-g_\textrm{min}(r)} \, ,
\end{equation}
which keeps $0 \leq g^{*} \leq 1$. We will solve a few examples for undirectional Stokes parameter $I$ with the following emissivity profile
\begin{equation}
    I\left(\frac{E_\textrm{obs}}{g},n_\textrm{em},r\right) = r^{-q}\bar{I}\left(\frac{E_\textrm{obs}}{g}\right) \, ,
\end{equation}
where $q=3$ is typically considered \citep{Zycki1998, Beckwith2004,Dovciak2014}. Then
\begin{equation}\label{to_be_solved}
\begin{aligned}
N_\textrm{E} &= \frac{\pi}{r_0^2} \int_{r_\textrm{in}}^{r_\textrm{out}} r^{1-q} \int_{0}^{1}\bar{I}\left(\frac{E_\textrm{obs}}{g^{*}[g_\textrm{max}-g_\textrm{min}]+g_\textrm{min}}\right) f(g^{*},r,\theta) \\
&\, \,\,\,\,\,\,\,\,\,\,\,\,\,\,\, \,\,\,\,\,\,\,\,\,\,\,\,\,\,\, \,\,\,\,\,\,\,\,\,\,\,\,\,\,\, \cdot  \frac{g^{*}[g_\textrm{max}-g_\textrm{min}]+g_\textrm{min}}{\sqrt{g^{*}}\sqrt{1-g^{*}}} \textrm{d}g^{*} \textrm{d}r \\
 &= \frac{\pi}{r_0^2} \int_{r_\textrm{in}}^{r_\textrm{out}} r^{1-q} \int_{-1}^{1} \bar{I} \left(\frac{E_\textrm{obs}}{\frac{1}{2}[\Tilde{g}^{*}+1][g_\textrm{max}-g_\textrm{min}]+g_\textrm{min}}\right)  \\
&\, \,\,\,\,\,\,\,\,\,\,\,\,\,\,\, \,\,\,\,\,\,\,\,\,\,\,\,\,\,\, \,\,\,\,\,\,\,\,\,\,\,\,\,\,\,  \cdot  f\left(\frac{1}{2}[\Tilde{g}^{*}+1],r,\theta\right) \Bigl\{ \frac{1}{2} [\Tilde{g}^{*}+1][g_\textrm{max}-g_\textrm{min}]+g_\textrm{min}\Bigr\}  \\  &\, \,\,\,\,\,\,\,\,\,\,\,\,\,\,\, \,\,\,\,\,\,\,\,\,\,\,\,\,\,\, \,\,\,\,\,\,\,\,\,\,\,\,\,\,\,  \cdot \frac{1}{\sqrt{1-\Tilde{g}^{*2}}} \textrm{d}\Tilde{g}^{*} \textrm{d}r \\
 &= \frac{\pi}{r_0^2} \int_{r_\textrm{in}}^{r_\textrm{out}} r^{1-q} \int_{-1}^{1} \bar{F}(\Tilde{g}^{*},r,\theta)\bar{S}(\Tilde{g}^{*}) \textrm{d}\Tilde{g}^{*} \textrm{d}r \, \, ,
\end{aligned}
\end{equation}
where $\bar{S}(\Tilde{g}^{*}) = [1-\Tilde{g}^{*2}]^{-\frac{1}{2}}$ is the diverging factor and $\bar{F}(\Tilde{g}^{*},r,\theta)$ is an analytical function on $g^{*} \in [-1;1]$. This allows the usage of Chebyshev-Gauss quadrature formula\footnote{\,Another option perhaps worth trying is splitting the inner integral of the first expression in~(\ref{to_be_solved}) into $\int_0^{\frac{1}{2}}\cdots \textrm{d}g^{*} + \int_{\frac{1}{2}}^1\cdots \textrm{d}g^{*} = \Hat{A} + \Hat{B}$. Then rescaling $\Hat{A}$ with $g^{*} = \frac{1}{2}\Tilde{g}^{*}$ substitution to~$\Tilde{g}^{*} \in [0;1]$ range and using Legendre polynomials $P_{2n}(\sqrt{\Tilde{g}^{*}})$ of the order $2n$ to solve for~singularity of type $\frac{1}{\sqrt{\Tilde{g}^{*}}}$. The other integral $\Hat{B}$ is to be rescaled with $g^{*} = \frac{1}{4}(\Tilde{g}^{*}+3)$ substitution to~$\Tilde{g}^{*} \in [-1;1]$ range and then using Jacobi polynomials $J_n(\Tilde{g}^{*},\alpha',\beta')$ of the order $n$ for $\alpha' = -\frac{1}{2}$, $\beta' = 0$ to solve for singularity of type $\frac{1}{\sqrt{1-\Tilde{g}^{*}}}$. See the discussion around table 4.4 in \cite{Ralston1965} for~general procedures and simplifications, also for two-dimensional integrals as a whole.} \citep{Ralston1965}
\begin{equation}
    \int_{-1}^{1} \frac{1}{\sqrt{1-x^2}} \bar{F}(x)\textrm{d}x \approx \sum_{j = 1}^{n} H_j\bar{F}(a_j) \, , \,\, H_j = \frac{\pi}{n}  \, , \,\, a_j = \cos \frac{[2j-1]\pi}{2n} \, .
\end{equation}
We obtain
\begin{equation}\label{polynomial}
\begin{aligned}
    N_\textrm{E} &= \frac{\pi^2}{nr_0^2} \int_{r_\textrm{in}}^{r_\textrm{out}} r^{1-q} \sum_{j = 1}^{n} \Bar{I}\left(\frac{E_\textrm{obs}}{\frac{1}{2}[\Tilde{g}_j^{*}+1][g_\textrm{max}-g_\textrm{min}]+g_\textrm{min}}\right)  \\
&\, \,\,\,\,\,\,\,\,\,\,\,\,\,\,\, \,\,\,\,\,\,\,\,\,\,\,\,\,\,\,  \cdot f\left(\frac{1}{2}[\Tilde{g}_j^{*}+1],r,\theta\right)\Bigl\{ \frac{1}{2} [\Tilde{g}_j^{*}+1][g_\textrm{max}-g_\textrm{min}]+g_\textrm{min}\Bigr\} \textrm{d}r \, ,  \\
&\, \,\,\,\,\,\,\,\,\,\,\,\,\,\,\, \,\,\,\,\,\,\,\,\,\,\,\,\,\,\,\,\, \Tilde{g}_j = \cos \frac{[2j-1]\pi}{n} \, ,
\end{aligned}
\end{equation}
which can be compared to the more classical result in $g$ coordinate:
\begin{equation}\label{option_classical}
\begin{aligned}
N_\textrm{E} &= \frac{\pi}{r_0^2} \int_{r_\textrm{in}}^{r_\textrm{out}} r^{1-q} \int_{g_\textrm{min}}^{g_\textrm{max}}\bar{I}\left(\frac{E_\textrm{obs}}{g}\right) f\left(\frac{g-g_\textrm{min}}{g_\textrm{max}-g_\textrm{min}},r,\theta\right)  \\
&\, \,\,\,\,\,\,\,\,\,\,\,\,\,\,\, \,\,\,\,\,\,\,\,\,\,\,\,\,\,\,  \cdot  \frac{g}{\sqrt{g-g_\textrm{min}}\sqrt{g_\textrm{max}-g}} \textrm{d}g \textrm{d}r \, \, .
\end{aligned}
\end{equation}

~\

\begin{figure}[!htb]\centering
	\includegraphics[width=\textwidth]{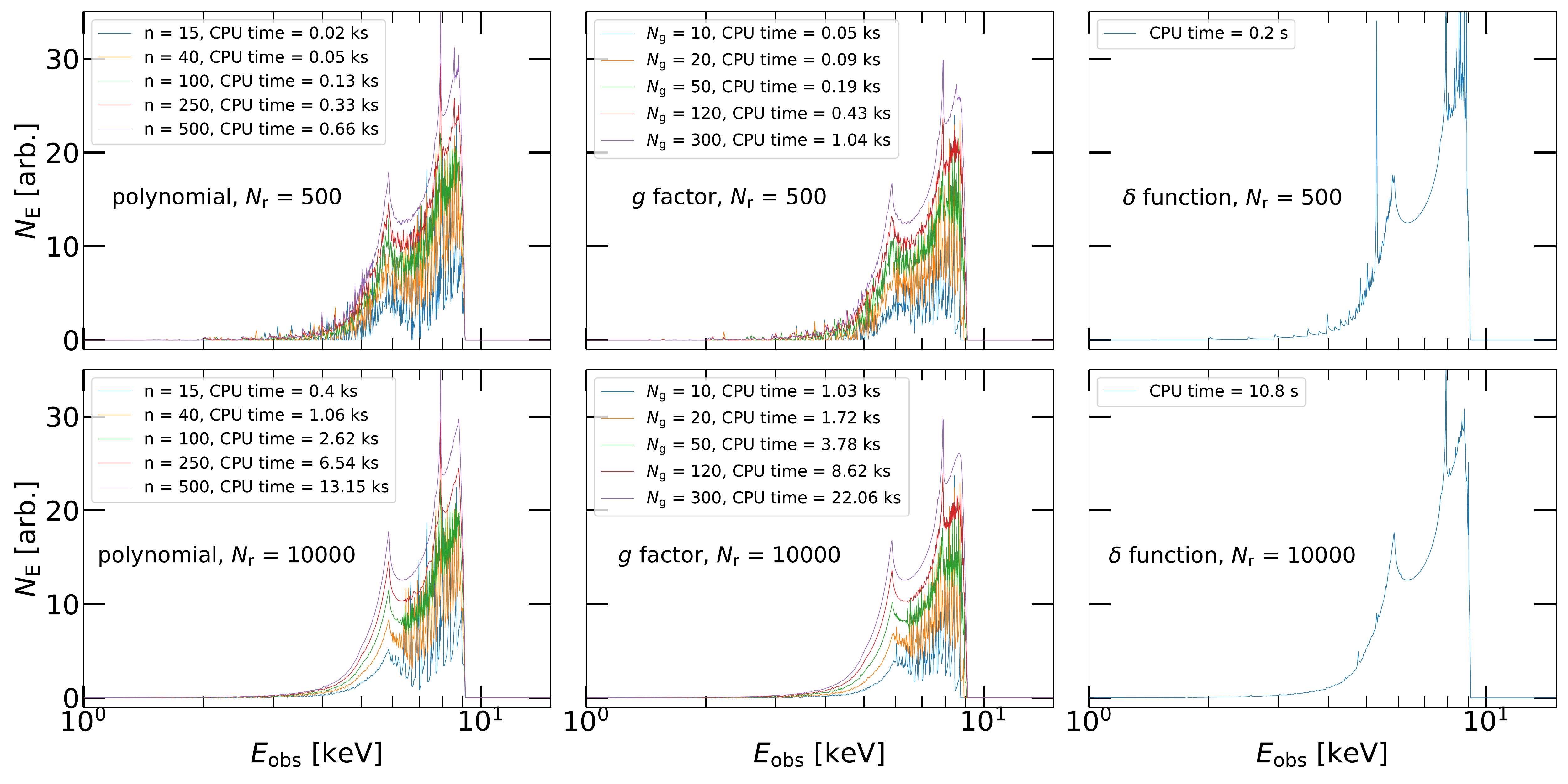}
	\caption{\footnotesize{Comparison of a narrow Gaussian line that is relativistically smeared for a distant observer, using the proposed polynomial integration technique (left) and using the classical integration in $g$ (middle) for various precision. Each case is renormalized to arbitrary value for~viewing clarity. The reference image of a smeared line profile given by $\delta$ function is also shown (right). The results are displayed for $N_\textrm{r} = 500$ (top) and $N_\textrm{r} = 10000$ (bottom). The~results are approximate, given significant simplifications (see text), as the main purpose is to test a~few examples of different integration methods.}}
	\label{singularities_comparison}
\end{figure}

Figure \ref{singularities_comparison} provides examples of spectral line broadening for a distant observer for both, polynomial and classical integration in $g$ in the left and middle panels, respectively. Computational processing unit (CPU) time for a simple script on~a~laptop is provided, but only relative difference between the results of~the~two methods is important. A Gaussian line profile at $E_\textrm{l} = 6.4$ keV was used 
\begin{equation}
\Bar{I}\left(\frac{E_\textrm{obs}}{g}\right) = \frac{1}{\omega \sqrt{2\pi}} e^{-\frac{\left[\frac{E_\textrm{obs}}{g}-E_\textrm{l}\right]^2}{2\omega^2}}  \, ,
\end{equation}
where $g = \frac{1}{2}[\Tilde{g}_j^{*}+1][g_\textrm{max}-g_\textrm{min}]+g_\textrm{min}$ for the case of (\ref{polynomial}). We used $\omega = 0.005$ for the line width and 500 linearly spaced bins in energy between $E_\textrm{min} = 1$ keV and $E_\textrm{max} = 15$ keV. In the right panels, we provide referential results for intensity given by $\delta$ function, which automatically solves for the inner integral and is hence the fastest solution:
\begin{equation}
\begin{aligned}
N_\textrm{E} &= \frac{\pi}{E_\textrm{obs}r_0^2} \int_{r_\textrm{in}}^{r_\textrm{out}} r^{1-q} \int_{\frac{1}{g_\textrm{min}}}^{\frac{1}{g_\textrm{max}}}\delta \left(x'-\frac{E_\textrm{l}}{E_\textrm{obs}}\right) f\left(\frac{\frac{1}{x'}-g_\textrm{min}}{g_\textrm{max}-g_\textrm{min}},r,\theta\right)  \\
&\, \,\,\,\,\,\,\,\,\,\,\,\,\,\,\, \,\,\,\,\,\,\,\,\,\,\,\,\,\,\, \,\,\,\,\,\,\,\,\,\,\,\,\,\,\, \cdot  \frac{1}{x'^2\sqrt{1-g_\textrm{min}x'}\sqrt{g_\textrm{max}x'-1}} \textrm{d}x' \textrm{d}r \\ 
&=
\begin{cases}
\frac{E_\textrm{obs}\pi}{E_\textrm{l}^2r_0^2} \int_{r_\textrm{in}}^{r_\textrm{out}} r^{1-q} f\left(\frac{\frac{E_\textrm{obs}}{E_\textrm{l}}-g_\textrm{min}}{g_\textrm{max}-g_\textrm{min}},r,\theta\right)&\cdot\frac{1}{\sqrt{1-g_\textrm{min}\frac{E_\textrm{l}}{E_\textrm{obs}}}\sqrt{g_\textrm{max}\frac{E_\textrm{l}}{E_\textrm{obs}}-1}} \textrm{d}r \ , \\ & \frac{E_\textrm{l}}{E_0} \in \left[\frac{1}{g_\textrm{max}};\frac{1}{g_\textrm{min}}\right] \ , \\
 0  \ , & \textrm{else} \, \, ,
\end{cases}
\end{aligned}
\end{equation}
where we used $x' = 1/g$. The results are provided for two linear binnings in $r$, given by $N_\textrm{r}$, and various binnings in $g$, given by $N_\textrm{g}$ (logarithmically compressed bins towards both integration edges starting from a mid-point), and various orders of $n$. For small $n$, we first observe $n$ nodes in the numerical defocusing, which gets smoothed to the exact result when $n$ approaches infinity. The unnatural features from realistic line profiles are given by the fact that we in addition imposed an approximation of the transfer function $\Bar{f}_\textrm{T}$ \citep[see][]{Cunningham1975} given by empirical law representing the curve $\Bar{f}_\textrm{T}(r)$ from figure 3a of \cite{Cunningham1975} and the same for $g_\textrm{min}(r)$ and $g_\textrm{max}(r)$ from figure 1a of \cite{Cunningham1975}, all for~$a = 0$ and $\cos\theta = 0.75$, as if these were constants in $\theta$. The remaining parameters were $q = 3$, $r_\textrm{in} = 2 \, r_\textrm{g}$, $r_\textrm{out} = 100 \, r_\textrm{g}$. The integrations in $r$ and $g$ were performed using the Simpson rule \citep[see e.g.][]{Ralston1965}.

Although some gain in efficiency is apparent, the polynomial approach did not seem to win dramatically over the integration in classical $g$ and $r$ coordinates, when also more detailed examples were tested than the above shown. We tried for interpolation of tabular emissivity profiles in discrete energy bins and precomputed transfer functions from \cite{Dovciak2004a}. We tried the usage of quadratures in $g$-integration compared to splitting the classical inner integral in (\ref{option_classical}) in three parts: the edges with singularities to be analytically integrated for quadratic interpolation of the non-diverging part of the integrand, and the mid part to be numerically integrated, using the Simpson rule. Various binning types, resolutions and line profiles were also tested. The intermediate results are disfavoring continuation of this work, but perhaps 
a negative result of an experiment also deserves to be recorded.

\chapter{General-relativistic change of the polarisation angle}\label{change_PA}

We will summarize the derivations of the change of the polarisation angle due to~GR effects in Kerr spacetime in between the lamp-post and the observer and~in~between the lamp-post and the disc. The original calculations were carried by Michal Dovčiak and in the scope of this dissertation only revisited and verified. These are now implemented in the {\tt KYNSTOKES} code \citep{Podgorny2023}.

~\

It is practical to use two Carter's constants of motion that are related to~the~impact parameters $\alpha$ and $\beta$. These will have a reduced form \mbox{$\bar{\lambda} = \alpha \sin{\theta}= 0$,} \mbox{$\bar{q} = \sqrt{\beta^2 + (\alpha^2-a^2)\cos^2{\!\theta}} = \sqrt{\beta^2 - a^2\cos^2{\!\theta}}$,} because we track only the photons from the rotation axis. The expressions for $p_\mu$ components are taken from \cite{Carter1968, Misner1973}: $|p_\theta|=\sqrt{\bar{q}^2+a^2\cos^2\!\theta}$, \mbox{$|p_r|=\sqrt{(r^2+a^2)^2-\Delta(\bar{q}^2+a^2)}/\Delta$} with $\Delta=r^2-2r+a^2$; $p_\phi = 0$ is trivial and~we normalize the conserved energy as $p_t = -1$. We define the unit polarisation \mbox{3-vector} $\Vec{f}$ in the polarisation plane (perpendicular to the photon's momentum) and a disc normal (or rotation axis) projected to that plane, which altogether set the polarisation angle $\Psi$ in any local frame of reference. We will first draft the general procedure and then solve for particular expressions of $\Delta \Psi$ between the lamp and the observer, denoted as $\chi_\textrm{p}$, and between the lamp and~the~disc, denoted as $\chi_\textrm{d}$.

~\

We assume the initial polarisation angle $\Psi_0 = 0$ at the lamp location. For~incident polarisation at the lamp in the local reference frame of the primary source (defined by the height, $h$, and constant of motion, $\bar{q}$) we write $\Vec{f_0}$. This local frame is given by an orthonormal tetrade $e_{\rm (a)}^\mu$, which sets four components of~a~polarisation 4-vector $f_0^{\rm (a)} = (0, \vec{f_0})$, the corresponding polarisation 4-vector being $f_0^\mu = e_{\!\rm (a)}^\mu f_0^{\rm (a)}$. This 4-vector can be parallelly transported in Kerr spacetime, using (\ref{geodesic}), between the lamp and a distant observer at spatial infinity (defined by the~observer inclination, $\theta$), or between the lamp and the disc (defined by the radius, $r$). At each receiving point, the polarisation 4-vector $f^\mu$ will, generally, have a non-zero time component in the local reference frame. We will thus subtract a multiple of a null vector after the parallel transport, so that the~properties of a polarisation vector are kept, i.e. the time component is set to~zero and the~space part is not rotated. This means transforming
\begin{equation}\label{reduction}
    f'^{\textrm{(a)}} = f^{\textrm{(a)}} - \frac{f^\textrm{(t)}}{p^\textrm{(t)}}p^{\textrm{(a)}} \, .
\end{equation}
Using the Walker-Penrose theorem and (\ref{KS}), we will express $\kappa_1$ and $\kappa_2$ by means of the initial $p^\mu$ and $f_0^\mu$ of each photon and the same for the three unknown components of the polarisation 4-vector $f^\mu$ in a given local reference frame at~the~receiving point, where we will construct the $\Psi$ with respect to the projected disc normal (or rotation axis).

~\

\begin{figure}[!htb]
	\centering
	\begin{minipage}[t]{0.49\textwidth}
		\includegraphics[width=\textwidth]{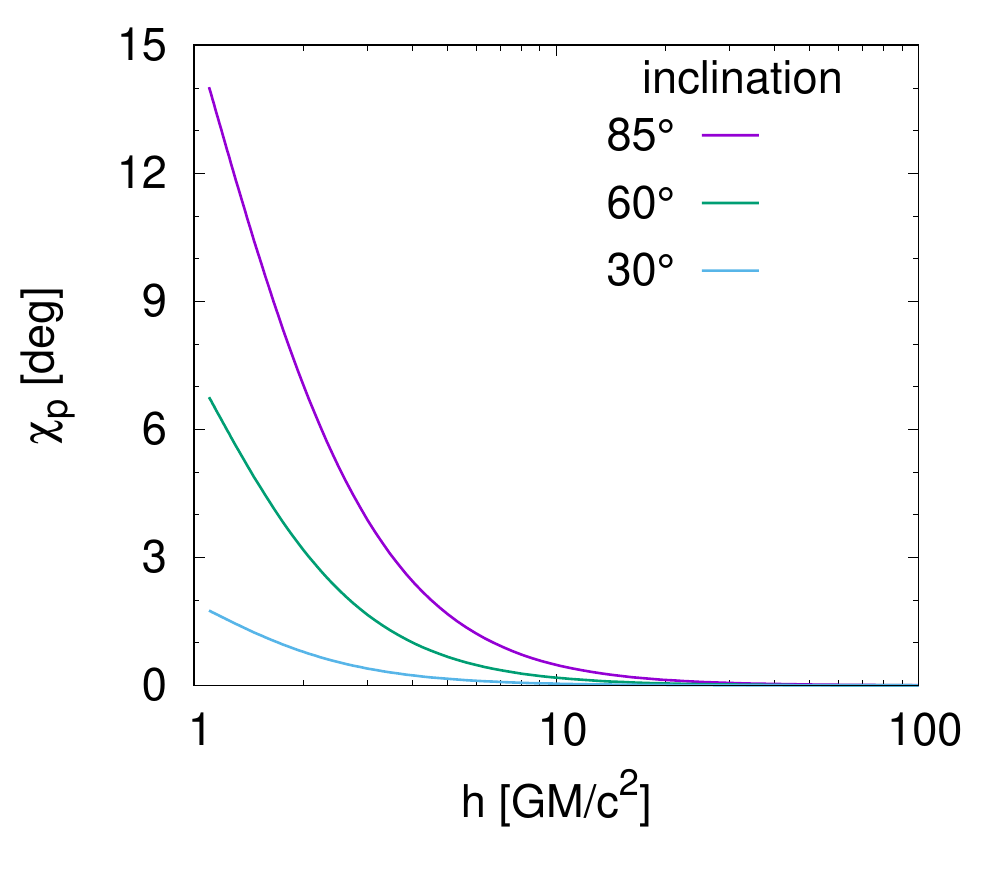}
		\caption{\footnotesize{The GR change of polarisation angle, $\chi_{\rm p}$, between the lamp-post and the observer versus lamp-post height $h$ for black-hole spin $a = 1$ and three different observer's inclinations $\theta = 30^{\circ},\,60^{\circ}$, and $85^{\circ}$. Image courtesy of Michal Dovčiak.}}
		\label{chip}
	\end{minipage}
	\hfill
	\begin{minipage}[t]{0.49\textwidth}
		\includegraphics[width=\textwidth]{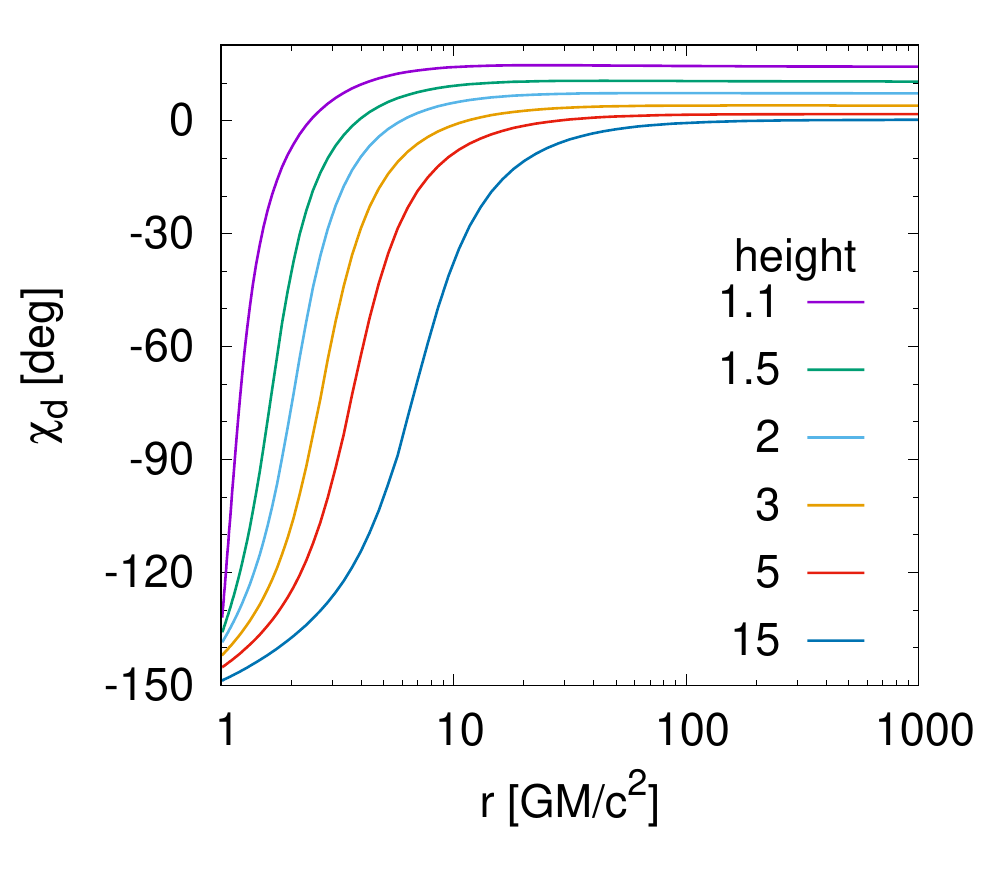}
		\caption{\footnotesize{The GR change of the polarisation angle, $\chi_{\rm d}$, between the lamp-post and the disc as a function of radius $r$ in the equatorial plane for black-hole spin $a = 1$ and several lamp-post heights. Image courtesy of Michal Dovčiak.}}
		\label{chid}
	\end{minipage}
\end{figure}

Let us first solve the case of transfer from the lamp to a distant observer. At~the~lamp it holds that
\begin{equation}\label{k10}
    \kappa_1 = h\sqrt{\frac{\bar{q}^2+a^2}{h^2+a^2}}\, ,
\end{equation}
\begin{equation}\label{k20}
    \kappa_2 = -a\sqrt{\frac{\bar{q}^2+a^2}{h^2+a^2}}\, ,
\end{equation}
where we set $f_{0\,t} = f_{0\,\phi} = 0$ and the other two components, $f_{0\,r}$ and $f_{0\,\theta}$ are obtained from general conditions of (i) polarisation vector being perpendicular to~the~momentum, $f^\mu p_\mu = 0$, and (ii) polarisation vector being normalized to~unity, $f_\mu f^\mu = 1$. Using the conditions $f^\mu p_\mu = p^\mu p_\mu = 0$, we perform the~transformation (\ref{reduction}) at spatial infinity
\begin{equation}\label{trafo1}
    f'^{\textrm{(t)}} = 0\, ,
\end{equation}
\begin{equation}
    f'^{\textrm{(r)}} = r^2p^\theta (p^\theta f^r - p^r f^\theta)\, ,
\end{equation}
\begin{equation}
    f'^{(\theta)} = rp^r (p^\theta f^r - p^r f^\theta )\, ,
\end{equation}
\begin{equation}\label{trafo4}
    f'^{(\phi)} = r\sin{\theta}f^{\phi}\, ,
\end{equation}
where $f^\mu$ and $p^\mu$ were transformed into the local reference frame defined by the~orthonormal tetrade $e^{\textrm{(t)}}_\mu = (-1,0,0,0)$, $e^{\textrm{(r)}}_\mu = (0,1,0,0)$, $e^{(\theta)}_\mu = -r\,(0,0,1,0)$, $e^{(\phi)}_\mu = r\sin{\theta}\,(0,0,0,1)$. Combining (\ref{trafo1})--(\ref{trafo4}) with (\ref{KS}) gives $\kappa_1$ and $\kappa_2$ for~each geodesic at spatial infinity
\begin{equation}\label{k1}
    \kappa_1 = \beta f'^{(\theta)} + a\sin{\theta}f'^{(\phi)}\, ,
\end{equation}
\begin{equation}\label{k2}
    \kappa_2 = -a\sin{\theta}f'^{(\theta)}+\beta f'^{(\phi)}\, .
\end{equation}
In the end we put (\ref{k1}) equal to (\ref{k10}) and (\ref{k2}) equal to (\ref{k20}), which brings the expression for the change of the polarisation angle $\chi_{\rm p}$ at the local reference frame of the observer
\begin{equation}
    \tan{\chi_{\rm p}} = -\frac{f'^{(\phi)}}{f'^{(\theta)}} = a\frac{\beta - h\sin{\theta}}{a^2\sin{\theta}+h\beta} \, .
\end{equation}
The dependence of $\chi_{\rm p}(h)$ on height is shown in Figure \ref{chip} for maximally rotating black hole and three observer inclinations.

~\

Now we will solve for the change of polarisation angle between the lamp and~the~disc. In the equatorial plane, (\ref{KS}) yields
\begin{equation}\label{kd1}
    \kappa_1 = \frac{r^2+a^2}{r}p_r f_t + \frac{r^2+a^2}{r} f_r + \frac{ap_r}{r}f_\phi \, ,
\end{equation}
\begin{equation}\label{kd2}
    \kappa_2 = \frac{ap_\theta}{r}f_t + \frac{a}{r}f_\theta + \frac{p_\theta}{r}f_\phi \, .
\end{equation}
A notable simplification of the computations comes from expressing $f_t$, $f_\theta$, $f_\phi$ as functions of $\kappa_1$, $\kappa_2$, $p^\theta$, $p^r$ and the metric coefficients, using the conditions \mbox{$p_\mu p^\mu = 0$,} $f^\mu p_\mu = 0$ and $f_\mu f^\mu = 1$. Instead of performing transformation in~the~form of~(\ref{reduction}), we will subtract the $r$-component of $f_\mu$ in the equatorial plane
\begin{equation}
    f'_\mu = f_\mu - \frac{f_r}{p_r}p_\mu \, ,
\end{equation}
because this also largely simplifies expressions. Since the polarisation vector is orthogonal to the photon momentum and definition of the polarisation angle is in~the~plane orthogonal to the photon momentum, any transformation \mbox{$f'_\mu = f_\mu + A p_\mu$} does not change its value (the tetrade vectors $e_{\rm (y)}^\mu$ and $e_{\rm (z)}^\mu$, defined below, into which we project the polarisation vector $f'^\mu$ in the polarisation angle definition at the disc, are perpendicular to the photon momentum). Hence, we obtain the~reduced components of $f'_\mu$ in the form of
\begin{equation}\label{dtrafo1}
    f'_t = -\frac{\Delta p^\theta}{ar(p^r)^2}\kappa_2 + \left[\frac{(r^2+a^2)}{a}\frac{\Delta (p^\theta)^2}{(p^r)^2} + g_{t\phi}\right]\frac{f_\phi}{g_{\phi \phi}}\, ,
\end{equation}
\begin{equation}\label{dtrafo2}
    f'_{r} = 0\, ,
\end{equation}
\begin{equation}\label{dtrafo3}
    f'_{\theta} = \frac{g_{\phi \phi}}{ar(p^r)^2}\kappa_2 - \frac{(r^2+a^2)p^\theta}{a(p^r)^2}f_\phi \, ,
\end{equation}
\begin{equation}\label{dtrafo4}
    f'_{\phi} = f_\phi = \frac{arp^r \kappa_1 + r(r^2+a^2)p^\theta \kappa_2}{q^2+a^2} \, ,
\end{equation}
where $g_{t\phi} = -2a/r$, $g_{\phi \phi} = [(r^2+a^2)^2-\Delta a^2]/r^2$ are taken from the~Kerr metric in the equatorial plane. The components of the 4-velocity are \mbox{$U^t=(r^2+a\sqrt{r})/r/\sqrt{r^2-3r+2a\sqrt{r}}$,} $U^r=U^\theta=0$ and $U^\phi=\Omega_{\rm K} U^t$ with relativistic Keplerian rotational velocity $\Omega_{\rm K}=(r^{3/2}+a)^{-1}$ \citep{Novikov1973}. We construct an orthonormal tetrade in the local comoving frame with the disc, in order to define the polarisation angle $\chi_\textrm{d}$. We choose the time-like tetrade vector to be given by the 4-velocity of the orbiting material, the first spatial tetrade vector to be directed in the direction of momentum of the incoming photon, the second spatial tetrade vector to be defined by the projection of the local disc normal into the plane orthogonal to the photon momentum, and~the~last spatial tetrade vector is constructed as the vector product of the others, so that the full orthonormal tetrade is formed. Thus, we set
\begin{equation}\label{de1}
    e_{\rm (t)}^\mu = U^\mu \, ,
\end{equation}
\begin{equation}
    e_{\rm (x)}^\mu = \frac{p^\mu+p^\nu U_\nu U^\mu}{-p_\sigma U^\sigma} \, ,
\end{equation}
\begin{equation}\label{de3}
    e_{\rm (y)}^\mu = \frac{n^\mu - \frac{n_\nu p^\nu}{-p_\sigma U^\sigma}e_{\rm (x)}^\mu}{\sqrt{1-\left(\frac{n_\rho p^\rho}{p_\iota U^\iota}\right)^2}} \, ,
\end{equation}
\begin{equation}\label{de4}
    e_{\rm (z)}^\mu = \frac{\varepsilon^{\alpha \beta \gamma \mu} U_\alpha p_\beta n_\gamma}{-p_\sigma U^\sigma\sqrt{1-\left(\frac{n_\rho p^\rho}{p_\iota U^\iota}\right)^2}} \, ,
\end{equation}
where $\varepsilon^{\alpha \beta \gamma \mu} = -\delta^{\alpha \beta \gamma \mu}/\sqrt{-\hat{g}}$ is the permutation tensor defined by the permutation symbol, $\delta^{\alpha \beta \gamma \mu}$, and the metric tensor determinant, $\sqrt{-\hat{g}} = r^2$, \mbox{and~$n^\mu = (0,0,-1/r,0)$} is the~local disc's normal. Using the relations $f^\mu p_\mu = 0$, $U_\theta = 0$, $p_\phi = 0$, (\ref{dtrafo2}), (\ref{de1})--(\ref{de4}) we obtain the expression for the change of~polarisation angle $\chi_\textrm{d}$ at the local reference frame of~the~observer comoving with the~disc
\begin{equation}
\label{eq:angle}
    \tan{\chi_\textrm{d}} = \frac{e_{\rm (z)}^\mu f'_\mu}{e_{\rm (y)}^\mu f'_\mu} = \frac{U_t p_r f'_\phi + U_r f'_\phi - U_\phi p_r f'_t}{p_\nu U^\nu f'_\theta-q U^\mu f'_\mu}\, ,
\end{equation}
where $U^\mu f'_\mu = U^t f'_t + U^\phi f'_\phi$, 
$p_\nu U^\nu = -U^t$, 
$f'_t$, $f'_\theta$ and $f'_\phi$ follow from (\ref{dtrafo1}), (\ref{dtrafo3}) and (\ref{dtrafo4}), respectively, where the constants $\kappa_1$ and $\kappa_2$ have to be taken from (\ref{k10}) and (\ref{k20}), respectively. The dependence of $\chi_{\rm d}(r)$ is shown in Figure \ref{chid} for maximally rotating black hole and several lamp-post heights.

\chapter{Axially symmetric reflection models for {\tt XSPEC}}\label{torus_model}

Both variants of the {\tt xsstokes} C routine that are presented in Section \ref{simple_models} operate on the same basis. We produce the Stokes parameters $I$, $Q$ and $U$ in particular reflection geometry for three independent primary polarisation states with unpolarized and $100\%$ horizontally ($\Psi_0 = \frac{\pi}{2}$) and $100\%$ diagonally ($\Psi_0 = \frac{\pi}{4}$) polarized irradiation, in order to use equation (\ref{interpolate_Stokes}). These tables were first produced in~the~American Standard Code for Information Interchange (ASCII) text file format and then converted to FITS tables in the OGIP format, using {\tt XSPEC Table Model Generator}\footnote{\,Available at \url{https://github.com/mbursa/xspec-table-models} on the day of submission, including documentation.}. This simple conversion script in Python 3 was developed by Michal Bursa, but also verified with a similar routine independently developed in~the~scope of this dissertation. We stress the usefulness of~the~{\tt XSPEC Table Model Generator} for obtaining any tabular models for {\tt XSPEC} version \textit{12.13.0} (and onwards) suitable for polarisation. A reduction of the code to a pure spectral FITS file generation (necessary for previous {\tt XSPEC} versions) is straight forward within. The {\tt xsstokes} code can be adapted to any precomputed non-relativistic static model of axially symmetric reflector, where the formula (\ref{interpolate_Stokes}) would apply and~where the three tabular dependencies could be converted to the right format using the {\tt XSPEC Table Model Generator}.

~\

In Section \ref{simple_models}, we already explained how we obtained the tabular dependencies for {\tt xsstokes\_disc}. For {\tt xsstokes\_torus} the precomputations were only a bit more demanding, requiring development of a Python 3 routine, called {\tt torus\_integrator}\footnote{\,Available at \url{https://github.com/jpodgorny/torus_integrator} on the day of submission, including documentation. The work is still in progress on the day of submission.}. We provide here the details of the numerical implementation not only for the sake of reproducibility of the results presented in Section \ref{simple_torus_reflection}, but because we also believe that the routine may be useful for other purposes. The mechanisms of integration of precomputed spectropolarimetric reflection tables over the two-dimensional surface are wavelength independent and~can be easily adapted to elliptical tori, for example. Different local reflection tables for different irradiation can be used, aiming at higher accuracy for~the~description of AGN tori, or reflecting media of completely different astrophysical origin. 

~\

Inside the {\tt torus\_integrator} routine, the user can define an isotropically emitting power-law point source located at the center of the coordinates (see Figure \ref{xsstokes_mo}) with $1.2 \leq \Gamma \leq 3.0$ and a state of incident polarisation, given by the~primary polarisation degree $p_0$ and primary polarisation angle $\Psi_0$. These parameters should preserve the values given by the local reflection tables, as the interpolation is not performed for these parameters for efficiency reasons. Then the~user selects any torus half-opening angle $1\degr < H' < 90\degr$ and any observer's inclination $0\degr < \theta < 90\degr$, both measured from the pole. In the current setup, the~results are independent of the distance of the torus from the center $r_\textrm{in}^\textrm{torus}$ (see below). The user also selects integration resolution in the coordinates $u \in (0, 2\pi]$ and~$v \in [\pi - H', \pi]$, where the grid is linearly divided in~$N_\textrm{v}$ and $2N_\textrm{u}$ bins in~the~corresponding ranges (the number of bins in $u$ covers only the $[\frac{\pi}{2},\frac{3\pi}{2}]$ interval, because the other half is symmetrically added). The energy binning of~the~output is preserved from the local reflection tables and the output is stored in ASCII file format with lower and upper energy bin edges and the corresponding renormalized Stokes parameters $I$, $Q$ and $U$. If a list of model parameter values is given, multiple files will be stored in the output directory under names of all parameters and their values.

~\

We utilize standard relations between the cartesian coordinate system \mbox{\{$x$, $y$, $z$\}} given by the~base vectors $\Vec{e}_\textrm{x} = (1,0,0)$, $\Vec{e}_\textrm{y} = (0,1,0)$, \mbox{$\Vec{e}_\textrm{z} = (0,0,1)$} and the torus surface coordinates \{$u$, $v$\} given by the base vectors \mbox{$\frac{\partial}{\partial u} = (-\sin{u}, \cos{u}, 0)$,} $\frac{\partial}{\partial v} = (-\sin{v}\cos{u}, -\sin{v}\sin{u}, \cos{v})$ (see Figure \ref{xsstokes_mo}):
\begin{equation}\label{coordtrafo}
	\begin{aligned}
		x &= (R'+r'\cos{v})\cos{u} \textrm{ ,}\\
		y &= (R'+r'\cos{v})\sin{u} \textrm{ ,}\\
            z &= r'\sin{v} \textrm{ ,}
	\end{aligned}
\end{equation}
where $R' = r' + r_\textrm{in}^\textrm{torus}$ is the distance between (0,0,0) and the center of the meridional section of the torus, and
\begin{equation}
    r' = r_\textrm{in}^\textrm{torus}\frac{\cos{H'}}{1-\cos{H'}}
\end{equation}
is the radius of the toroidal section circles in the meridional plane, following equation (\ref{opening_angle}). The surface is given by another well-known implicit formula
\begin{equation}
\Phi = (x^2 + y^2 + z^2 + R'^2 - r'^2)^2 - 4R'^2(x^2+y^2) = 0 \textrm{ .}
\end{equation}

\begin{figure}[!htb]
	\includegraphics[trim={2.1cm 1.1cm 1.1cm 1.1cm},clip,width=0.55\columnwidth]{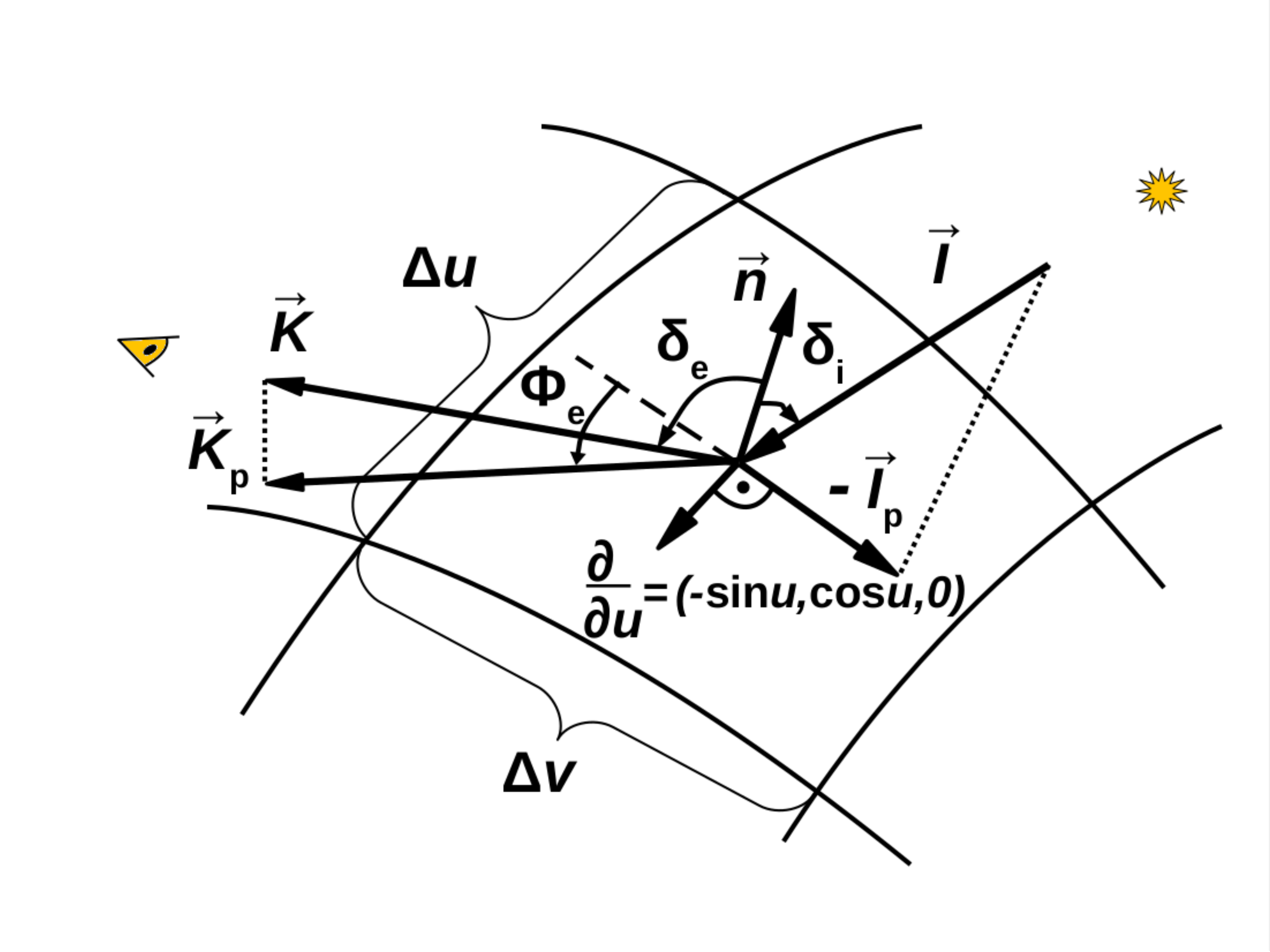}
        \includegraphics[width=0.43\columnwidth]{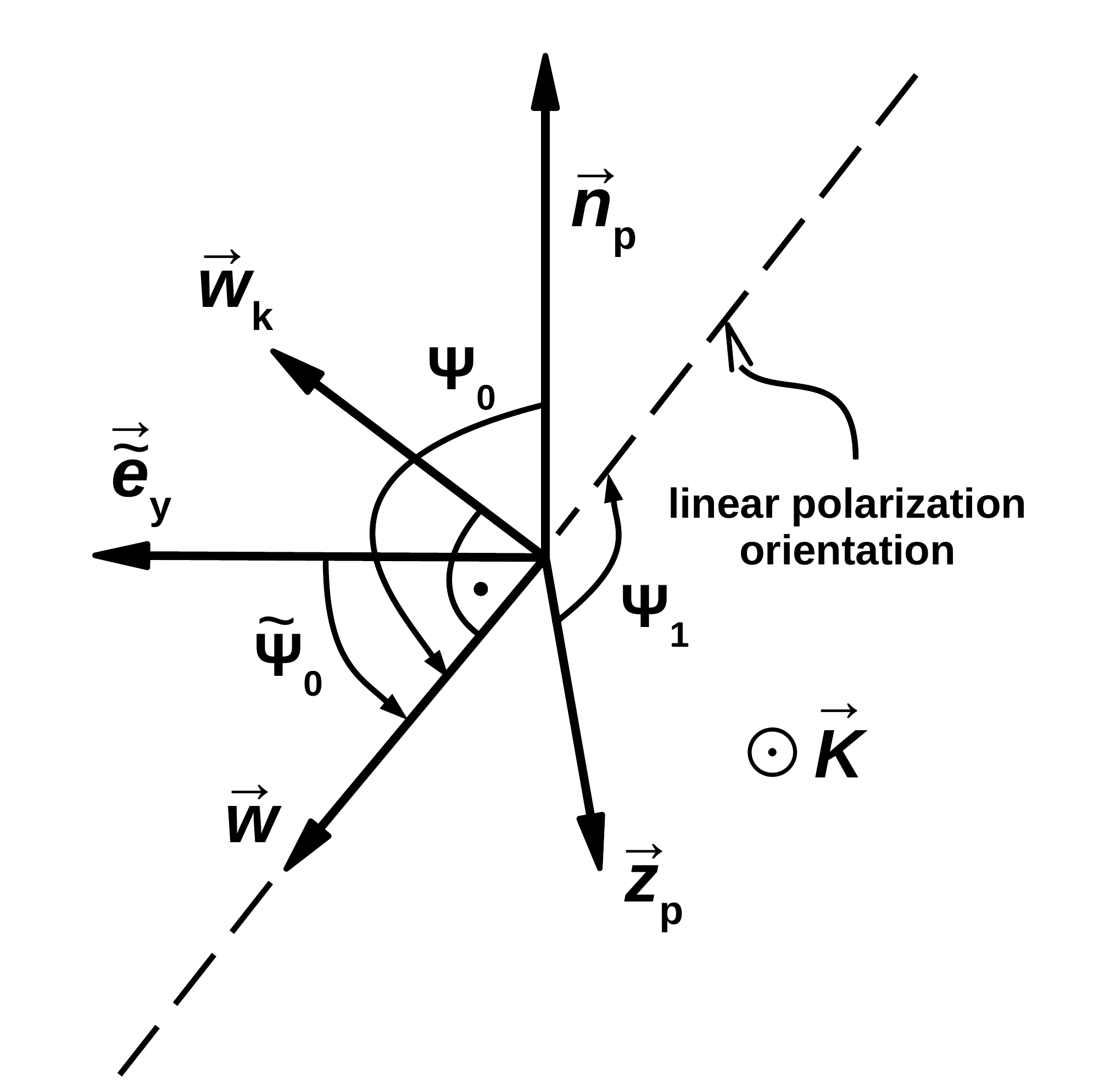}
	\caption{{\footnotesize Left: the definitions of angles and vectors in the locally reflecting plane, tangent to the torus surface. Right: the definitions of angles and vectors in the polarisation plane, perprendicular to the photon momentum direction. These are needed for the calculations of~rotation of the polarisation angle by changing frame of reference. Image adapted from \cite{Podgorny2023e}.}}
	\label{local_angles}
\end{figure}

The local reflection tables from \cite{Podgorny2021} are linearly interpolated at each at each point in the \{$u$, $v$\} grid in $\mu_\textrm{i} = \cos{\delta_\textrm{i}}$, $\mu_\textrm{e} = \cos{\delta_\textrm{e}}$, and $\Phi_\textrm{e}$, which are schematically shown in Figure \ref{local_angles} (left). We first precompute all three angles at each grid point in the local frame, given by the local surface normal $\Vec{n} = \frac{\nabla \Phi}{\lvert \nabla \Phi \rvert}$, the incident vector $\Vec{I} = (x, y, z)$, the emission vector $\Vec{K} = (0, \sin \theta, \cos \theta)$, and~their respective projections $\Vec{I}_\textrm{p}$ and $\Vec{K}_\textrm{p}$ to the local tangent plane through scalar products.

The code secures that the grid points with $\delta_\textrm{i} \notin [0, \frac{\pi}{2}]$ and $\delta_\textrm{e} \notin [0, \frac{\pi}{2}]$ are not accounted for. We limit $\pi - H' \leq v \leq \pi$, i.e. we do not account for those regions that lie below the equatorial plane and for any $v$ smaller than the shadow boundary constrained by $H'$. The region of reflection has to be also directly visible by the observer, which is for each $u$ and $\theta$ given by the boundary curve of the zero-transparency condition $0 = \Vec{n} \cdot \Vec{K}$. Therefore, after a few simple steps we arrive at a prescription for this curve
\begin{equation}
    v_\textrm{limit}(u,\theta) = -\arctan ( 2 \sin u \tan \theta) + \pi \textrm{ .} 
\end{equation}
Next, we have to account for self-obscuration of the reflected rays by the opposite side of the torus, closer to the observer. This is the scenario for highly inclined observers. To select such region, we define a grazing vector $\Vec{G}$ that begins \mbox{at~$\Vec{X}_1=$ ($x_1$, $y_1$, $z_1$)} at~the~opposite half-space from the observer given by $\pi \leq u \leq 2\pi$ and some $v_\textrm{self-obs}$ per each $u$. The grazing vector $\Vec{G}$ points in~the~$\Vec{K}$ direction towards the observer and ends at a point $\Vec{X}_2=$ ($x_2$, $y_2$, $z_2$) tangential to the torus surface, i.e. defined by $v_\textrm{limit}$ for unknown $0 \leq u_\textrm{t} \leq \pi$. Thus, parametrically for $\Vec{X}_1 + t\Vec{K} = \Vec{X}_2$ we write
\begin{equation}\label{cond1}
	\begin{aligned}
		(R' + r'\cos v_\textrm{self-obs})\cos u + tK_1 &= (R' + r'\cos v_\textrm{limit})\cos u_\textrm{t} \textrm{ ,}\\
		(R' + r'\cos v_\textrm{self-obs})\sin u + tK_2  &=  (R' + r'\cos v_\textrm{limit})\sin u_\textrm{t} \textrm{ ,}\\
            r'\sin v_\textrm{self-obs} + tK_3 &=  r'\sin v_\textrm{limit} \textrm{ .}
	\end{aligned}
\end{equation}
Then, the limiting curve $v_\textrm{self-obs}(u, \theta)$ is found as a solution \{$v_\textrm{self-obs}, t, u_\textrm{t}$\} of a set of equations
\begin{equation}\label{cond2}
	\begin{aligned}
		0 &= (R' + r'\cos v_\textrm{self-obs})\cos u - (R' + r'\cos v_\textrm{limit}(u_\textrm{t},\theta) )\cos u_\textrm{t} \textrm{ ,}\\
		0 &= (R' + r'\cos v_\textrm{self-obs})\sin u + t\sin \theta - (R' + r'\cos v_\textrm{limit}(u_\textrm{t},\theta) )\sin u_\textrm{t} \textrm{ ,}\\
            0 &= r'\sin v_\textrm{self-obs} + t\cos \theta - r'\sin v_\textrm{limit}(u_\textrm{t},\theta) \textrm{ .}
	\end{aligned}
\end{equation}

~\

Each local scattering surface in the user-defined grid is given by boundaries $v_1(u,v)$, $v_2(u,v)$, $u_1(u,v)$, $u_2(u,v)$. Or the local surface is set to zero, if~the~central bin point \{$u$, $v$\} does not fall within the restricting boundary conditions above, which is valid for sufficiently high resolution. We then approximate the~contributing local scattering surfaces by a tangent rectangle with area
\begin{equation}
\begin{split}
    A_\textrm{u,v} & = \int_{u_1}^{u_2} \int_{v_1}^{v_2} r'(R'+r'\cos v) dv du \\
    & \approx r'(R'+r'\cos v)(v_2-v_1)(u_2-u_1) \textrm{ .}
\end{split}
\end{equation}
At each \{$u,v$\} we perform the tri-linear interpolation in the local reflection angles $\mu_\textrm{i}$, $\mu_\textrm{e}$, and $\Phi_\textrm{e}$ of the tabulated Stokes parameters $I$, $Q$ and $U$, commonly denoted as $S(\mu_\textrm{i},\mu_\textrm{e},\Phi_\textrm{e}; E, \Gamma, p_0, \Psi_0)$. The interpolated result, which we denote as $\Bar{S}_\textrm{u,v}(\mu_\textrm{i}(u,v),\mu_\textrm{e}(u,v),\Phi_\textrm{e}(u,v); E, \Gamma, p_0, \Psi_0)$, is further transformed.

The polarisation vector of the interpolated reflected rays needs to be rotated from the local frame to conform to the photon momentum direction in~the~global system. In global coordinates, the Stokes parameters are defined with respect to~the~system axis of symmetry. The incident electric vector orientation does not have to be rotated to conform to the polarisation definition in~the~local frame due~to~the axial symmetry of the reflecting surface and due to~the~placement of~the~point-source in the central point. The polarisation fraction after the rotation $p_1$ is equal to the $p_0$ in the local reflection frame. Similarly for~the~intensity, $I_1 = I_0$. Hence, for~the~remaining Stokes parameters we have $Q_1 = \sqrt{Q_0^2 + U_0^2}\cos{(2\Psi_1)}$ and $U_1 = \sqrt{Q_0^2 + U_0^2}\sin{(2\Psi_1)}$. In order to obtain the $\Psi_1(\Psi_0)$ function, we will go back to the adopted definitions (\ref{psi_def}), which are illustrated in Figure \ref{local_angles} (right) in the polarisation plane. We assume $\Psi_0$ with respect to the projected normal $\Vec{n}_\textrm{p} = \Vec{n} - (\Vec{n} \cdot \Vec{K})\Vec{K}$ and $\Psi_1$ with respect to~the~projected global axis of symmetry $\Vec{z}_\textrm{p} = \Vec{e}_\textrm{z} - (\Vec{e}_\textrm{z} \cdot \Vec{K})\Vec{K}$. Then $\Vec{\Tilde{e}}_\textrm{y} = \Vec{K} \times \frac{\Vec{n}_\textrm{p}}{\lvert \Vec{n}_\textrm{p} \rvert}$, $\Vec{\Tilde{e}}_\textrm{x} = \frac{\Vec{n}_\textrm{p}}{\lvert \Vec{n}_\textrm{p} \rvert}$ and $\Vec{\Tilde{e}}_\textrm{z} = \Vec{K}$ are the base vectors in the polarisation plane and the polarisation vector is locally $\Vec{w}_\textrm{loc} = (\cos{\Psi_0}, \sin{\Psi_0}, 0)$. The conditions $\cos{\Psi_0} = \Vec{\Tilde{e}}_\textrm{x} \cdot \Vec{w}$, $\cos{\Tilde{\Psi}_0} = \sin{\Psi_0} = \Vec{\Tilde{e}}_\textrm{y} \cdot \Vec{w}$ and $\Vec{w} \cdot \Vec{K} = 0$ provide
\begin{gather}
 \Vec{w}_\textrm{loc}
 = 
  \begin{pmatrix}
   \Vec{\Tilde{e}}_\textrm{x}, &
   \Vec{\Tilde{e}}_\textrm{y}, & \Vec{\Tilde{e}}_\textrm{z}
   \end{pmatrix} \cdot \Vec{w} = B \cdot \Vec{w} \textrm{ .}
\end{gather}
Therefore, we solve for this matrix equation to get $\Vec{w} = B^{-1} \cdot \Vec{w}_\textrm{loc}$, in order to~obtain $\Vec{w}_\textrm{k} = - \Vec{K} \times \Vec{w}$. In the end,
\begin{equation}
    \Psi_1 = 
	\begin{cases}
		\arccos{ \left( \dfrac{\Vec{z}_\textrm{p} \cdot \Vec{w}}{\lvert \Vec{z}_\textrm{p} \rvert \lvert \Vec{w} \rvert} \right) } \ , & \textrm{if} \ \arccos{\left( \dfrac{\Vec{z}_\textrm{p} \cdot \Vec{w}_\textrm{k}}{\lvert \Vec{z}_\textrm{p} \rvert \lvert \Vec{w}_\textrm{k} \rvert} \right)} \leq \dfrac{\pi}{2} \textrm{ ,}\\
		-\arccos{ \left( \dfrac{\Vec{z}_\textrm{p} \cdot \Vec{w}}{\lvert \Vec{z}_\textrm{p} \rvert \lvert \Vec{w} \rvert} \right) }  \ , & \textrm{if} \ \arccos{\left( \dfrac{\Vec{z}_\textrm{p} \cdot \Vec{w}_\textrm{k}}{\lvert \Vec{z}_\textrm{p} \rvert \lvert \Vec{w}_\textrm{k} \rvert} \right)} > \dfrac{\pi}{2} \ .
	\end{cases}
\end{equation}

~\

The code results are then produced by numerically integrating the rotated Stokes parameters from locally tangent planes projected to the line of sight across all \{$u,v$\} that fulfill the aforementioned shadow and self-obscuration conditions:
\begin{equation}
    S_\textrm{tot}(\theta, H', p_0, \Psi_0, \Gamma, E) = K_0 \sum_\textrm{u,v} A_\textrm{u,v} \mu_\textrm{e}(u,v) \mu_\textrm{i}(u,v) \frac{1}{\varrho(u,v)^{2}} \Bar{S}_\textrm{u,v} \textrm{ ,}
\end{equation}
with weighting according to the locally illuminating flux that is dependent on~distance from the center $\varrho(u,v)$ as $\sim \frac{1}{\varrho(u,v)^{2}}$. Because of the weighting factor $\varrho(u,v)$ that scales linearly with $r_\textrm{in}^\textrm{torus}$, the results are independent of $r_\textrm{in}^\textrm{torus}$. Because we take for all surface points the most neutral version of the local reflection tables, the ionization parameter is not dependent on the local flux received in our approximation. The output $S_\textrm{tot}$ is renormalized via $K_0$ constant for storage convenience before being saved in a user-defined directory in the ASCII file format. Given the axial symmetry of our toroidal setup (the incident polarisation angle does not change with respect to the local reflection frame), neglect of the relativistic effects, and the linearity of Stokes parameters, we precompute only three tables in $S_\textrm{tot}$ for three independent incident states of polarisation. Such $S_\textrm{tot}$ tables converted to FITS format are then ready to be used inside {\tt xsstokes\_torus}, following the~interpolation relation (\ref{interpolate_Stokes}).

\begin{landscape}
\end{landscape}

\end{sloppypar}

\includepdf{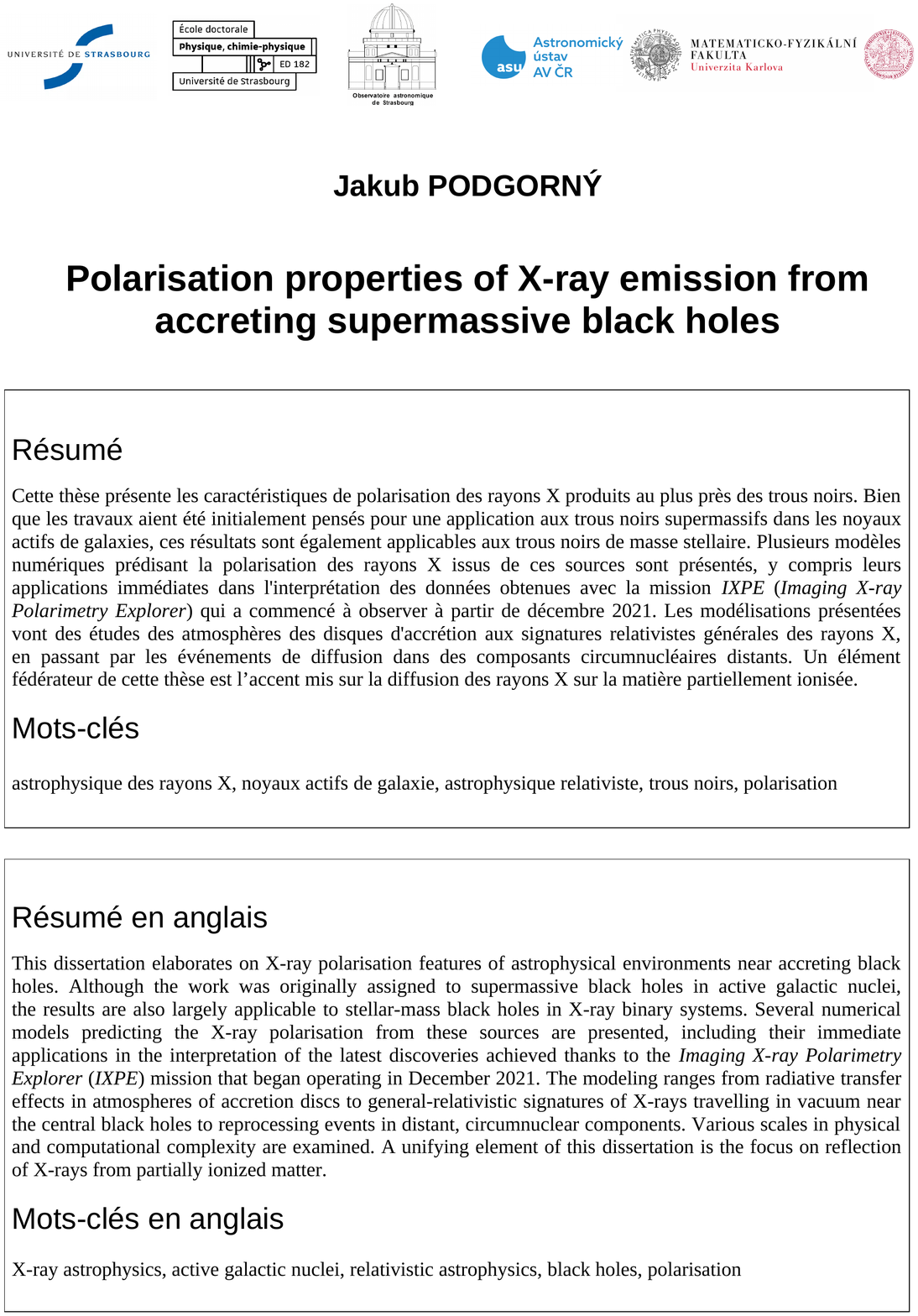}

\openright
\end{document}